\documentclass[a4paper,fleqn,usenatbib,useAMS]{mnras}

\usepackage{graphicx}	
\usepackage{amsmath}	
\usepackage{amssymb}	
\usepackage{bm}		
\usepackage{pdflscape}	
\usepackage{natbib}
\usepackage{times}
\usepackage{calc}
\usepackage{rotating}
\usepackage{lscape}
\usepackage{hyperref}
\bibpunct{(}{)}{;}{a}{}{,} 
\usepackage{longtable}
\usepackage{enumitem} 
\usepackage{hyperref}
\usepackage{pifont}
\usepackage{tikz}

\usepackage{tikz,xcolor,hyperref}

\definecolor{lime}{HTML}{A6CE39}
\DeclareRobustCommand{\orcidicon}{
	\begin{tikzpicture}
	\draw[lime, fill=lime] (0,0) 
	circle [radius=0.14] 
	node[white] {{\fontfamily{qag}\selectfont \tiny ID}};
	\draw[white, fill=white] (-0.0625,0.095) 
	circle [radius=0.007];
	\end{tikzpicture}
	\hspace{-2mm}
}

\foreach \x in {A, ..., Z}{\expandafter\xdef\csname orcid\x\endcsname{\noexpand\href{https://orcid.org/\csname orcidauthor\x\endcsname}
			{\noexpand\orcidicon}}
}

\def\oversim#1#2{\lower0.5pt\vbox{\baselineskip0pt \lineskip-0.5pt
     \ialign{$\mathsurround0pt #1\hfil##\hfil$\crcr#2\crcr\sim\crcr}}}

\usepackage[T1]{fontenc}
\usepackage{ae,aecompl}

\usepackage{mathptmx}

\def\mnras {{MNRAS}}
\def\aas  {{A\&A}}

\title[IPHAS Imaging catalogue]{First deep images catalogue of extended IPHAS PNe}

\author[L.\,Sabin et al.]{L. Sabin\thanks{E-mail:lsabin@astro.unam.mx (LS)}$^{1\orcidA{}}$, 
M.A. Guerrero$^{2\orcidB{}}$, 
G. Ramos-Larios$^{3\orcidC{}}$, 
P. Boumis$^{4\orcidD{}}$, 
A. A. Zijlstra$^{5,6\orcidE{}}$,  
D. N. F. Awang Iskandar$^{7\orcidF{}}$, \newauthor M. J. Barlow$^{8\orcidG{}}$, J. A. Toal\'{a}$^{9\orcidH{}}$, Q. A. Parker$^{10\orcidI{}}$,  R. M. L. Corradi$^{11,12\orcidJ{}}$ and R. A. H. Morris$^{13}$ \\
$^1$Instituto de Astronom\'{i}a,  Universidad Nacional Aut\'onoma de M\'exico, Apdo. Postal 106, 22800 Ensenada, B.C., Mexico\\ 
$^2$Instituto de Astrof\'{i}sica de Andaluc\'{i}a (IAA-CSIC), Glorieta de la Astronom\'{i}a S/N, 18008 Granada, Spain\\
$^3$Instituto de Astronom\'{i}a y Meteorolog\'{i}a, CUCEI,  Universidad de Guadalajara, Av. Vallarta 2602, Arcos Vallarta, 44130 Guadalajara, Mexico\\
$^4$Institute for Astronomy, Astrophysics, Space Applications and Remote Sensing, National Observatory of Athens, 15236 Penteli, Athens, Greece \\
$^5$Jodrell Bank Centre for Astrophysics, The University of Manchester, Oxford Road, Manchester M13 9PL, UK\\
$^6$Laboratory for Space Research, University of Hong Kong, Pokfulam Road, Hong Kong\\
$^7$Faculty of Computer Science and Information Technology, Universiti Malaysia Sarawak, Sarawak 94300, Malaysia\\
$^8$Dept. of Physics \& Astronomy, University College London, Gower St, London WC1E 6BT,UK \\
$^9$Instituto de Radioastronom\'{i}a y Astrof\'{i}sica, UNAM, Campus Morelia, C.P. 58089, Morelia, Mexico \\
$^{10}$Department of Physics, The University of Hong Kong, Chong Yuet Ming Physics Building Pokfulam Road, Hong Kong \\
$^{11}$GRANTECAN, Cuesta de San Jos\'e s/n, E-38712, Bre\~na Baja, La Palma, Spain \\
$^{12}$Instituto de Astrof\'\i sica de Canarias, E-38200 La Laguna, Tenerife, Spain \\
$^{13}$School of Physics, Bristol University, Tyndall Avenue, Bristol, BS8 1TL, UK 
}

\pubyear{2021}

\begin{document}
\label{firstpage}
\pagerange{\pageref{firstpage}--\pageref{lastpage}}
\maketitle

\begin{abstract}
We present the first instalment of a deep imaging catalogue containing 58 True, Likely and Possible extended PNe detected with the Isaac Newton Telescope Photometric H$\alpha$ Survey (IPHAS). 
The three narrow-band filters in the emission lines of H$\alpha$, [N~{\sc ii}] $\lambda$6584 \AA, and [O~{\sc iii}] $\lambda$5007 \AA\ used for this purpose allowed us to improve our description of the morphology and dimensions of the nebulae. 
In some cases even the nature of the source has been reassessed. 
We were then able to unveil new macro- and micro-structures, which will without a doubt contribute to a more accurate analysis of these PNe. 
It has been also possible to perform a primary classification of the targets based on their ionization level. A Deep Learning classification tool has also been tested.
We expect that all the PNe from the IPHAS catalogue of new extended planetary nebulae will ultimately be part of this deep H$\alpha$, [N~{\sc ii}] and [O~{\sc iii}] imaging catalogue.

\end{abstract}





\section{Introduction}

\begin{table*}
\footnotesize\addtolength{\tabcolsep}{-3.5pt}
\caption{\label{tab:setup} Observational setup}
\begin{center}
\begin{tabular}{l|l|l|l|l|c|c|l|l|l|r|}
\hline
\multicolumn{1}{c}{Setup} & 
\multicolumn{1}{c}{Date} & 
\multicolumn{1}{c}{Telescope} & 
\multicolumn{1}{c}{Instrument} & 
\multicolumn{1}{c}{CCD} & 
\multicolumn{1}{c}{Plate scale} & 
\multicolumn{1}{c}{FoV} & 
\multicolumn{1}{c}{Filter 1} & 
\multicolumn{1}{c}{Filter 2} & 
\multicolumn{1}{c}{Filter 3} & 
\multicolumn{1}{c}{Exposure Time} \\
\multicolumn{5}{c}{} & 
\multicolumn{1}{c}{(arcsec)} & 
\multicolumn{1}{c}{(arcmin)} & 
\multicolumn{3}{c}{} & 
\multicolumn{1}{c}{(s)} \\
\hline
N1  & 2007-09 & NOT & ALFOSC & E2V 42-40 2k$\times$2k     & 0.189 & 6.5 & [O~{\sc iii}] \#40 NOT & H$\alpha$ \#21 NOT & [N~{\sc ii}] \#78 NOT & 1$\times$1200~~~~~ \\
N2  & 2015-07 & NOT & ALFOSC & E2V 42-40 2k$\times$2k     & 0.189 & 6.5 & [O~{\sc iii}] \#40 NOT & H$\alpha$ \#21 NOT & [N~{\sc ii}] \#78 NOT & 3$\times$450~~~~~ \\
N3  & 2015-09 & NOT & ALFOSC & E2V 42-40 2k$\times$2k     & 0.189 & 6.5 & [O~{\sc iii}] \#90 NOT & H$\alpha$ \#21 NOT & [N~{\sc ii}] \#78 NOT & 3$\times$600~~~~~ \\
N4  & 2016-05 & NOT & ALFOSC & E2V CCD231-42 2k$\times$2k & 0.211 & 7.1 & [O~{\sc iii}] \#90 NOT & H$\alpha$ \#21 NOT & [N~{\sc ii}] \#78 NOT & 3$\times$600~~~~~ \\
N4  & 2016-11 & NOT & ALFOSC & E2V CCD231-42 2k$\times$2k & 0.211 & 7.1 & [O~{\sc iii}] \#90 NOT & H$\alpha$ \#21 NOT & [N~{\sc ii}] \#78 NOT & 3$\times$600~~~~~ \\
N5  & 2018-06 & NOT & ALFOSC & E2V CCD231-42 2k$\times$2k & 0.211 & 7.1 & [O~{\sc iii}] \#90 NOT & H$\alpha$ OSN & [N~{\sc ii}] OSN & 3$\times$600~~~~~ \\
N5  & 2018-09 & NOT & ALFOSC & E2V CCD231-42 2k$\times$2k & 0.211 & 7.1 & [O~{\sc iii}] \#90 NOT & H$\alpha$ OSN & [N~{\sc ii}] OSN & 3$\times$600~~~~~ \\
N5  & 2020-01 & NOT & ALFOSC & E2V CCD231-42 2k$\times$2k & 0.211 & 7.1 & [O~{\sc iii}] \#90 NOT & H$\alpha$ OSN & [N~{\sc ii}] OSN & 3$\times$600~~~~~ \\
A1  & 2016-11 & AT & LN2CCD & E2V CCD42-40 2k$\times$2k & 0.16 & 5.5 & [O~{\sc iii}] AT & H$\alpha$ AT & [N~{\sc ii}] AT & 3$\times$800~~~~~ \\
A2  & 2016-08 & AT & LN2CCD & E2V CCD42-40 2k$\times$2k & 0.16 & 5.5 & [O~{\sc iii}] AT & H$\alpha$ AT & [N~{\sc ii}] AT & 1$\times$1200~~~~~ \\
A2  & 2016-09 & AT & LN2CCD & E2V CCD42-40 2k$\times$2k & 0.16 & 5.5 & [O~{\sc iii}] AT & H$\alpha$ AT & [N~{\sc ii}] AT & 1$\times$1200~~~~~ \\
A1  & 2017-08 & AT & LN2CCD & E2V CCD42-40 2k$\times$2k & 0.16 & 5.5 & [O~{\sc iii}] AT & H$\alpha$ AT & [N~{\sc ii}] AT & 3$\times$800~~~~~ \\
A1  & 2017-10 & AT & LN2CCD & E2V CCD42-40 2k$\times$2k & 0.16 & 5.5 & [O~{\sc iii}] AT & H$\alpha$ AT & [N~{\sc ii}] AT & 3$\times$800~~~~~ \\
A1  & 2018-06 & AT & LN2CCD & E2V CCD42-40 2k$\times$2k & 0.16 & 5.5 & [O~{\sc iii}] AT & H$\alpha$ AT & [N~{\sc ii}] AT & 3$\times$800~~~~~ \\
\hline
\hline
\end{tabular}
\end{center}
\end{table*}

The Isaac Newton Telescope Photometric H$\alpha$ Survey of the Northern Galactic Plane \citep[IPHAS;][]{Drew2005} has allowed the detection of a significant number of new planetary nebulae (PNe) candidates. More than 150 of those have already been confirmed in follow-up spectroscopic observations \citep{Sabin2014} and more are will be presented in upcoming papers (Ritter et al., in preparation).  

The observational strategy of IPHAS was optimized to guaranty the completion of the survey in a reasonable amount of time with the acquisition of images in the Sloan $i'$ and $r'$ filters and through a 95 \AA\ wide H$\alpha$ filter with fixed exposure times of 10~s, 30~s and 120~s, respectively\footnote{
We note that a database of images from the INT Galactic Plane Survey (IGAPS, \citealt{Monguio2020}) will soon be published (Greimel et al. submitted).}. 
The shallow IPHAS images thus provided only limited information on the morphologies of those new PNe.  
Furthermore, IPHAS does not provide information on the ionization structure because the used filter includes both the H$\alpha$ and [N~{\sc ii}] $\lambda$6583\AA\ emission lines. 
The PN morphology, however, depends strongly on the excitation of the ion emitting a particular line.  
In particular, the [O {\sc iii}] $\lambda$5007\AA\ emission line  preferentially reveals the morphology of high-excitation nebular regions, whereas the [N~{\sc ii}] $\lambda$6584 \AA\ emission line reveals low-excitation regions and features (see for example the catalogues by \citealt{Manchado1996} and \citealt{Stan2016}).

It is thus very likely that the faintest nebular structures and low-excitation features such as outflows/jets or knots can still elude us in IPHAS images.  
Not only the detailed morphology of the IPHAS PNe is compromised, but also their exact sizes might require a deeper and sharper imaging.  
This information can be expected to give us more insights into the formation process and more generally the characteristics of these PNe. 
Therefore we focus on the morphological and nebular excitation analysis in this new instalment on our study of IPHAS PNe presenting a multi-wavelength optical H$\alpha$, [N~{\sc ii}] and [O~{\sc iii}] imaging survey of 58 True, Likely and Possible extended planetary nebulae selected mainly from the IPHAS Catalogue of New Extended Planetary Nebulae \citep[][hereafter Paper I]{Sabin2014}, but also from the follow-up spectroscopy program (Ritter et al., in prep). The article is organized as follows. The observations are presented in section \S\ref{Obs}, the list of objects and images are described in section \S\ref{Res}, and our discussion in section \S\ref{Disc}.  The conclusions are presented in section \S\ref{Conc}.

\begin{table}
\footnotesize\addtolength{\tabcolsep}{-4pt}
\caption{\label{tab:filter} Properties of the narrow-band filters}
\begin{tabular}{l|l|l|r|c|}
\hline
\hline
Telescope & ~~~~Filter & Central Wavelength  &  FWHM    & Transmission \\
          &            & ~~~~~~~~~~~~~~(\AA) & (\AA)~~~ &     (\%)     \\
\hline
AT   & H$\alpha$ 6563 \AA     & ~~~~~~~~~~~~6567   & 17.0~~ & 69.6 \\
AT   & [O {\sc iii}] 5007 \AA & ~~~~~~~~~~~~5011   & 30.0~~ & 83.9 \\
AT   & [N {\sc ii}] 6583 \AA  & ~~~~~~~~~~~~6568   & 17.0~~ & 79.6 \\
OSN  & H01  H$\alpha$         & ~~~~~~~~~~~~6564.6 & 12.7~~ & 80.2 \\
OSN  & E16 [N {\sc ii}]       & ~~~~~~~~~~~~6583.3 & 12.6~~ & 79.2 \\
NOT  & H$\alpha$ \#21         & ~~~~~~~~~~~~6564   &  3.3~~ & 66.0 \\
NOT  & [O {\sc iii}] \#40     & ~~~~~~~~~~~~5010   &  4.3~~ & 56.0 \\
NOT  & [N {\sc ii}] \#78      & ~~~~~~~~~~~~6583   &  3.6~~ & 74.0 \\
\hline
\hline
\end{tabular}
\end{table}

\section{NOT and Helmos Observations}\label{Obs}

We have acquired narrow-band images in the [O~{\sc iii}] $\lambda$5007 \AA, H$\alpha$, and [N~{\sc ii}] $\lambda$6584 \AA\ emission lines using the ALhambra Faint Object Spectrograph and Camera (ALFOSC) installed on the 2.56-m Nordic Optical Telescope (NOT) at the Observatorio de El Roque de los Muchachos (ORM) in La Palma (Spain) and the LN2CCD imaging camera at the 2.3-m (f/8) Aristarchos telescope (AT) of Helmos Observatory in Peloponese (Greece).
The instrumental setup has been changing between the different observational runs and we summarize the information regarding the detectors, plate scale, field-of-view (FoV), filters and number of frames and integration time of individual exposures in Table~\ref{tab:setup} and Table~\ref{tab:filter}.
A code is assigned to each instrumental setup in the first column of Table~\ref{tab:setup}.

All the data were bias-subtracted and then flat-fielded using twilight flats employing standard {\sc {\sc iraf}} \citep{Tody1993} routines\footnote{{\sc iraf}, the Image Reduction and Analysis Facility, is distributed by the National Optical Astronomy Observatory, which is operated by the Association of Universities for Research in Astronomy (AURA) under cooperative agreement with the National Science Foundation.}.

\section{Results}\label{Res}

The complete list of objects studied here is presented in Table~\ref{tab:catalogue} and their individual H$\alpha$, [N~{\sc ii}] and [O~{\sc iii}] narrow-band images in addition to colour-composite are presented in Figures~\ref{1.img}--\ref{12.img} of Appendix~A. For simplicity, each object listed in Table~\ref{tab:catalogue} is identified by its identification number as defined in Paper I (and other papers of this series), therefore adopting the Sab~\# denomination hereafter\footnote{But the last source in Table~\ref{tab:catalogue}, named J191104.8, which is not listed in Paper~I.}, as well as by the standard and official IAU PNG and IPHAS identifications.  
For each object we list next the code describing the instrumental setup used (as defined in the first column of Table~\ref{tab:setup}), the observing date and the average seeing during the observation. 
In the following section, we present the main characteristics and features that can be derived from the deep images.

It is important to emphasize that the main goal of the observing program presented here is the acquisition of morphological information on different emission lines for IPHAS PNe. 
Therefore the observing program did not include standard stars for flux calibration (and surface brightness measurement). 
A proper calibration of the images taking into account the characteristics of the different filters and guided by spectroscopic data for some of the sources in our sample will be presented in a forthcoming paper.

\section{Discussion}\label{Disc}

\begin{table*}
\footnotesize\addtolength{\tabcolsep}{-4pt}
\caption{\label{tab:catalogue} IPHAS sources observed with the Nordic Optical Telescope (NOT) and Aristarchos telescope (AT, Helmos Observatory). \textit{Sab} numbers, designation, coordinates and the observation date are listed. Notes: The asterisks "*" indicate a change in total size (from the originally reported dimensions) $\geq 10\arcsec$, $\dagger$ the full size is taken from \citealt{Mampaso2006}. Note that Sab\,15 and \,50 are now rejected as PNe. }
\begin{tabular}{r|l|l|c|c|c|r|c|}
\hline
\multicolumn{1}{c}{Sab~\#} & 
\multicolumn{1}{c}{IAU PNG} & 
\multicolumn{1}{c}{IPHAS ID:} & 
\multicolumn{1}{c}{Setup} & 
\multicolumn{1}{c}{Obs.\ date} & 
\multicolumn{1}{c}{Seeing} & 
\multicolumn{1}{c}{Major Diam.} & 
\multicolumn{1}{c}{Morphology} \\
 & 
\multicolumn{1}{c}{designation} & 
\multicolumn{1}{c}{IPHASX} & 
\multicolumn{1}{c}{}    & 
\multicolumn{1}{c}{yyyy-mm-dd} & 
\multicolumn{1}{c}{(arcsec)} & 
\multicolumn{1}{c}{(arcsec)} & 
\multicolumn{1}{c}{ Old$\rightarrow$New } \\ 

\hline
\hline
  2~~~~~ & G119.2$+$04.6	      & ~~J001333.8$+$671803     & N4 & 2016-11-28 &      0\farcs8      &  $^*$46~~~~~~ & B    $\rightarrow$ Bprs    \\ 
  4~~~~~ & G126.6$+$01.3          & ~~J012507.9$+$635652$^a$ & N3 & 2015-10-02 &      0\farcs7      & $\dagger$105~~~~~~ & Bamp $\rightarrow$ Bampr \\ 
  6~~~~~ & G129.6$+$03.4          & ~~J015624.9$+$652830     & A1 & 2016-11-05 &      1\farcs0      &     196~~~~~~ & Rar  $\rightarrow$ Rms     \\ 
  7~~~~~ & G132.8$+$02.0          & ~~J022045.0$+$631134     & N3 & 2015-10-02 &      0\farcs6      &	  33~~~~~~ & Eas  $\rightarrow$ Es      \\ 
 11~~~~~ & G144.1$-$00.5	      & ~~J033105.3$+$553851     & N4 & 2016-11-27 &      0\farcs7      &  $^*$36~~~~~~ & Rar  $\rightarrow$ Ras	  \\ 
 13~~~~~ & G150.0$-$00.3	      & ~~J040329.5$+$520825     & N5 & 2020-01-26 & 0\farcs8--1\farcs3 &  $^*$34~~~~~~ & Bp   $\rightarrow$ Bp      \\ 
 15~~~~~ & G159.4$+$02.0	      & ~~J045358.6$+$465842     & N5 & 2020-01-26 &      1\farcs0      & $^*$126~~~~~~ & Am   $\rightarrow$ I \\ 
 19~~~~~ & G183.0$+$00.0	      & ~~J055242.8$+$262116$^b$ & N5 & 2020-01-26 &      1\farcs0      &      16~~~~~~ & Rar  $\rightarrow$ Ram     \\ 
 21~~~~~ & G190.7$-$01.3	      & ~~J060412.2$+$190031     & N5 & 2020-01-26 &      1\farcs1      &	     72~~~~~~ & Ba   $\rightarrow$ Bp      \\ 
 22~~~~~ & G195.4$-$04.0	      & ~~J060416.2$+$133250     & N4 & 2016-11-28 &      1\farcs1      &  $^*$75~~~~~~ & Ba   $\rightarrow$ Br	  \\ 
 23~~~~~ & G204.3$-$01.6	      & ~~J062937.8$+$065220     & N4 & 2016-11-28 &      0\farcs8      &  31~~~~~~ & Ba   $\rightarrow$ Br      \\ 
 31~~~~~ & G038.4$+$03.6	      & ~~J184834.6$+$063302     & A2 & 2016-08-05 &      2\farcs0      &      25~~~~~~ & Ba   $\rightarrow$ A/Ba    \\ 
 36~~~~~ & G032.7$-$00.5	      & ~~J185312.9$-$002529     & A2 & 2016-08-05 &      1\farcs6      &	     26~~~~~~ & E    $\rightarrow$ E       \\ 
 39~~~~~ & G032.9$-$01.4	      & ~~J185640.0$-$003804     & A2 & 2016-08-06 &      1\farcs6      &	     21~~~~~~ & Bp   $\rightarrow$ Br      \\ 
 40~~~~~ & G043.3$+$03.5	      & ~~J185744.4$+$105053     & A2 & 2016-08-05 &      1\farcs9      &     110~~~~~~ & Eams $\rightarrow$ Eas     \\ 
 42~~~~~ & G040.6$+$01.5	      & ~~J185957.0$+$073544     & N5 & 2018-06-05 & 0\farcs8--1\farcs1 &	      5~~~~~~ & B    $\rightarrow$ B/S	  \\ 
 44~~~~~ & G044.4$+$03.1	      & ~~J190115.5$+$114150     & A2 & 2016-08-06 &      1\farcs5      &	     30~~~~~~ & Ear  $\rightarrow$ Ra      \\ 
 50~~~~~ & G038.7$-$00.5	      & ~~J190401.5$+$045433     & N2 & 2015-07-17 &      0\farcs6      &    >140~~~~~~ & Ims  $\rightarrow$ I	      \\ 
 53~~~~~ & G036.4$-$01.9	      & ~~J190438.6$+$021424     & N4 & 2016-05-02 &      0\farcs8      &  	 21~~~~~~ & Bmp  $\rightarrow$ Bmpw    \\ 
 54~~~~~ & G045.4$+$02.6	      & ~~J190447.9$+$121844     & N5 & 2018-09-08 &      0\farcs7      &	     12~~~~~~ & Er   $\rightarrow$ Br      \\ 
 55~~~~~ & G043.6$+$01.7	      & ~~J190454.0$+$101801     & A2 & 2016-08-06 &      1\farcs6      &	     20~~~~~~ & Ears $\rightarrow$ Ra      \\ 
 61~~~~~ & G038.9$-$01.3	      & ~~J190718.1$+$044056     & N4 & 2016-05-01 & 0\farcs9 &	  12~~~~~~ & I    $\rightarrow$ Bmpsw   \\ 
 63~~~~~ & G045.7$+$01.4	      & ~~J190954.7$+$120455     & N4 & 2016-05-01 & 1\farcs1 &	  18~~~~~~ & B    $\rightarrow$ Es      \\ 
 65~~~~~ & G044.9$+$00.8	      & ~~J191022.1$+$110538     & N5 & 2018-06-07 & 1\farcs0 &	  11~~~~~~ & Ear  $\rightarrow$ Ems     \\ 
 69~~~~~ & G047.6$+$01.0	      & ~~J191445.1$+$133219     & N4 & 2016-05-01 & 1\farcs5 &	  16~~~~~~ & Ears $\rightarrow$ Em      \\ 
 79~~~~~ & G051.7$+$01.3	      & ~~J192146.7$+$172055     & N2 & 2015-07-19 & 1\farcs5--1\farcs8 &	  34~~~~~~ & Ears $\rightarrow$ Ia      \\ 
 81~~~~~ & G049.2$+$00.0	      & ~~J192153.9$+$143056     & N5 & 2018-09-09 & 0\farcs9 & $^*$60~~~~~~ & Ear  $\rightarrow$ Br  \\ 
 86~~~~~ & G049.5$-$01.4	      & ~~J192751.3$+$140127     & N5 & 2018-09-10 & 0\farcs9 &      14~~~~~~ & Rars $\rightarrow$ Ia      \\ 
 88~~~~~ & G045.7$-$03.8	      & ~~J192847.2$+$093436     & N2 & 2015-07-18 & 0\farcs7 &	  72~~~~~~ & Br   $\rightarrow$ Brs     \\ 
 99~~~~~ & G056.1$-$00.4	      & ~~J193718.6$+$202102$^c$ & N4 & 2016-11-27 & 0\farcs9 &  $^*$64~~~~~~ & Bas  $\rightarrow$ Brs     \\ 
105~~~~~ & G063.3$+$02.2	      & ~~J194240.5$+$275109     & A1 & 2017-08-20 & 2\farcs4 &	  76~~~~~~ & Rar  $\rightarrow$ Ra      \\ 
106~~~~~ & G054.2$-$03.4	      & ~~J194359.5$+$170901$^d$ & N1 & 2007-09-03 & 0\farcs7 & $^*$120~~~~~~ & Bamps$\rightarrow$ Bmprs   \\ 
108~~~~~ & G057.8$-$01.7	      & ~~J194533.6$+$210808     & A1 & 2016-11-05 & 1\farcs8 & $^*$118~~~~~~ & Ear  $\rightarrow$ Is      \\ 
109~~~~~ & G059.8$-$00.6	      & ~~J194556.2$+$232833     & A1 & 2018-06-06 & 1\farcs0 &	  58~~~~~~ & Bas  $\rightarrow$ Es      \\ 
111~~~~~ & G059.7$-$01.0	      & ~~J194727.5$+$230816     & N2 & 2015-07-19 & 1\farcs3 &	  40~~~~~~ & Rars $\rightarrow$ Rs      \\ 
113~~~~~ & G066.8$+$02.9	      & ~~J194751.9$+$311818     & N5 & 2018-09-08 & 0\farcs8 &	  13~~~~~~ & B    $\rightarrow$ Bpw     \\ 
116~~~~~ & G062.7$+$00.0	      & ~~J194940.9$+$261521     & N1 & 2007-09-04 & 0\farcs6 &	  22~~~~~~ & Bps  $\rightarrow$ Bmr     \\ 
117~~~~~ & G063.5$+$00.0	      & ~~J195126.5$+$265838     & N5 & 2018-06-06 & 1\farcs0 &	  14~~~~~~ & B    $\rightarrow$ Es      \\ 
118~~~~~ & G067.9$+$02.4	      & ~~J195221.6$+$315859     & A1 & 2017-10-20 & 0\farcs9 &  $^*$57~~~~~~ & Ear  $\rightarrow$ Es      \\ 
119~~~~~ & G062.7$-$00.7	      & ~~J195248.8$+$255359     & A2 & 2016-08-05 & 1\farcs7 &  $^*$50~~~~~~ & B    $\rightarrow$ Bw      \\ 
125~~~~~ & G064.1$-$00.9	      & ~~J195657.6$+$265713     & N2 & 2015-07-18 & 1\farcs0 &  $\sim^*$55~~~~~~ & Bars $\rightarrow$ Brs \\ 
130~~~~~ & G068.0$+$00.0	      & ~~J200224.3$+$304845     & N3 & 2015-09-29 & 0\farcs8 &	  30~~~~~~ & Bar  $\rightarrow$ Br      \\ 
135~~~~~ & G077.6$+$04.3	      & ~~J200940.9$+$411442     & A1 & 2017-08-25 & 1\farcs7 &	  48~~~~~~ & Rars $\rightarrow$ R       \\ 
138~~~~~ & G077.4$-$04.0	      & ~~J204414.1$+$360737     & N4 & 2016-11-26 & 1\farcs3--1\farcs6 &	  25~~~~~~ & Es   $\rightarrow$ Es      \\ 
139~~~~~ & G079.5$-$03.8	      & ~~J205002.8$+$375315     & A2 & 2016-08-06 & 1\farcs7 &	  42~~~~~~ & Ears $\rightarrow$ I/E     \\ 
140~~~~~ & G086.5$+$01.8	      & ~~J205013.6$+$465515     & A1 & 2017-08-24 & 2\farcs5 &     332~~~~~~ & Ea   $\rightarrow$ Ems     \\ 
141~~~~~ & G081.0$-$03.9	      & ~~J205527.2$+$390359     & N1 & 2007-09-04 & 0\farcs7 &      31~~~~~~ & B    $\rightarrow$ Bmpr    \\ 
144~~~~~ & G090.5$-$01.7	      & ~~J212151.8$+$473301     & A2 & 2016-09-04 & 1\farcs8 &      30~~~~~~ & Bs   $\rightarrow$ Es      \\ 
145~~~~~ & G095.9$+$03.5	      & ~~J212200.9$+$550430     & A2 & 2016-08-05 & 1\farcs6 &      56~~~~~~ & Bams $\rightarrow$ Bs      \\ 
146~~~~~ & G091.6$-$01.0	      & ~~J212335.3$+$484717     & N3 & 2015-09-29 & 0\farcs7 &      20~~~~~~ & B    $\rightarrow$ Br      \\ 
147~~~~~ & G095.8$+$02.6	      & ~~J212608.3$+$542015     & N3 & 2015-10-01 & 0\farcs7 &      15~~~~~~ & Bas  $\rightarrow$ Eps      \\ 
148~~~~~ & G095.5$+$00.5	      & ~~J213423.2$+$523727     & N4 & 2016-11-27 & 0\farcs9 &       8~~~~~~ & E    $\rightarrow$ Em      \\ 
150~~~~~ & G098.9$-$01.1	      & ~~J215842.3$+$533003     & N4 & 2016-11-26 & 0\farcs7 &      31~~~~~~ & Rar  $\rightarrow$ Ra      \\ 
151~~~~~ & G101.5$-$00.6	      & ~~J221118.0$+$552841     & N3 & 2015-09-30 & 0\farcs7 &      78~~~~~~ & Bas  $\rightarrow$ Brs     \\ 
153~~~~~ & G114.2$+$03.7	      & ~~J232713.2$+$650923     & N3 & 2015-10-04 & 1\farcs6 &      21~~~~~~ & Rs   $\rightarrow$ R	      \\ 
154~~~~~ & G114.4$+$00.0	      & ~~J233841.2$+$614146     & A2 & 2016-09-04 & 1\farcs4 &      61~~~~~~ & Ears $\rightarrow$ E       \\ 
157~~~~~ & G114.7$-$01.2	      & ~~J234403.8$+$603242     & N4 & 2016-11-26 & 0\farcs9 &      24~~~~~~ & Ras  $\rightarrow$ Rs      \\ 
J191104.8 & G040.6$-$01.5         & ~~J191104.8$+$060845$^e$ & N5 & 2018-06-06 & 0\farcs8 &      32~~~~~~ & Brs    	                  \\ 
\hline
\end{tabular}
\begin{minipage}[b]{15cm}
 See also ${^a}$\citealt{Mampaso2006},${^b}$\citealt{Guerrero2021}, ${^c}$\citealt{Sabin2021}, ${^d}$\citealt{Corradi2011},${^e}$ \citealt{Rodriguez2021}.   \ \
\end{minipage} 
\end{table*}

\begin{figure*} 
\centering 
\includegraphics[height=2.5in]{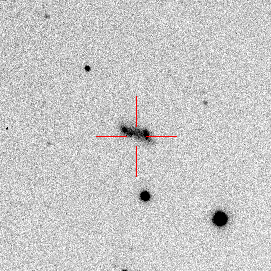}
\includegraphics[height=2.5in]{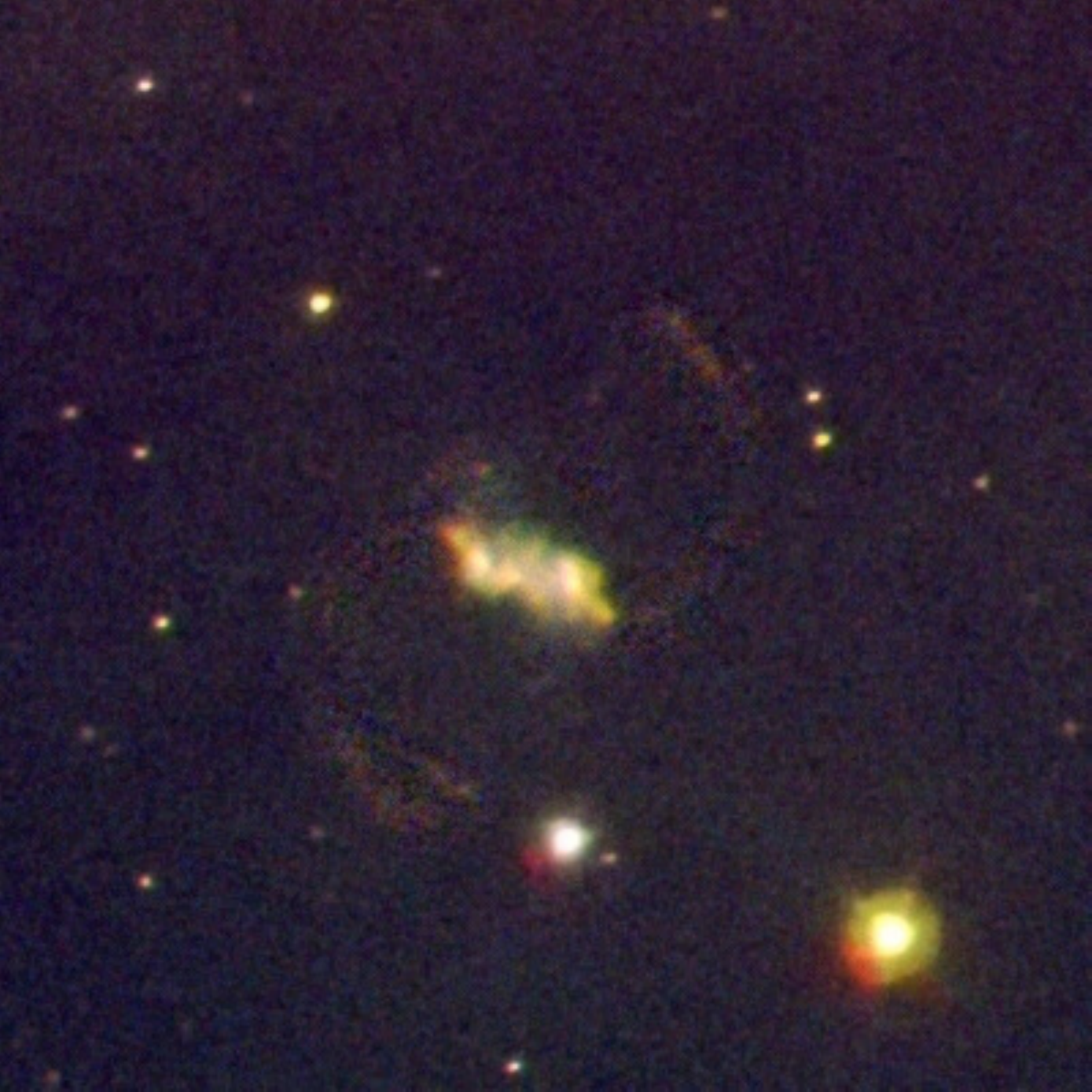}
\caption{
The 120s exposure original IPHAS image of Sab~2 in the H$\alpha$ filter (left) and the deep 1800s red (R), green (G) and blue (B) colour-composite picture with R=[N~{\sc ii}], G=H$\alpha$ and B=[O~{\sc iii}] (right). 
The comparison underlines the bipolar [N~{\sc ii}]-dominated outflow in the IPHAS image. 
North is up, East is left.  
The FoV is 90$\arcsec$.
} 
\label{new_outflow} 
\end{figure*}

\begin{figure*} 
\centering 
\includegraphics[height=2.5in]{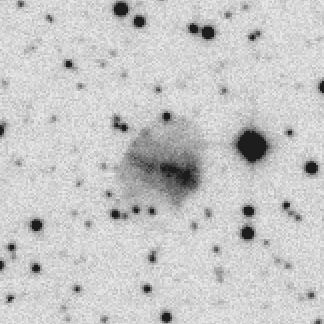}
\includegraphics[height=2.5in]{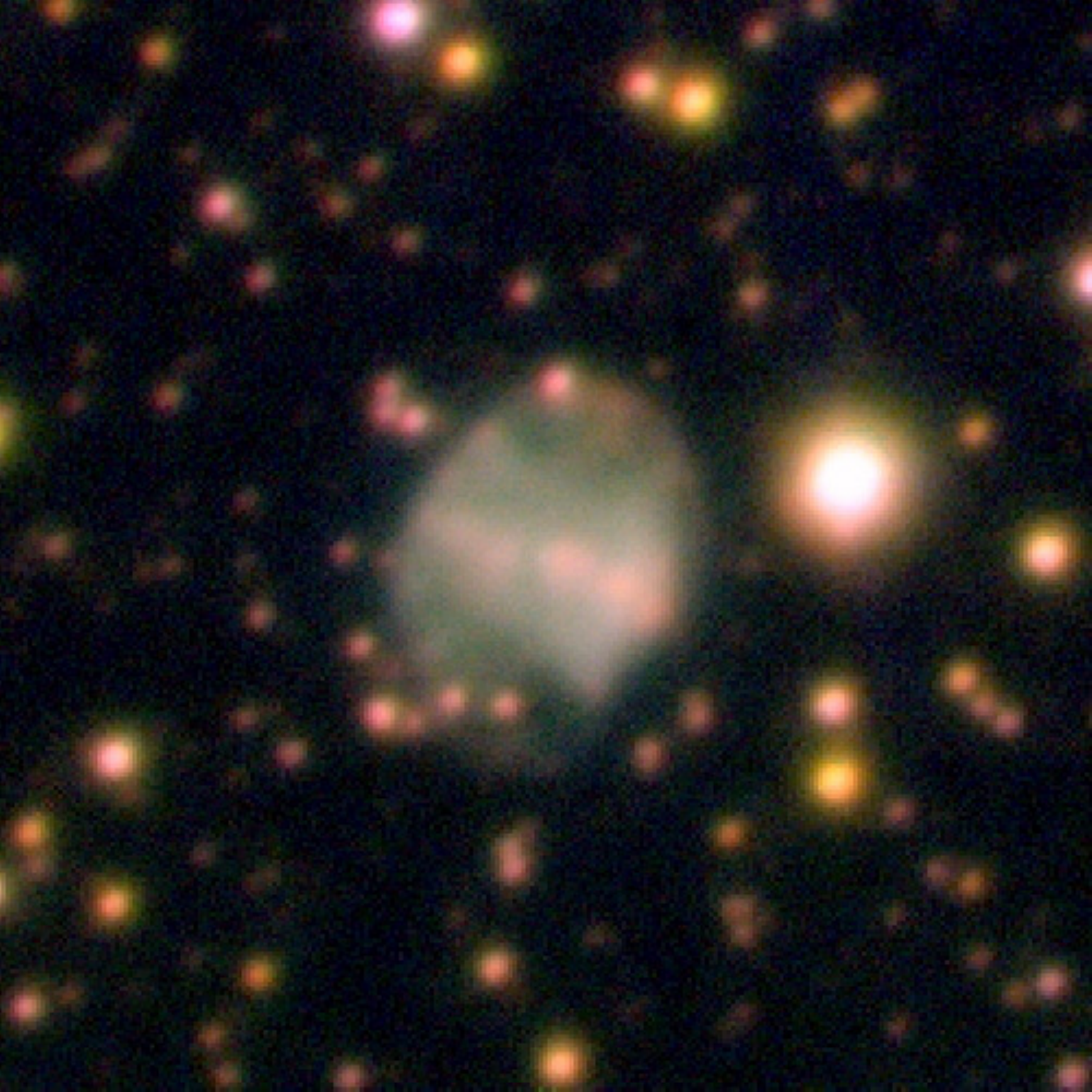}
\caption{ 
The bright homogeneous band seen in the original H$\alpha$ IPHAS image of Sab 63, turned out to be a string of individual knots in the deep 1800s colour-composite picture (same color code as Fig.\ref{new_outflow}, right). 
North is up, East is left. 
The FoV is 50$\arcsec$.
} 
\label{new_knot} 
\end{figure*}

\begin{figure*} 
\centering 
\includegraphics[height=2.5in]{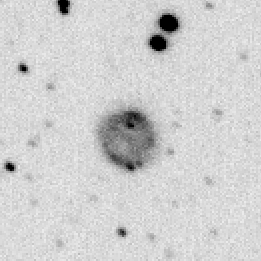}
\includegraphics[height=2.5in]{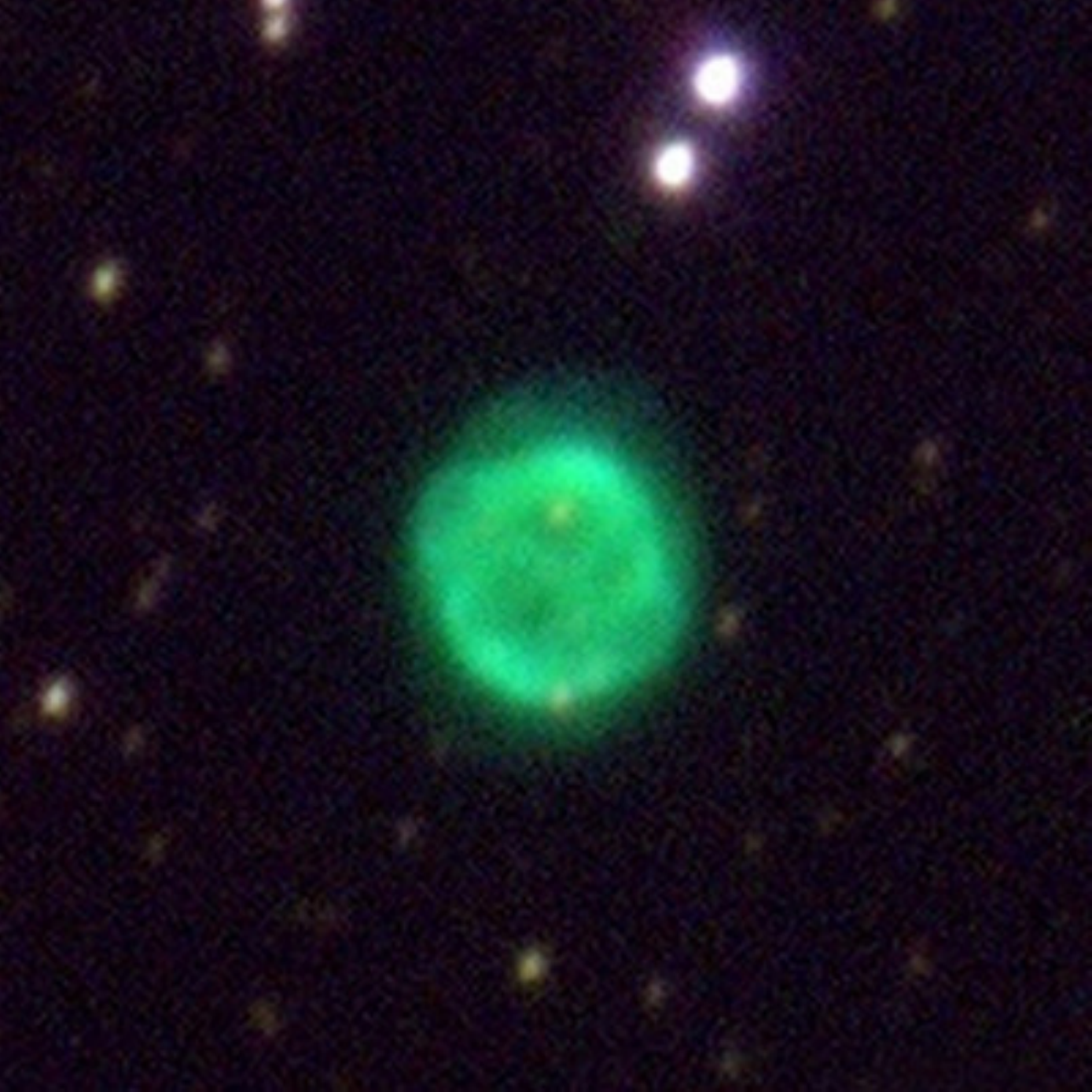}
\caption{ 
The original H$\alpha$ IPHAS image of Sab 65 (left) indicates a plain round morphology without any obvious internal or external structures. 
The deep NOT colour-composite picture (same color code as Fig.\ref{new_outflow}) reveals a filamentary PN (right). 
North is up, East is left. 
The FoV is 40$\arcsec$.
} 
\label{new_filament} 
\end{figure*}

\subsection{The power of deep imaging}

The first gain provided by the new imagery is a better estimate of the
nebular sizes.  Indeed, a significant increase ($\geq$10\arcsec) in
the previously stated dimensions (Paper I) is found for 13 objects
when measuring their angular extent in the new H$\alpha$ and/or
[N~{\sc ii}] images as listed in Table~\ref{tab:catalogue}.  In most
cases, the increase in the measured size arises from the detection of
faint bipolar outflows, outer structures or ansae previously missed in
the IPHAS images.  In the former case, the nebular emission formally
considered as the whole PN is revealed to correspond to the brightest
equatorial regions as illustrated in Figure~\ref{new_outflow} for
Sab~2.

Then the combination of depth and improved spatial resolution reached
by the new images allows the identification of small-scale features
previously undetected or not obvious such as knots and filaments.
Relatively compact knots are revealed in 11 objects, namely Sab\,2
(Fig.~\ref{new_outflow}), Sab\,7, Sab\,54, Sab\,61, Sab\,63, Sab\,79,
Sab\,86, Sab\,99, Sab\,106, Sab\,144, and J191104.8).  Knots are
mostly prominent in [N~{\sc ii}], except for highly ionized PNe where
the knots emit mostly in [O~{\sc iii}] as is the case of Sab~63
(Fig.~\ref{new_knot}).  Filaments are also detected in a number of
sources, namely Sab\,11, Sab\,15, Sab\,19, Sab\,50, Sab\,65
(Fig.~\ref{new_filament}), Sab\,140, Sab\,150, and
Sab\,157.

Actually, we have discarded Sab\,15 (Fig.~\ref{2.img}, 2$^{nd}$ row)
and Sab\,50 {\bf (Fig.~\ref{4.img}, 3$^{rd}$ row)} as genuine PNe
based on their morphology.  These are unlikely PNe, as their extended,
irregular and highly filamentary nature reveal a morphology which
mirrors more closely that of the shell of supernova remnants (SNRs)
than PNe. They were first classified as possible PNe (the lowest grade
of our classification scheme) based on the difficulty to have a clear
detection from the spectroscopy.  These two sources will not be
discussed further.


More generally, the shallow IPHAS images provided limited information
on the morphology of those PNe within the primary \emph{ERBIAS}
(Elliptical, Round, Bipolar, Irregular, Asymmetric and quasi-Stellar)
and secondary \emph{amspr} (a:one sided enhancement/asymmetries, m:
multiple shells or external structure, p: point symmetry, r: well
defined ring structure or annulus and s: resolved, internal structure)
classification schemes (see \citealt{Parker2006}).  We have revised
this morphology in the last column of Table~\ref{tab:catalogue}, with
most sources showing changes in their secondary \emph{amspr}
morphology and some of them even in their primary \emph{ERBIAS}
morphology.  Here we note the introduction of the secondary
classification key \emph{``w''} to denote sources with a pinched,
unresolved waist.

\subsection{Ionization state of the PNe}

The relative intensity of the H$\alpha$, [O~{\sc iii}], and [N~{\sc ii}] emission lines can be used to assess the excitation degree of a PN.  
The images presented in Figures~\ref{1.img}--\ref{12.img} are not flux-calibrated, however, and therefore they cannot be used to obtain a quantitative estimate of the nebular excitation.  
A qualitative assessment of nebular excitation can still be inferred from the comparison between the images in different emission lines in these figures as all images were obtained under photometric conditions through narrow-band filter sets with similar transmissions and using CCDs with similar responses.  
The relative contribution of these emission lines is immediately revealed in the colour-composite pictures obtained using similar intensity levels and contrast for the different emission lines: $R$ = [N~{\sc ii}], $G$ = H$\alpha$, and $B$ = [O~{\sc iii}].  
We shall define \emph{``red''}, \emph{``green''}, and \emph{``blue''} PNe according to the prevalent colour in their colour-composite pictures.

The \emph{blue} PNe are those completely dominated by [O~{\sc iii}] emission.  
The most obvious case is maybe Sab~7 (Fig.~\ref{1.img}), but also Sab~118. 
We can add here PNe with a \emph{turquoise} (\emph{blue}+\emph{green}) colour where H$\alpha$ has a non negligible contribution (e.g., Sab~44, Fig.~\ref{4.img}).   
These  \emph{blue} and \emph{turquoise} PNe are highly ionized sources that are expected to harbour the hottest PNe central stars (CSPNe) in this sample of sources.  
In a few cases, the \emph{blue} and \emph{turquoise} colours are dominant in the innermost regions, but \emph{green} and mostly \emph{red} (or their combination \emph{orange}) prevail in the outer regions, thus revealing a notable ionization stratification.  
The most obvious case is possibly Sab~88 (Fig.~\ref{6.img}), but also Sab~6, Sab~99 (\citealt{Sabin2021}), Sab~106, Sab~ 135, Sab~141, Sab~150, and Sab~151.  
This ionization stratification is suggestive of a hot CSPN with a limited ionizing flux, most likely an evolved CSPN already in the cooling track of white dwarfs.

On the opposite corner there is a number of \emph{red} PNe, i.e., sources that are dominated by [N~{\sc ii}] emission.  
Among them, we can list Sab~11, Sab~23, Sab~36, Sab~39, Sab~81, Sab~109, Sab~130, Sab~139, and J191104.8 (\citealt{Rodriguez2021}).  
These objects can be associated with low excitation nebulae. 
An enhanced nitrogen content can be suggested, linking them to evolved Type I PNe, which is further supported by their bipolar or ring-like morphology.  

We can add here also sources with an \emph{orange} colour resulting from the combination of H$\alpha$ and [N~{\sc ii}] emissions, e.g., Sab\,22, Sab\,53, Sab\,111, Sab\,113, Sab\,119, Sab\,125, and Sab~138.  
Alternatively, \emph{orange} sources may be heavily extinguished, thus reducing significantly the observed flux of the [O~{\sc iii}] emission line.  
Indeed, the much smaller number of field stars in the [O~{\sc iii}] images of some of them (e.g., Sab\,53 and Sab\,111) with respect to that in the H$\alpha$ and [N~{\sc ii}] images may be indicative of high extinction along the line of sight. 
The same can apply to some \emph{red} sources (e.g., Sab\,11), although in those cases it is not straightforward to disentangle the effects of low excitation and extinction, as both tend to diminish the relative intensity of the emission in the [O~{\sc iii}] line.

There is also a small group of \emph{green} PNe (not to be confused in our scheme with those harboring the [O~{\sc iii}] ''green'' line) whose emission is dominated by H$\alpha$ with nearly no [N~{\sc ii}] and little [O~{\sc iii}] emission, such as Sab~13 and Sab~44 and to some extent Sab~65. 
The significant shortfall of emission in the forbidden emission lines while H$\alpha$ is bright may be indicative of a recombining nebula.

Finally, we identify a \emph{white} PN, namely Sab~148 (Fig.~\ref{11.img}), whose colour-picture implies similar contributions of the three emission lines.  
Interestingly, this double shell PN shows low-ionization knots protruding along the major axis of the inner shell.

\subsection{Classification and Deep Learning}

Although the objects in this survey were selected as PNe, some may not be. Indeed, in Paper I some objects are classified as Likely or Possible. 
The HASH catalogue \citep{Parker2016} classifies 51 of our targets as 'True PN', four as 'Likely PN', one as 'Possible PN', and the two remaining objects are not classified as PN.  The classification is based mainly on available survey images, aided by spectra.\\

\citet{Iskandar2020}  applied Deep Transfer Learning (DTL) to assign confidence levels to the HASH classifications. 
The network was trained on three sets of images available in HASH: optical H$\alpha$ images, quotient images (H$\alpha$ divided by the continuum image) and \emph{WISE} images combining bands 2, 3, and 4 at 4.6$\mu$m, 12$\mu$m and 22$\mu$m, respectively.  
Each object was classified using each of the three resources.  
An object was classified as a true PN if at least two resources classified it accordingly.  
The DTL classified the majority of Likely PNe as PN (64\%), but a smaller fraction of the Possible PNe (41\%). 
This agreed with the expected higher level of confidence for the Likely PNe.  
In comparison, the model classified 94\% of the True PN as PN.  

We can now test how well the DTL output agrees with the deeper images presented here. 
The  trained network of \citet{Iskandar2020} with the {\sc Densenet201} algorithm was used. Densenet201 uses 201 layers, and differs from other tested implementations by making the input feature map reusable and accessible to later layers \citep{Huang2017}. In the tests of \citet{Iskandar2020} it was found to be among the most effective algorithms for the current purpose. We applied this to the HASH images of the 58 objects in this catalogue. An object was classified as a PN if at least two image types returned this classification. 
The three methods agree on a PN classification for 40 out of the 58 objects in this sample and it was questioned only by one criterion for seven objects, thus leading to a PN classification for 47 sources.  
Eleven sources were thus rejected as PN, in 3 cases by all three criteria.

\begin{table*}
\footnotesize\addtolength{\tabcolsep}{-4pt}
\caption{\label{tab:class} Objects for which the classification as PN is not certain }
\begin{tabular}{r|l|l|l|l|}
\hline
\multicolumn{1}{c}{Sab~\#} & 
\multicolumn{1}{l}{~~HASH~~~} & 
\multicolumn{1}{l}{DTL~~~~} & 
\multicolumn{1}{l}{deep image~~~~} &
\multicolumn{1}{l}{comment} \\ 
\hline
\hline
2   & ~~true     & not & true \\
13  & ~~true     & not & true &  \\
15  & ~~possible & not & not & supernova remnant\\
19  & ~~true     & true & possible & unusual structure; very low luminosity for PN star\\
22  & ~~likely   & true & true \\
31  & ~~true     & not & true & very faint \\
36  & ~~likely   & true & possible & too faint to classify \\
42  & ~~true     & not & possible & no morphological information\\
50  & ~~not      & not & not & supernova remnant \\
53  & ~~not      & not & possible & Star too red to be a PN central star\\
61  & ~~true     & not & possible & very elongated. Jet, PN or YSO?\\
81  & ~~true     & not & true \\
99   & ~~true     & true & possible & asymmetric \\
108  & ~~likely   & true & possible & lack of symmetry suggestive of ISM\\
119  & ~~likely   & true & true \\
125 & ~~true     & not & possible & not fully symmetric\\
138 & ~~true     & not & true \\
140   & ~~true     & true & possible & Irregular halo \\
\hline
\hline
\end{tabular}
\end{table*}

The 18 objects for which either the HASH database, the DTL or the current images leaves some doubt about the identification as PN are listed in Table \ref{tab:class}. \\
Most of them are likely to be PNe, but we cannot be fully confident about it. 
A few nebulae are too faint to confidently classify them. 
Here deeper images would certainly help. 
The HASH and deep image classification agree for five objects, but the DTL does not. 
In most of those cases the deep images reveal structure which make a PN classification very likely. 
Irregular structures are likely to lead to a DTL classification as non-PN, although PNe can become irregular during late stages of their evolution due to ISM interactions.  
Anyhow it casts lower confidence in the classification as other ISM structures also often are irregular. 

Two objects are confirmed as non-PN (15, 50), both very likely supernova remnants. 
Sab\,53 is unlikely to be a PN and may be related to symbiotic stars. 
This might also be the case of Sab\,61, which has a very unusual structure for a PN and needs further study. 
Finally, Sab\,19 is classified as a true PN in the detailed study of \citet{Guerrero2021}.

\section{Conclusion}\label{Conc}

This paper presents a new set of deep images for 58 True, Likely and Possible extended PNe detected with IPHAS. 
These new images have been obtained using H$\alpha$, [N~{\sc ii}] $\lambda$6584 \AA, and [O~{\sc iii}] $\lambda$5007 \AA\ narrow-band filters.

The depth of these images in different emission lines has allowed us to detect macro- and micro-structures in these PNe such as outflows, filaments as well as knots and ansae. 
These new features has led us to the morphological reclassification of some sources. 
Two sources originally catalogued as \emph{Probable PNe} have even been discarded as PNe based on their filamentary nature most likely implying a SNR nature. 
In some cases we were able to uncover the full extent of the optical nebulae with an update on the total size.

The images presented here also indicate that these PNe span over a large excitation or ionization range from the Type-I like PNe (N-enhanced) to the highly ionized PNe, highlighting the wide and rich range of IPHAS objects. 
Indeed, our {\it blue} ([O~{\sc iii}] dominated), {\it red} ([N~{\sc ii}] dominated), {\it green} (H$\alpha$ dominated) and {\it white} (similar contribution in all filters) qualitative classification of PNe is indicative of the different characteristics of the ionizing central stars as well as their evolutionary status. 
We note, however, that images acquired through the [O~{\sc iii}] filter are more prone to extinction effects, which may affect the above classification. 

We also used Deep Transfer Learning for the automated classification of the objects owing that our sample includes True, Likely and Possible PNe. These results were compared to the classification performed using the HASH catalogue as well as our deep images. Most of the objects are thus classified as PNe by all three methods and amongst the 18 doubtful cases it appears that our deep images are very useful, if not critical, mostly in the case of the faintest and more irregular PNe. 

This first imaging catalogue is therefore a valuable tool to support further more exhaustive studies of these PNe \citep[hence the analysis of a group of such PNe has already been carried on by][]{Corradi2011,Miszalski2013,Guerrero2021,Rodriguez2021,Sabin2021} and we expect to proceed to the deep imaging of the full catalogue presented in our Paper I.

\section*{Acknowledgements}
We thank the referee for the his/her comments which help improving the paper.
LS acknowledges support from PAPIIT grant IN-101819 (Mexico). 
MAG acknowledges support from grant AYA  PGC2018-102184-B-I00  co-funded  with  FEDER funds. 
GR-L acknowledges support from CONACyT (grant 263373) and PRODEP (Mexico). JAT acknowledges funding by Direcci\'{o}n
General de Asuntos del Personal Academico of the Universidad Nacional Aut\'{o}noma de M\'{e}xico (DGAPA, UNAM) project IA100720 and the Marcos Moshinsky Fundation (Mexico).
AAZ acknowledges funding from STFC under grant number ST/T000414/1 and Newton grant ST/R006768/1, and from a Hung Hing Ying visiting professorship at the University of Hong Kong. DNFAI acknowledges funding under `Deep Learning for Classification of Astronomical Archives' from the Newton-Ungku Omar Fund: F08/STFC/1792/2018.

This work is partially based on observations made with the Nordic Optical Telescope, operated by the Nordic Optical Telescope Scientific Association installed in the Spanish Observatorio  de El  Roque  de  los  Muchachos  of  the  Instituto  de  Astrof\'isica de Canarias in La Palma, Spain, and the Aristarchos telescope, which is operated at the Helmos Observatory by the IAASARS of the National Observatory of Athens. 
This paper makes use of data obtained as part  of  the  INT  Photometric  H$\alpha$ Survey  of  the  Northern Galactic Plane (IPHAS, www.iphas.org) carried out at the Isaac Newton  Telescope (INT). The INT is operated on the island of La Palma by the Isaac Newton Group in the Spanish Observatorio del Roque de los Muchachos of the Instituto de Astrof\'isica de Canarias. All IPHAS data are processed by the Cambridge Astronomical Survey Unit, at the Institute of Astronomy in Cambridge. The bandmerged DR2 catalogue was assembled at the Centre for Astrophysics Research, University of Hertfordshire, supported by STFC grant ST/J001333/1. This work has made extensive use of NASA’s Astrophysics
Data System.

\section*{Data availability}

The images here presented will be included in due time in the HASH database  (\url{http://202.189.117.101:8999/gpne/index.php}), once an option currently under development to download the best available optical images of PNe will become available. 
In the meantime the data presented in this work are available on request to the first author.






\appendix

\section{Images}
\label{sec:all_images}

\begin{figure*} 
\centering 
\includegraphics[height=1.7in]{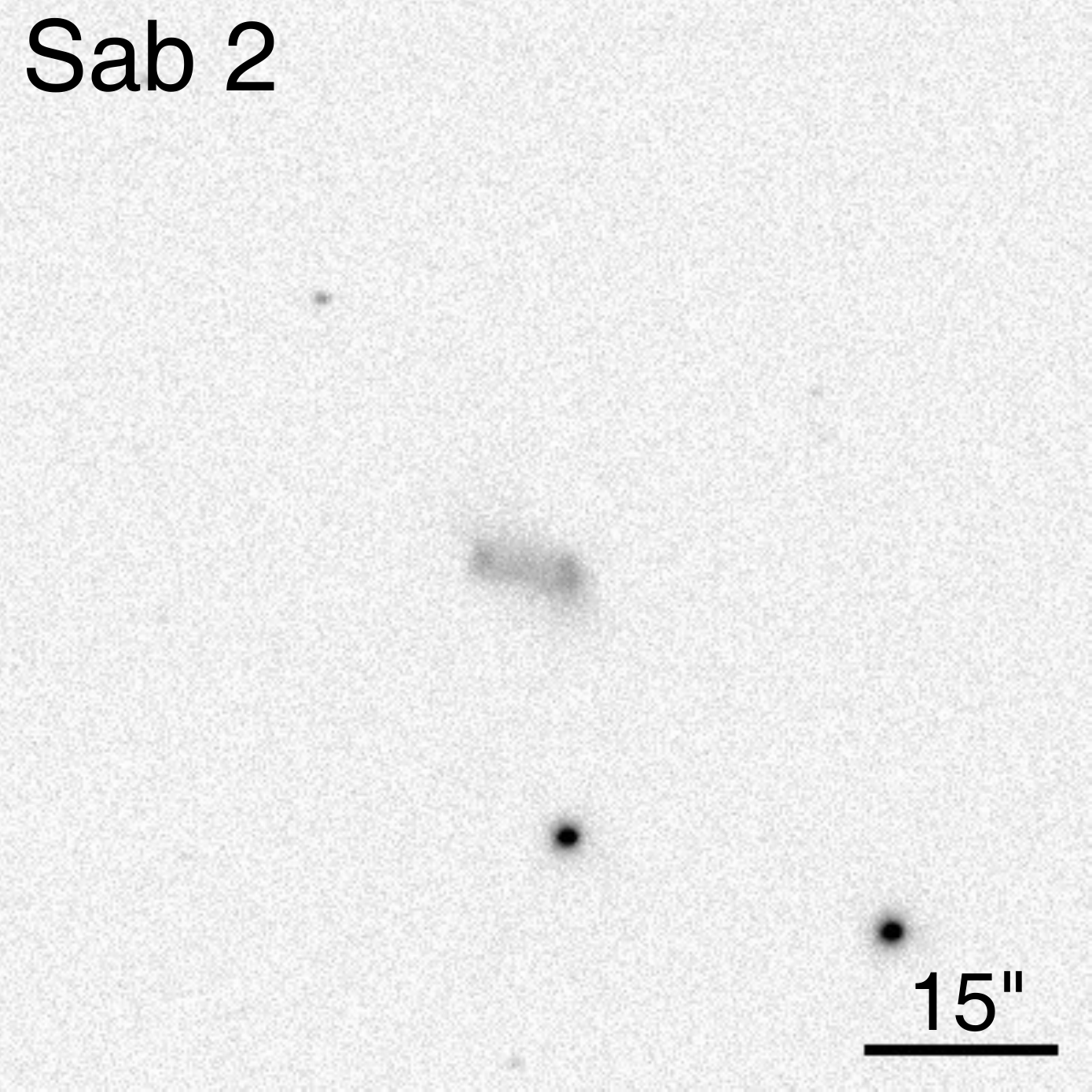} 
\includegraphics[height=1.7in]{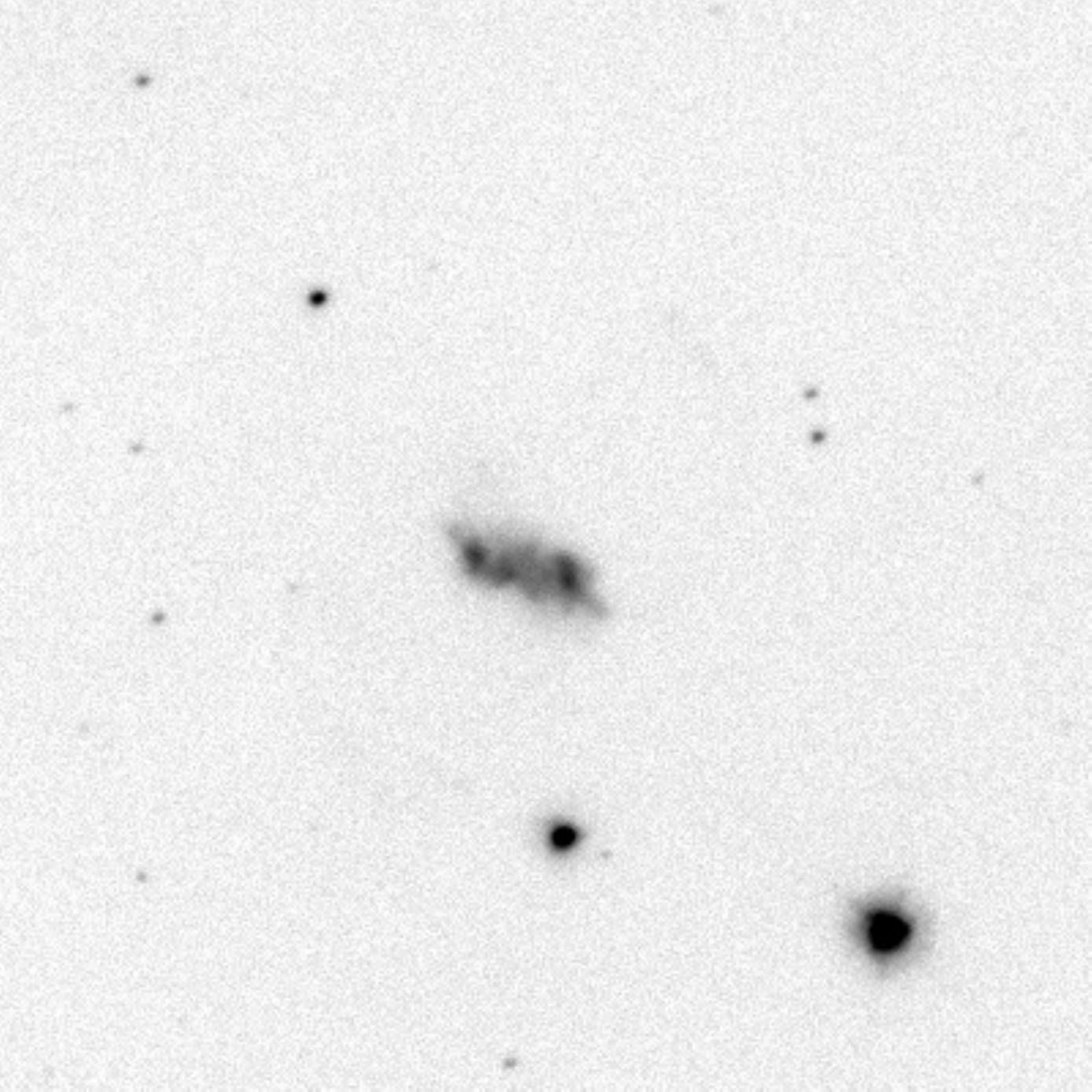}
\includegraphics[height=1.7in]{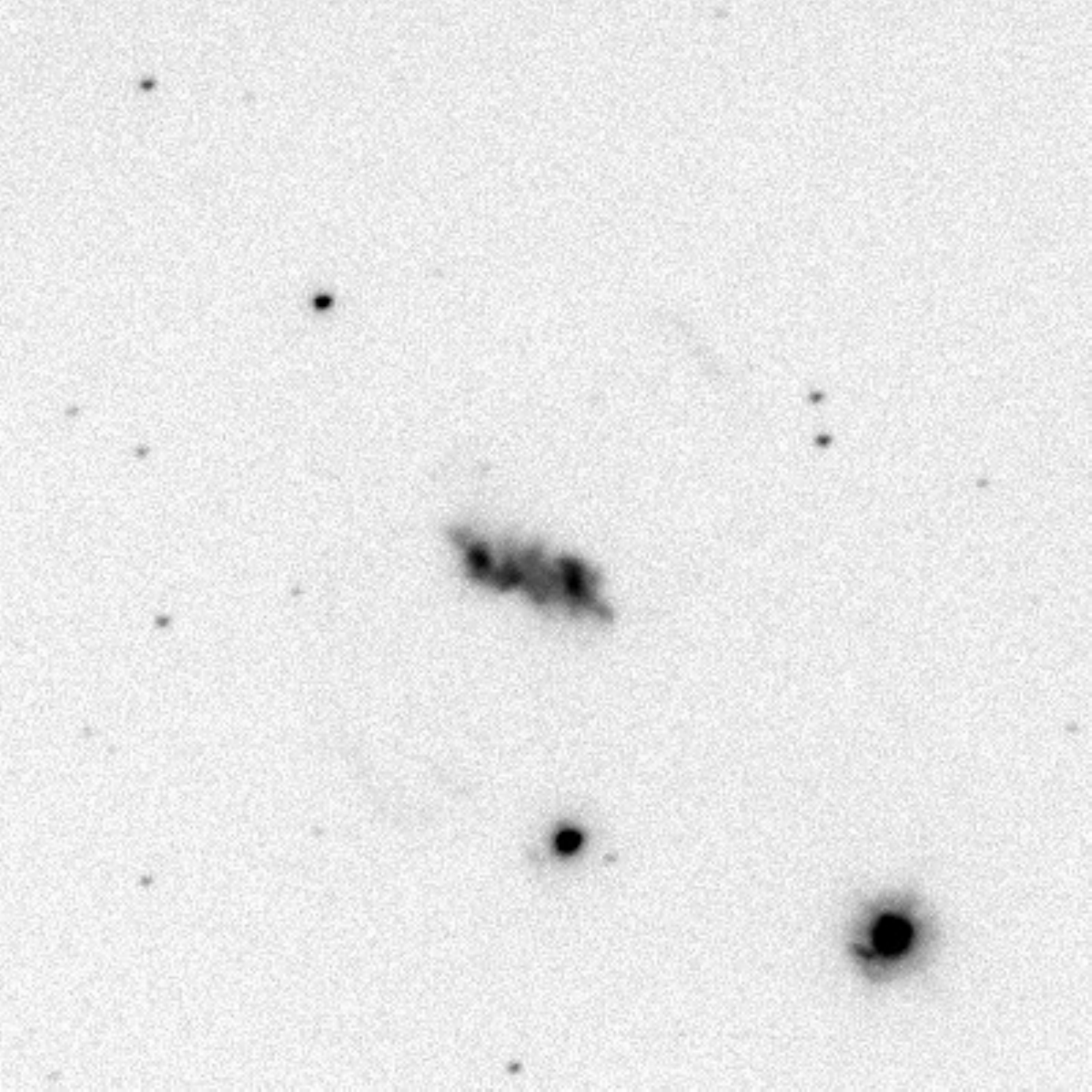}
\includegraphics[height=1.7in]{2_G119.eps}
\vskip .1in 
\includegraphics[height=1.4in]{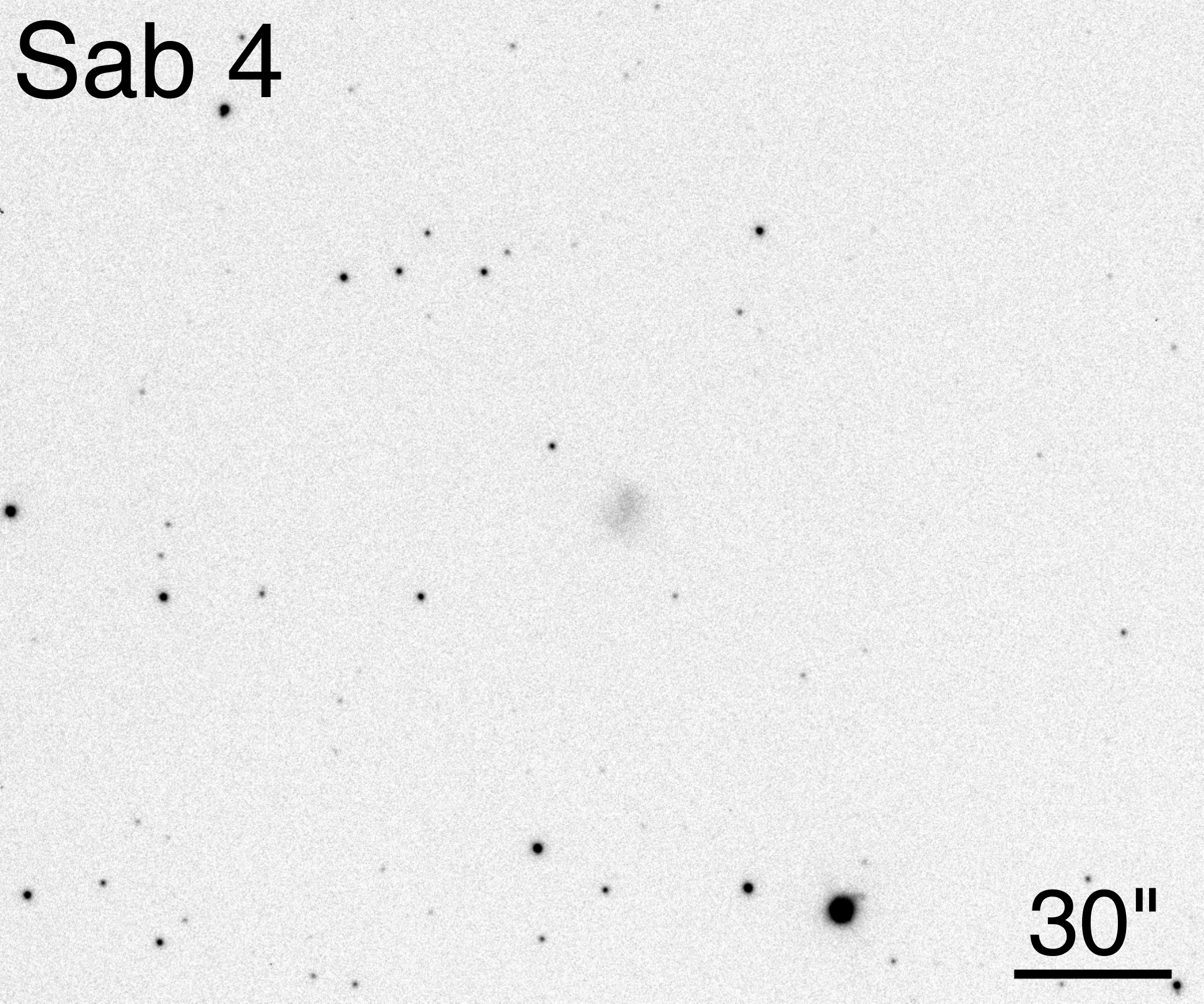} 
\includegraphics[height=1.4in]{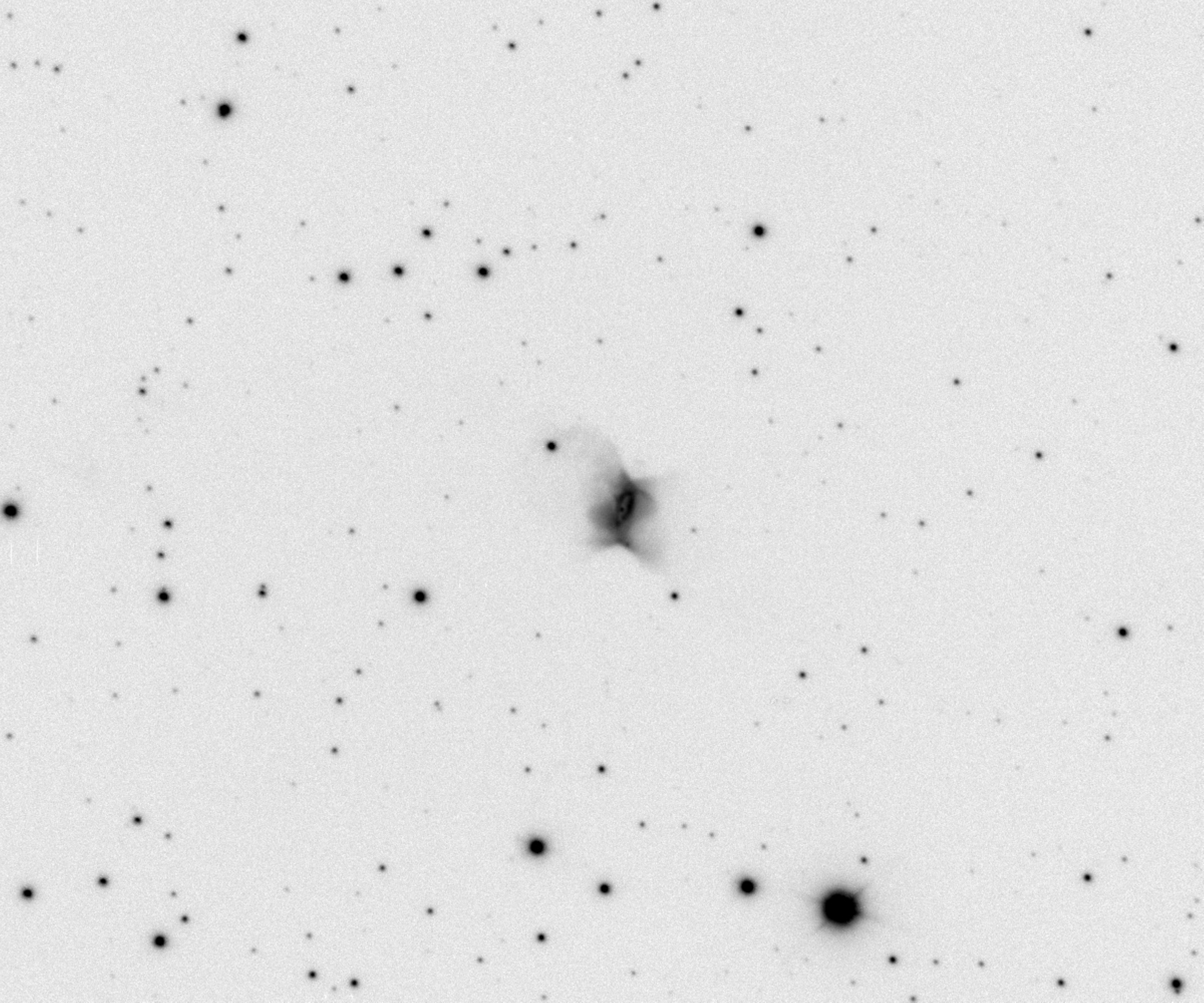}
\includegraphics[height=1.4in]{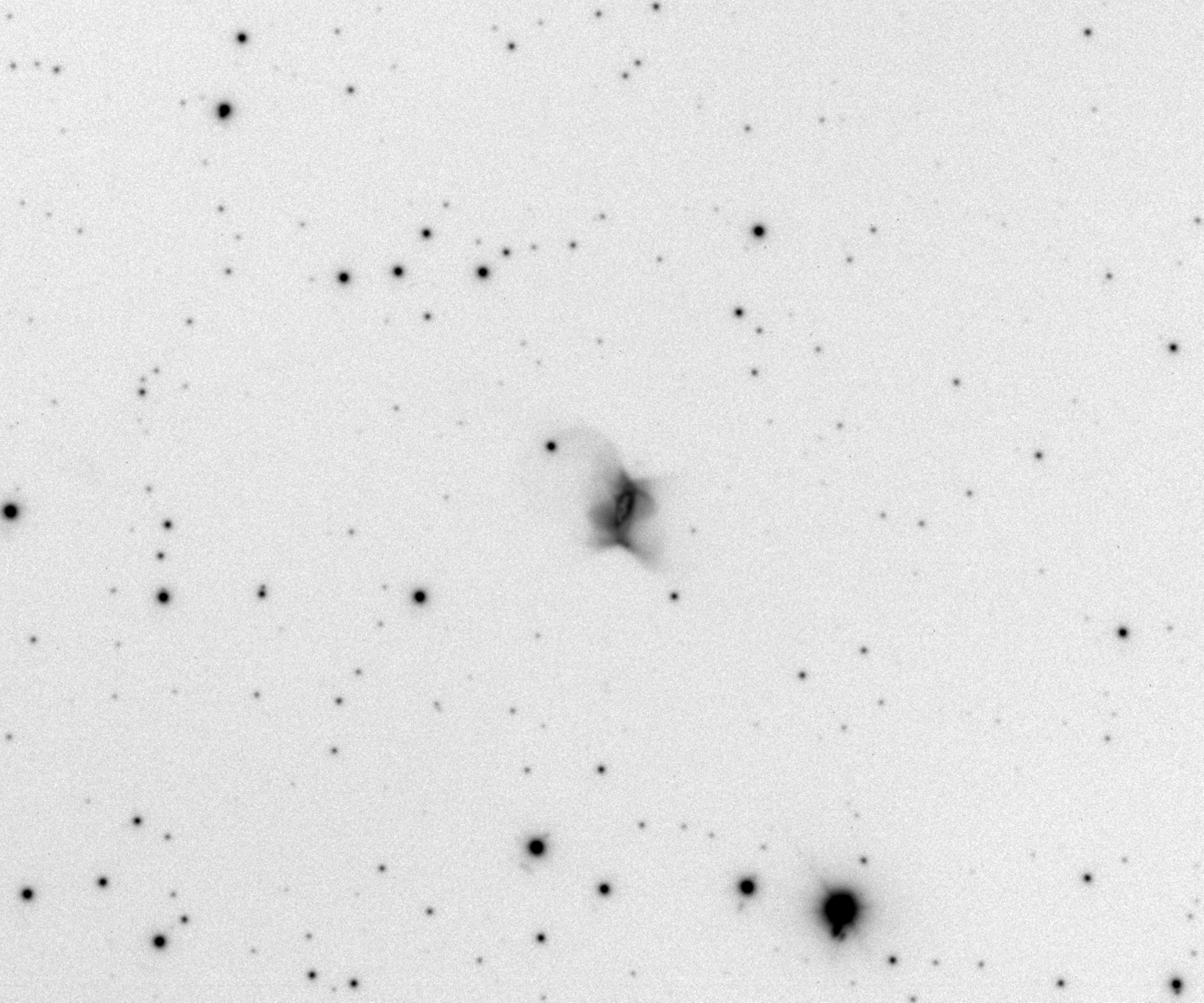}
\includegraphics[height=1.4in]{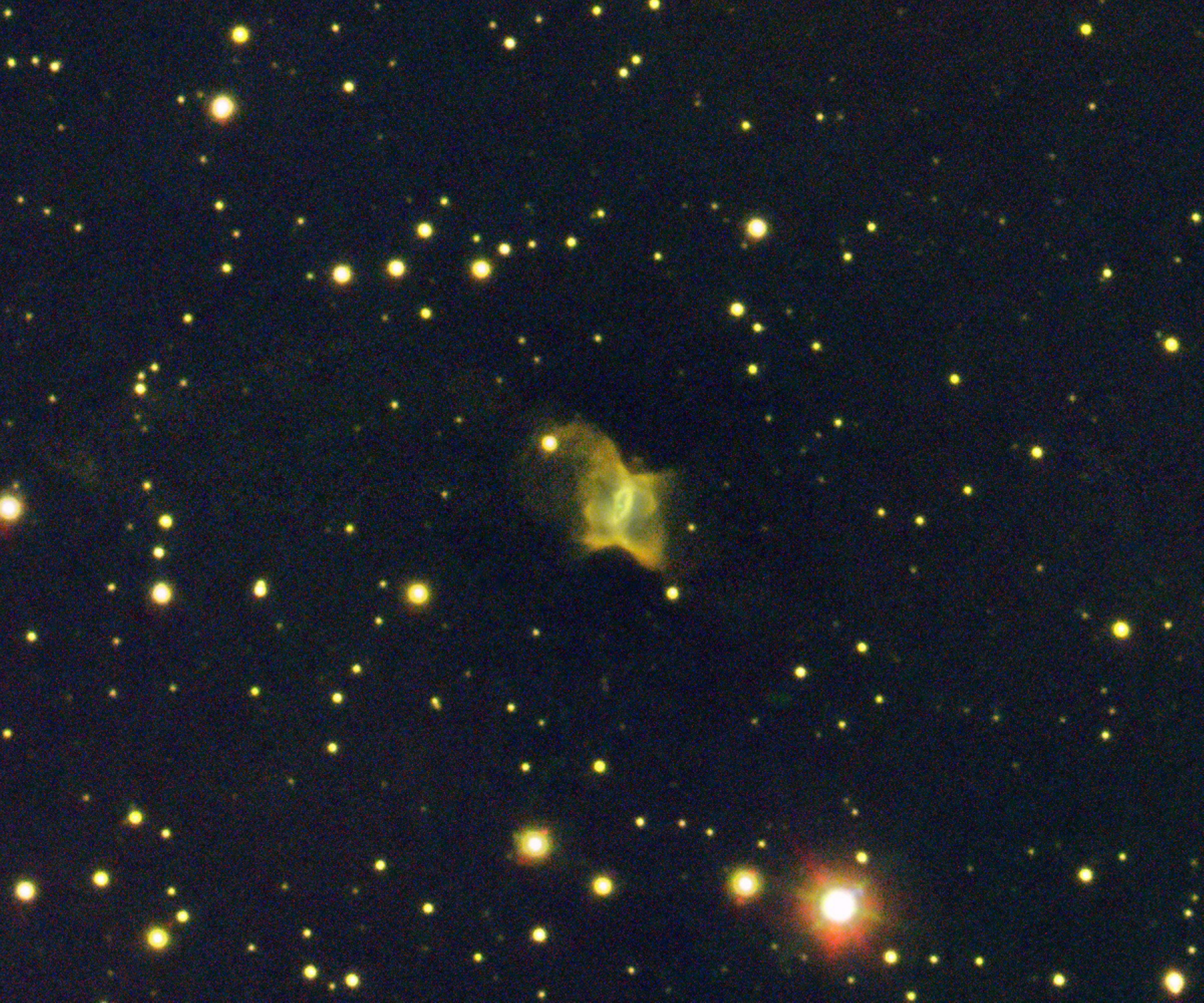}
\vskip .1in 
\includegraphics[height=1.7in]{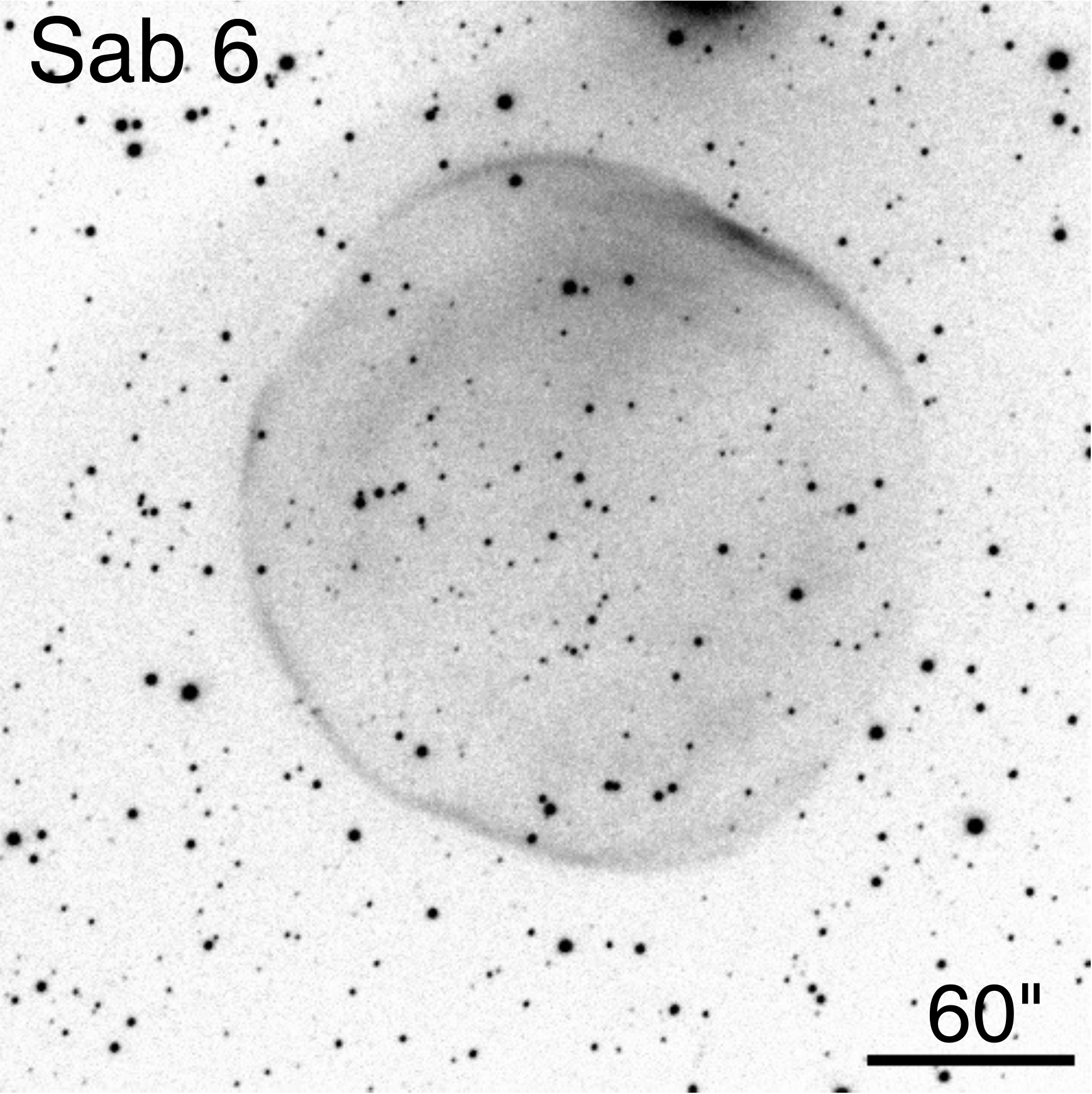} 
\includegraphics[height=1.7in]{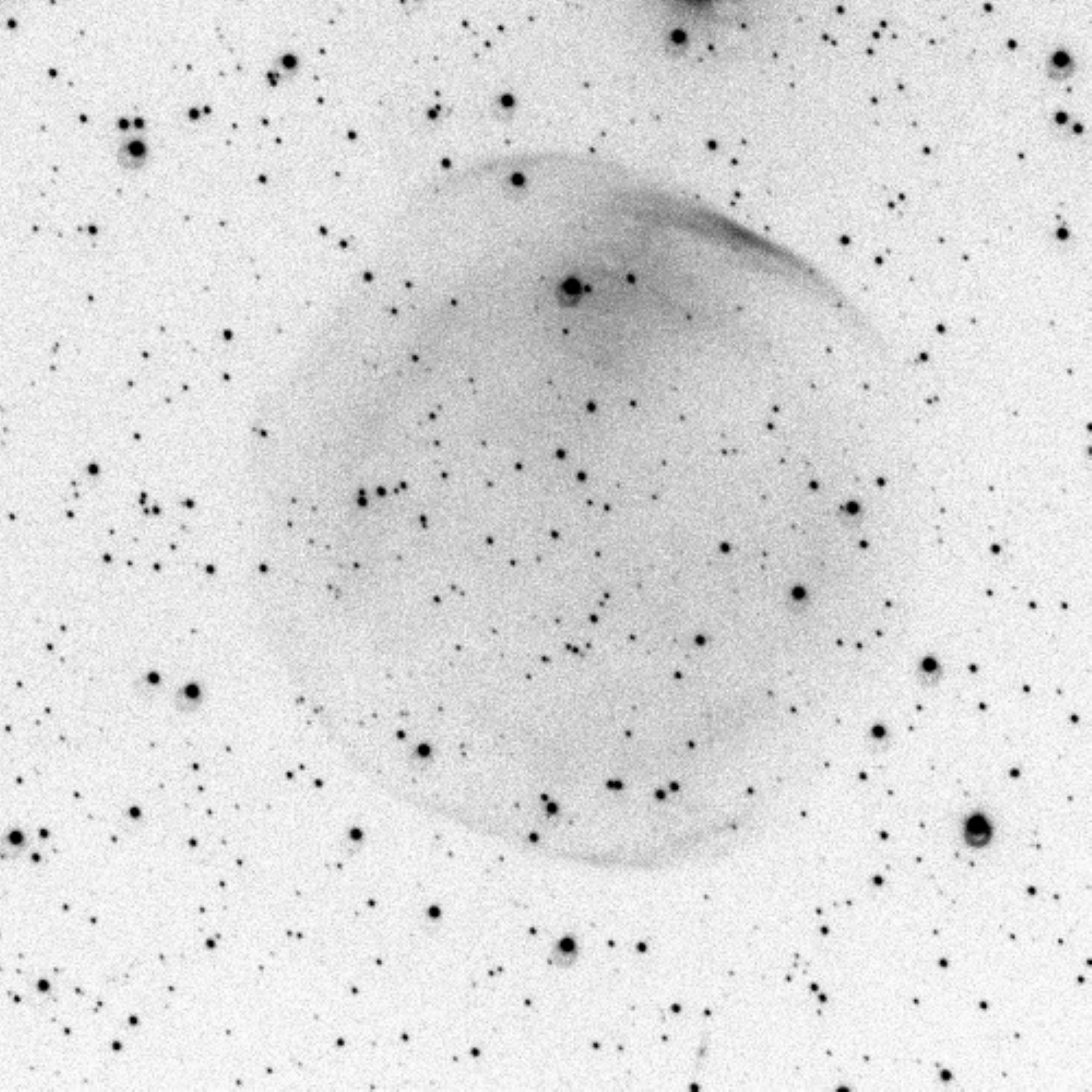}
\includegraphics[height=1.7in]{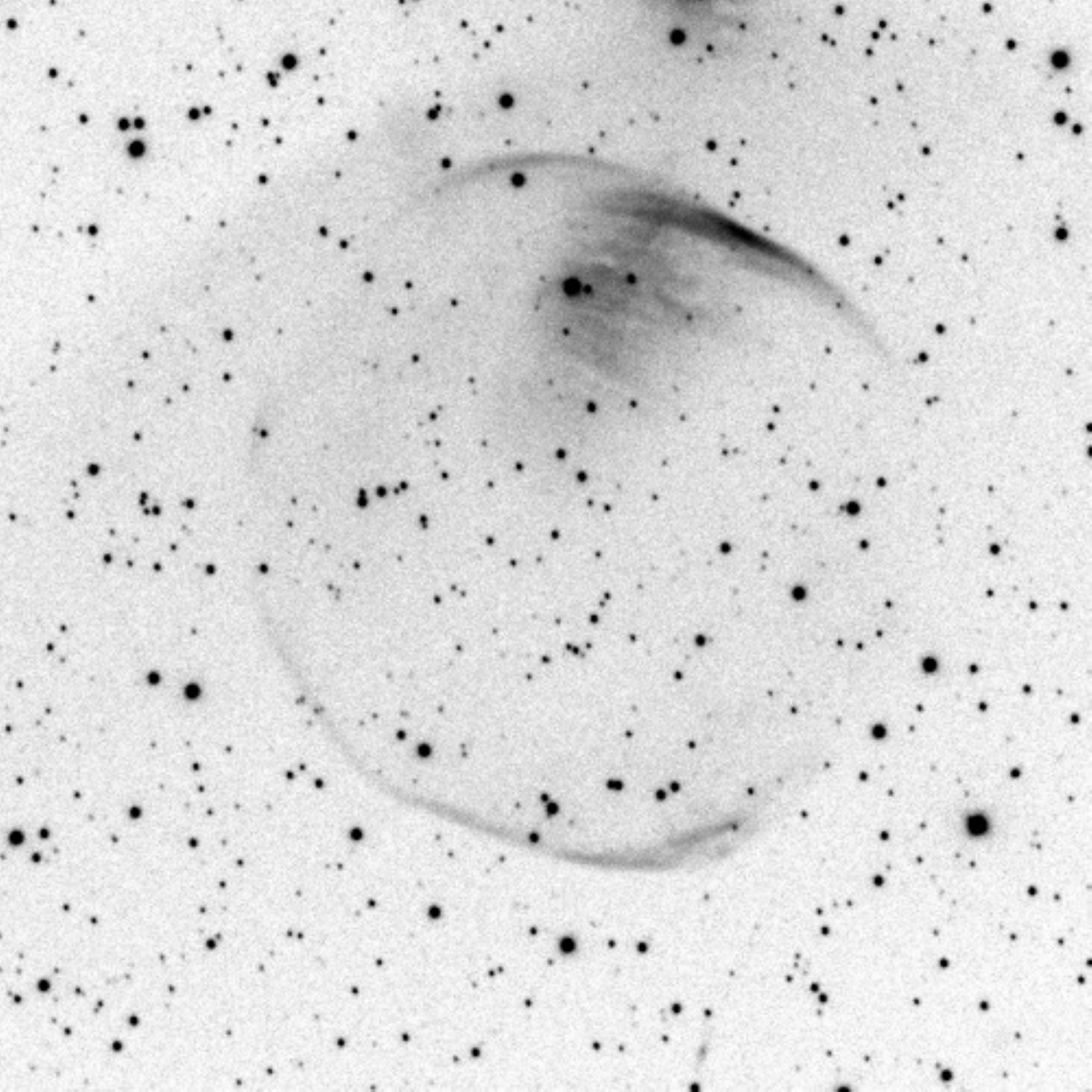}
\includegraphics[height=1.7in]{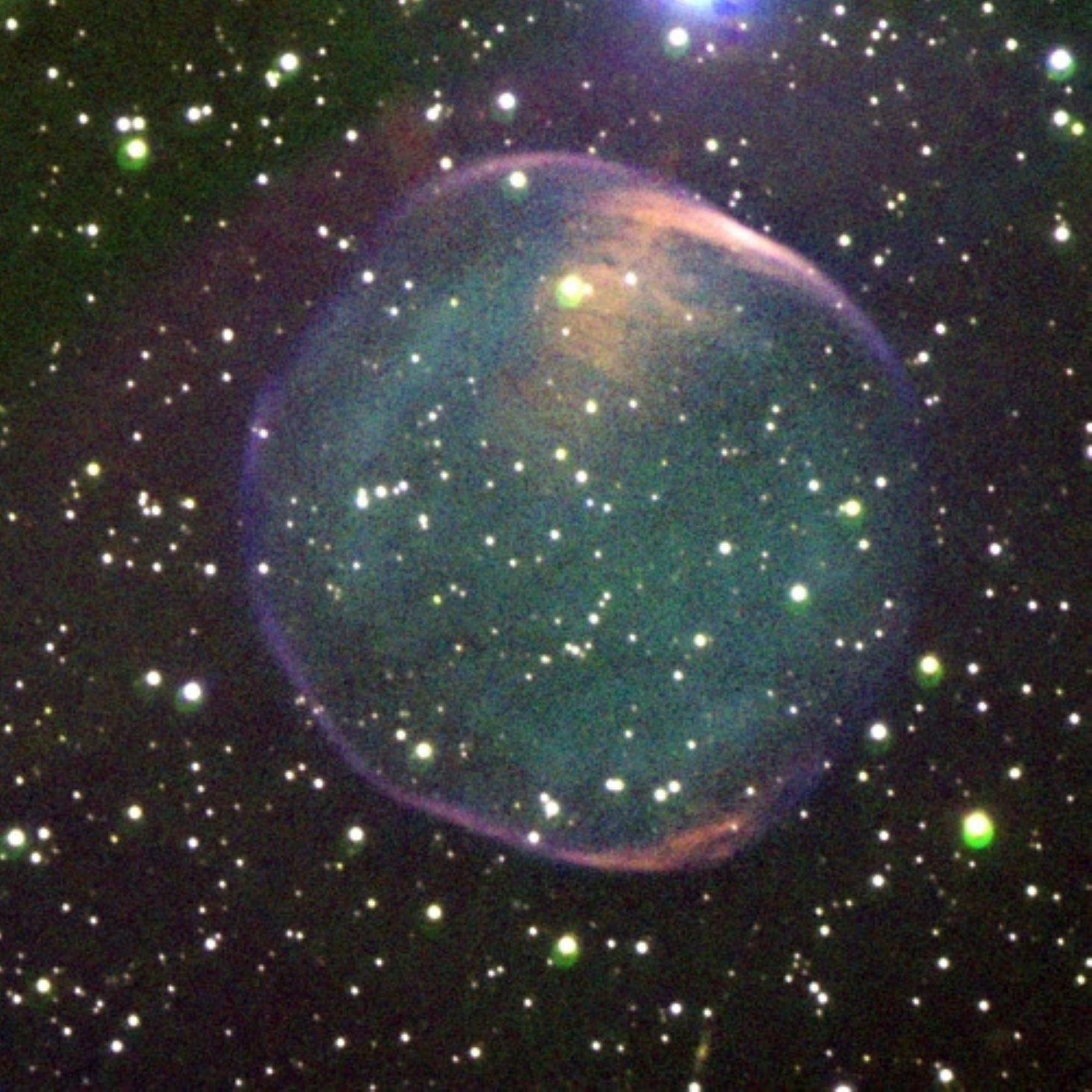}
\vskip .1in 
\includegraphics[height=1.7in]{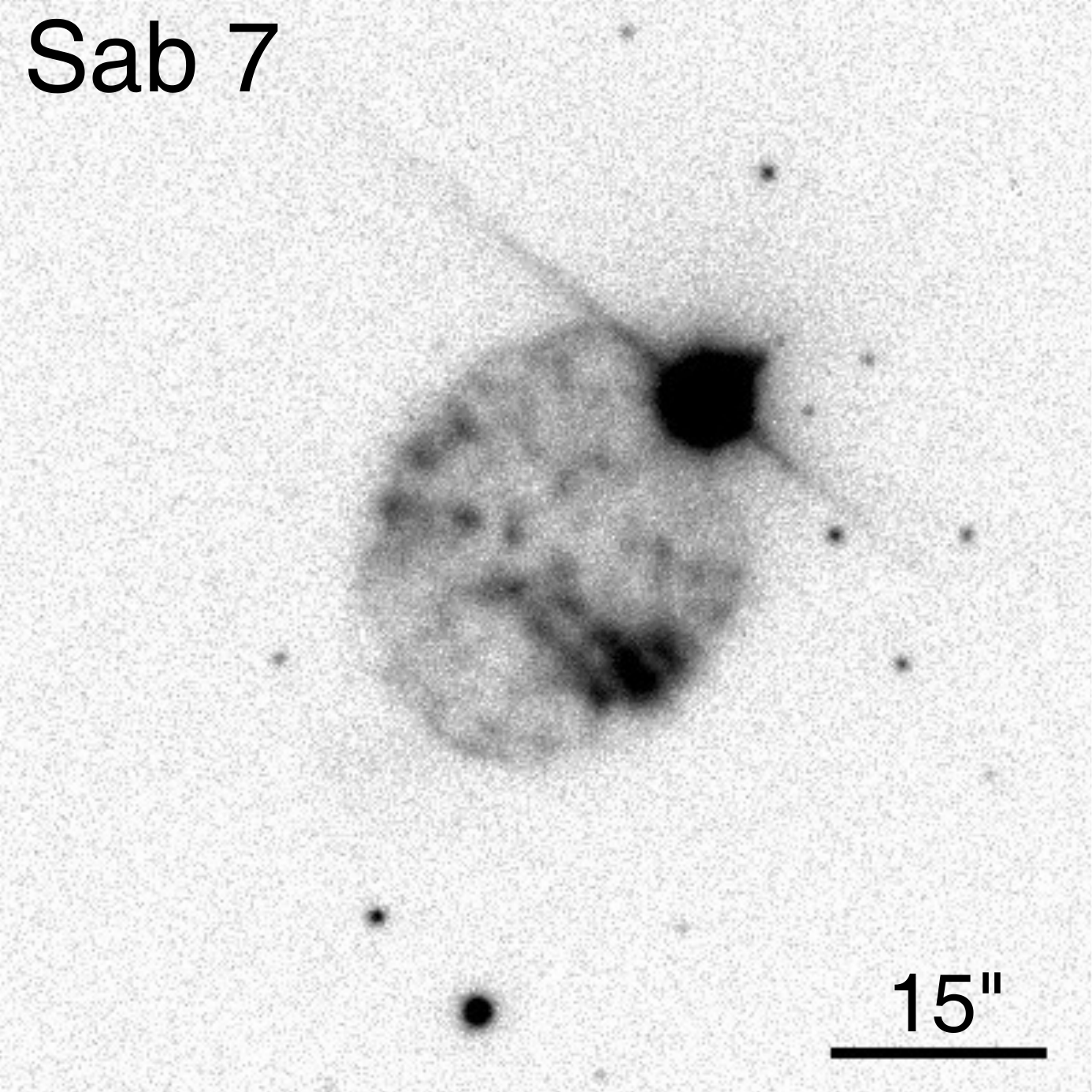} 
\includegraphics[height=1.7in]{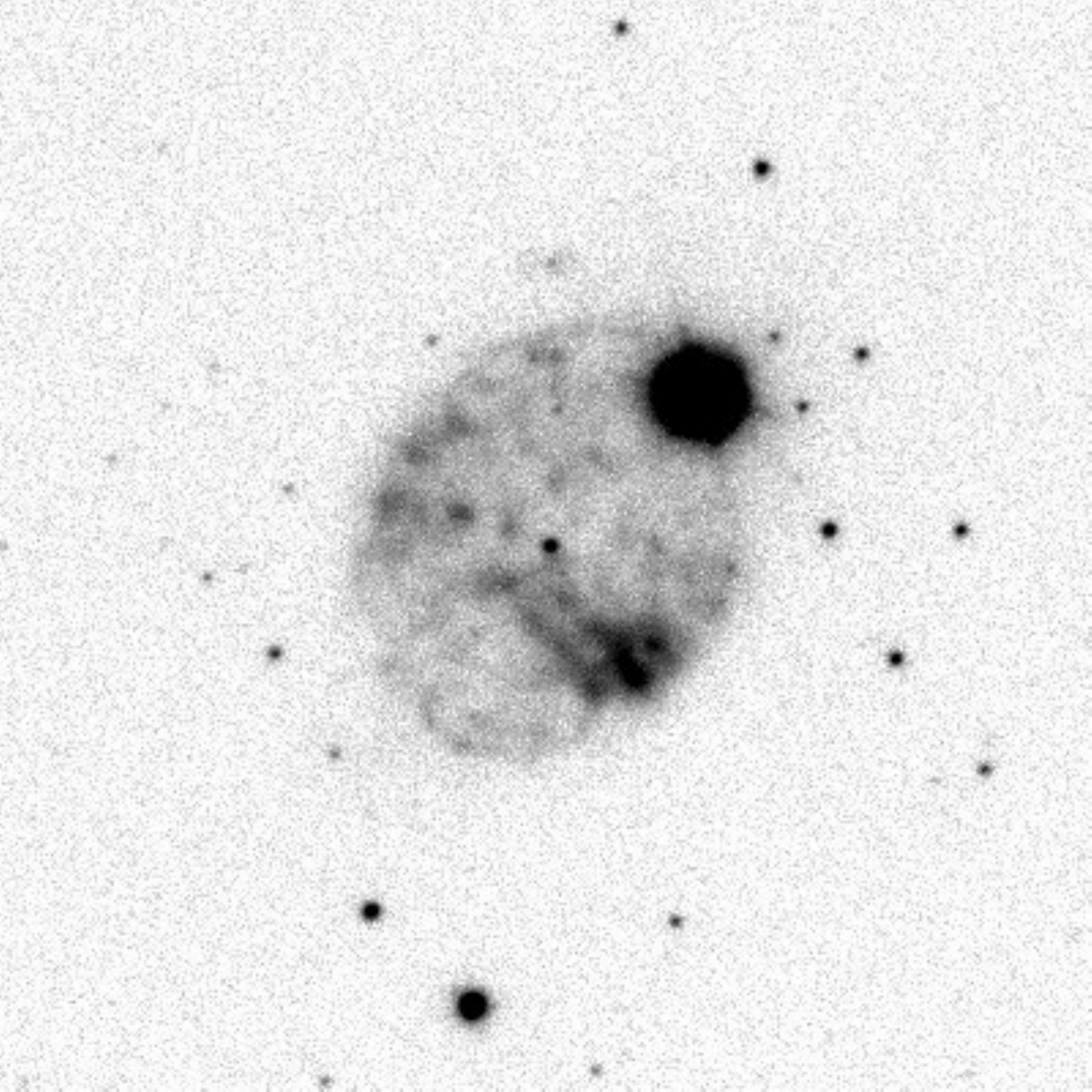}
\includegraphics[height=1.7in]{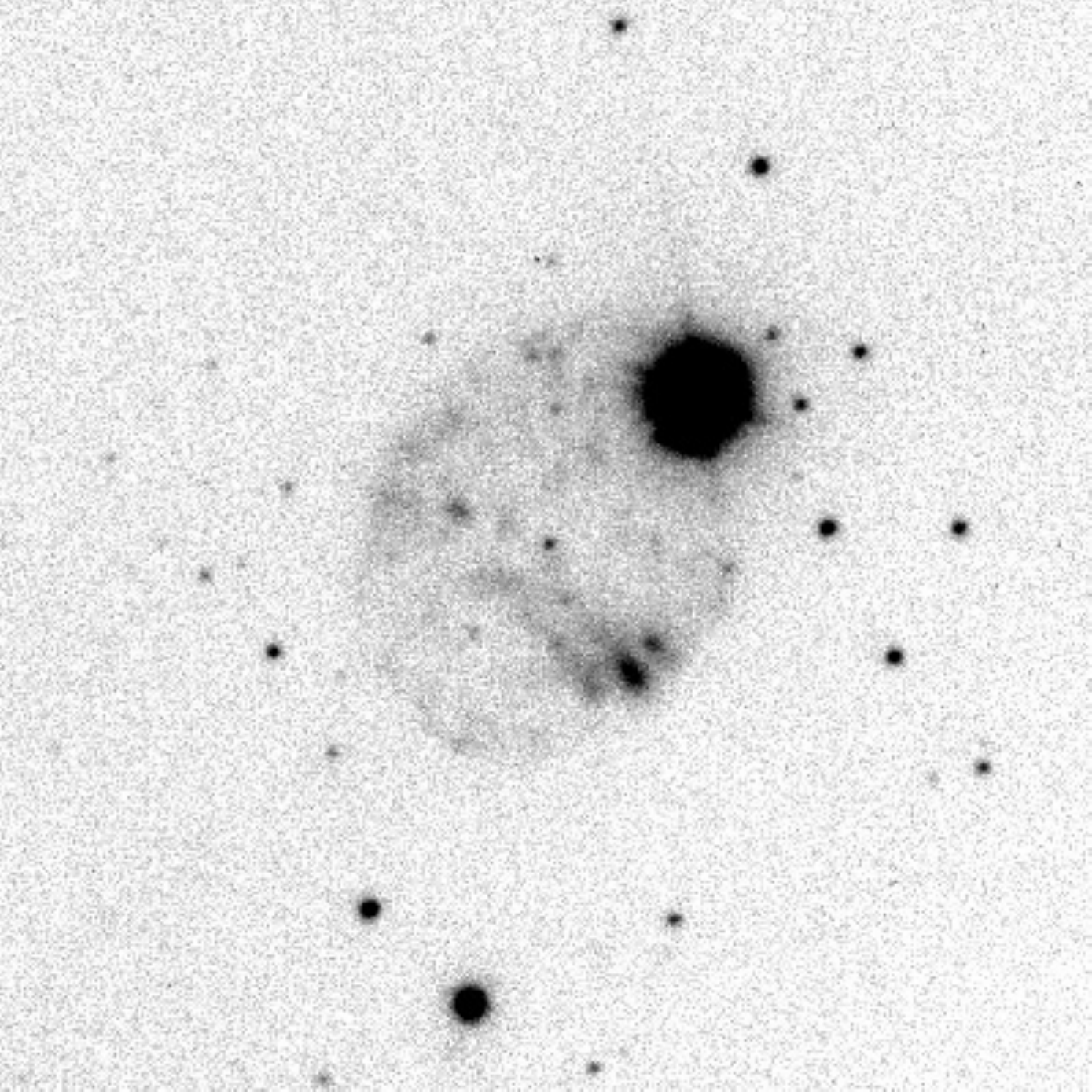}
\includegraphics[height=1.7in]{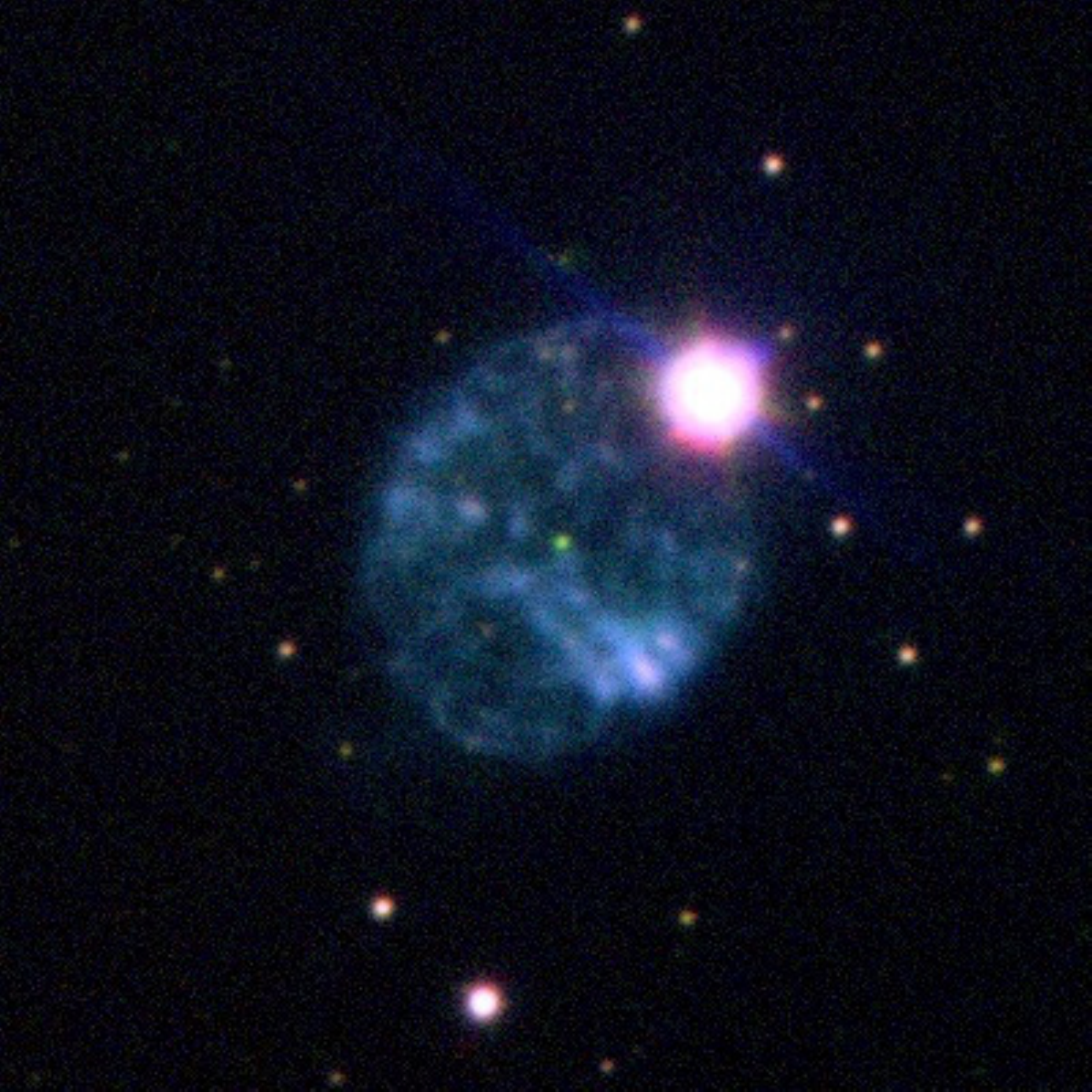}
\vskip .1in 
\includegraphics[height=1.7in]{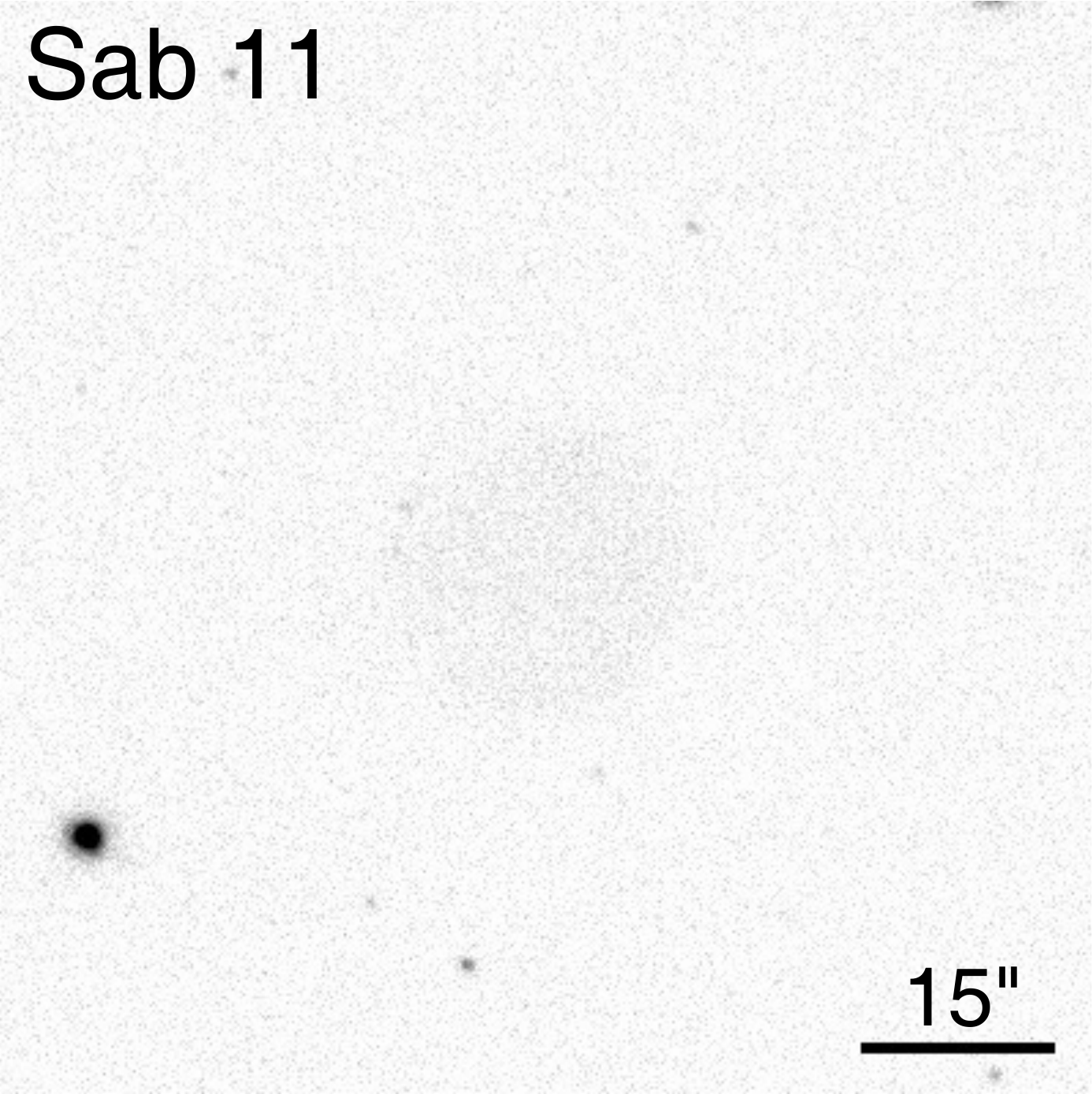} 
\includegraphics[height=1.7in]{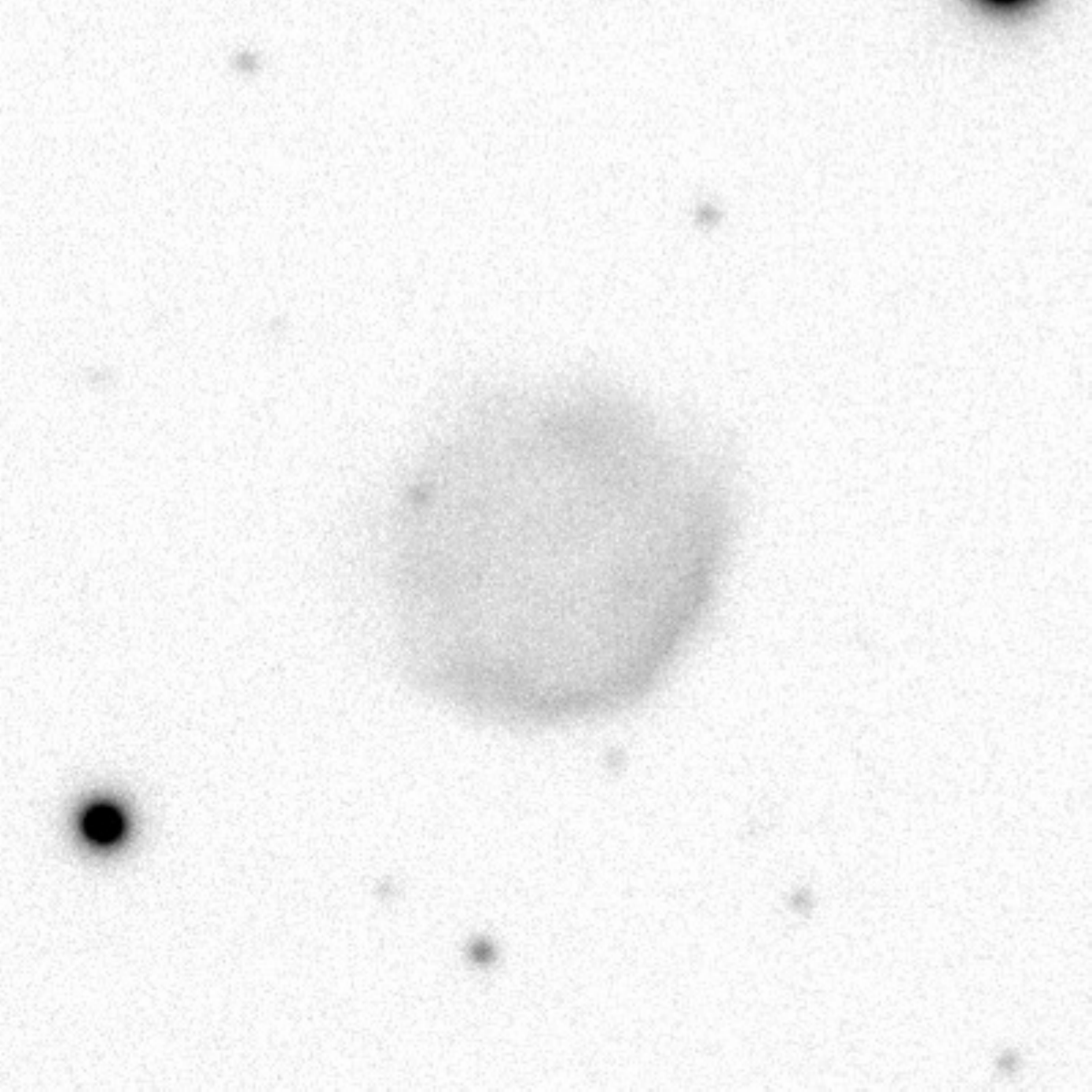}
\includegraphics[height=1.7in]{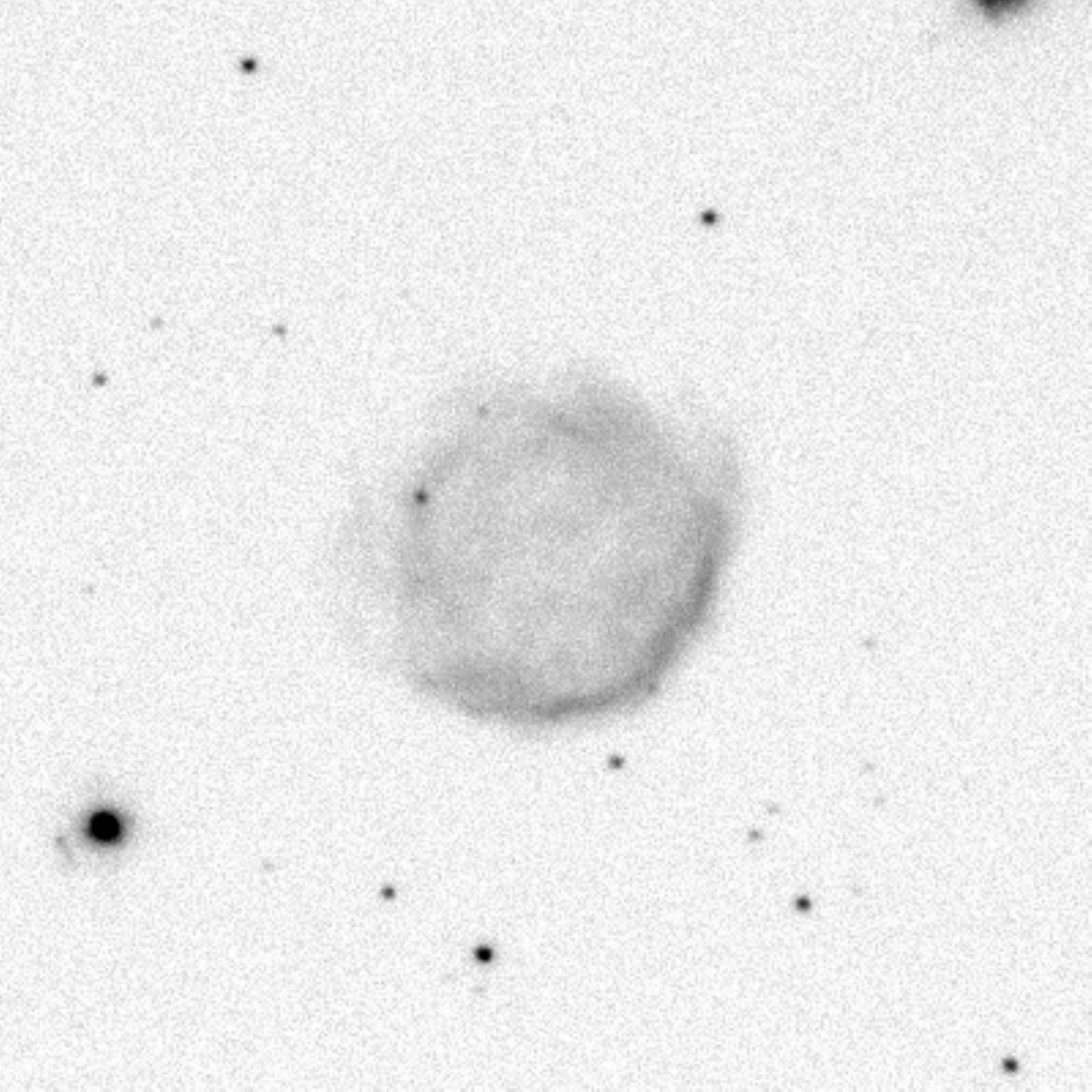}
\includegraphics[height=1.7in]{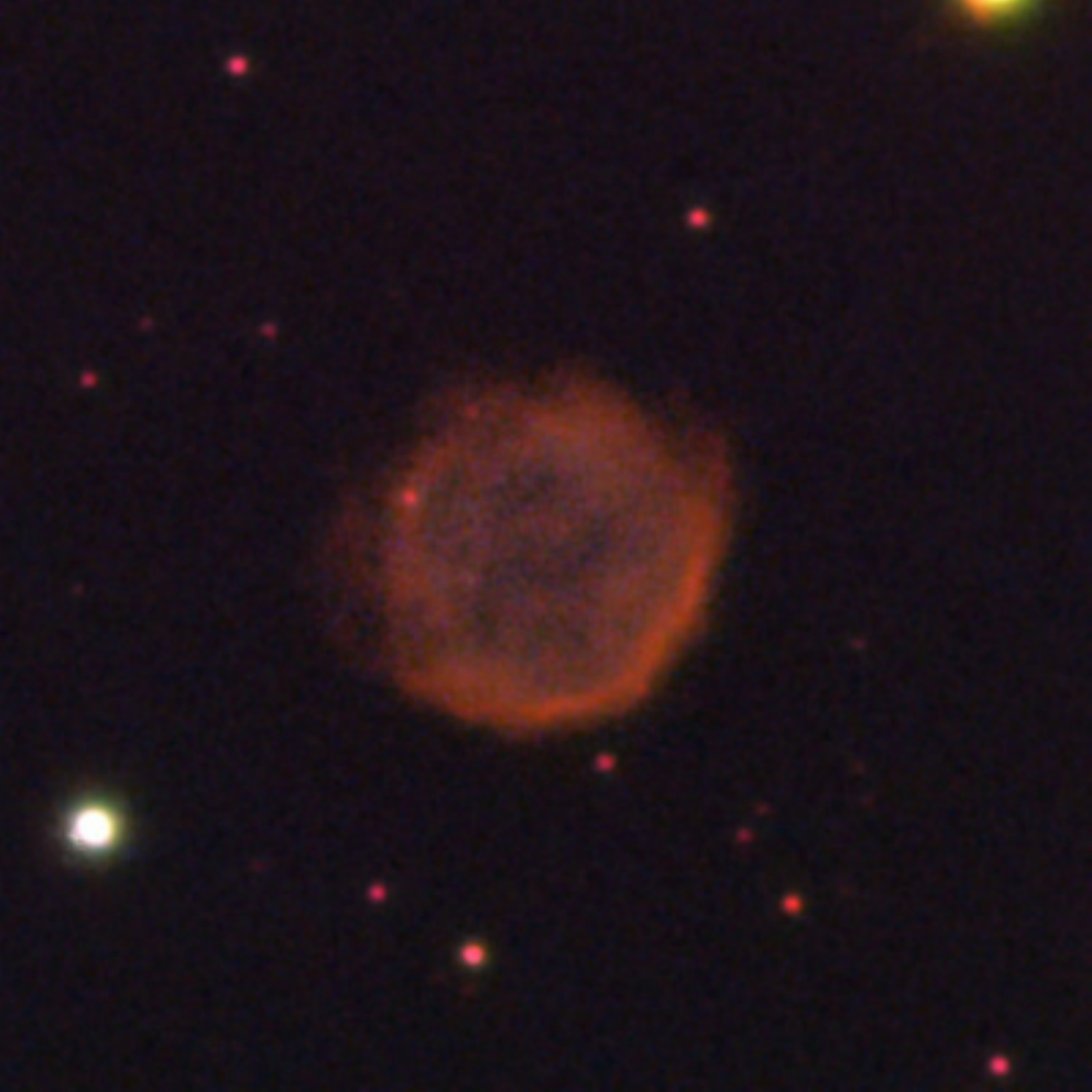}
\caption{NOT and HELMOS [O~{\sc iii}] $\lambda$5007 (left), H$\alpha$ $\lambda$6563 (center-left), and [N~{\sc ii}] $\lambda$6583 (center-right) images and \textit{RGB}-composite pictures (right) with R=[N~{\sc ii}], G=H$\alpha$ and B=[O~{\sc iii}]. 
North is up, East is left. 
} 
\label{1.img} 
\end{figure*}

\begin{figure*} 
\centering 
\includegraphics[height=1.7in]{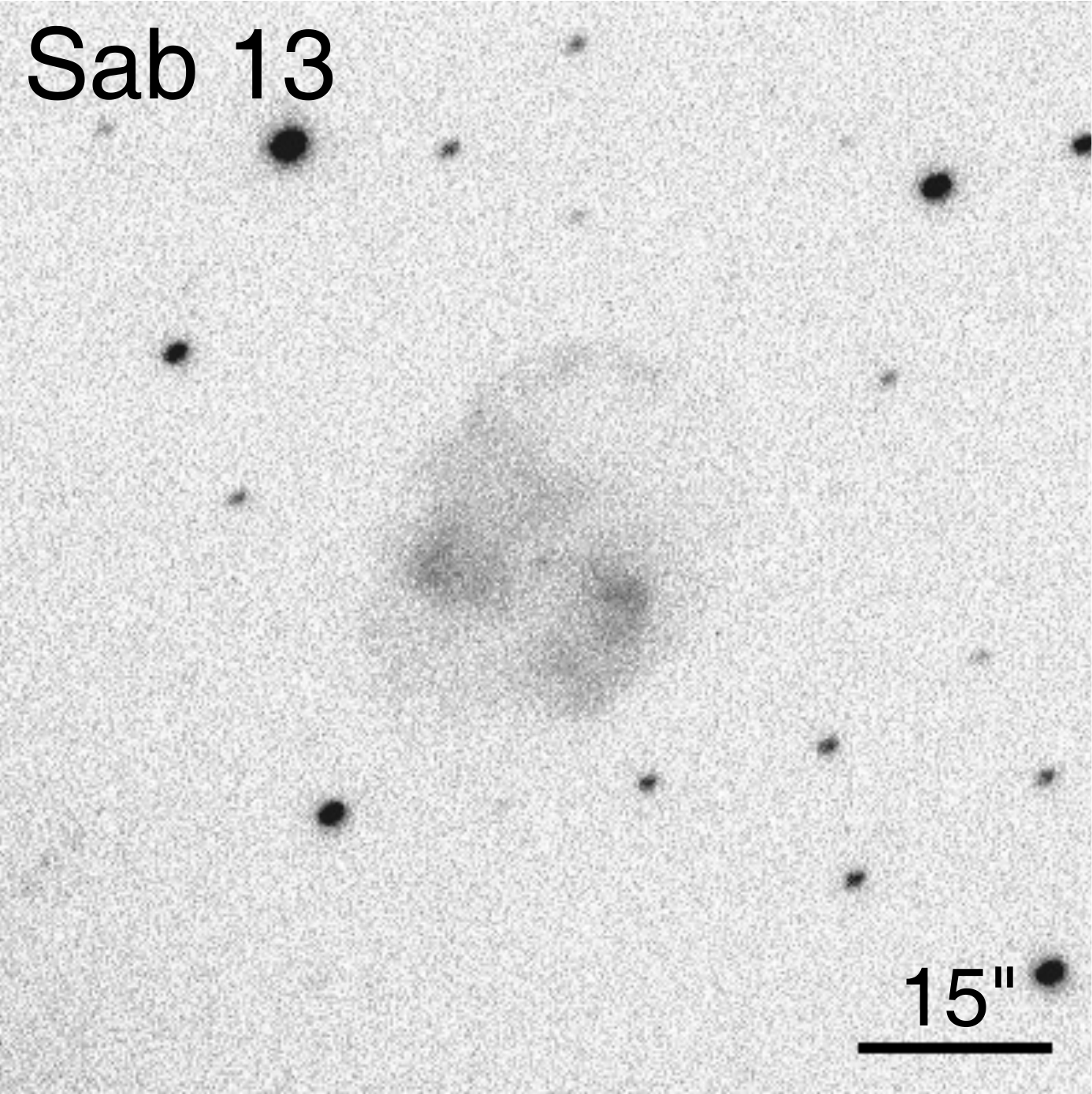} 
\includegraphics[height=1.7in]{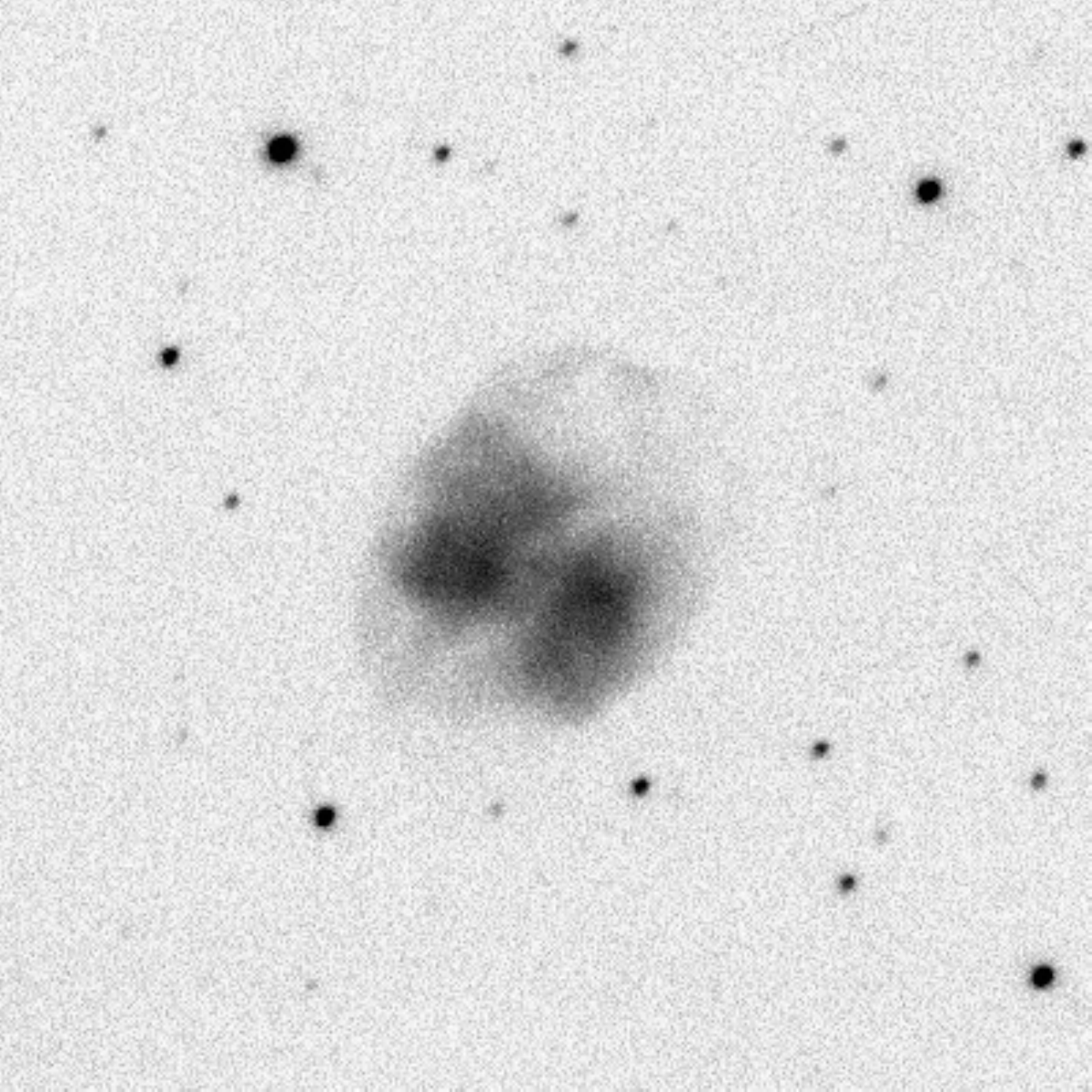}
\includegraphics[height=1.7in]{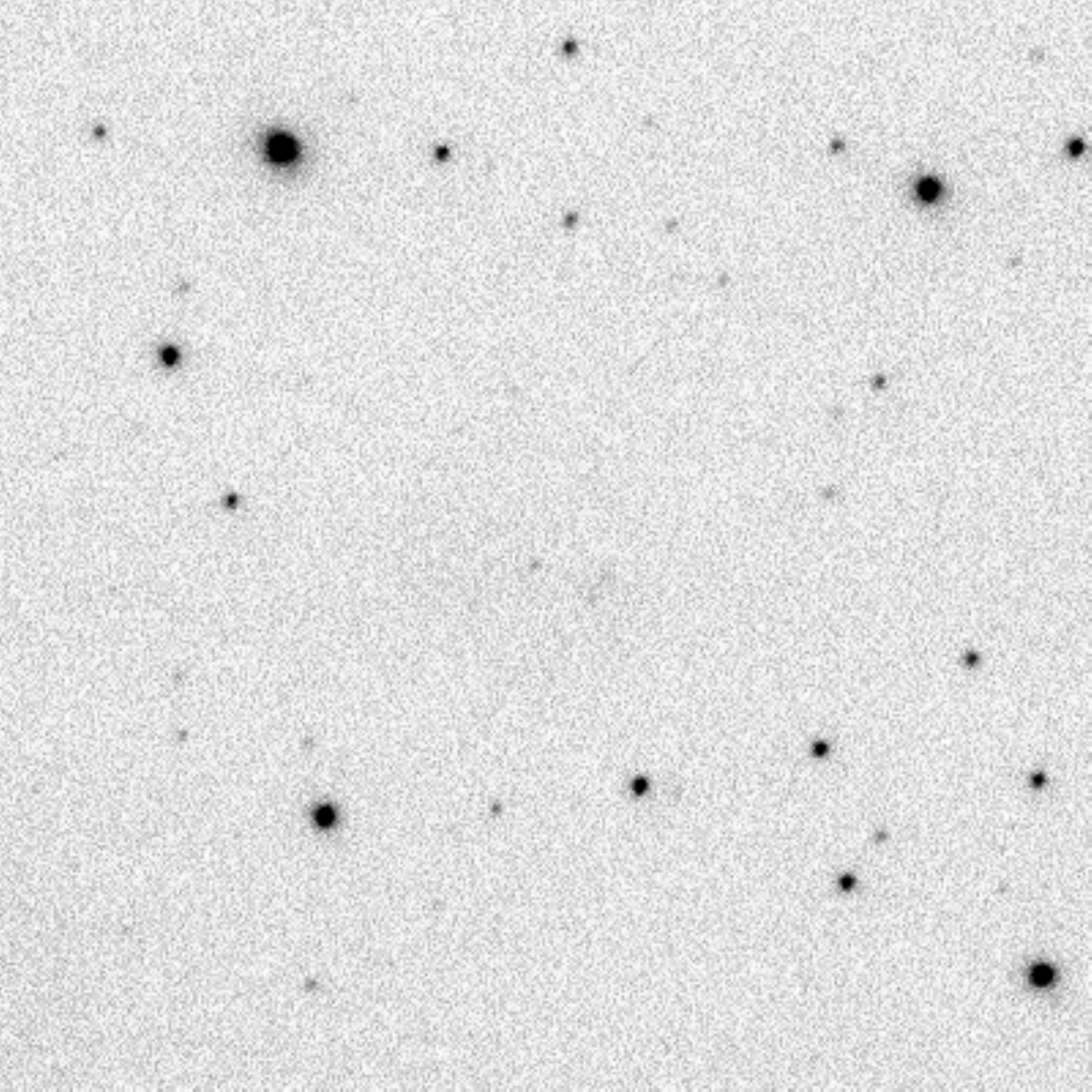}
\includegraphics[height=1.7in]{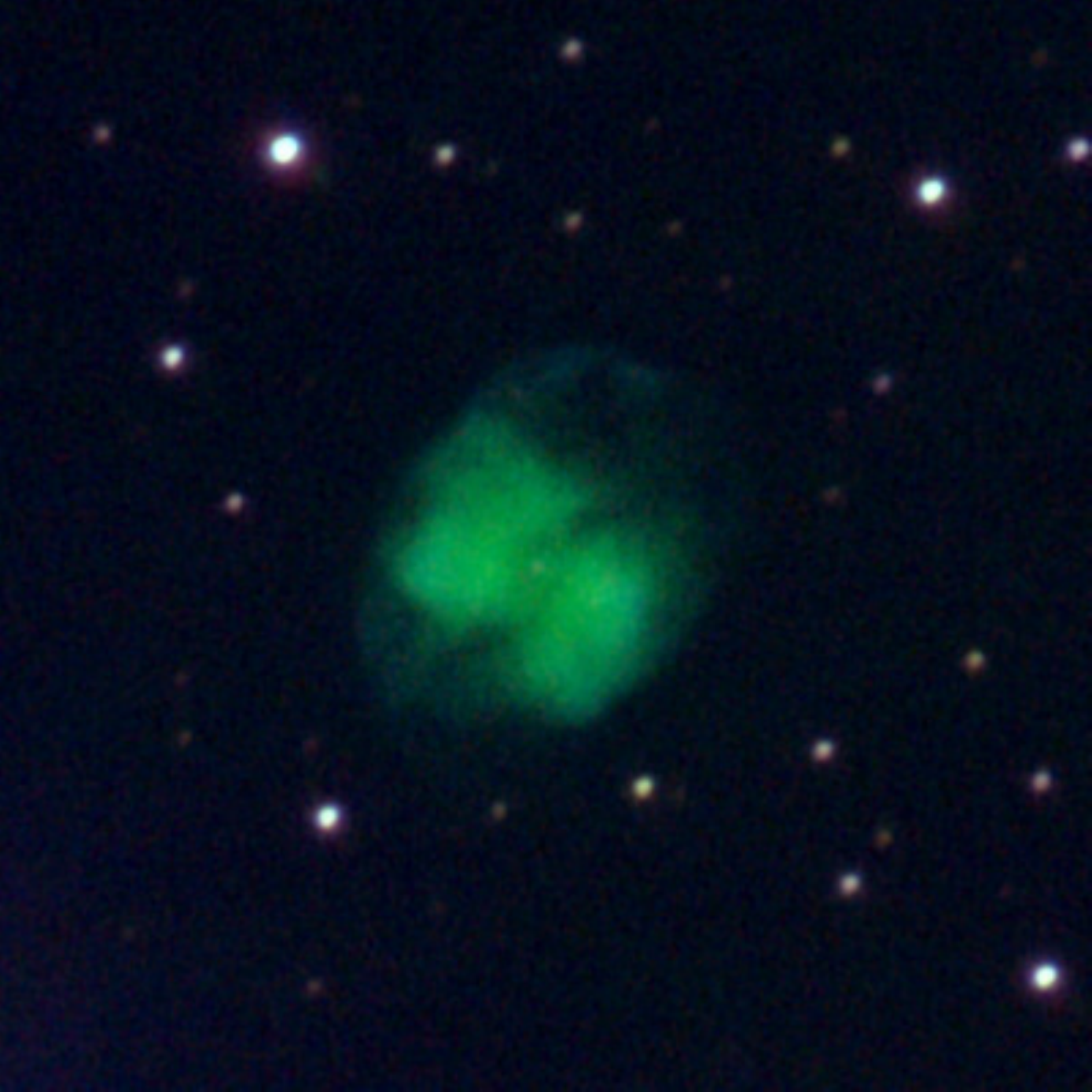}
\vskip .1in 
\includegraphics[height=1.7in]{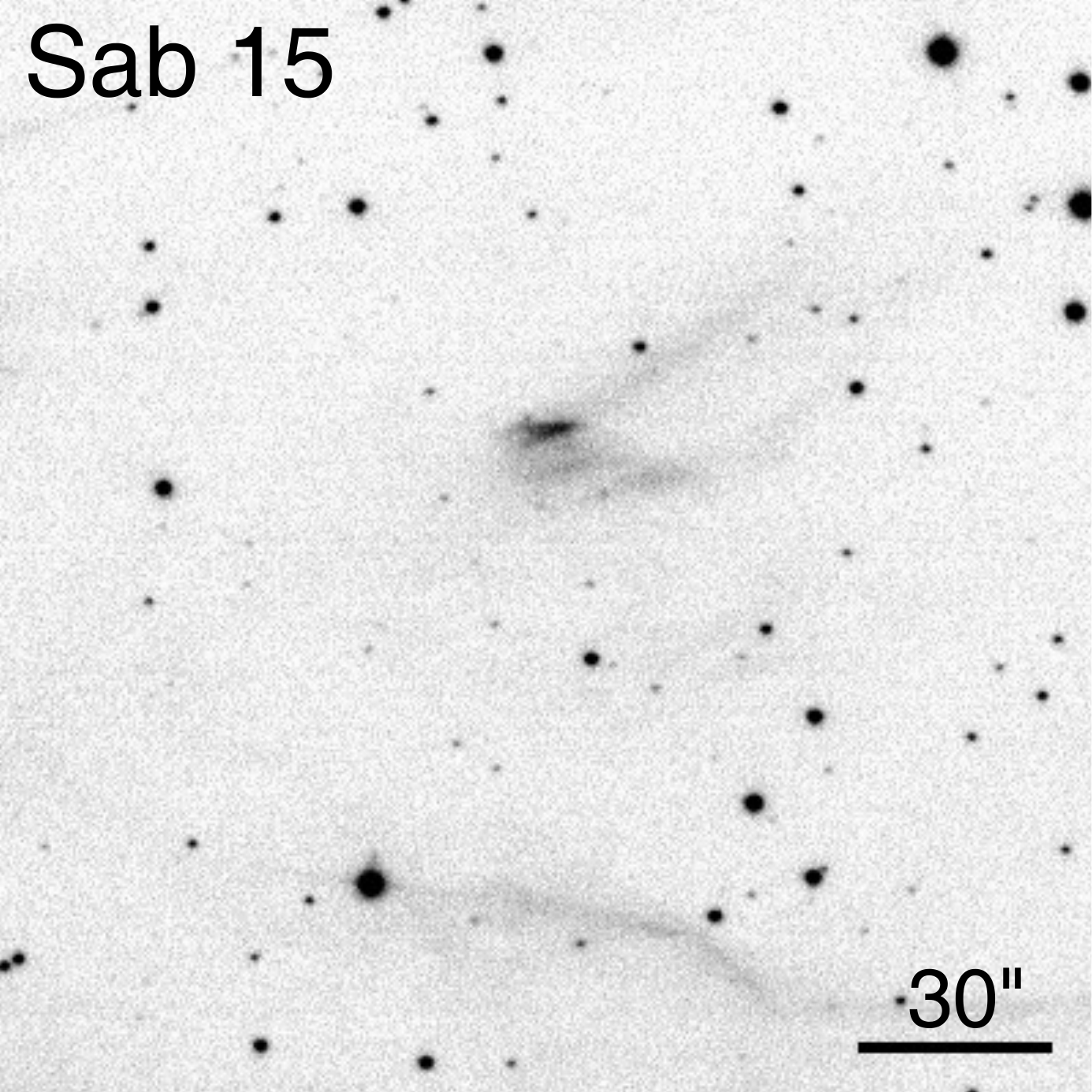} 
\includegraphics[height=1.7in]{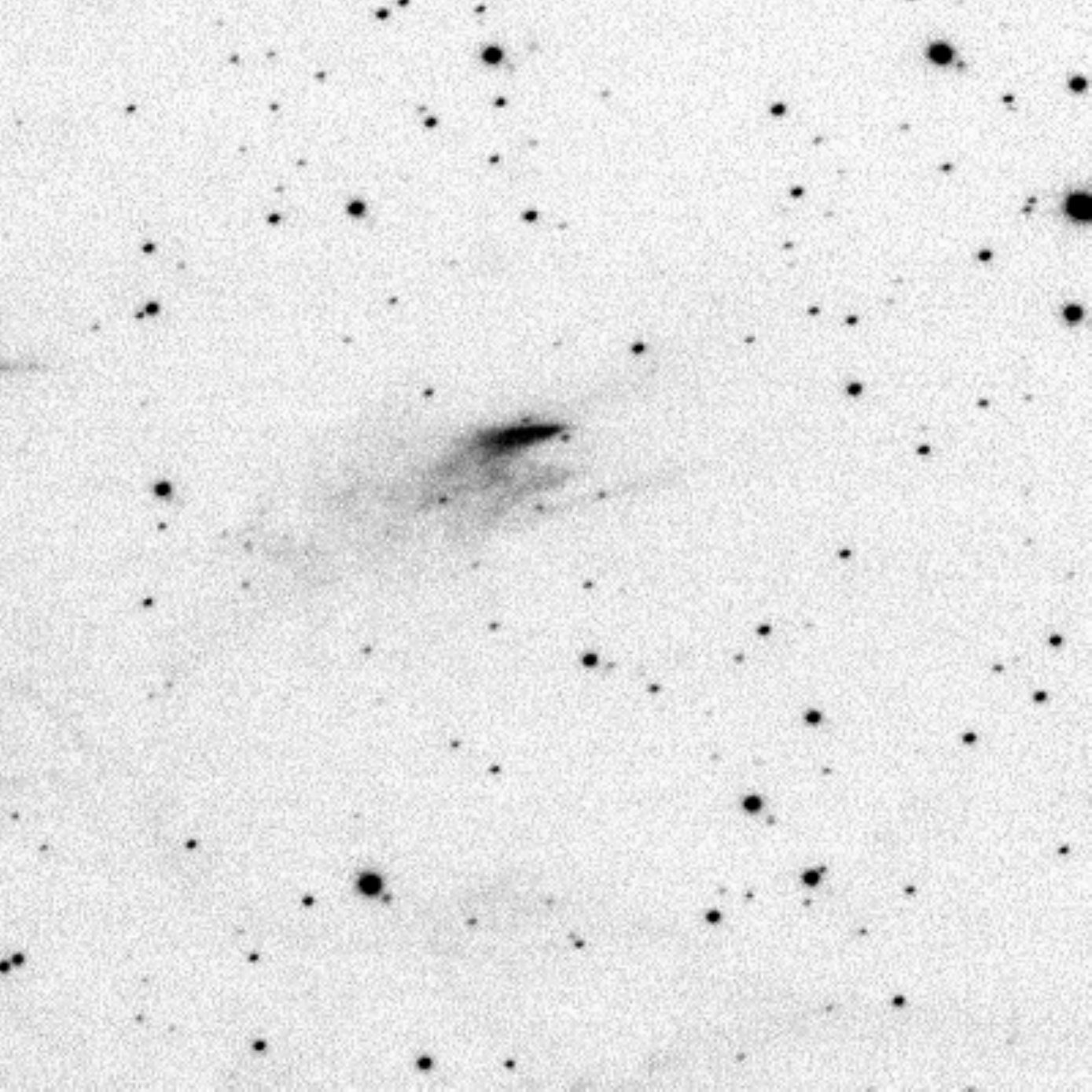}
\includegraphics[height=1.7in]{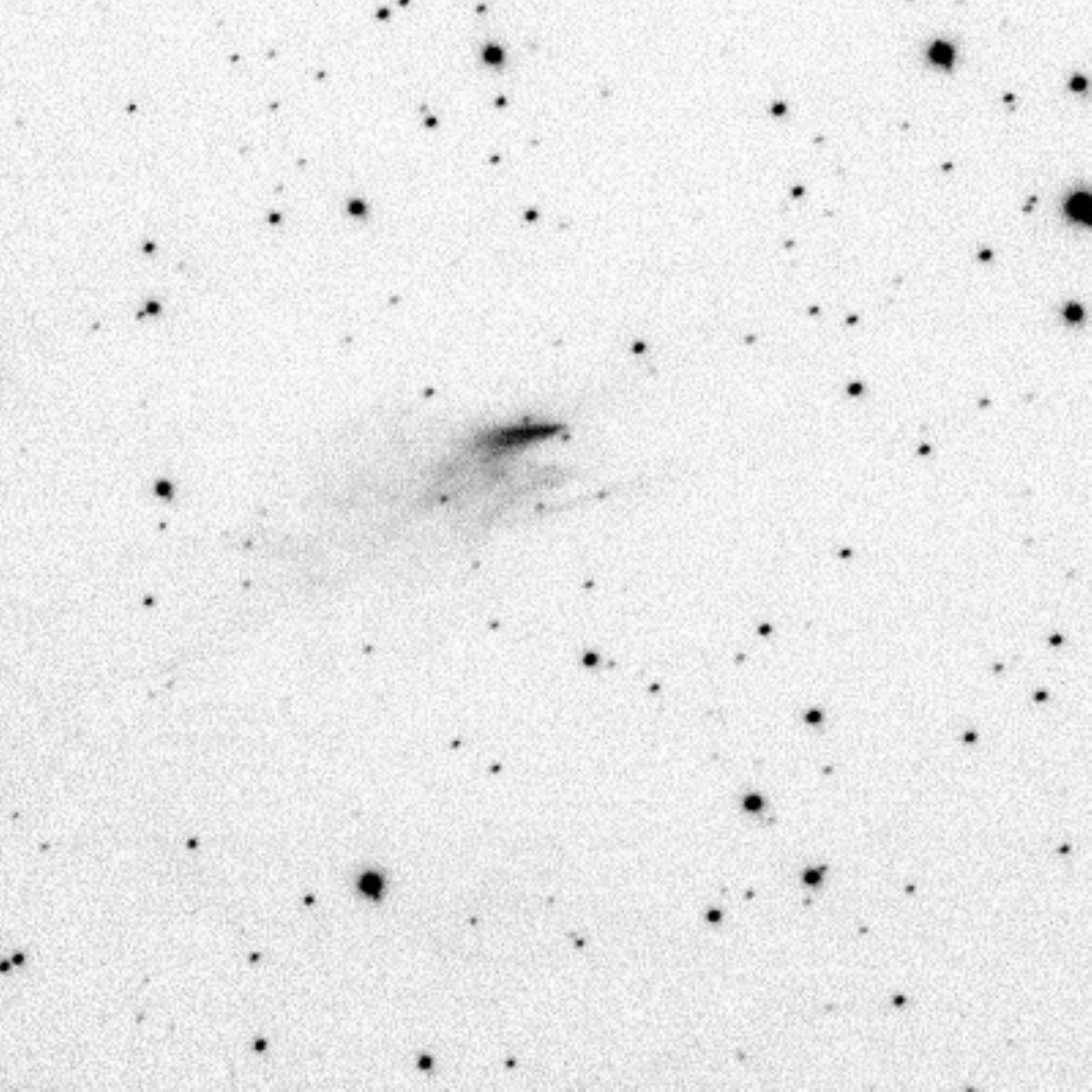}
\includegraphics[height=1.7in]{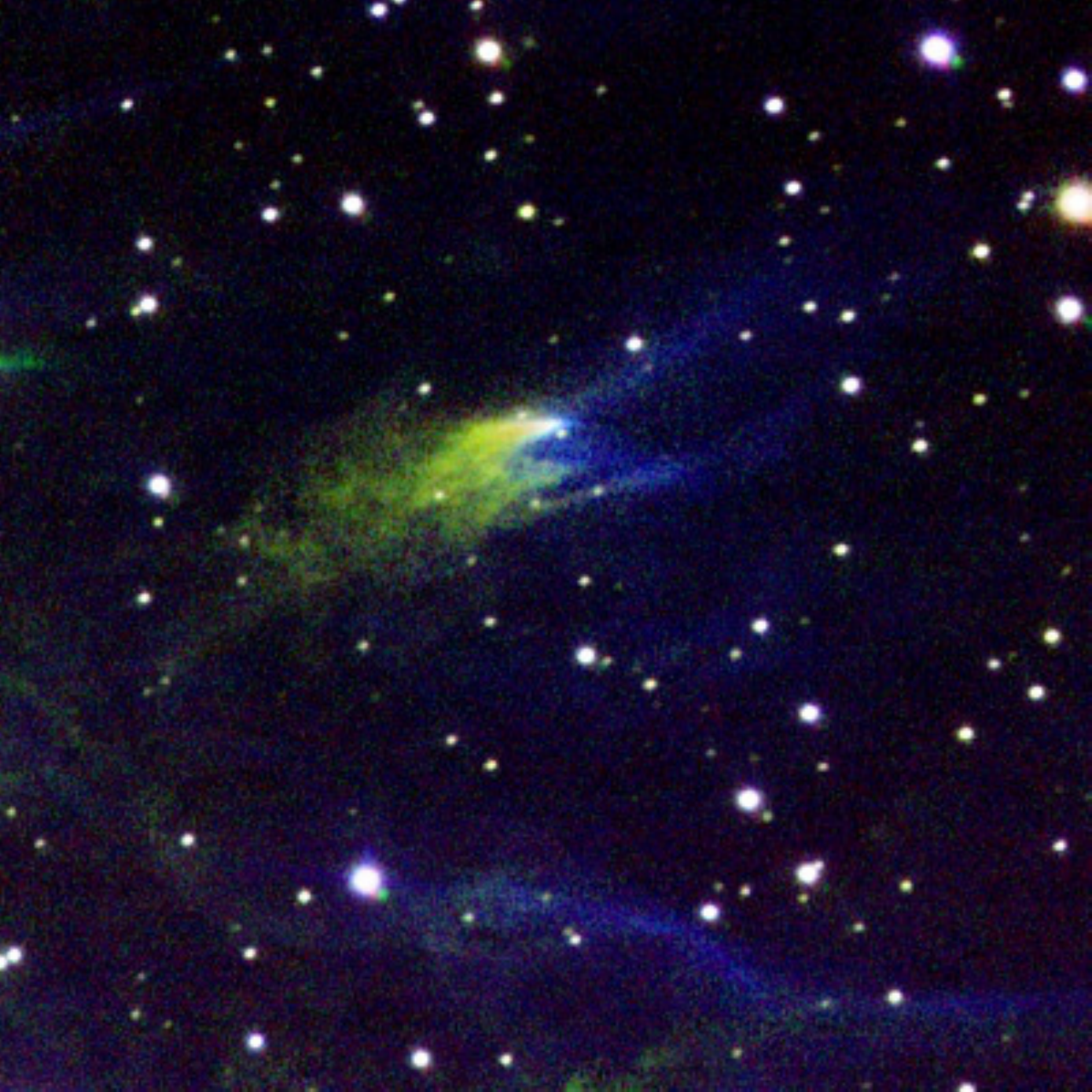}
\vskip .1in 
\includegraphics[height=1.7in]{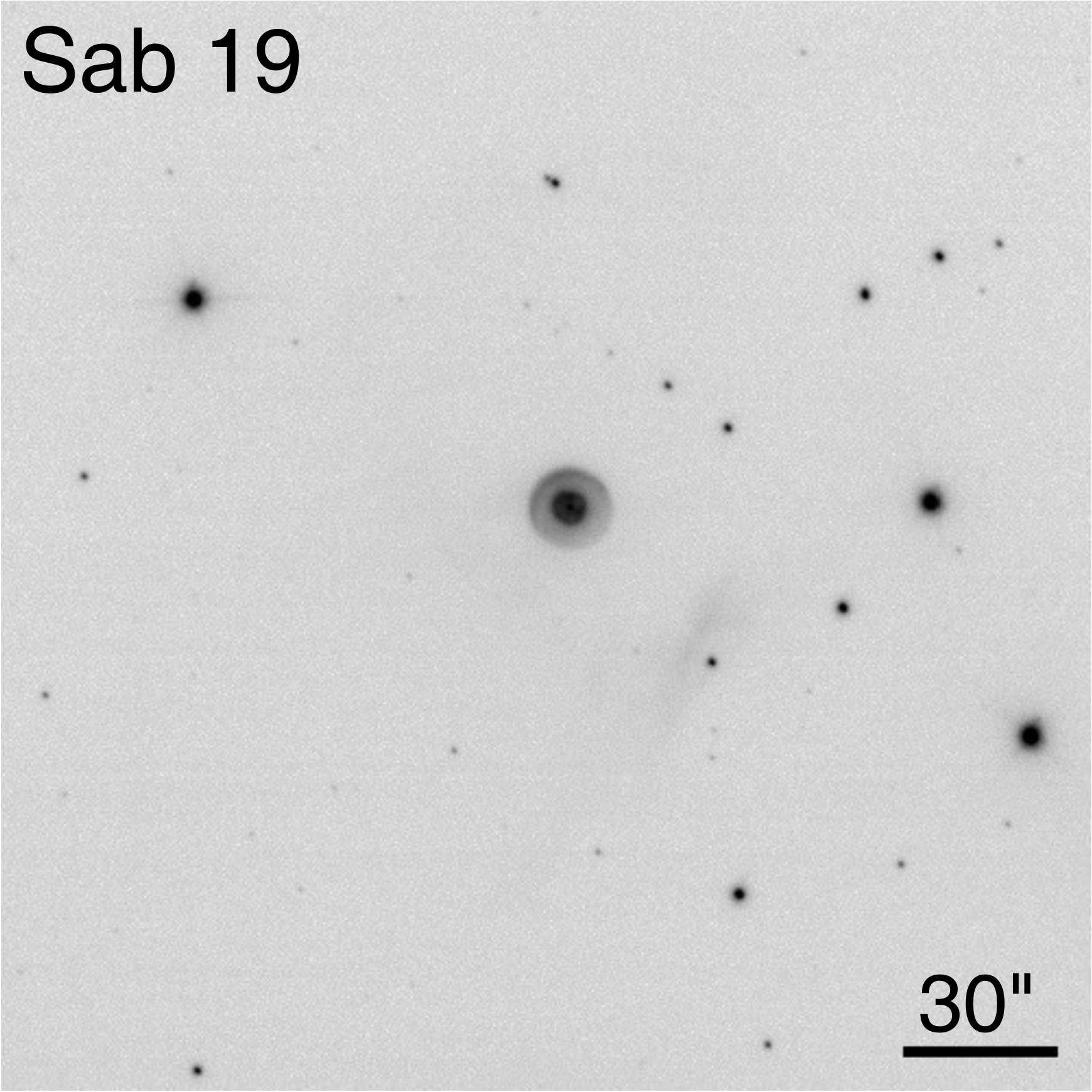} 
\includegraphics[height=1.7in]{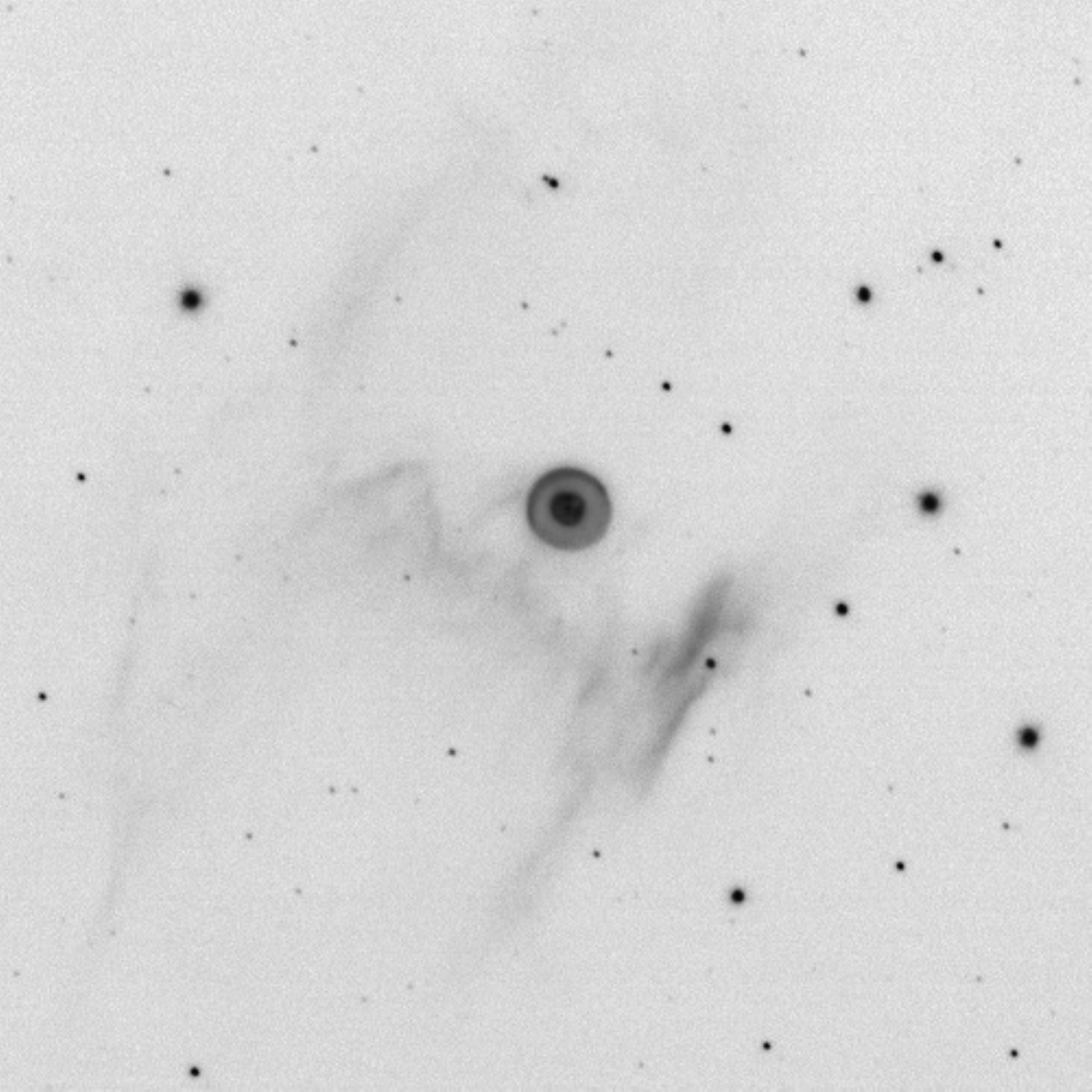}
\includegraphics[height=1.7in]{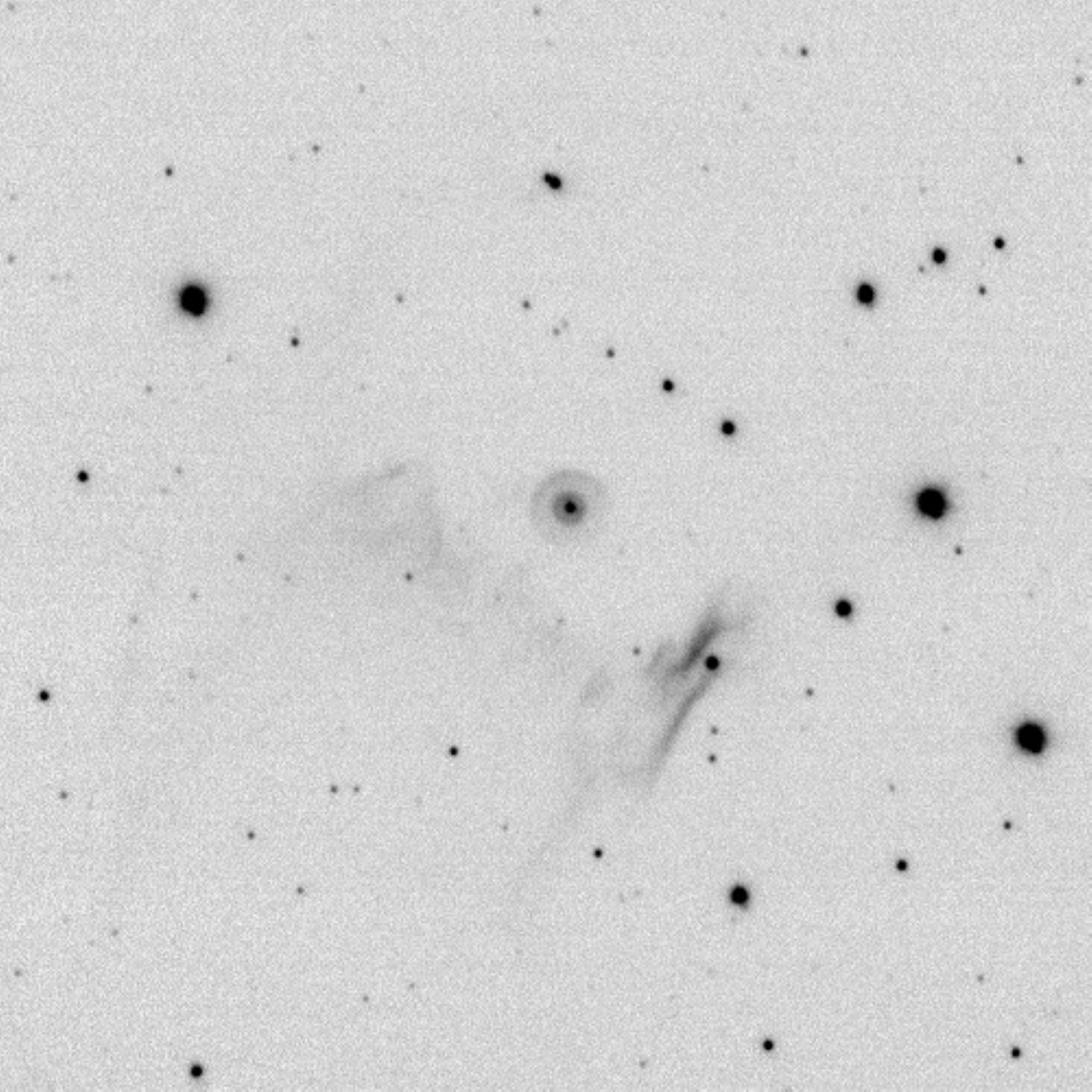}
\includegraphics[height=1.7in]{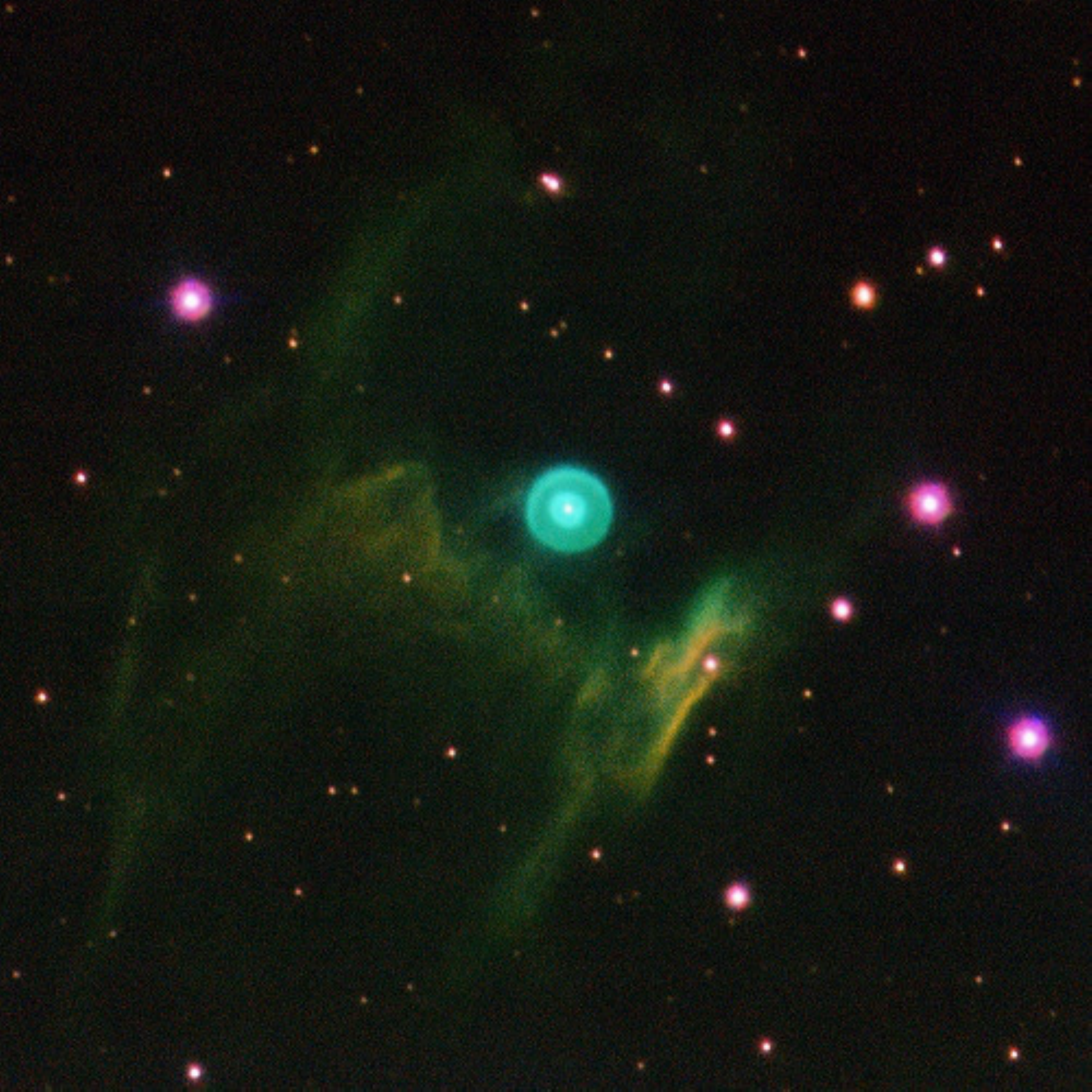}
\vskip .1in 
\includegraphics[height=1.7in]{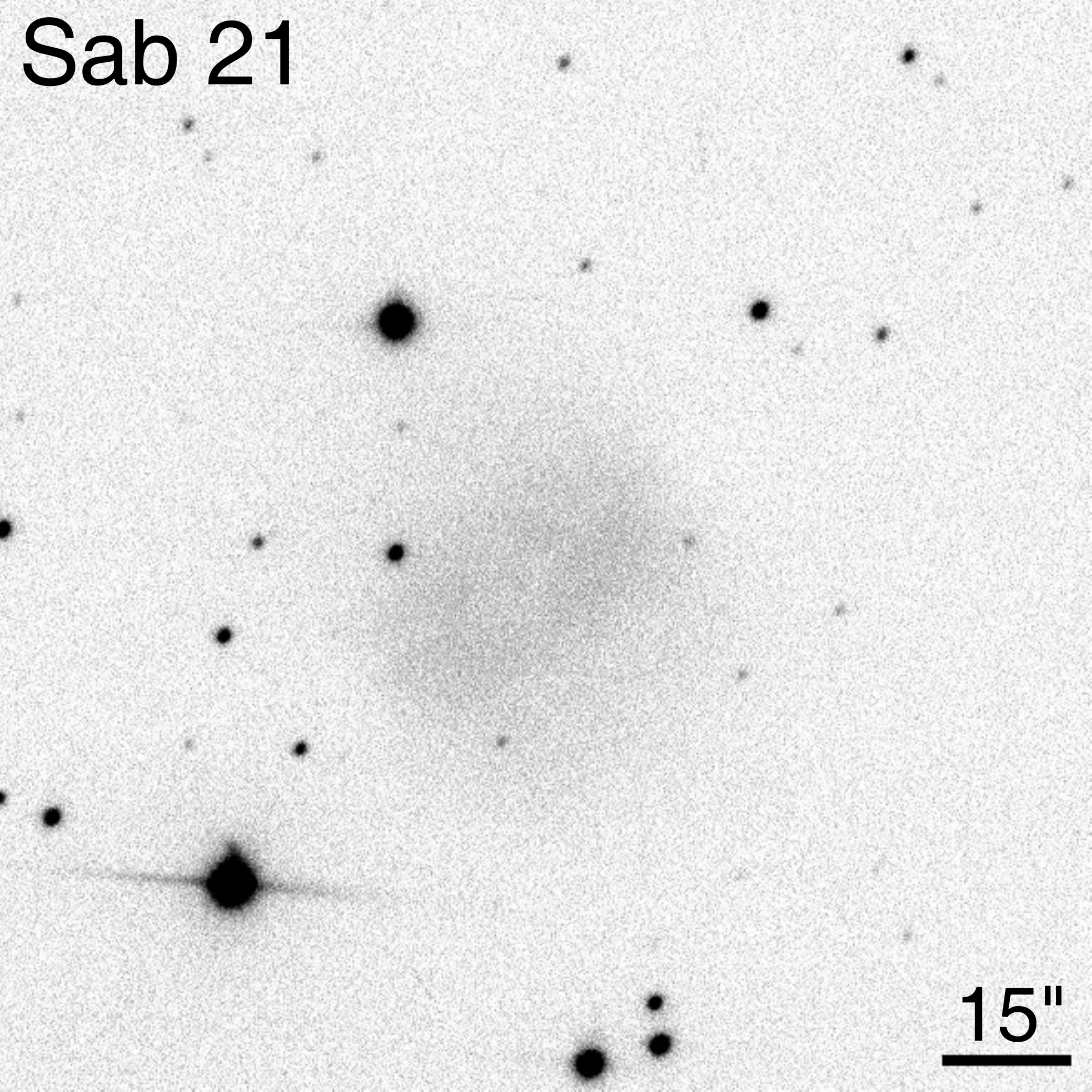} 
\includegraphics[height=1.7in]{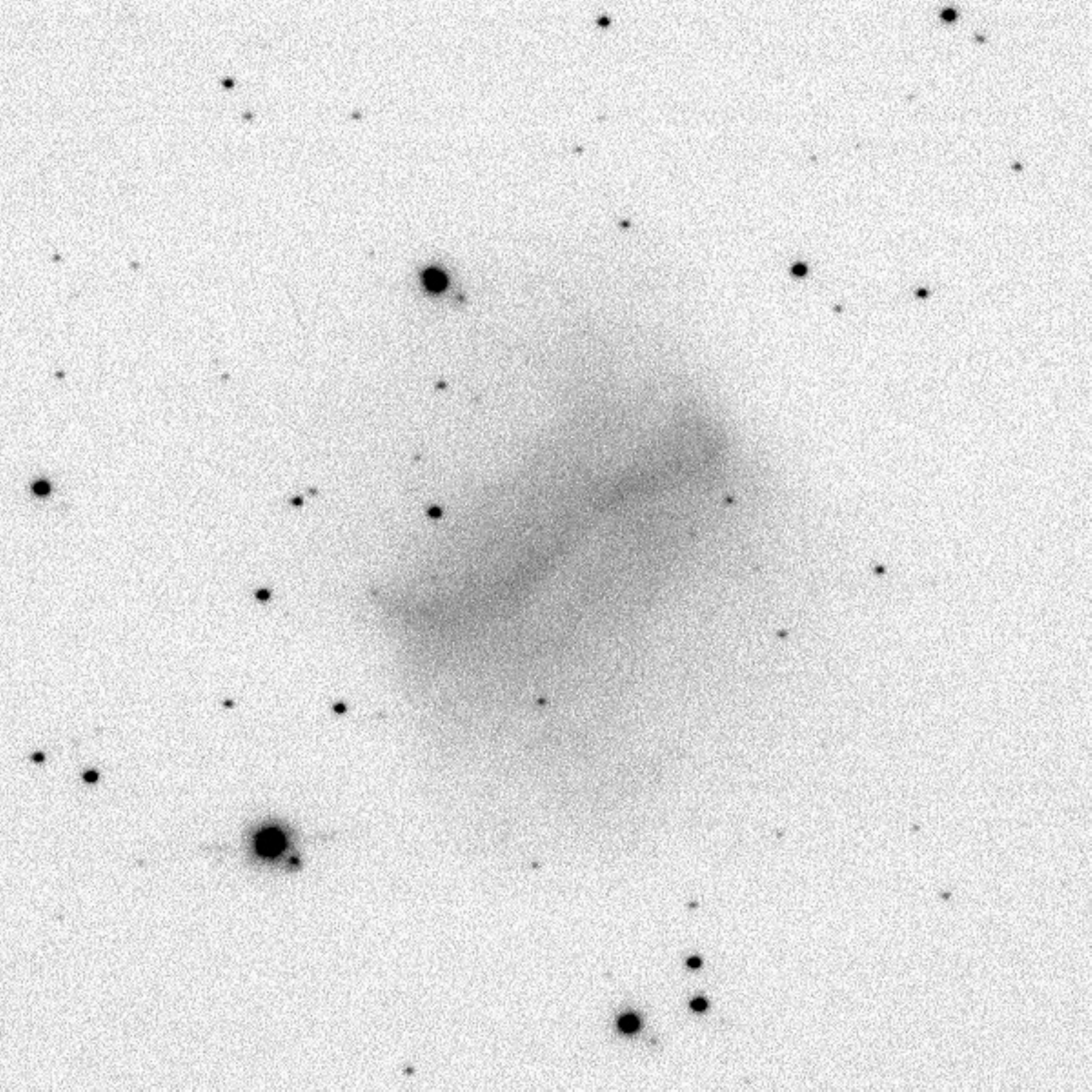}
\includegraphics[height=1.7in]{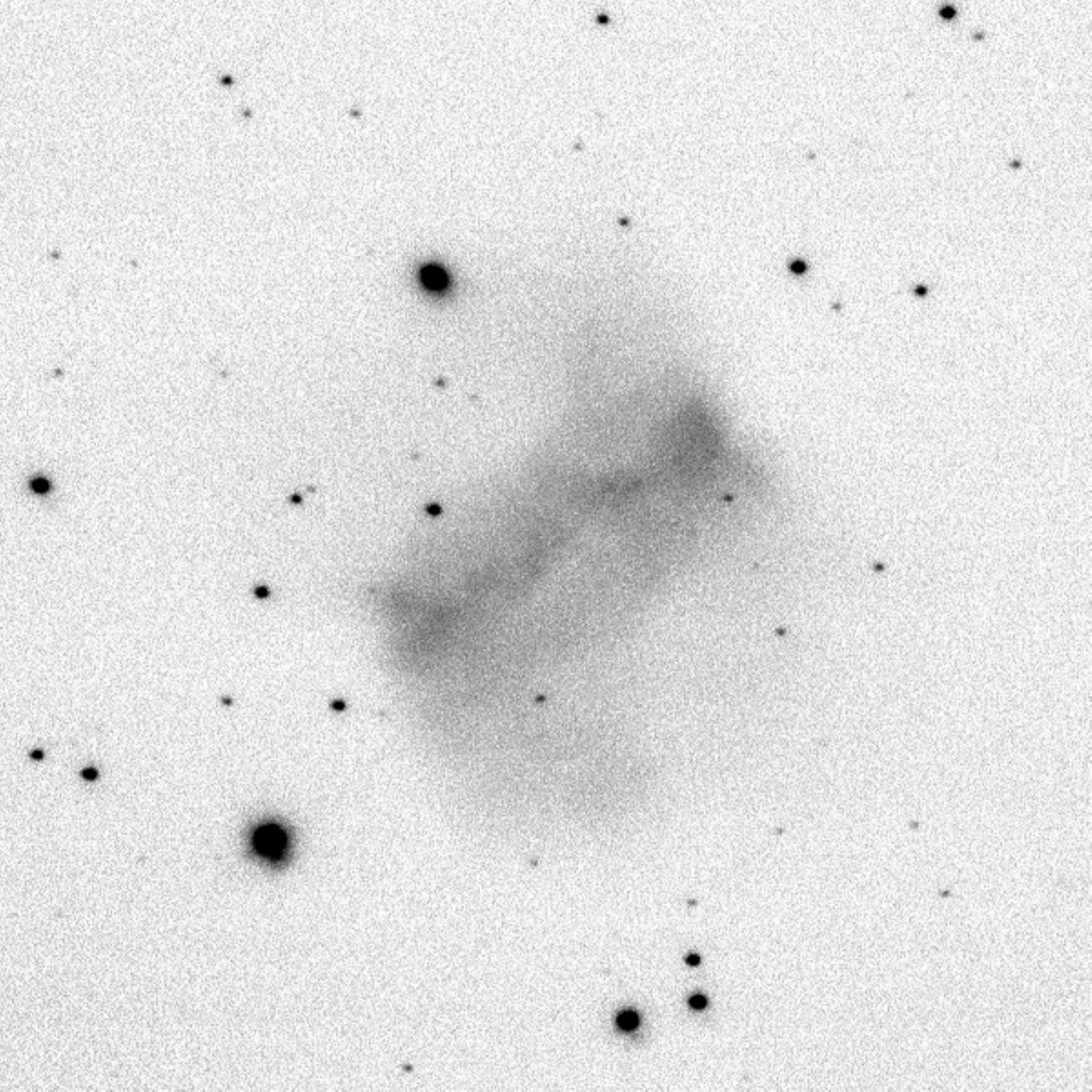}
\includegraphics[height=1.7in]{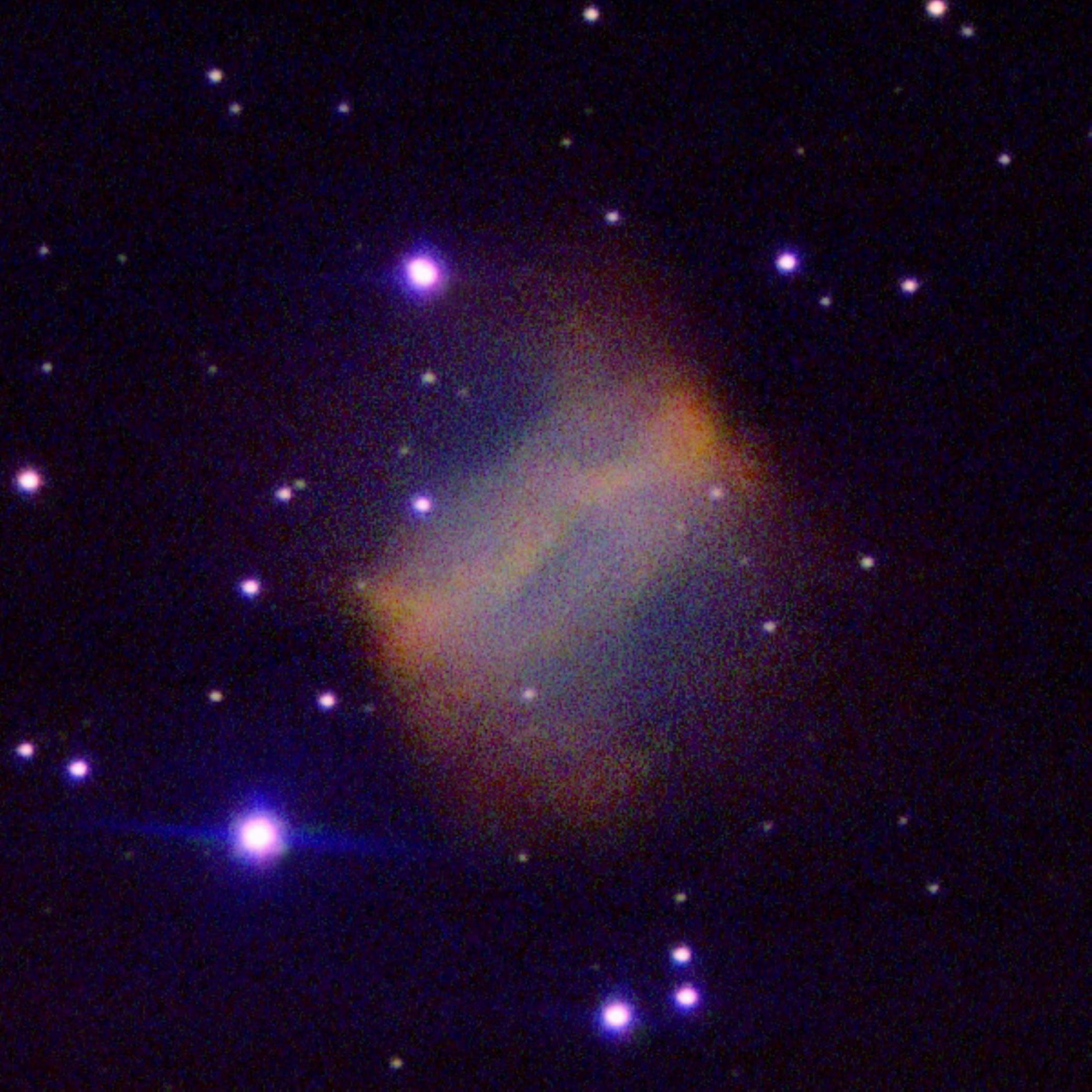}
\vskip .1in
\includegraphics[height=1.7in]{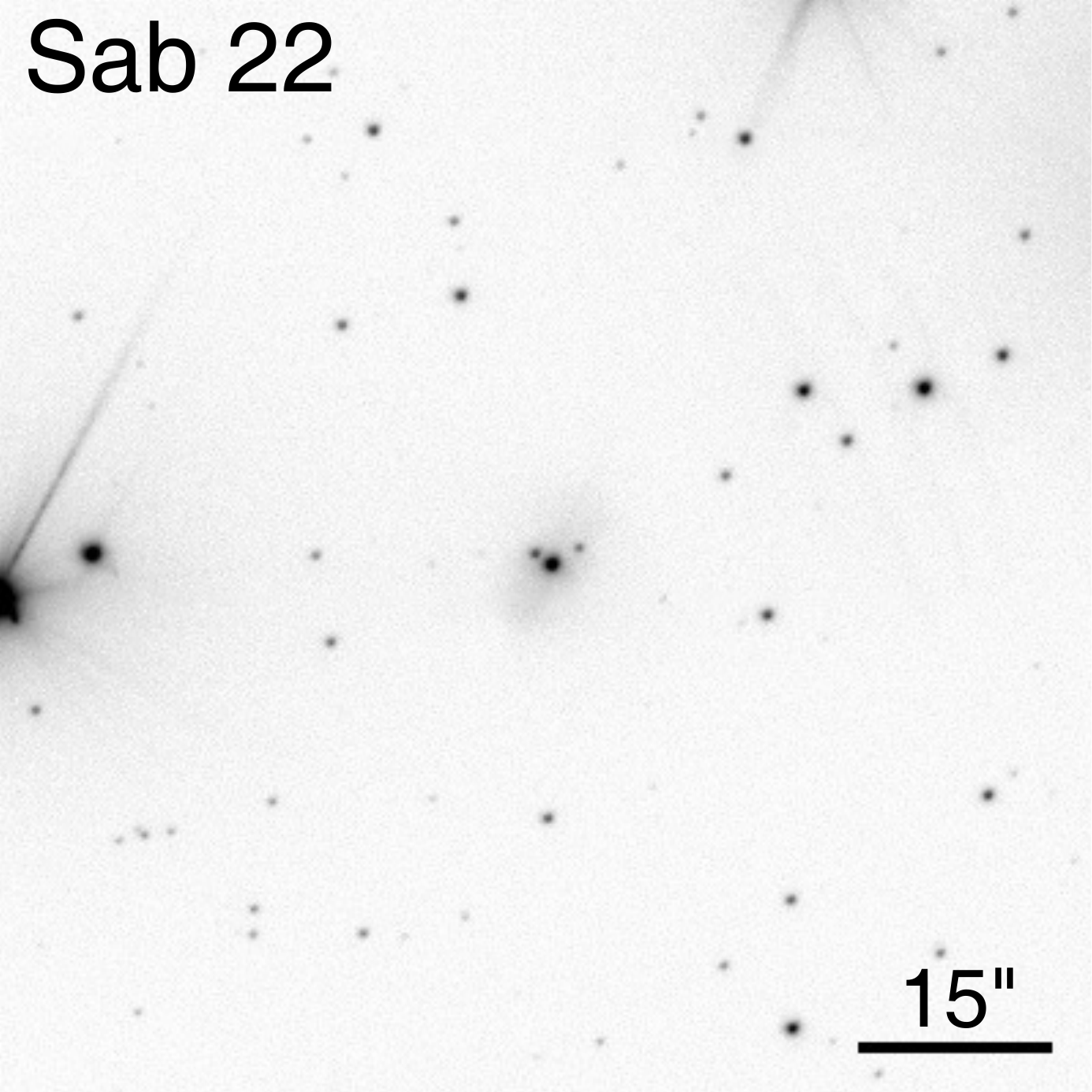} 
\includegraphics[height=1.7in]{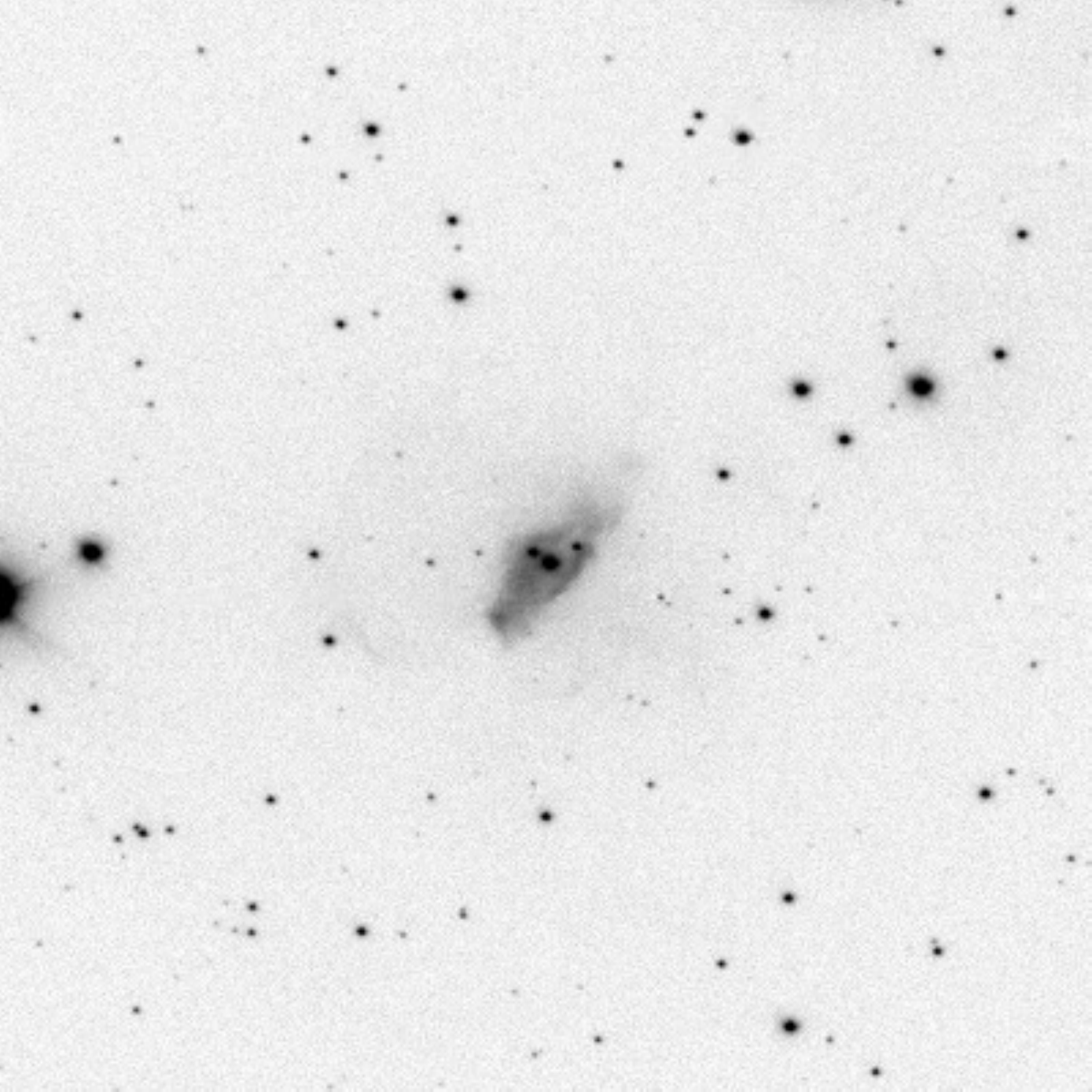}
\includegraphics[height=1.7in]{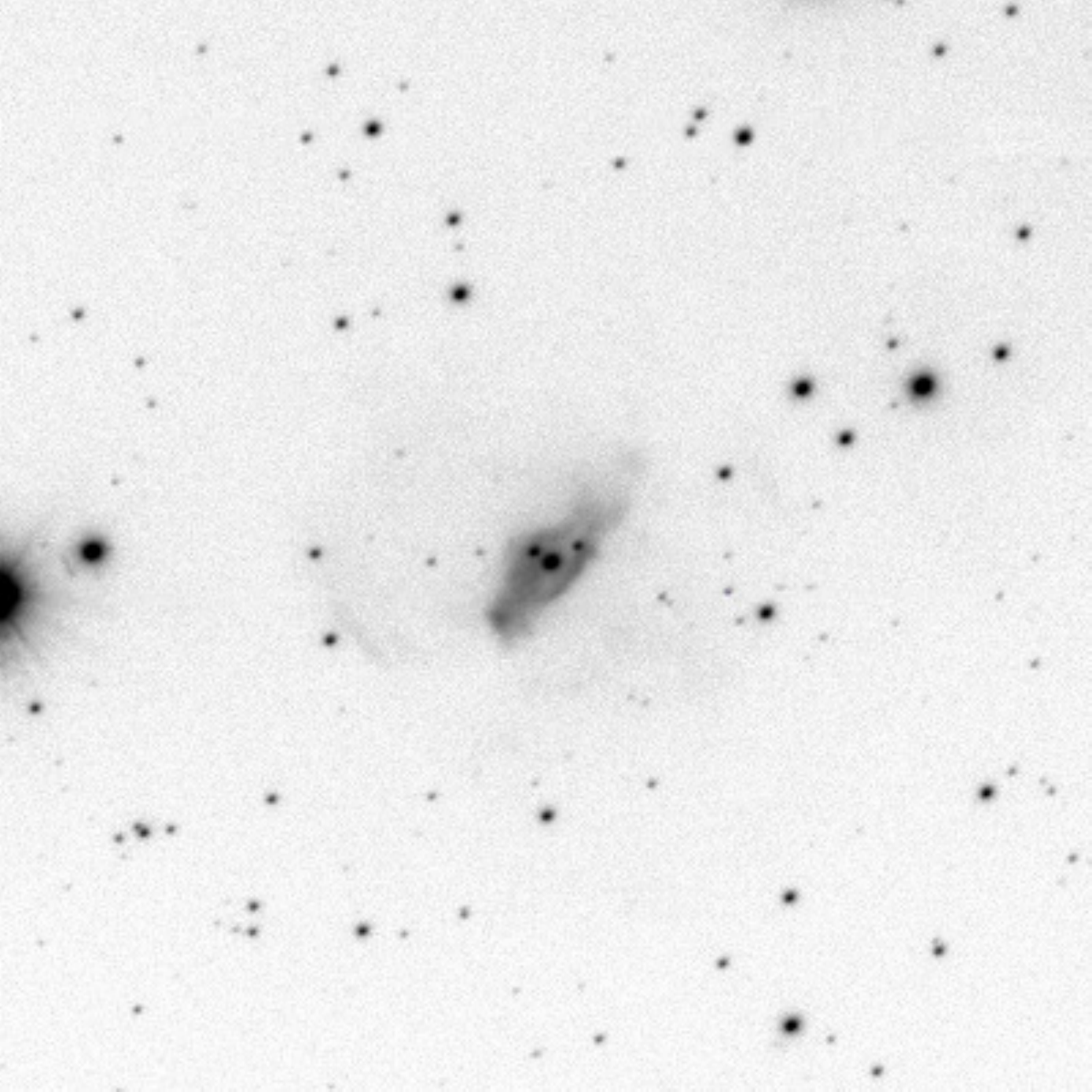}
\includegraphics[height=1.7in]{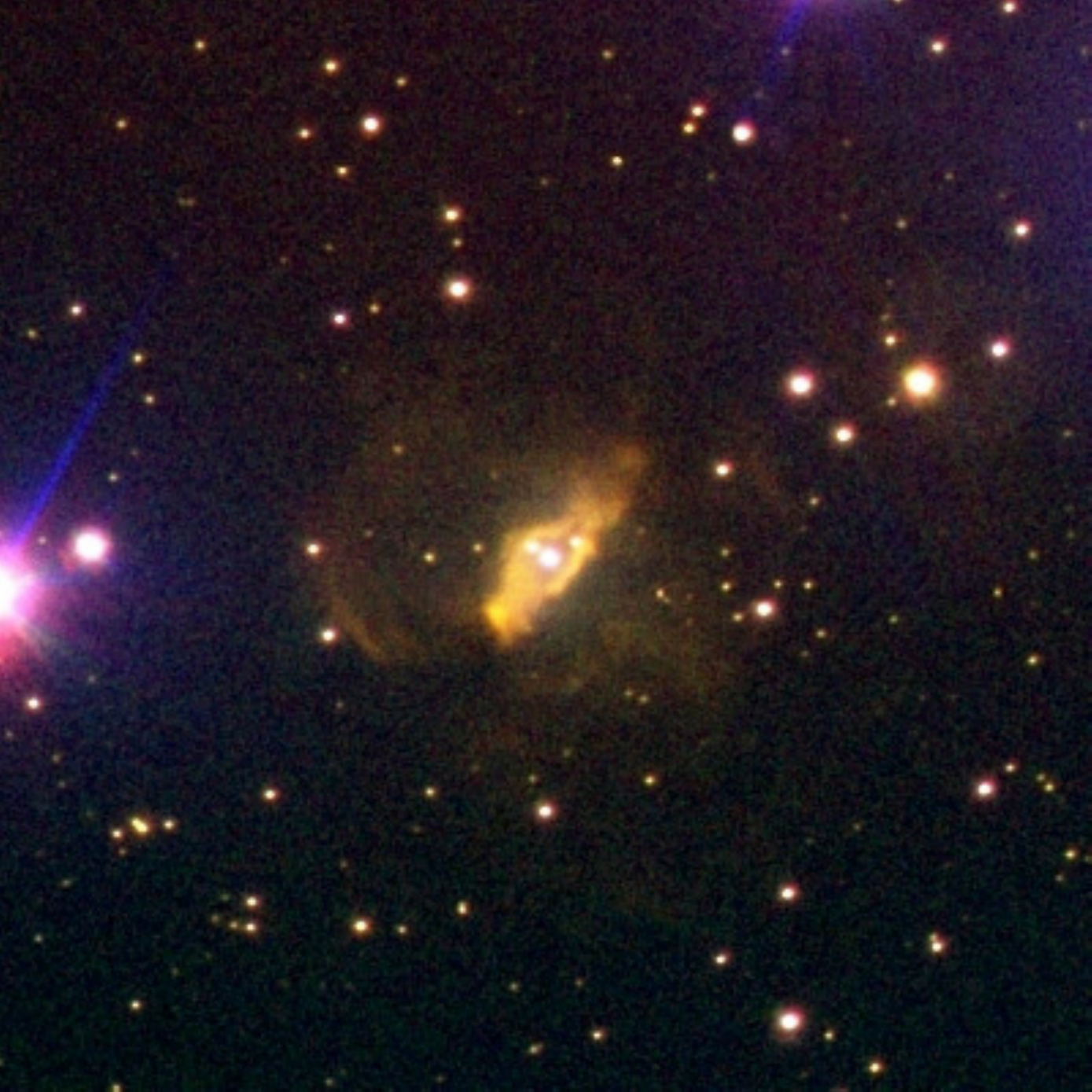}
\caption{Same as Figure~\ref{1.img}.} 
\label{2.img} 
\end{figure*}

\begin{figure*} 
\centering 
\includegraphics[height=1.7in]{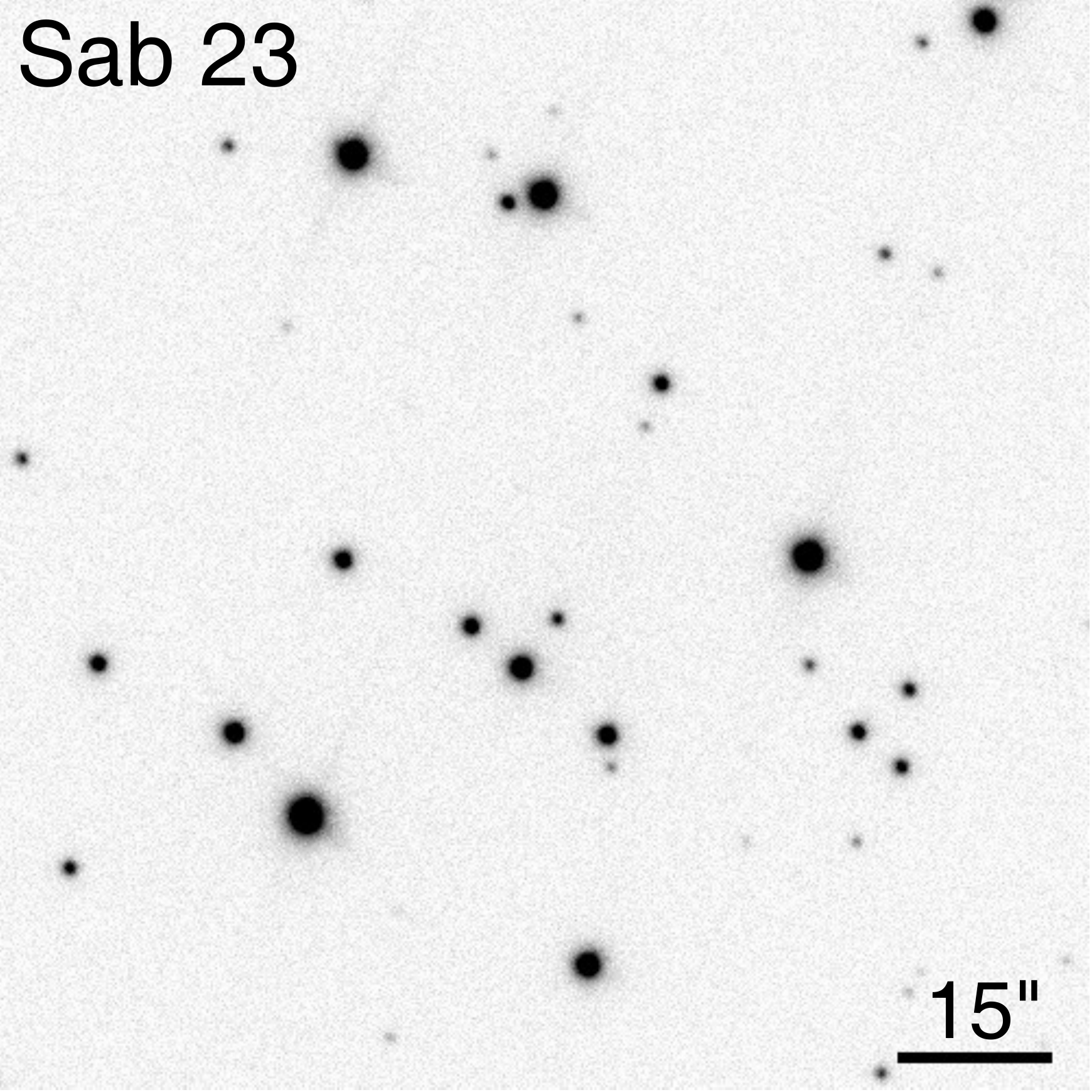} 
\includegraphics[height=1.7in]{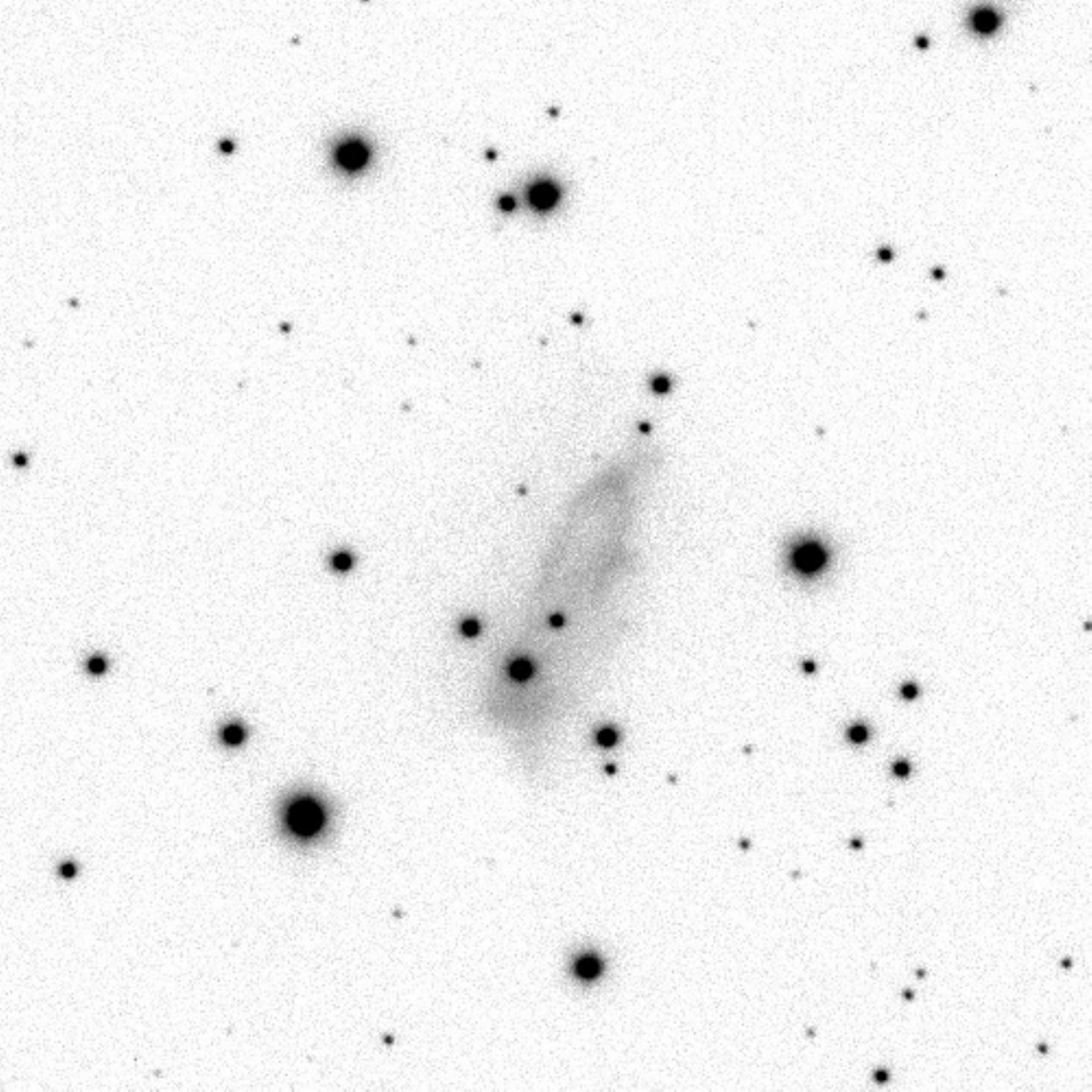}
\includegraphics[height=1.7in]{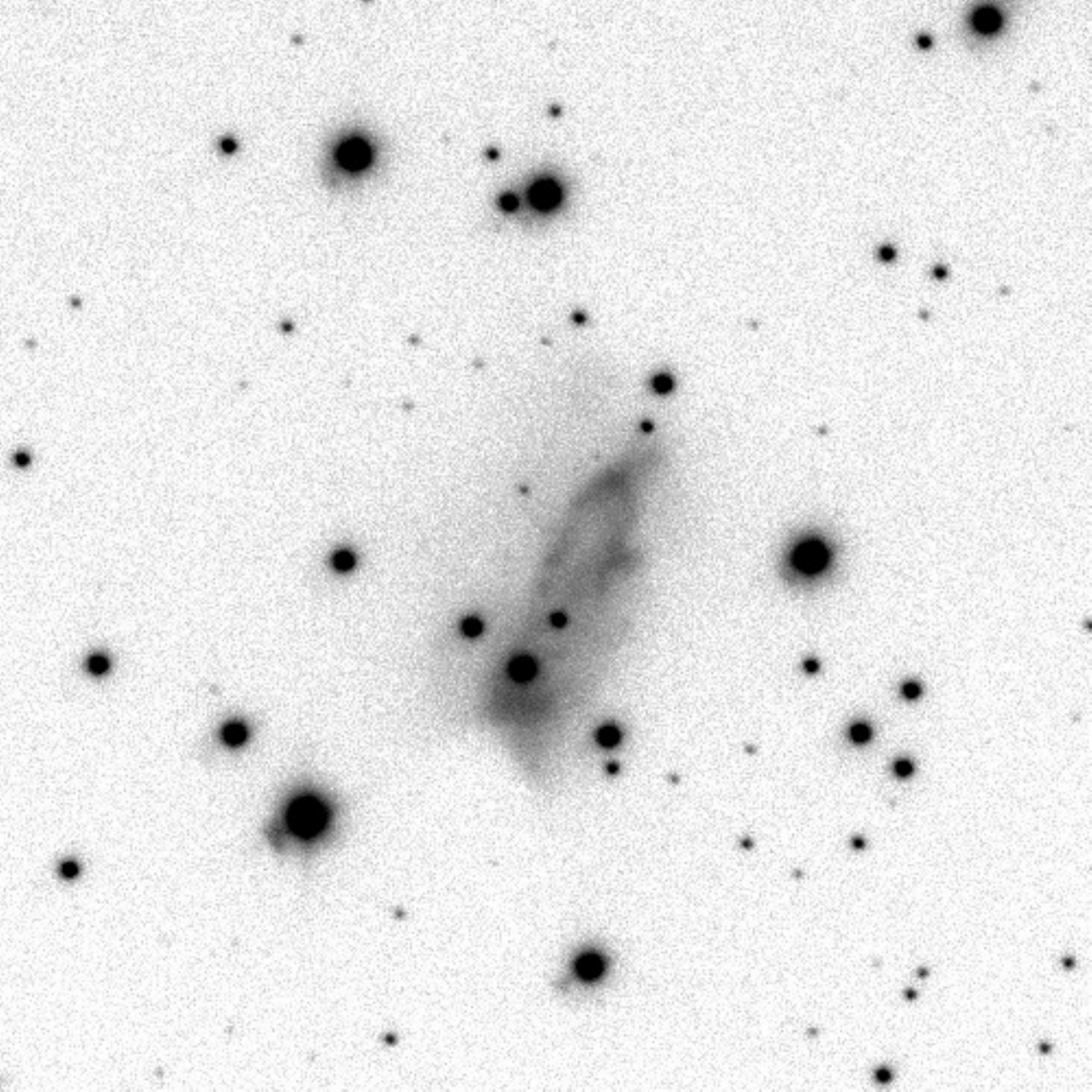}
\includegraphics[height=1.7in]{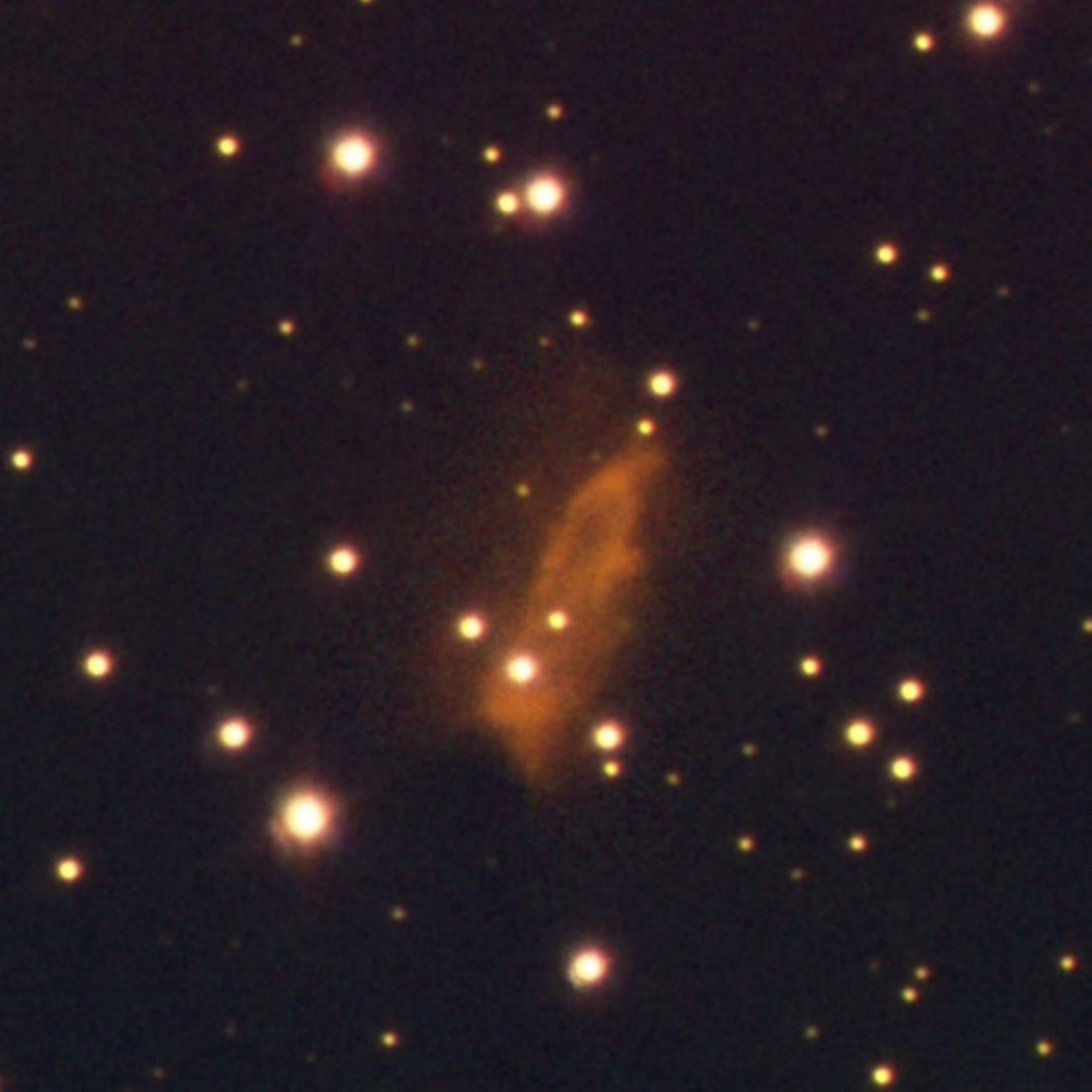}
\vskip .1in 
\includegraphics[height=1.7in]{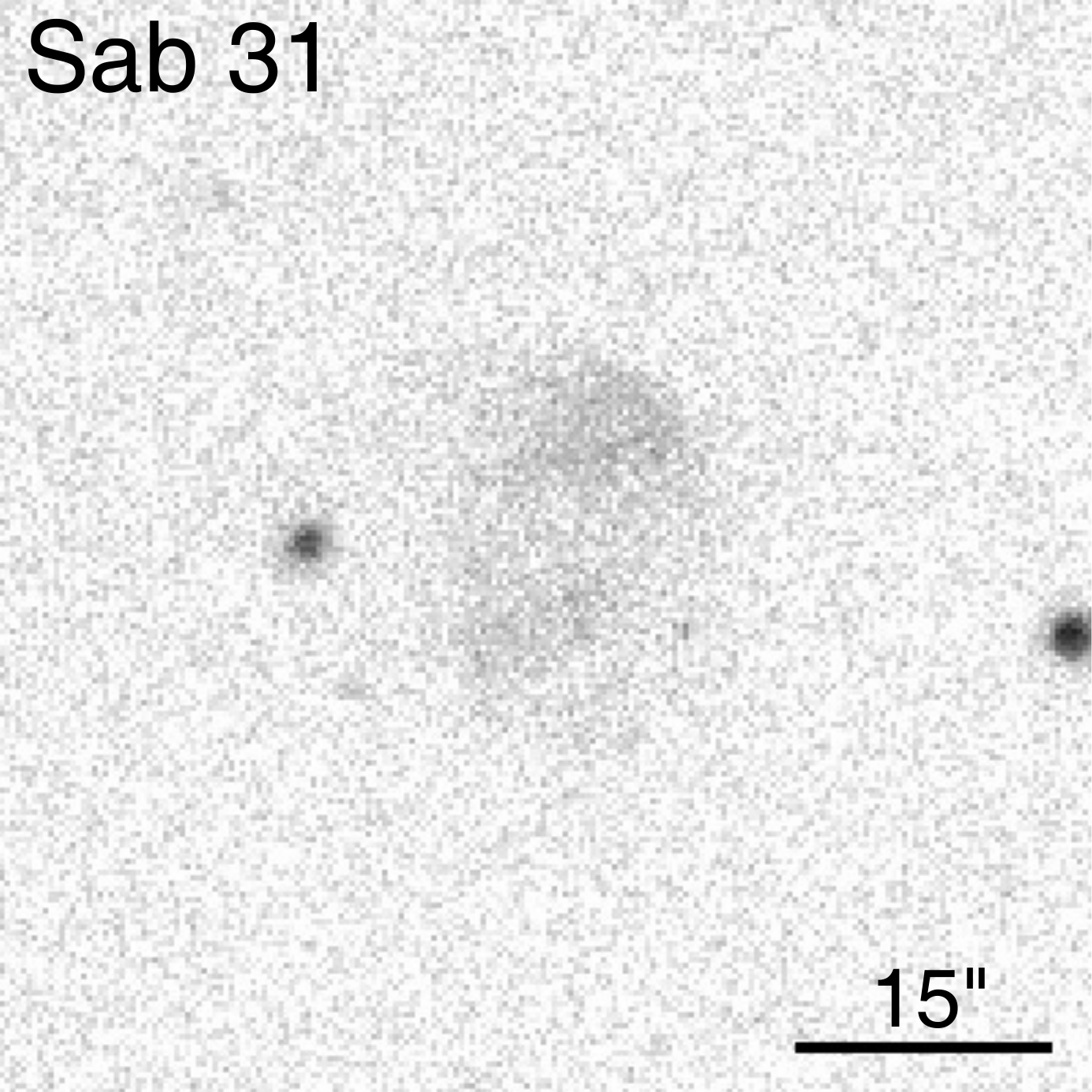} 
\includegraphics[height=1.7in]{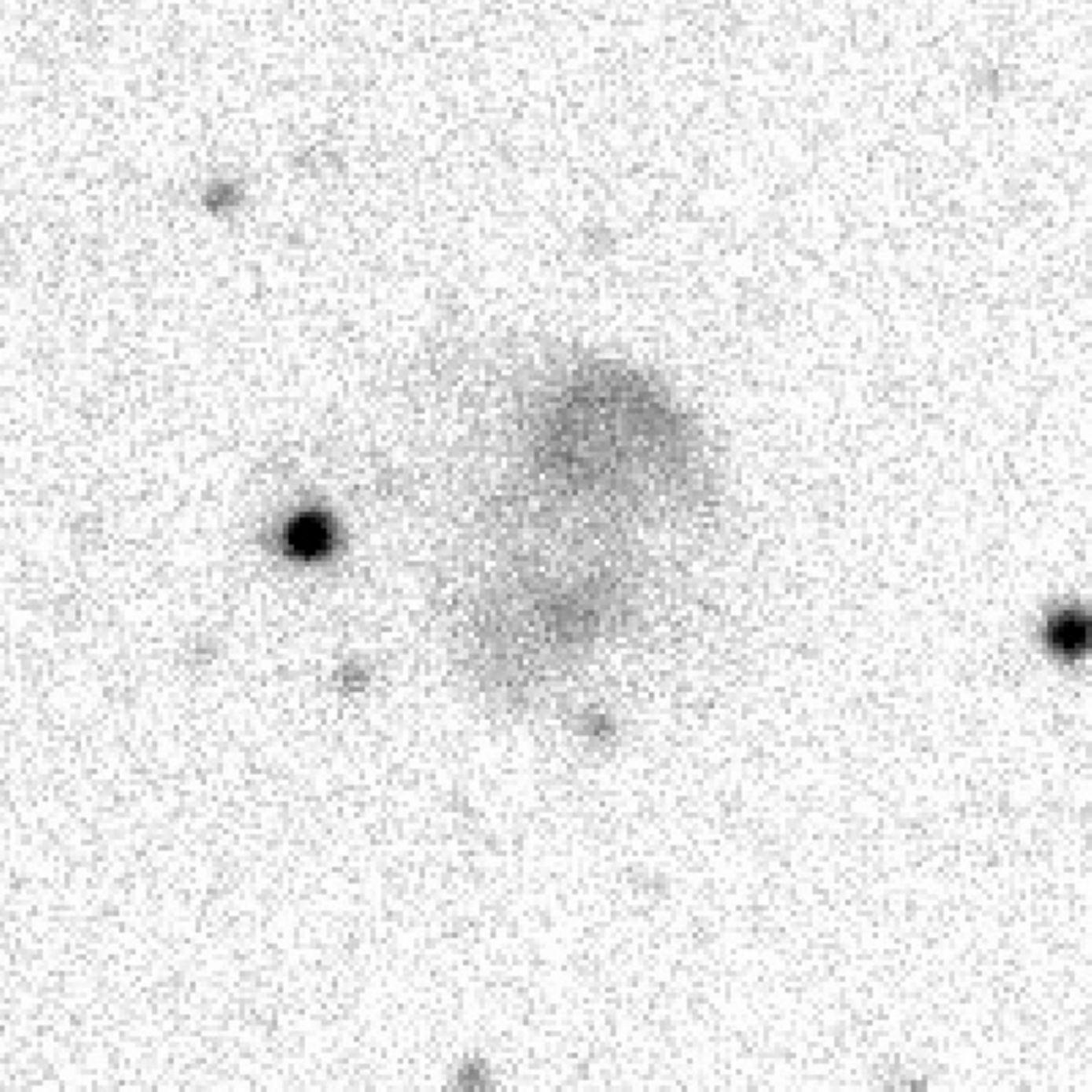}
\includegraphics[height=1.7in]{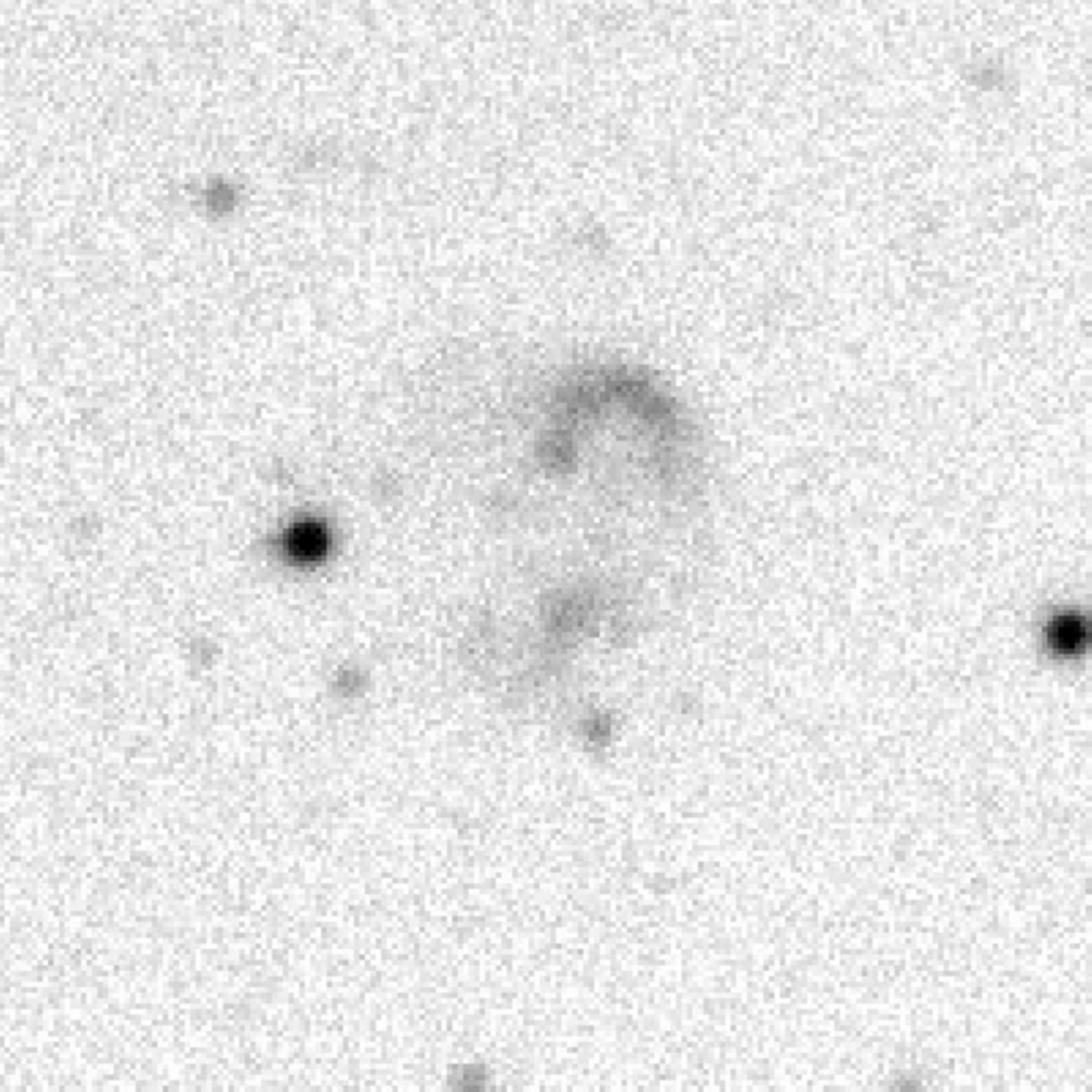}
\includegraphics[height=1.7in]{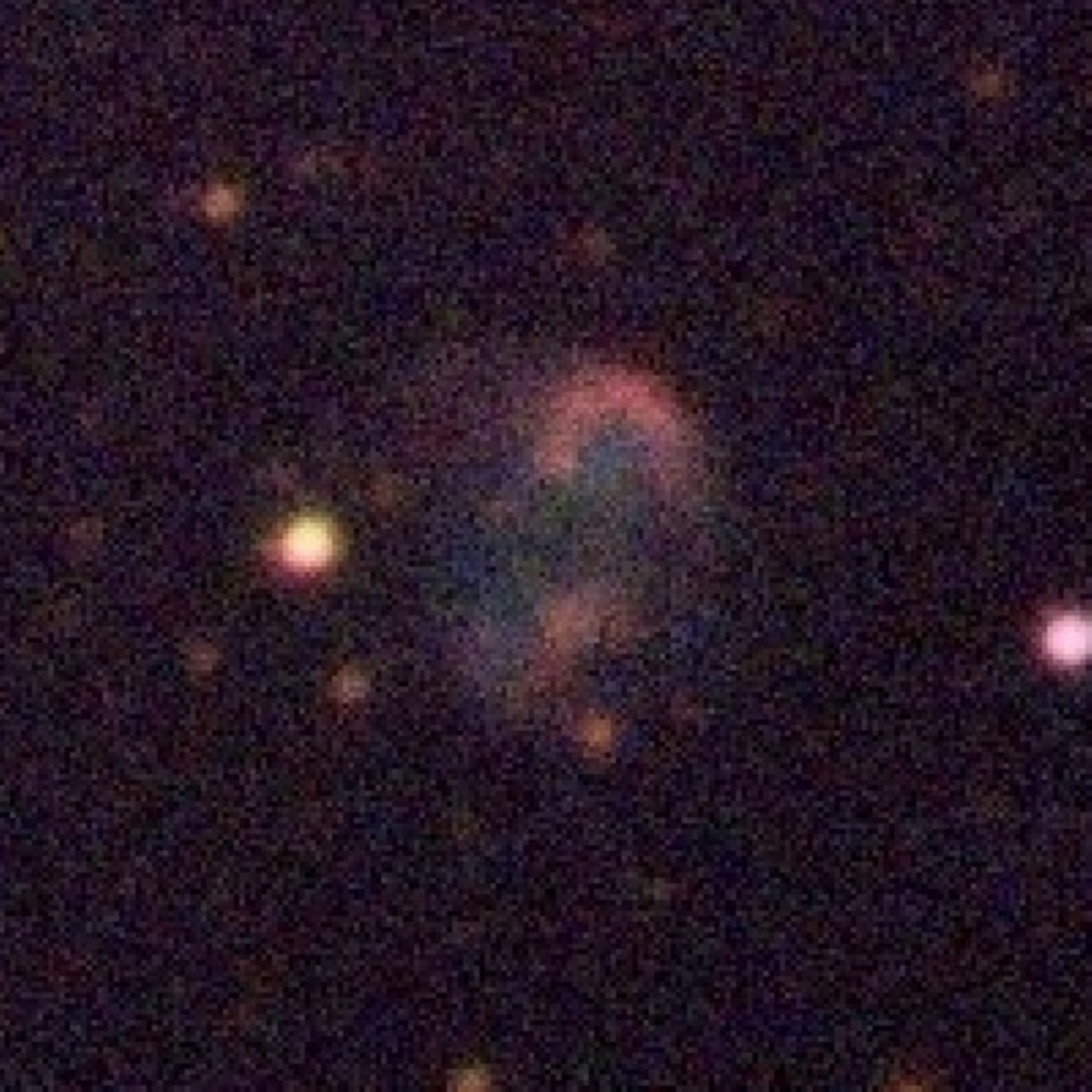}
\vskip .1in 
\includegraphics[height=1.7in]{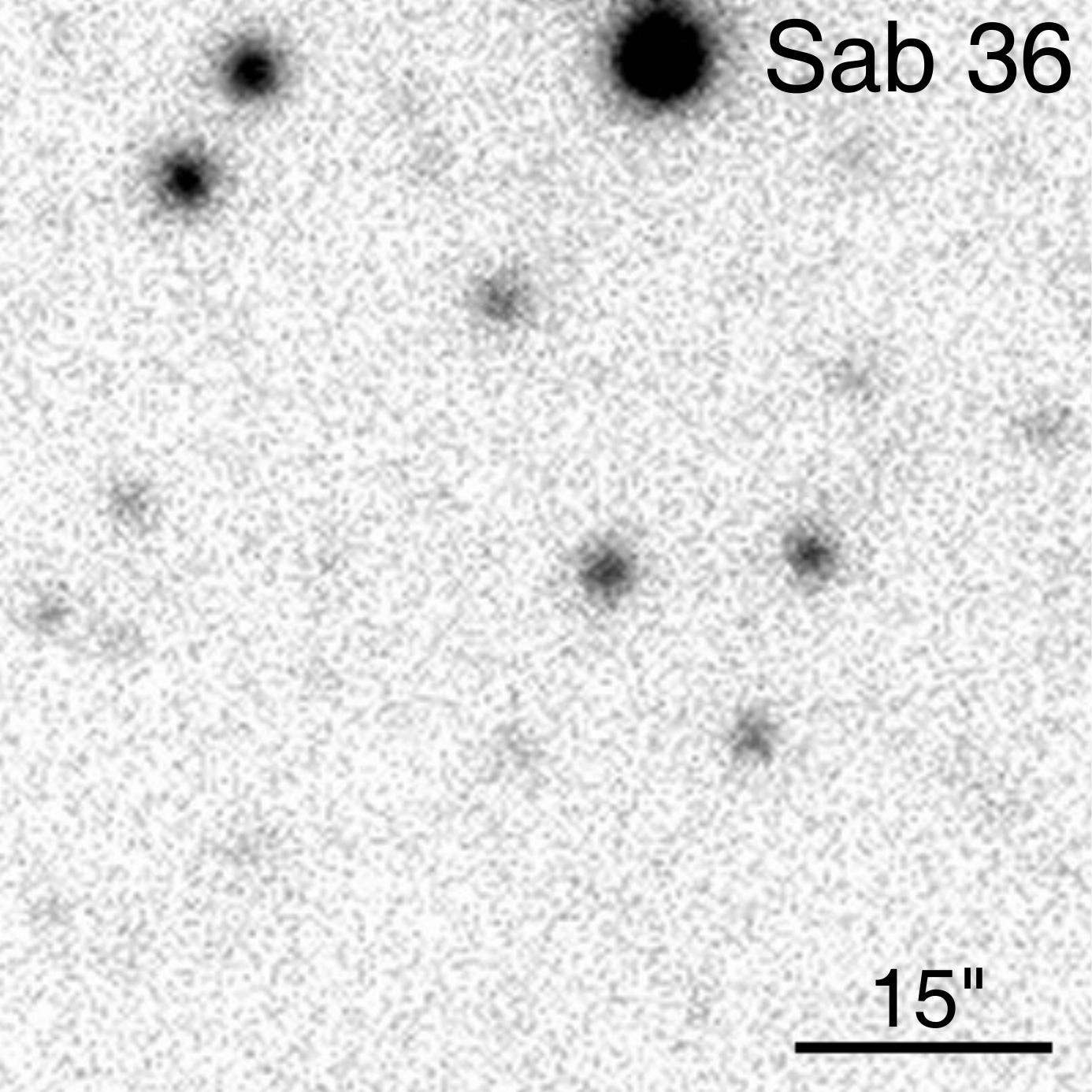} 
\includegraphics[height=1.7in]{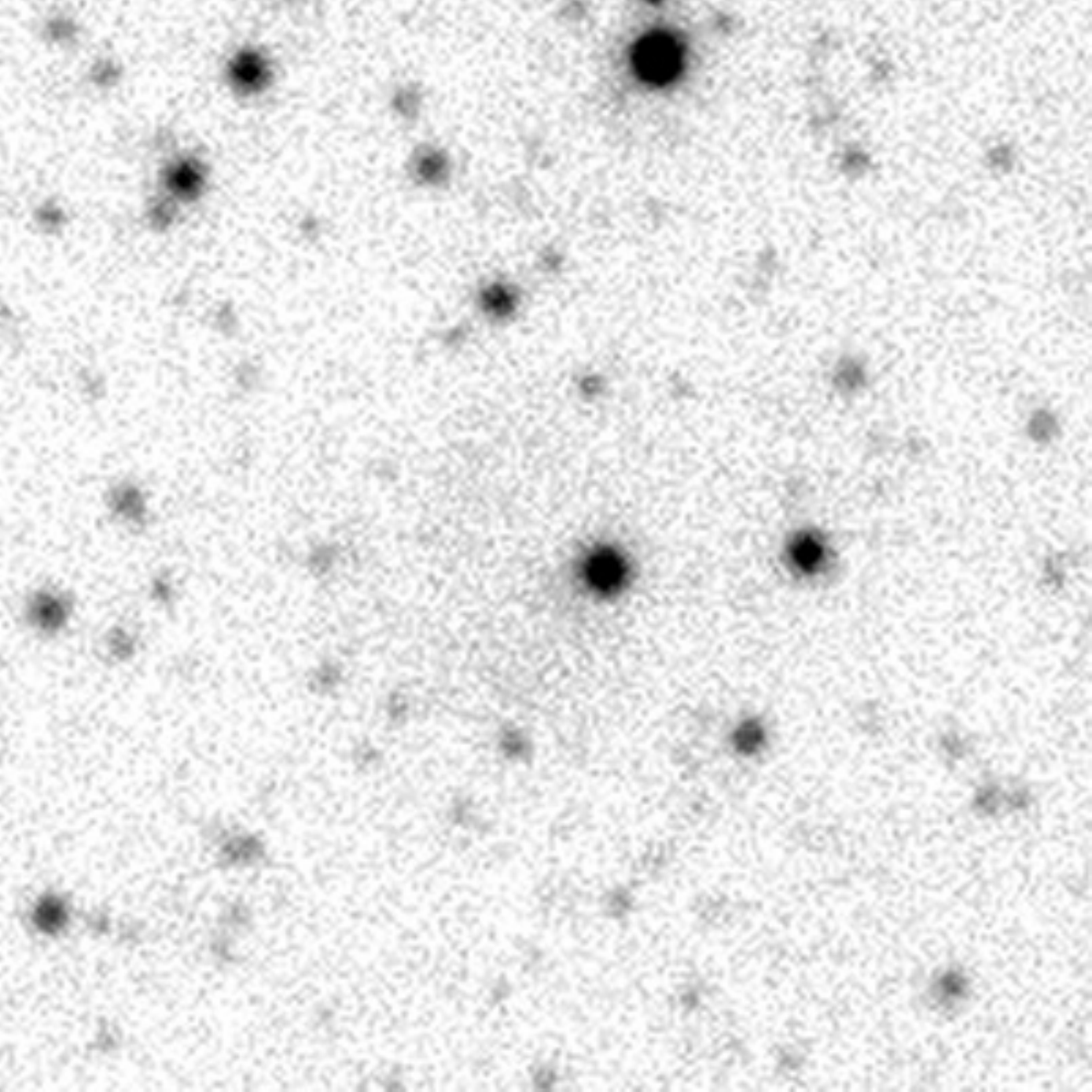}
\includegraphics[height=1.7in]{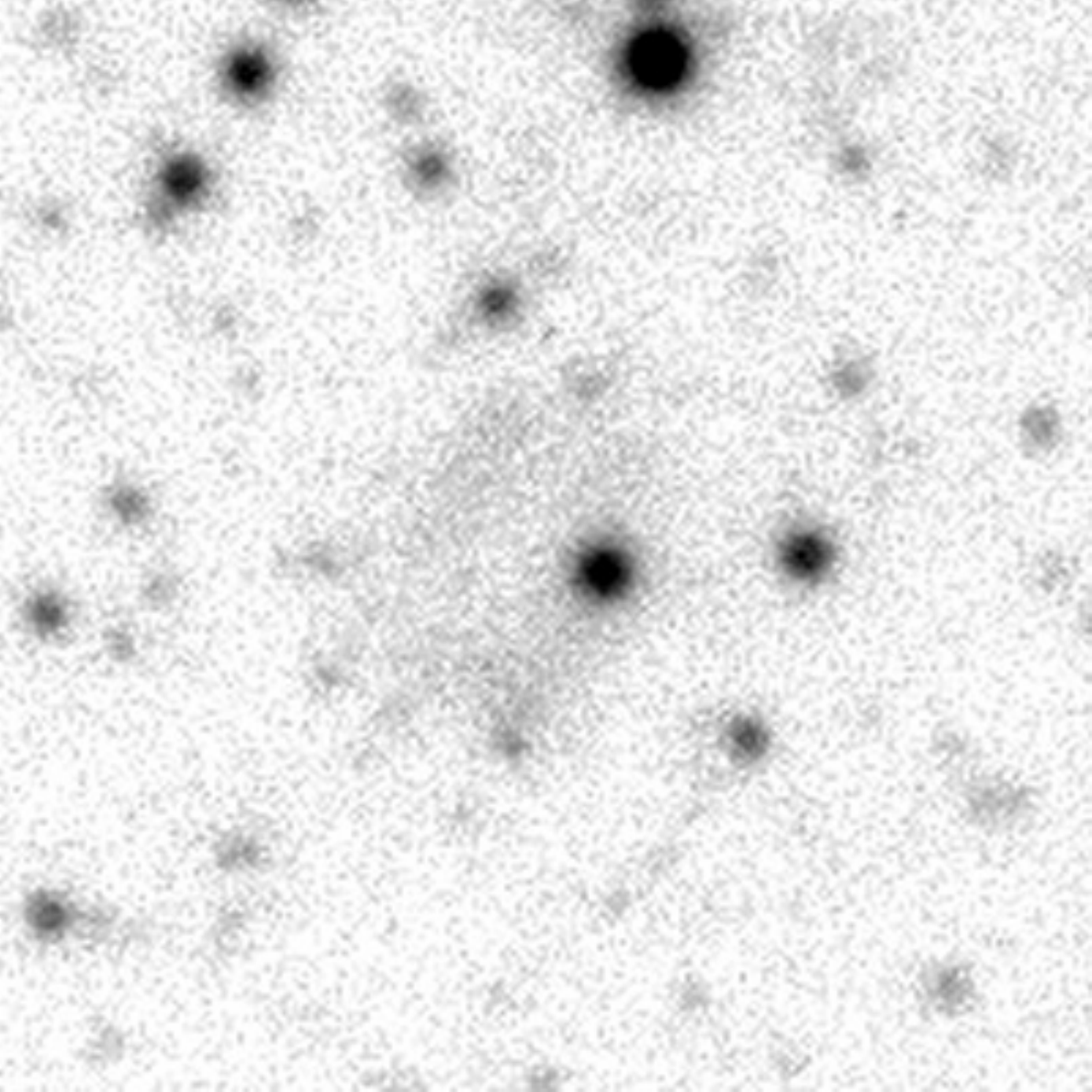}
\includegraphics[height=1.7in]{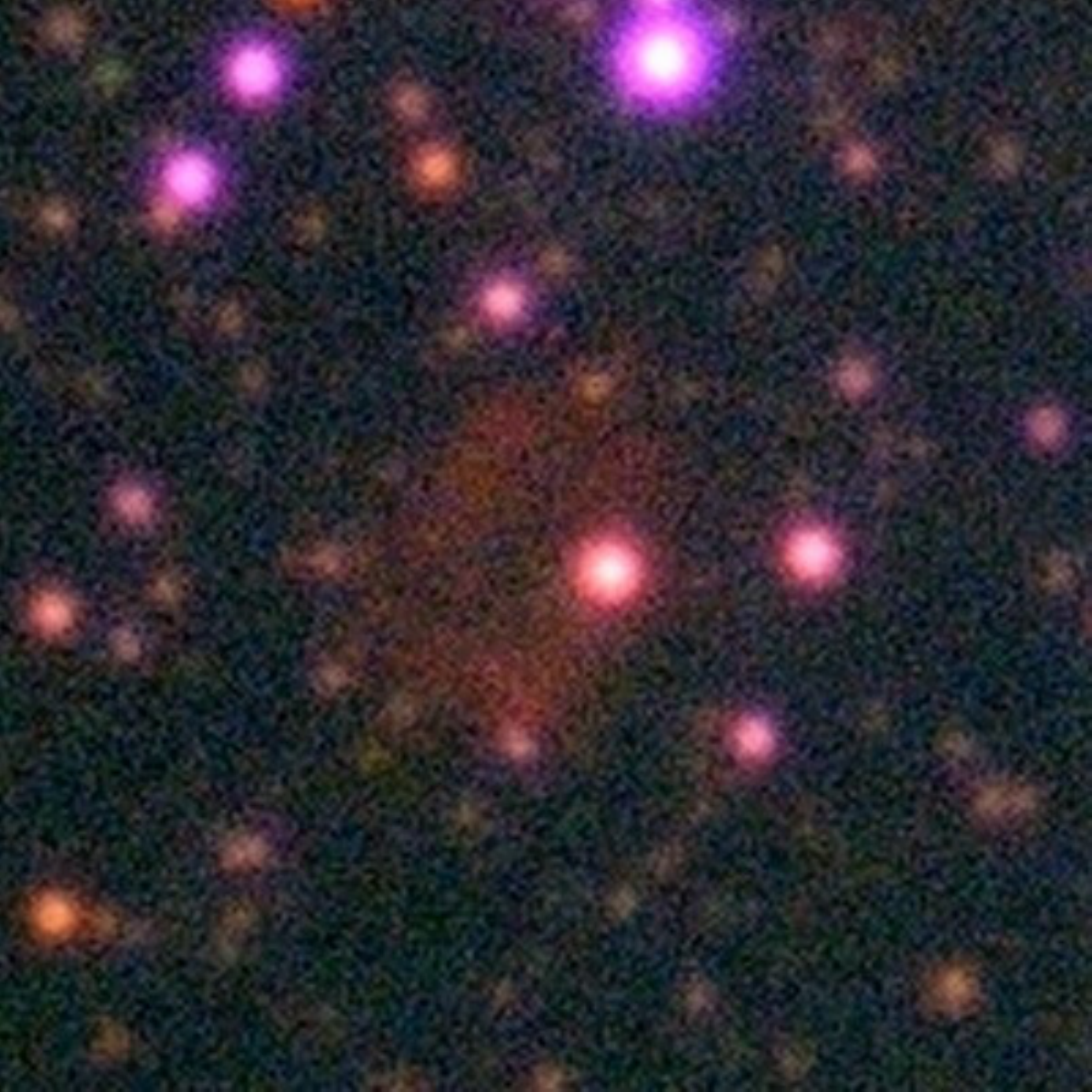}
\vskip .1in 
\includegraphics[height=1.7in]{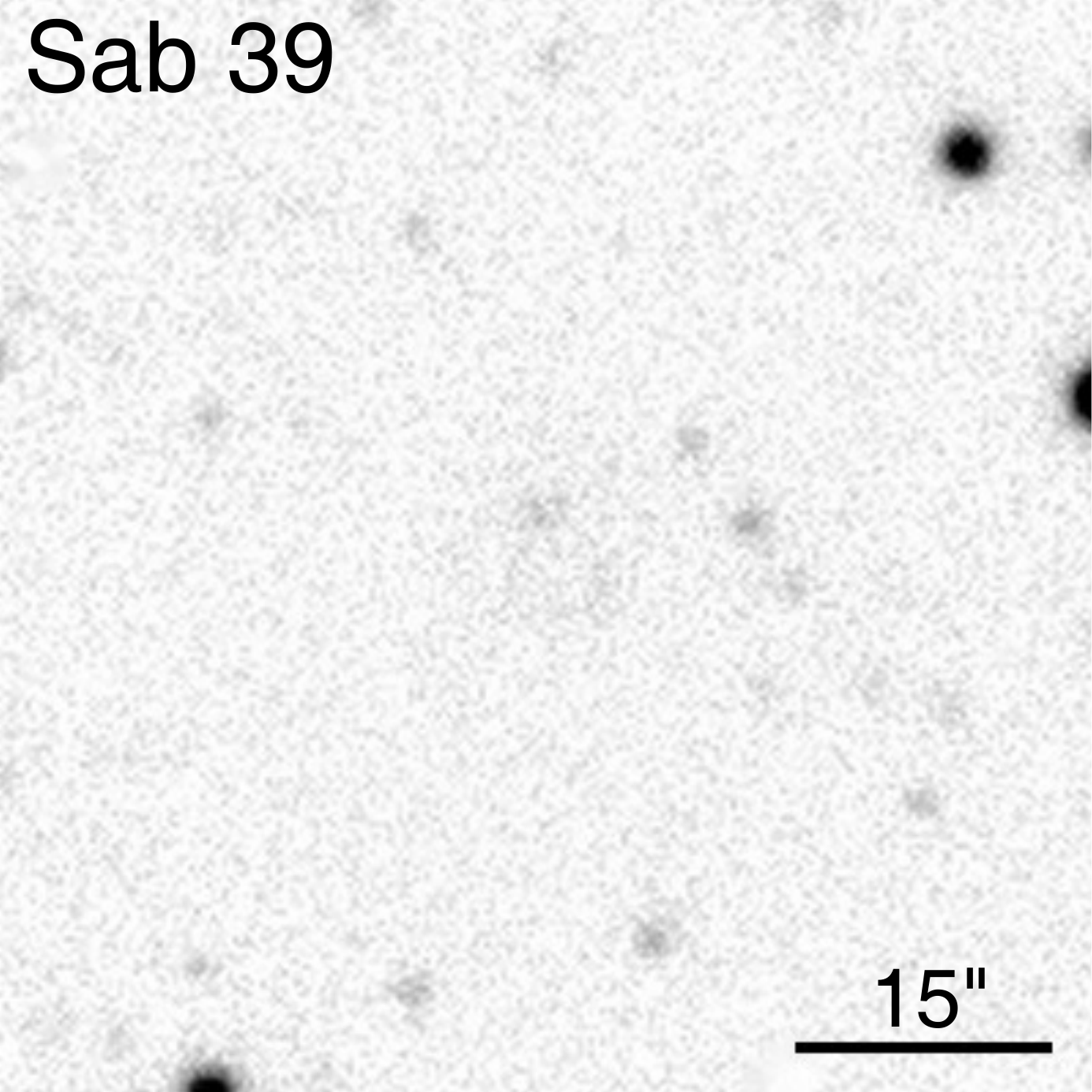} 
\includegraphics[height=1.7in]{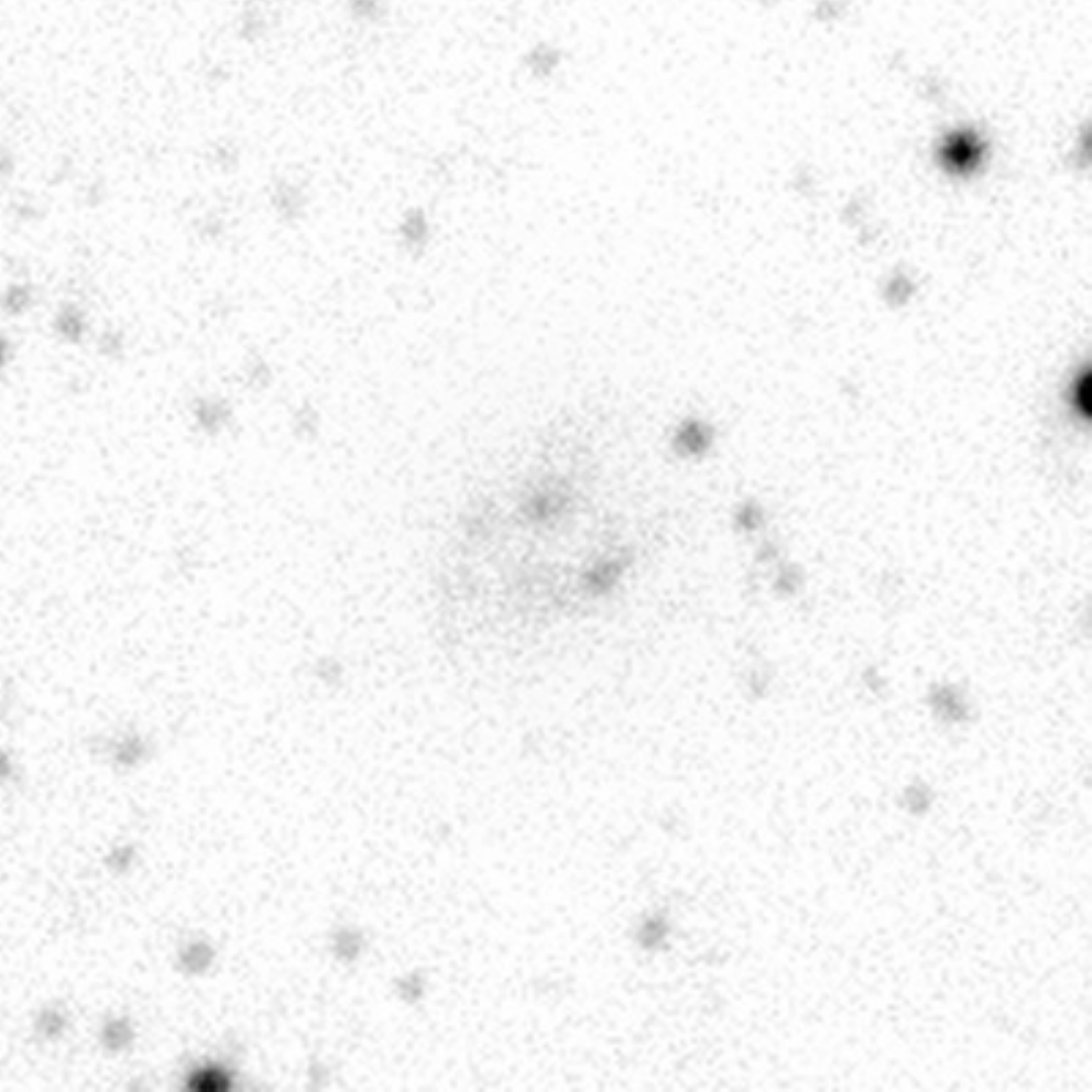}
\includegraphics[height=1.7in]{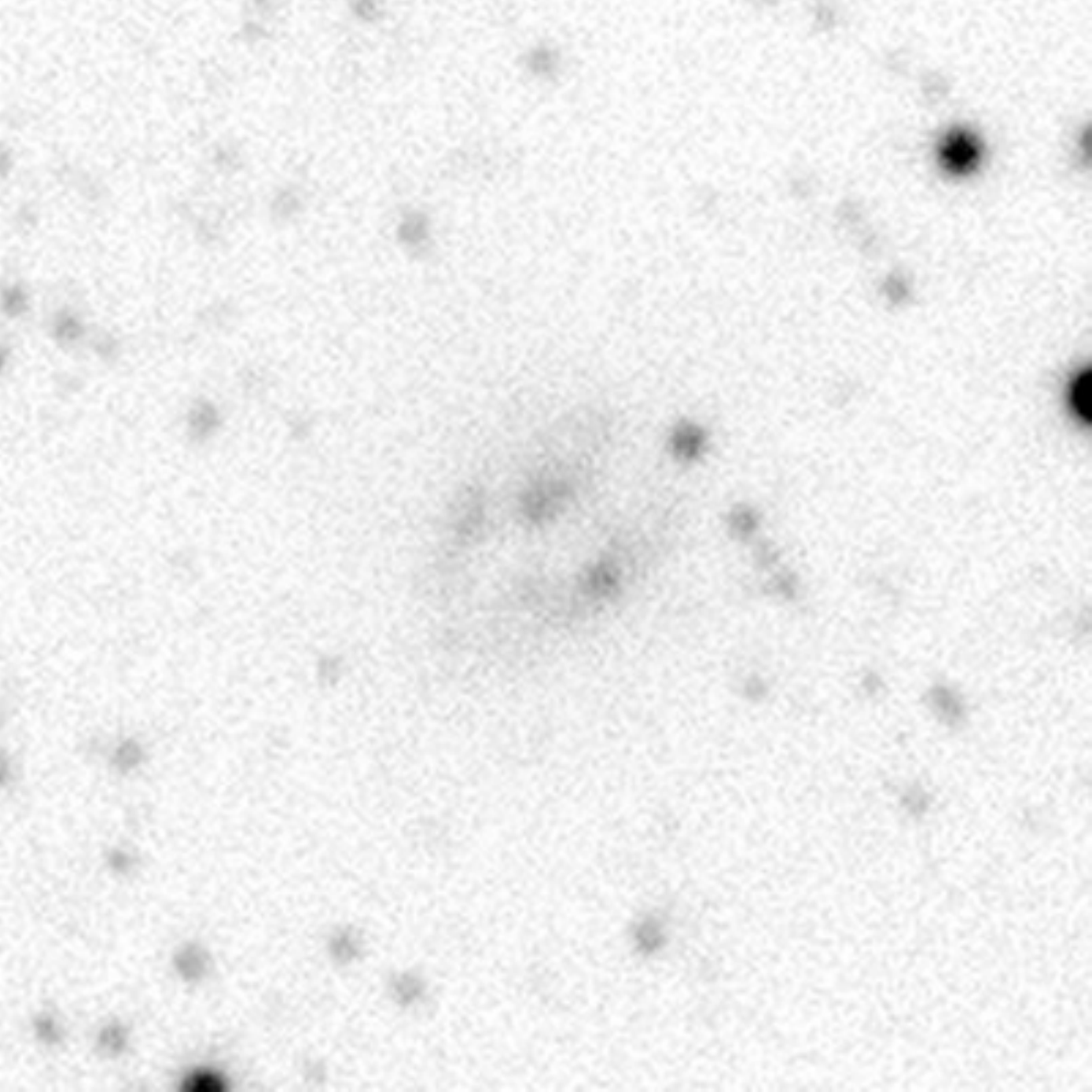}
\includegraphics[height=1.7in]{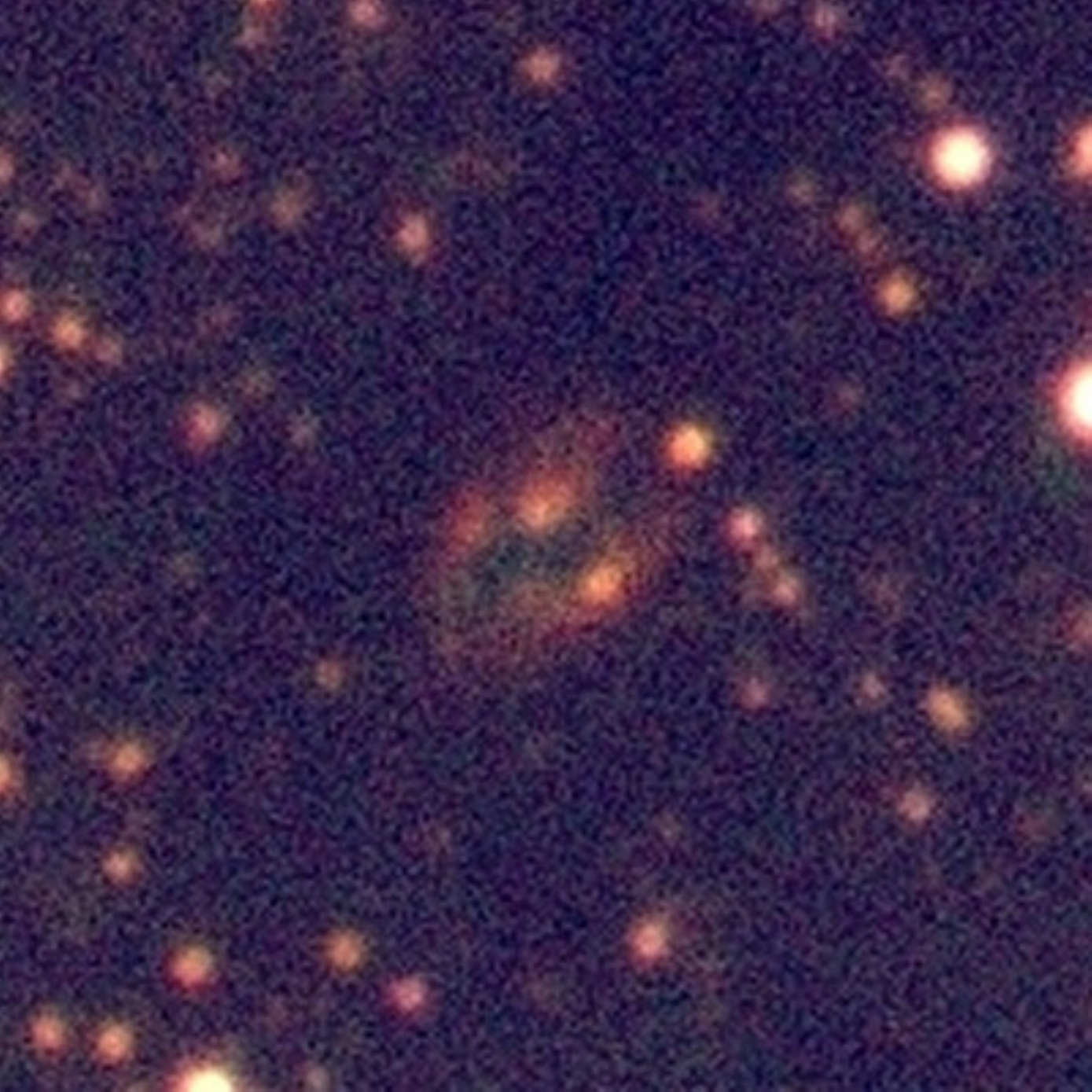}
\vskip .1in 
\includegraphics[height=1.7in]{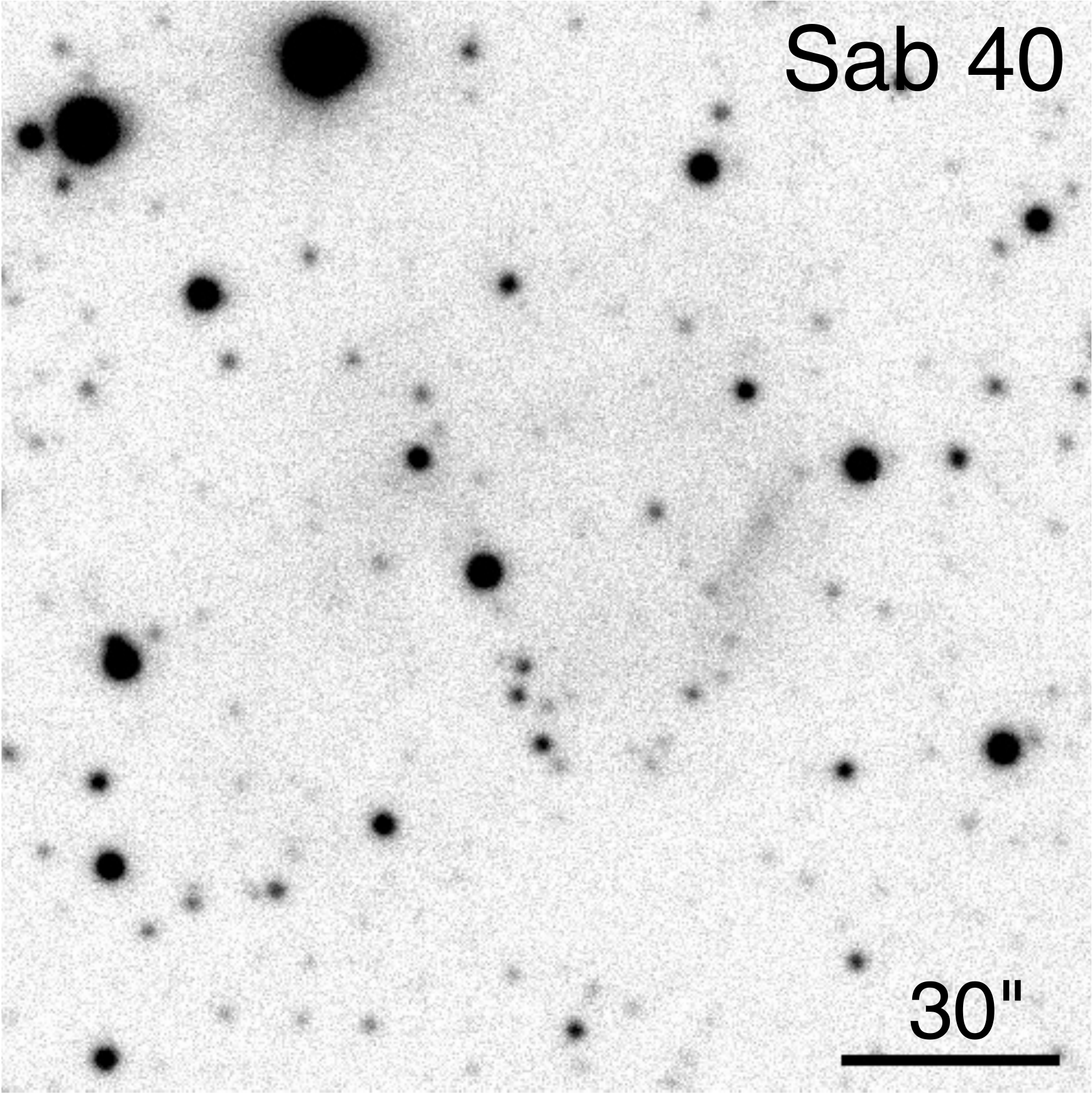} 
\includegraphics[height=1.7in]{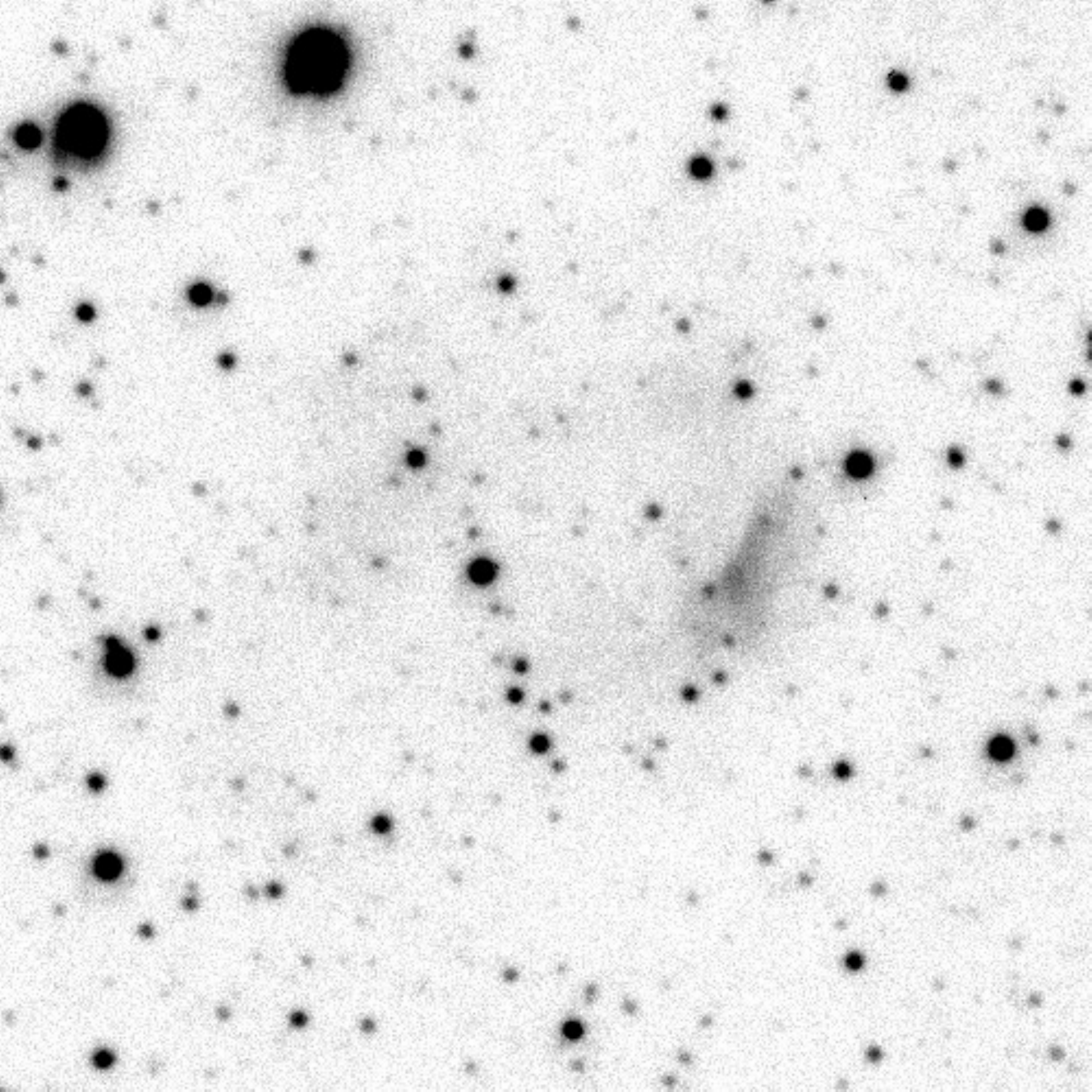}
\includegraphics[height=1.7in]{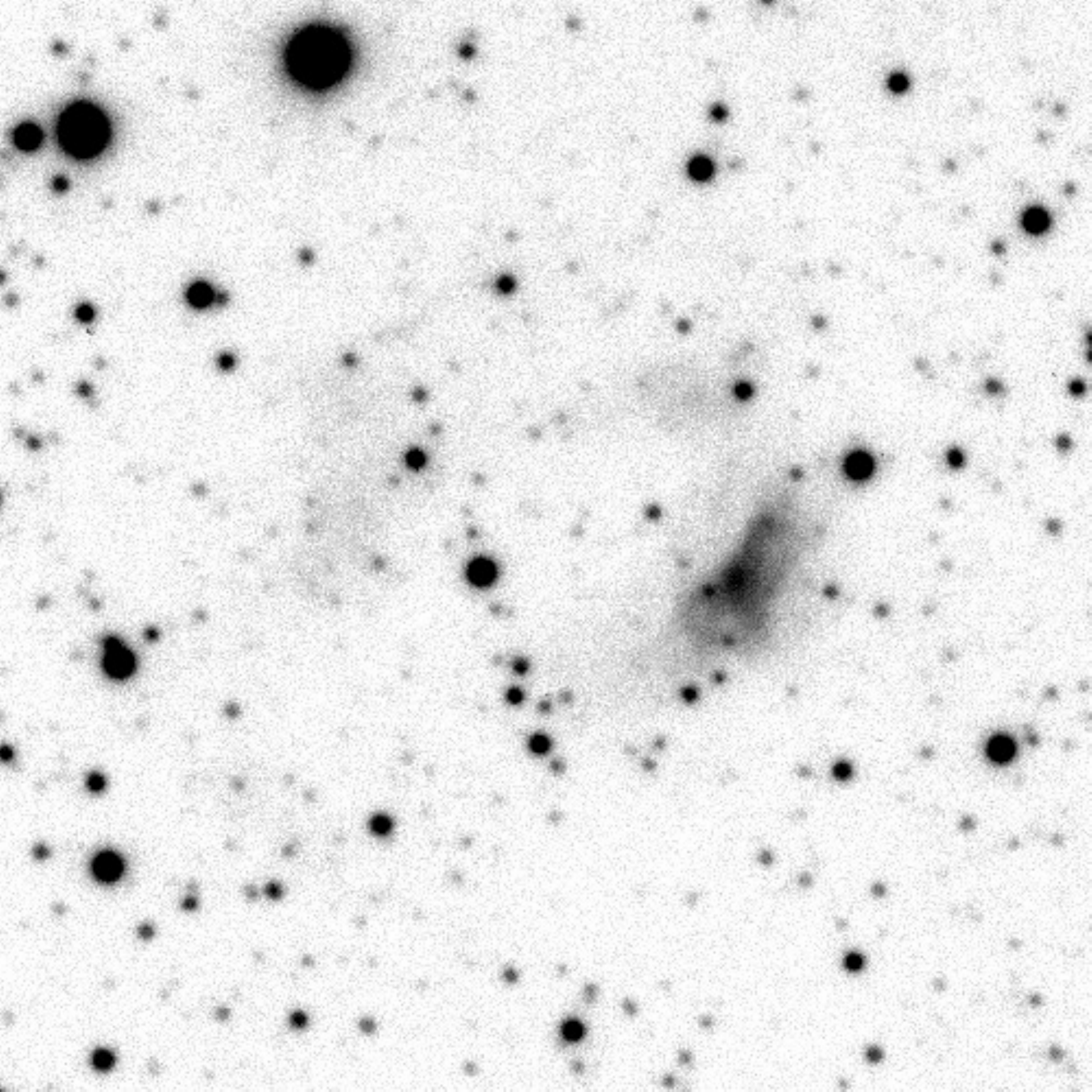}
\includegraphics[height=1.7in]{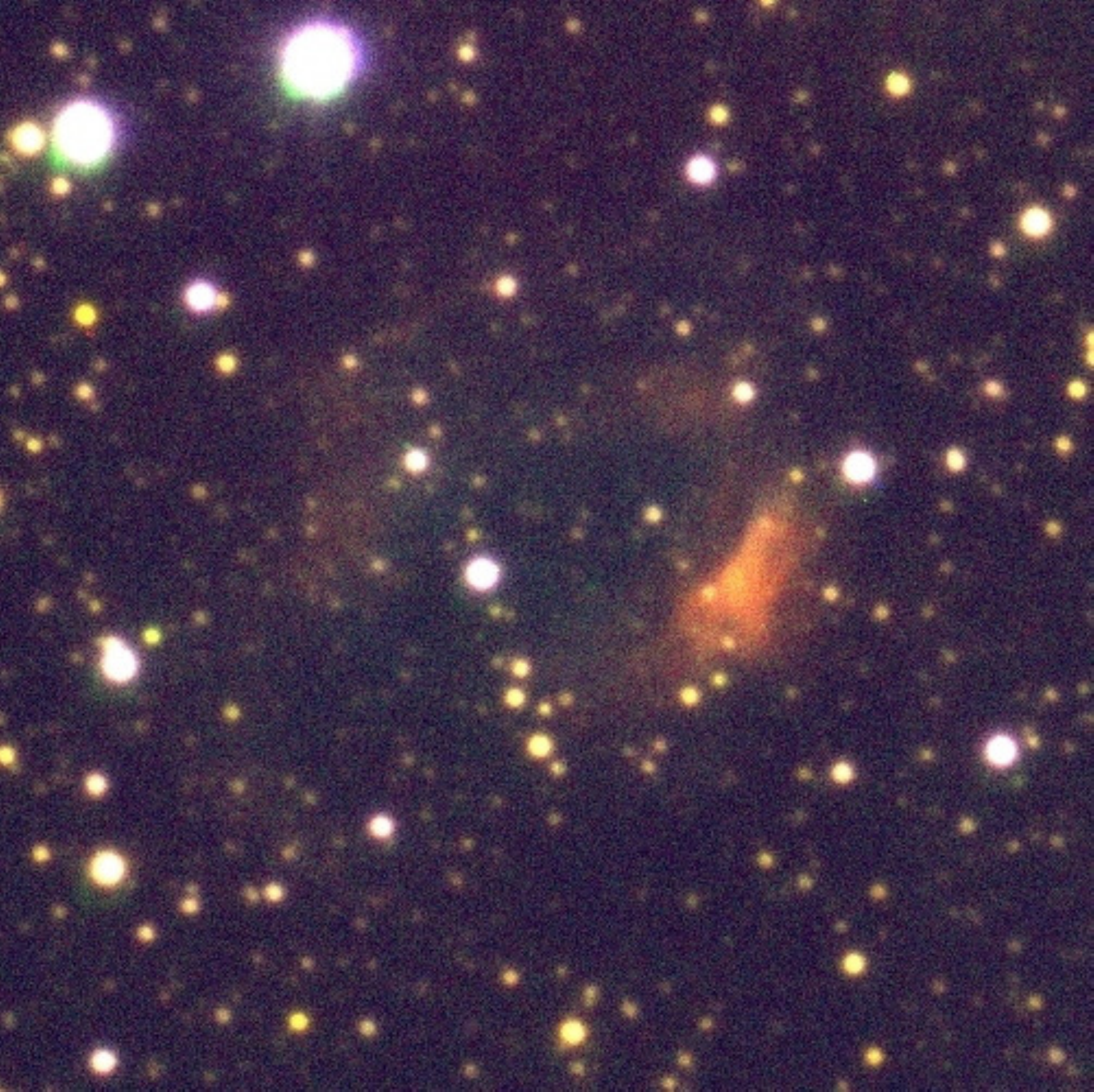}
\caption{Same as Figure~\ref{1.img}. } 
\label{3.img} 
\end{figure*}

\begin{figure*} 
\centering 
\includegraphics[height=1.7in]{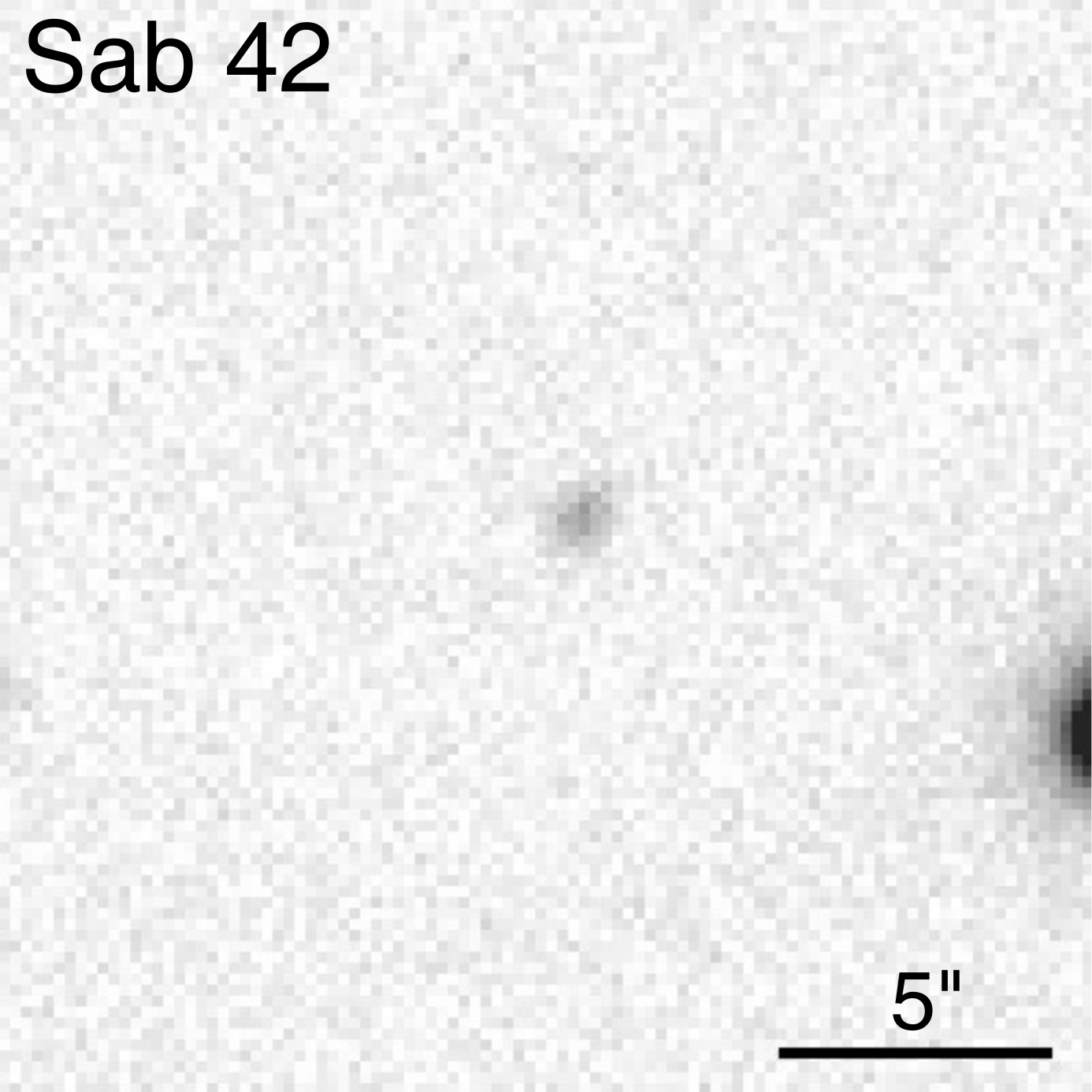} 
\includegraphics[height=1.7in]{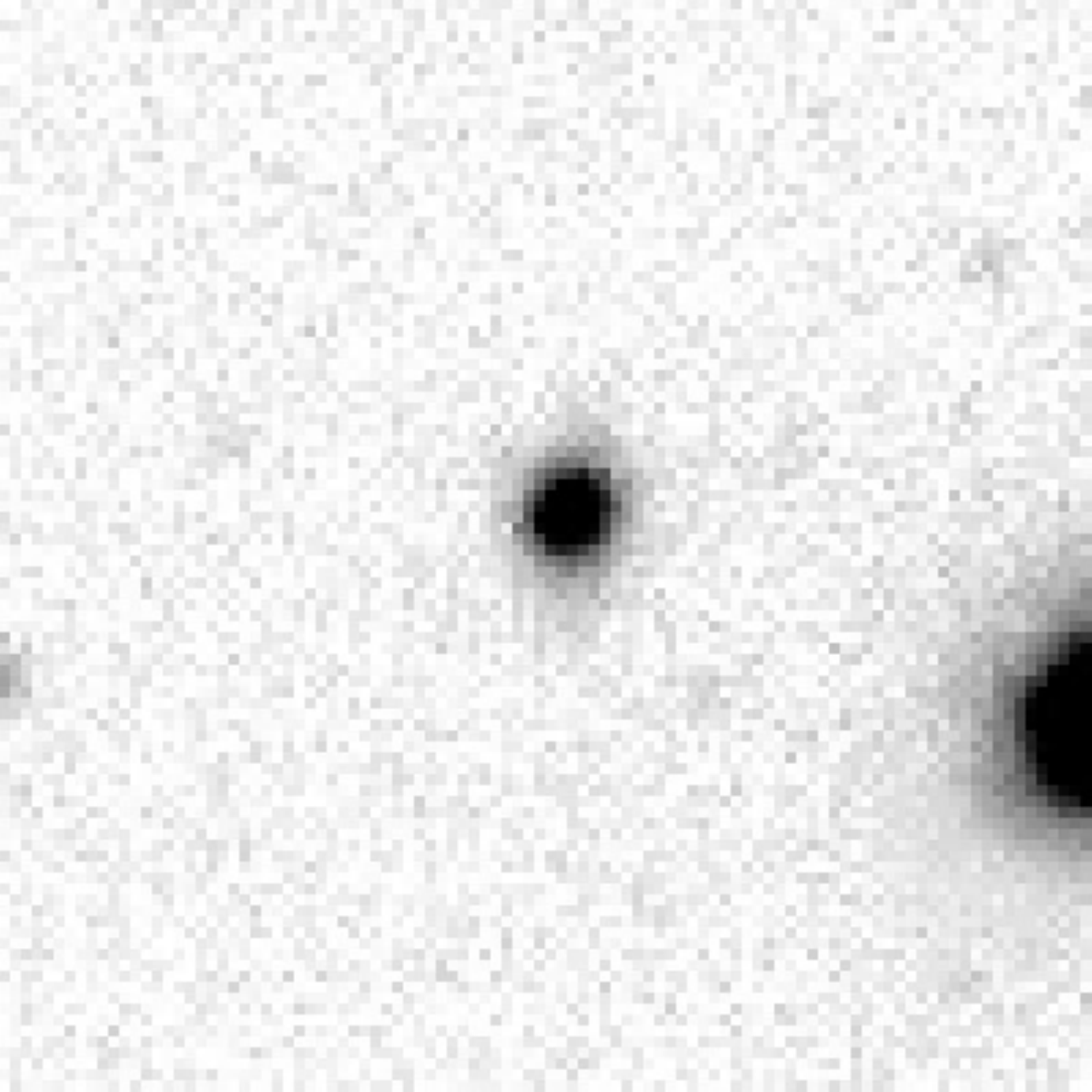}
\includegraphics[height=1.7in]{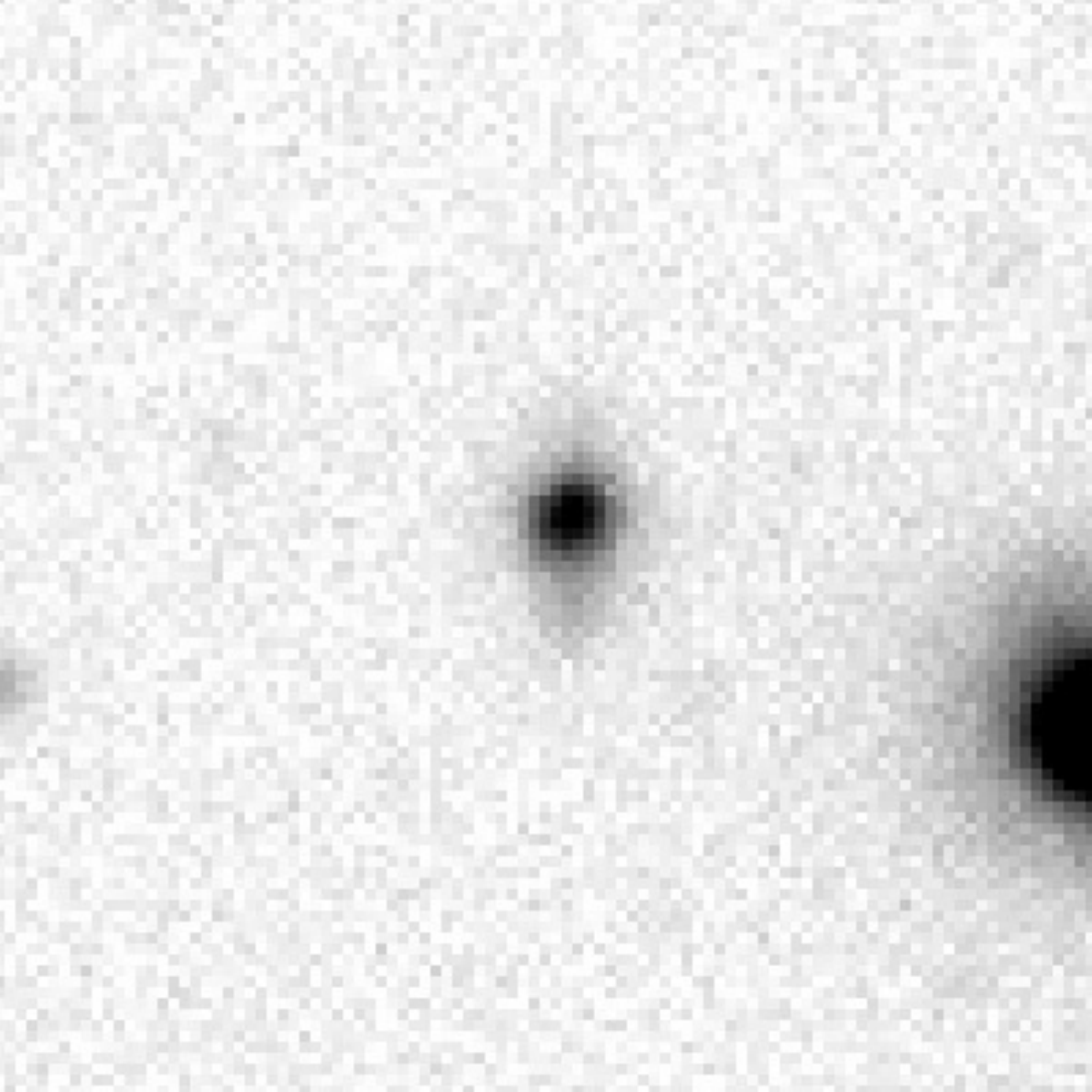}
\includegraphics[height=1.7in]{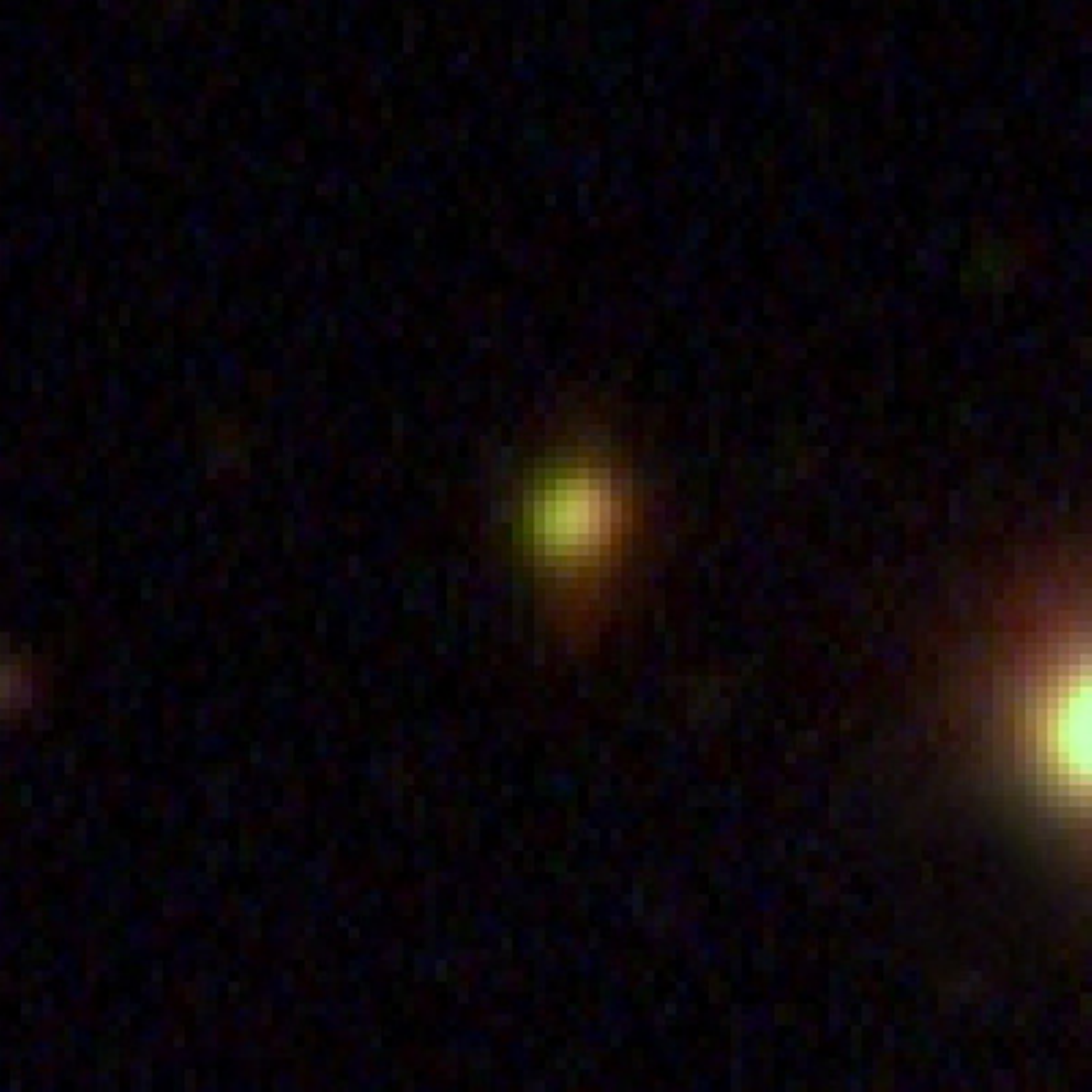}
\vskip .1in 
\includegraphics[height=1.7in]{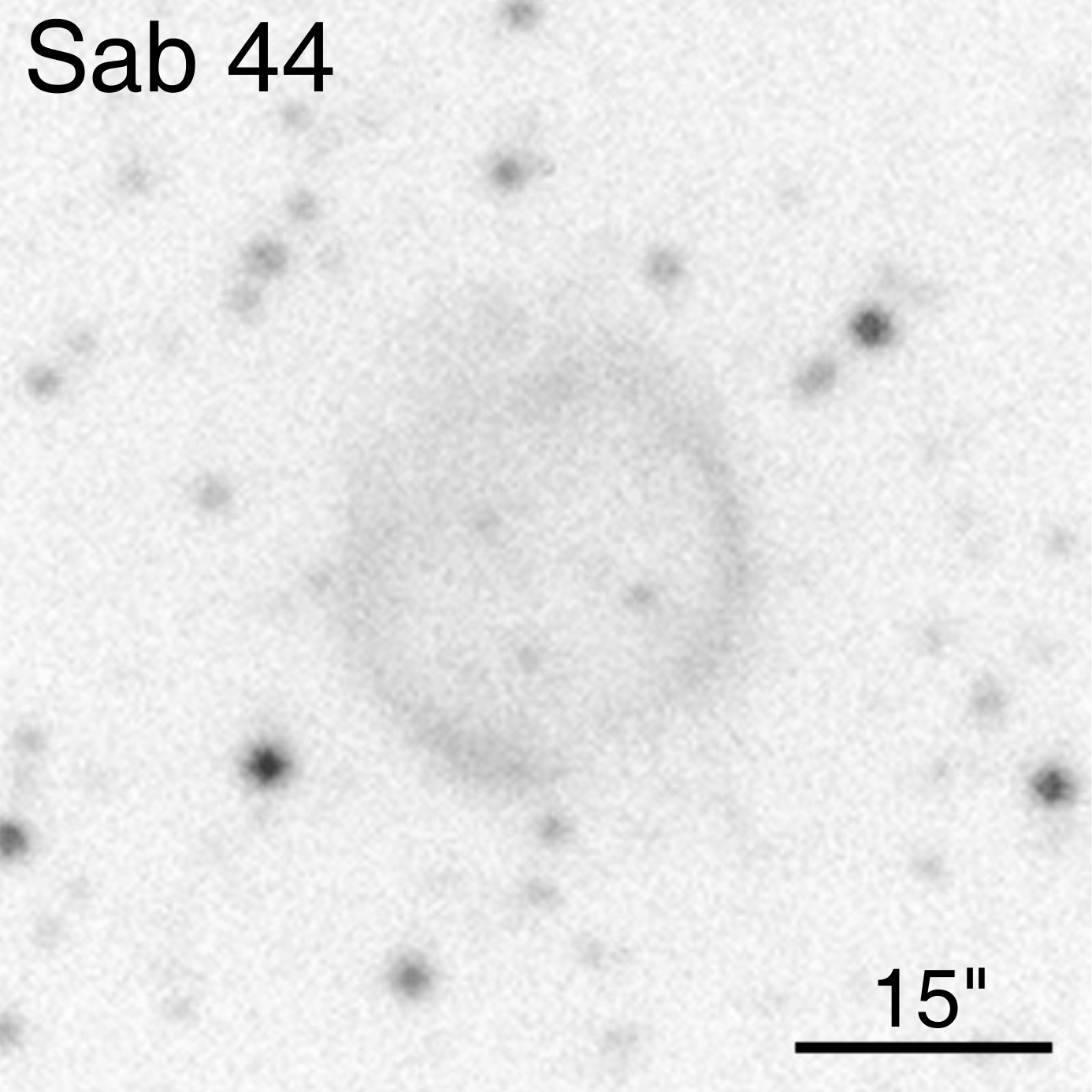} 
\includegraphics[height=1.7in]{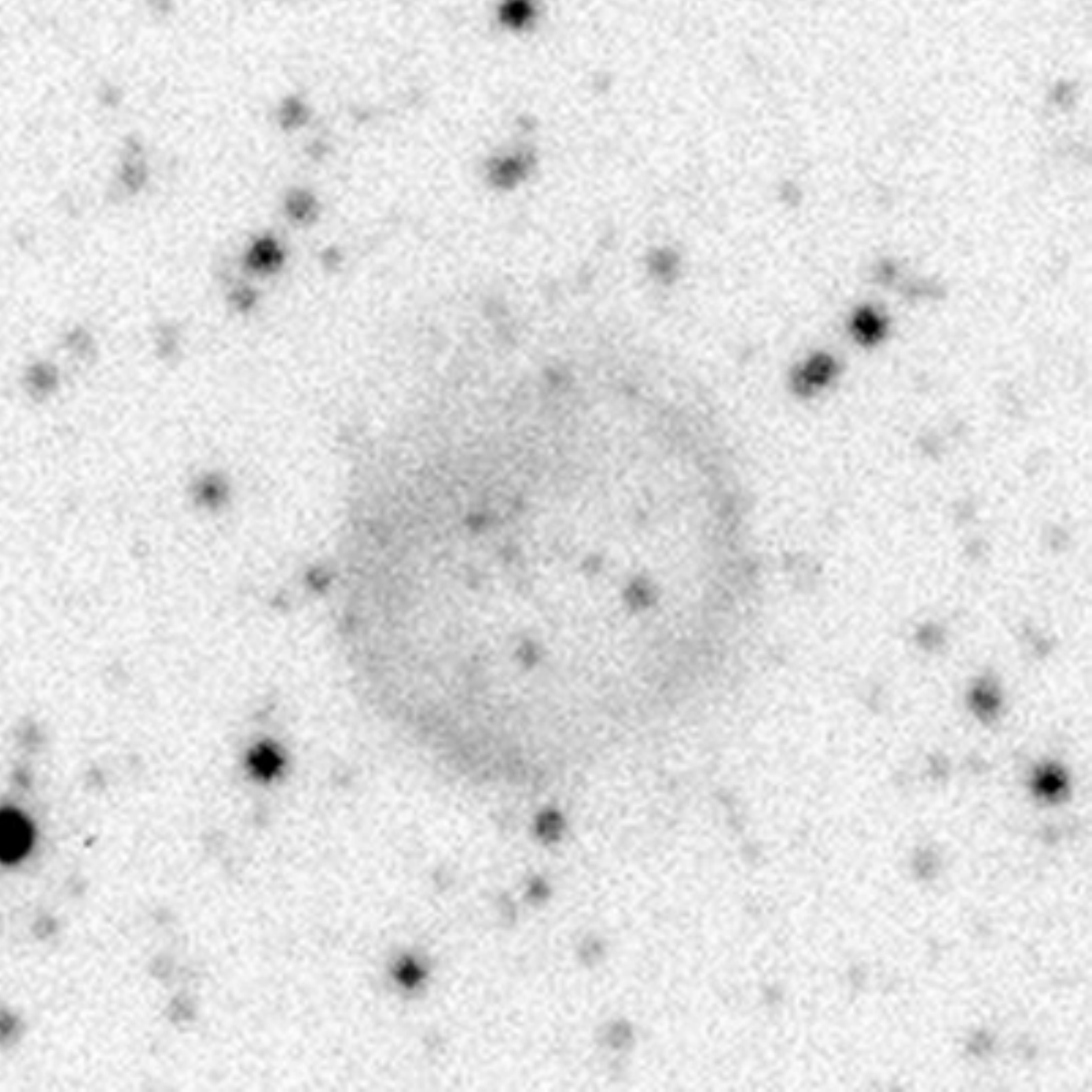}
\includegraphics[height=1.7in]{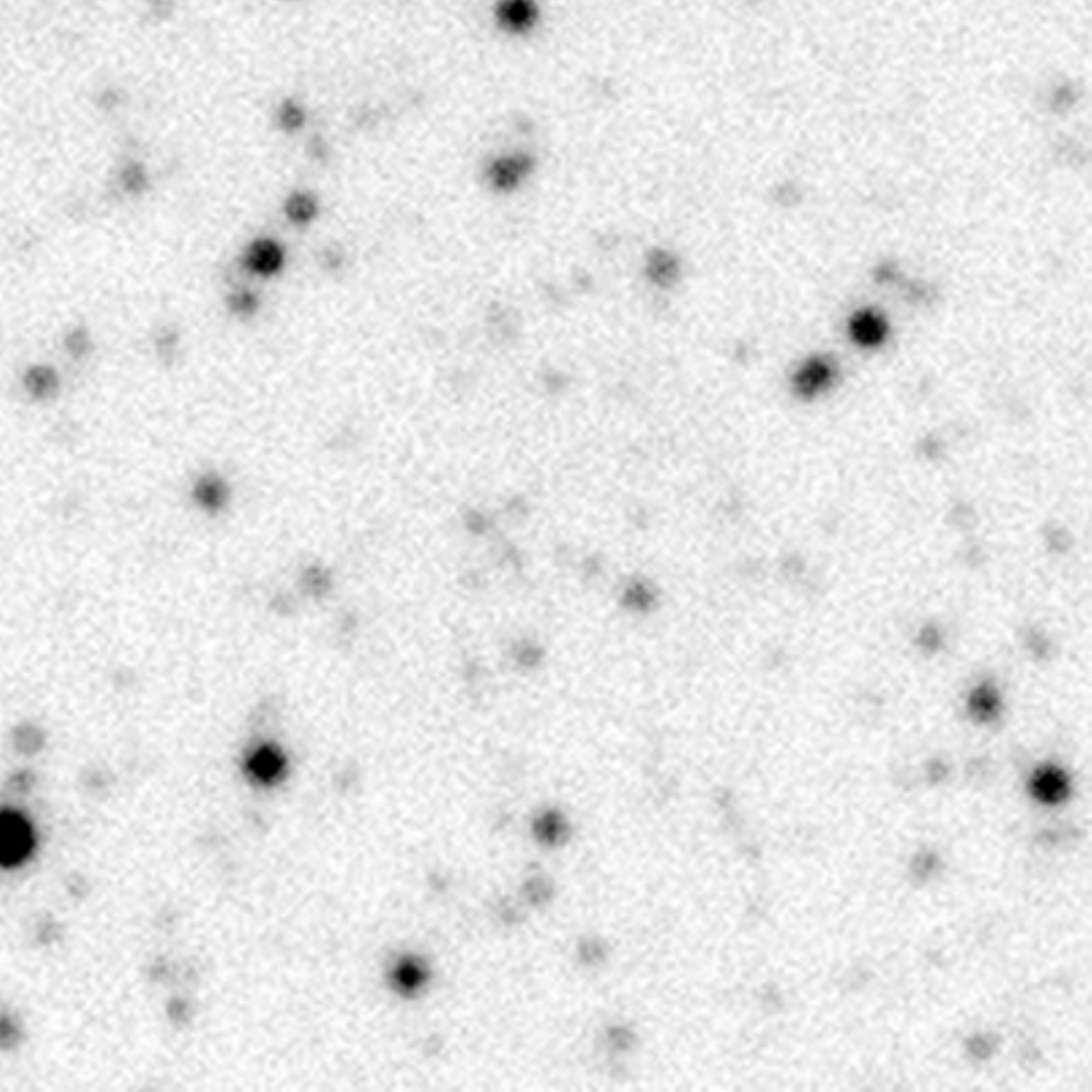}
\includegraphics[height=1.7in]{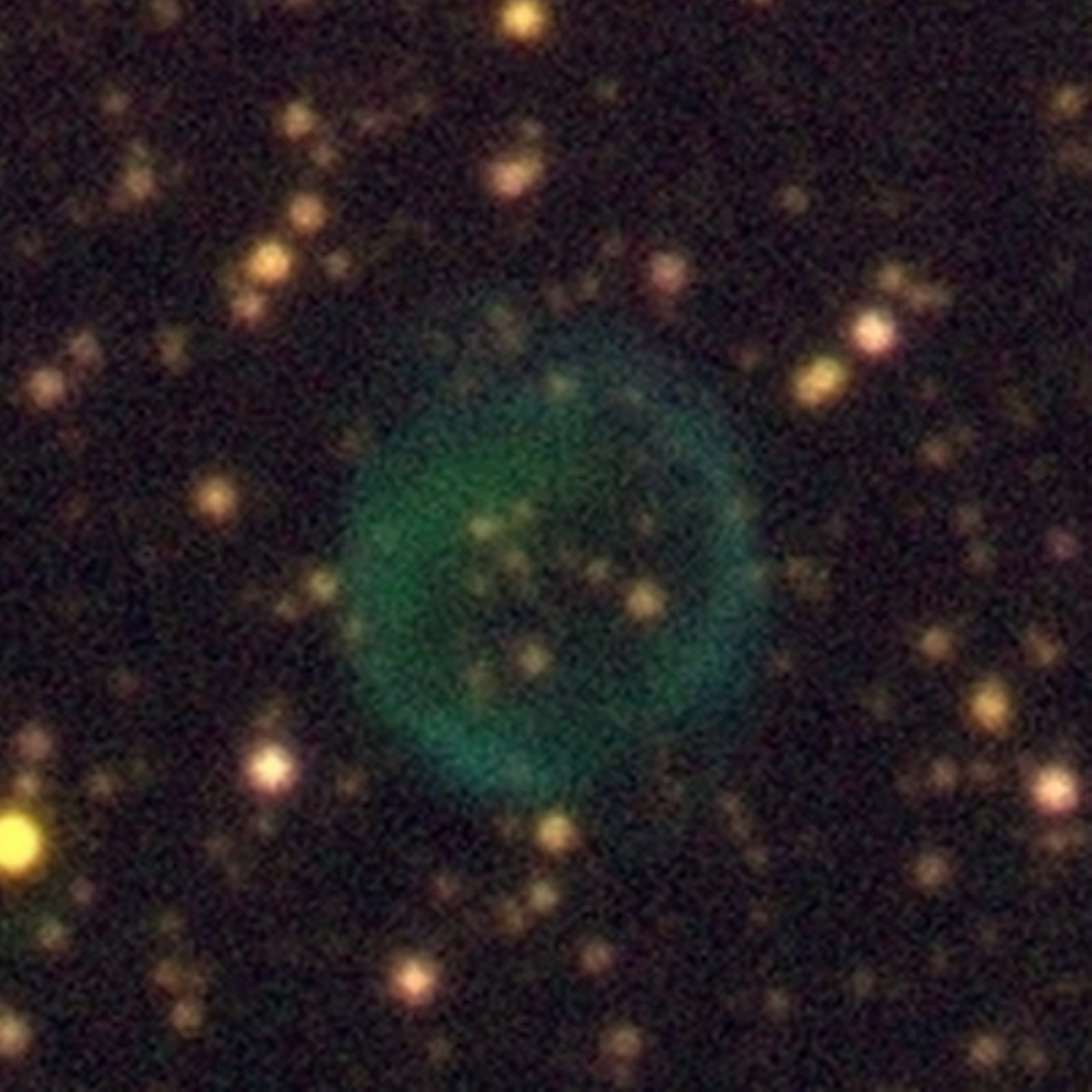}
\vskip .1in 
\includegraphics[height=1.7in]{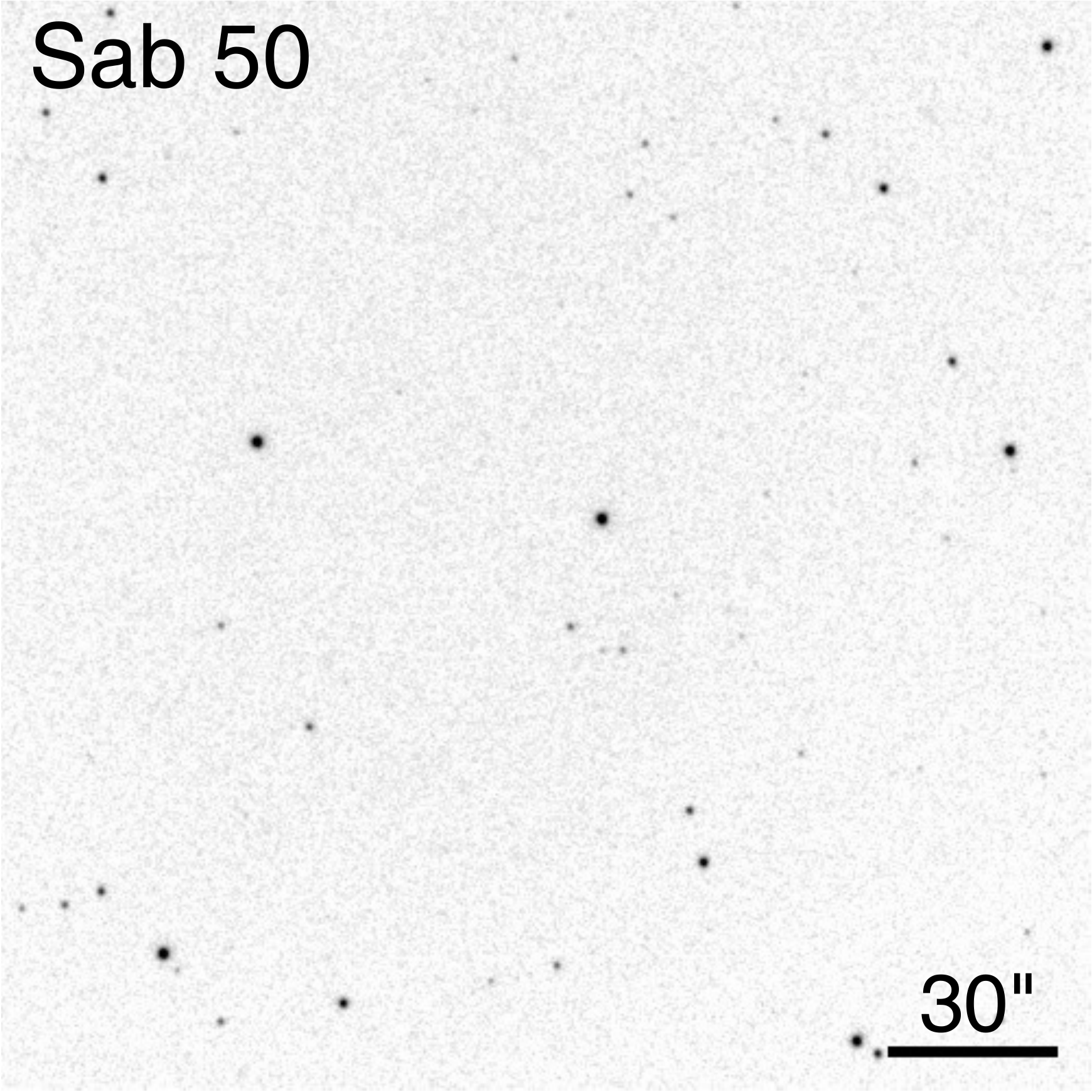} 
\includegraphics[height=1.7in]{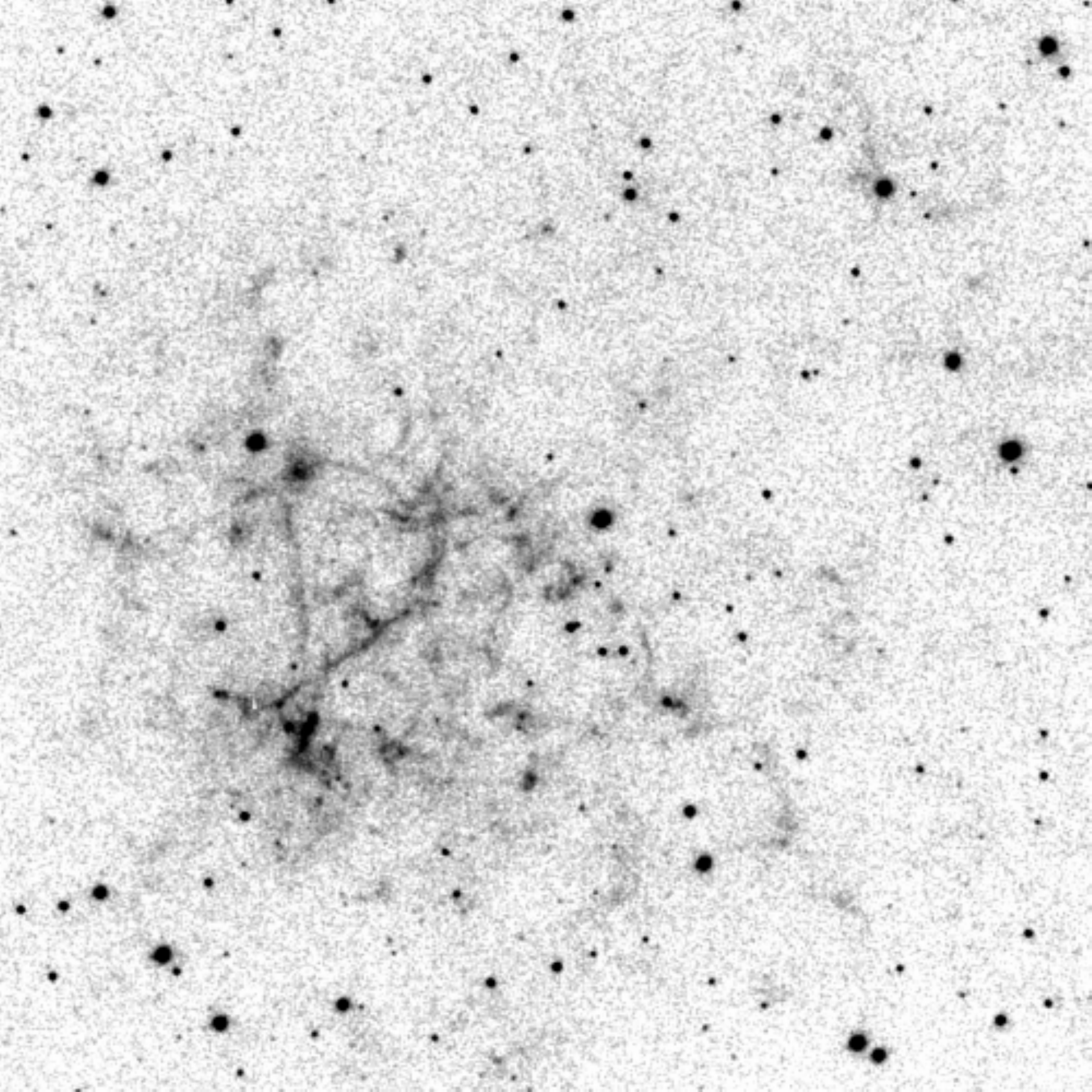}
\includegraphics[height=1.7in]{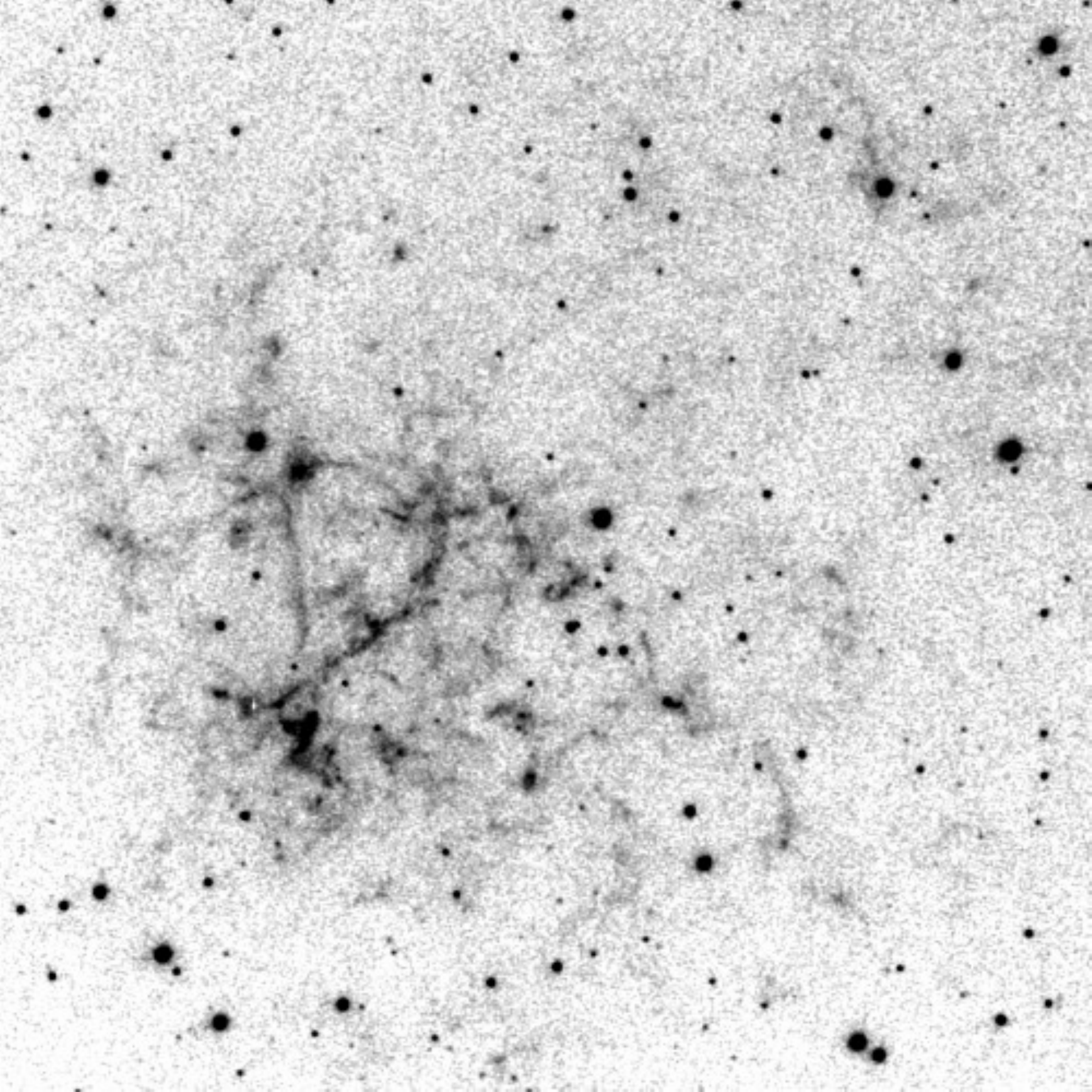}
\includegraphics[height=1.7in]{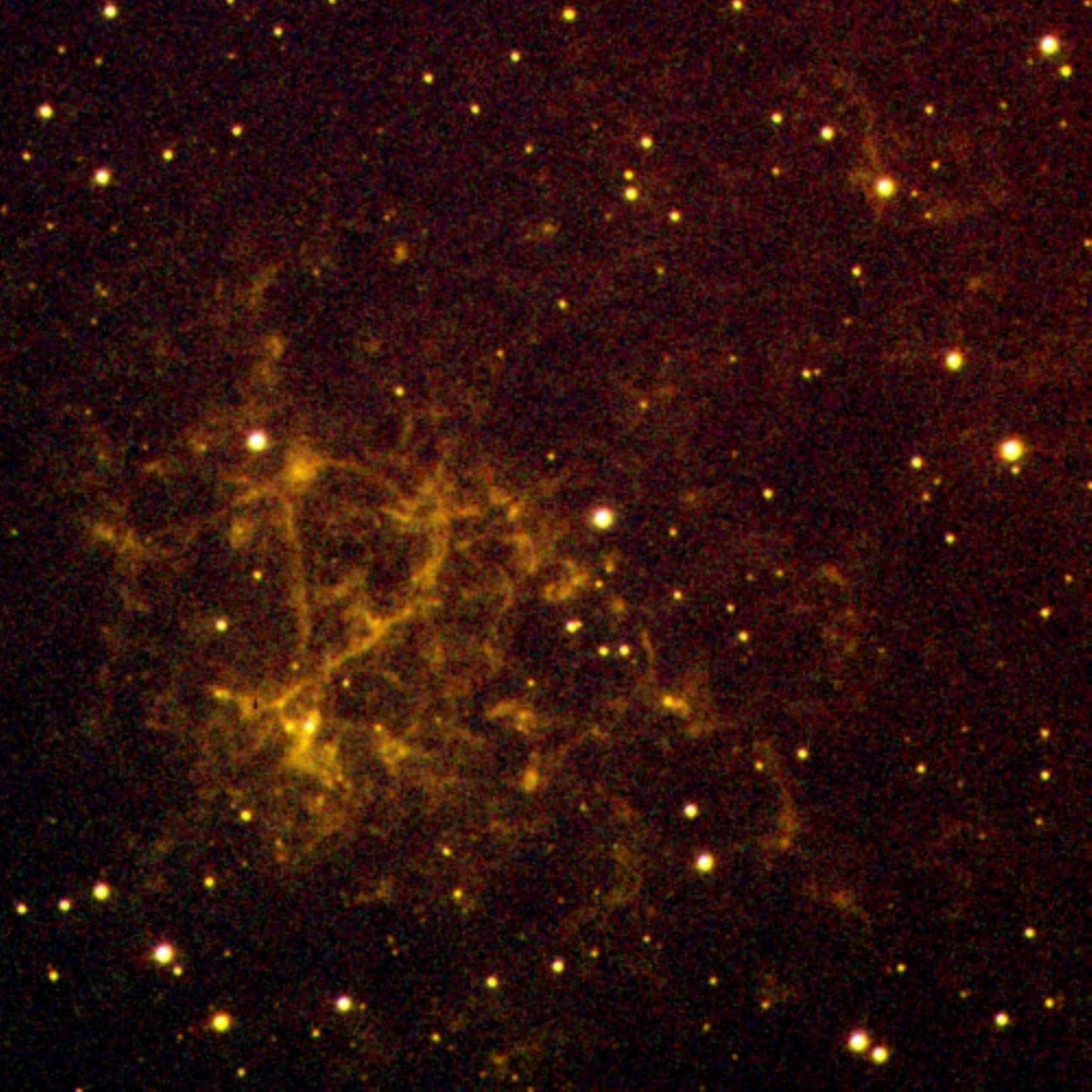}
\vskip .1in 
\includegraphics[height=1.7in]{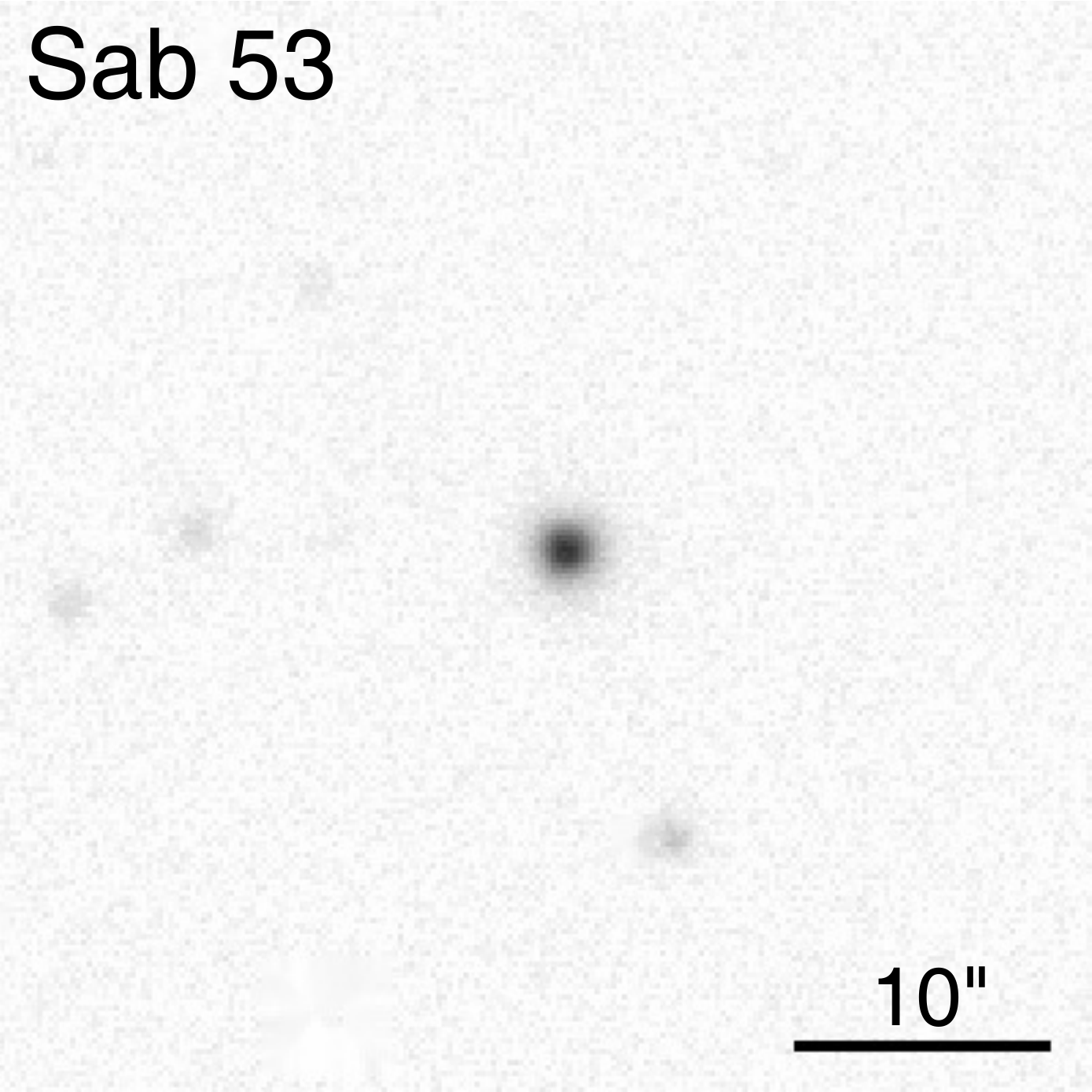} 
\includegraphics[height=1.7in]{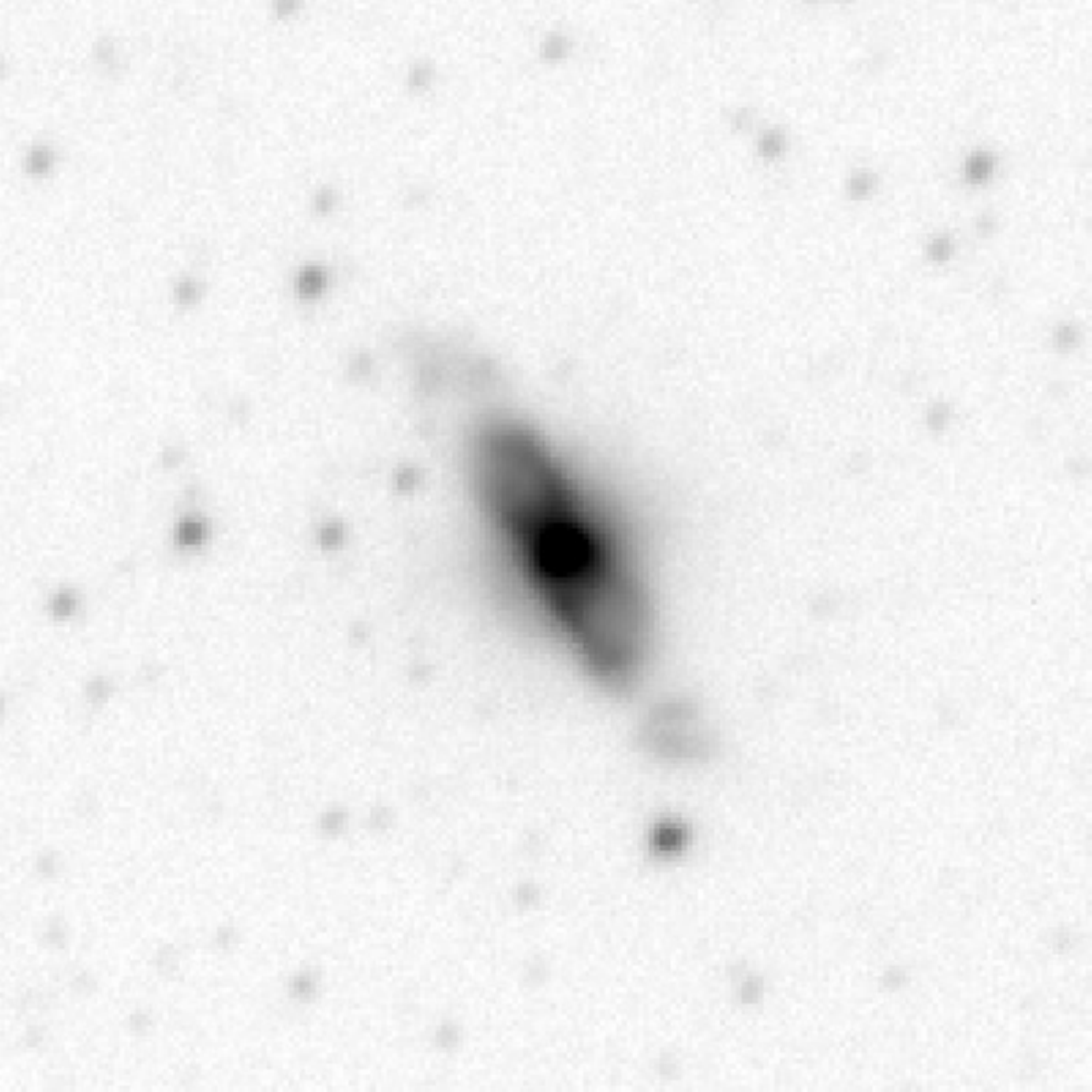}
\includegraphics[height=1.7in]{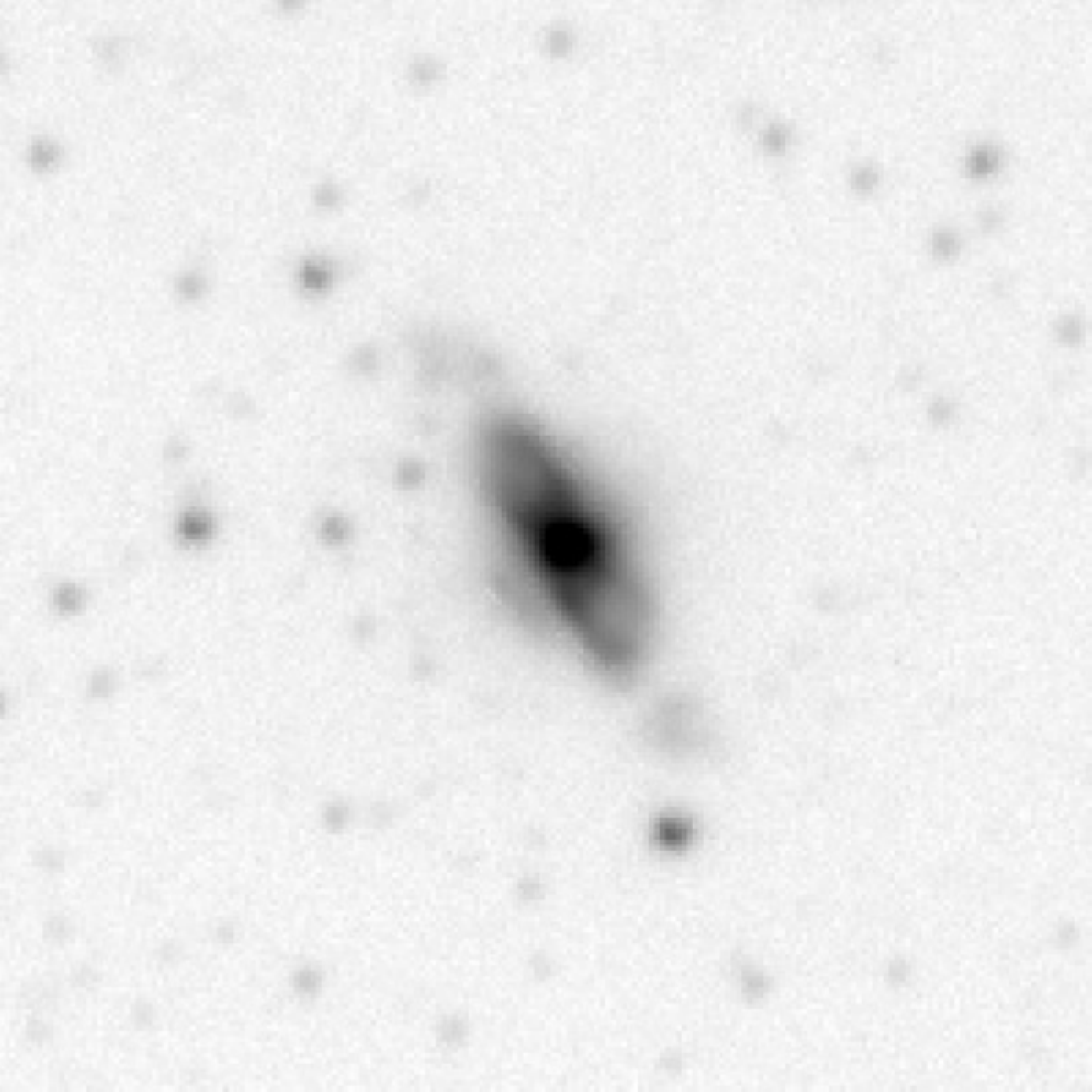}
\includegraphics[height=1.7in]{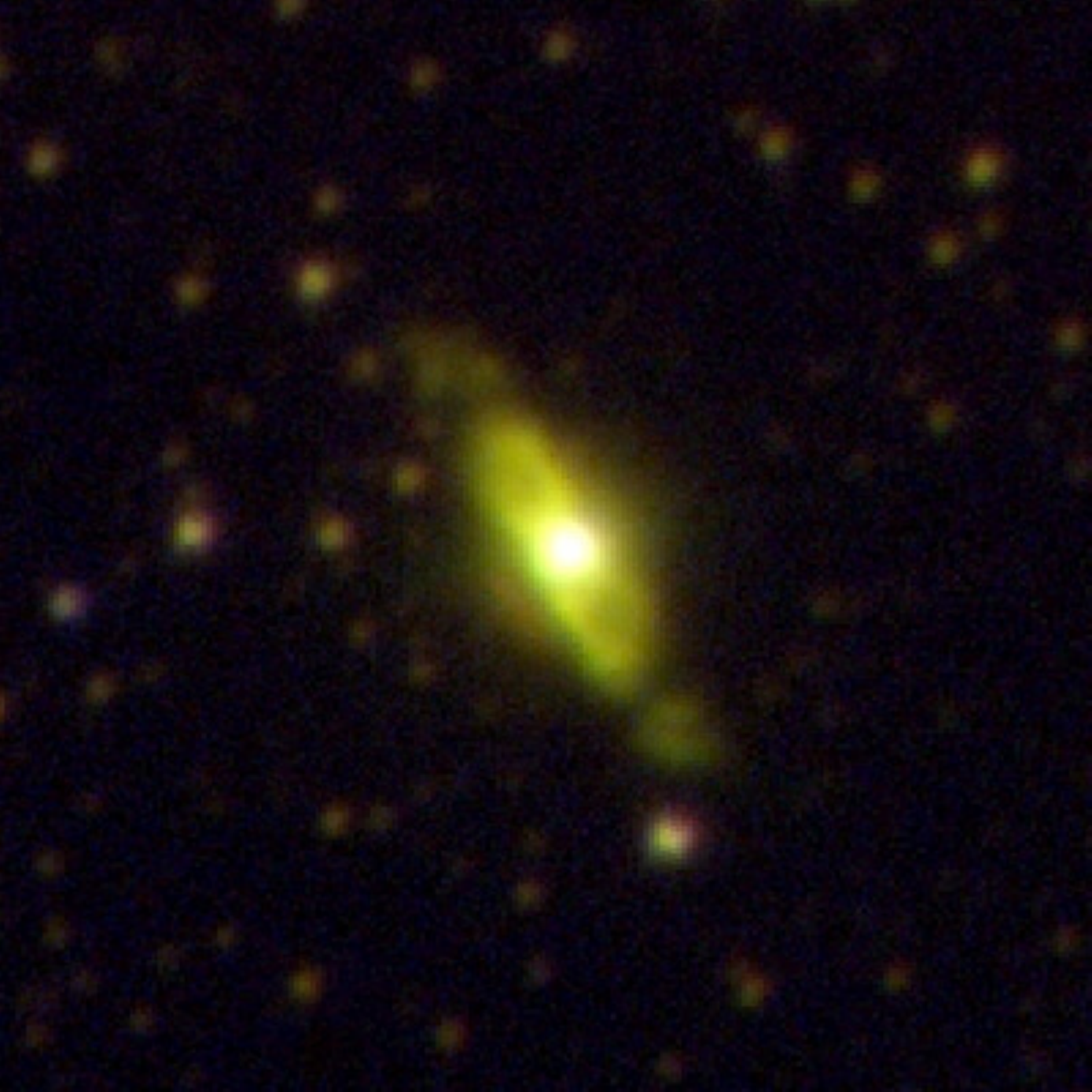}
\vskip .1in
\includegraphics[height=1.7in]{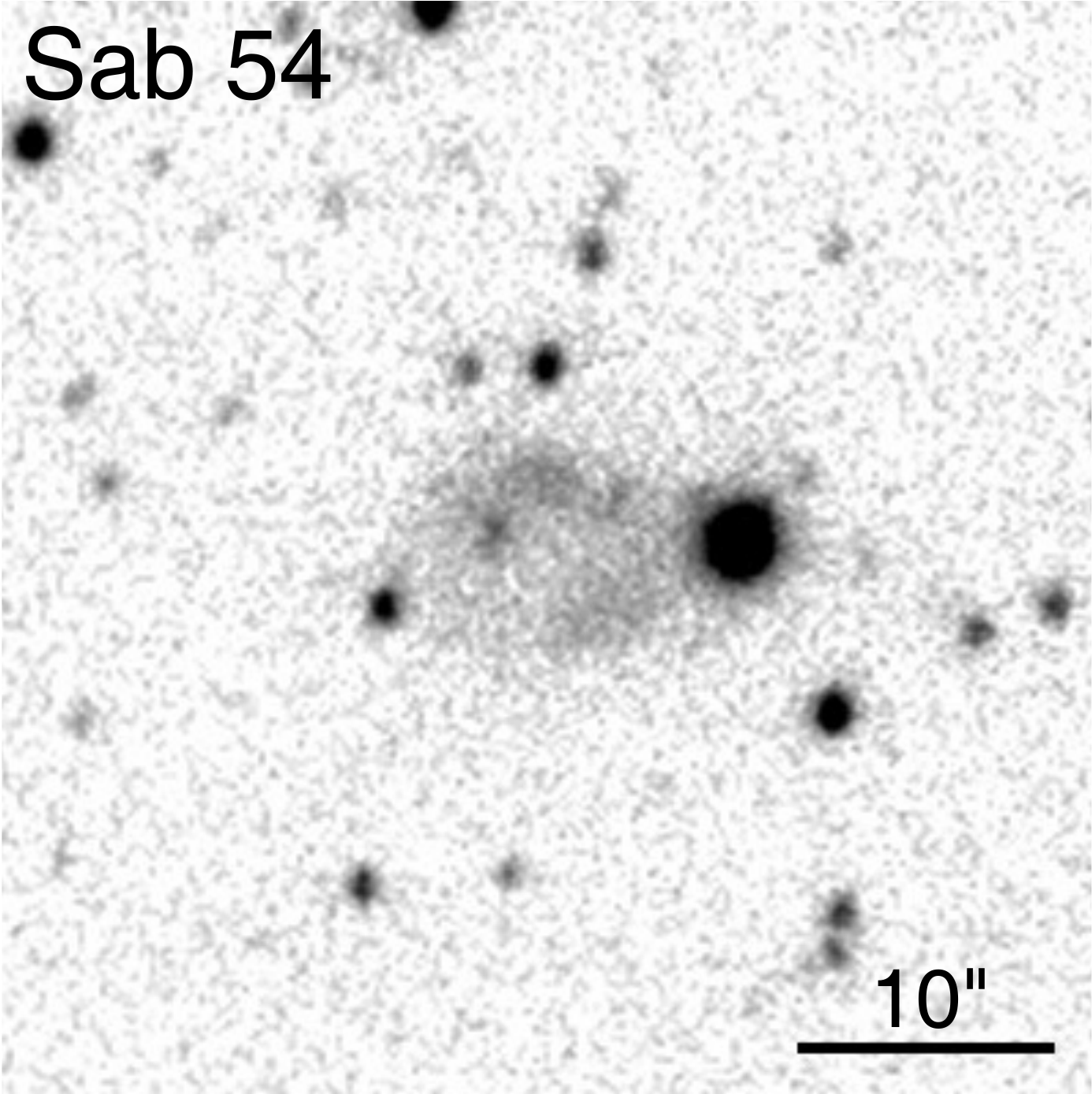} 
\includegraphics[height=1.7in]{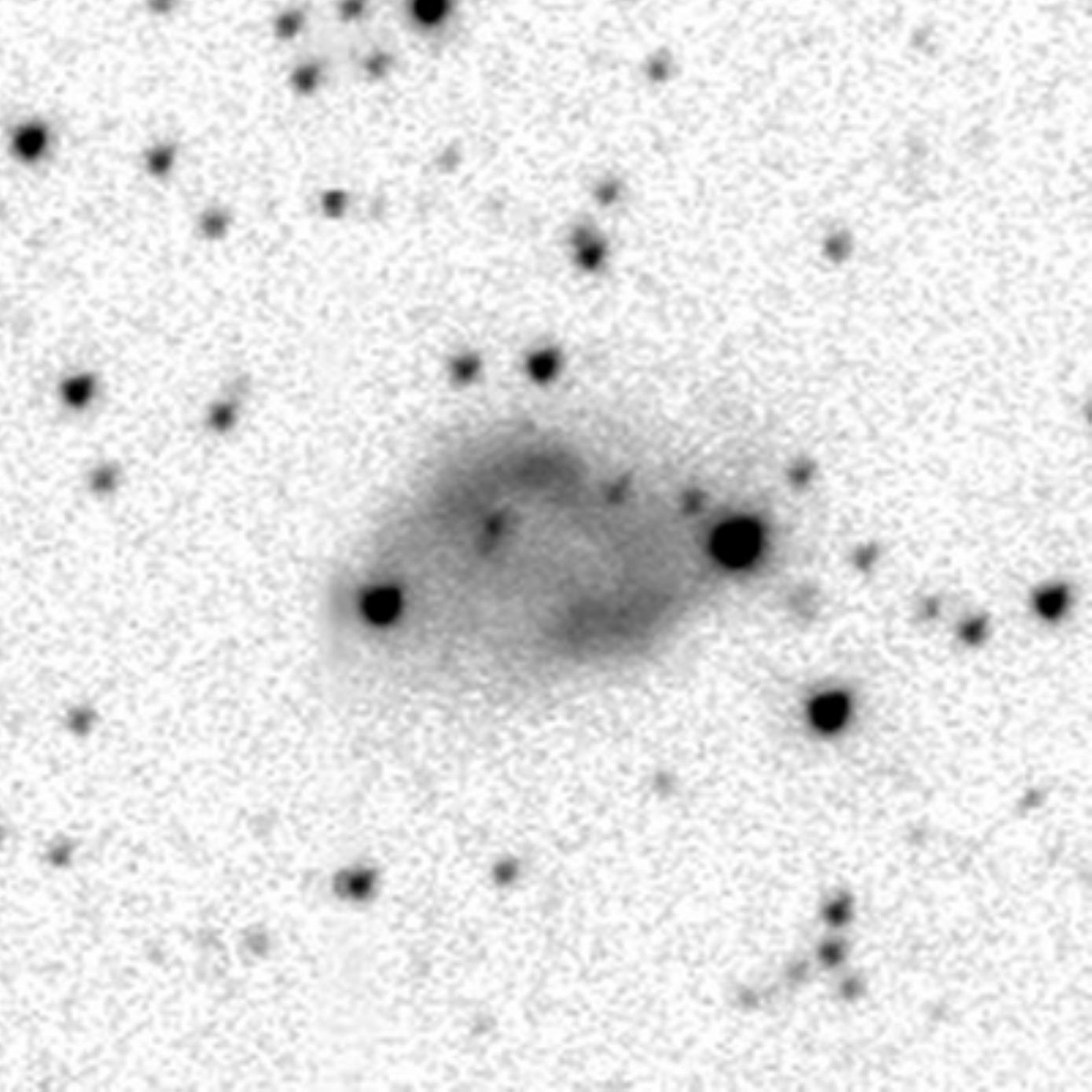}
\includegraphics[height=1.7in]{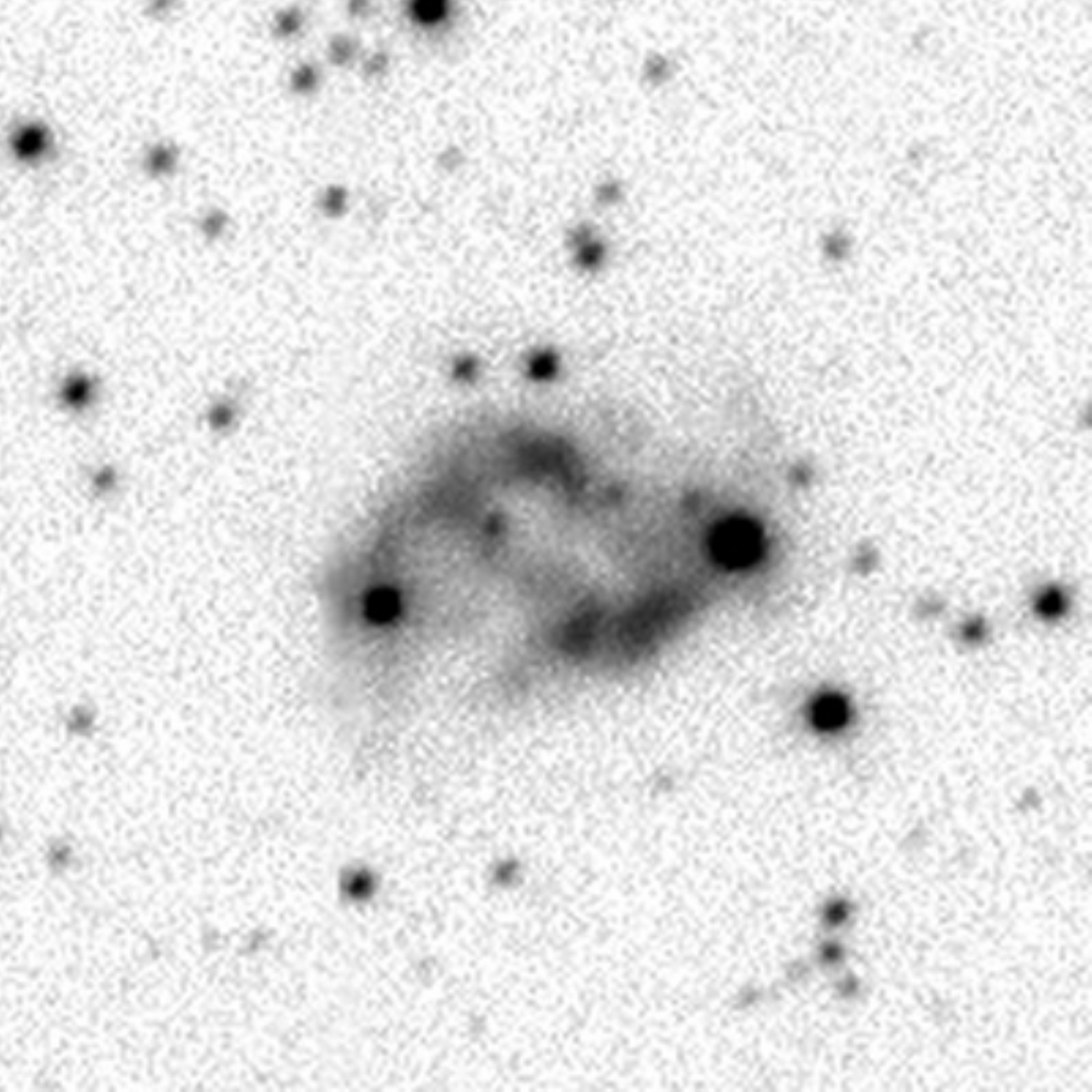}
\includegraphics[height=1.7in]{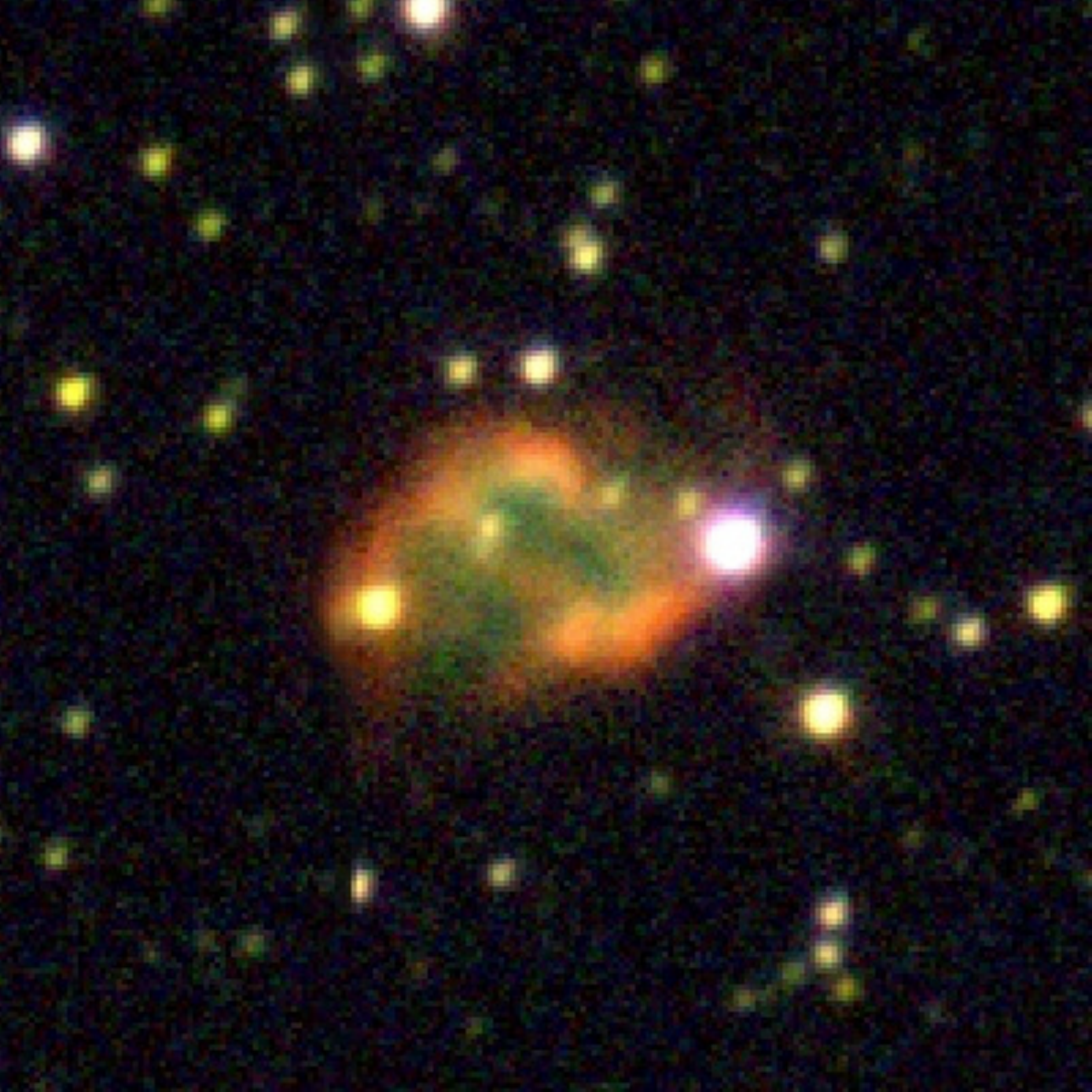}
\caption{Same as Figure~\ref{1.img}. } 
\label{4.img} 
\end{figure*}

\begin{figure*} 
\centering 
\includegraphics[height=1.7in]{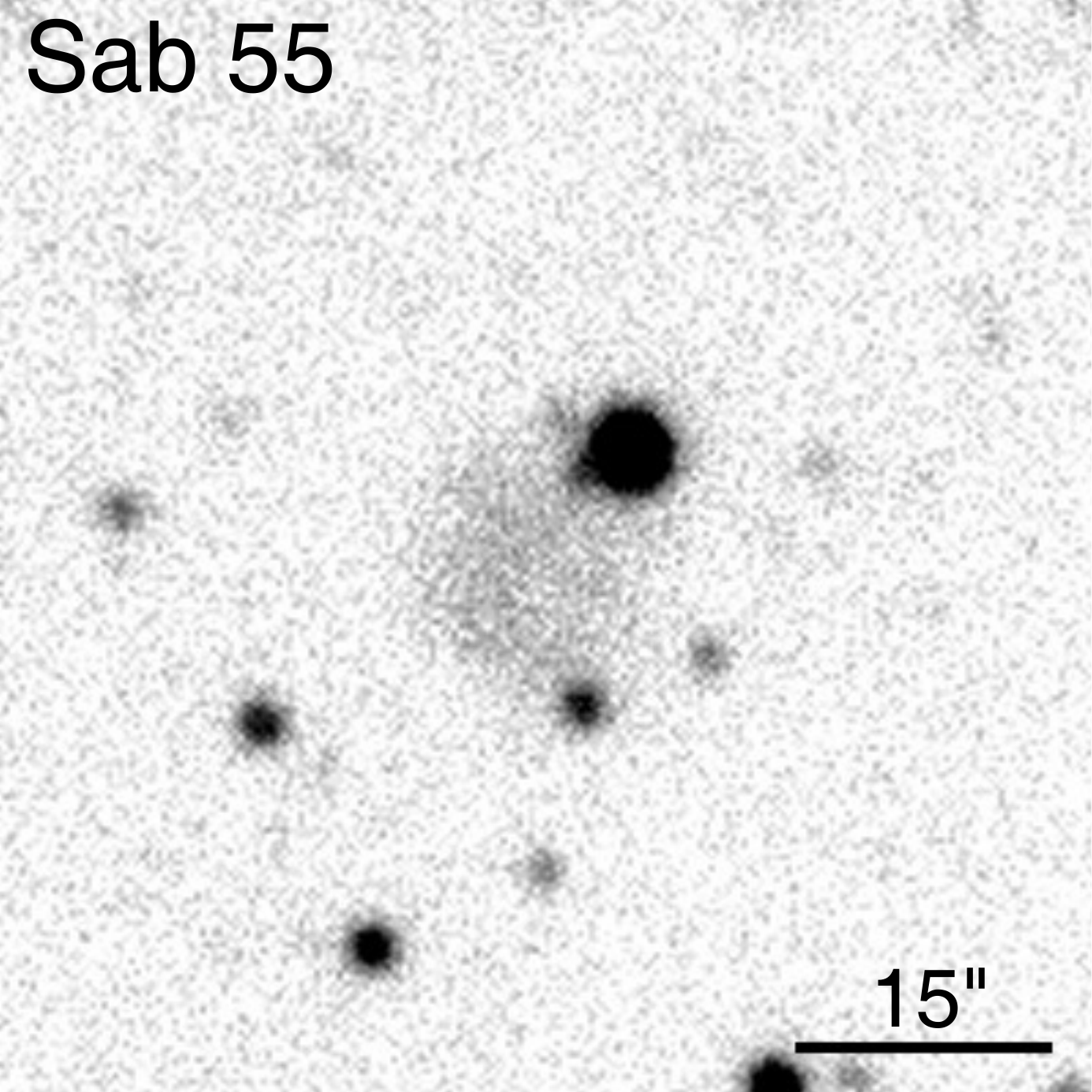} 
\includegraphics[height=1.7in]{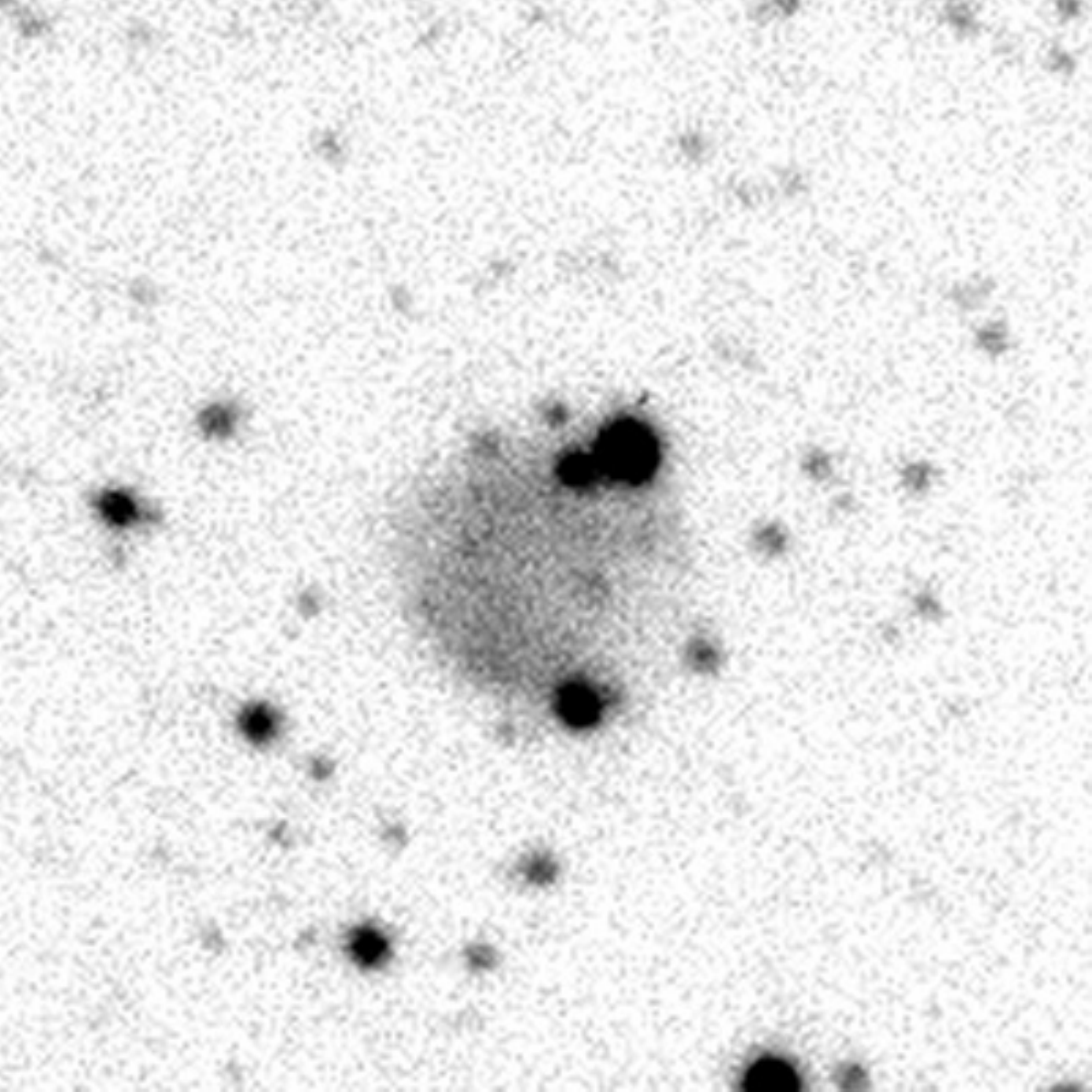}
\includegraphics[height=1.7in]{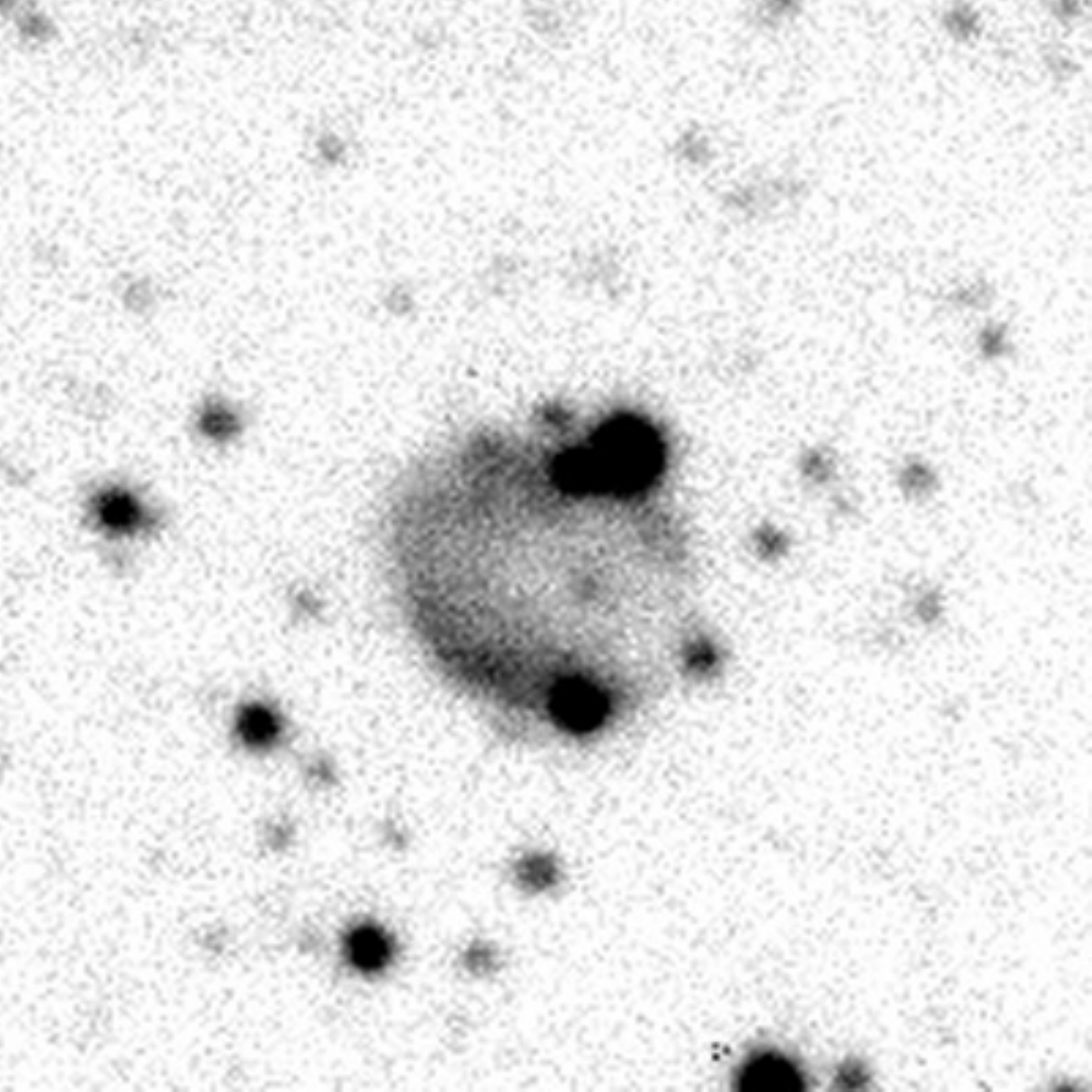}
\includegraphics[height=1.7in]{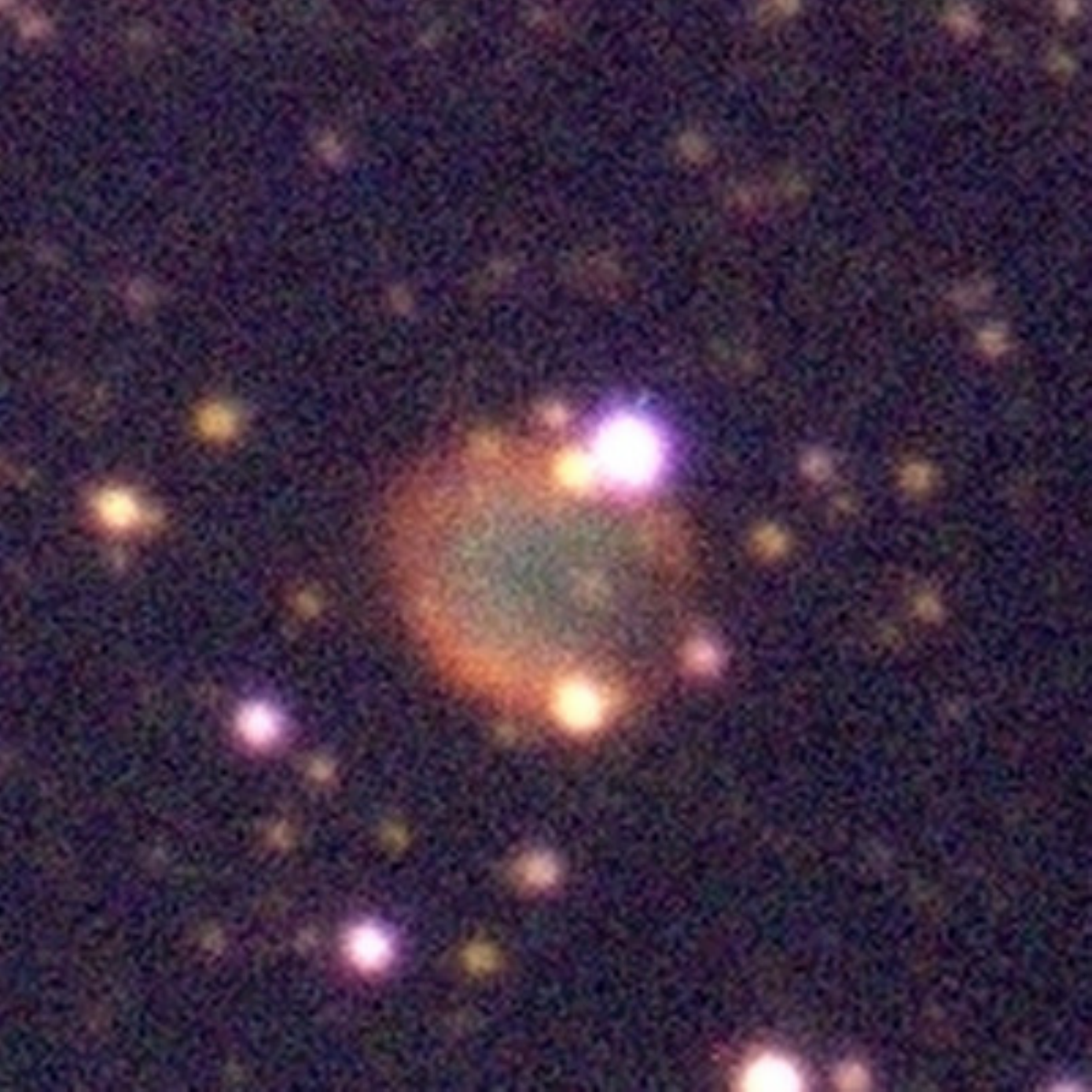}
\vskip .1in 
\includegraphics[height=1.7in]{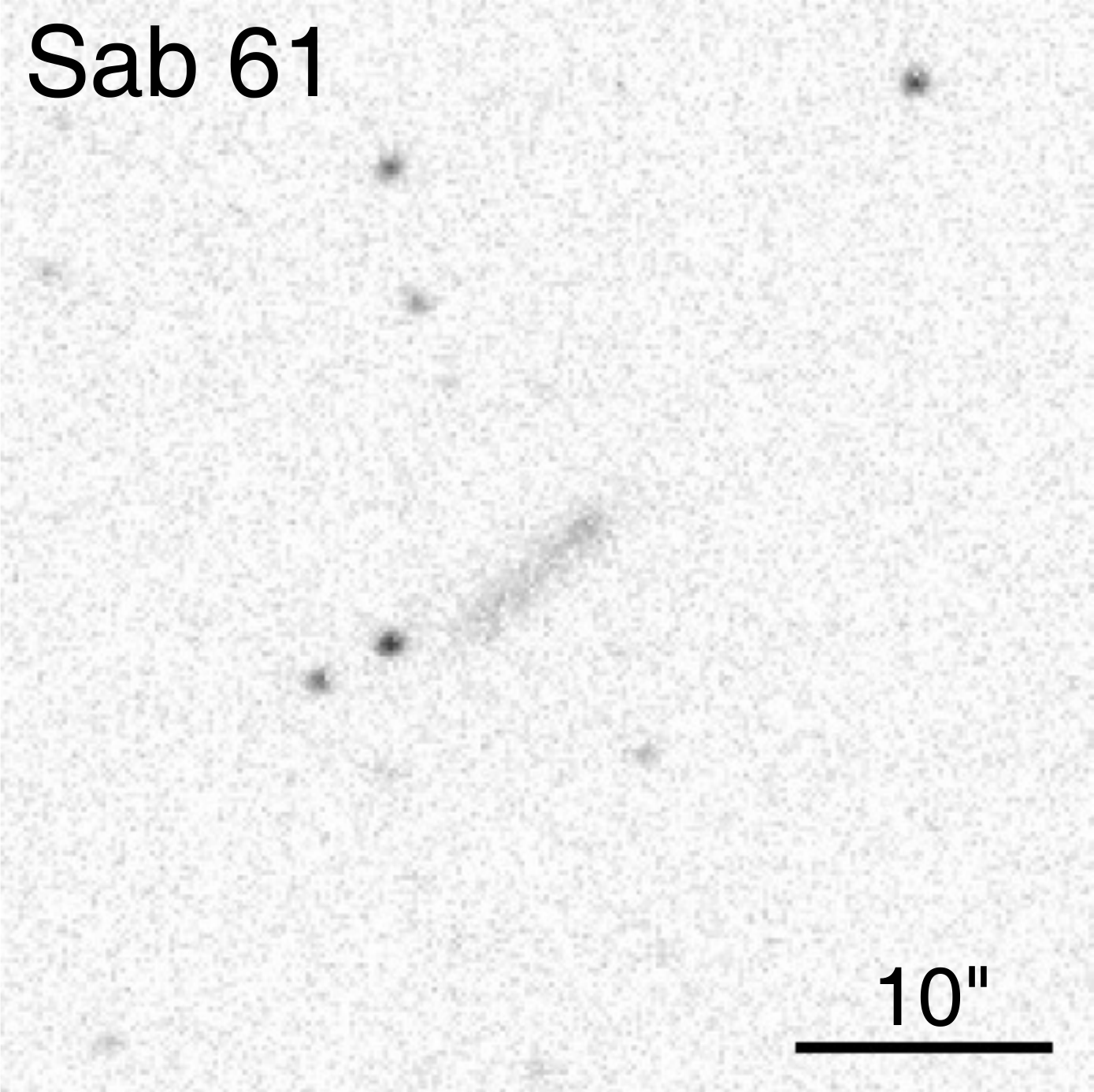} 
\includegraphics[height=1.7in]{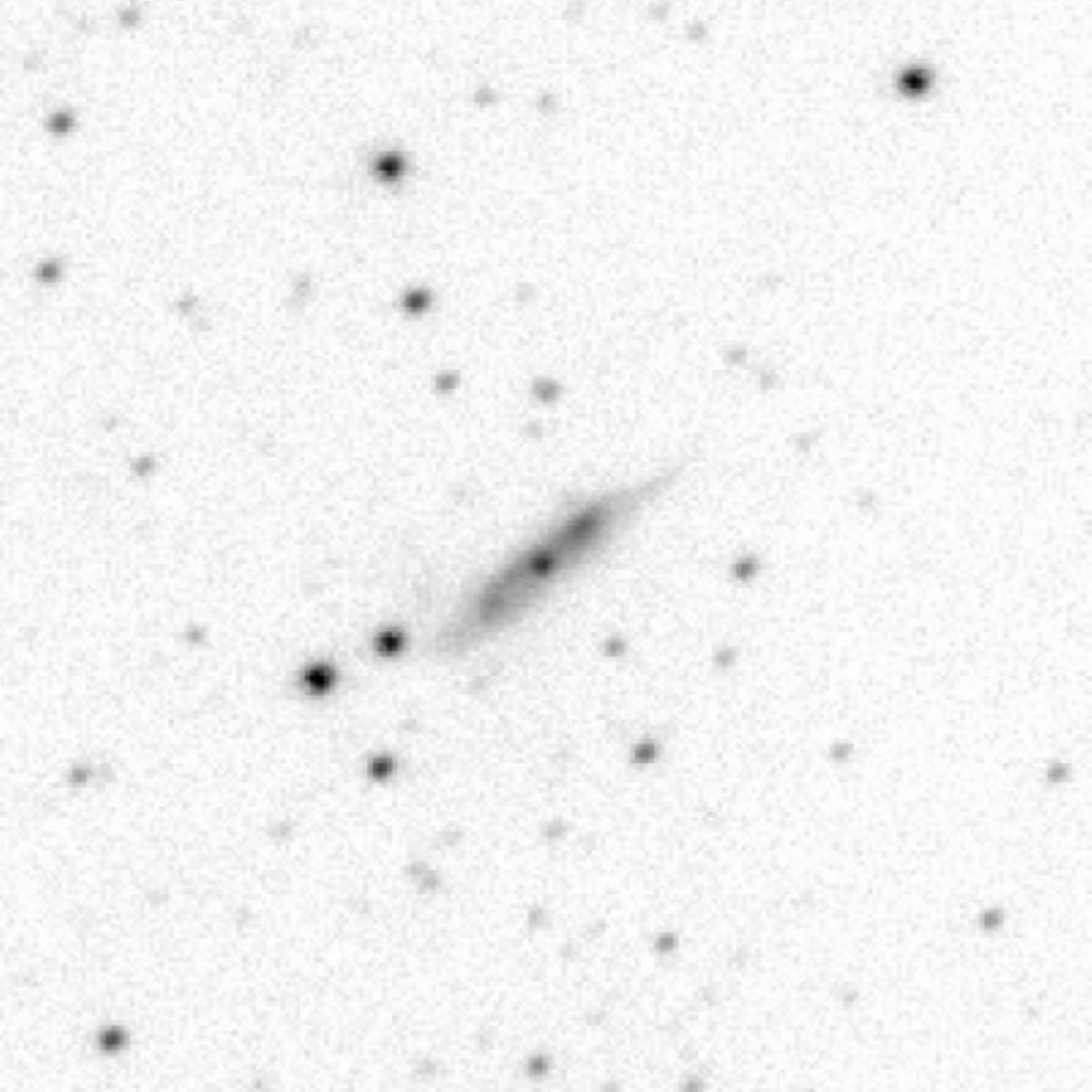}
\includegraphics[height=1.7in]{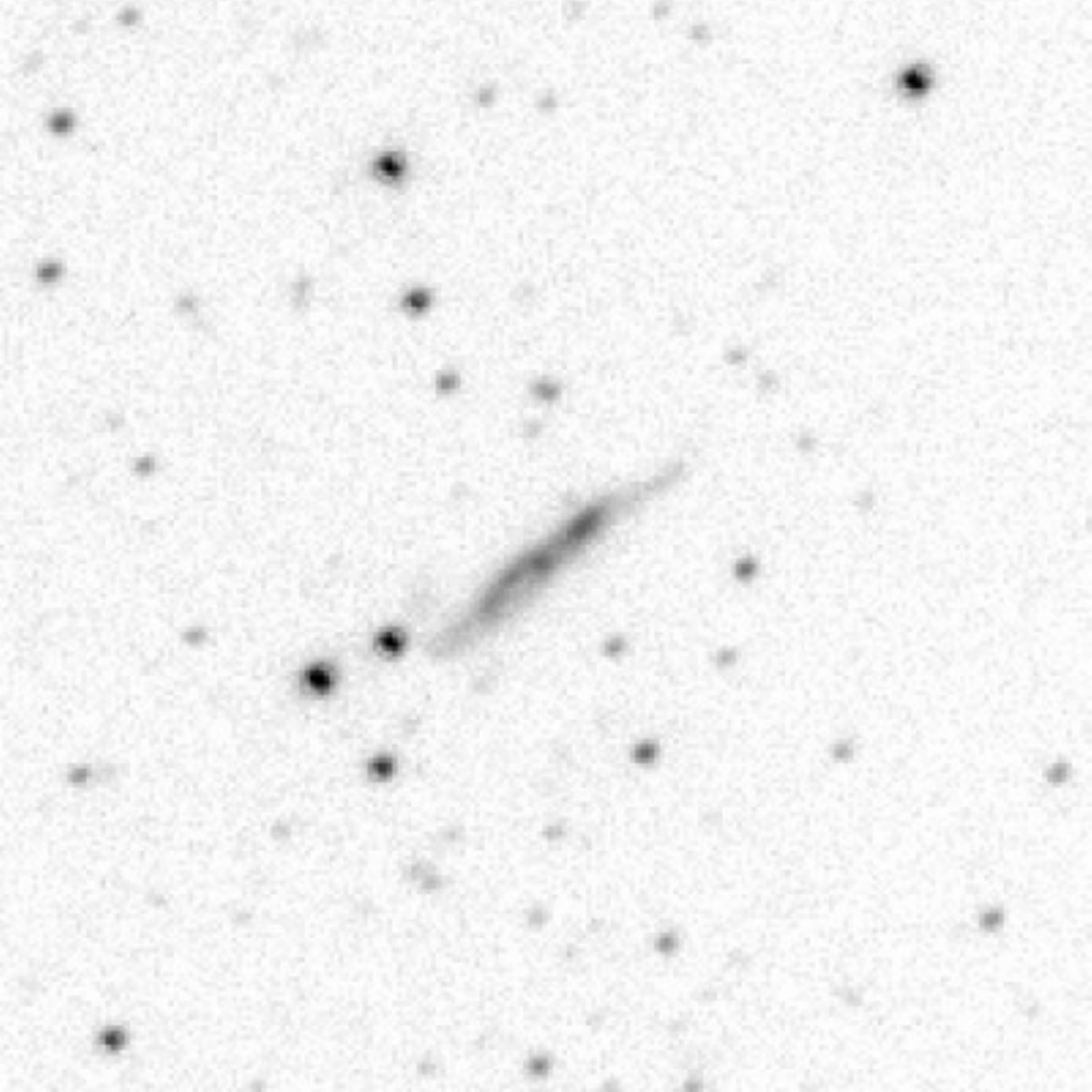}
\includegraphics[height=1.7in]{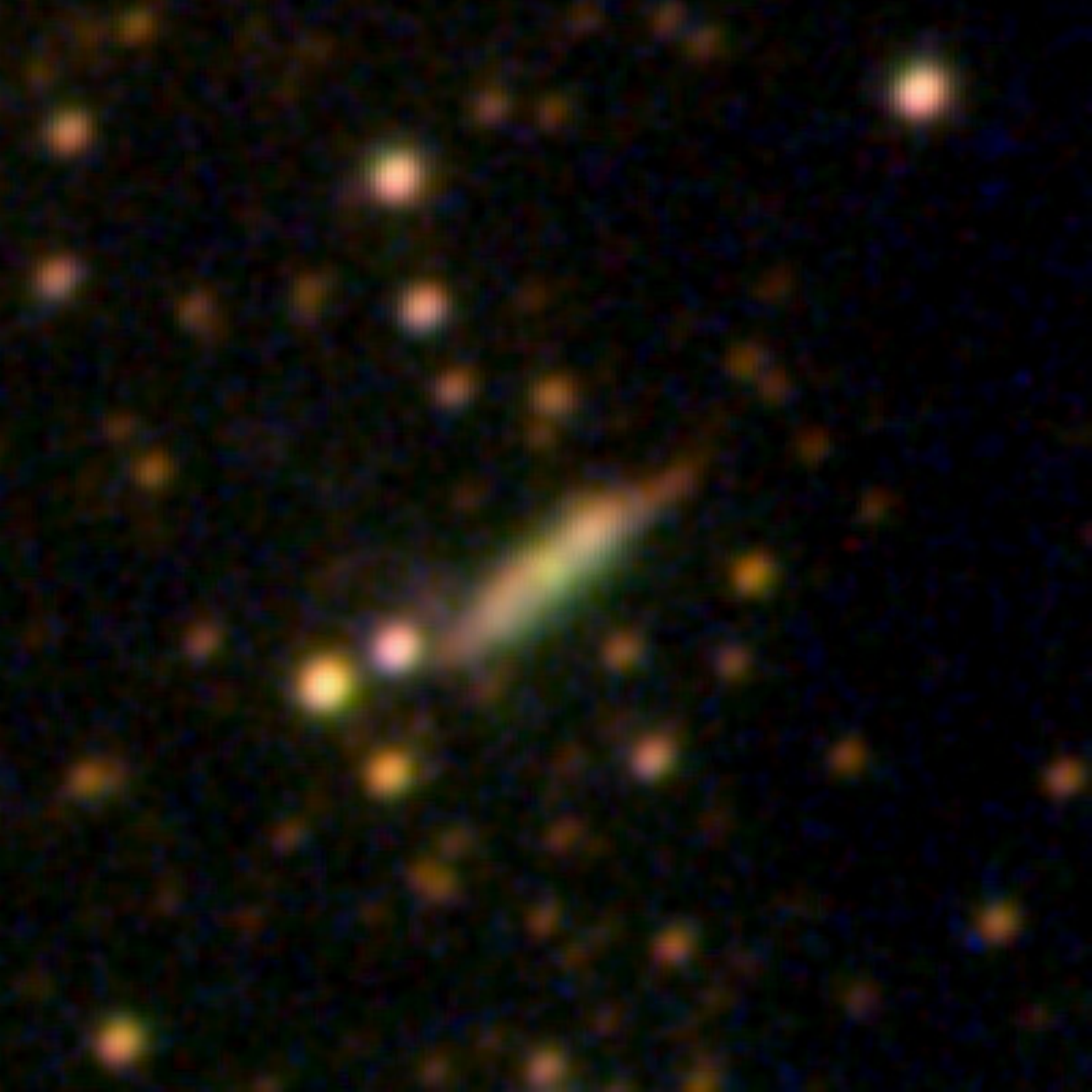}
\vskip .1in 
\includegraphics[height=1.7in]{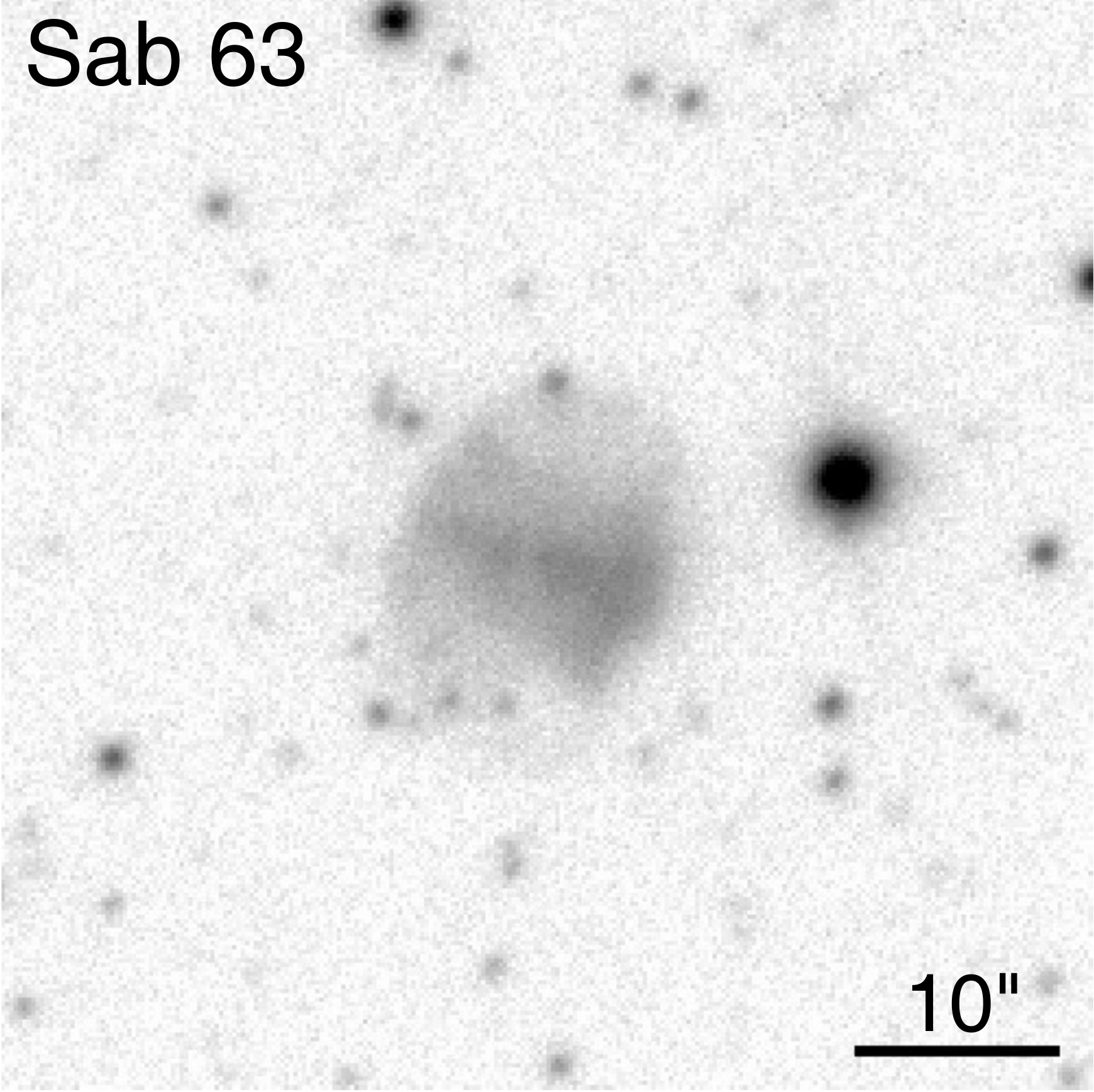} 
\includegraphics[height=1.7in]{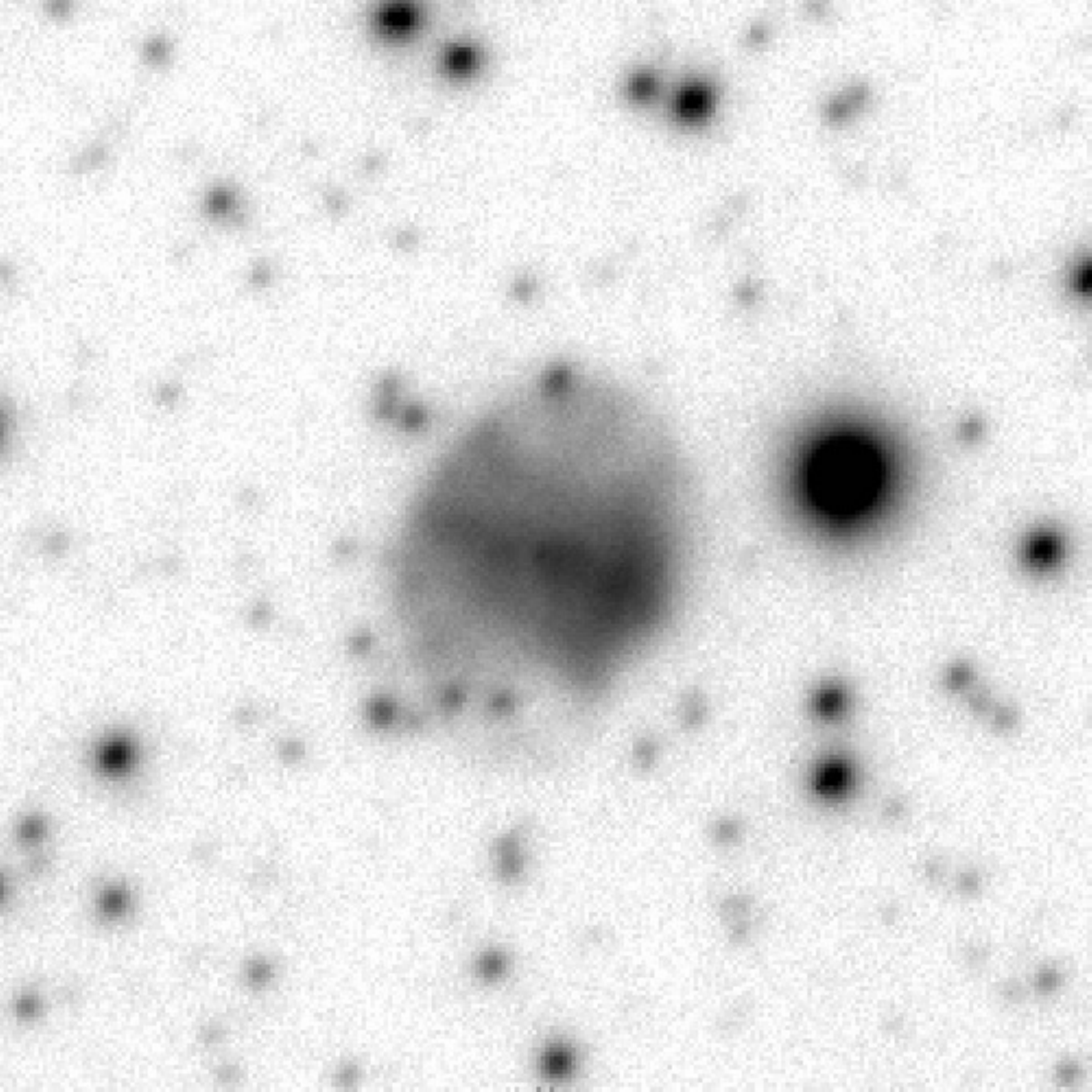}
\includegraphics[height=1.7in]{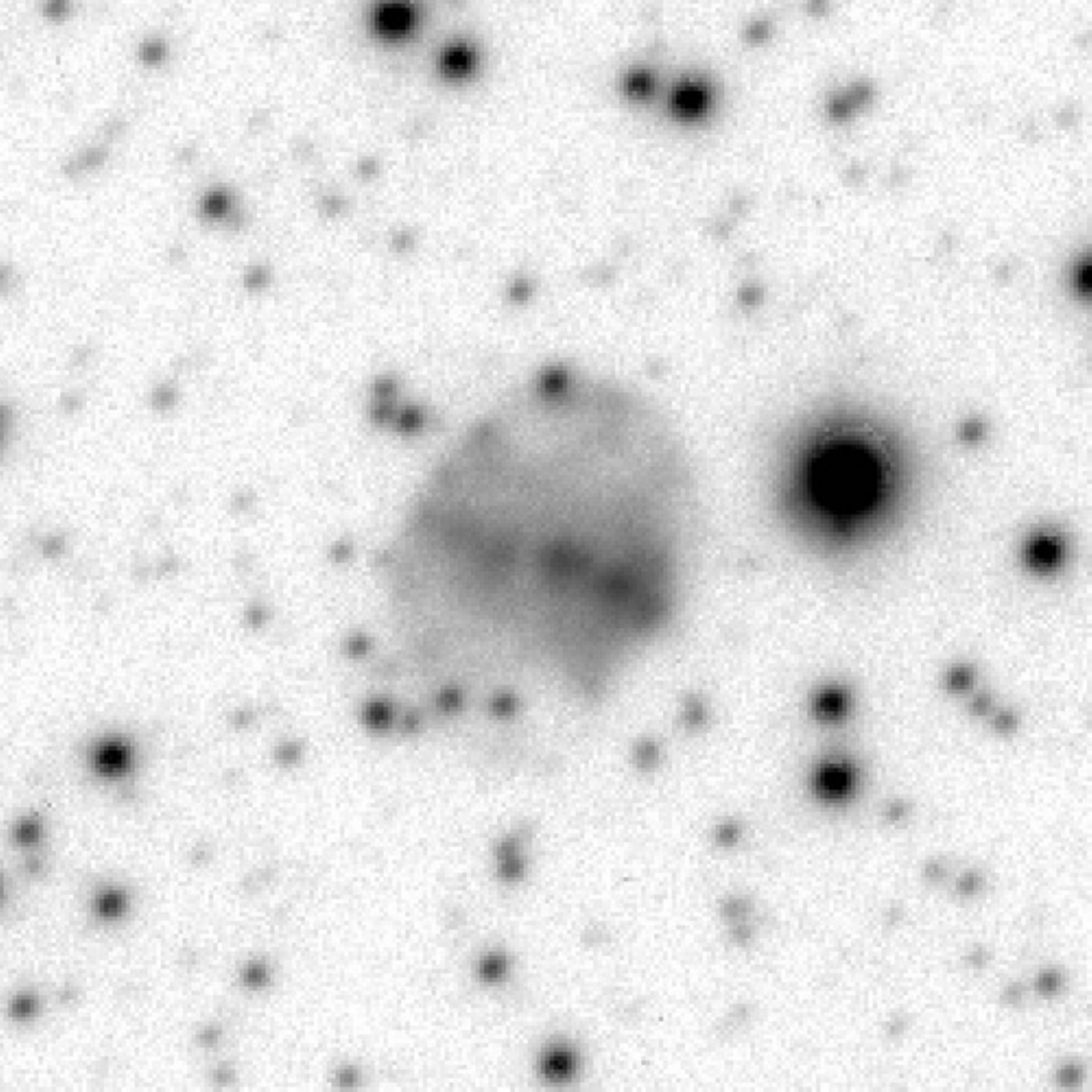}
\includegraphics[height=1.7in]{63_G045.eps}
\vskip .1in 
\includegraphics[height=1.7in]{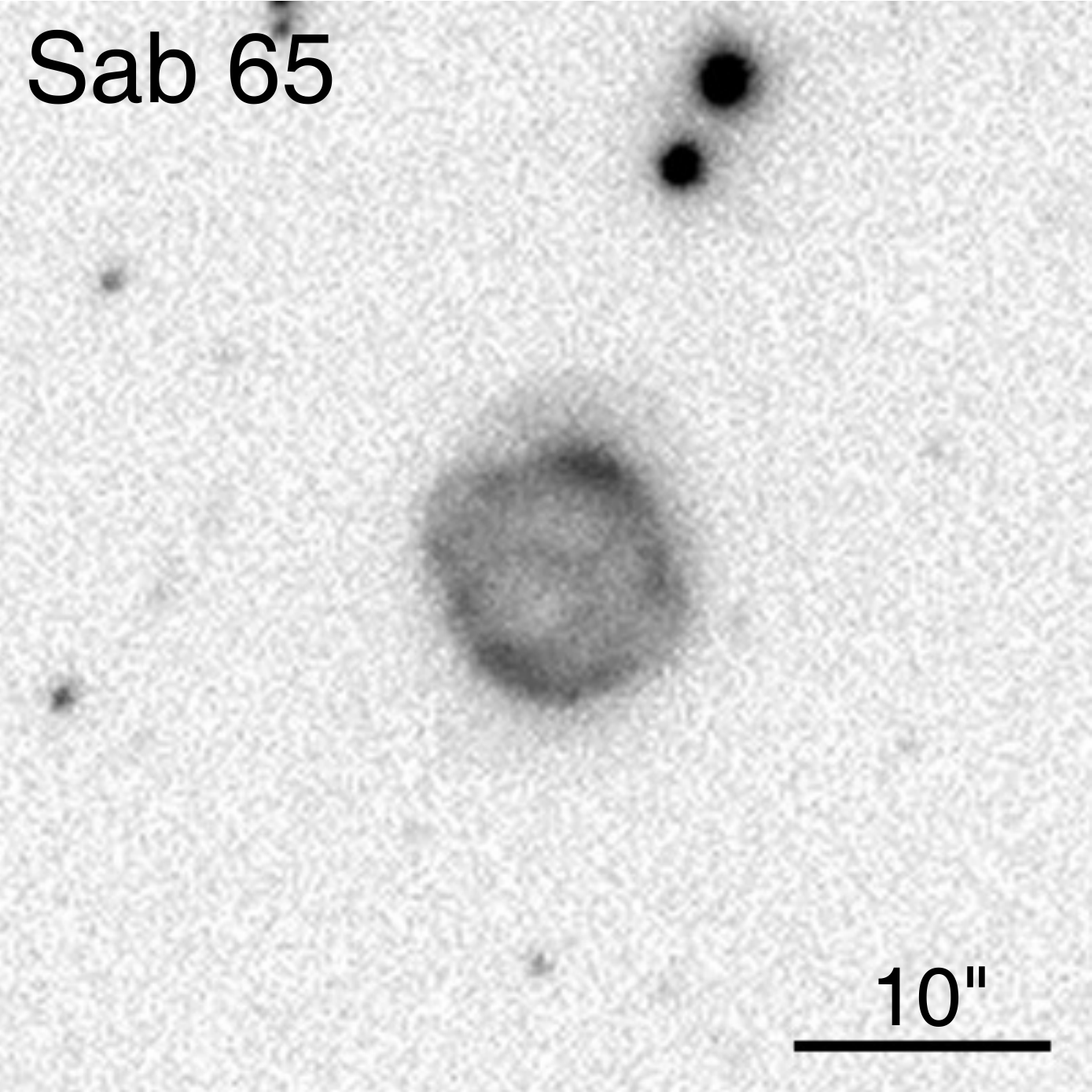} 
\includegraphics[height=1.7in]{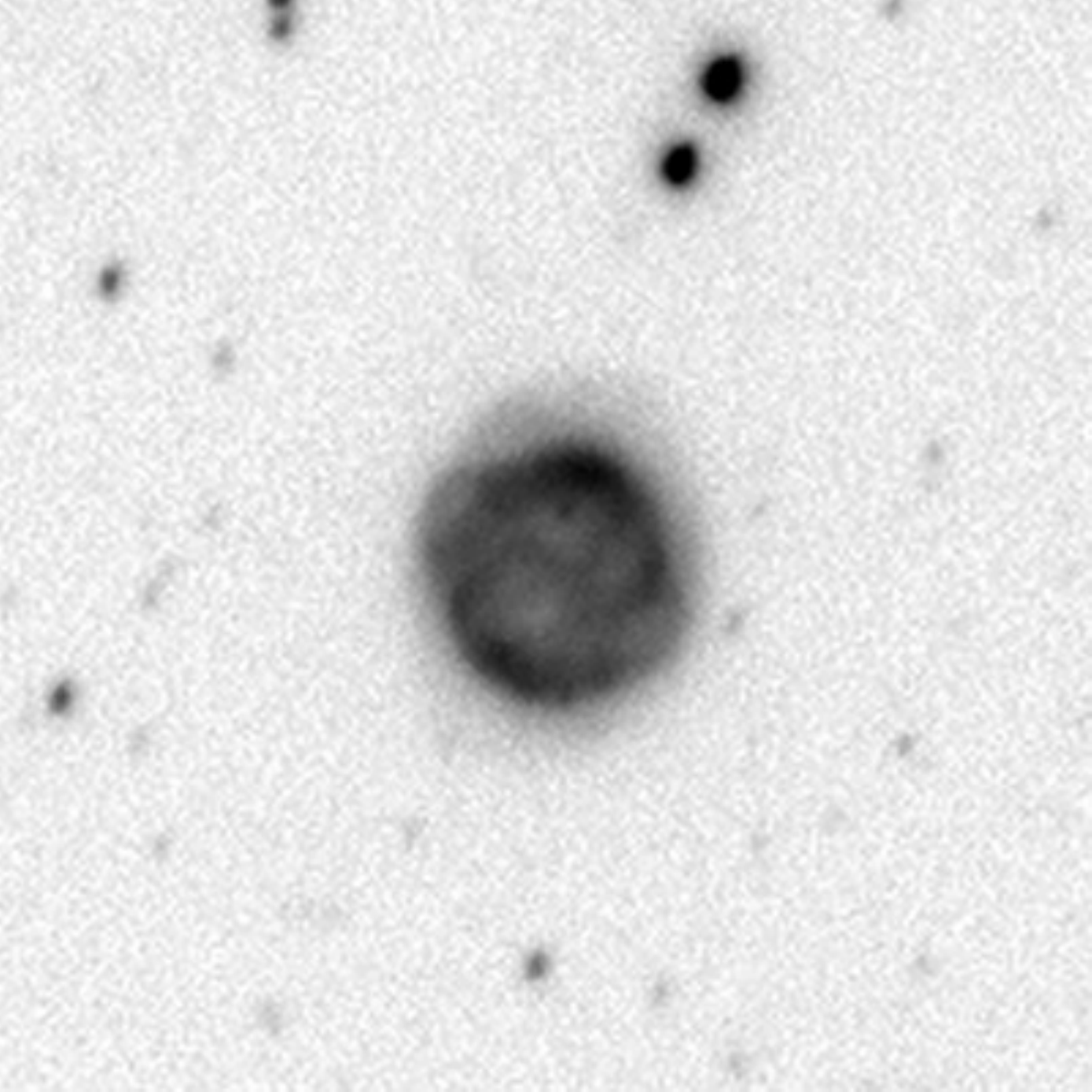}
\includegraphics[height=1.7in]{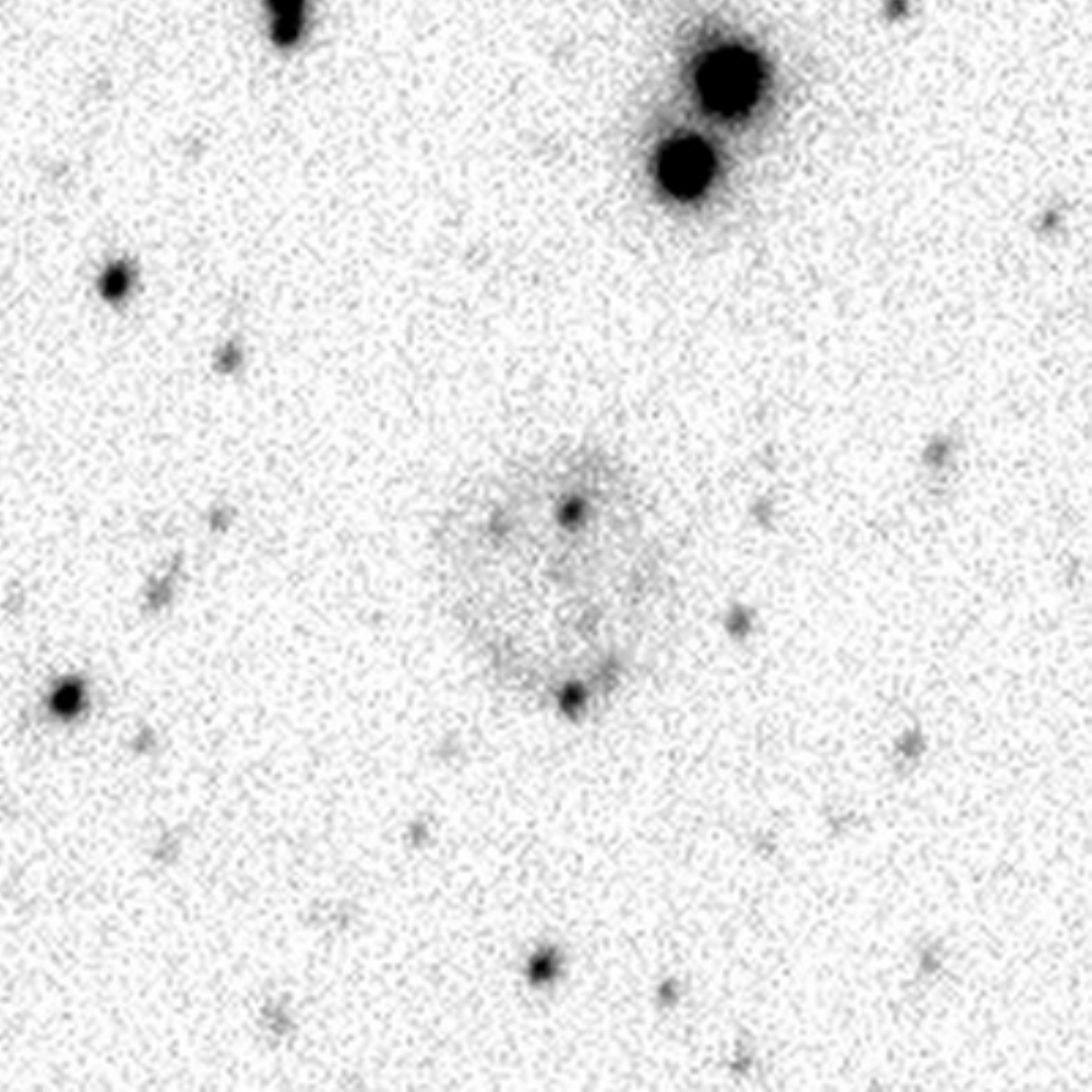}
\includegraphics[height=1.7in]{65_G044.eps}
\vskip .1in 
\includegraphics[height=1.7in]{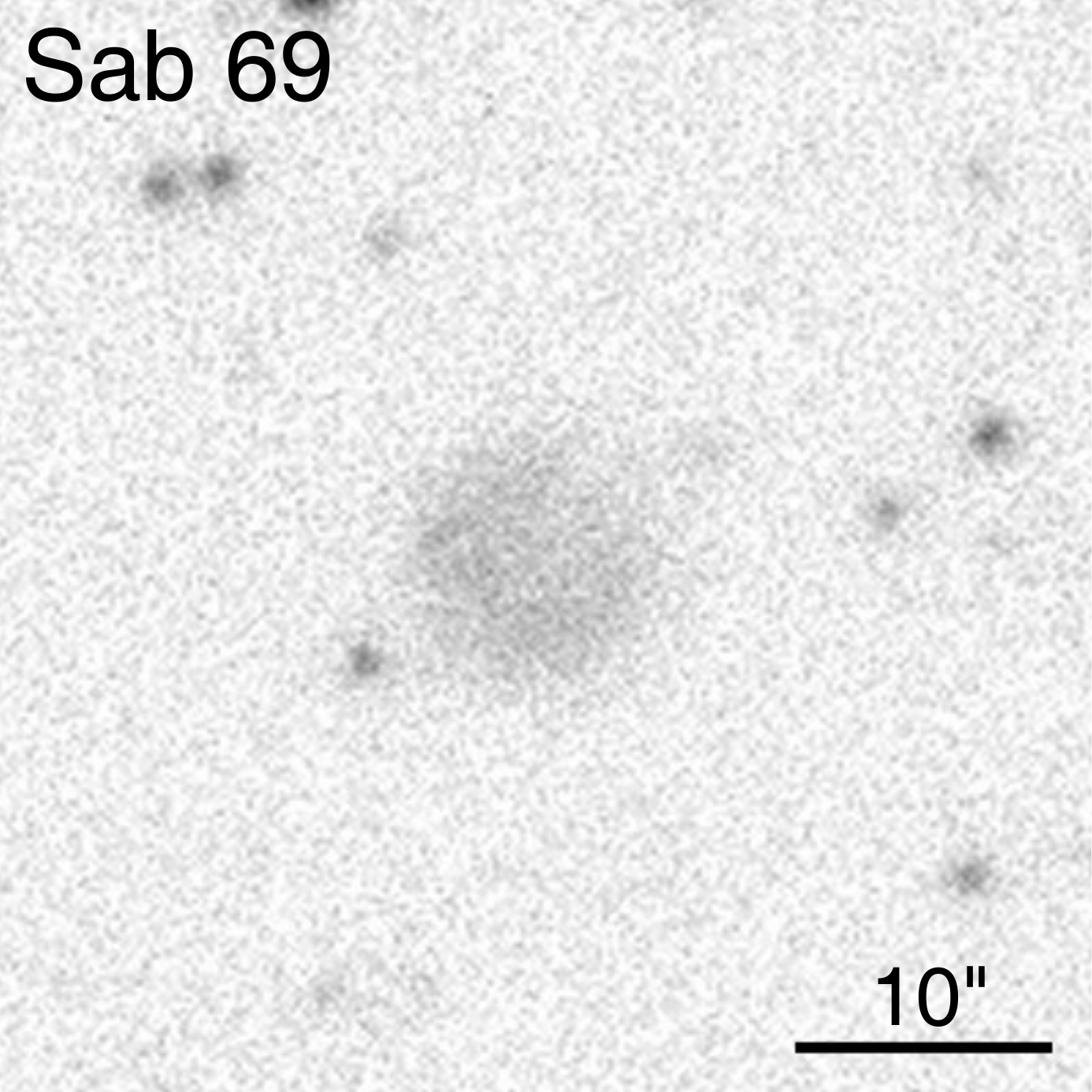} 
\includegraphics[height=1.7in]{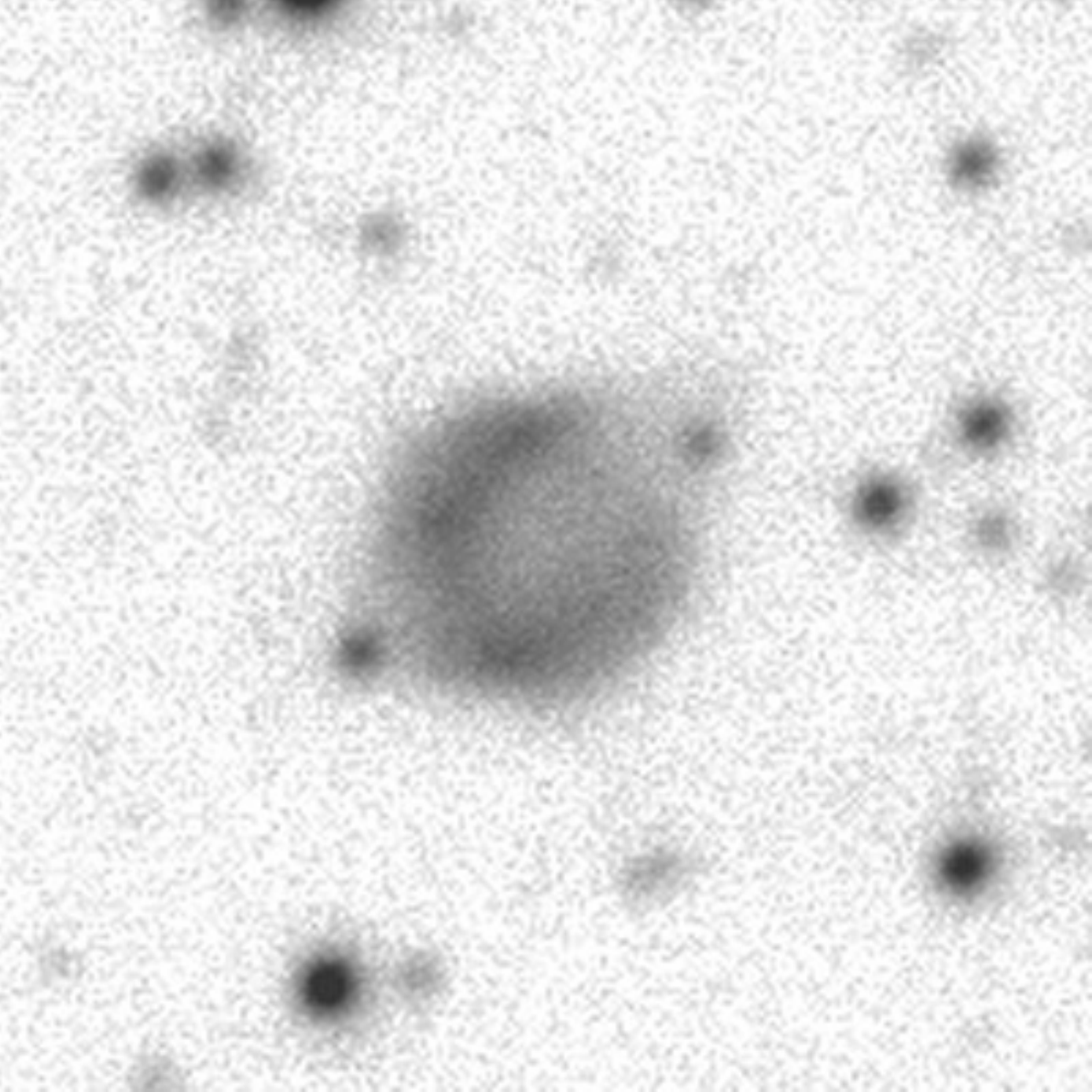}
\includegraphics[height=1.7in]{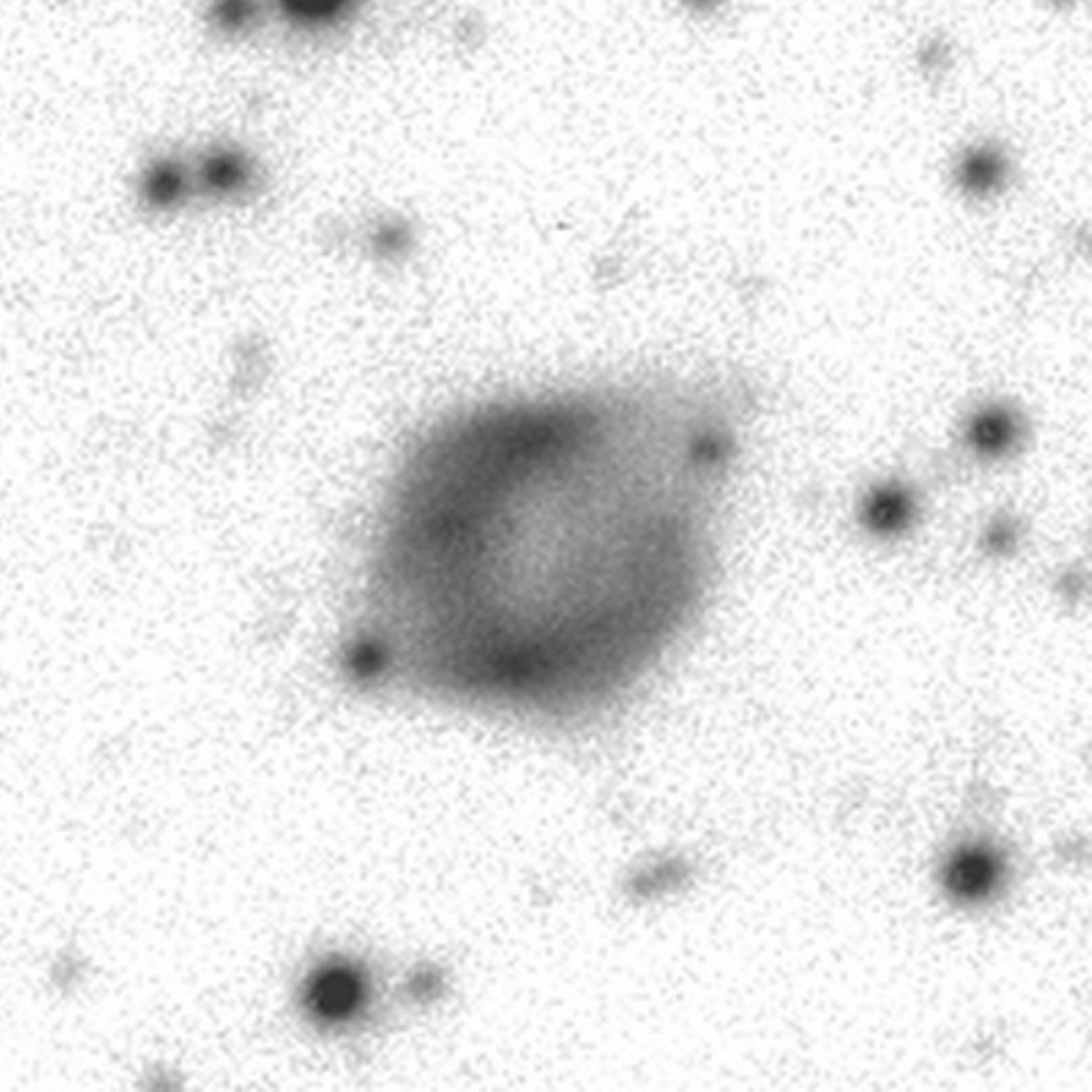}
\includegraphics[height=1.7in]{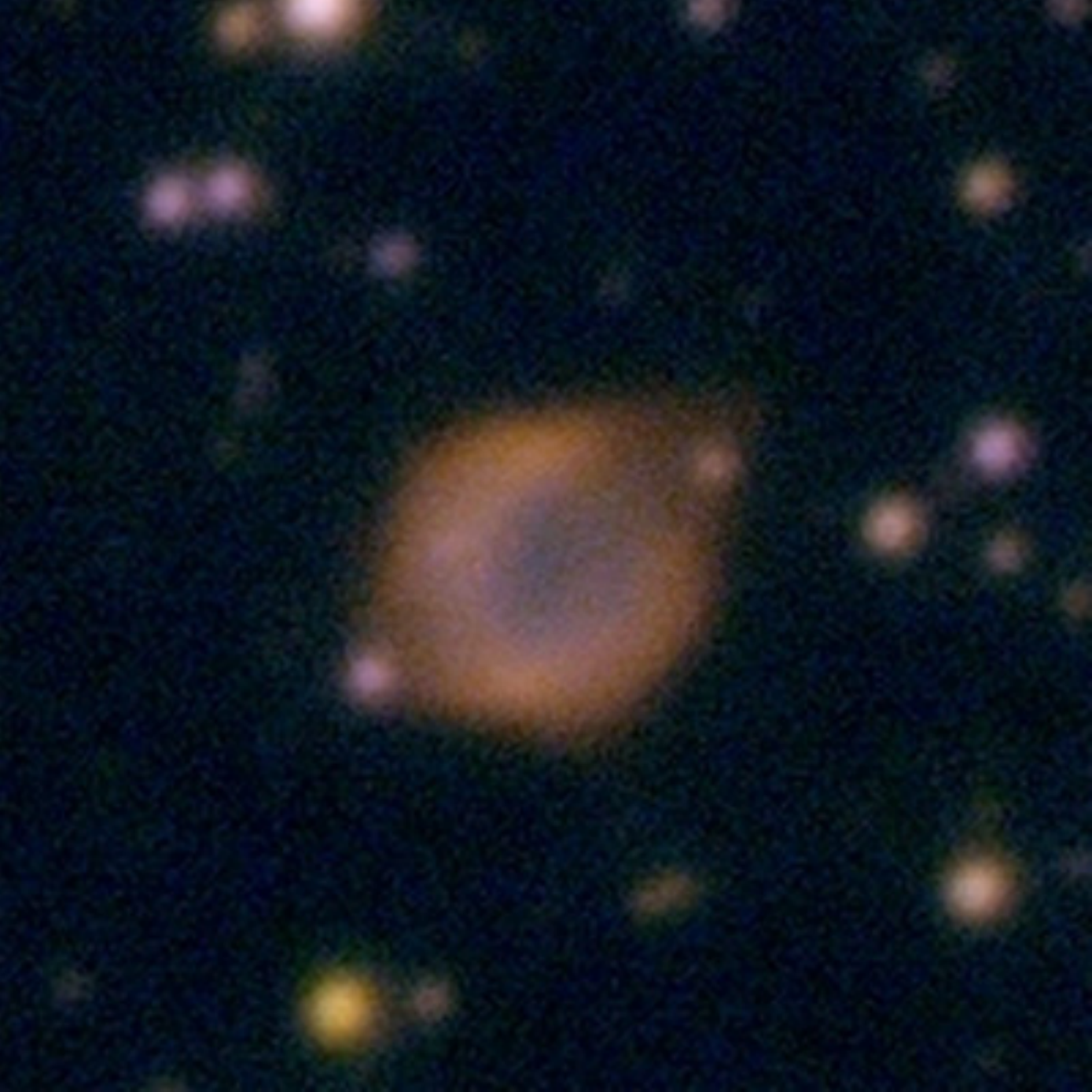}
\caption{Same as Figure~\ref{1.img}. } 
\label{5.img} 
\end{figure*}

\begin{figure*} 
\centering 
\includegraphics[height=1.7in]{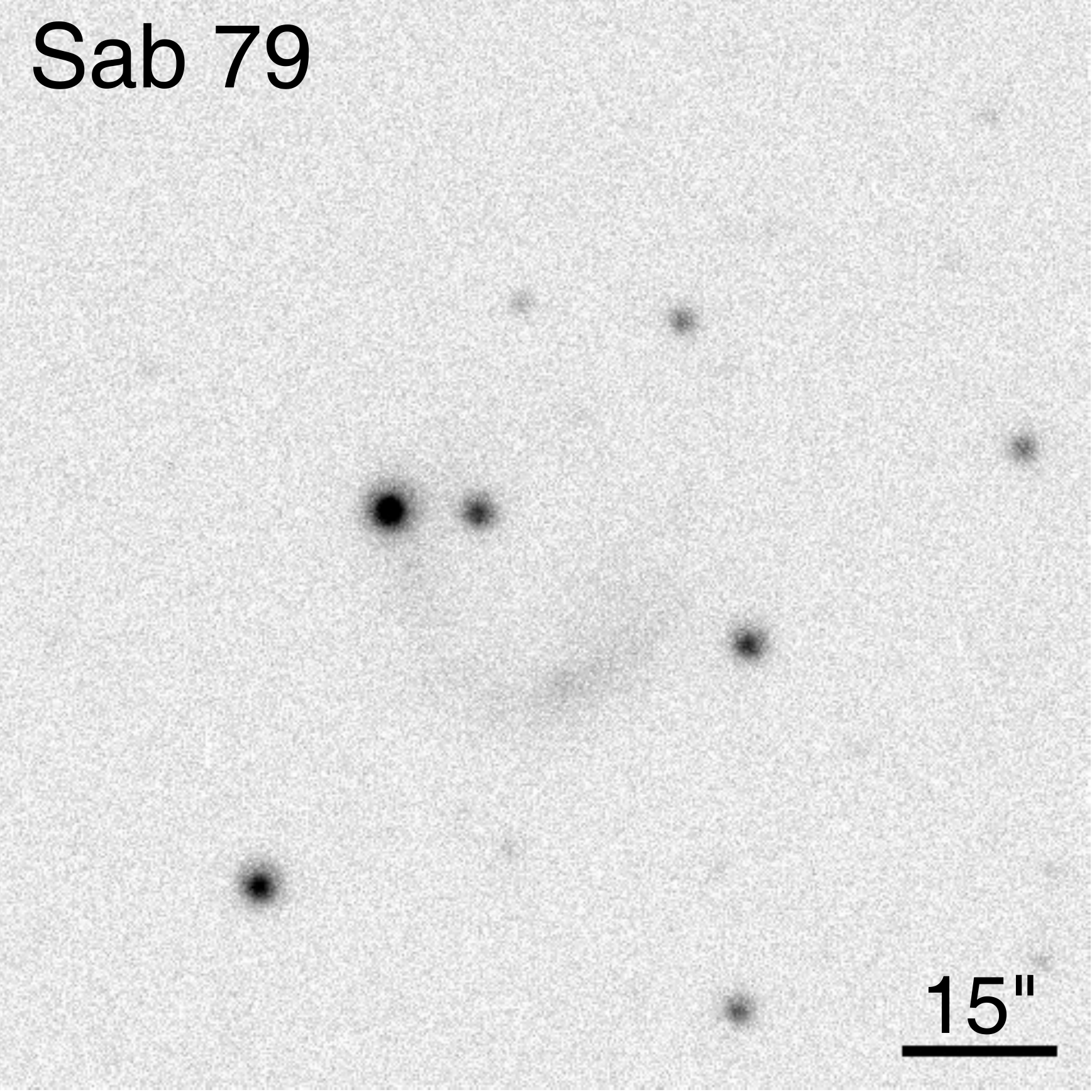} 
\includegraphics[height=1.7in]{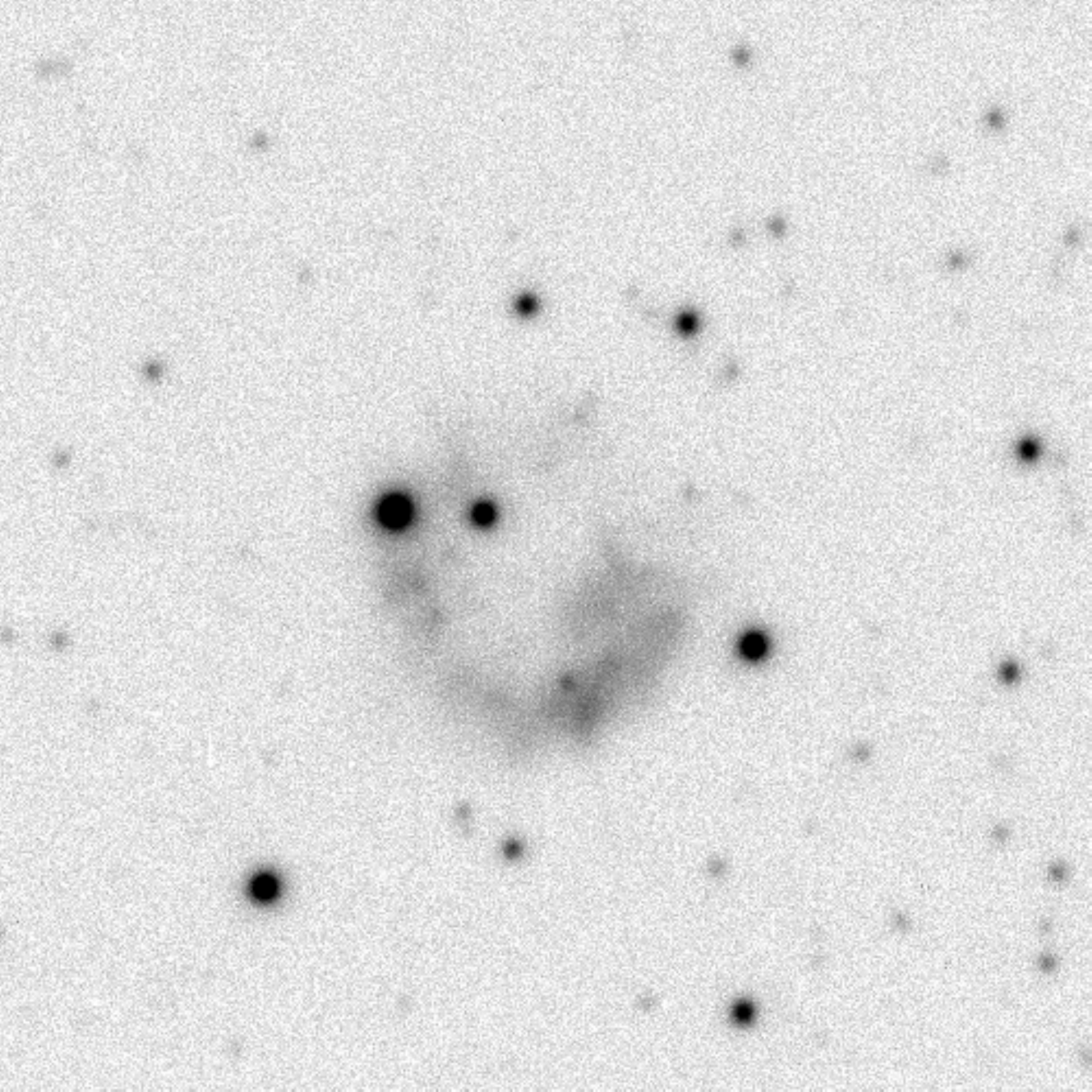}
\includegraphics[height=1.7in]{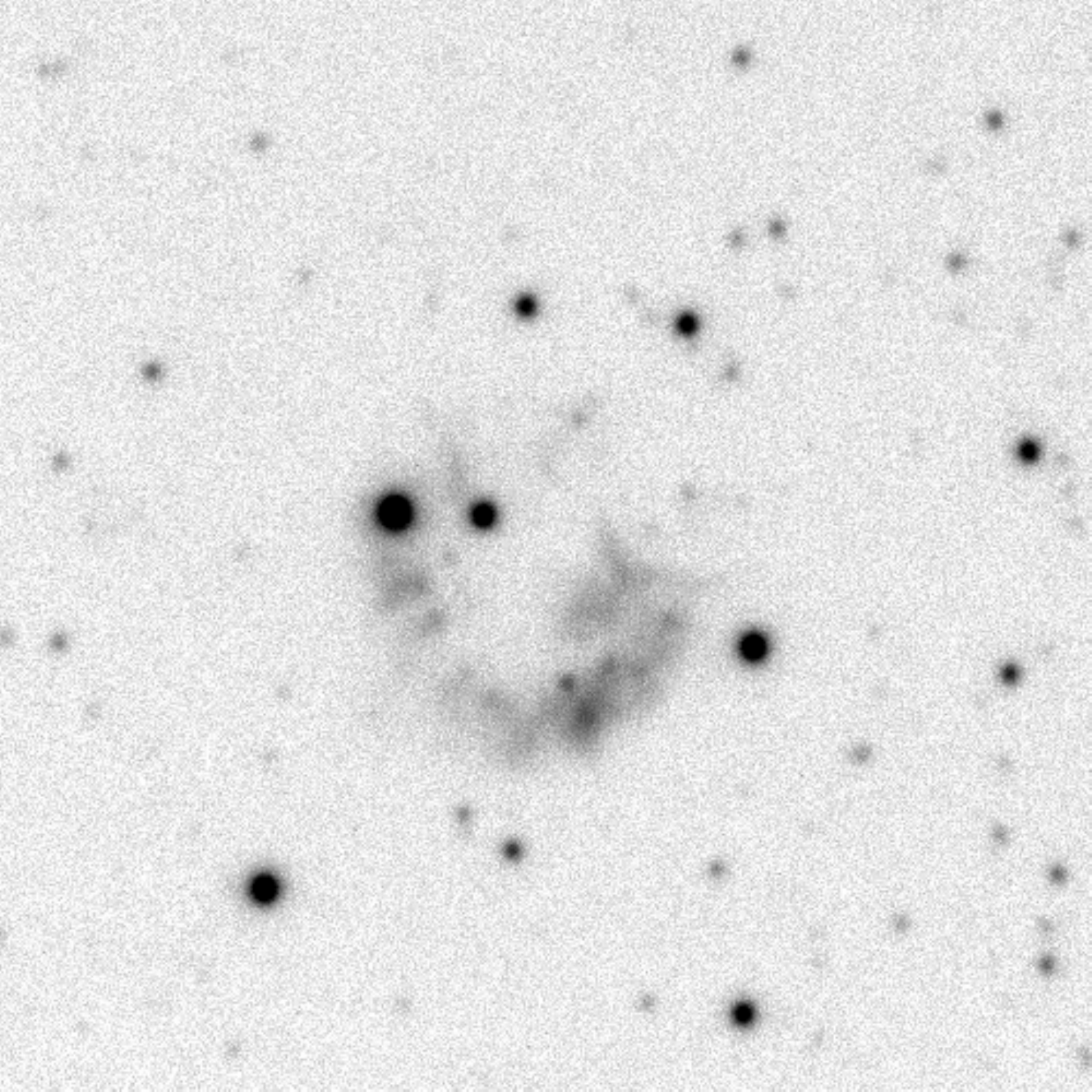}
\includegraphics[height=1.7in]{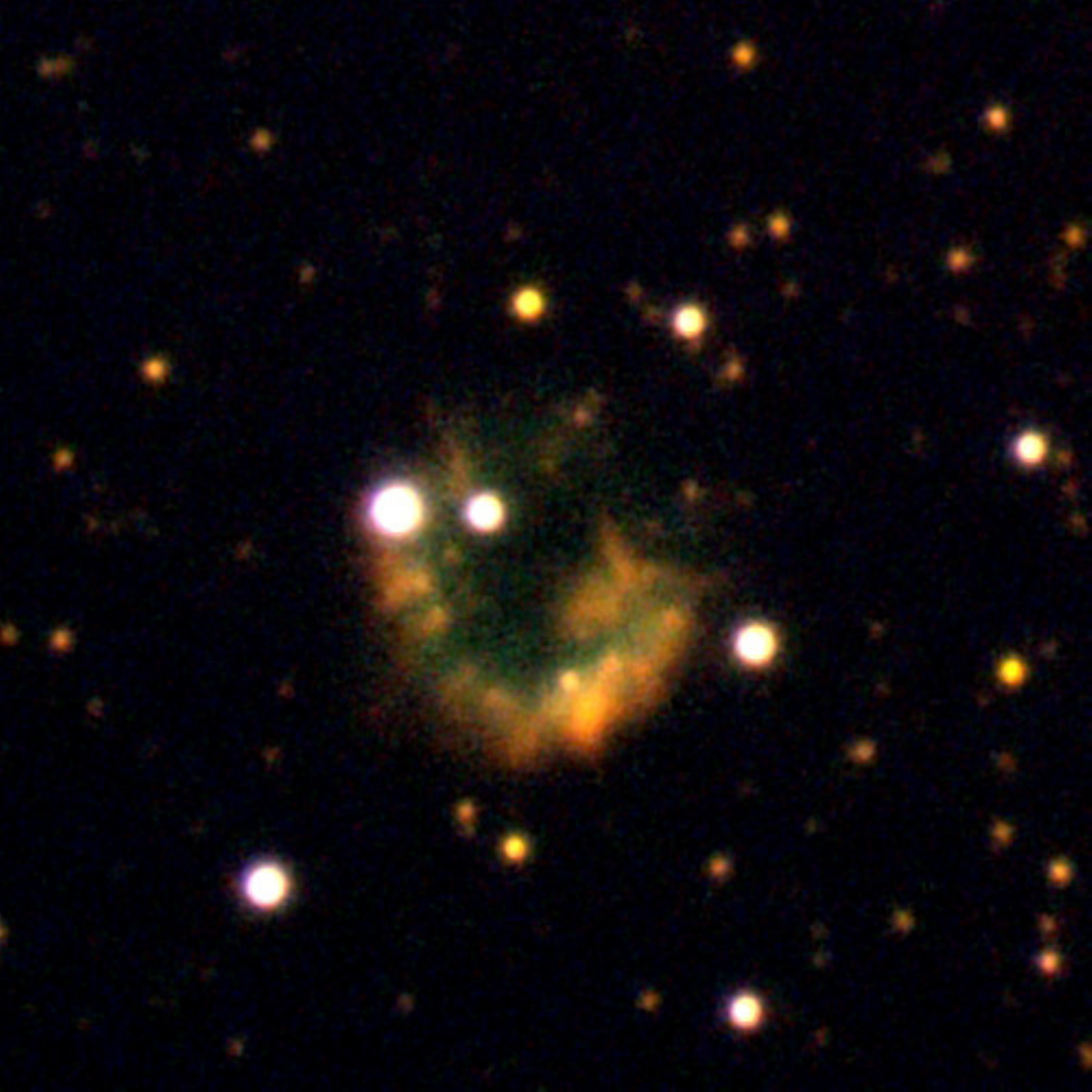}
\vskip .1in 
\includegraphics[height=1.7in]{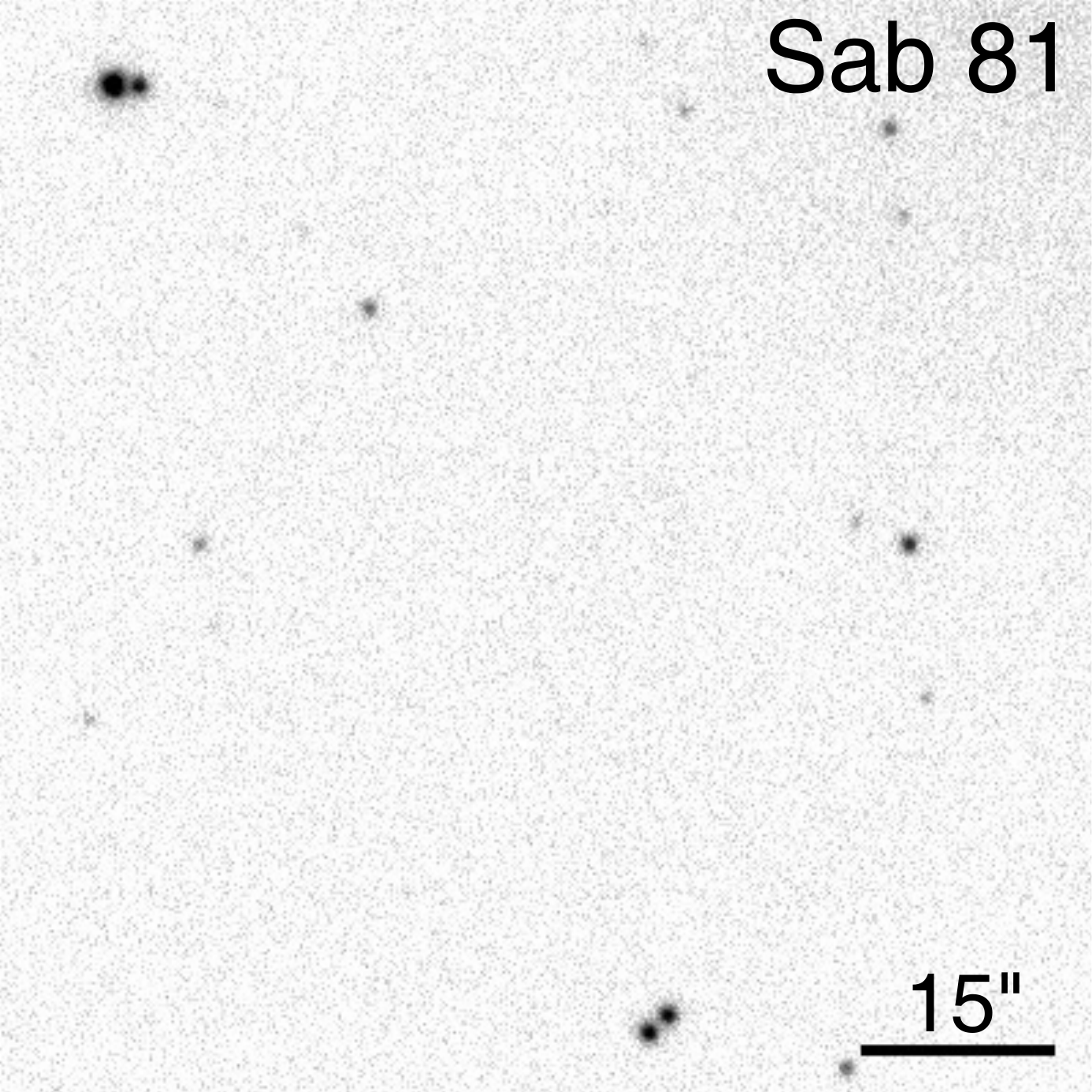} 
\includegraphics[height=1.7in]{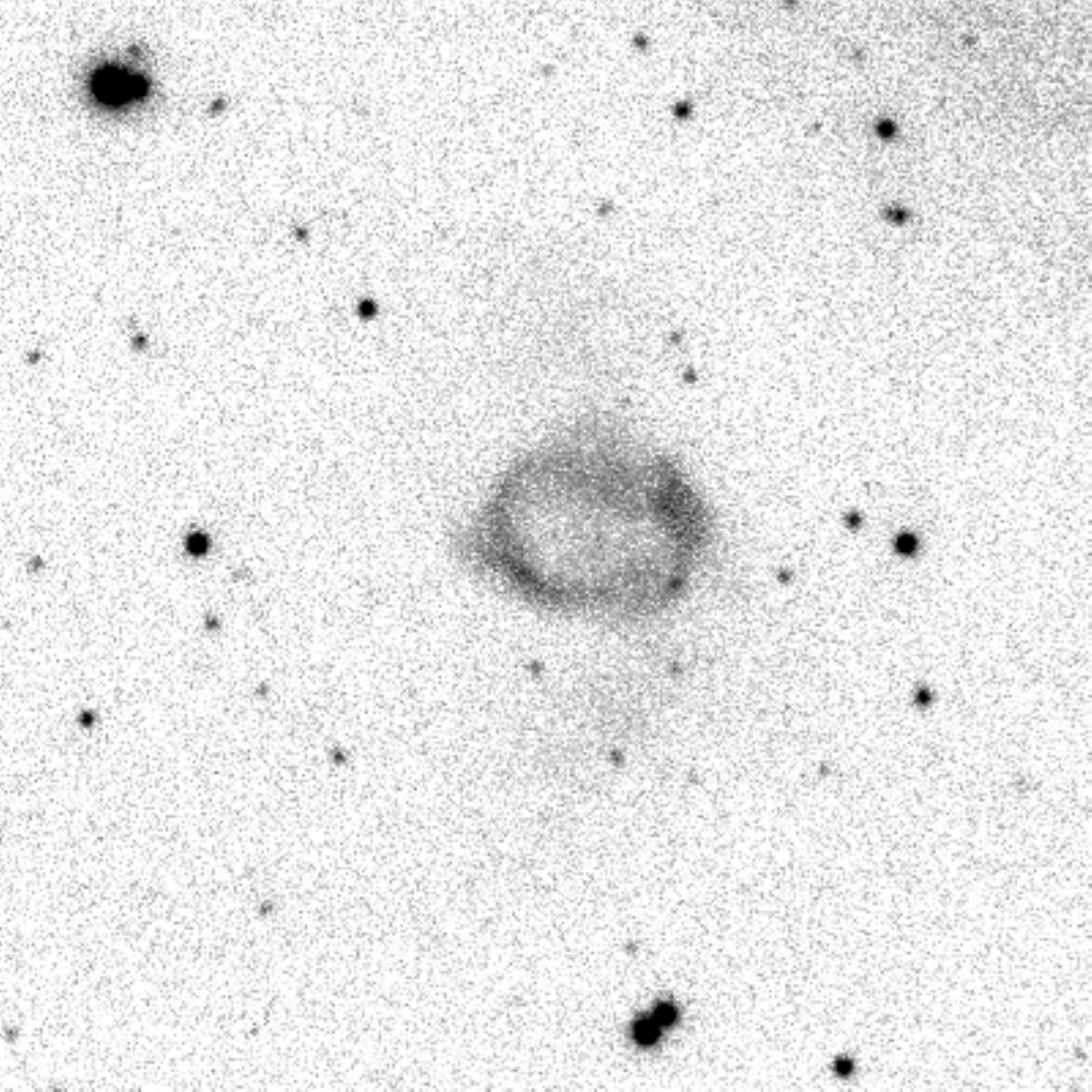}
\includegraphics[height=1.7in]{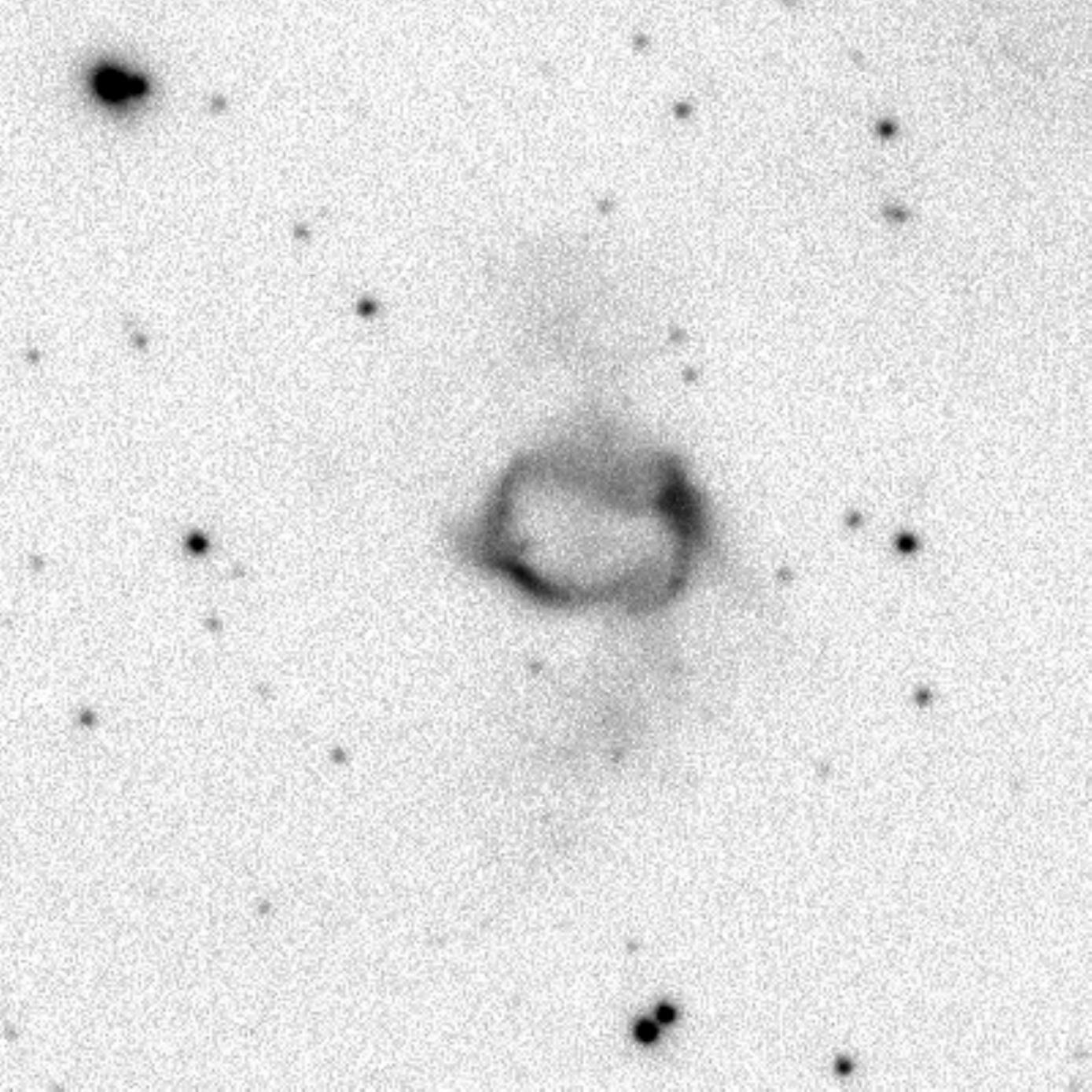}
\includegraphics[height=1.7in]{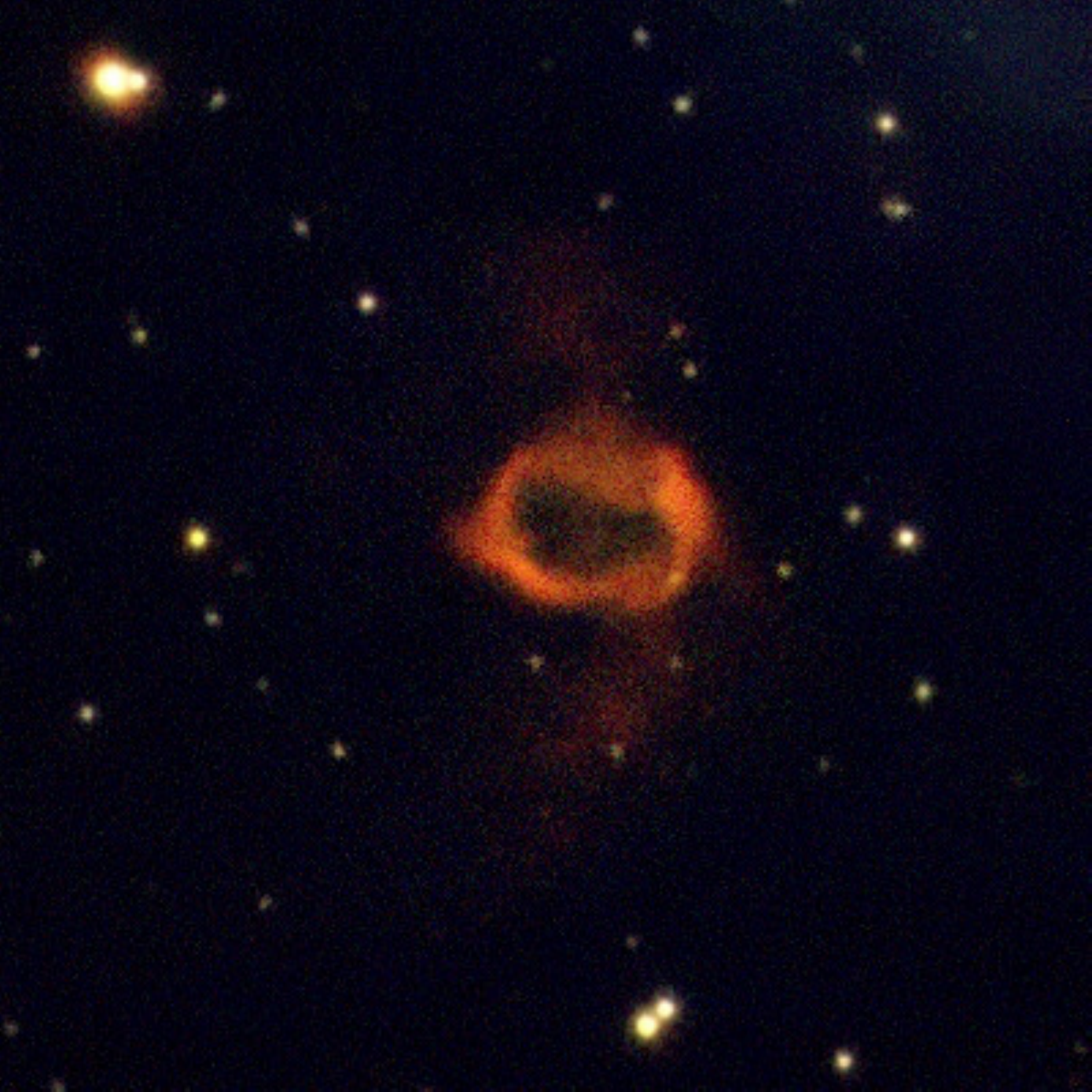}
\vskip .1in 
\includegraphics[height=1.7in]{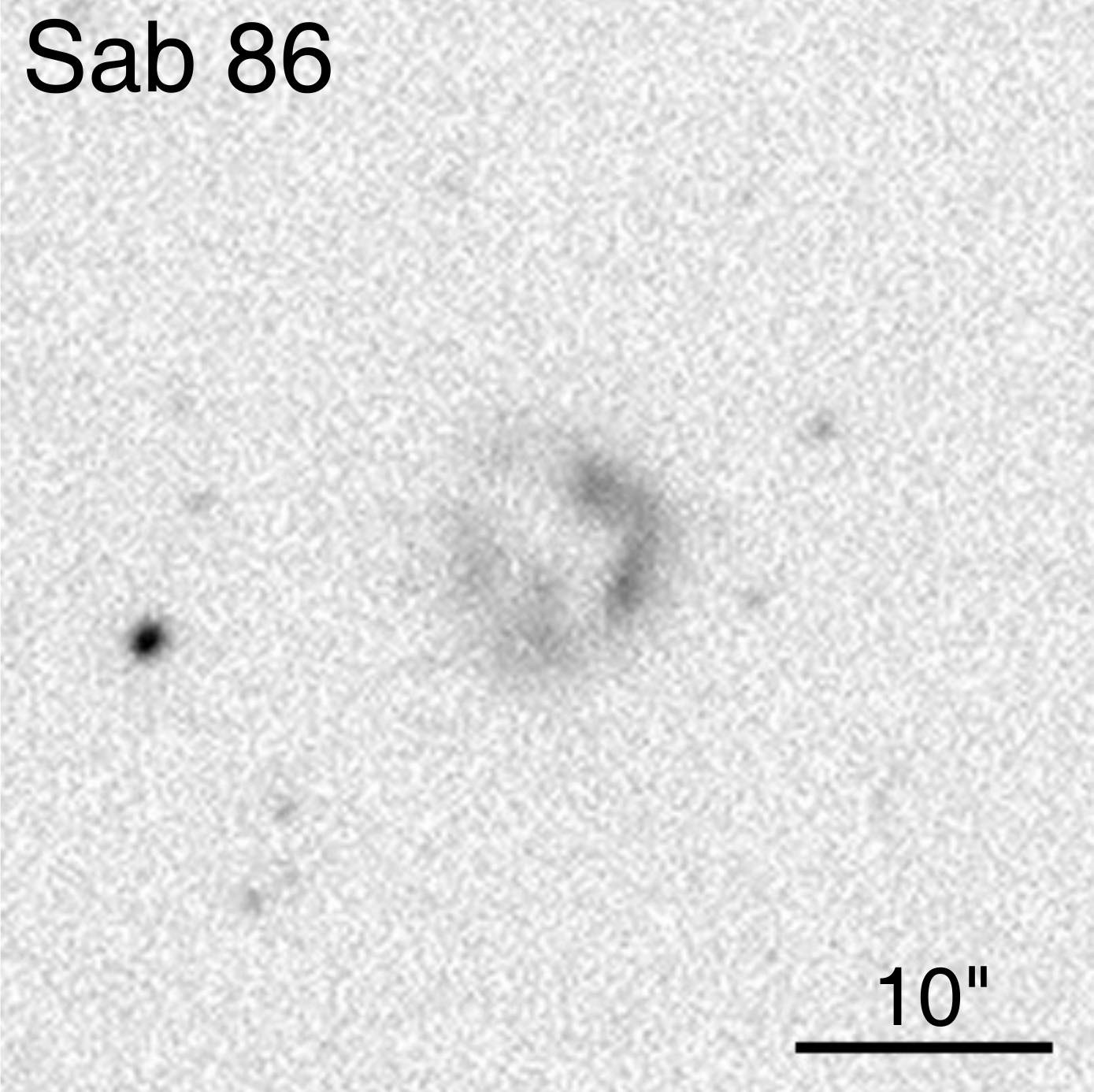} 
\includegraphics[height=1.7in]{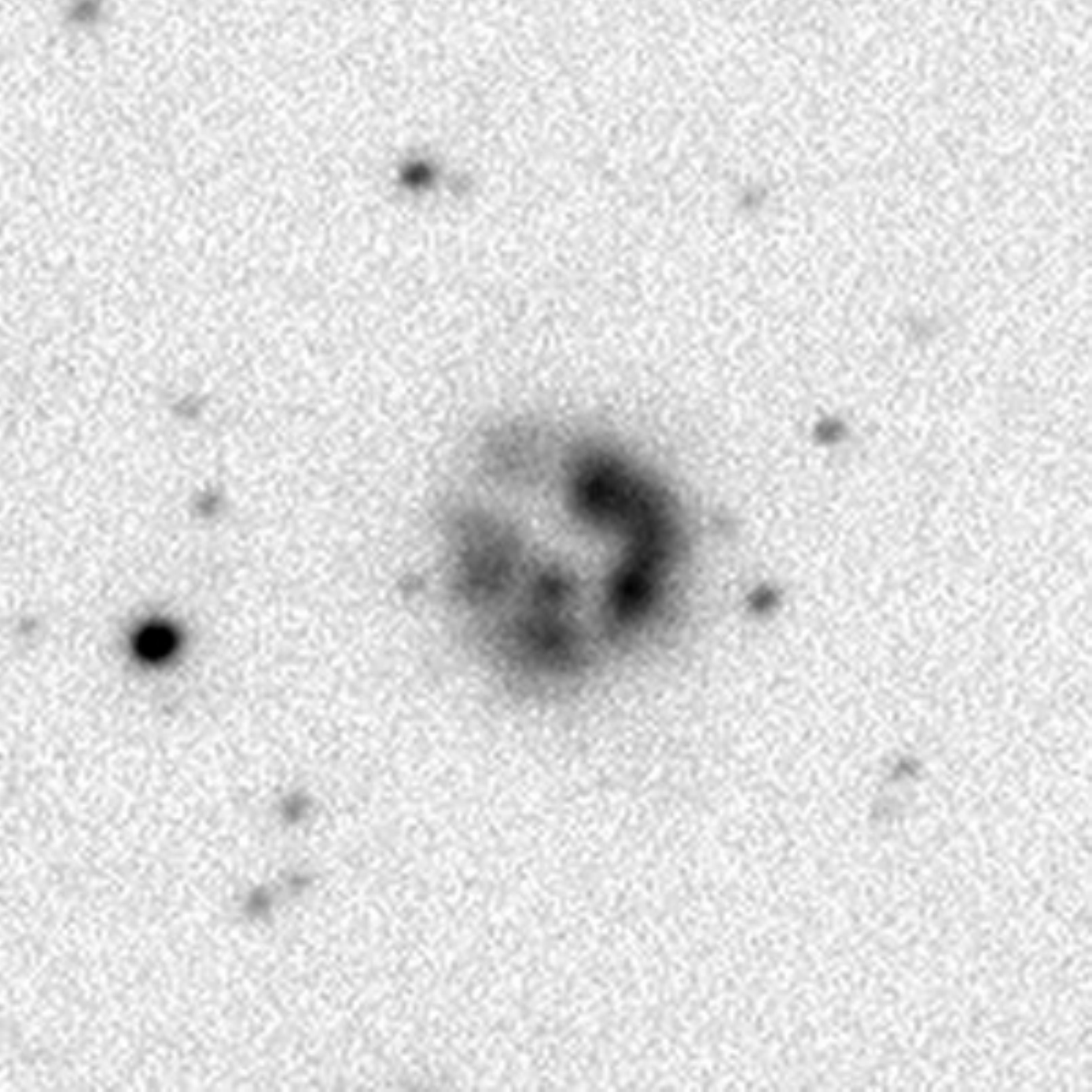}
\includegraphics[height=1.7in]{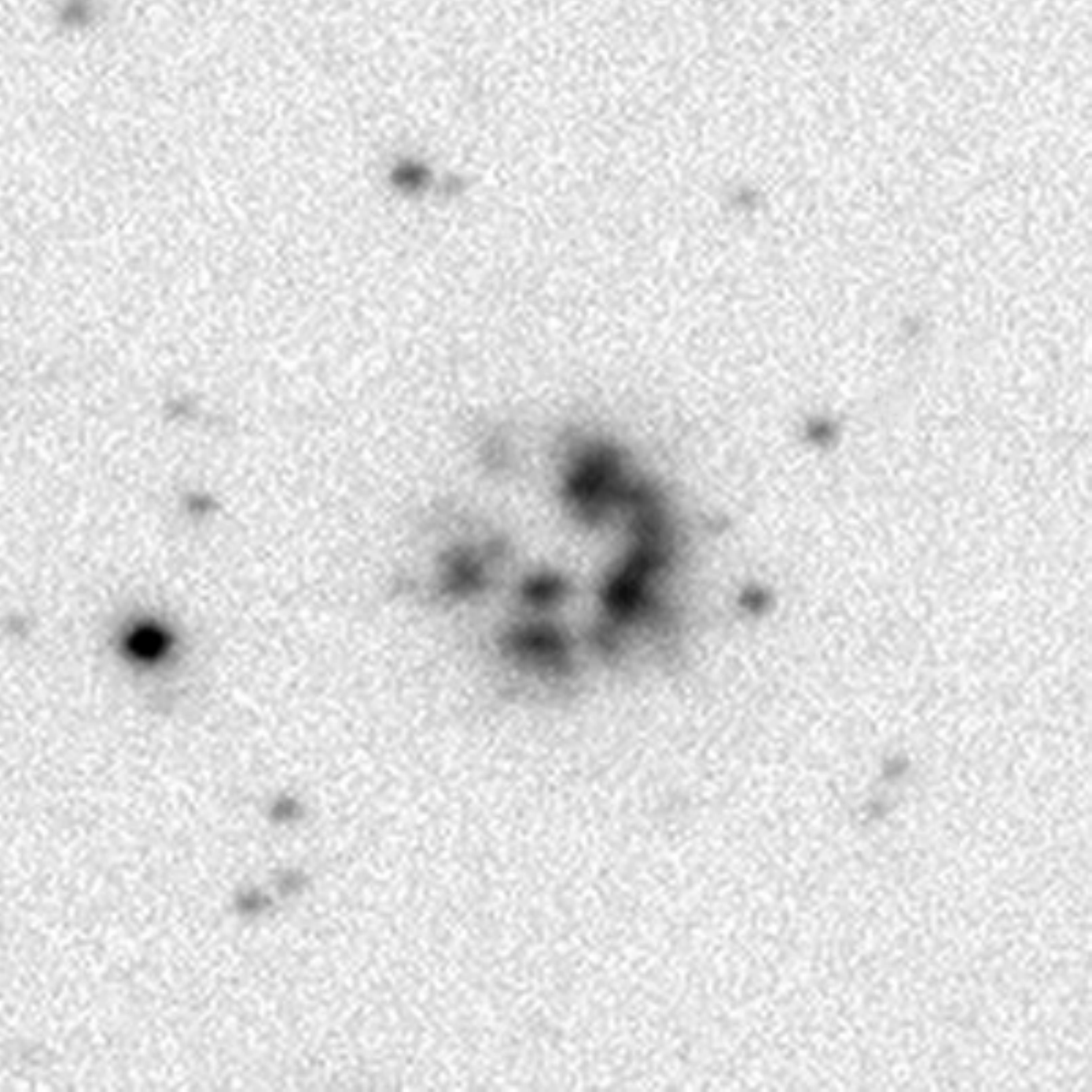}
\includegraphics[height=1.7in]{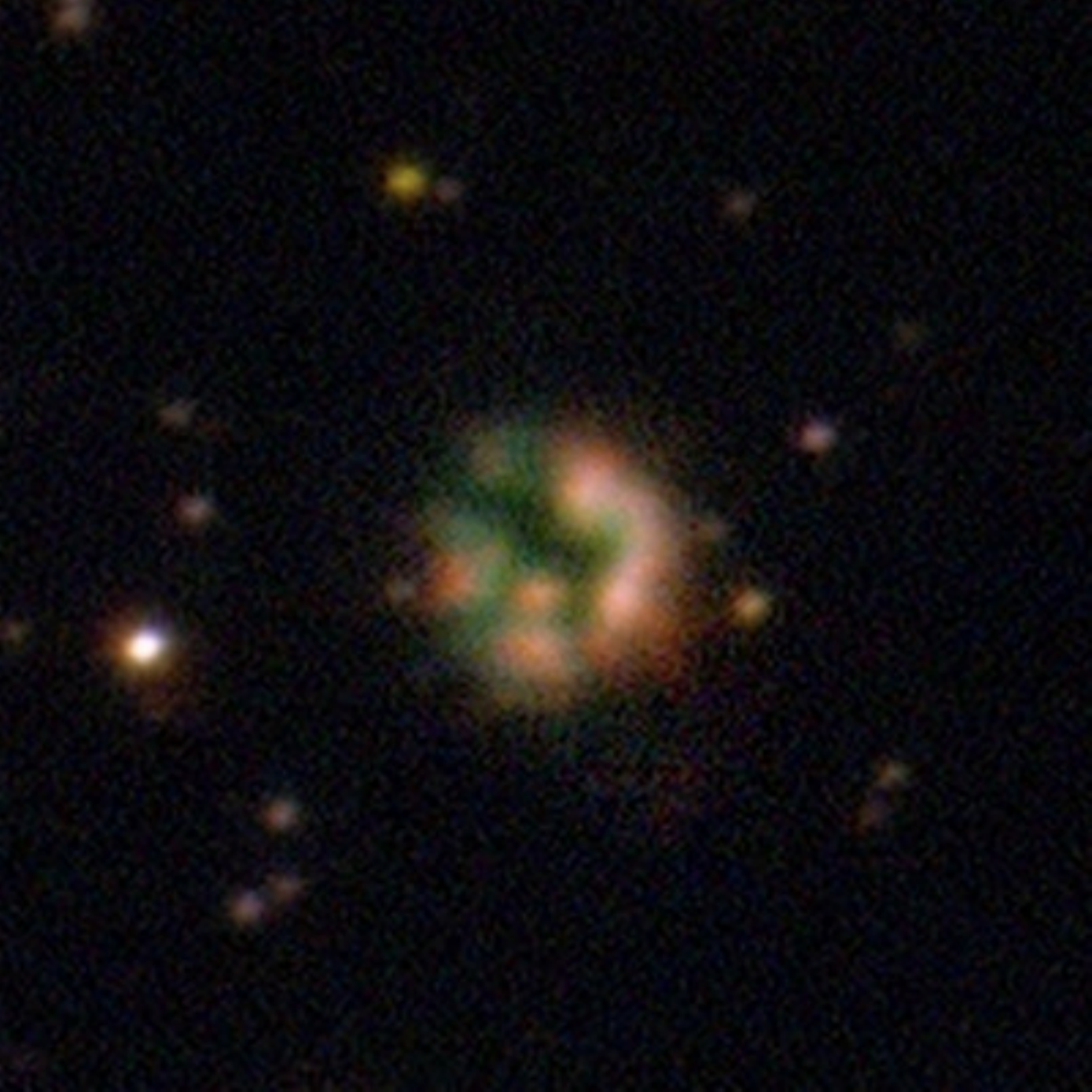}
\vskip .1in
\includegraphics[height=1.7in]{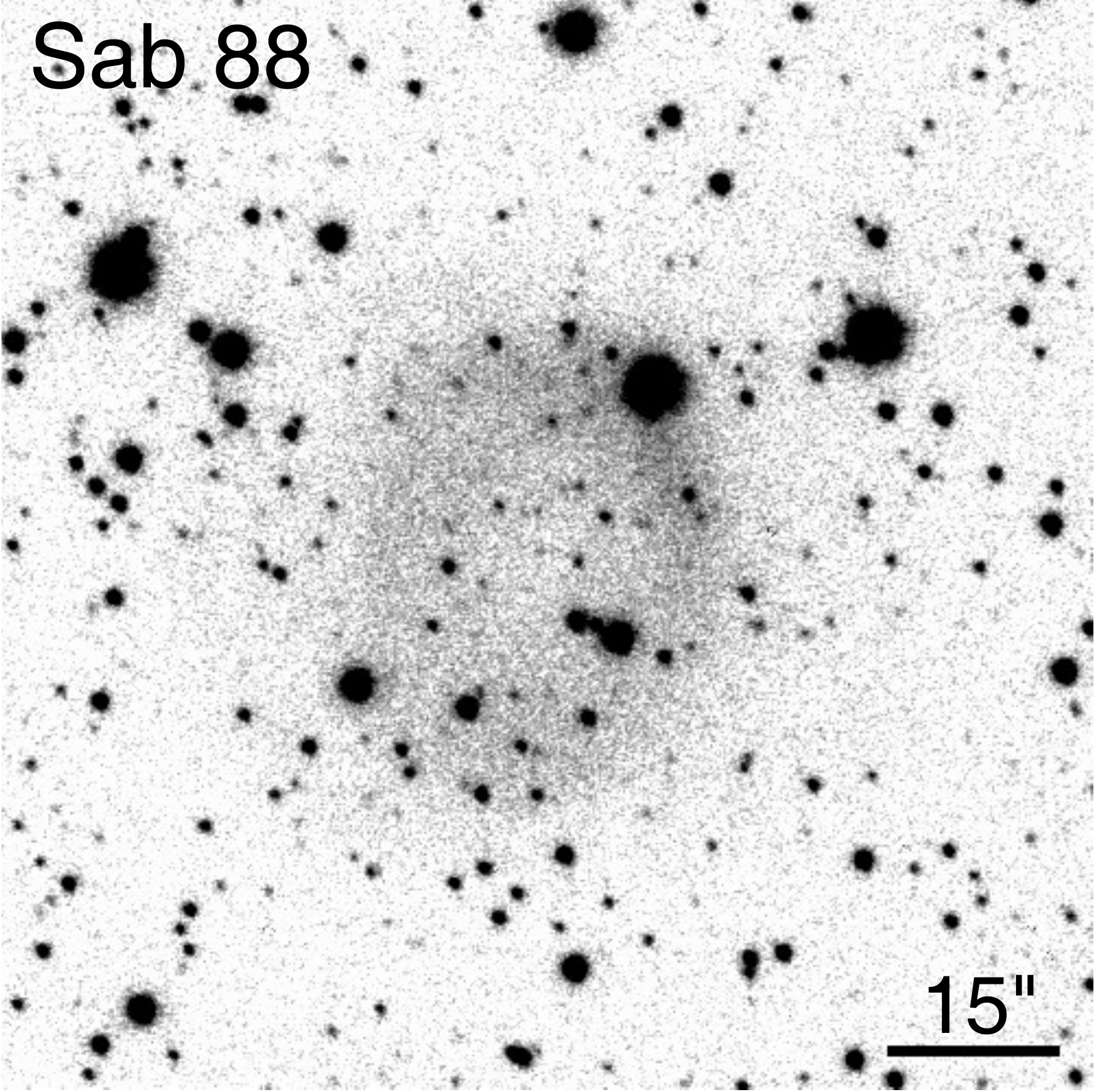} 
\includegraphics[height=1.7in]{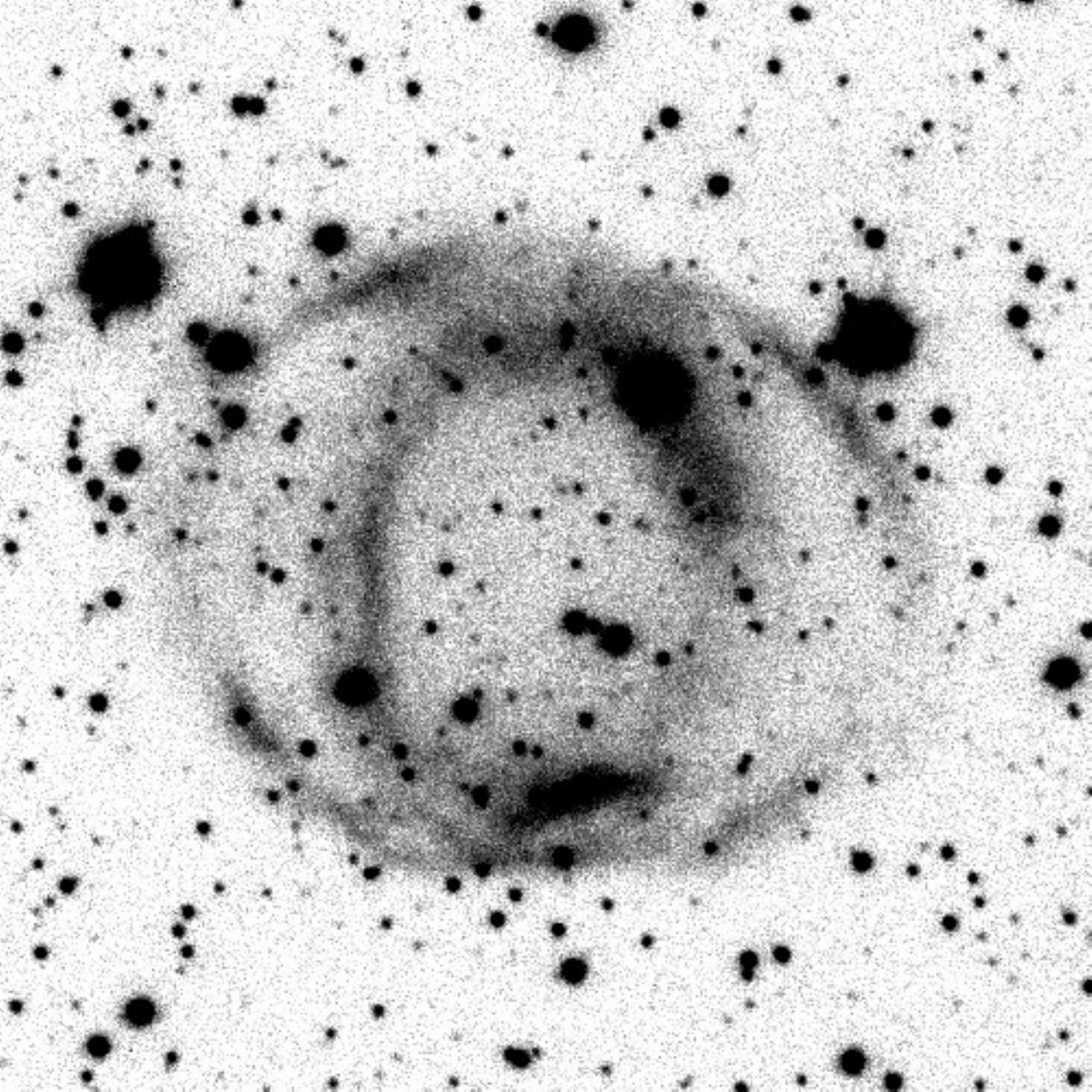}
\includegraphics[height=1.7in]{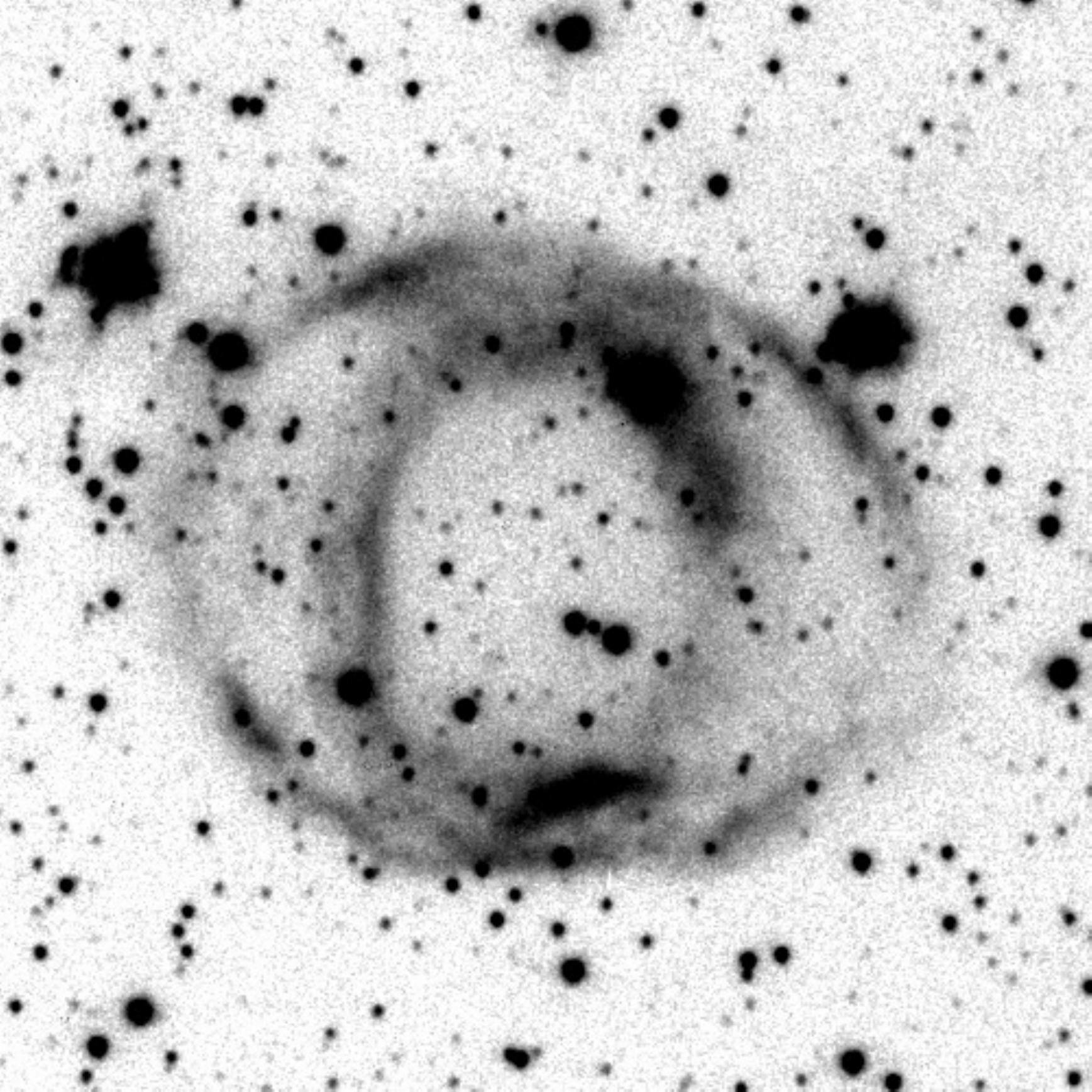}
\includegraphics[height=1.7in]{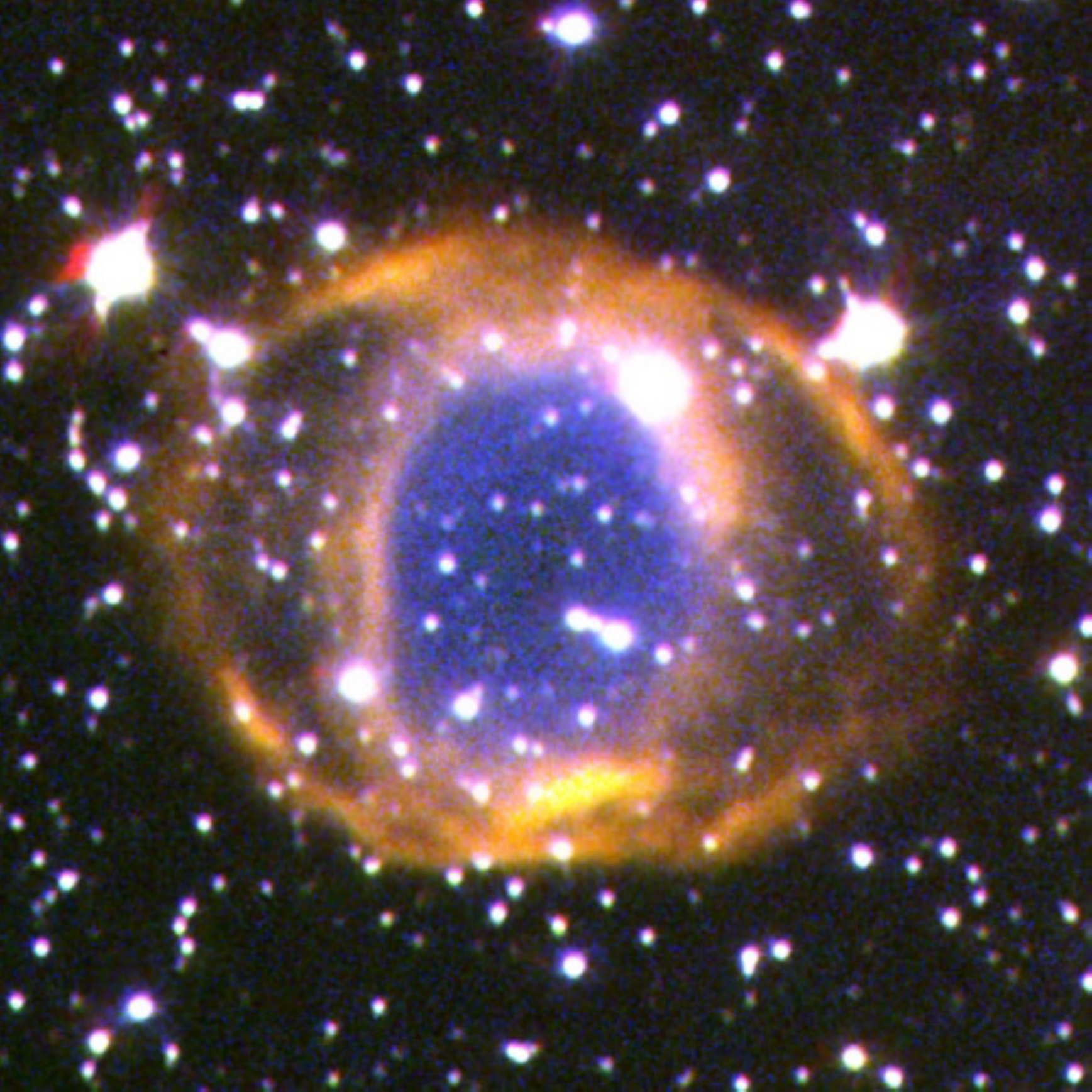}
\vskip .1in
\includegraphics[height=1.7in]{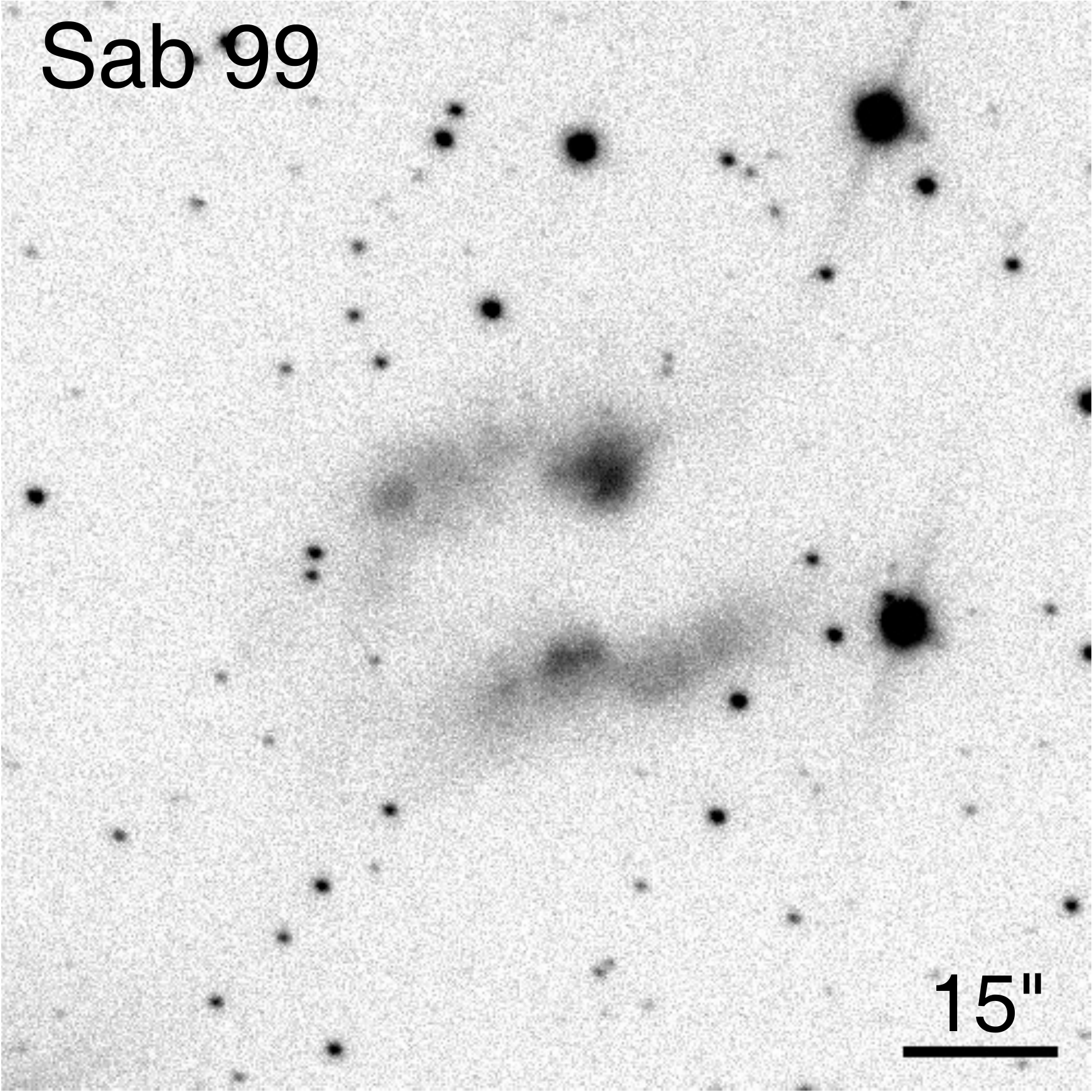} 
\includegraphics[height=1.7in]{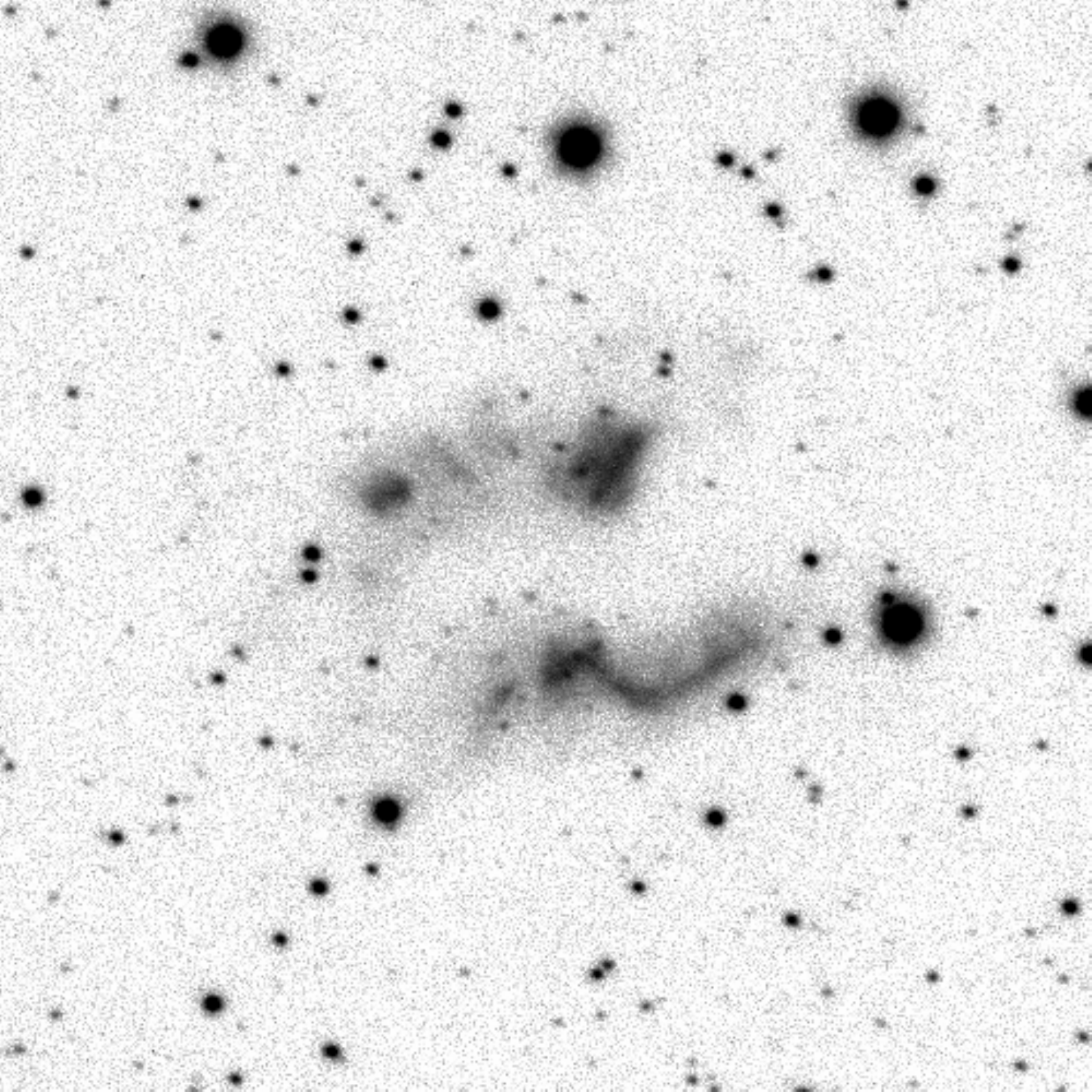}
\includegraphics[height=1.7in]{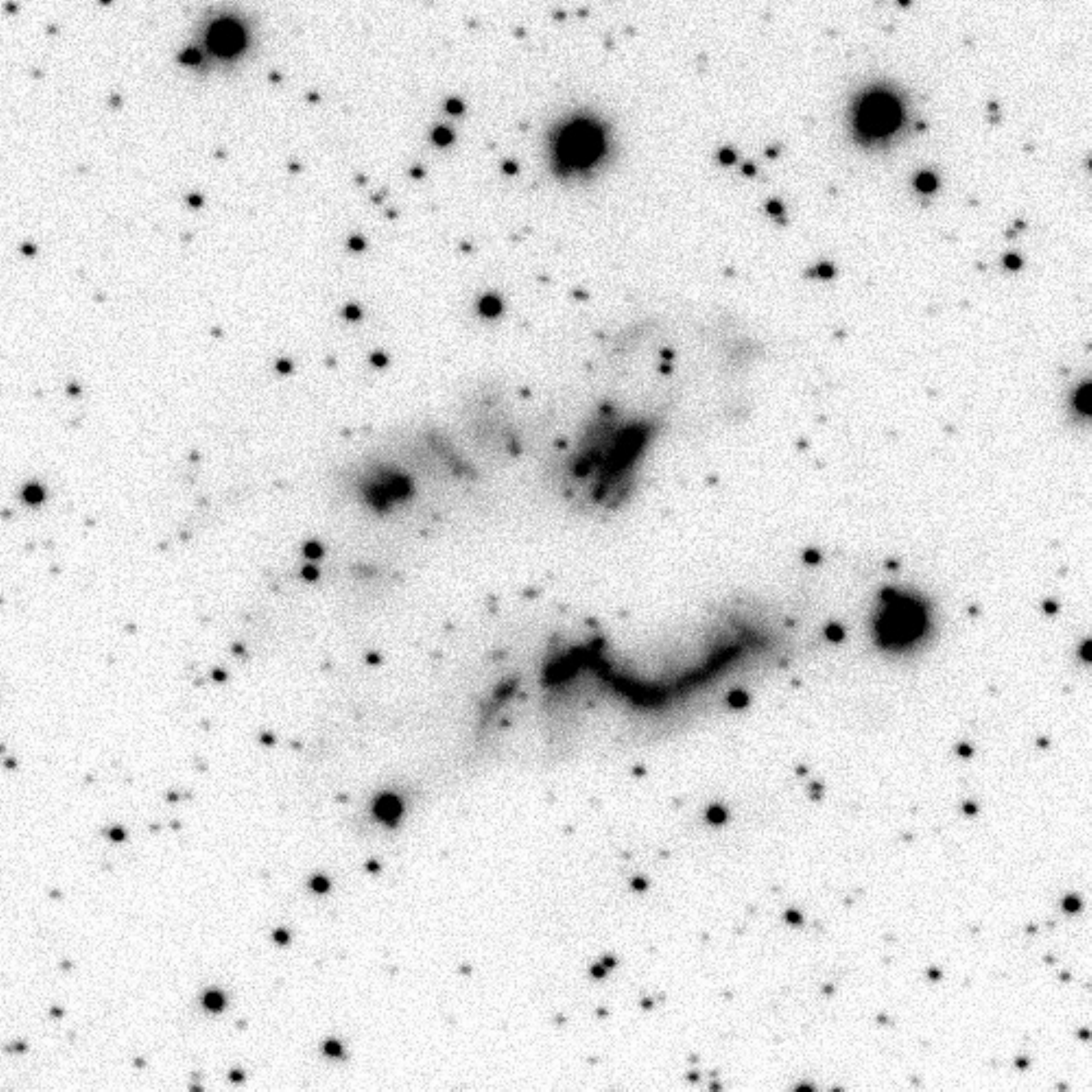}
\includegraphics[height=1.7in]{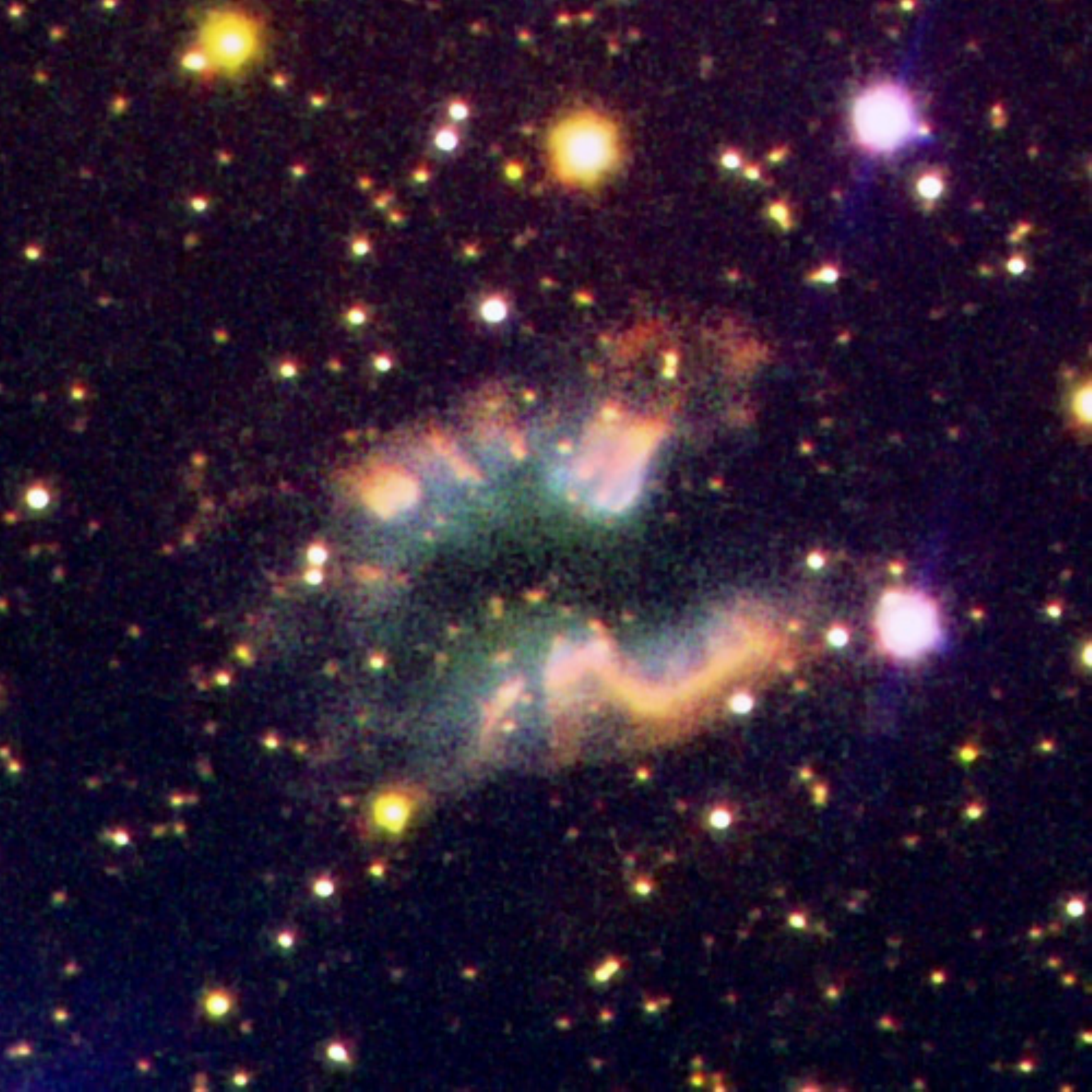}
\caption{Same as Figure~\ref{1.img}. } 
\label{6.img} 
\end{figure*}

\begin{figure*} 
\centering 
\includegraphics[height=1.7in]{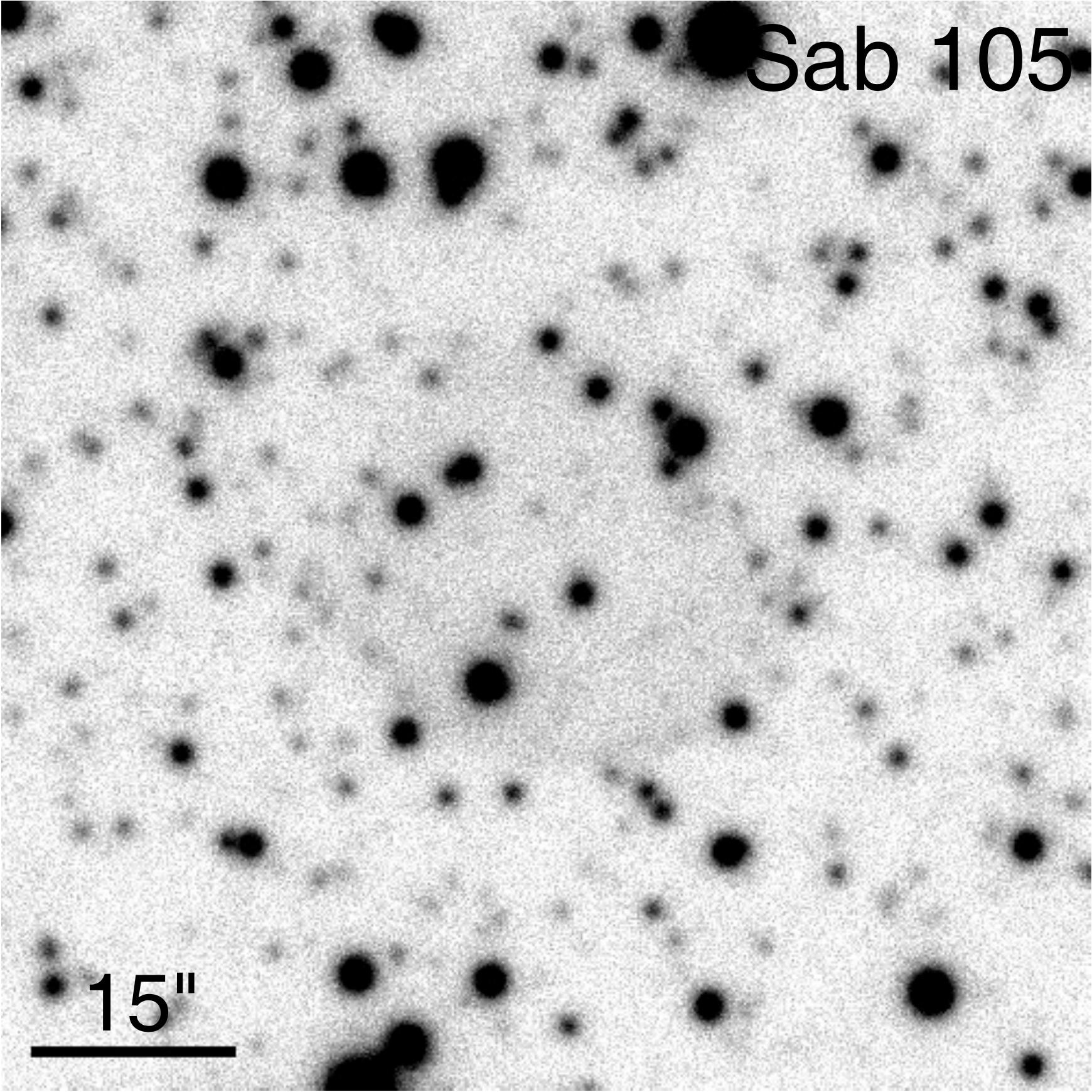} 
\includegraphics[height=1.7in]{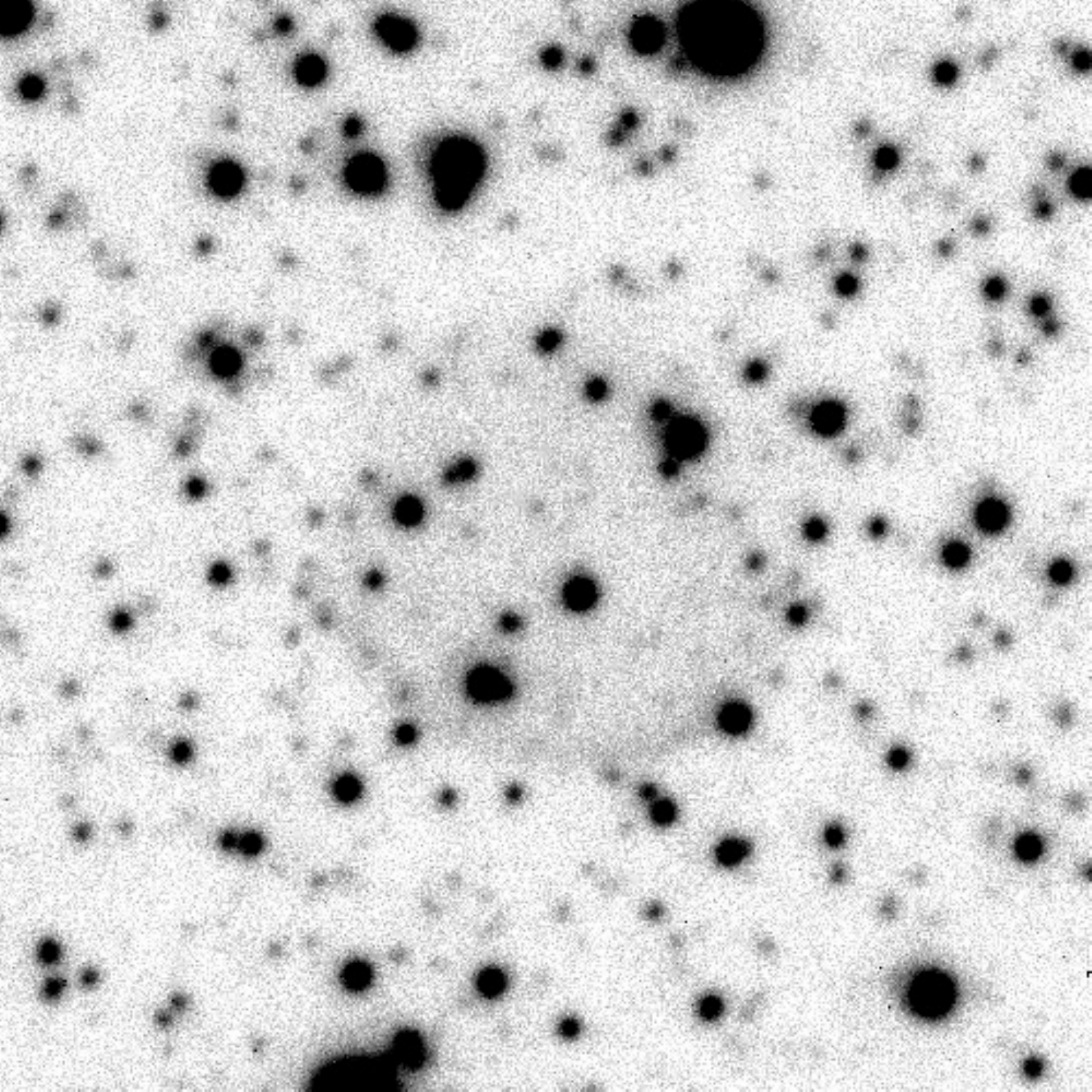}
\includegraphics[height=1.7in]{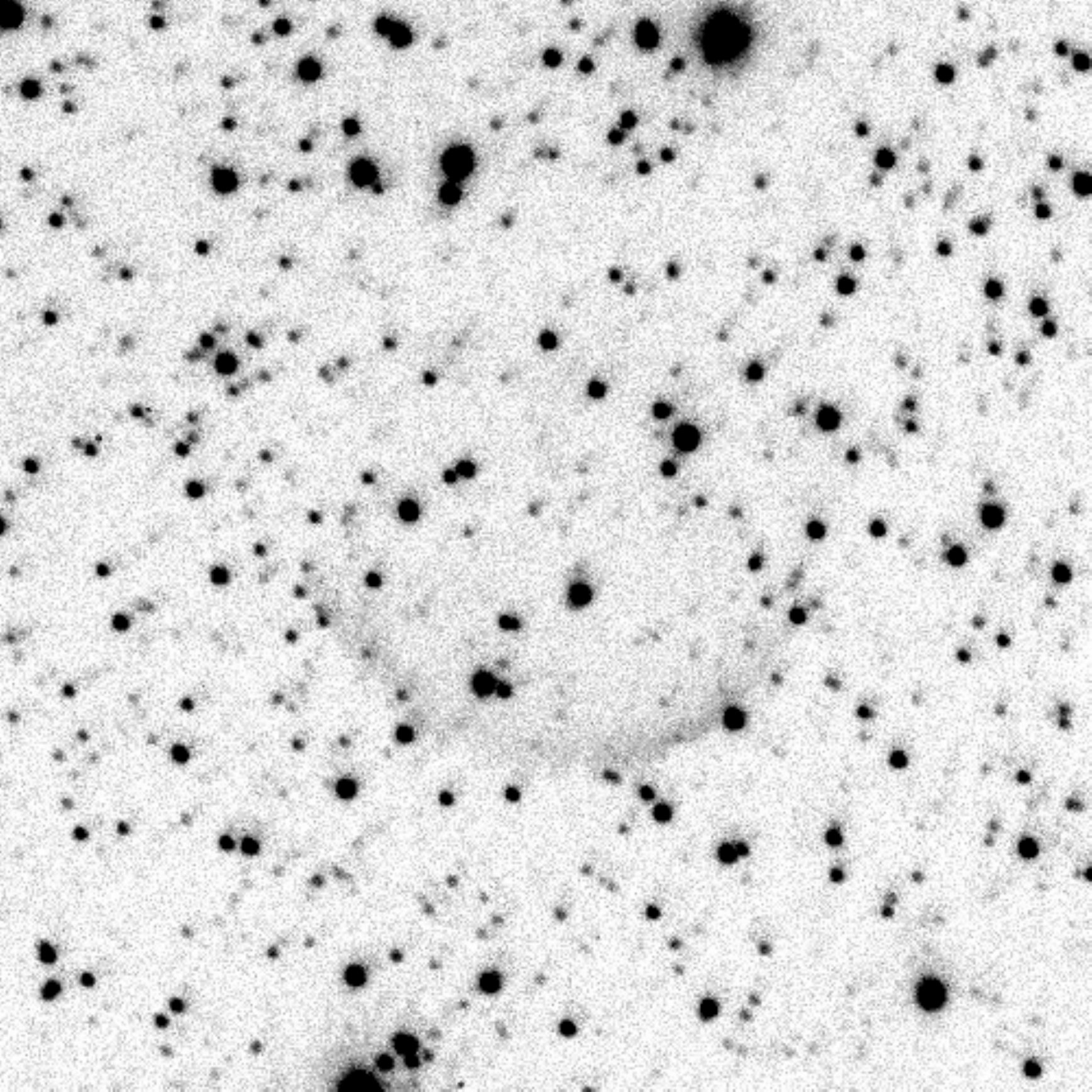}
\includegraphics[height=1.7in]{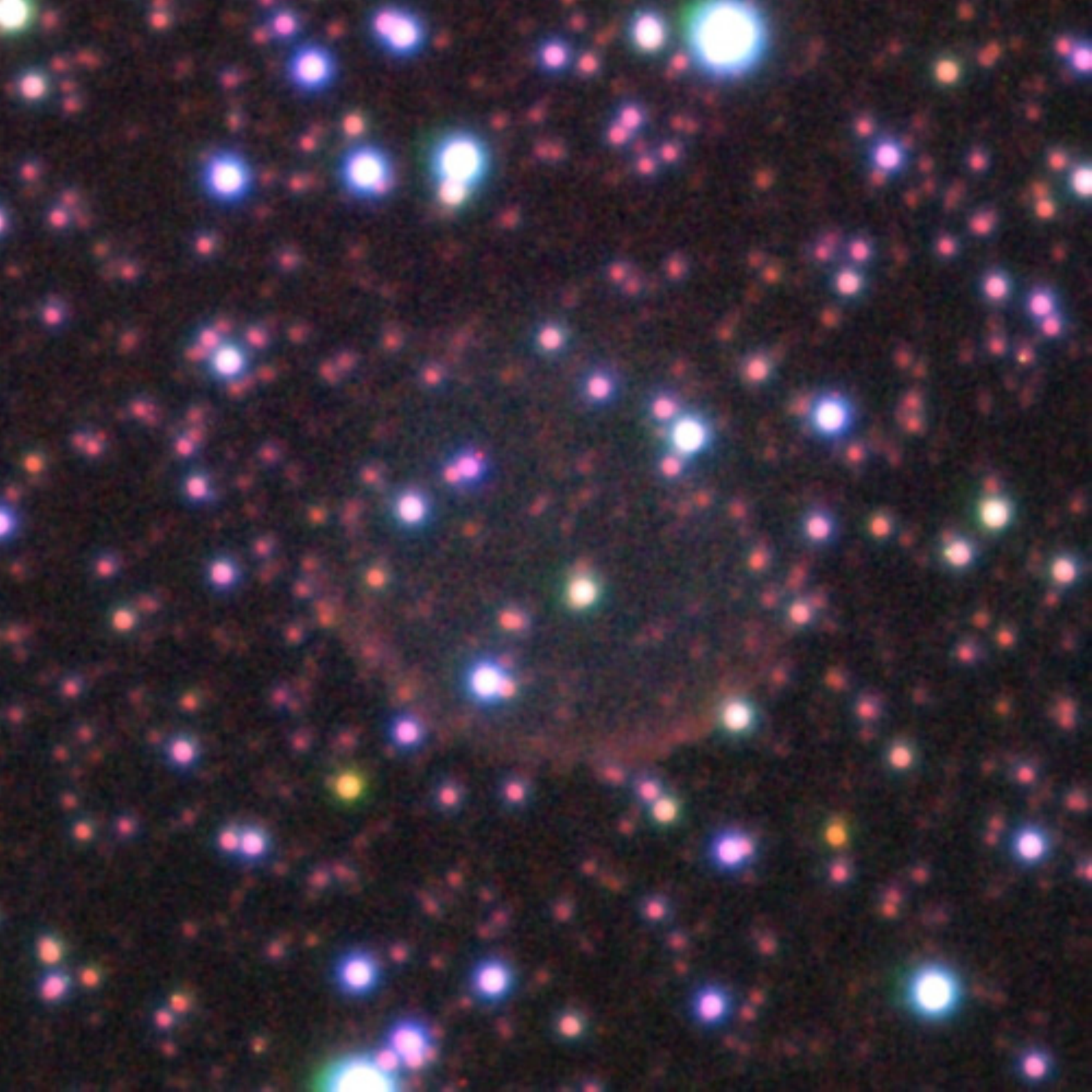}
\vskip .1in
\includegraphics[height=1.7in]{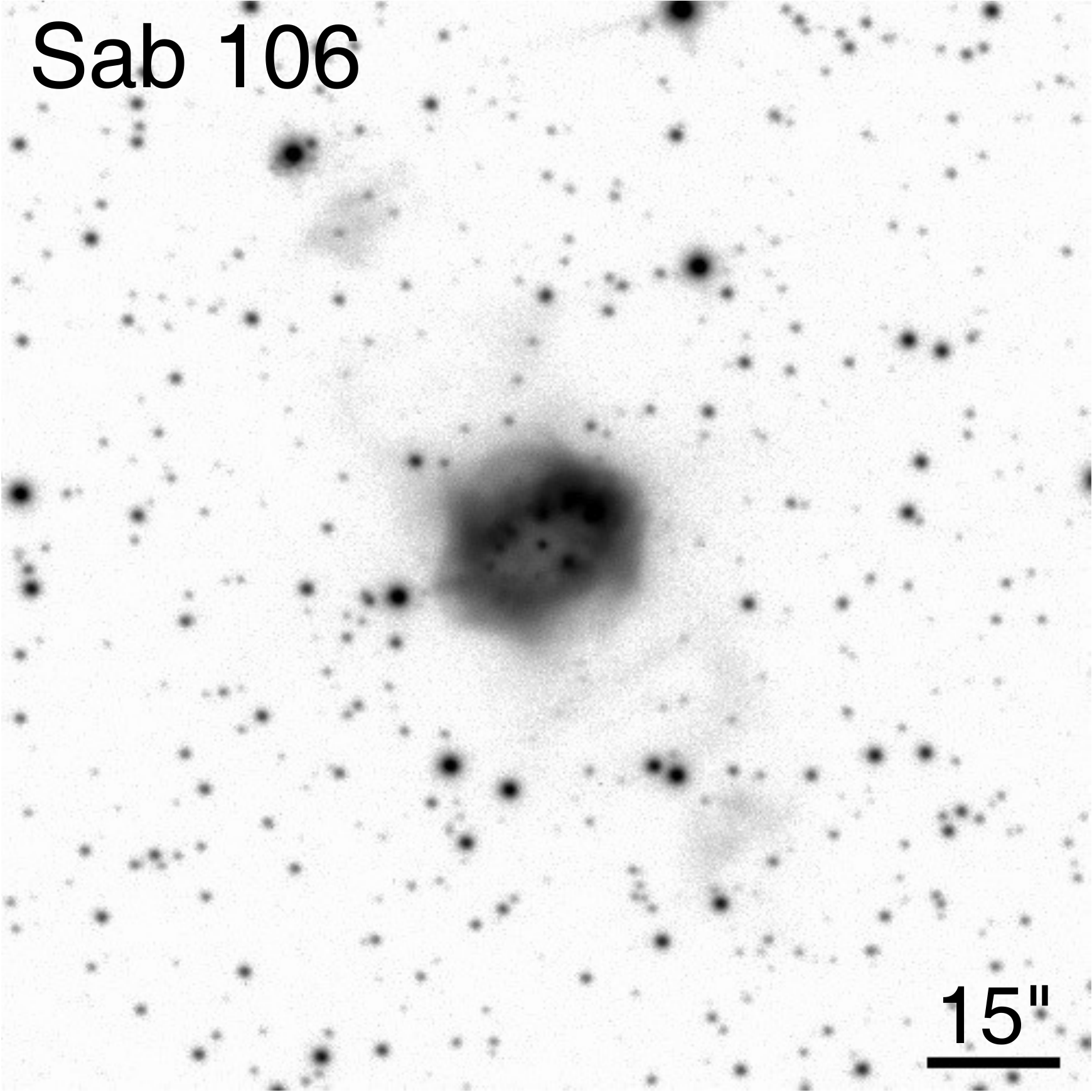} 
\includegraphics[height=1.7in]{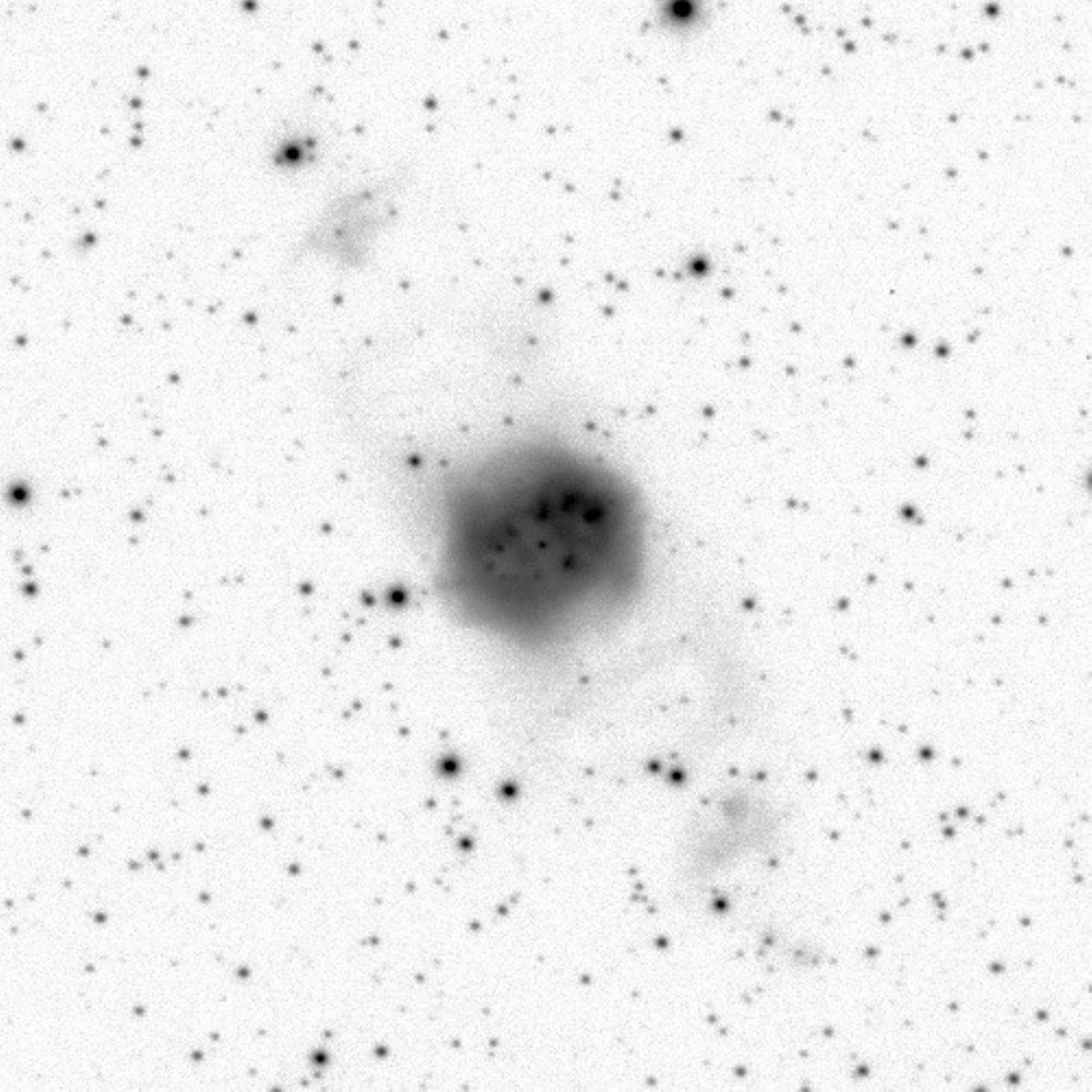}
\includegraphics[height=1.7in]{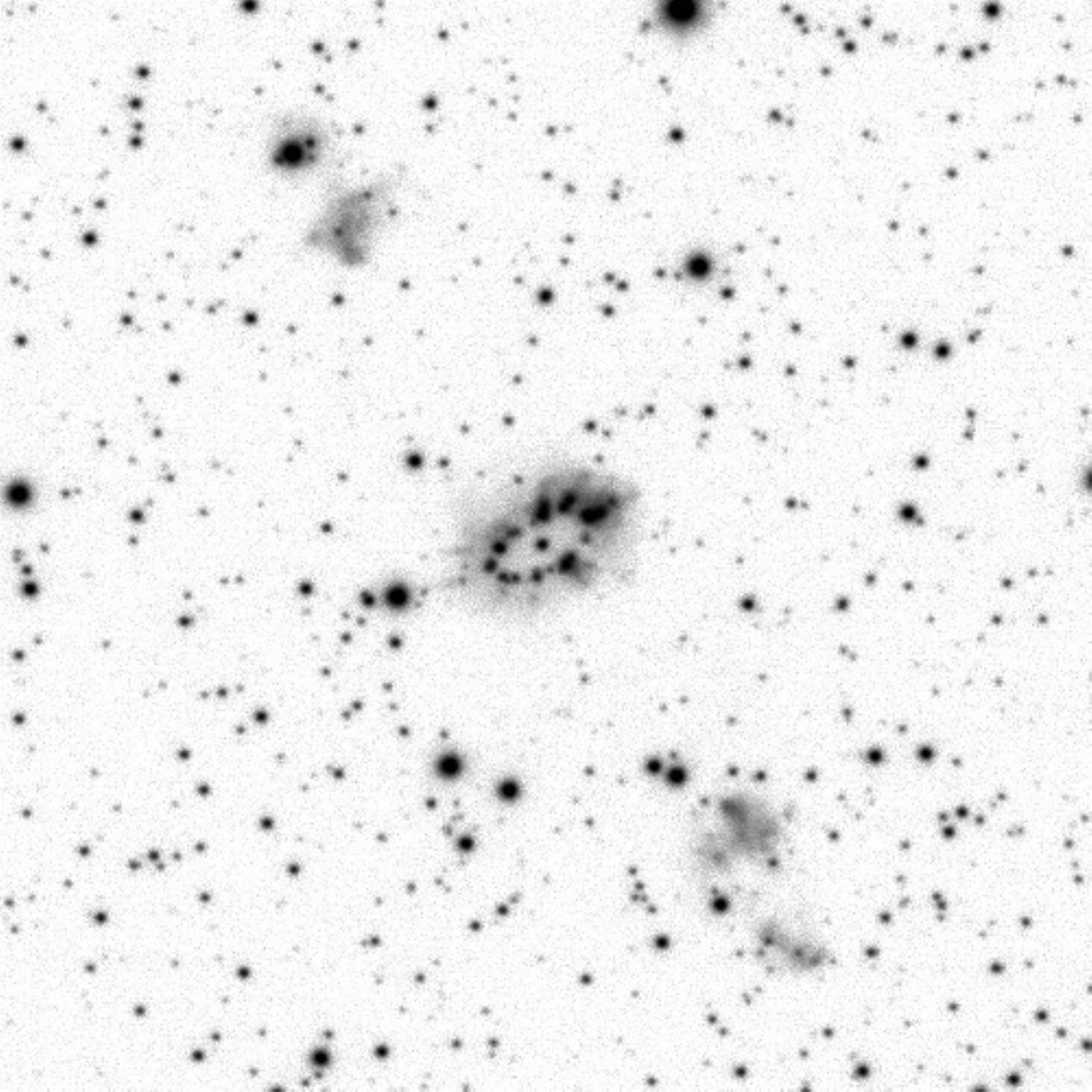}
\includegraphics[height=1.7in]{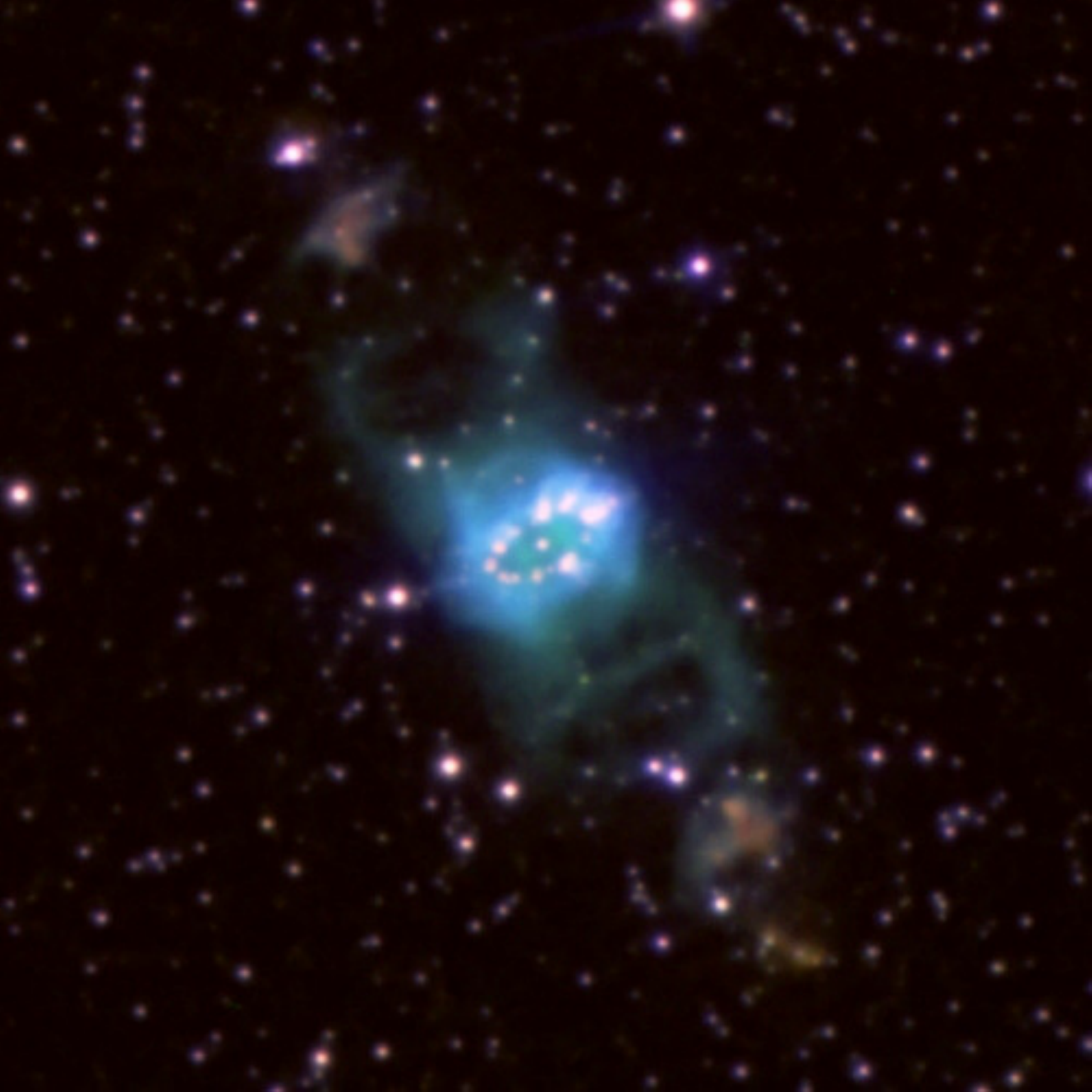}
\vskip .1in 
\includegraphics[height=1.7in]{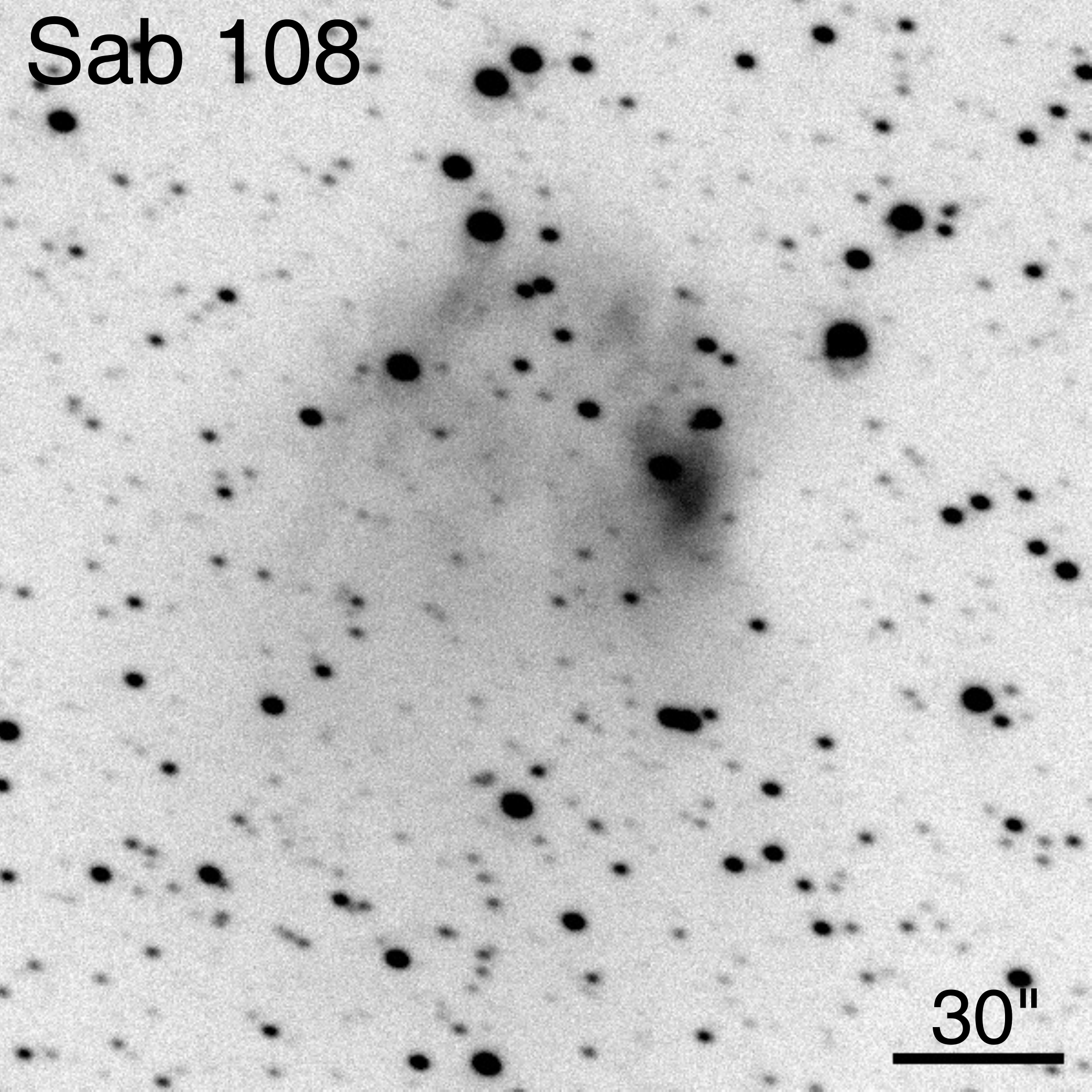} 
\includegraphics[height=1.7in]{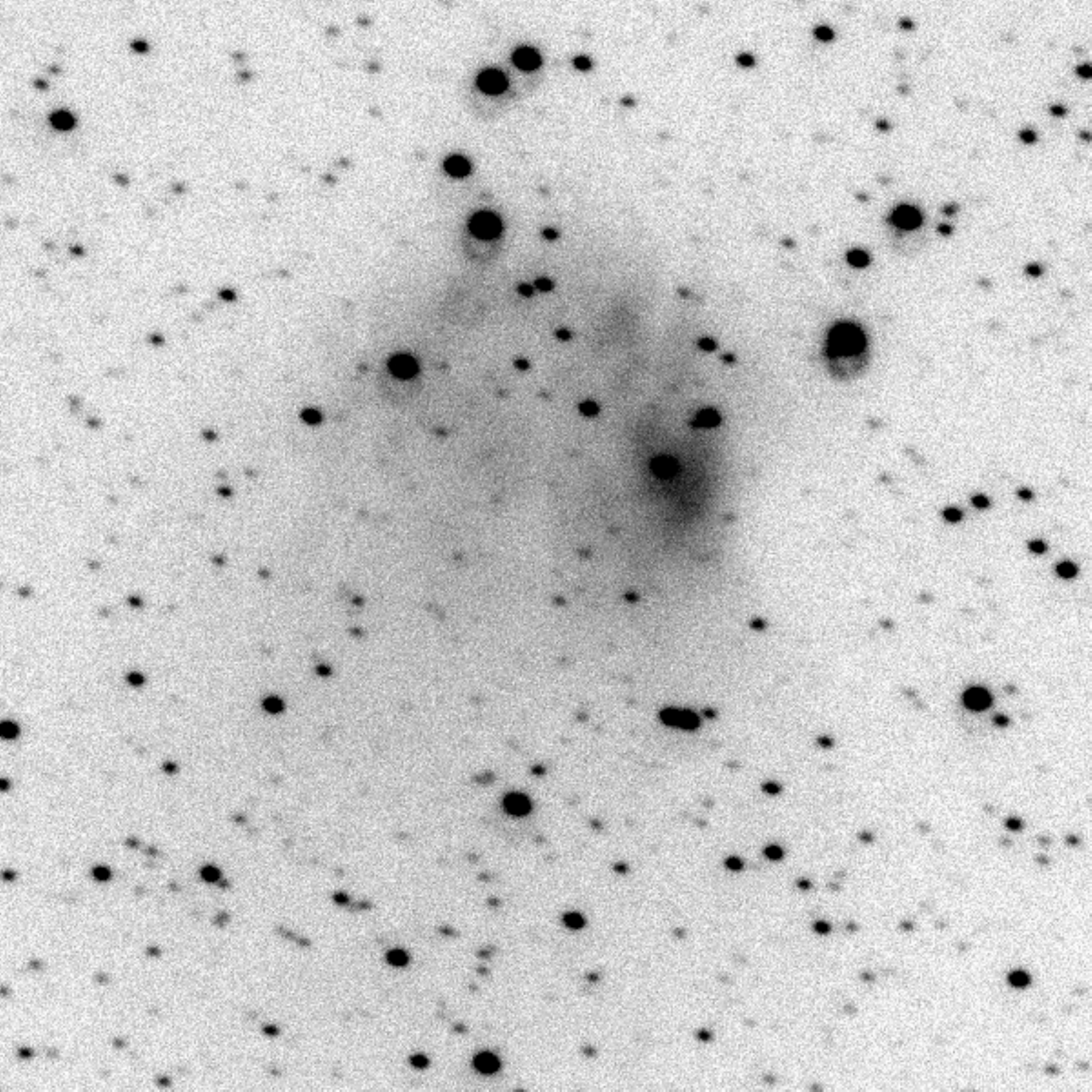}
\includegraphics[height=1.7in]{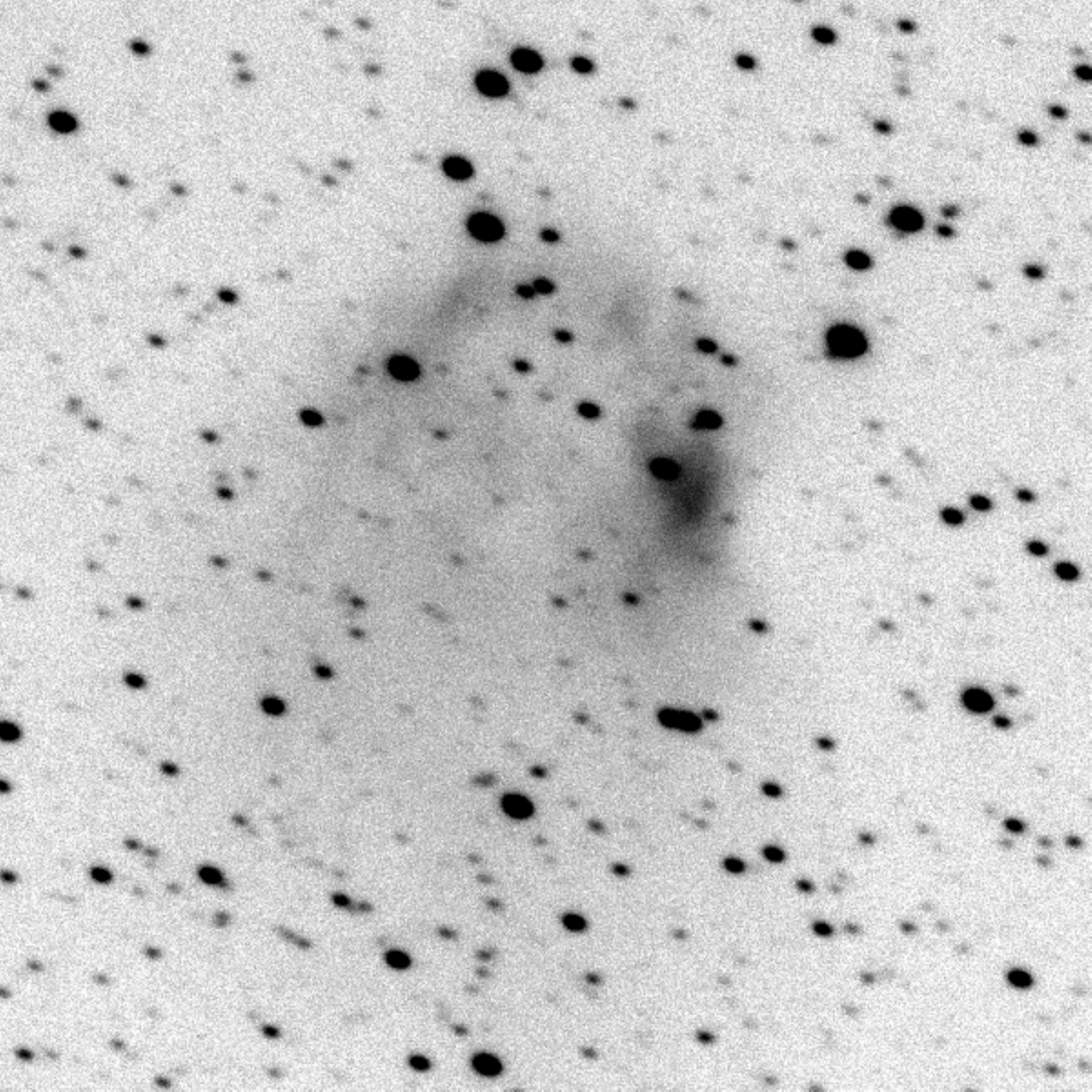}
\includegraphics[height=1.7in]{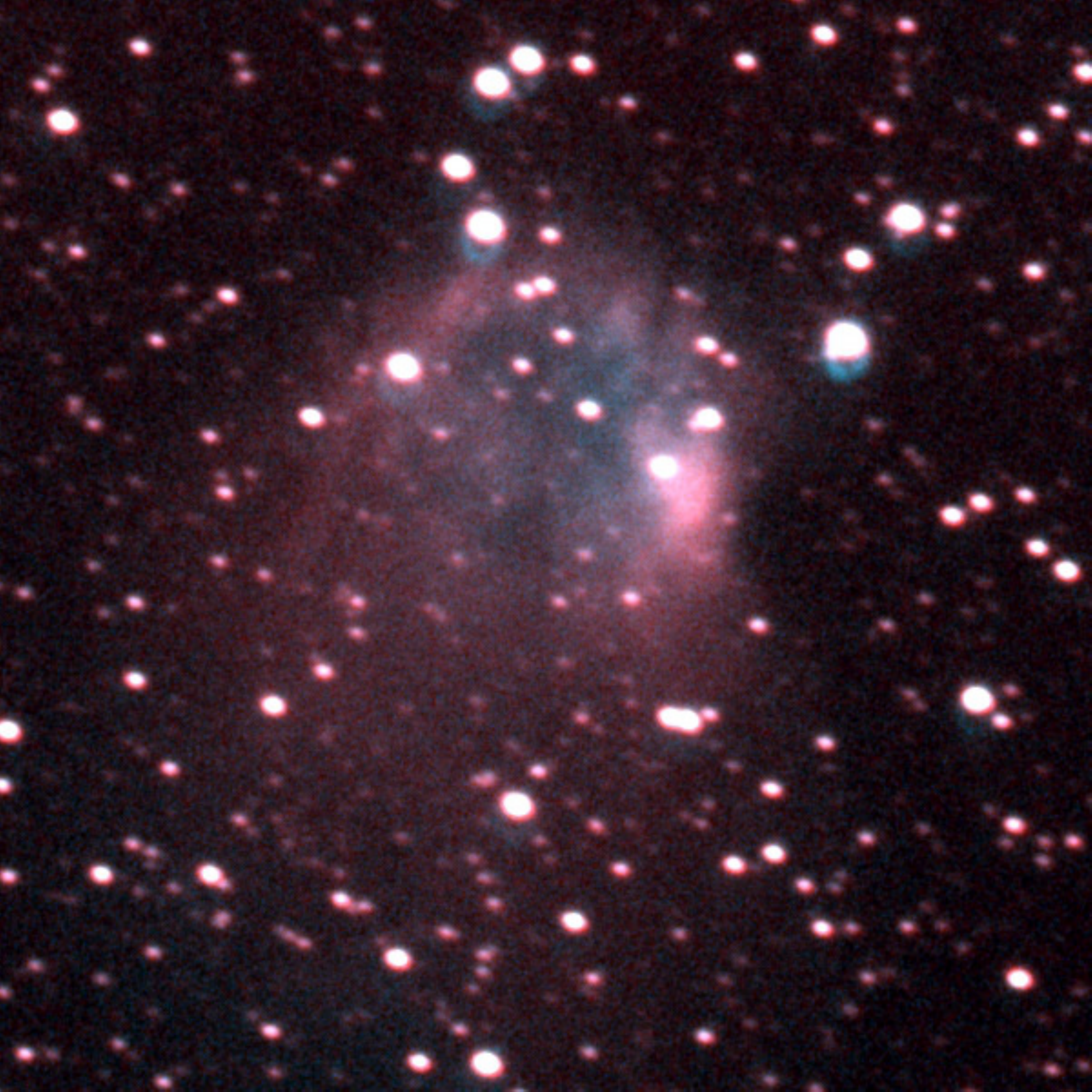}
\vskip .1in 
\includegraphics[height=1.7in]{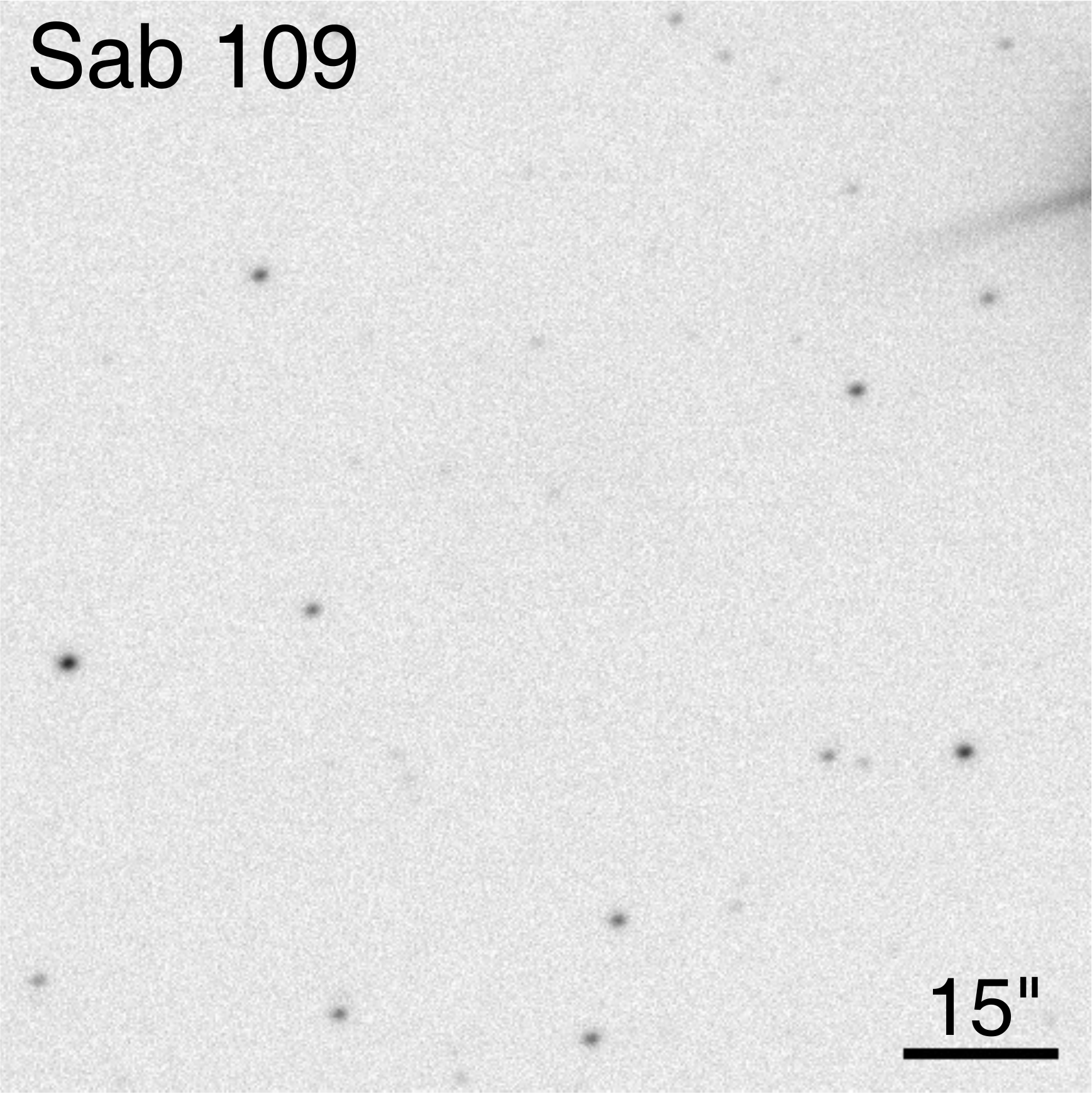} 
\includegraphics[height=1.7in]{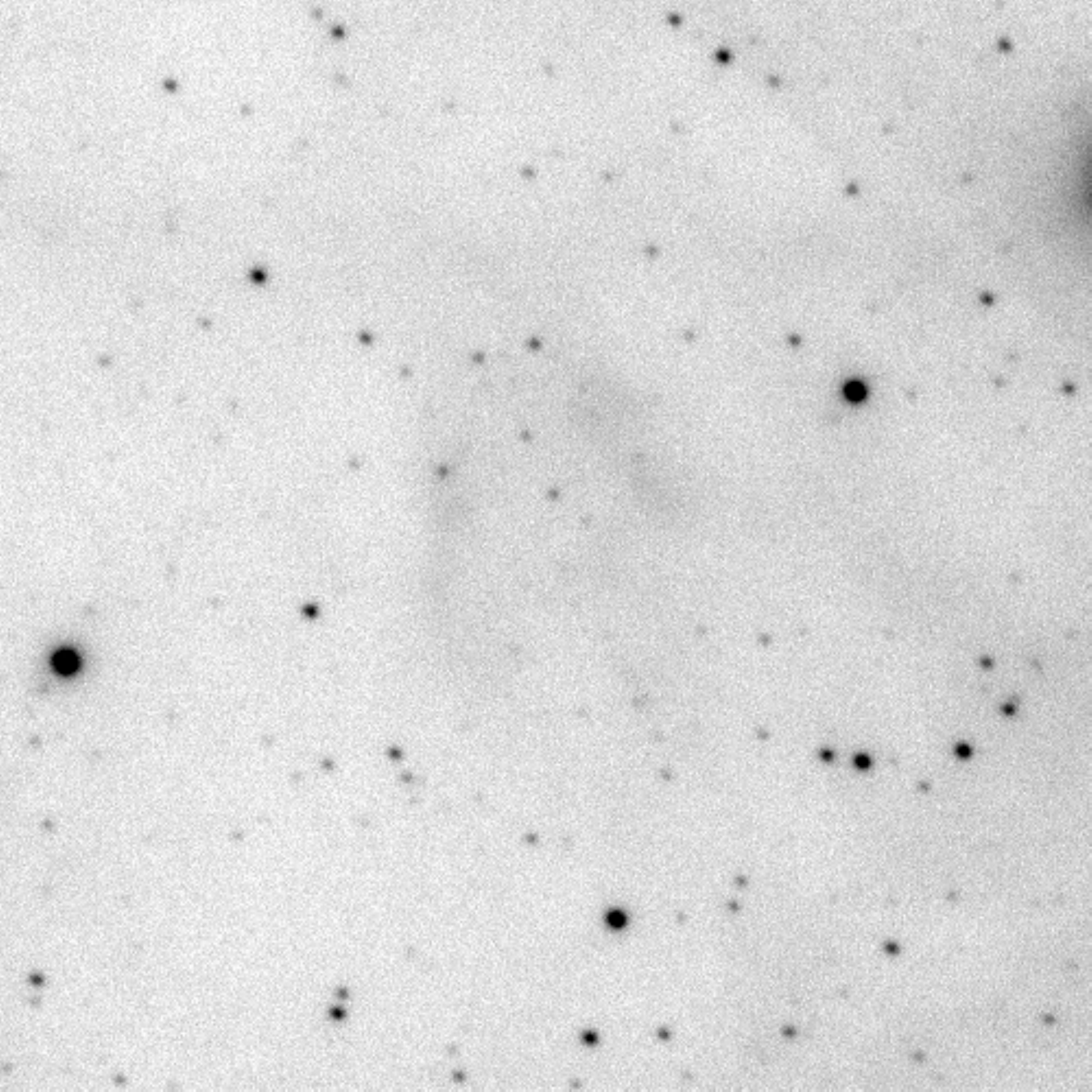}
\includegraphics[height=1.7in]{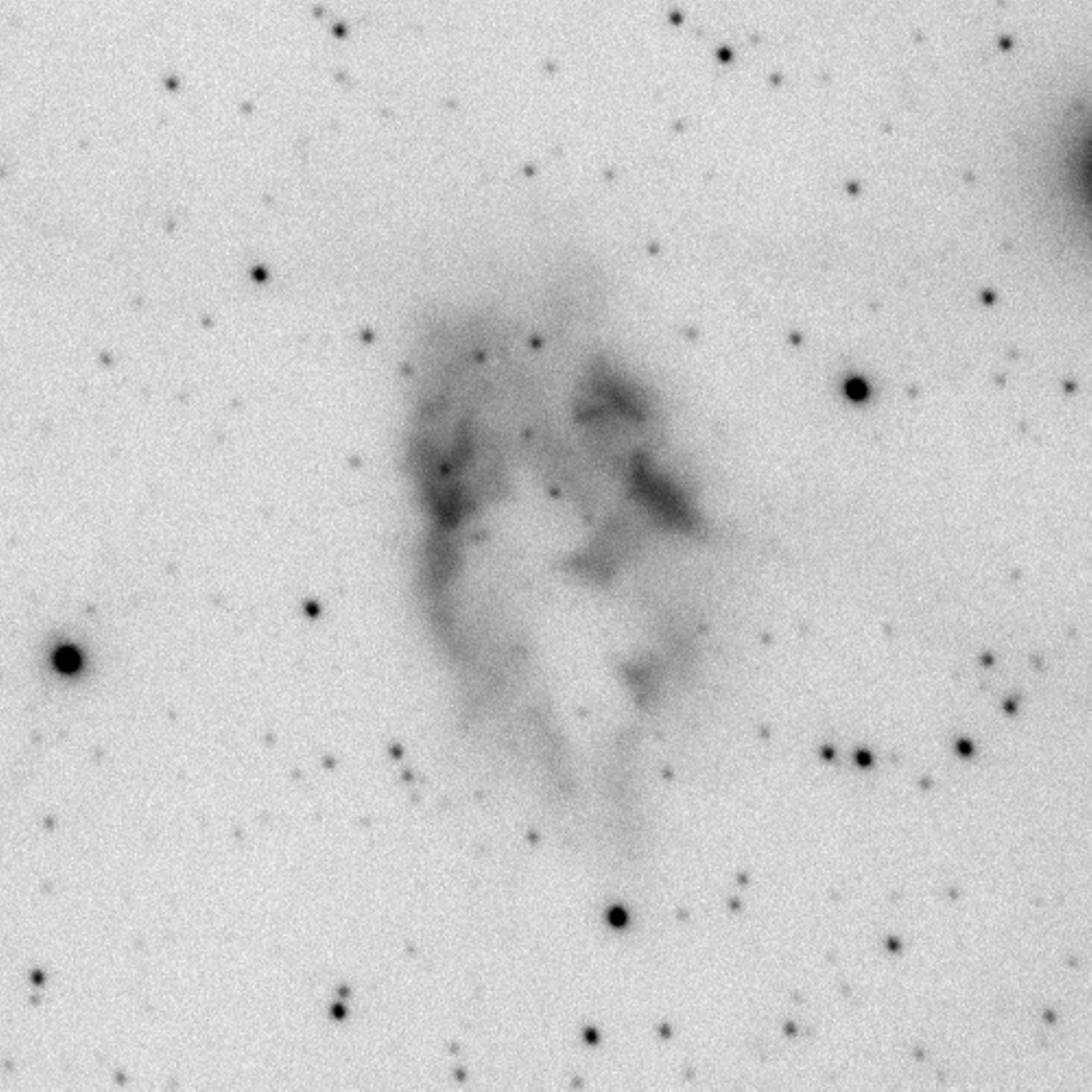}
\includegraphics[height=1.7in]{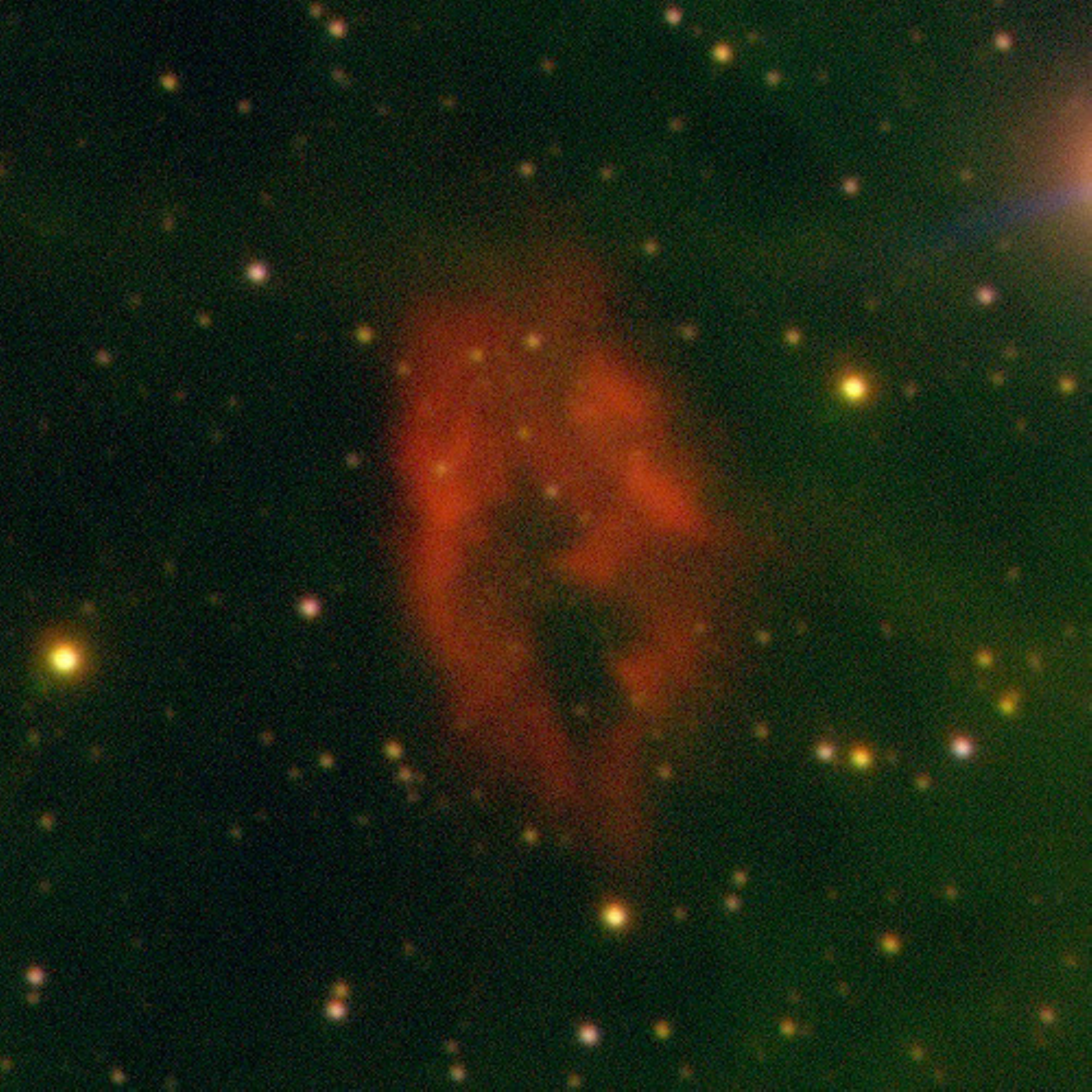}
\vskip .1in 
\includegraphics[height=1.7in]{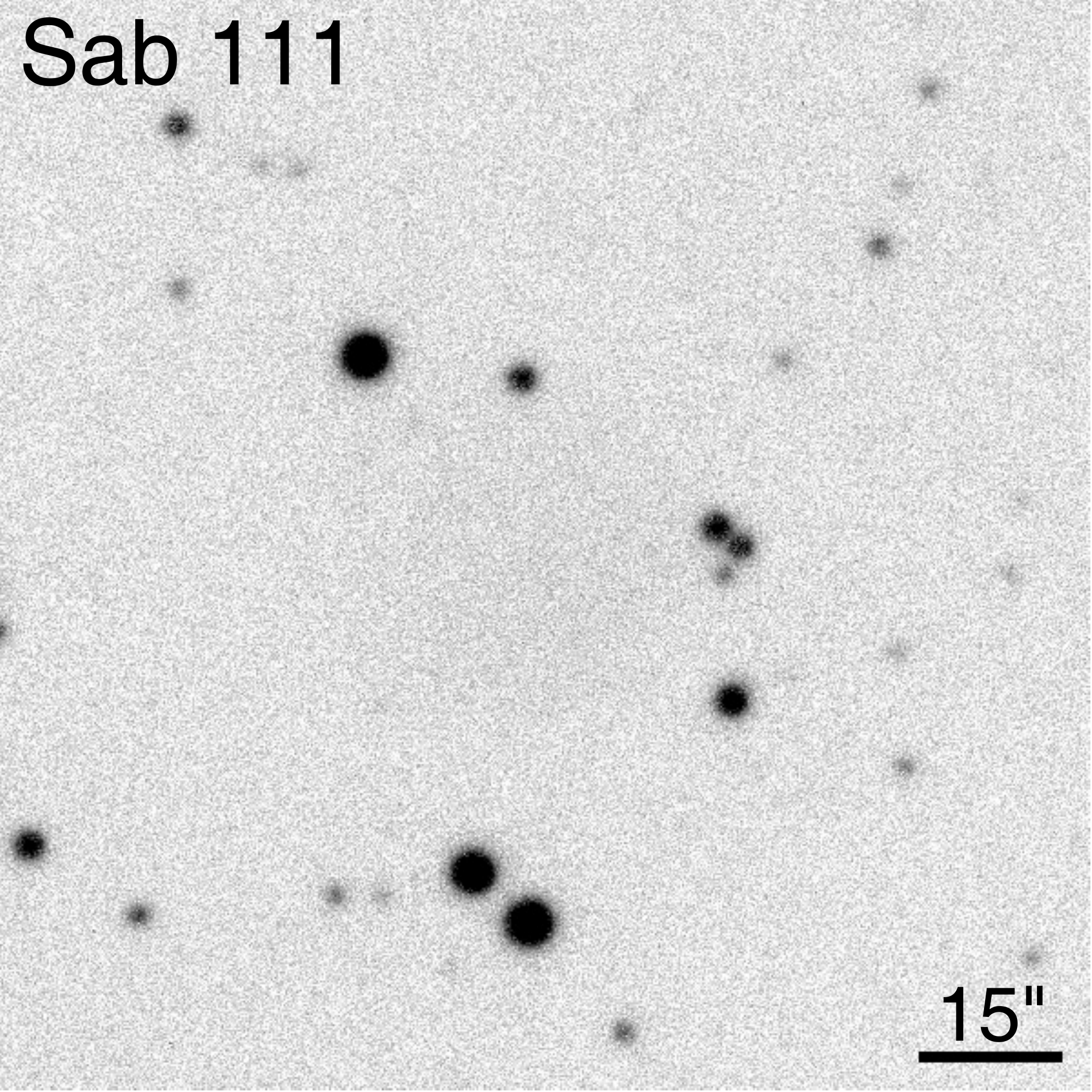} 
\includegraphics[height=1.7in]{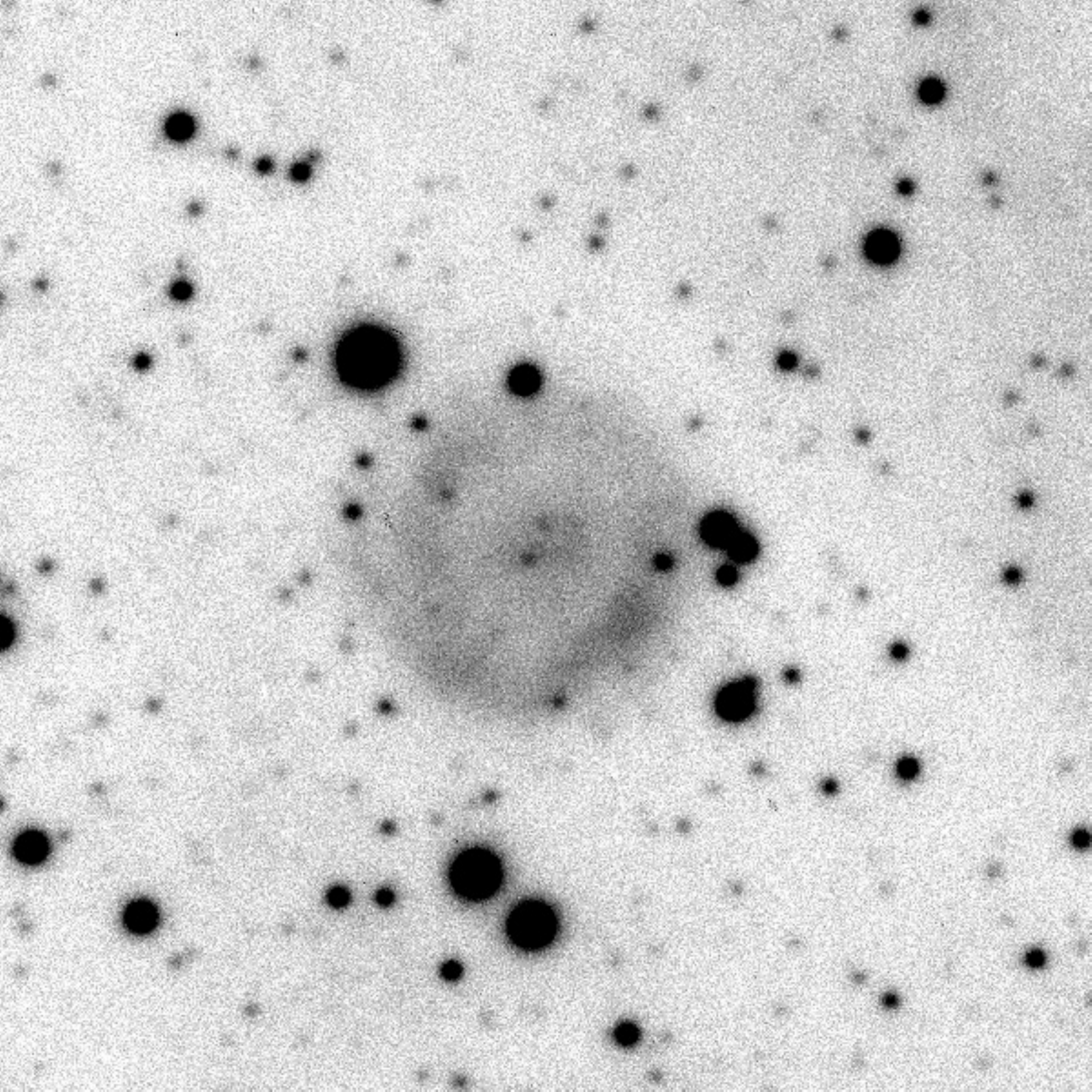}
\includegraphics[height=1.7in]{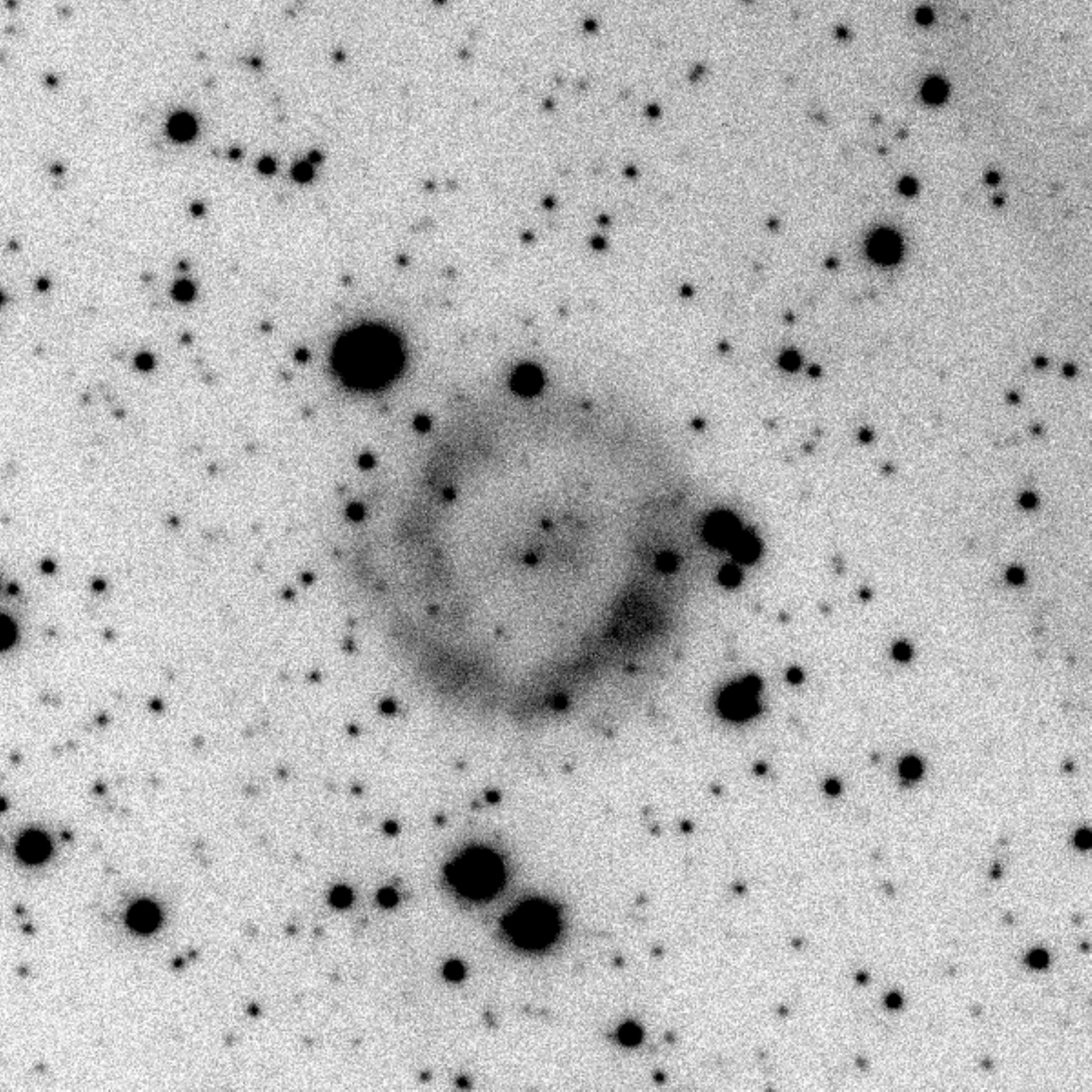}
\includegraphics[height=1.7in]{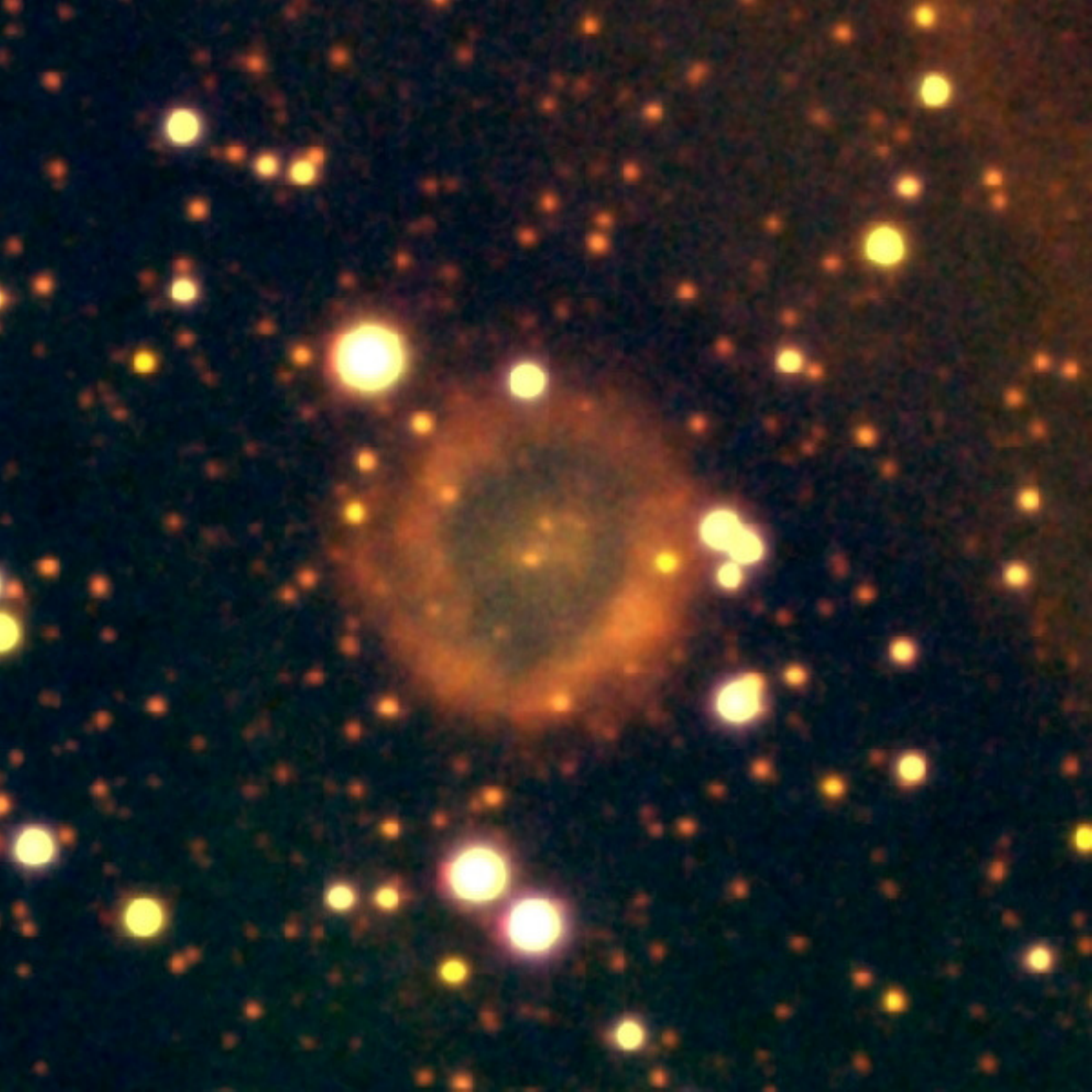}
\caption{Same as Figure~\ref{1.img}. } 
\label{7.img} 
\end{figure*}

\begin{figure*} 
\centering
\includegraphics[height=1.7in]{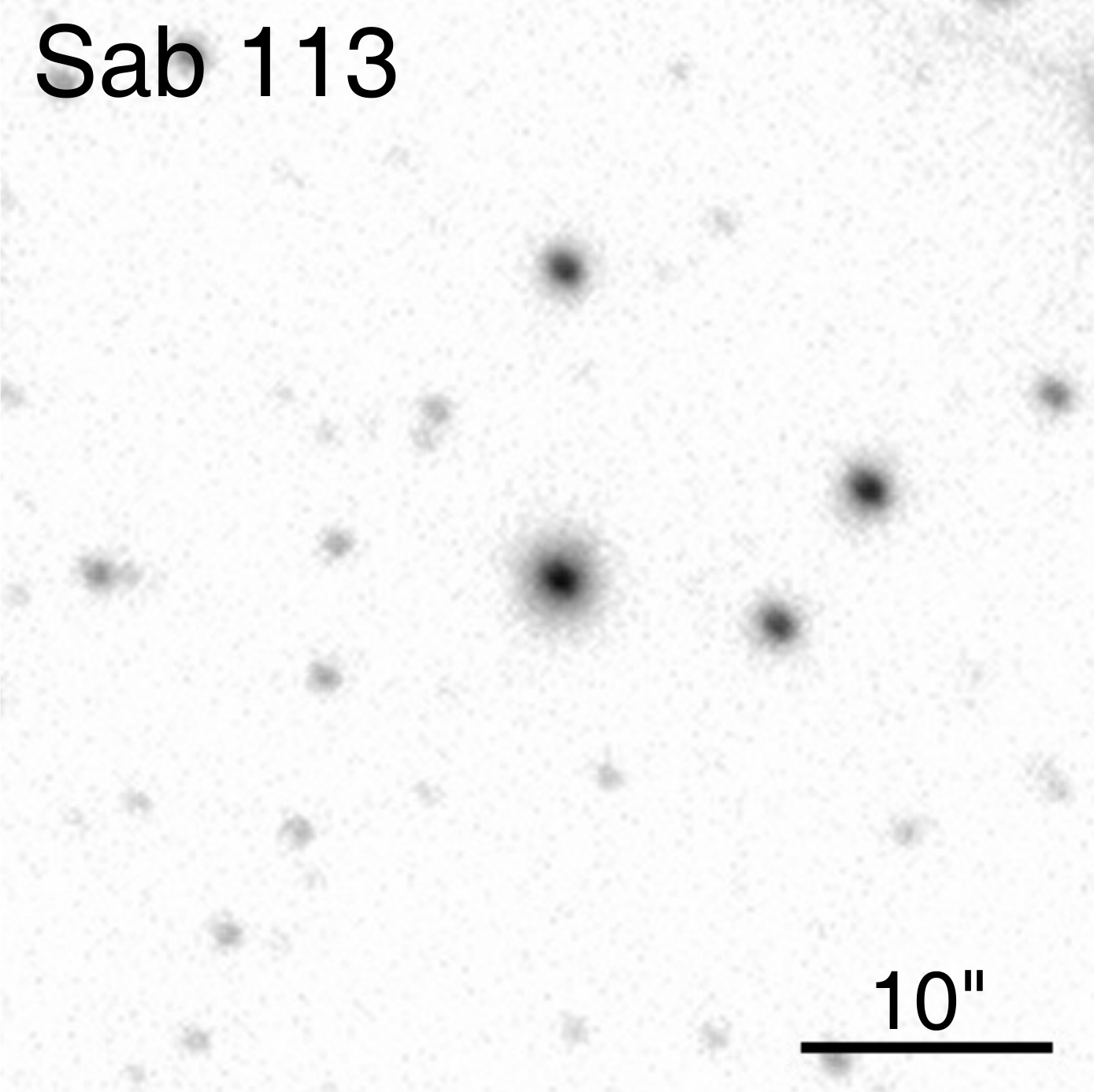} 
\includegraphics[height=1.7in]{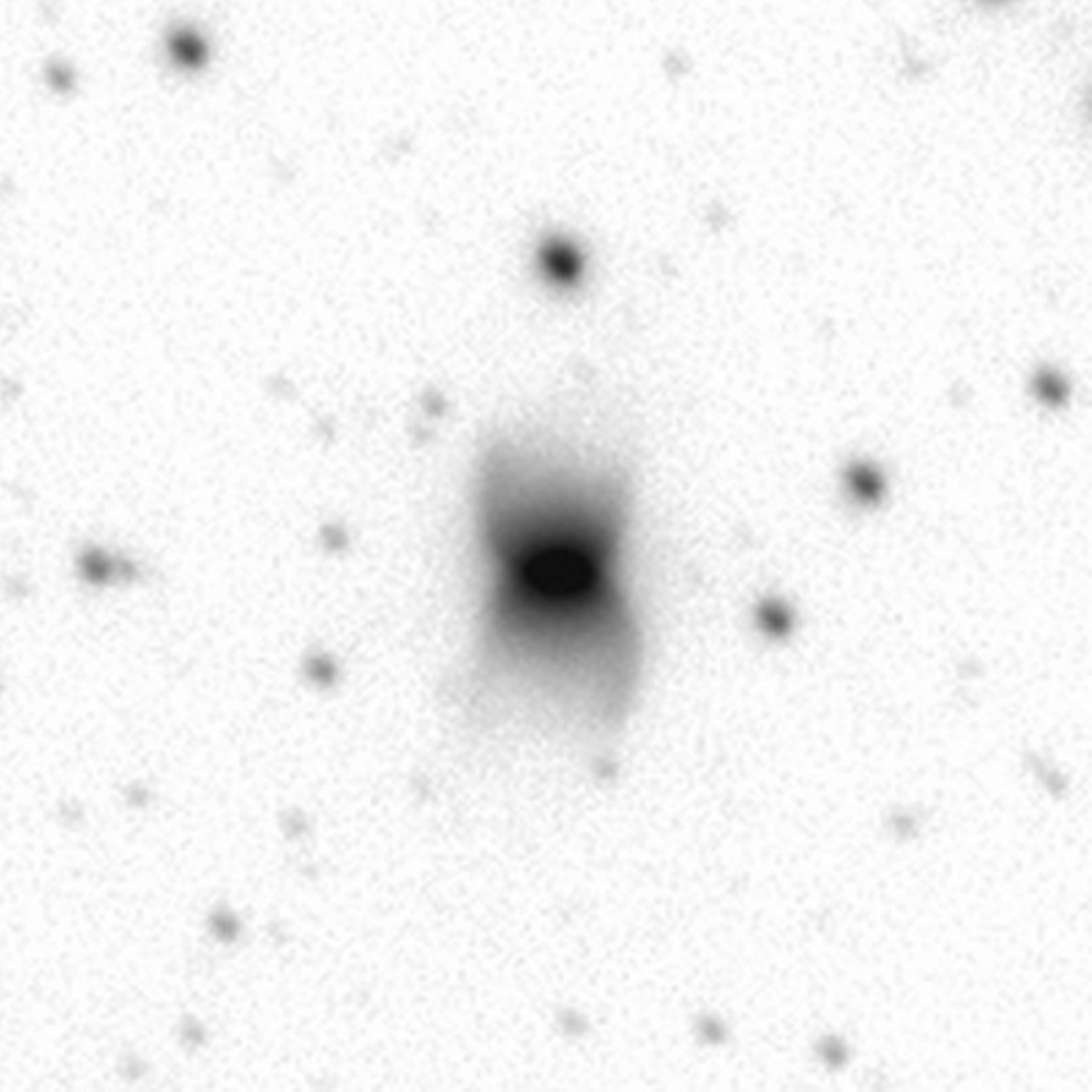}
\includegraphics[height=1.7in]{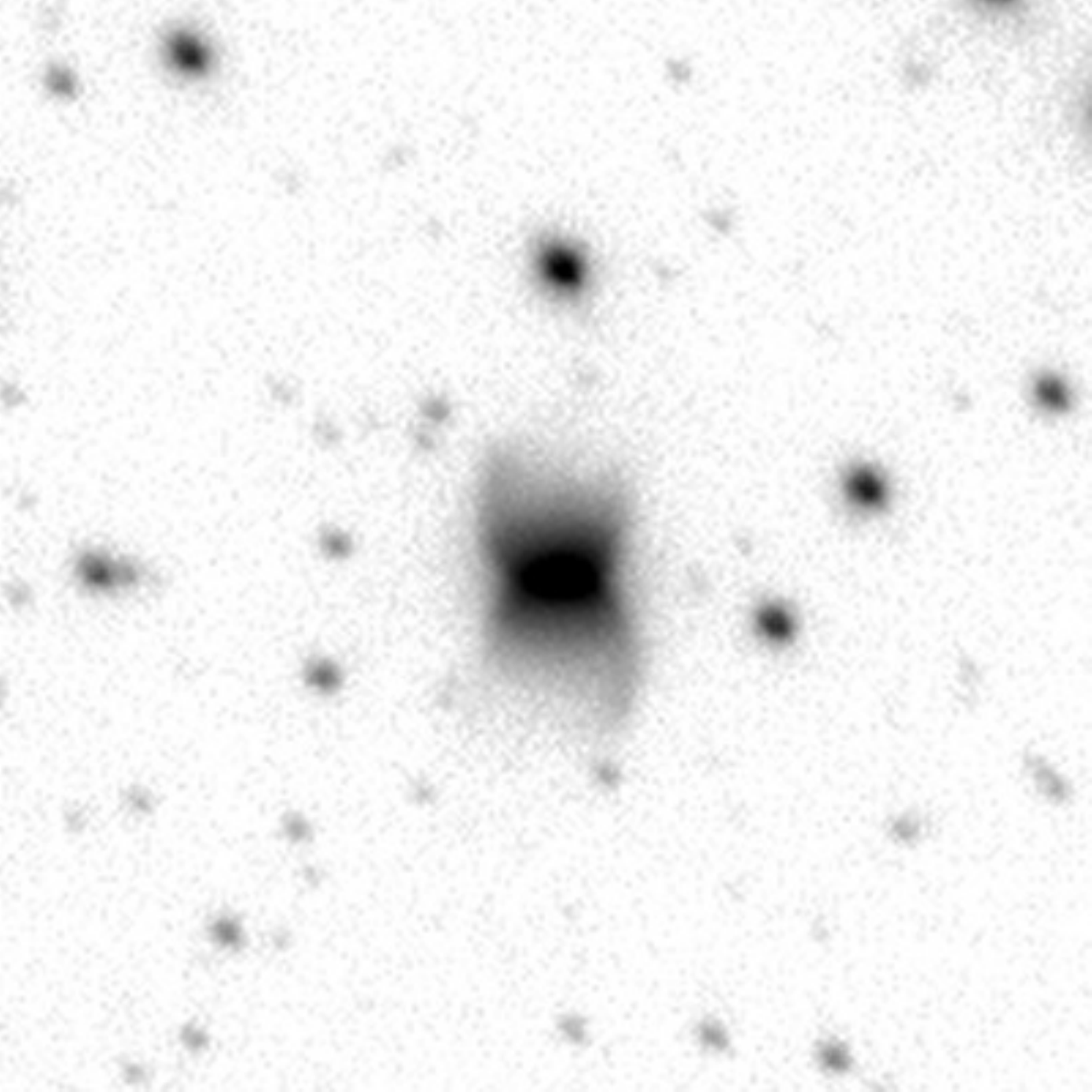}
\includegraphics[height=1.7in]{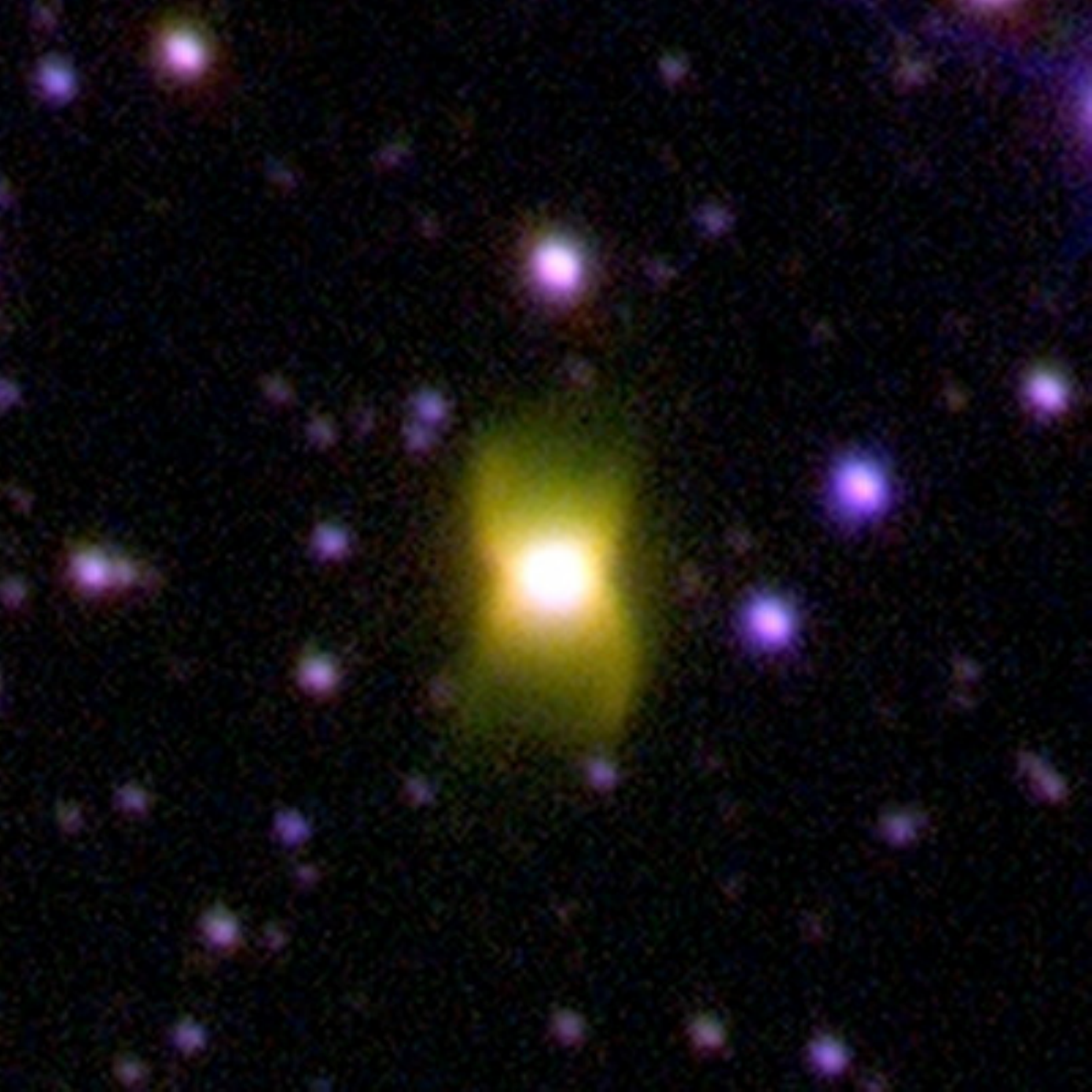}
\vskip .1in 
\includegraphics[height=1.7in]{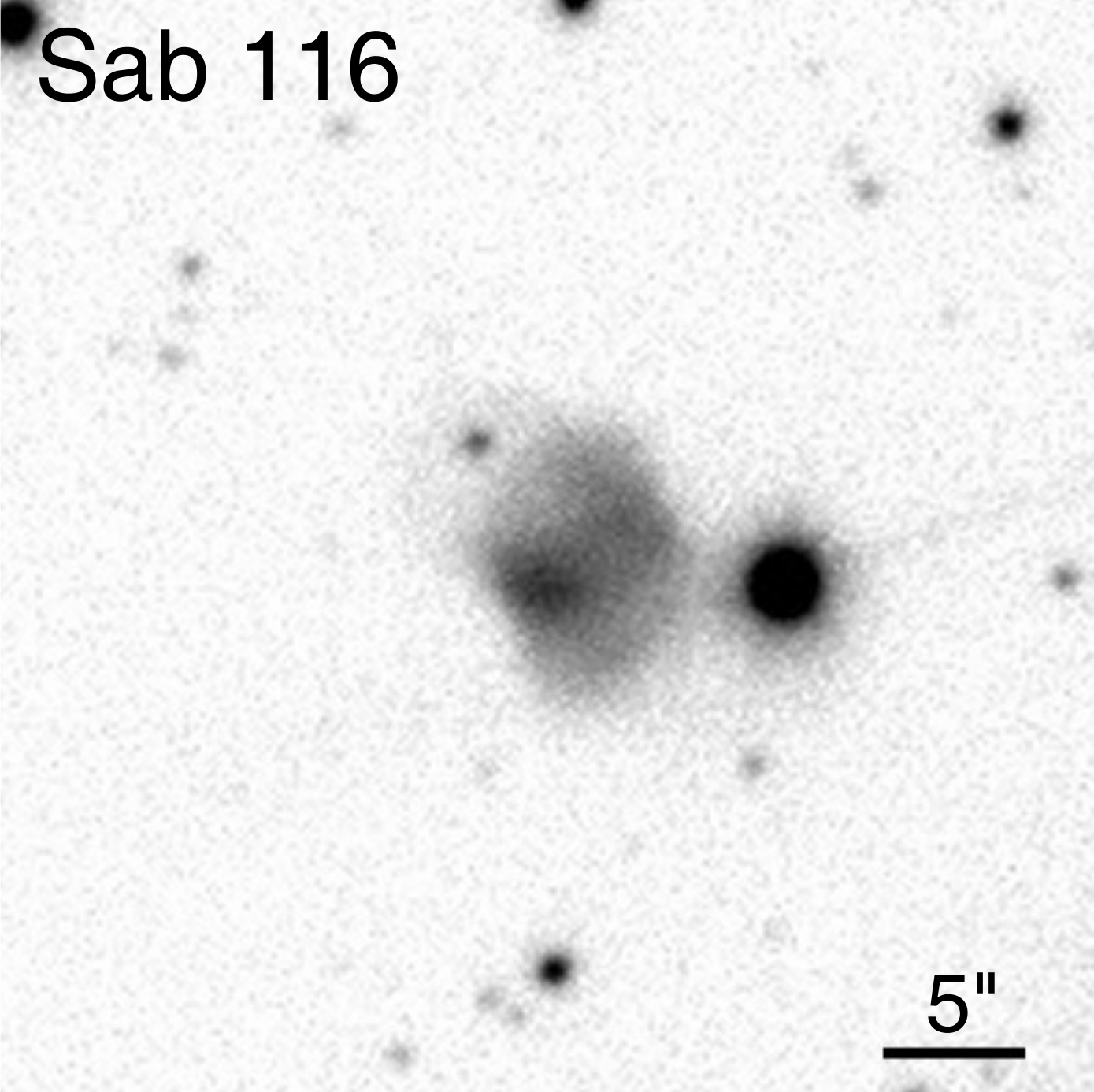} 
\includegraphics[height=1.7in]{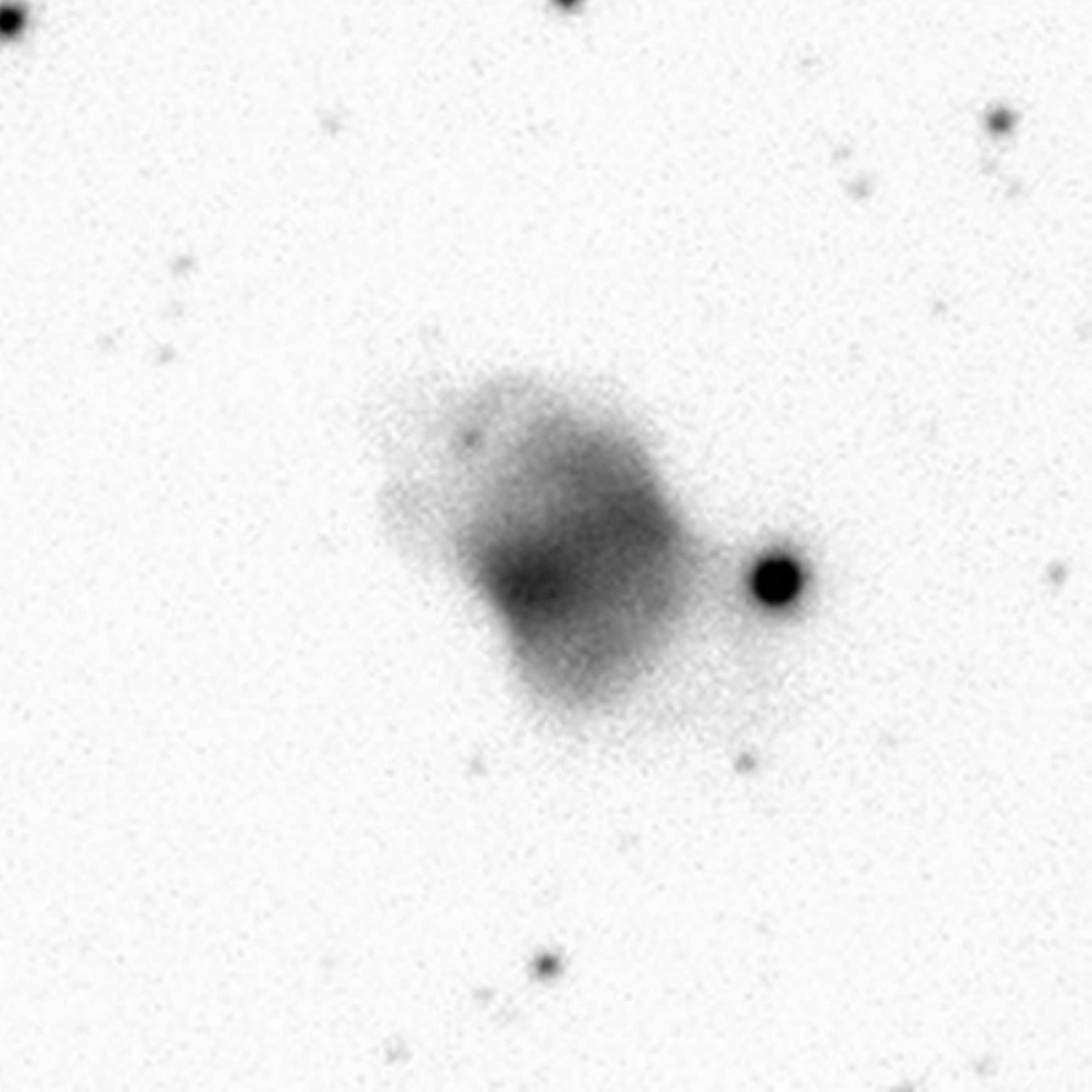}
\includegraphics[height=1.7in]{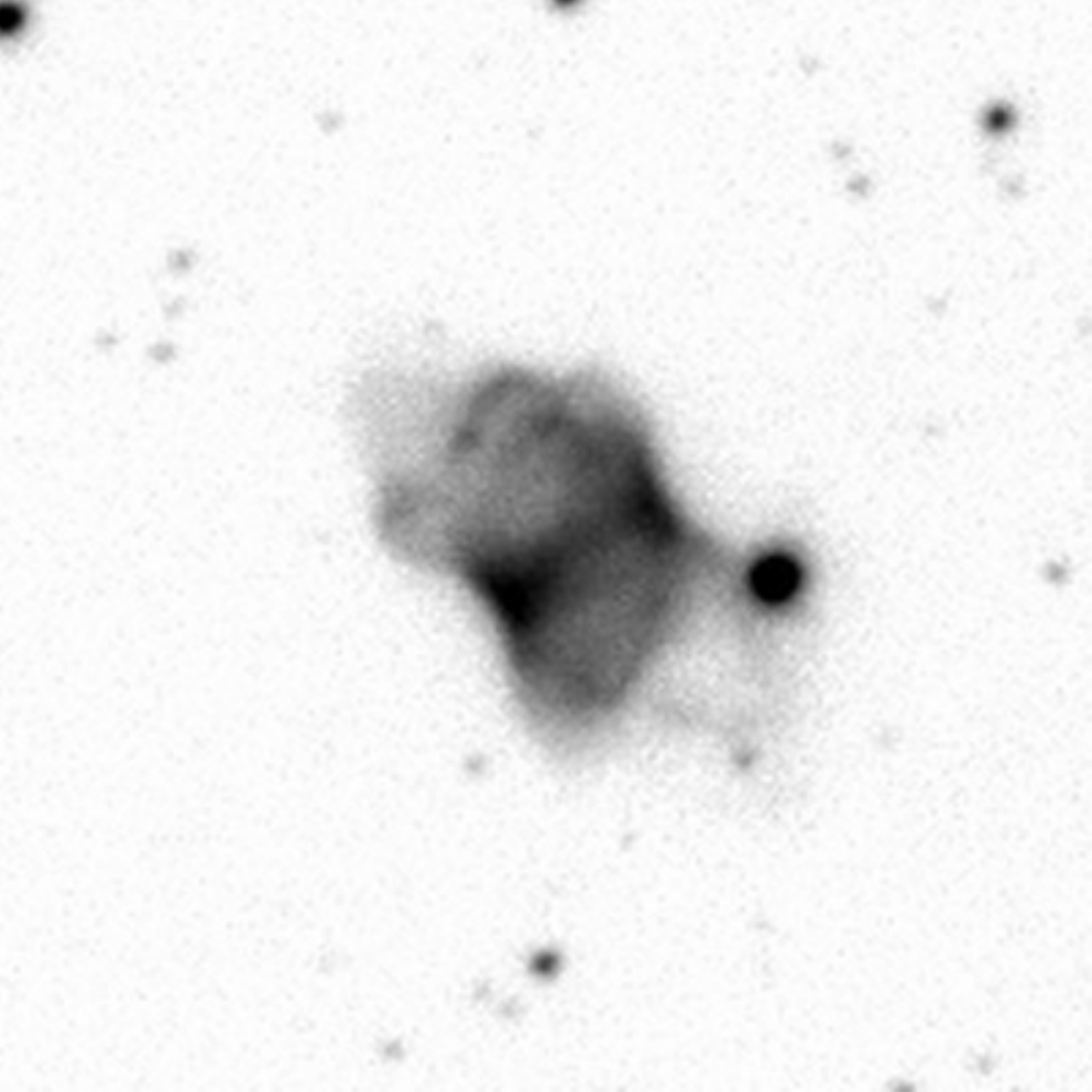}
\includegraphics[height=1.7in]{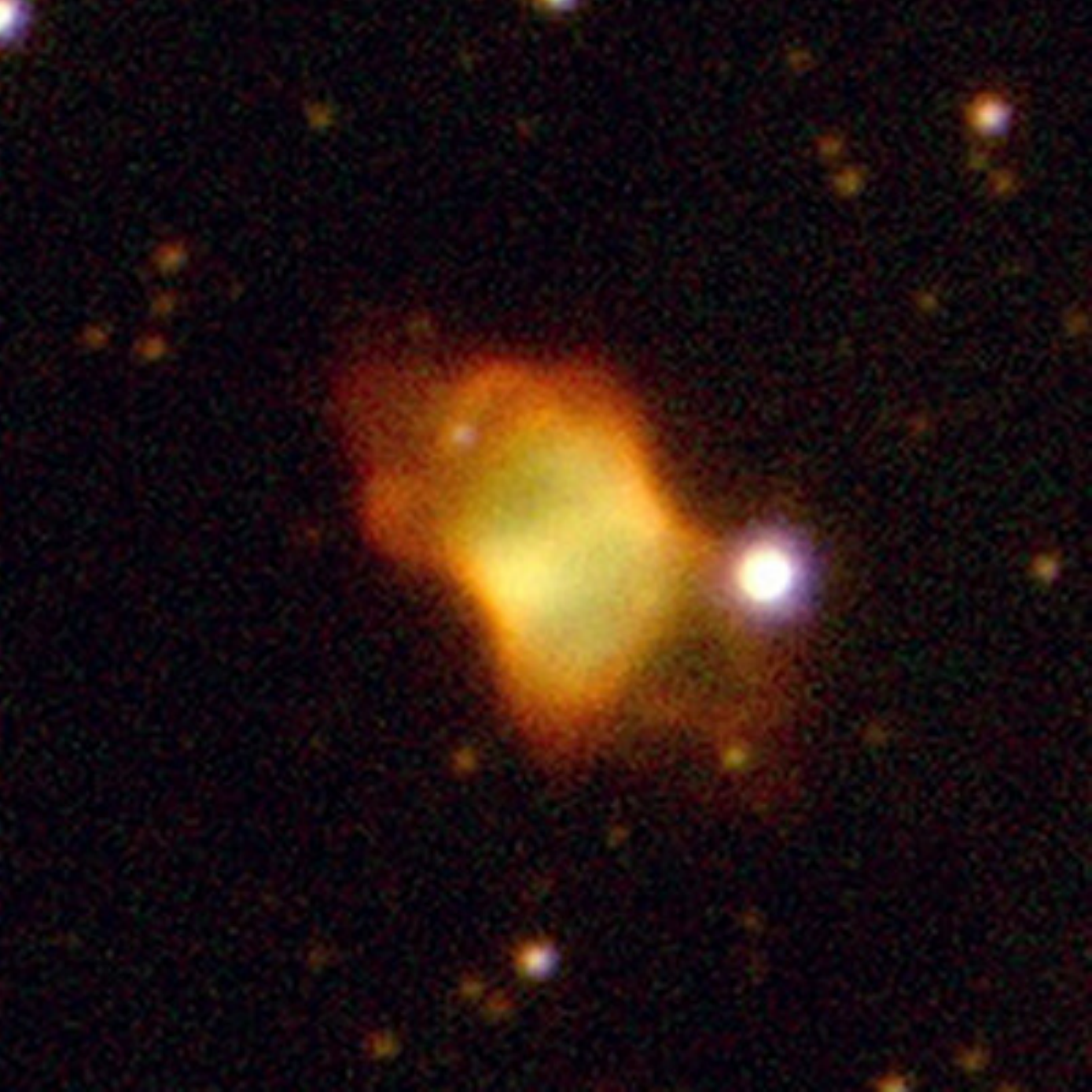}
\vskip .1in 
\includegraphics[height=1.7in]{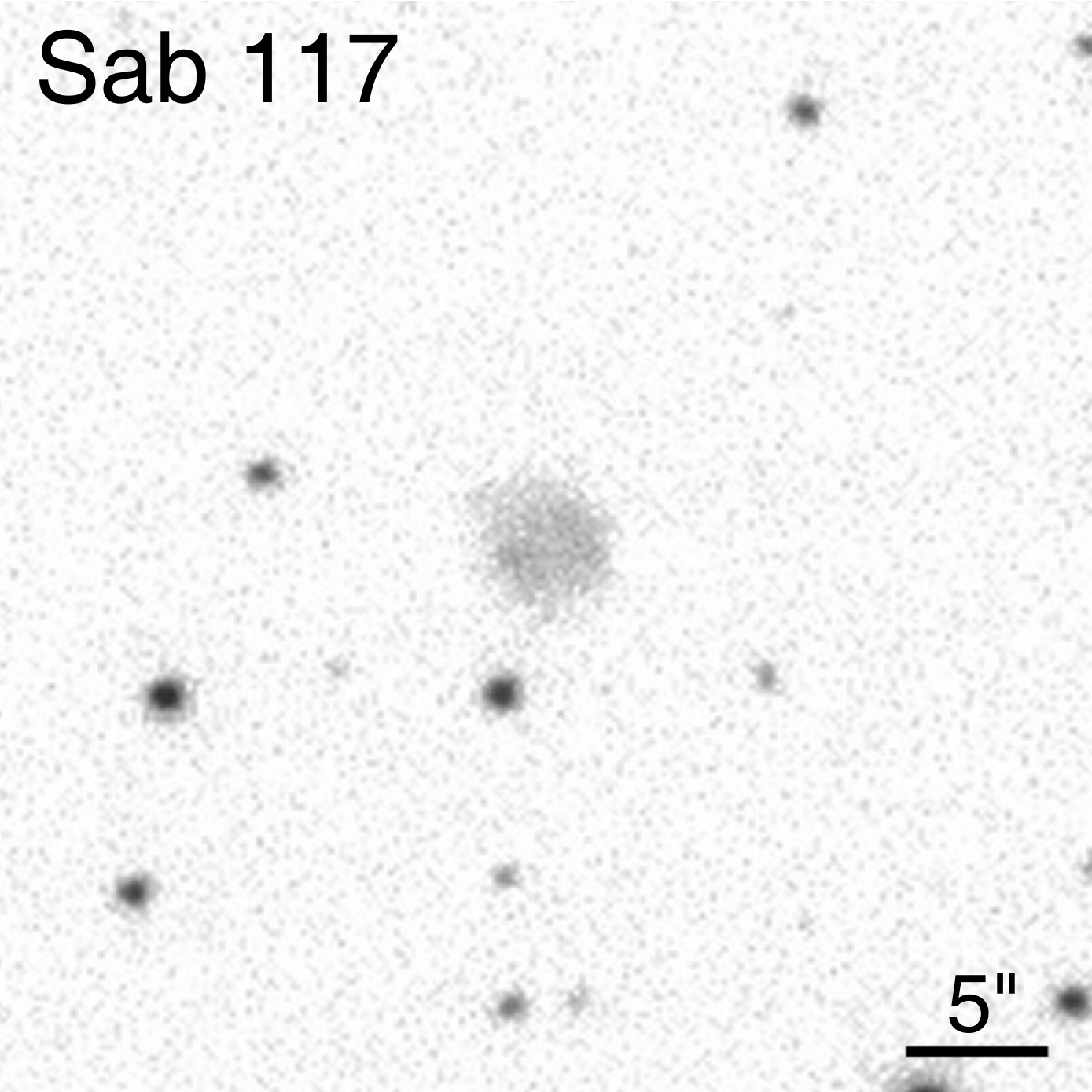} 
\includegraphics[height=1.7in]{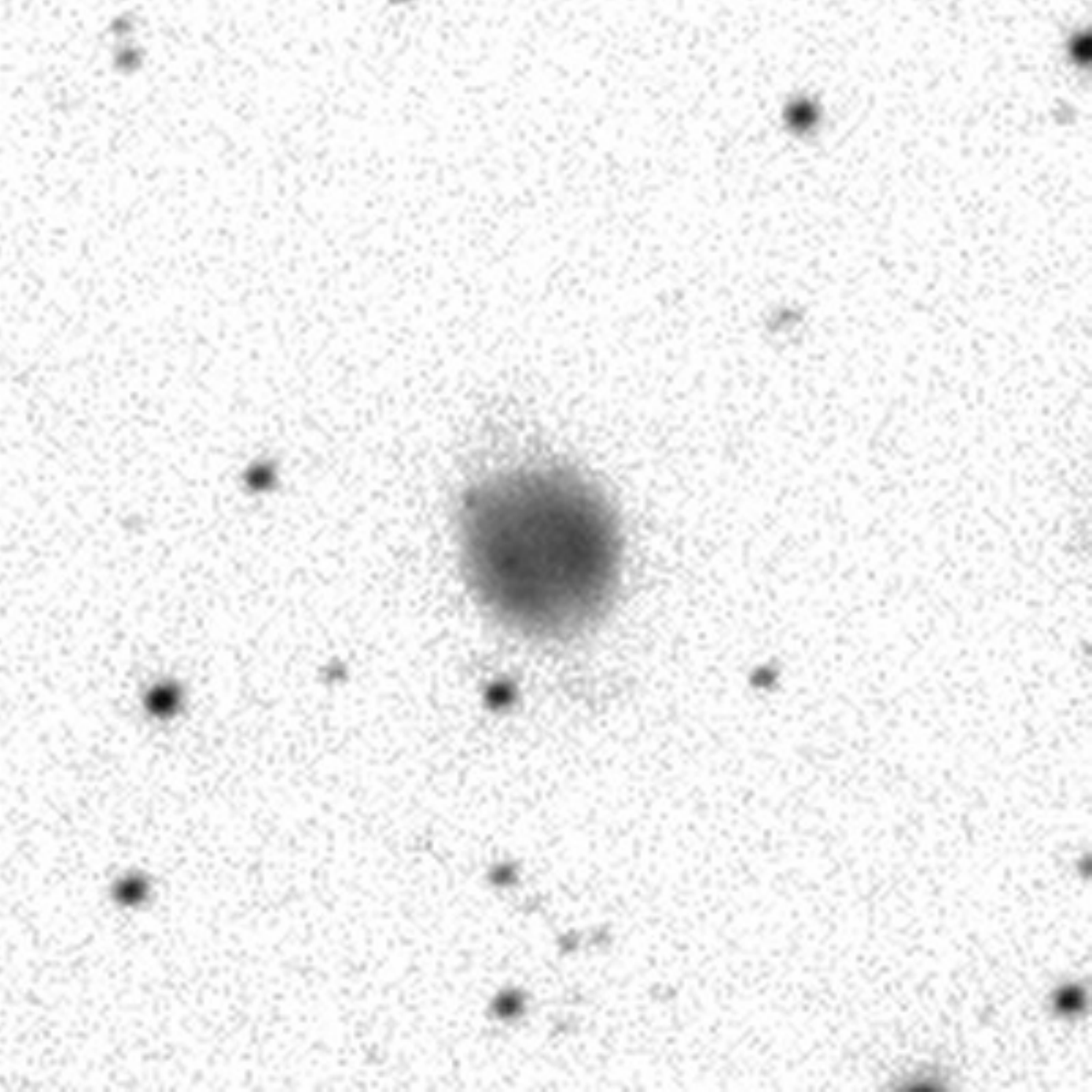}
\includegraphics[height=1.7in]{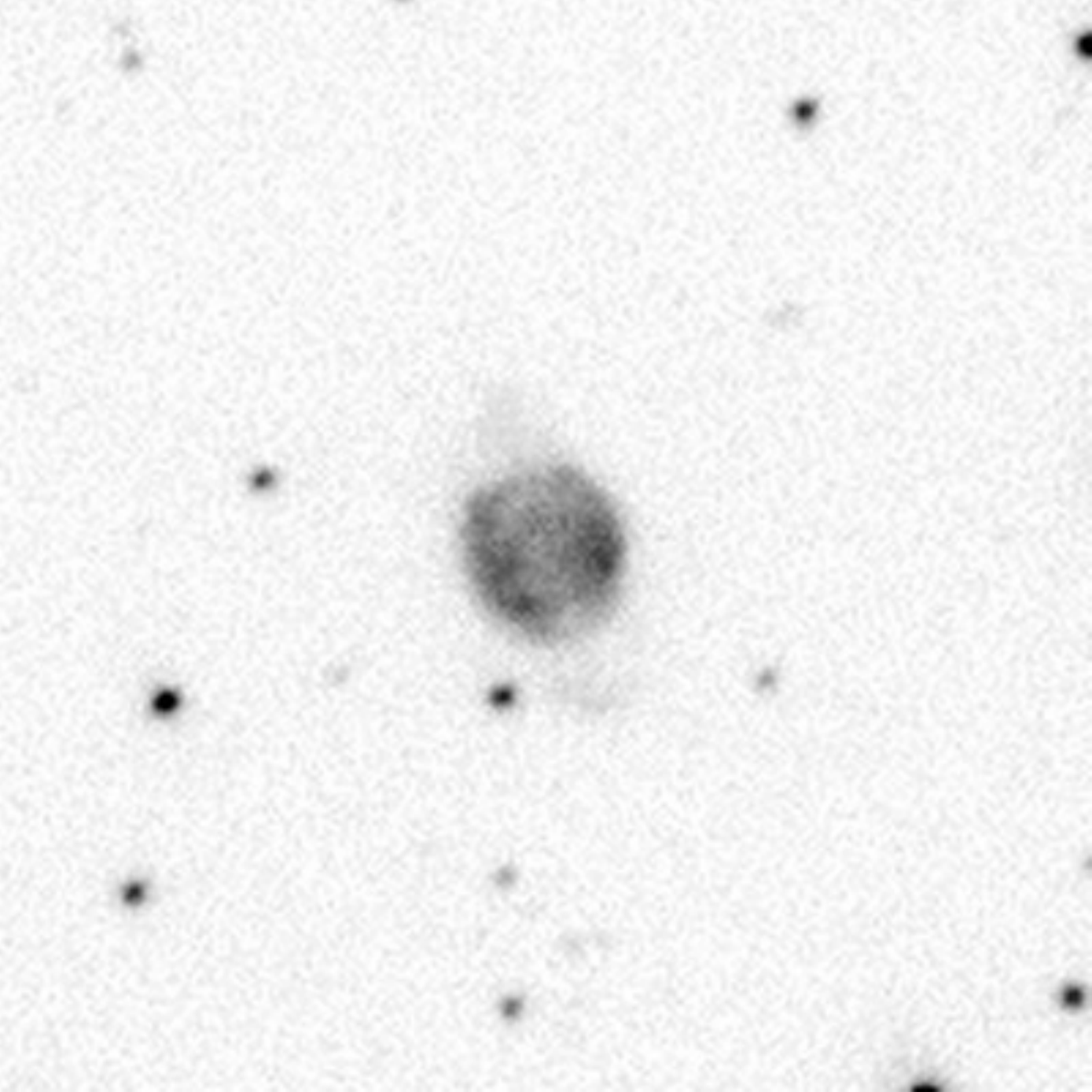}
\includegraphics[height=1.7in]{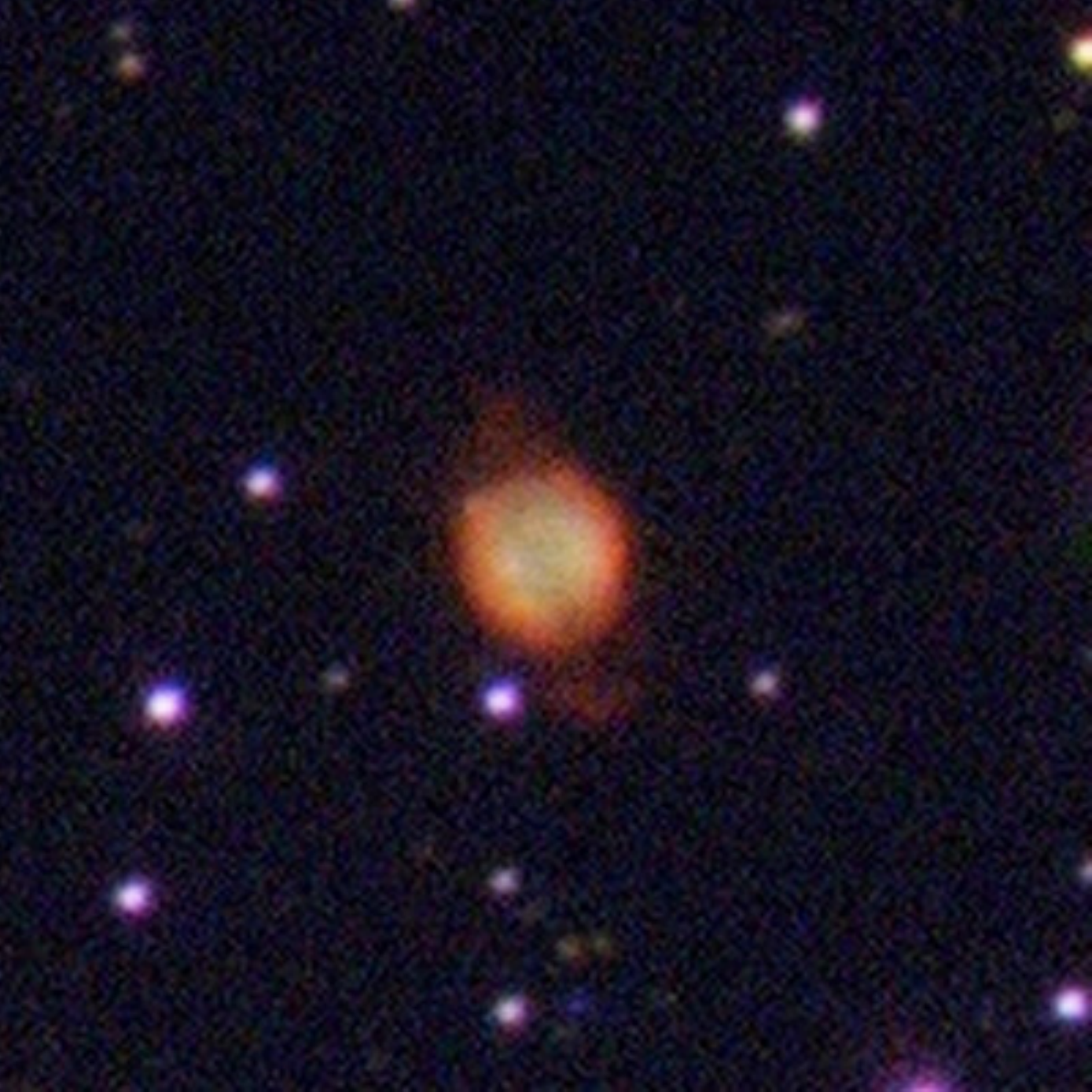}
\vskip .1in 
\includegraphics[height=1.7in]{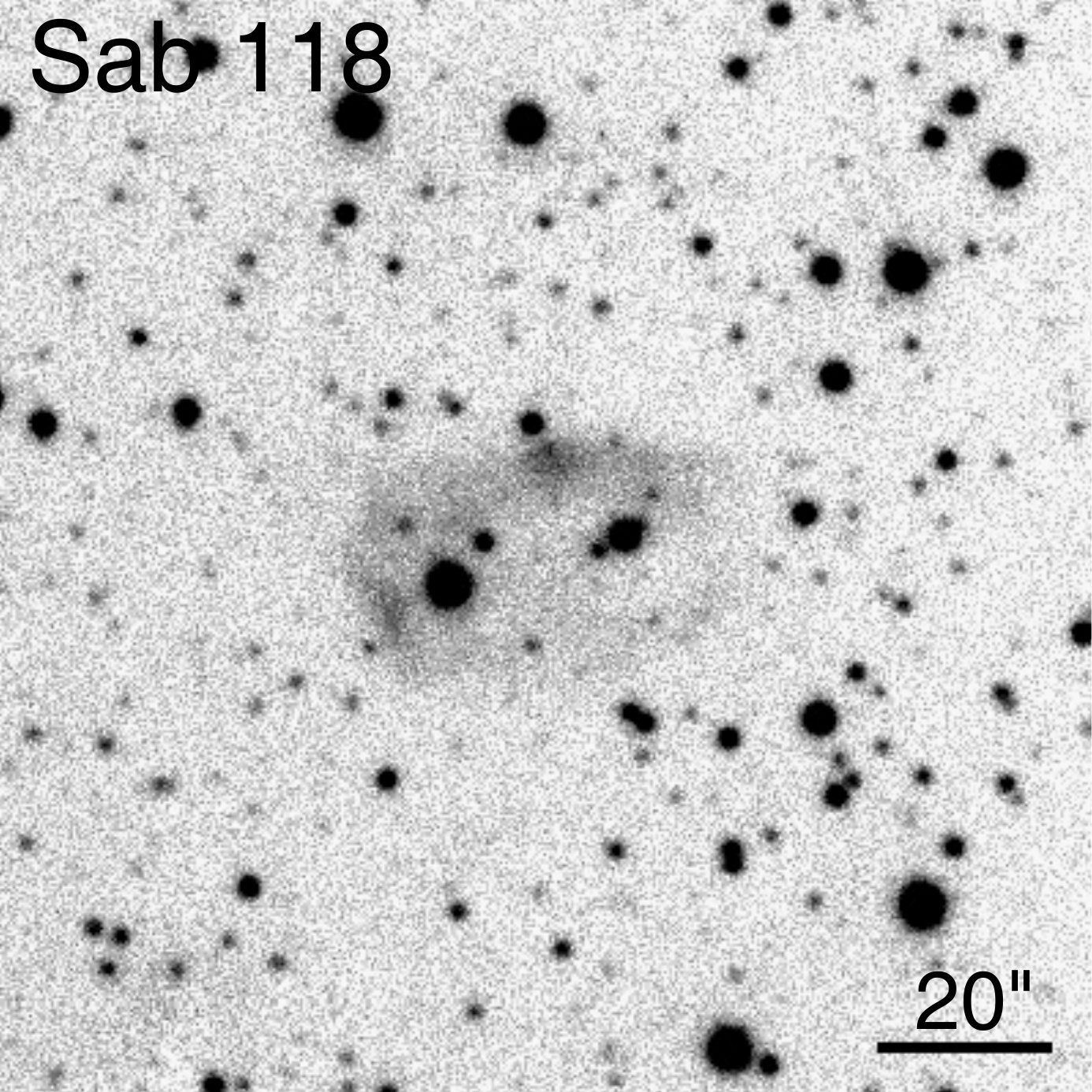} 
\includegraphics[height=1.7in]{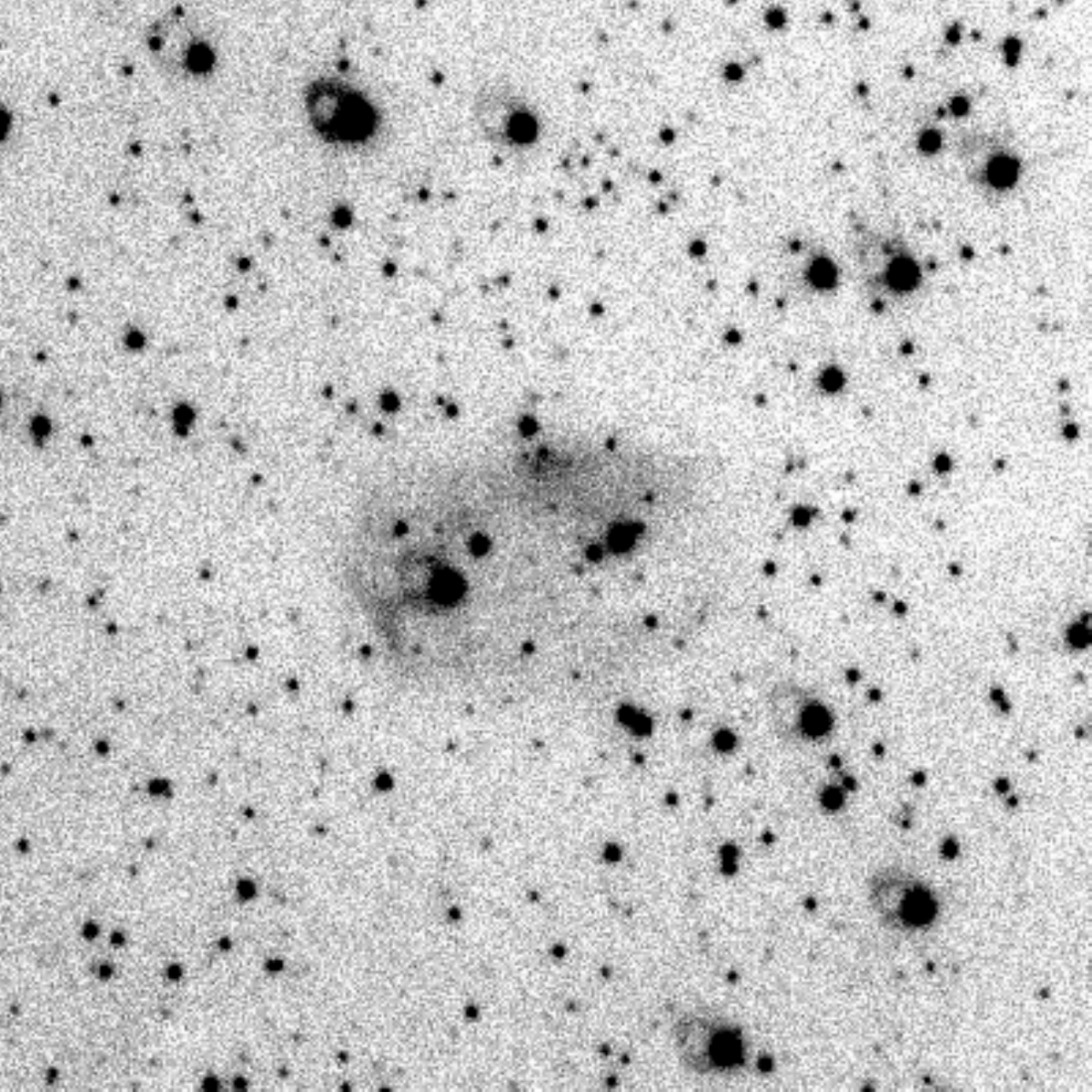}
\includegraphics[height=1.7in]{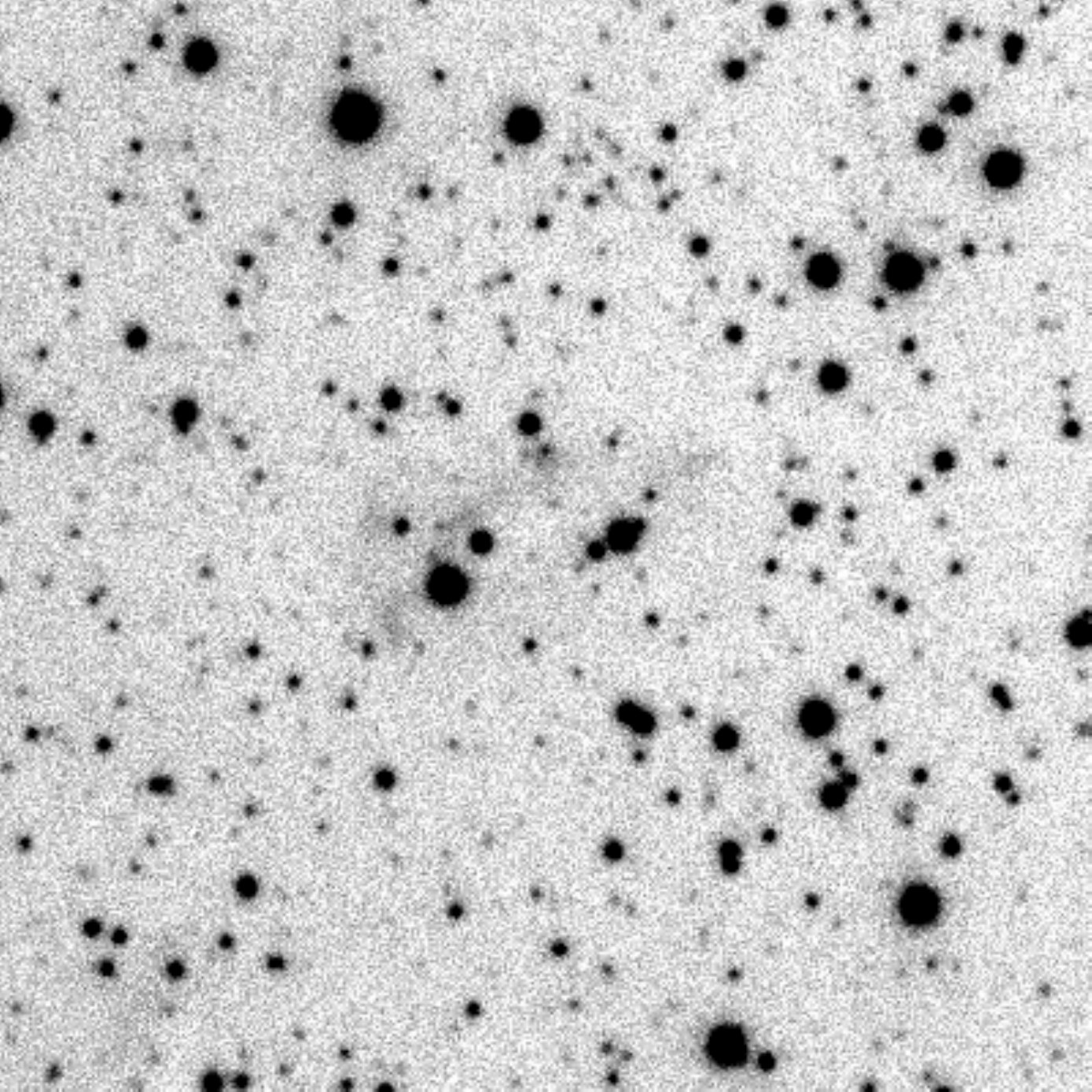}
\includegraphics[height=1.7in]{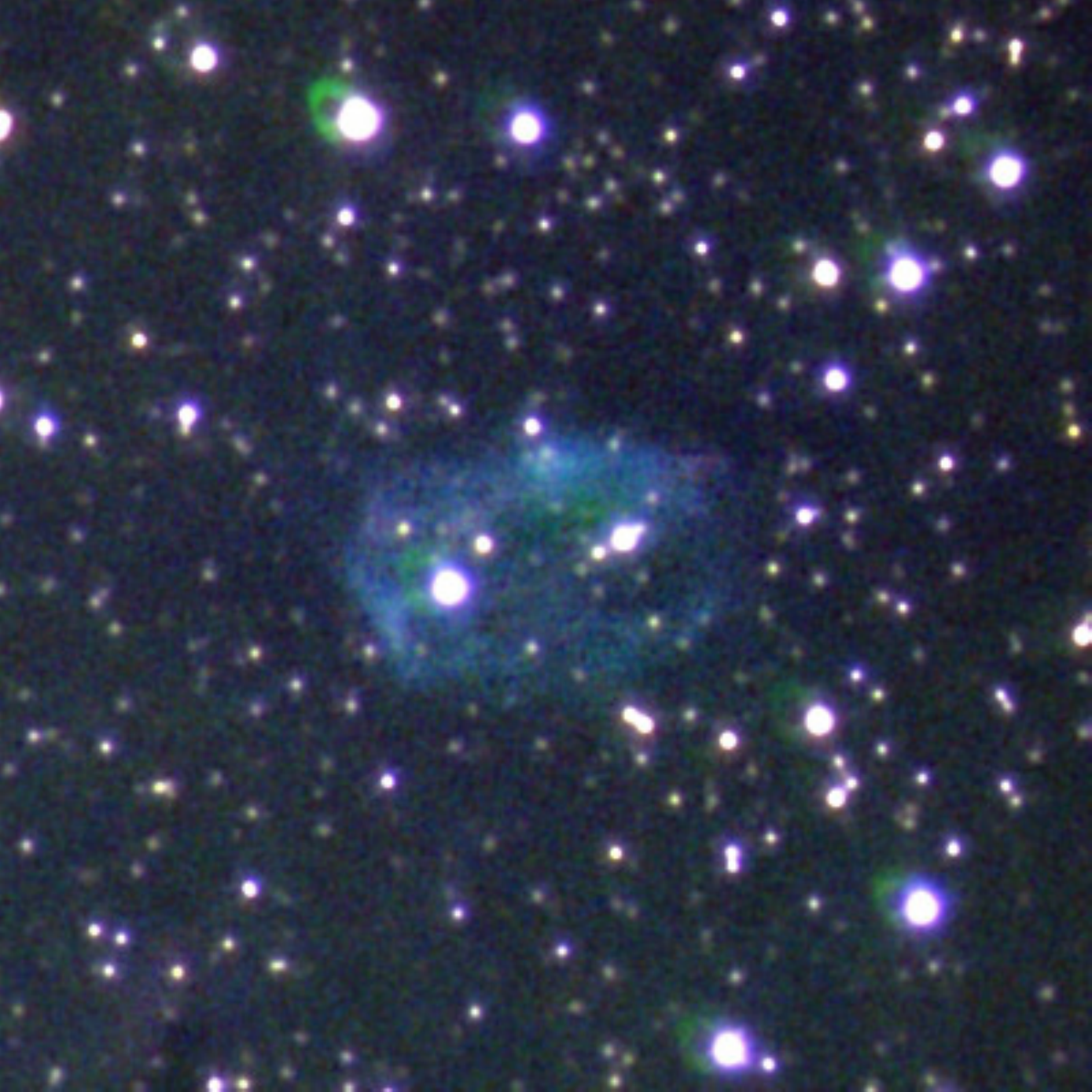}
\vskip .1in
\includegraphics[height=1.7in]{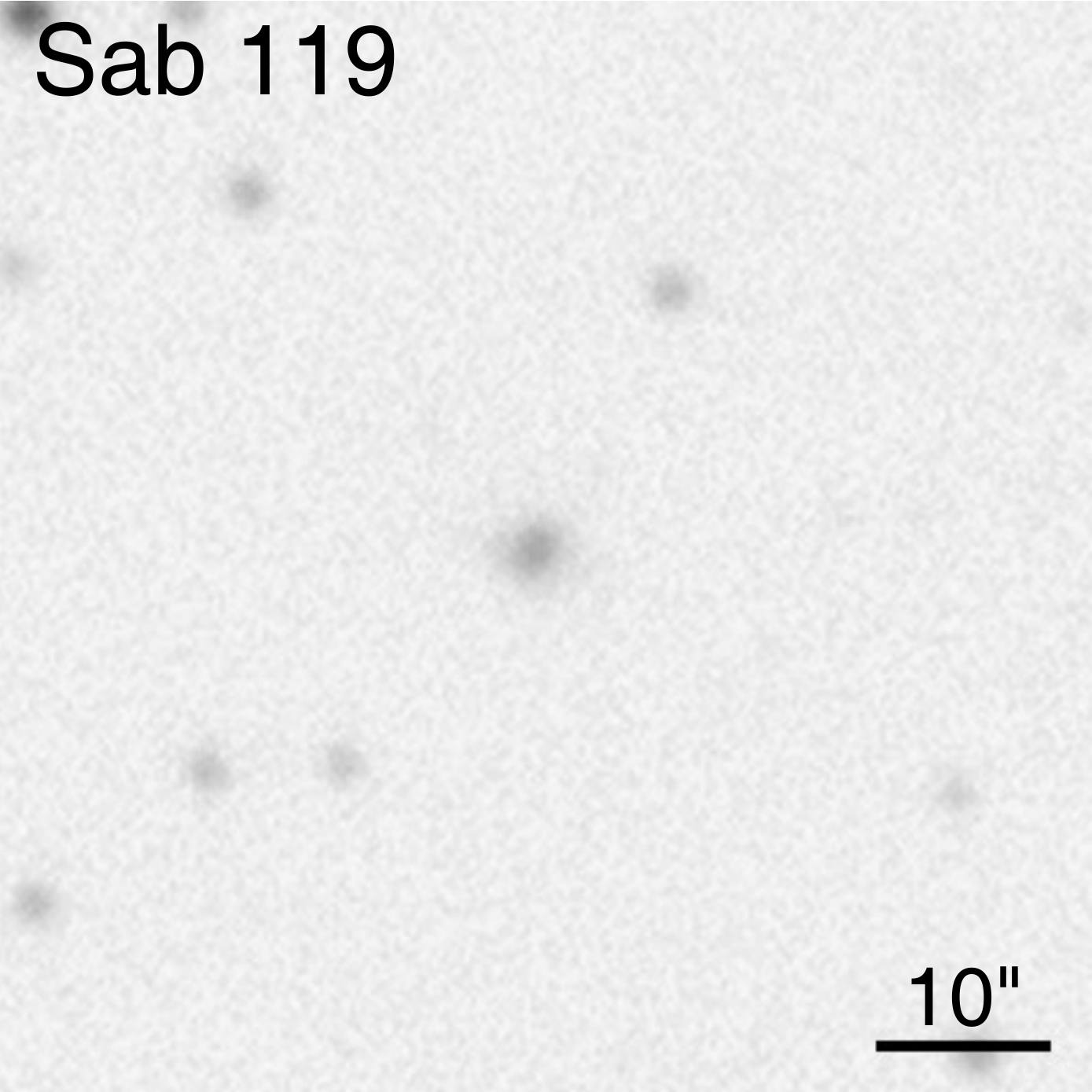} 
\includegraphics[height=1.7in]{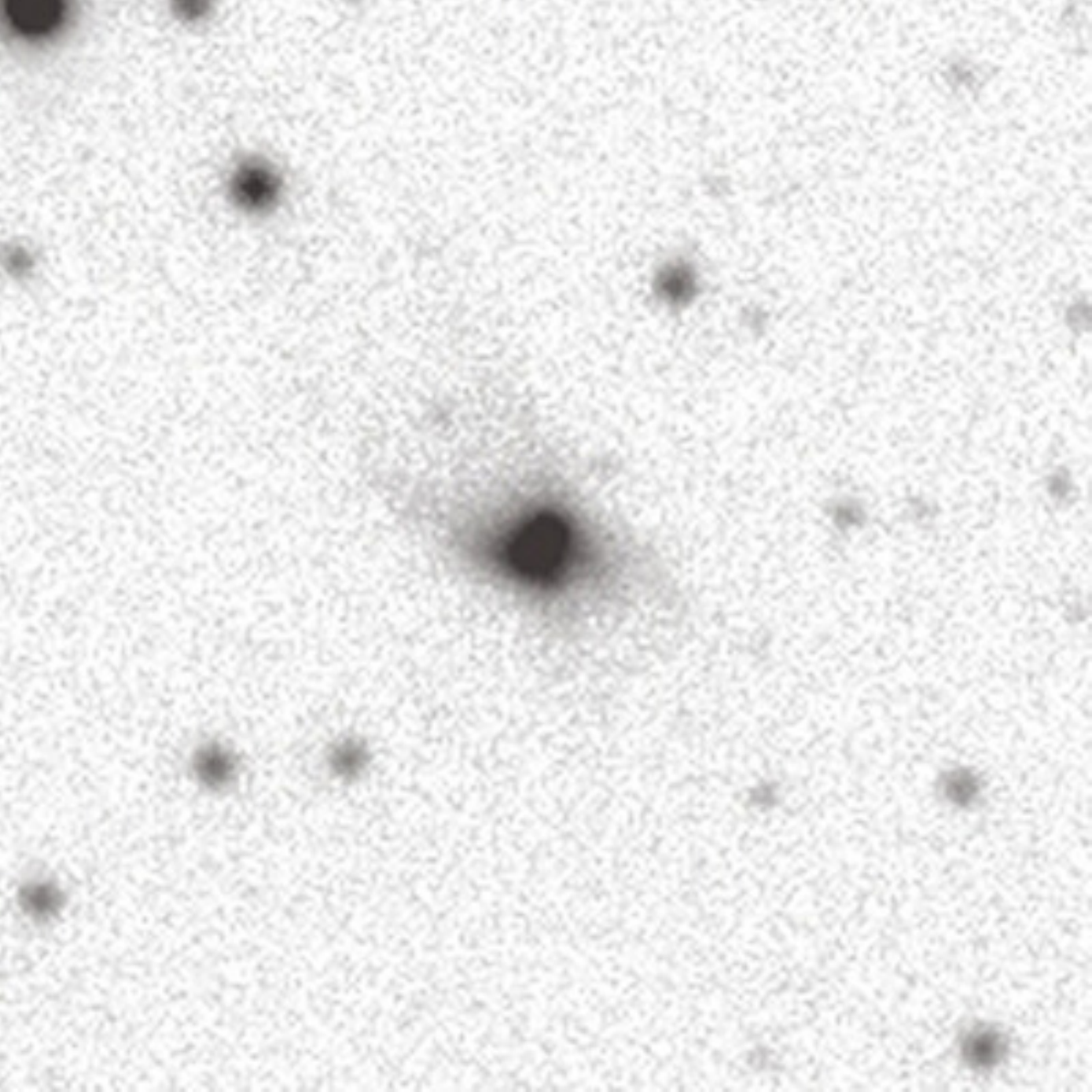}
\includegraphics[height=1.7in]{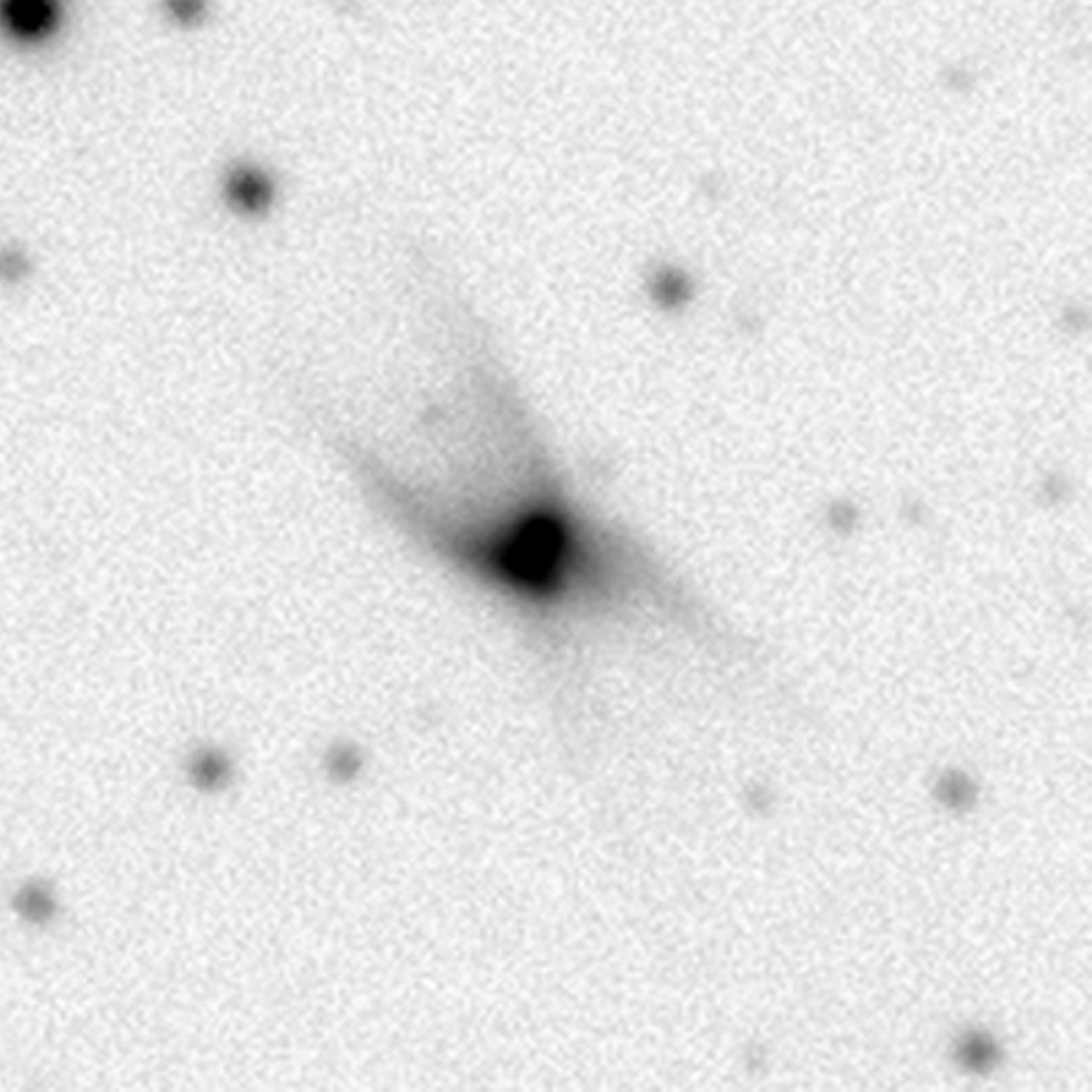}
\includegraphics[height=1.7in]{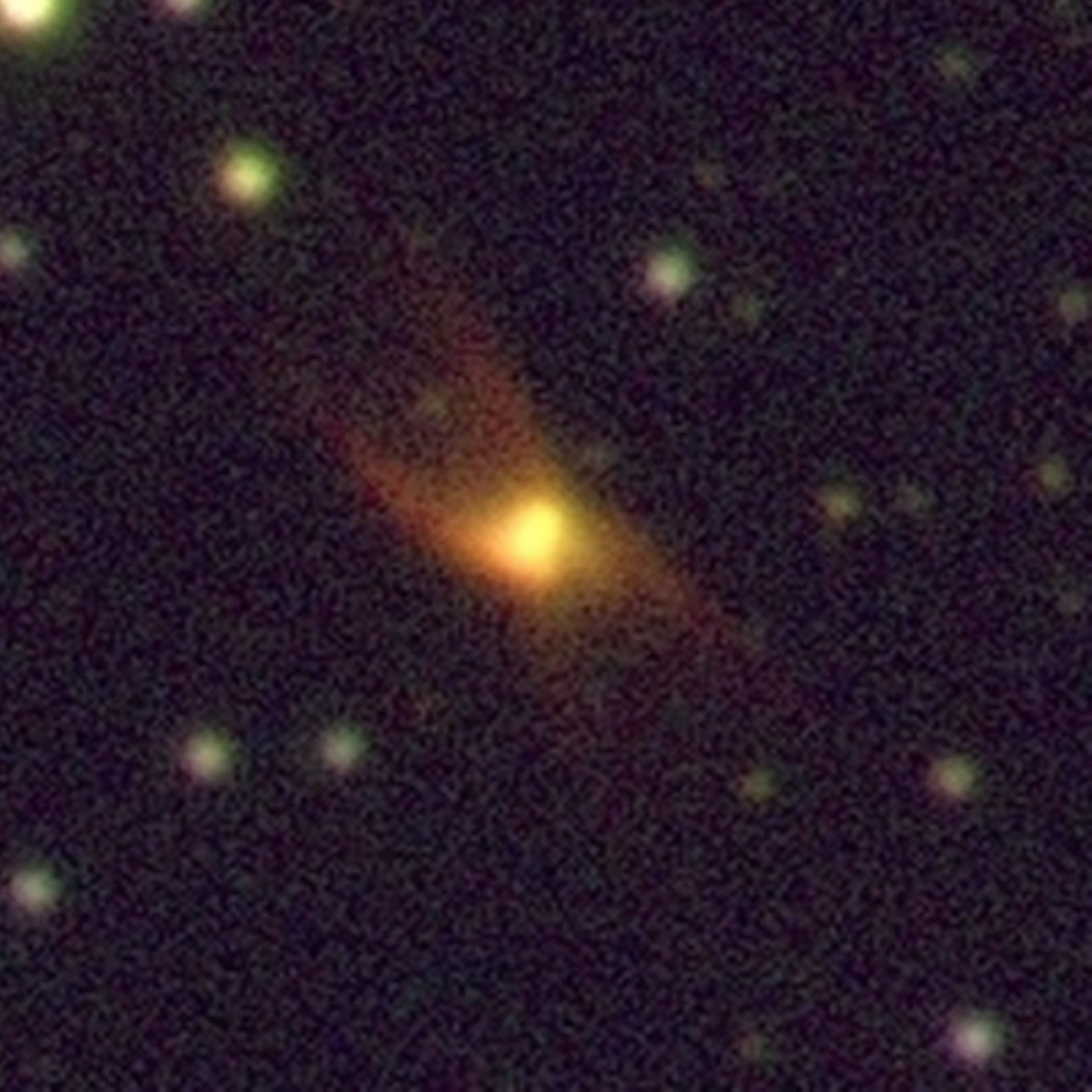}
\caption{Same as Figure~\ref{1.img}. } 
\label{8.img} 
\end{figure*}

\begin{figure*} 
\centering 
\includegraphics[height=1.7in]{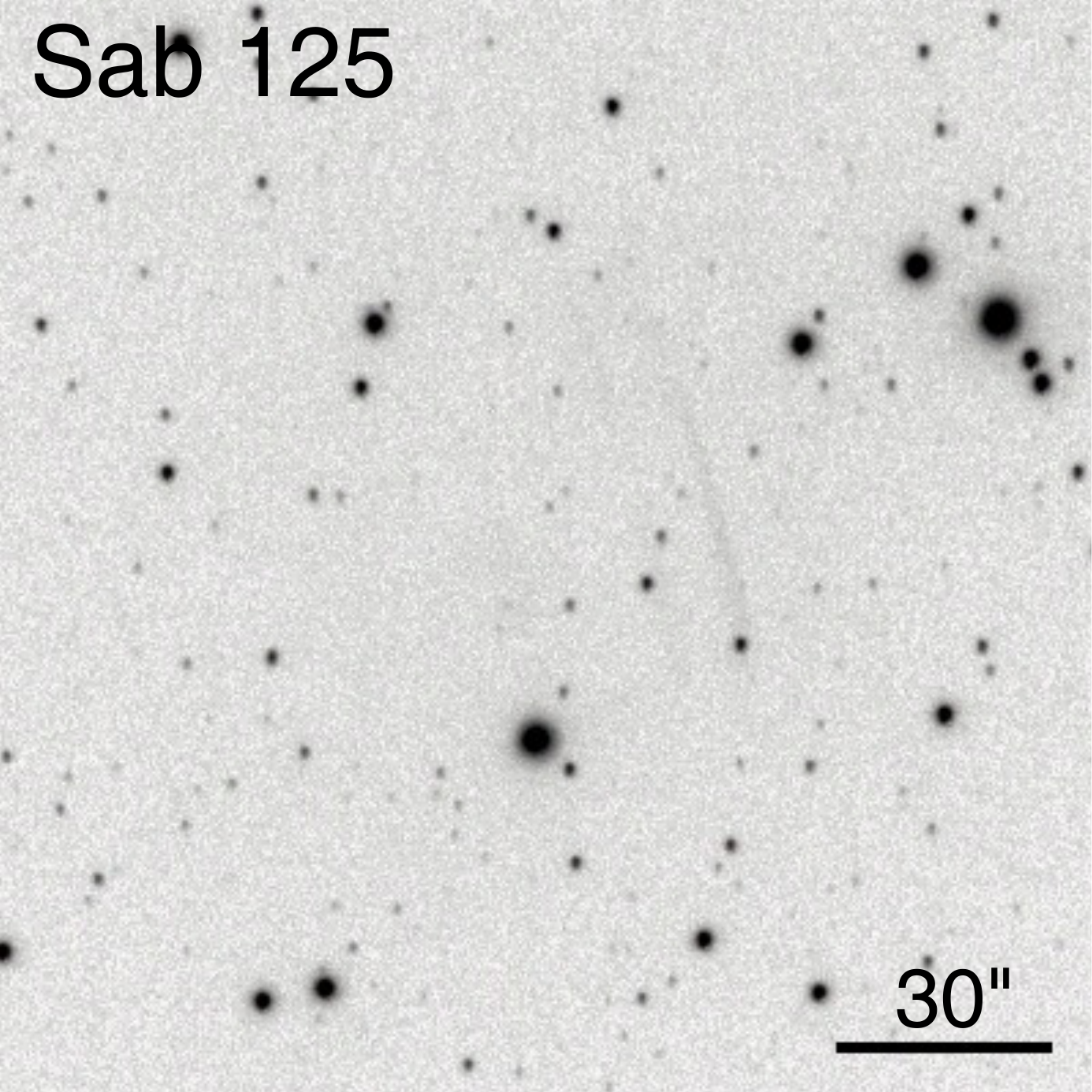} 
\includegraphics[height=1.7in]{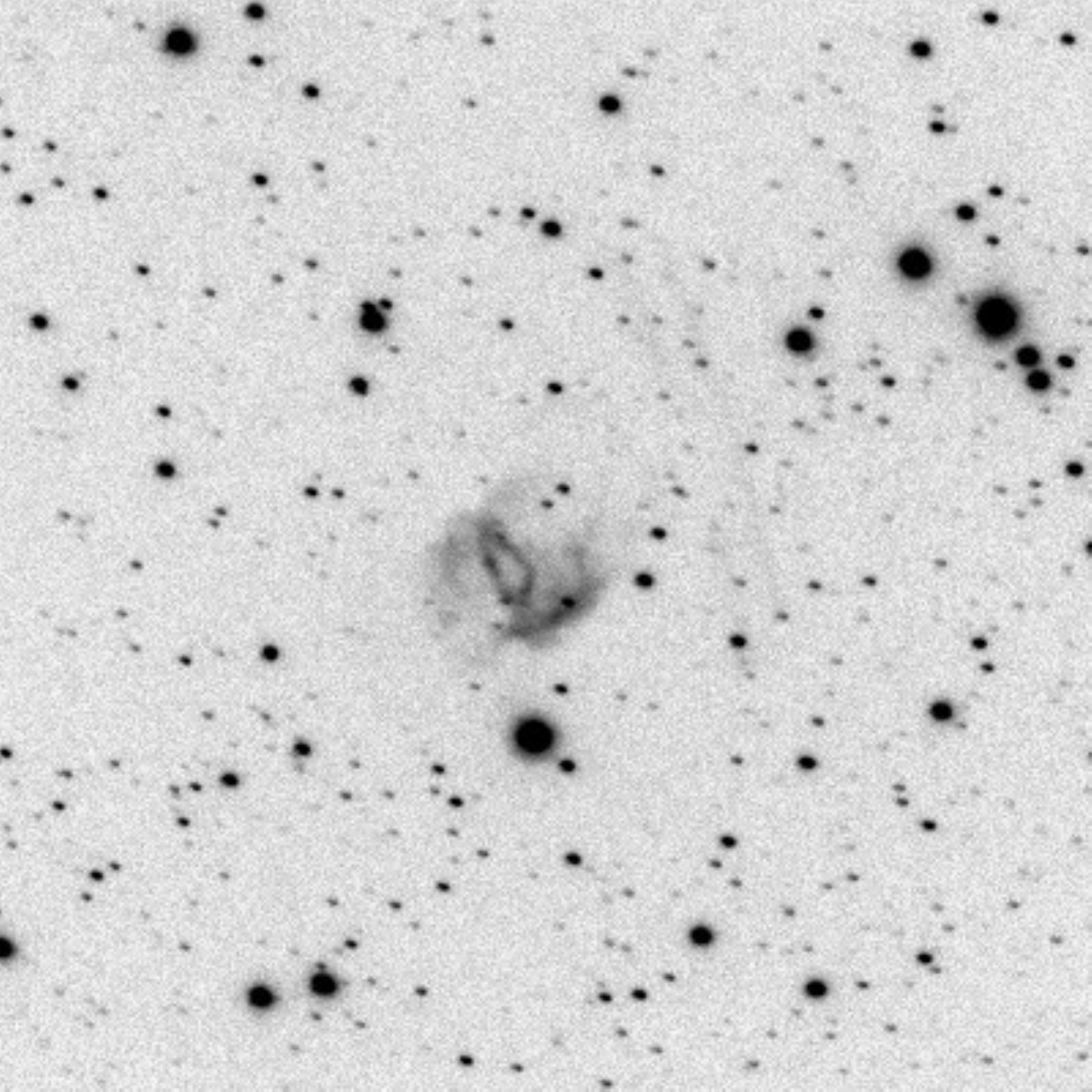}
\includegraphics[height=1.7in]{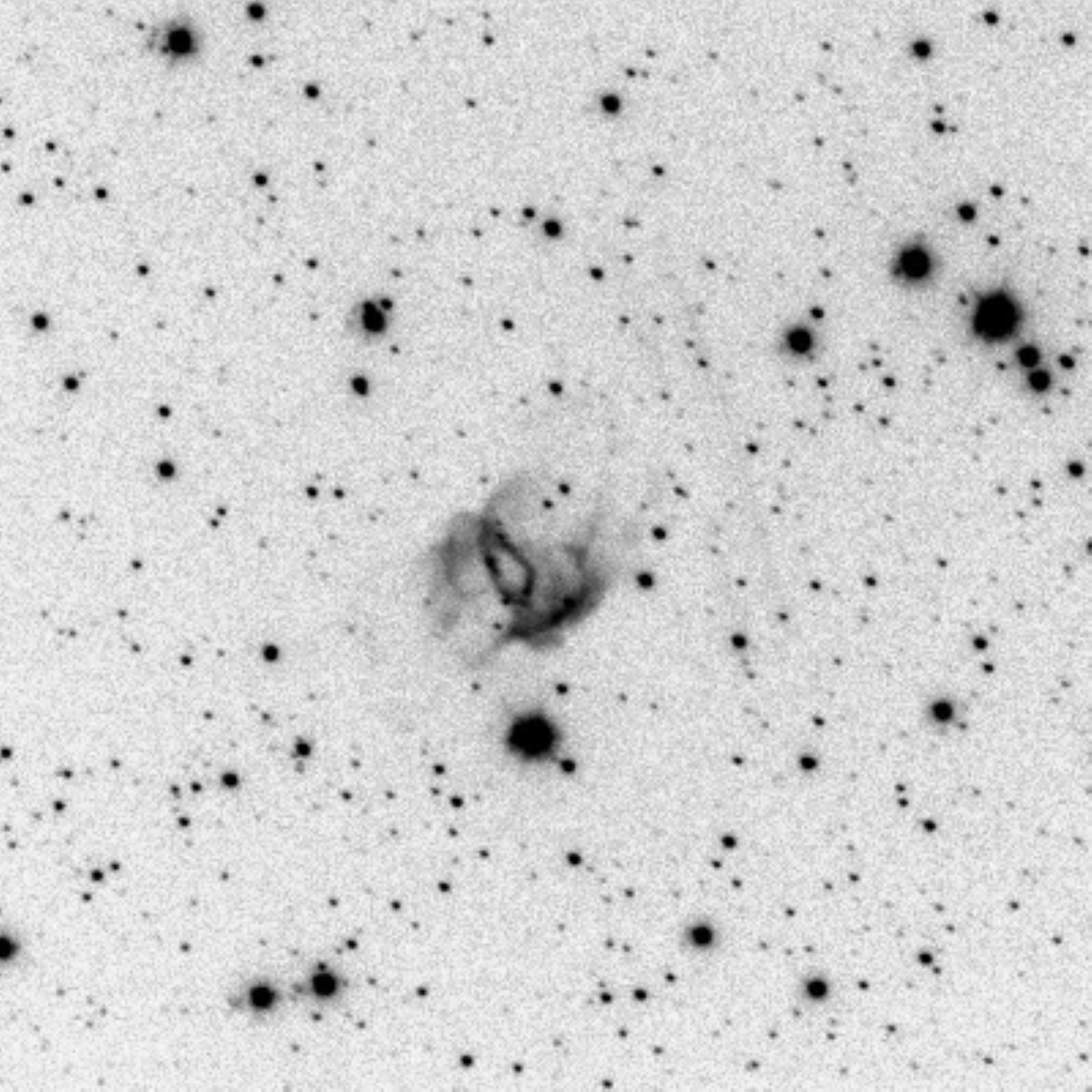}
\includegraphics[height=1.7in]{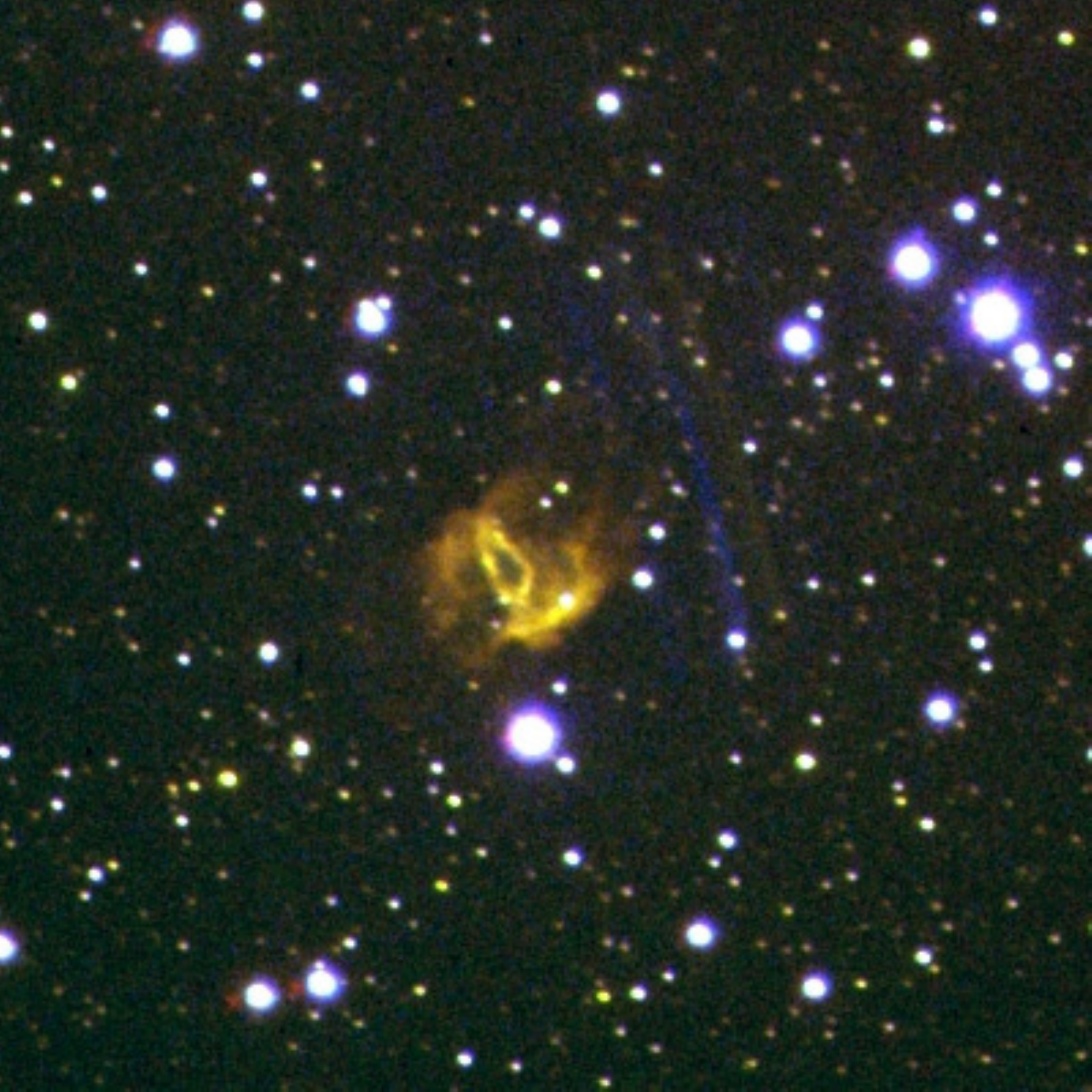}
\vskip .1in 
\includegraphics[height=1.7in]{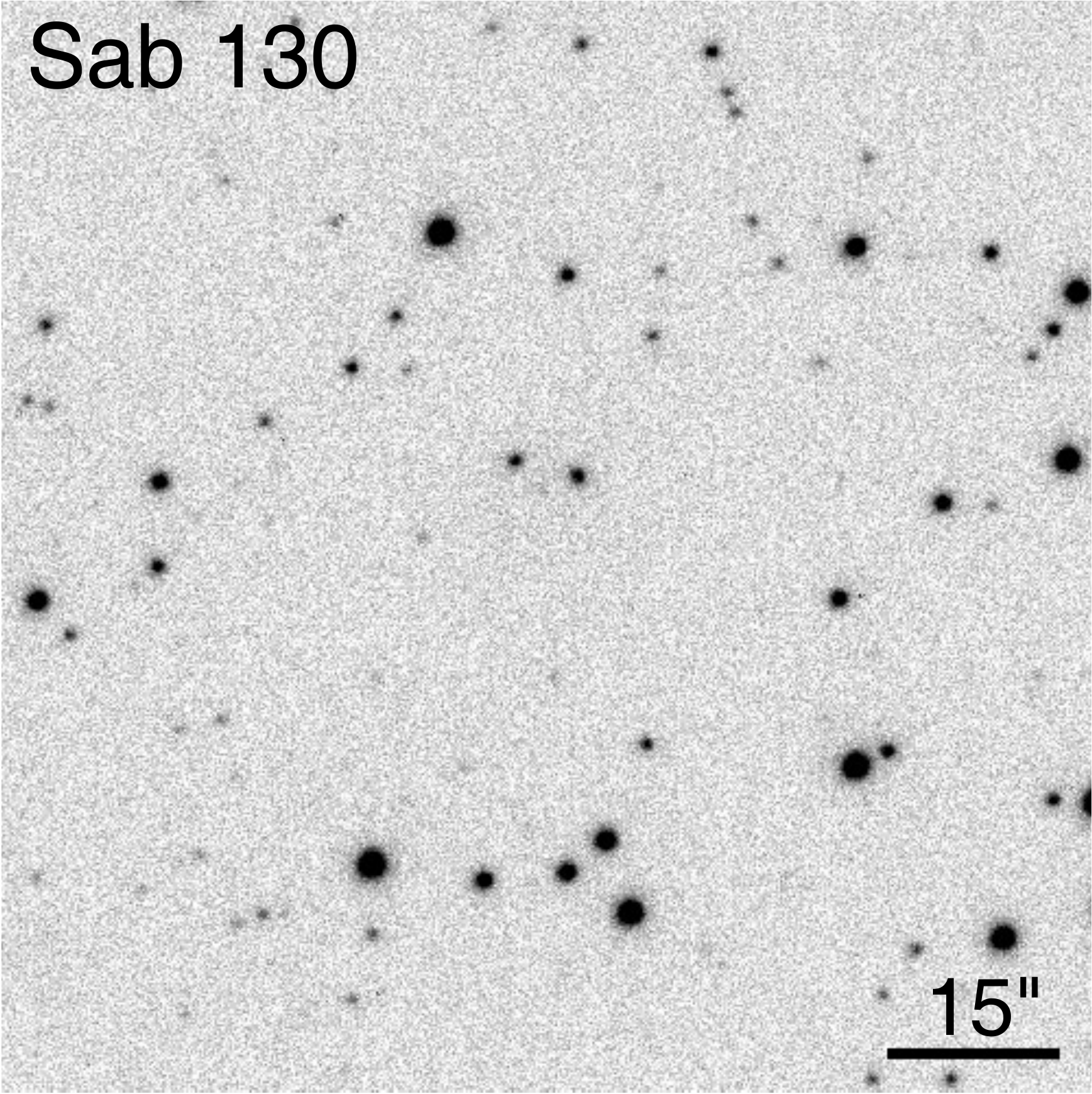} 
\includegraphics[height=1.7in]{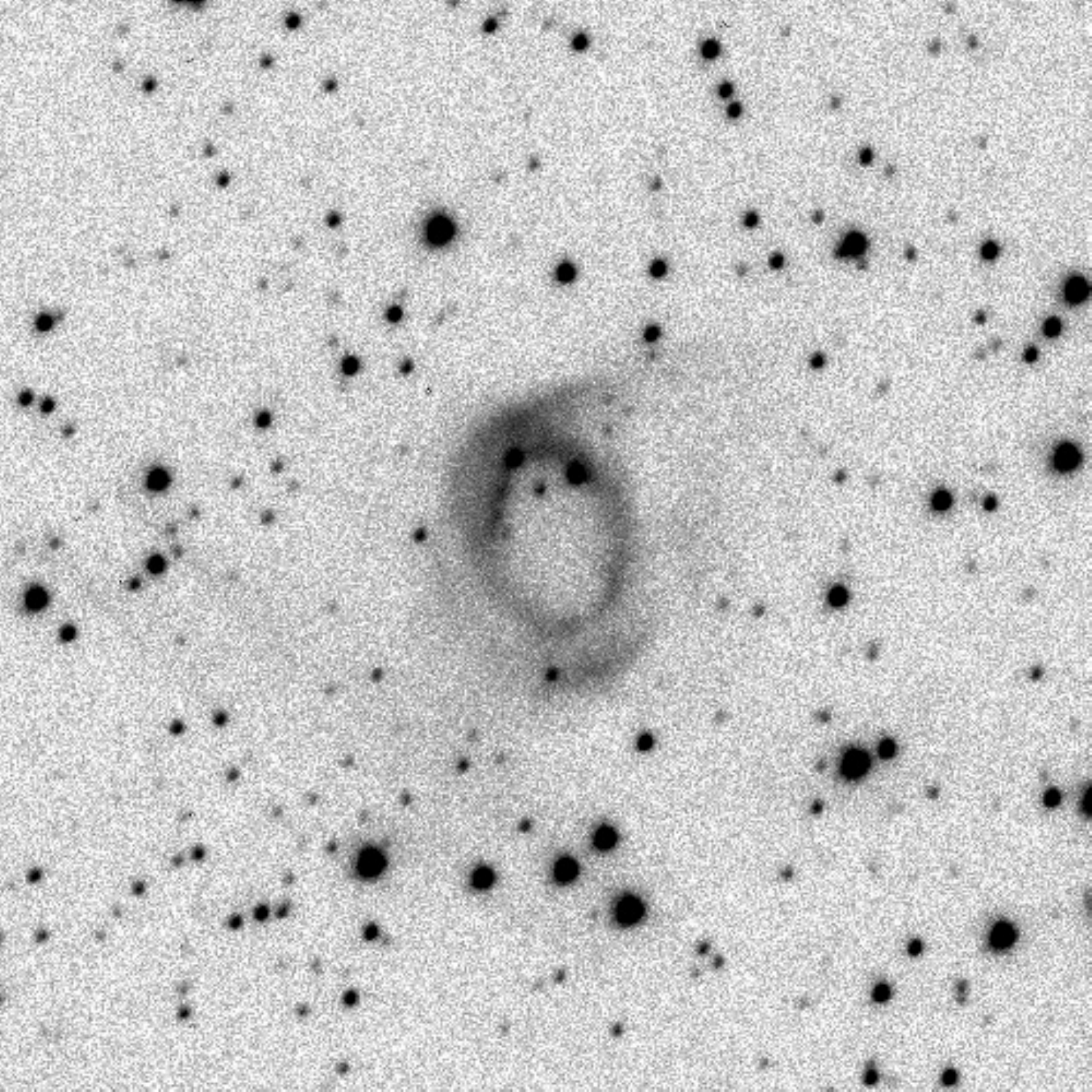}
\includegraphics[height=1.7in]{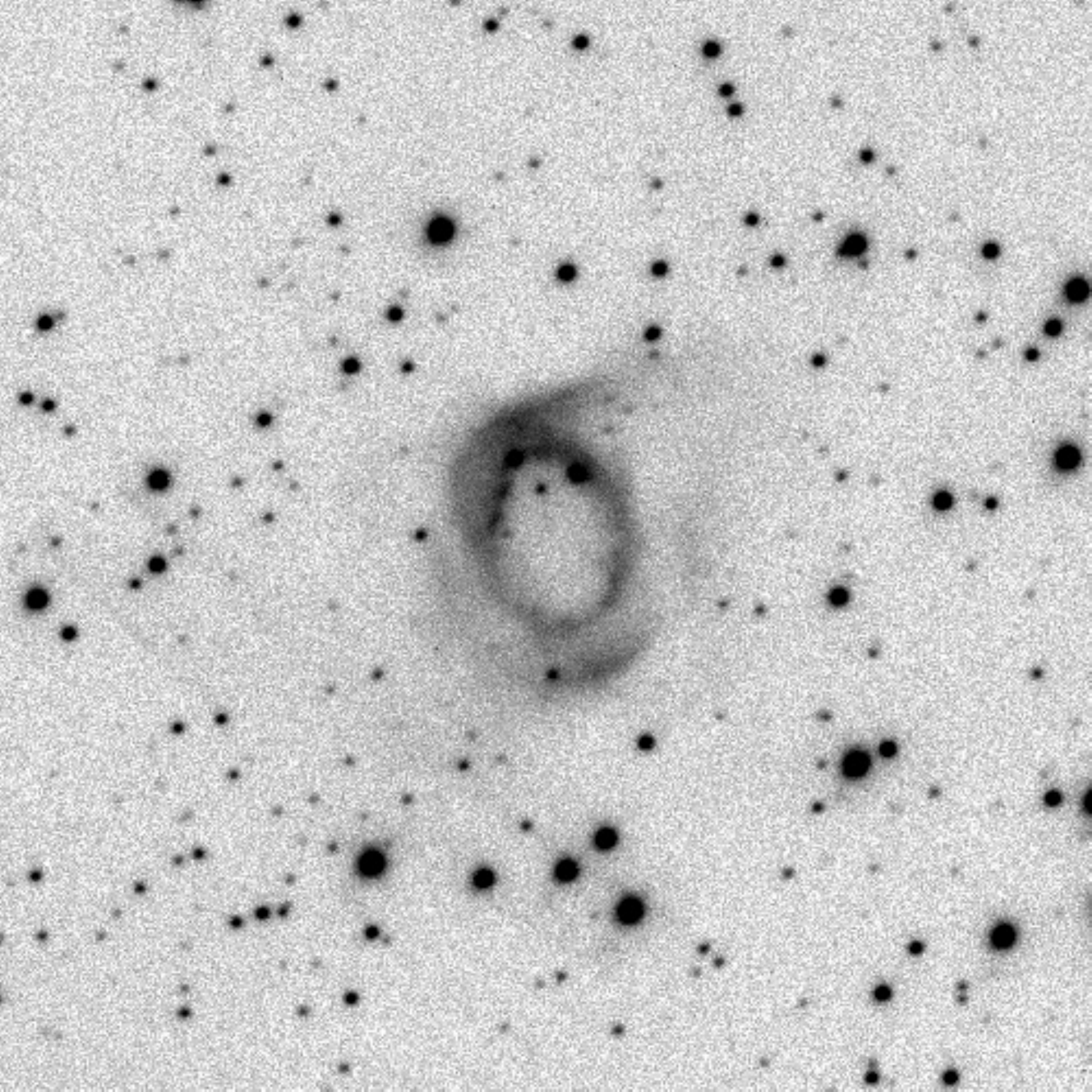}
\includegraphics[height=1.7in]{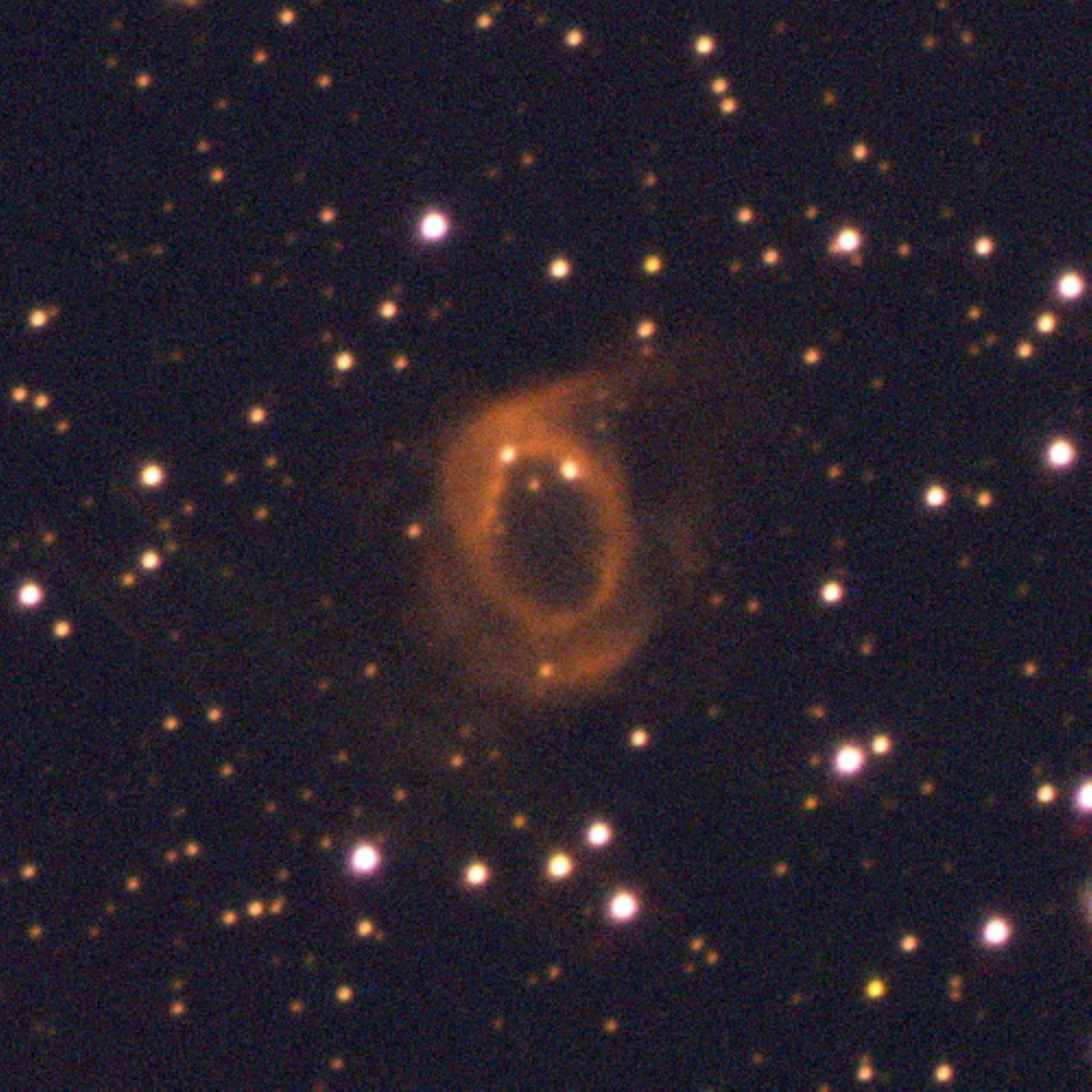}
\vskip .1in 
\includegraphics[height=1.7in]{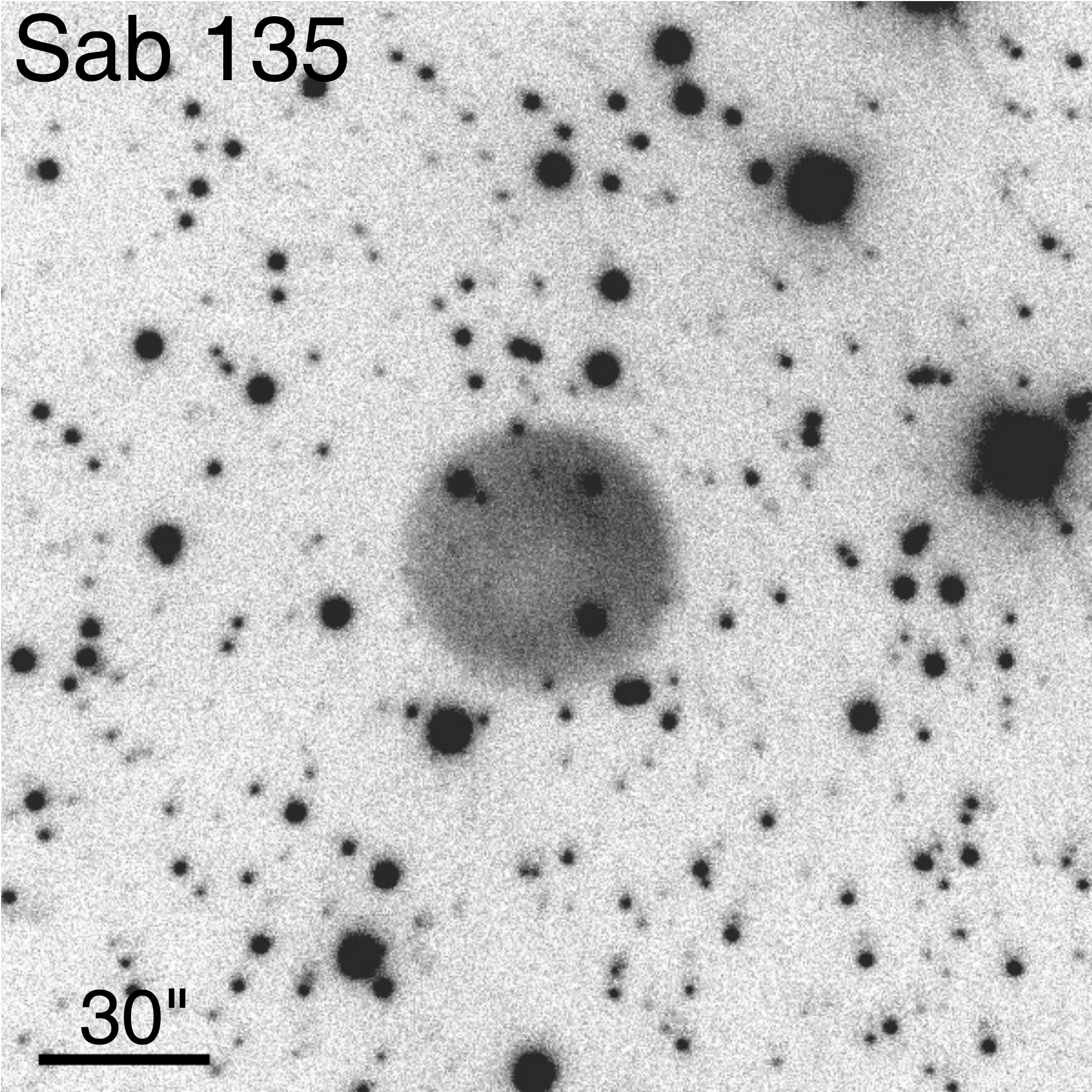} 
\includegraphics[height=1.7in]{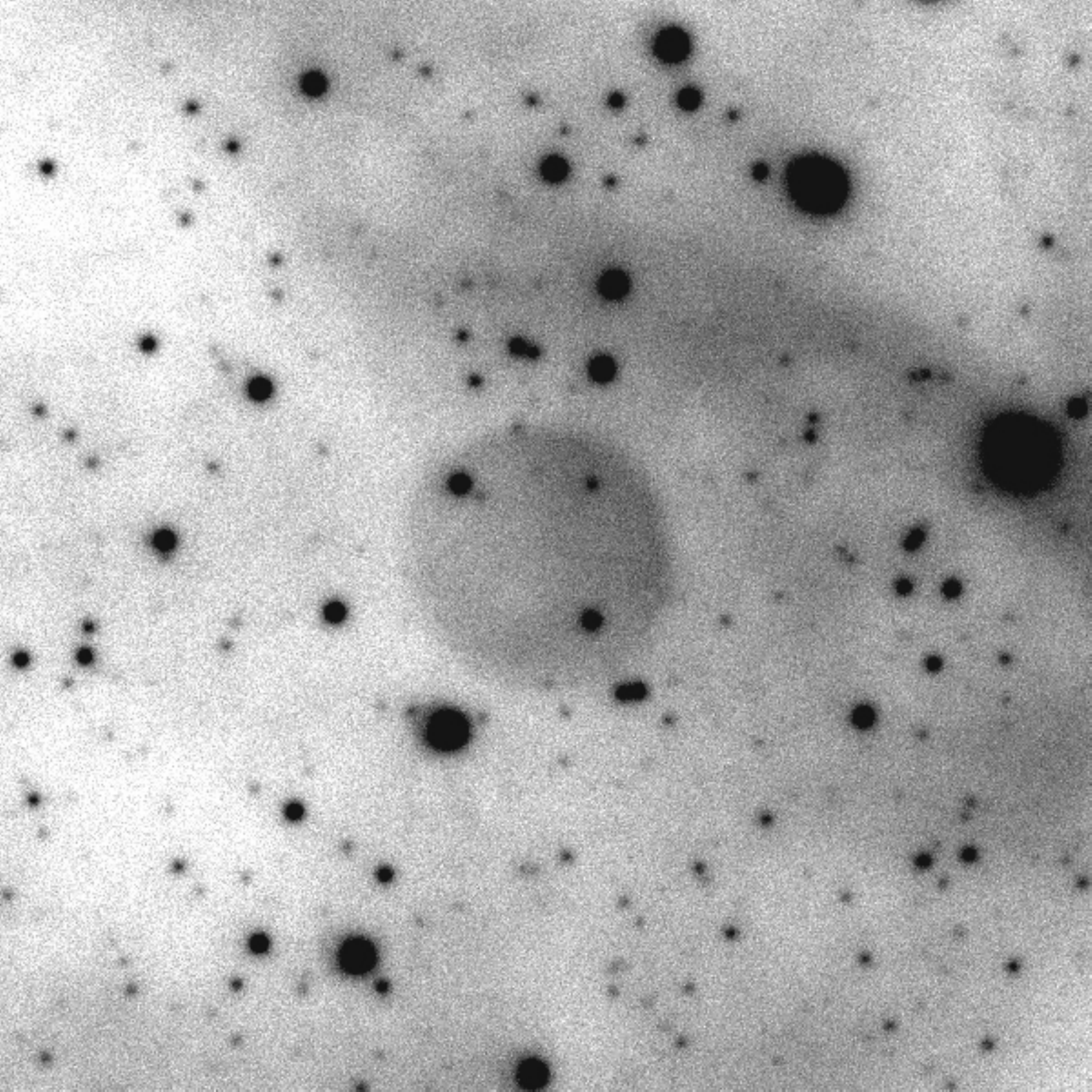}
\includegraphics[height=1.7in]{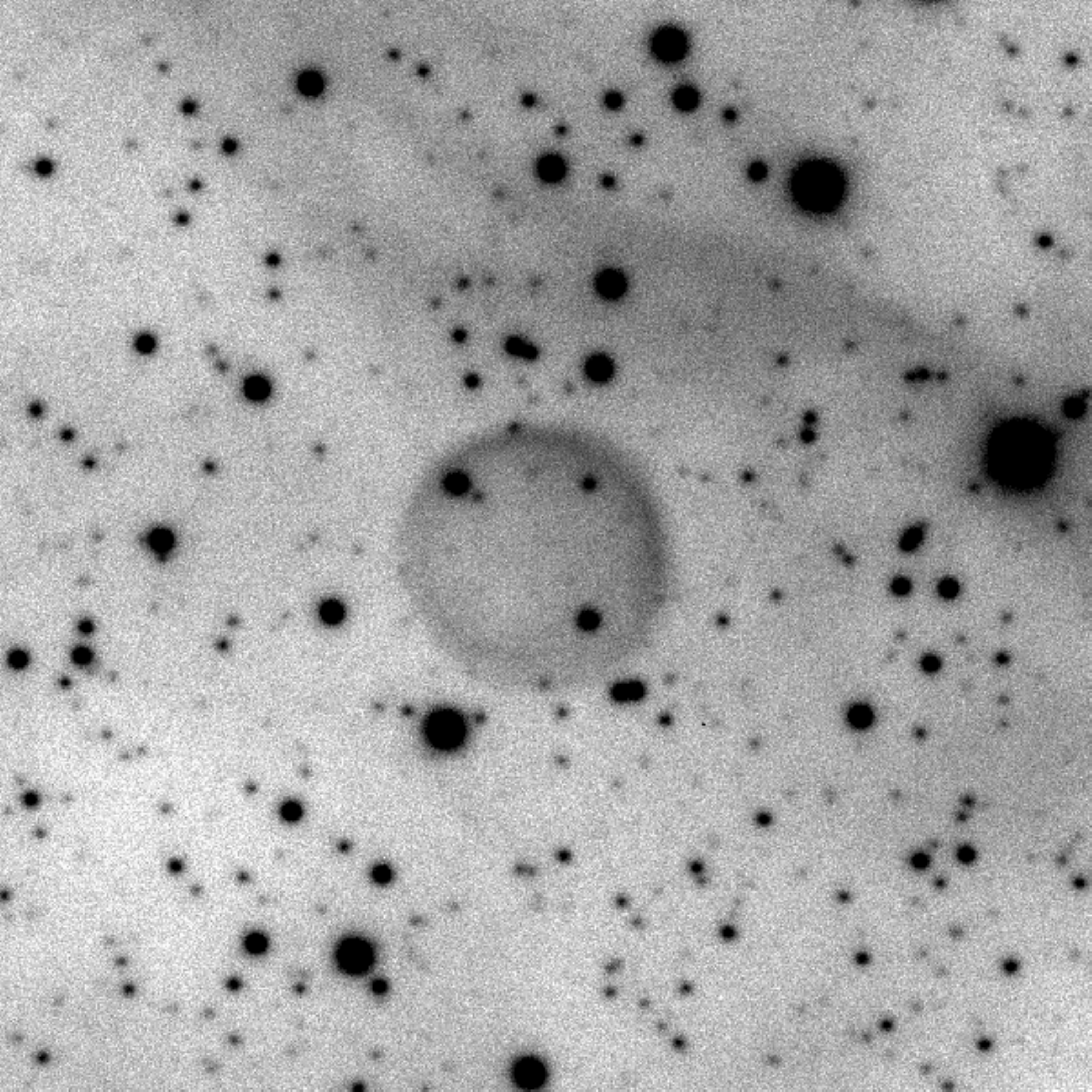}
\includegraphics[height=1.7in]{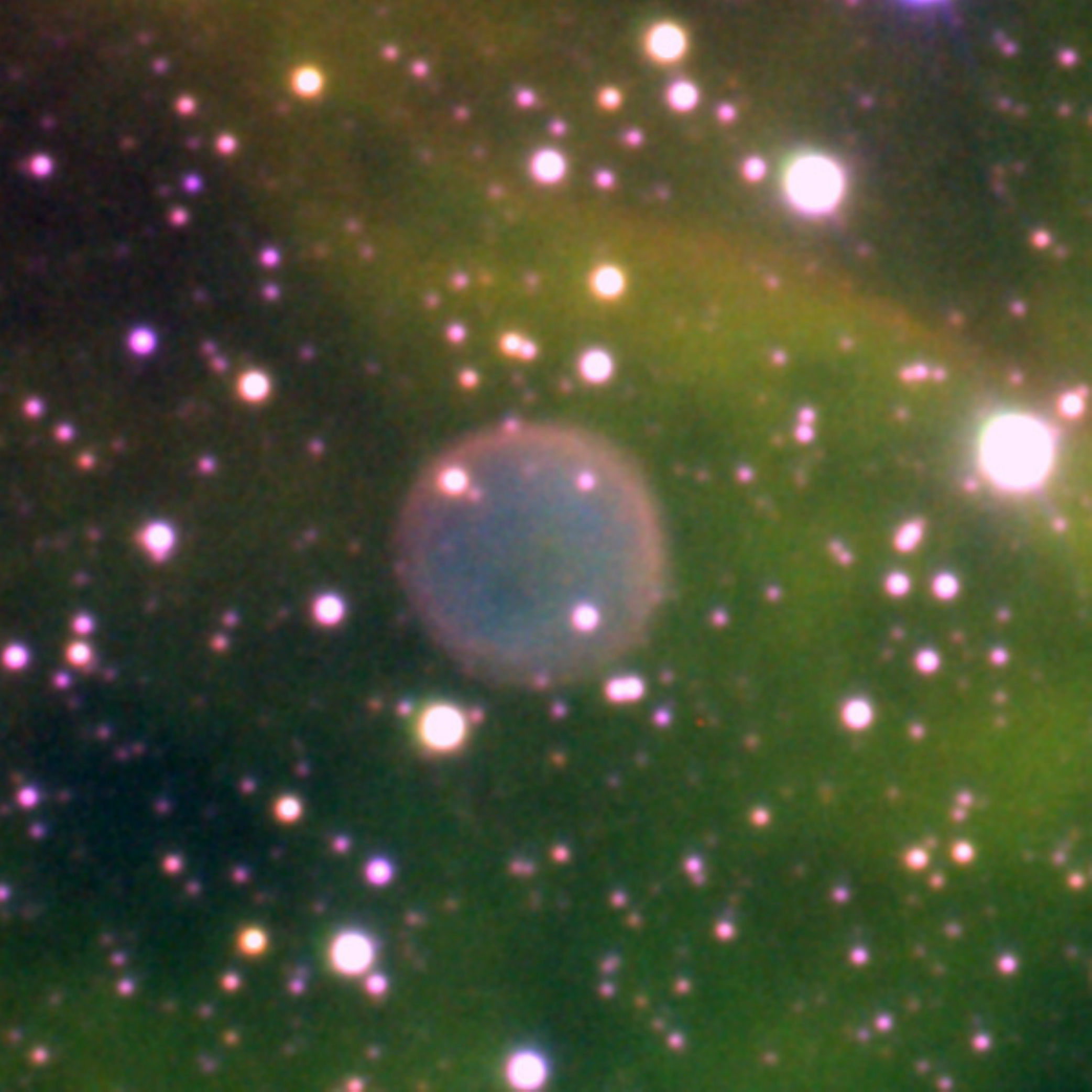}
\vskip .1in 
\includegraphics[height=1.7in]{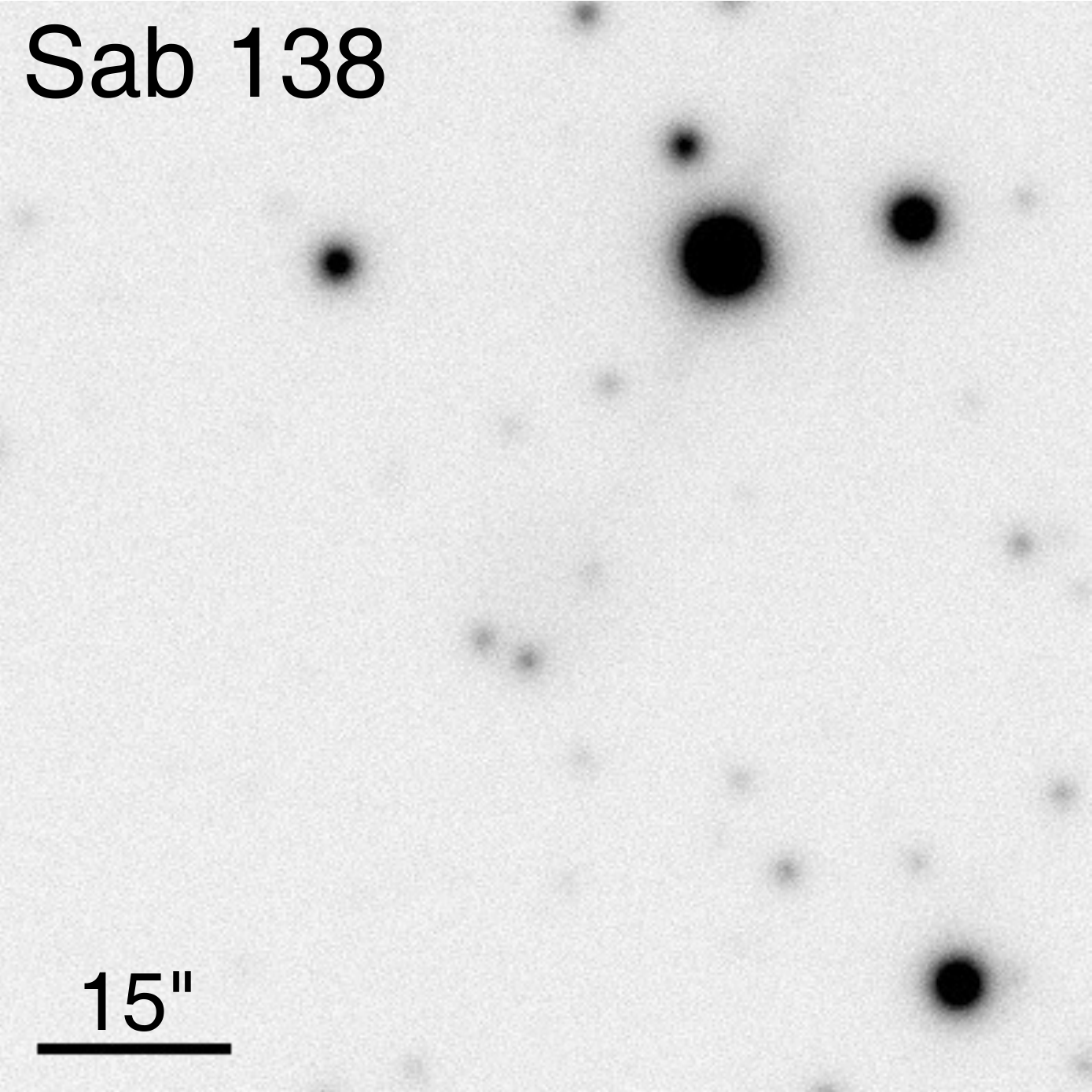} 
\includegraphics[height=1.7in]{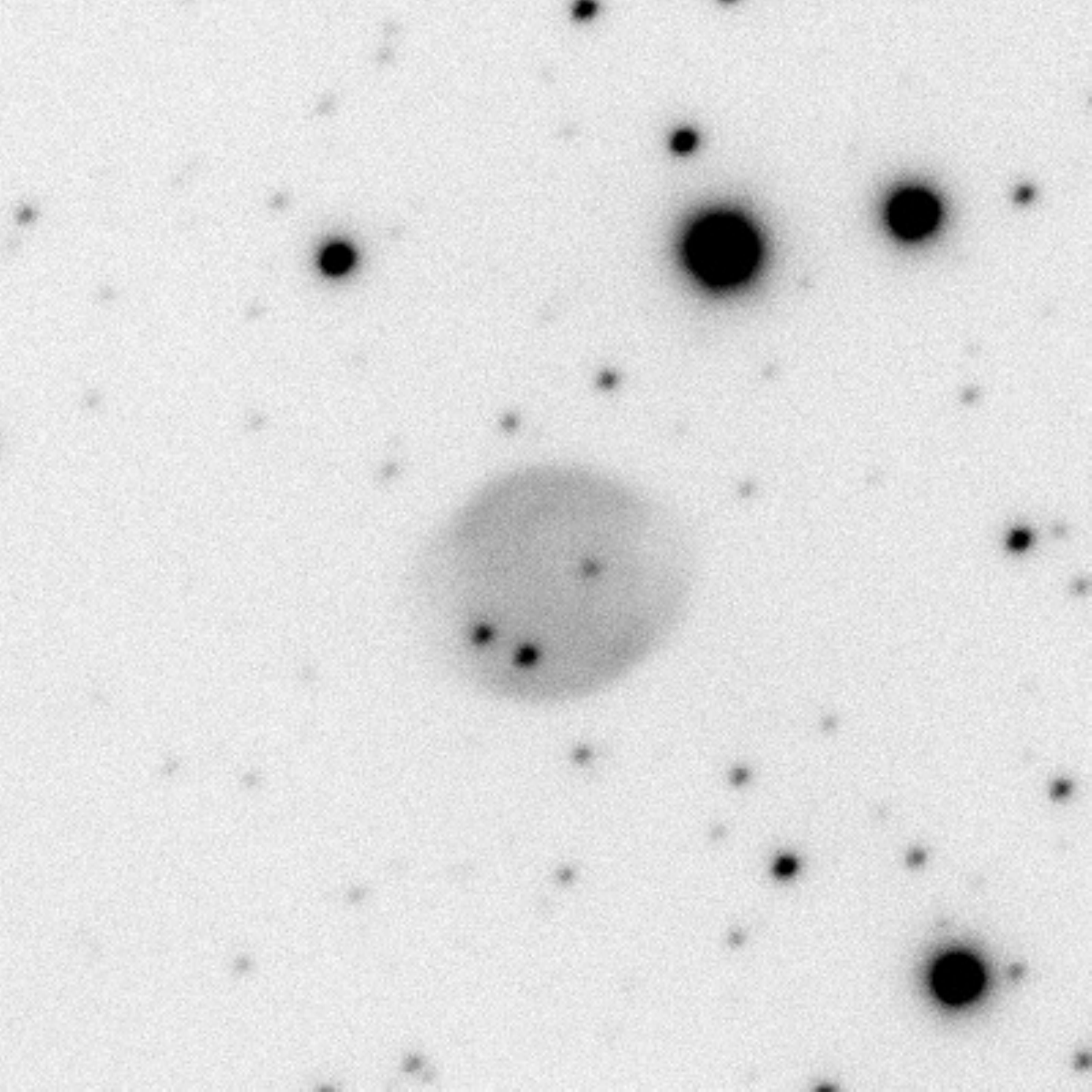}
\includegraphics[height=1.7in]{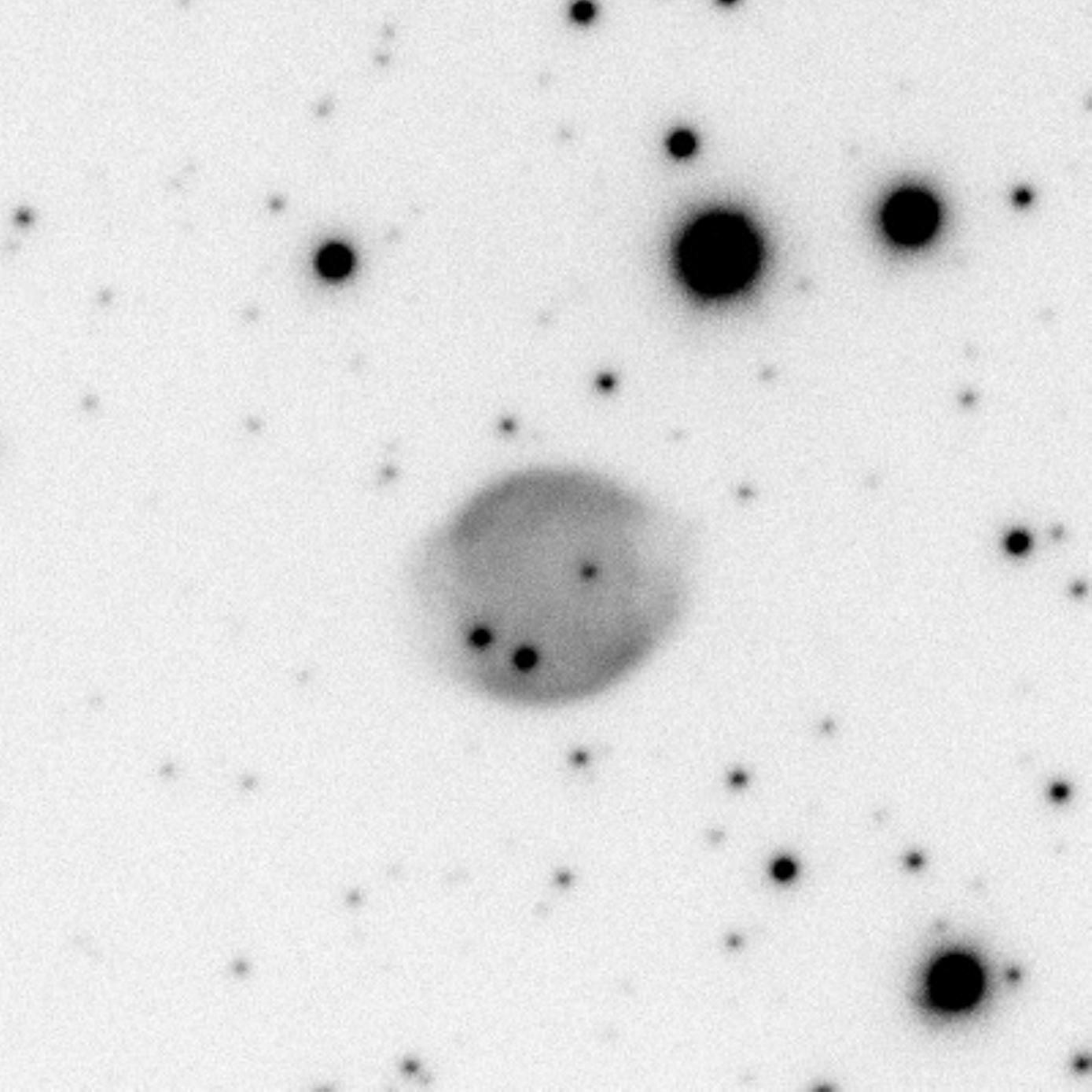}
\includegraphics[height=1.7in]{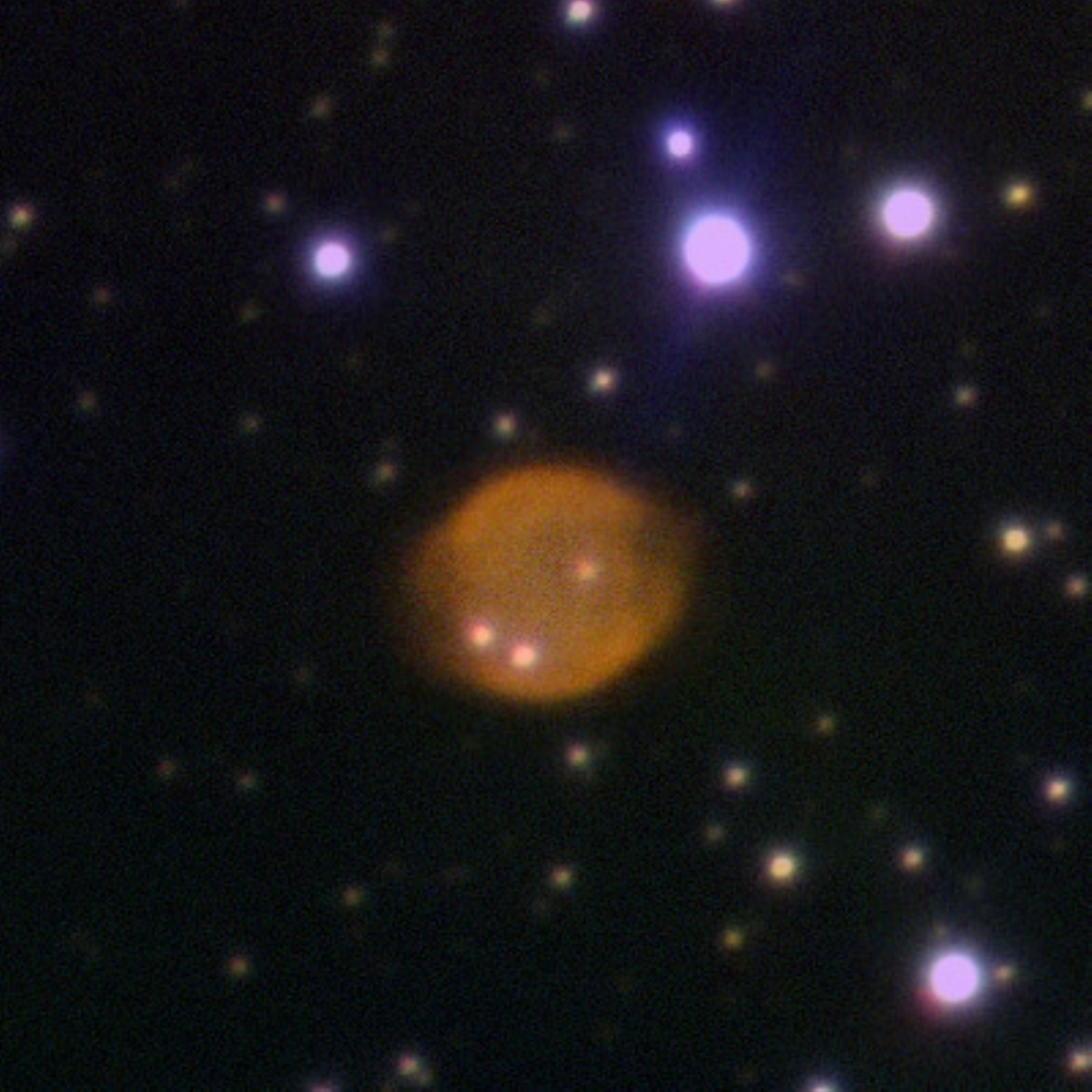}
\vskip .1in
\includegraphics[height=1.7in]{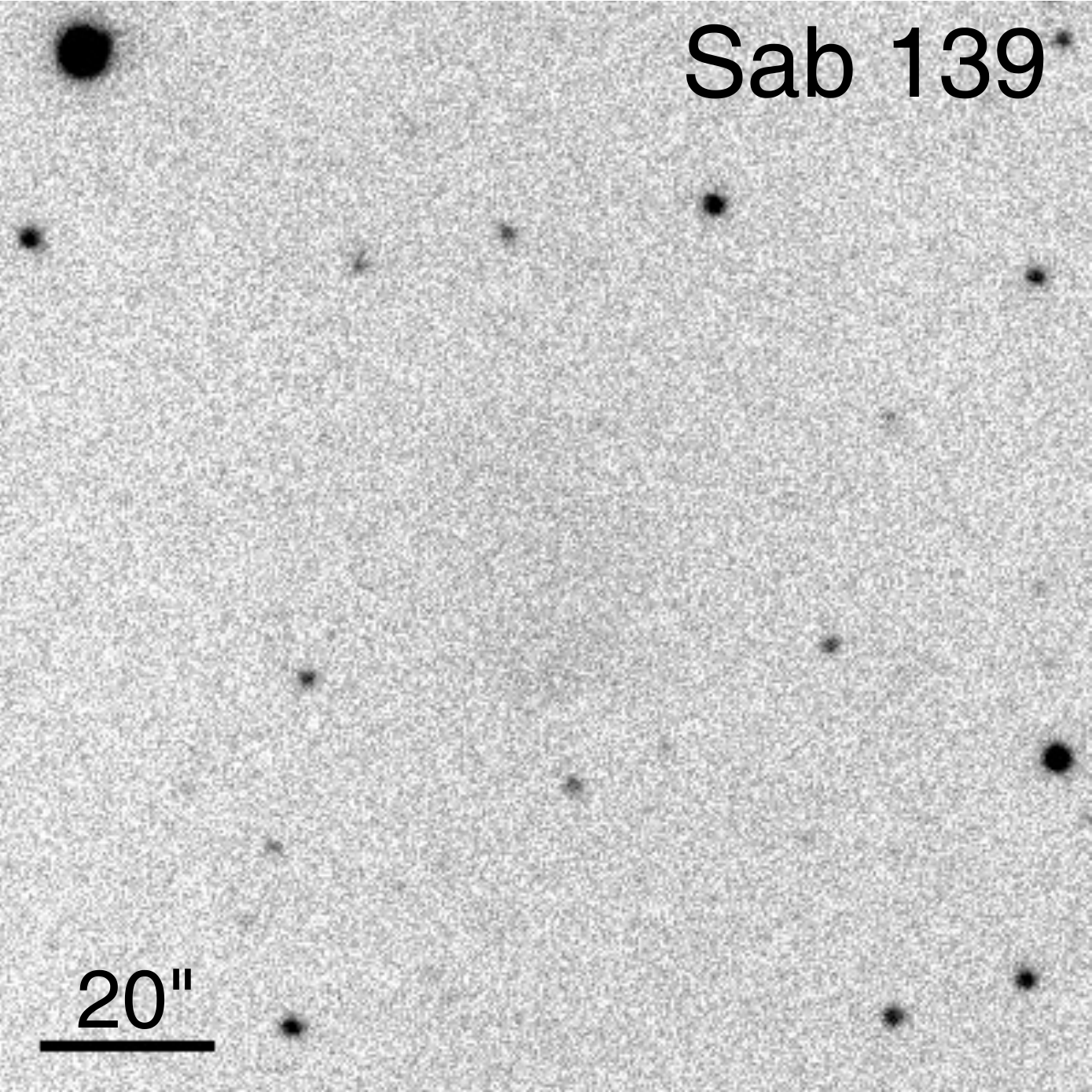} 
\includegraphics[height=1.7in]{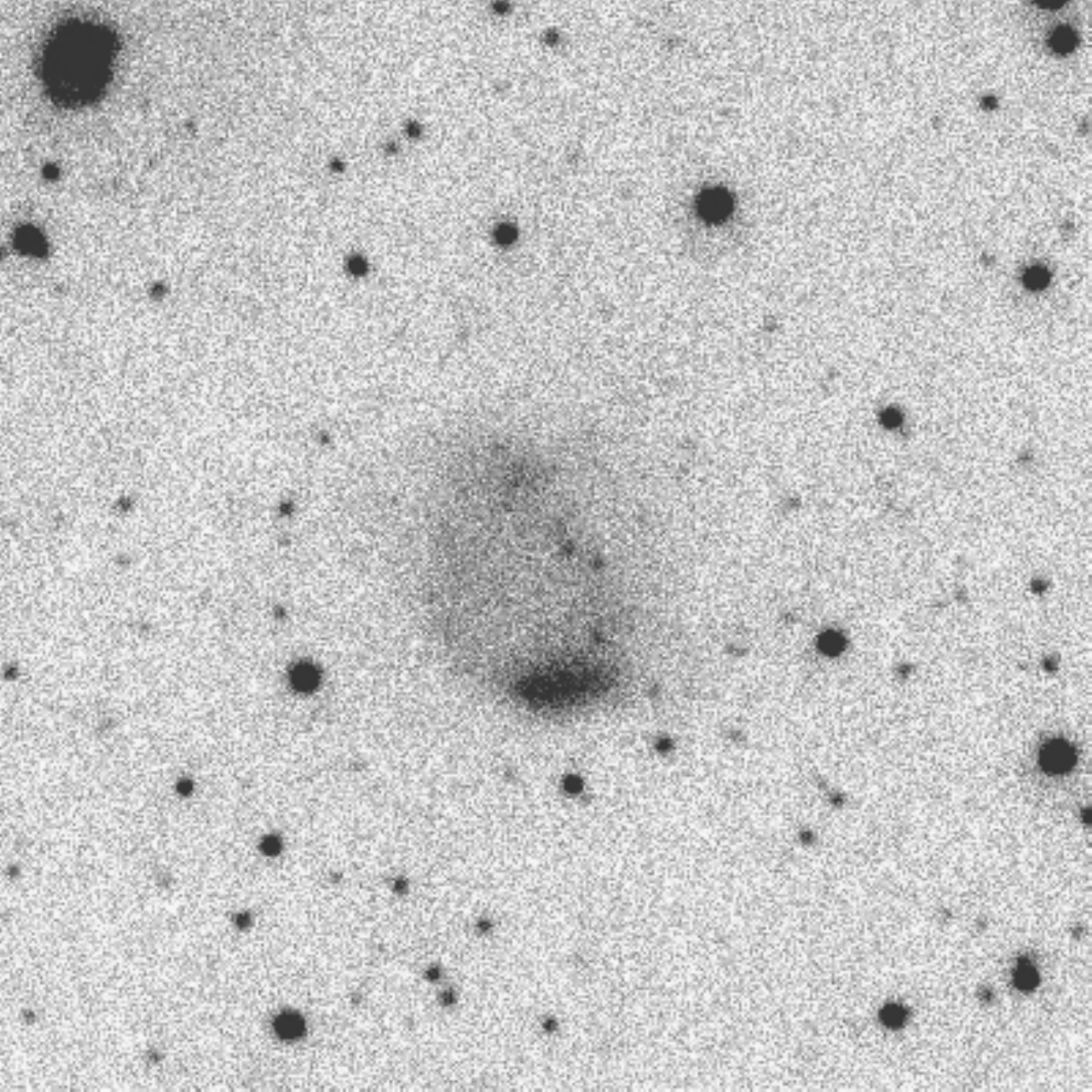}
\includegraphics[height=1.7in]{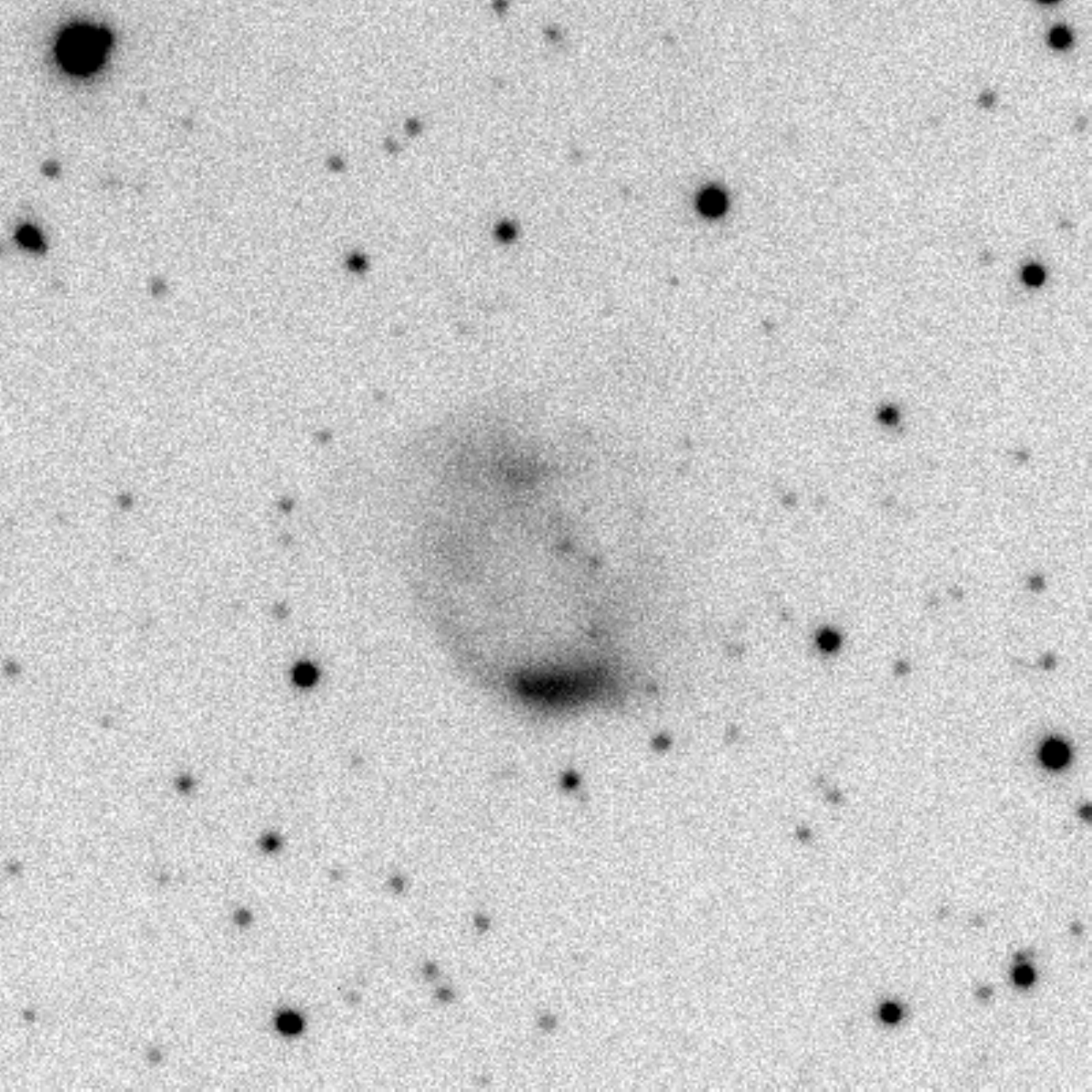}
\includegraphics[height=1.7in]{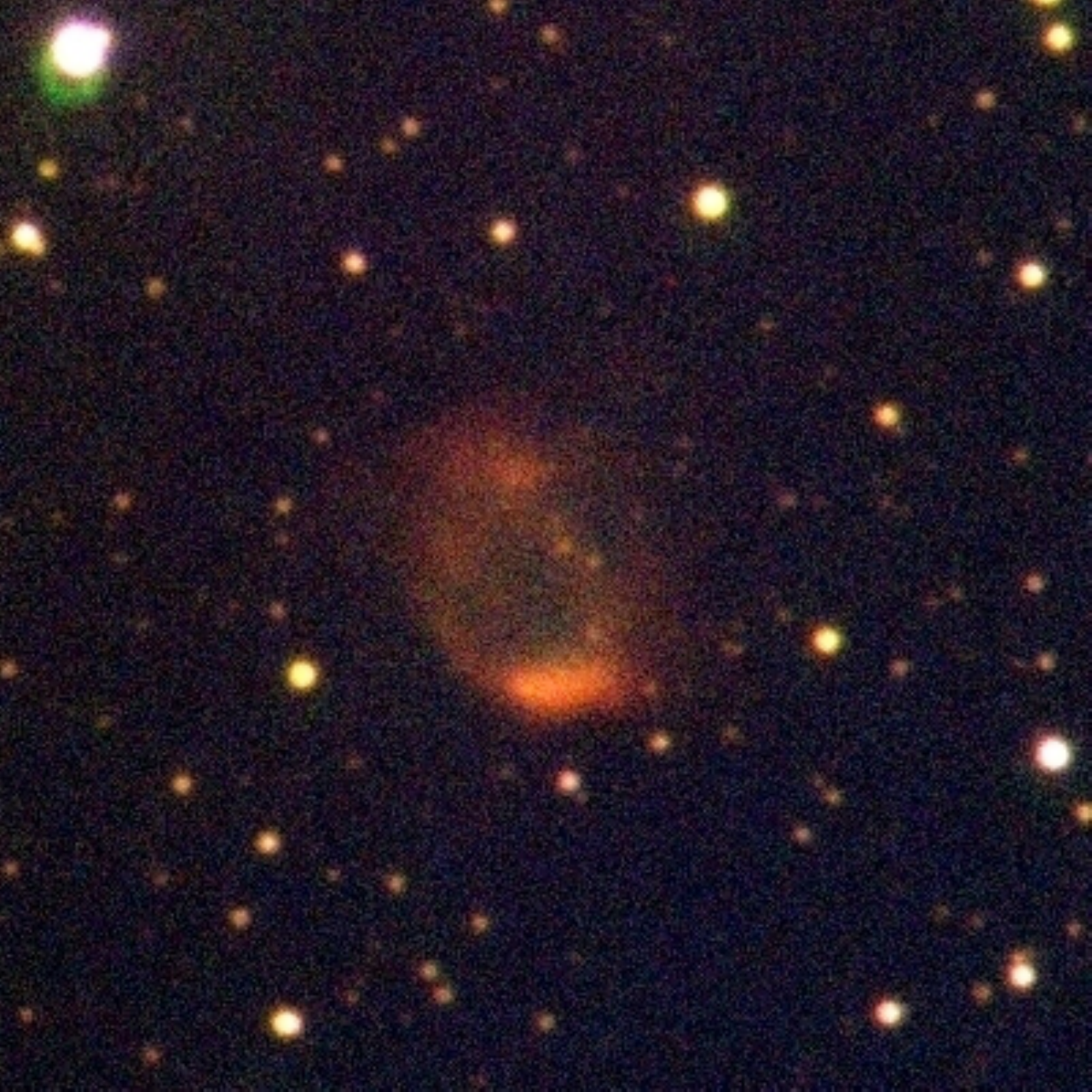}
\caption{Same as Figure~\ref{1.img}. } 
\label{9.img} 
\end{figure*}

\begin{figure*} 
\centering 
\includegraphics[height=1.7in]{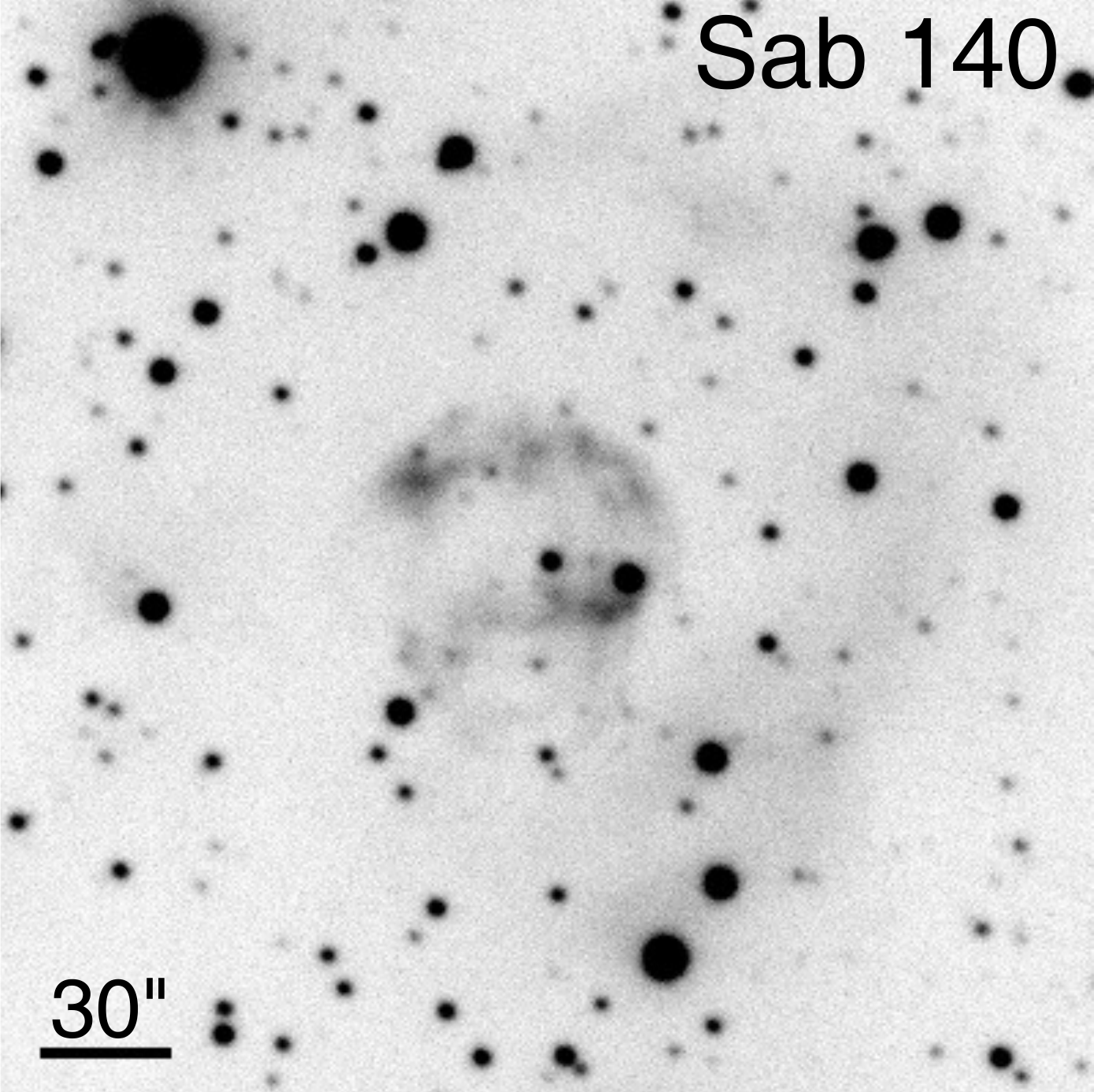} 
\includegraphics[height=1.7in]{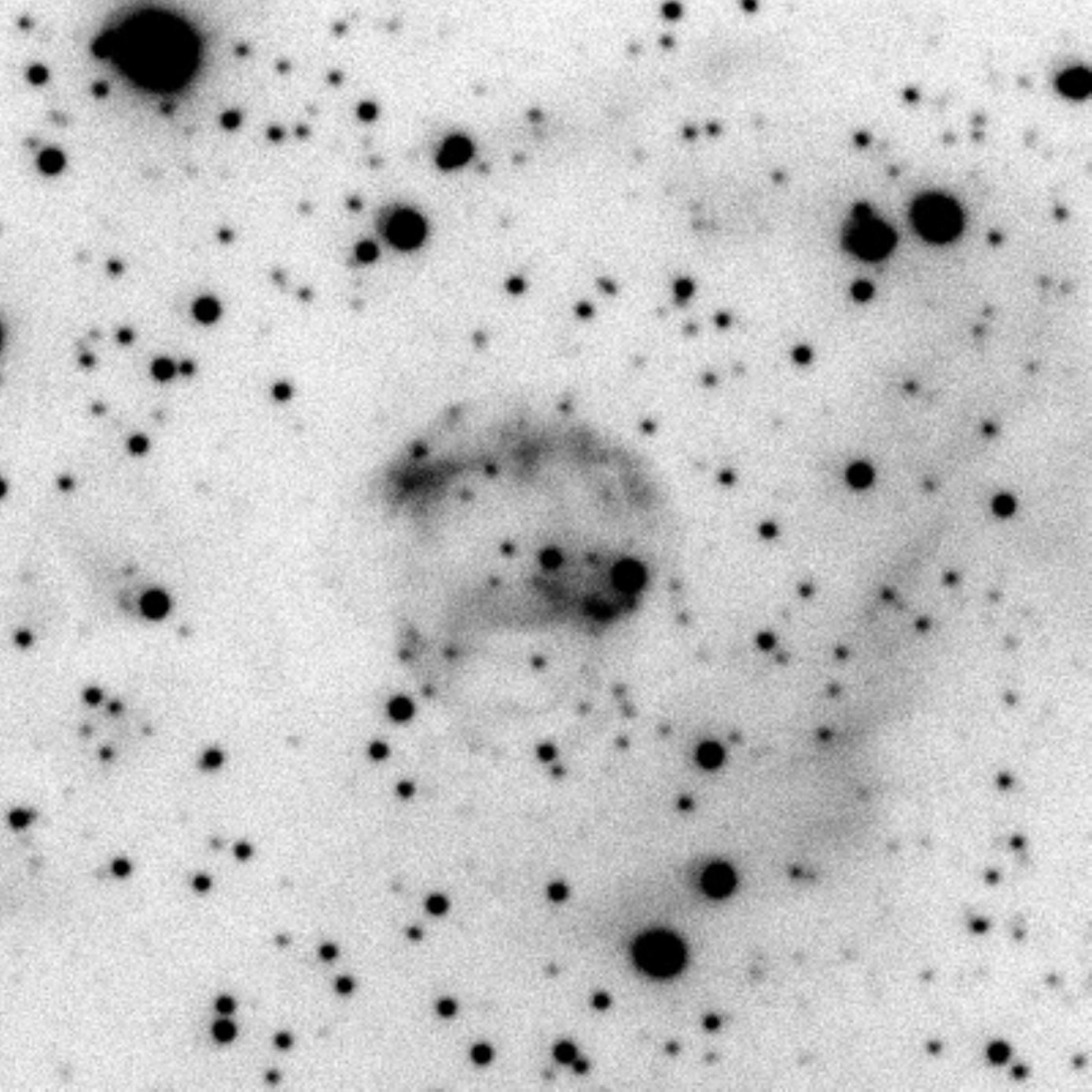}
\includegraphics[height=1.7in]{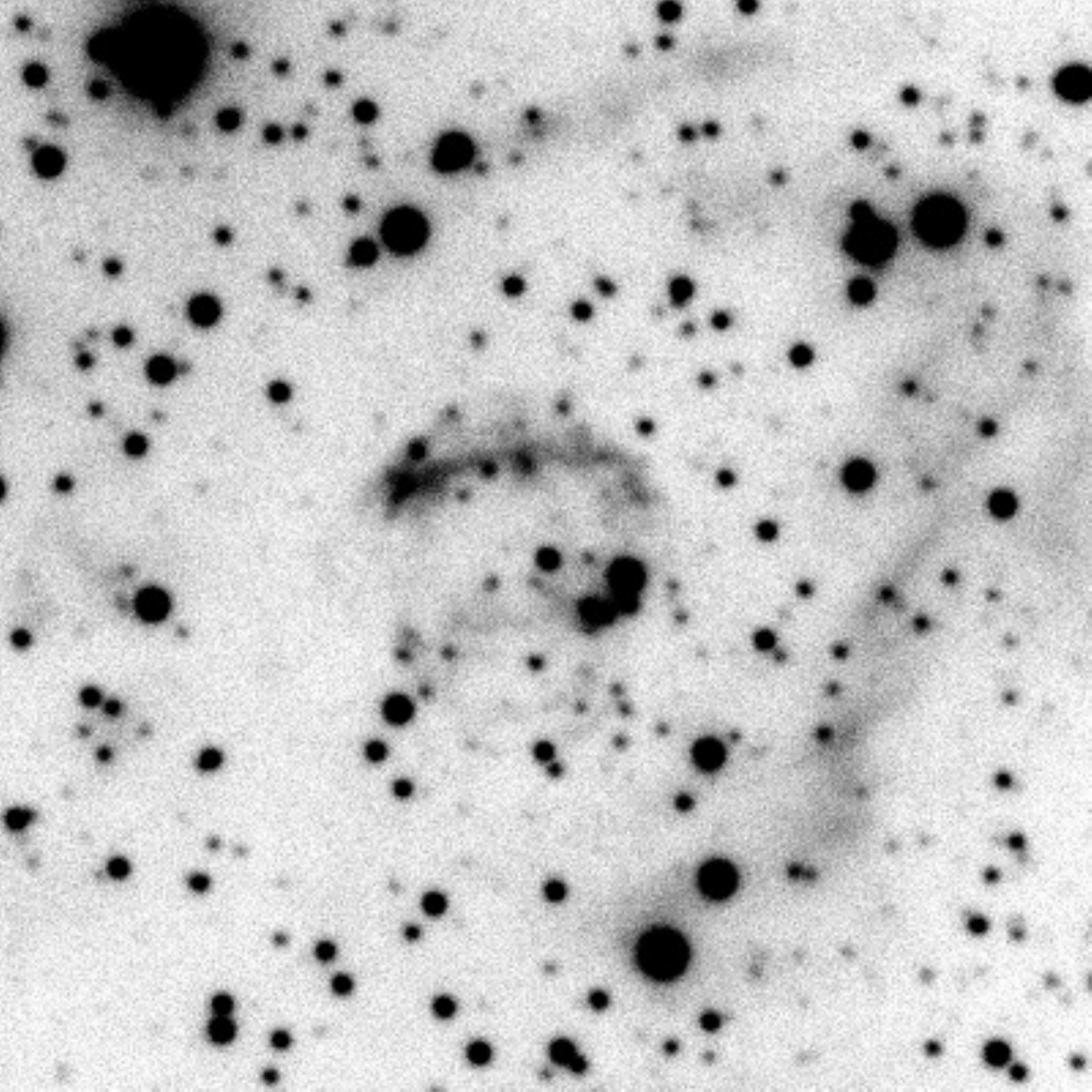}
\includegraphics[height=1.7in]{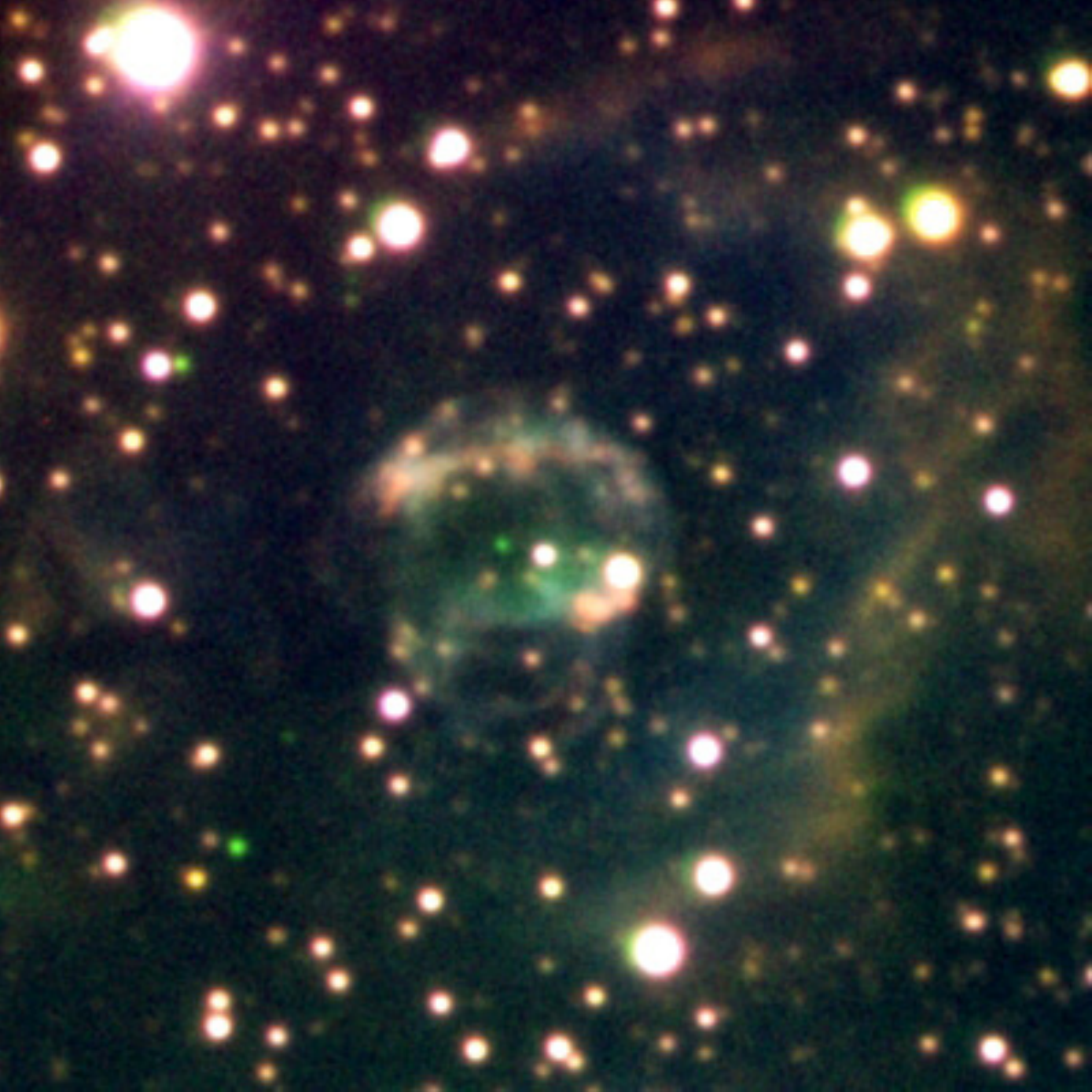}
\vskip .1in 
\includegraphics[height=1.7in]{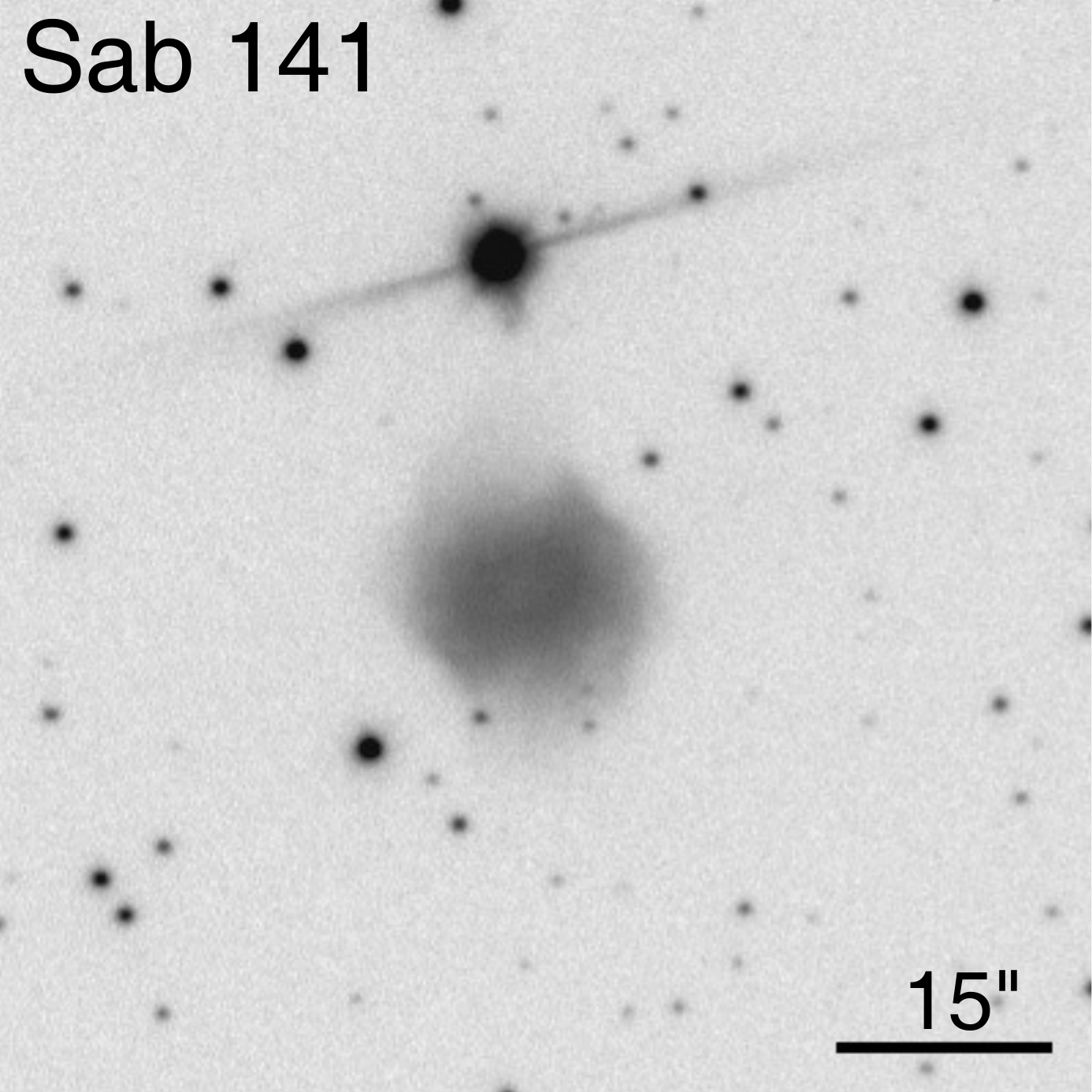} 
\includegraphics[height=1.7in]{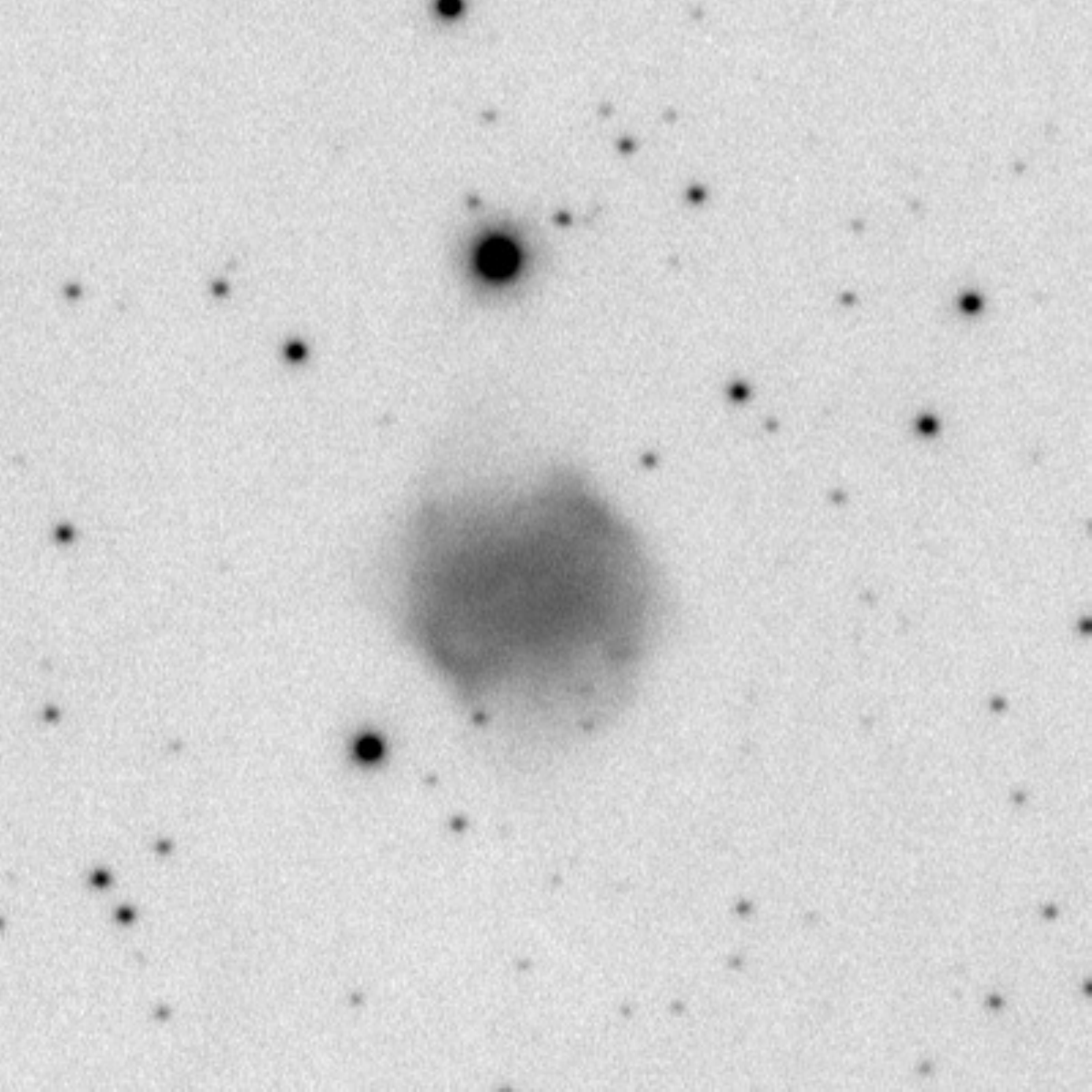}
\includegraphics[height=1.7in]{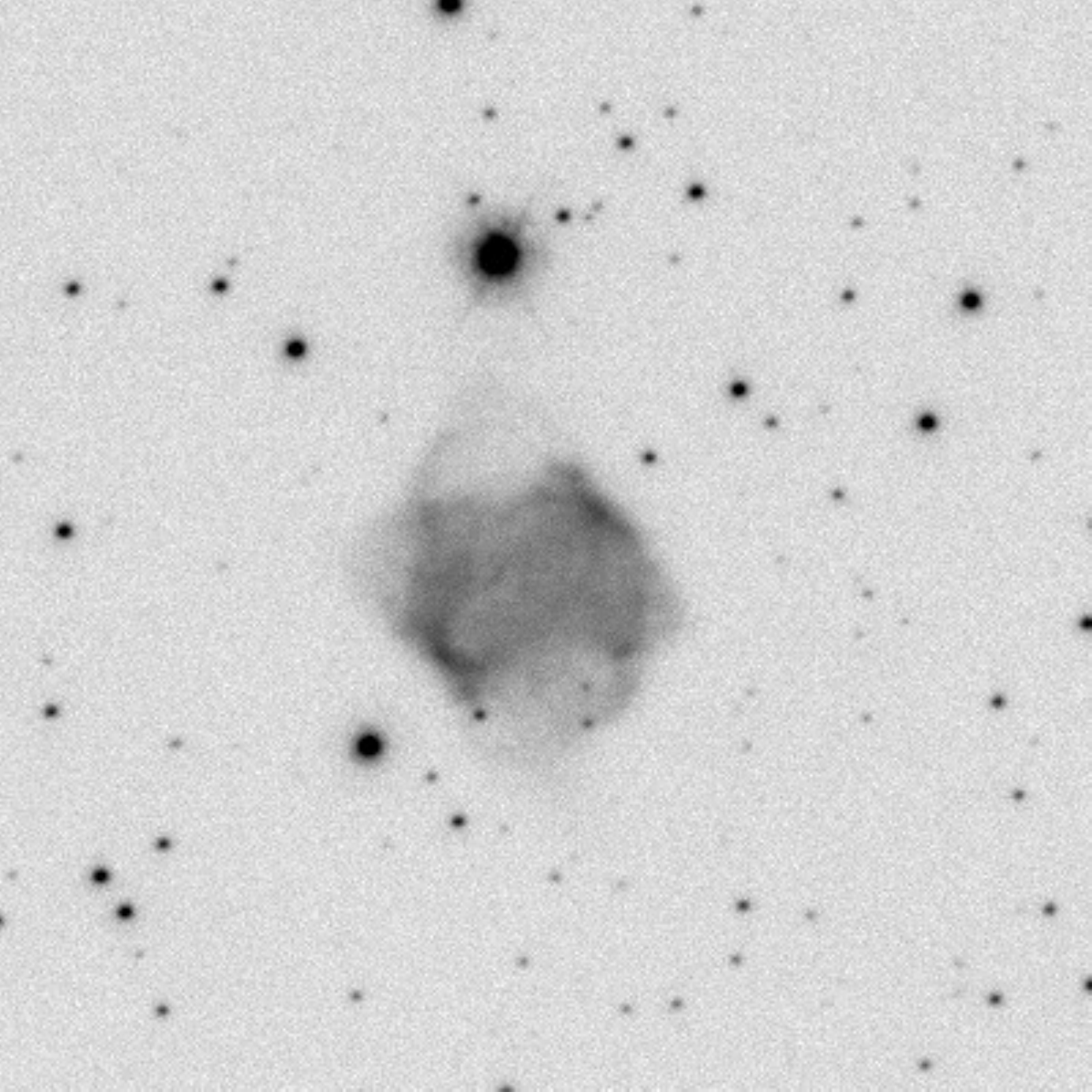}
\includegraphics[height=1.7in]{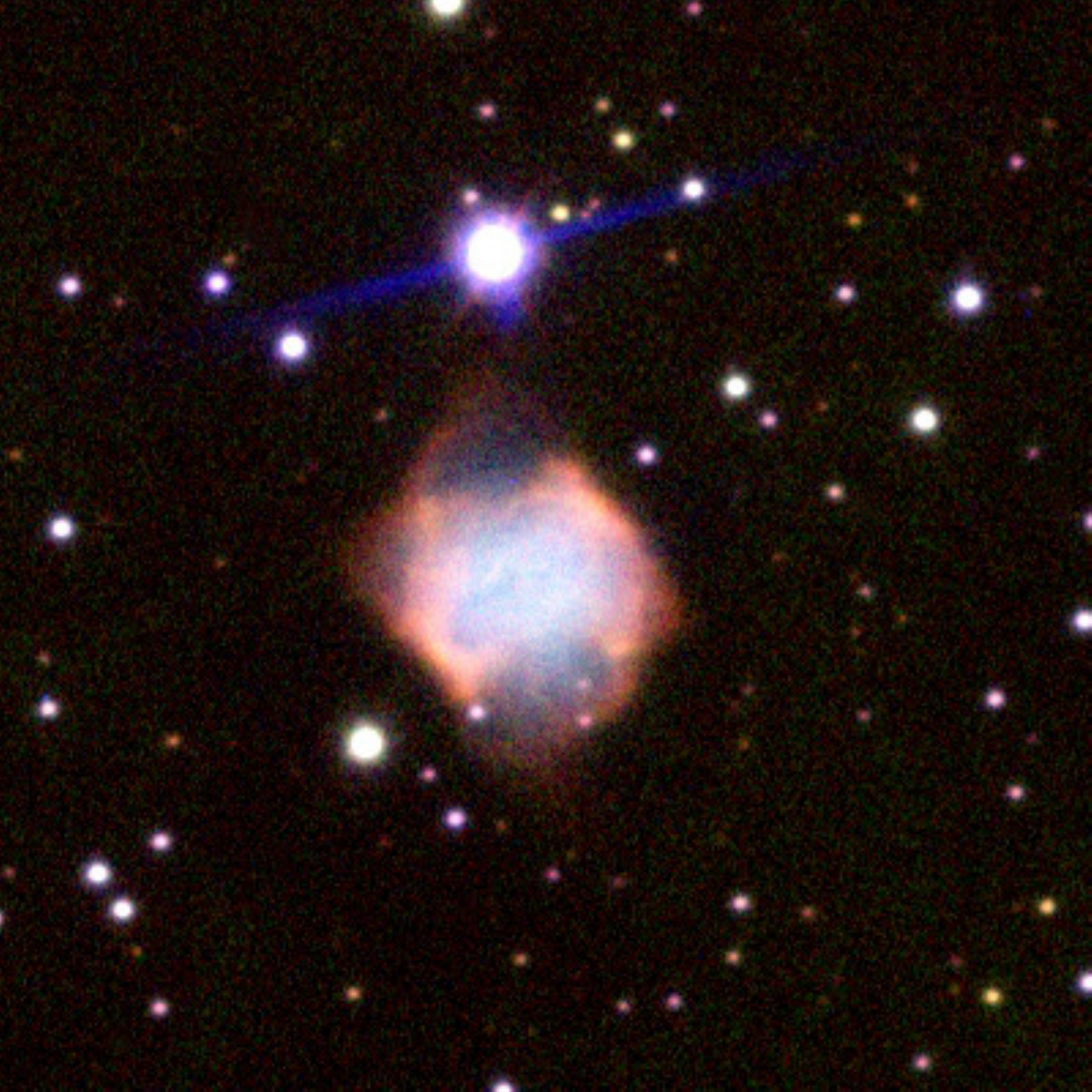}
\vskip .1in 
\includegraphics[height=1.7in]{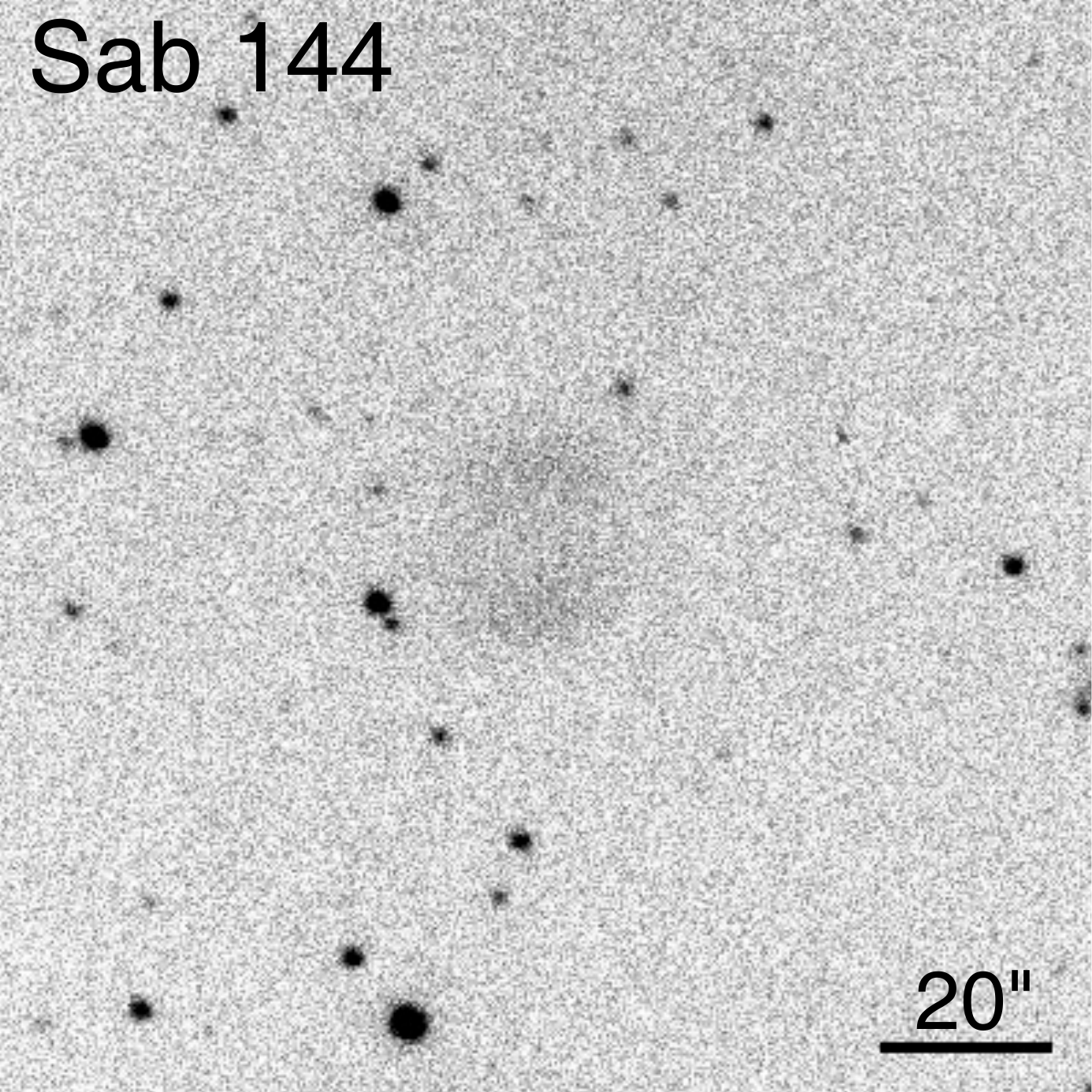} 
\includegraphics[height=1.7in]{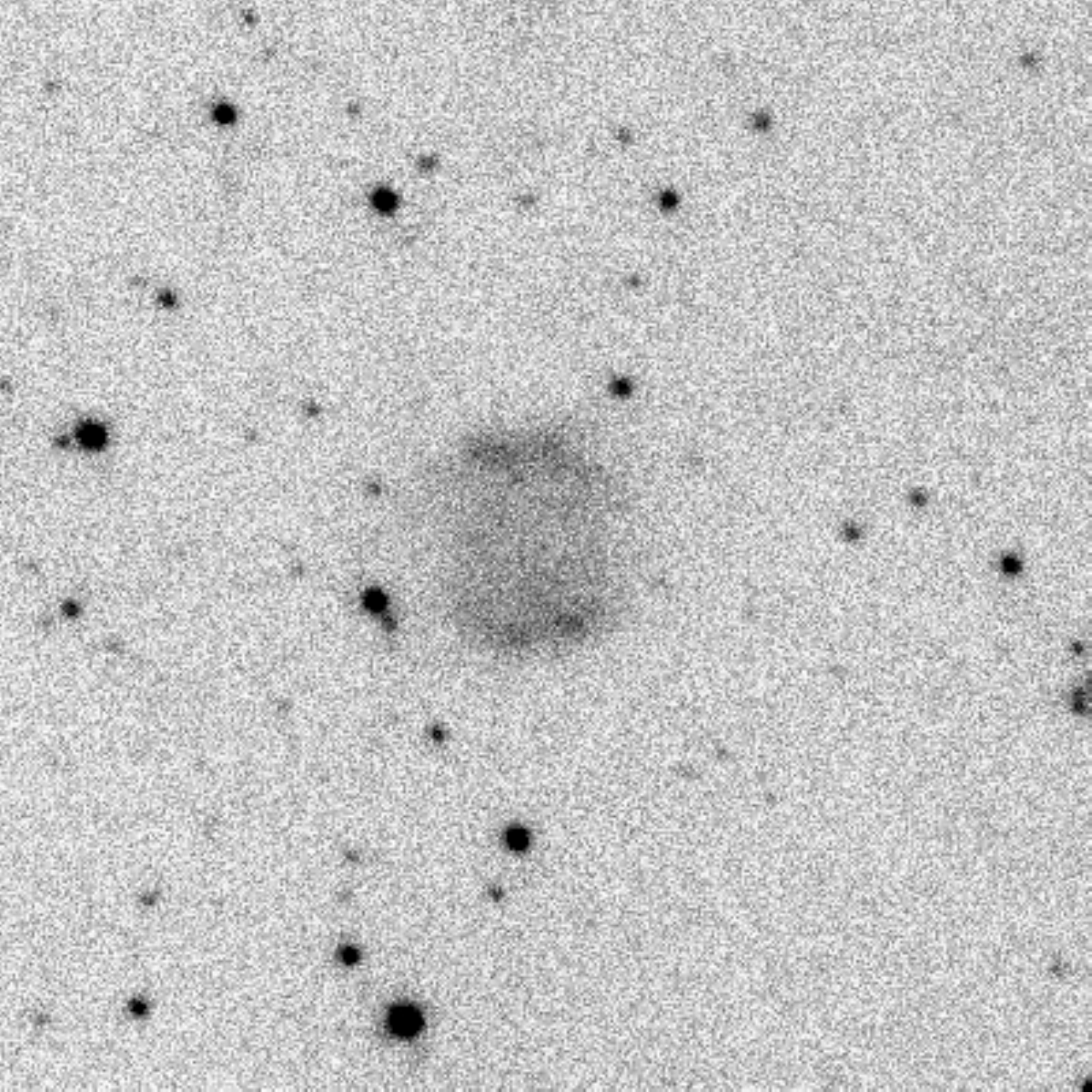}
\includegraphics[height=1.7in]{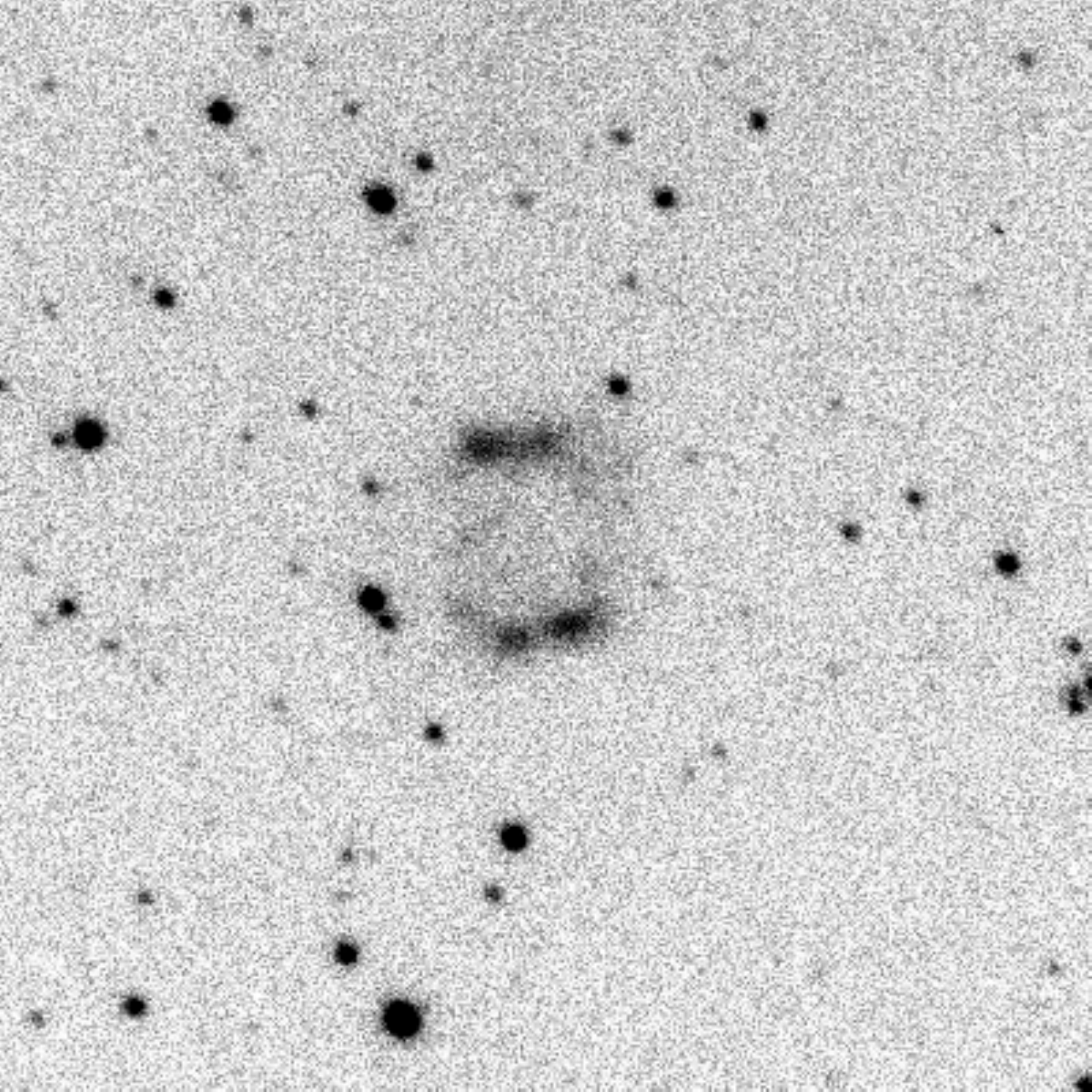}
\includegraphics[height=1.7in]{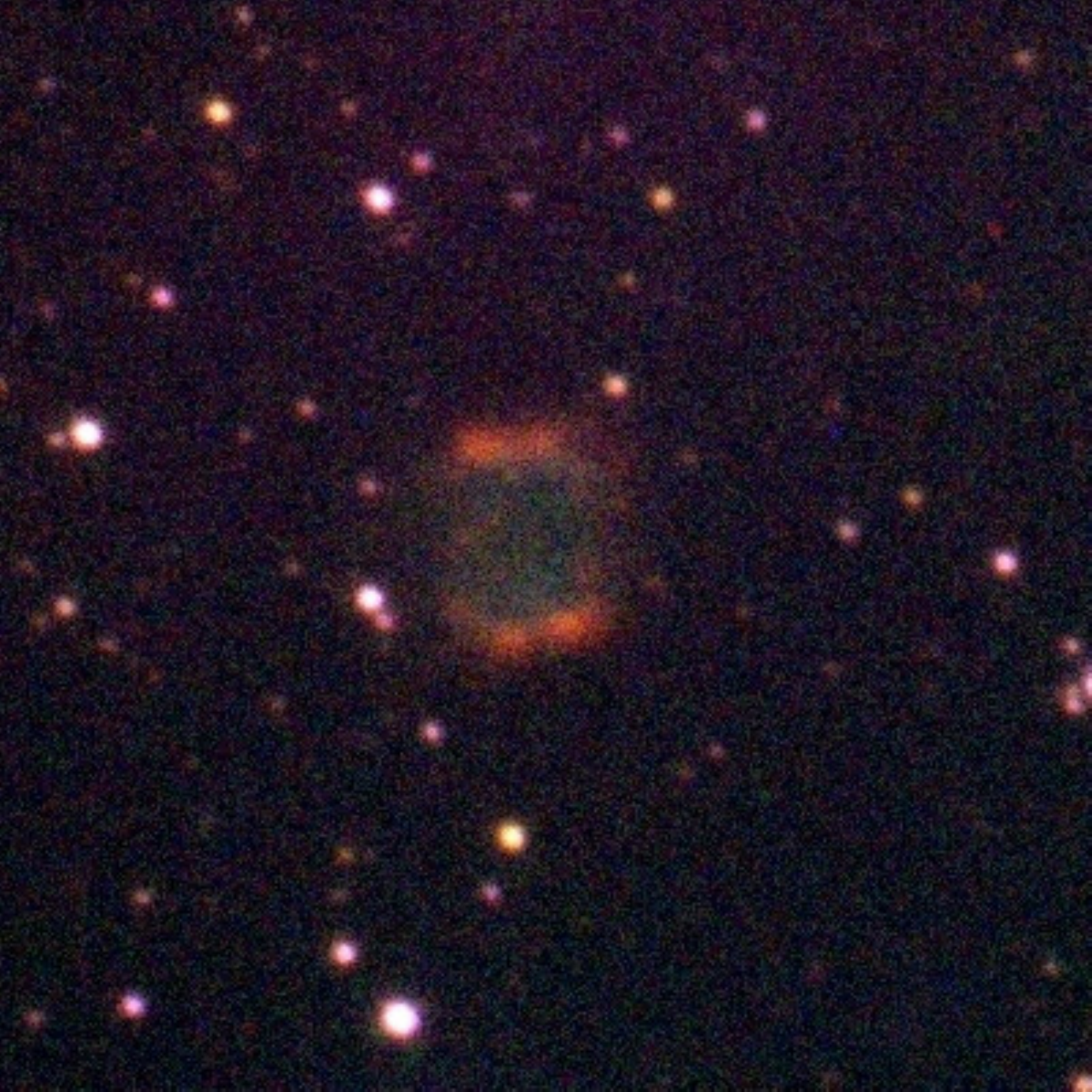}
\vskip .1in 
\includegraphics[height=1.7in]{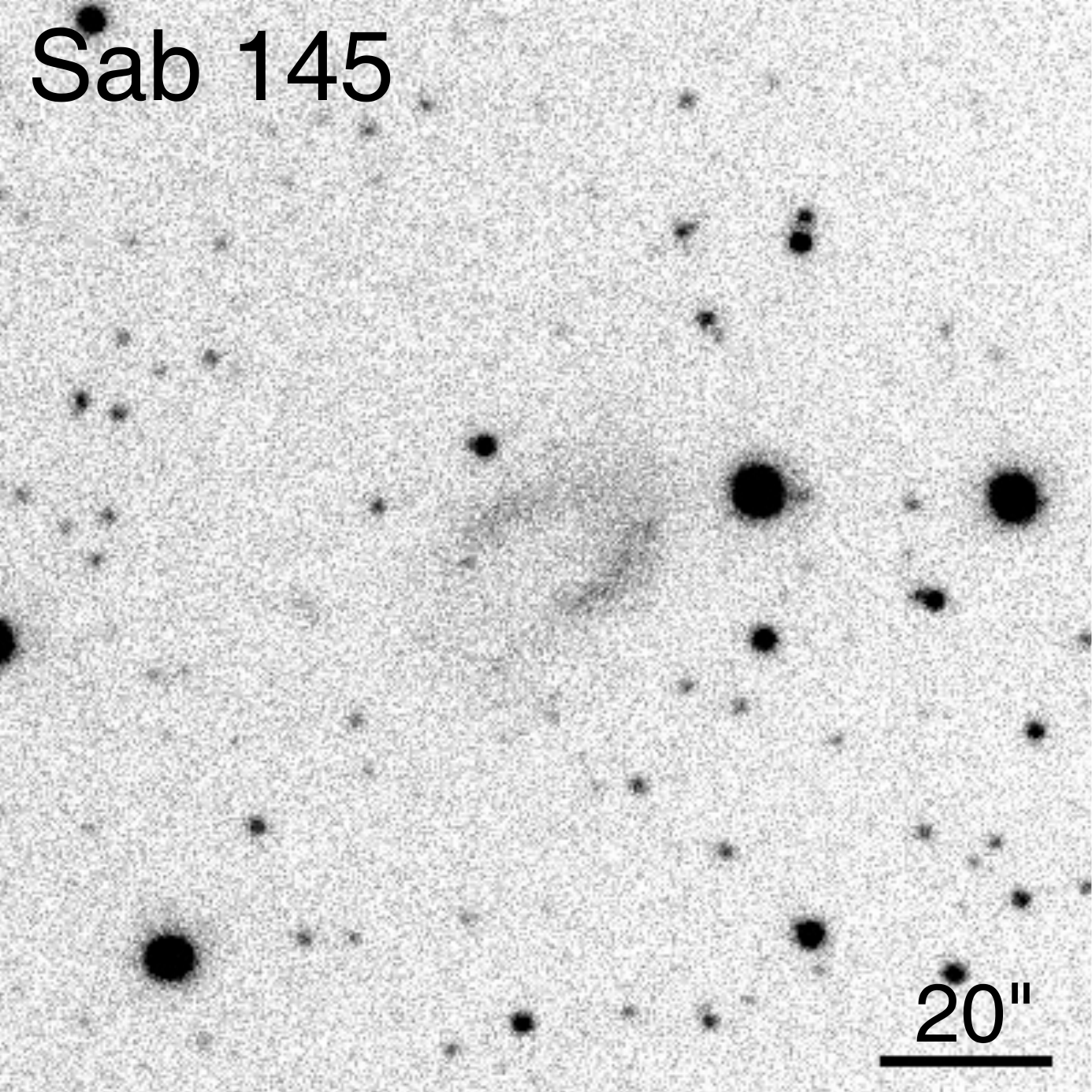} 
\includegraphics[height=1.7in]{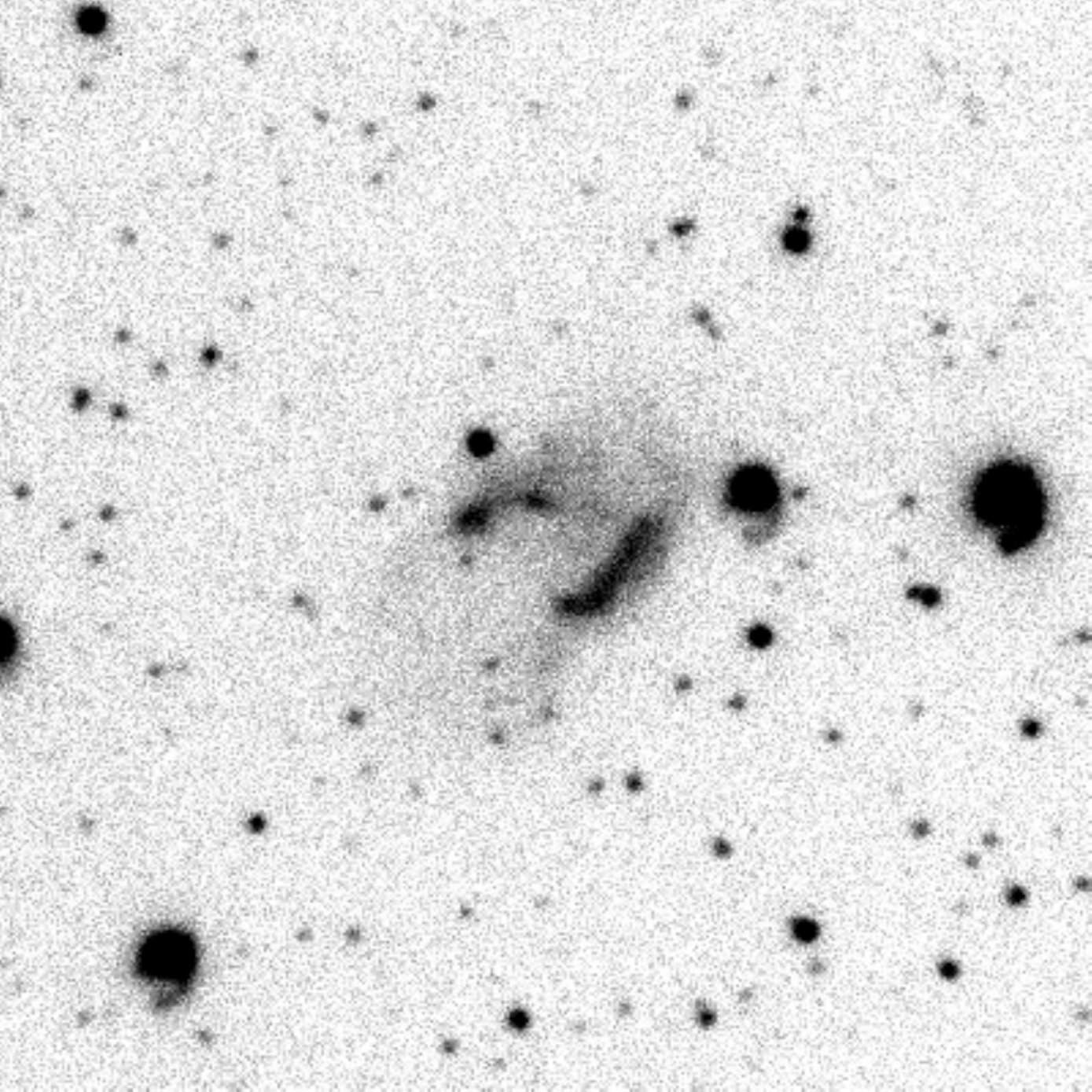}
\includegraphics[height=1.7in]{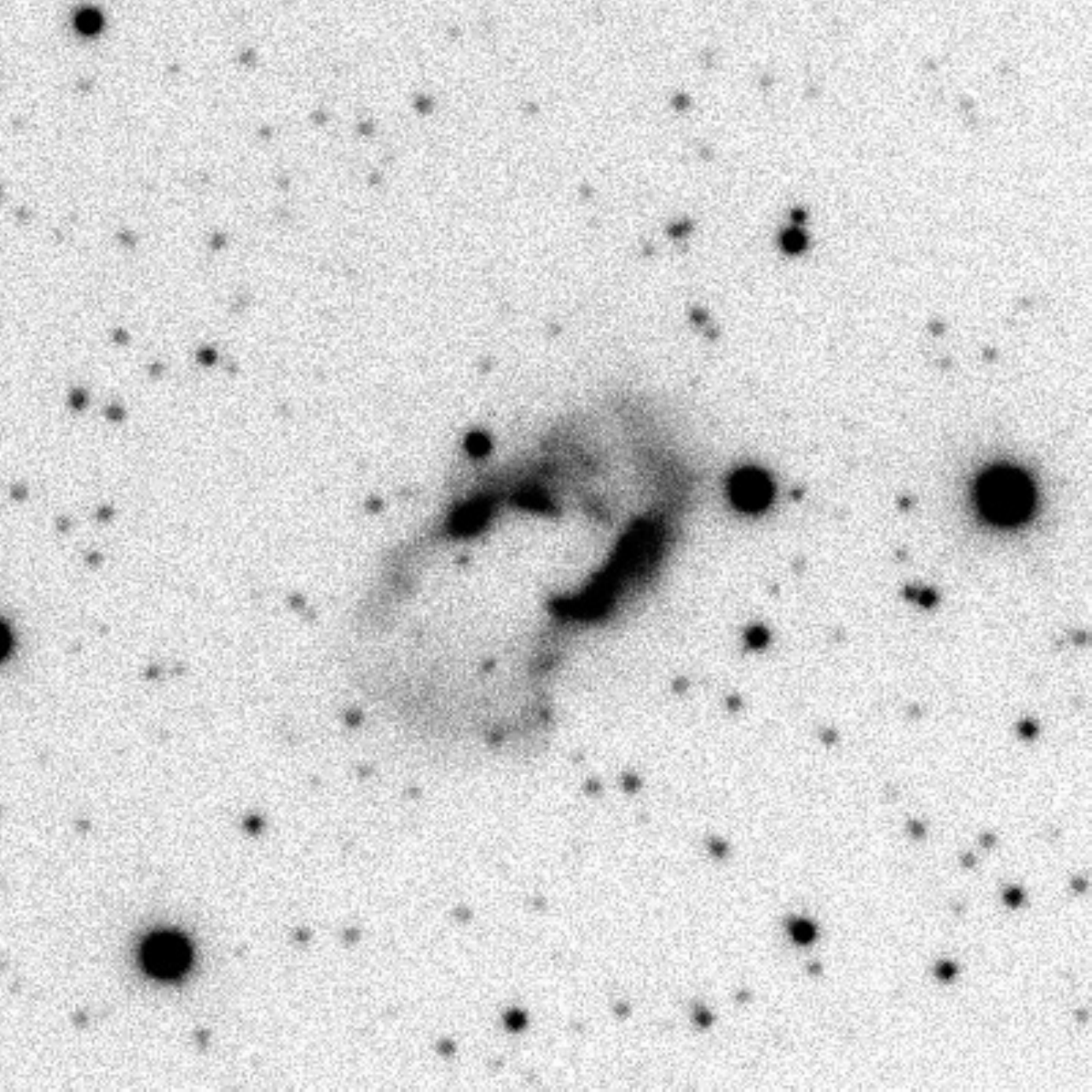}
\includegraphics[height=1.7in]{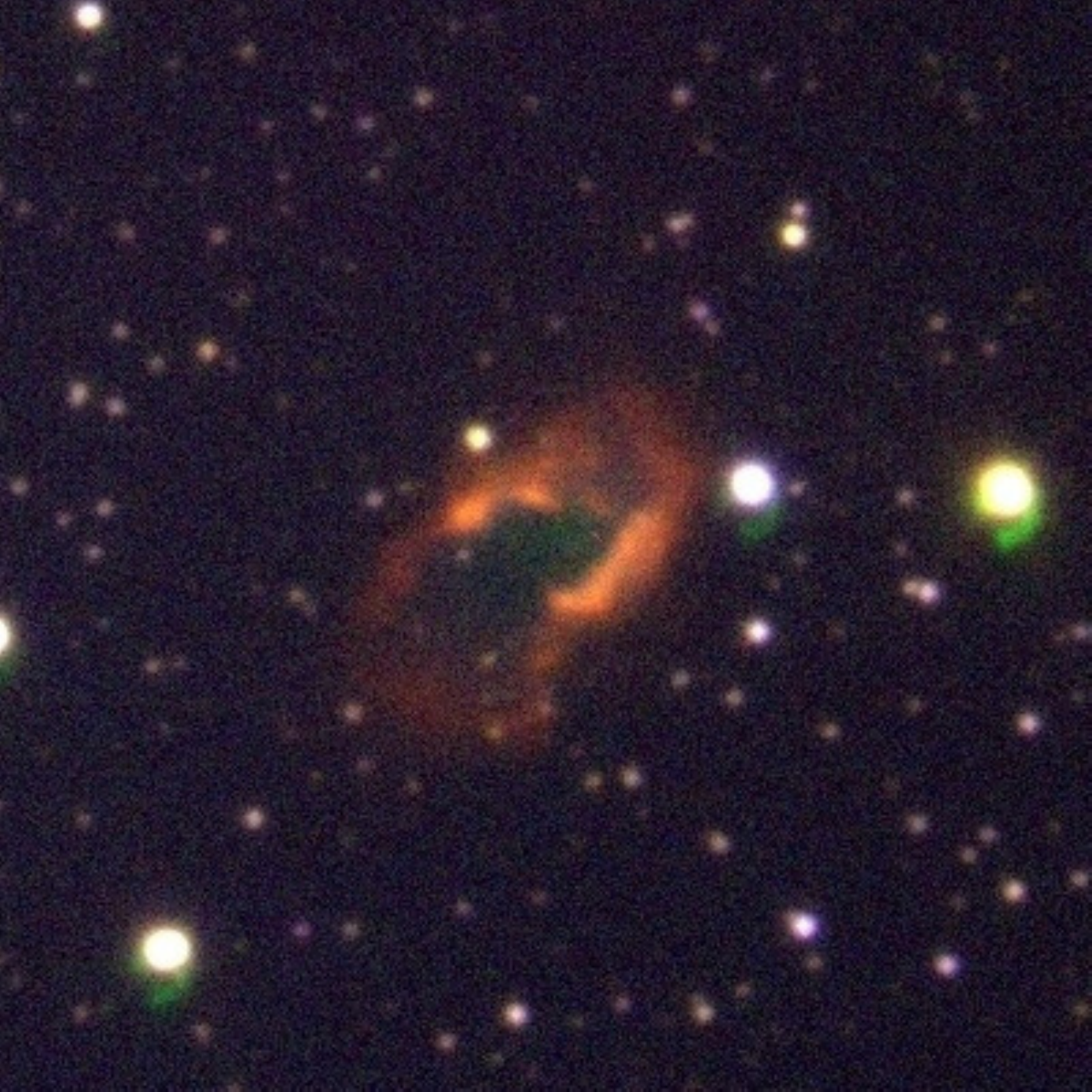}
\vskip .1in
\includegraphics[height=1.7in]{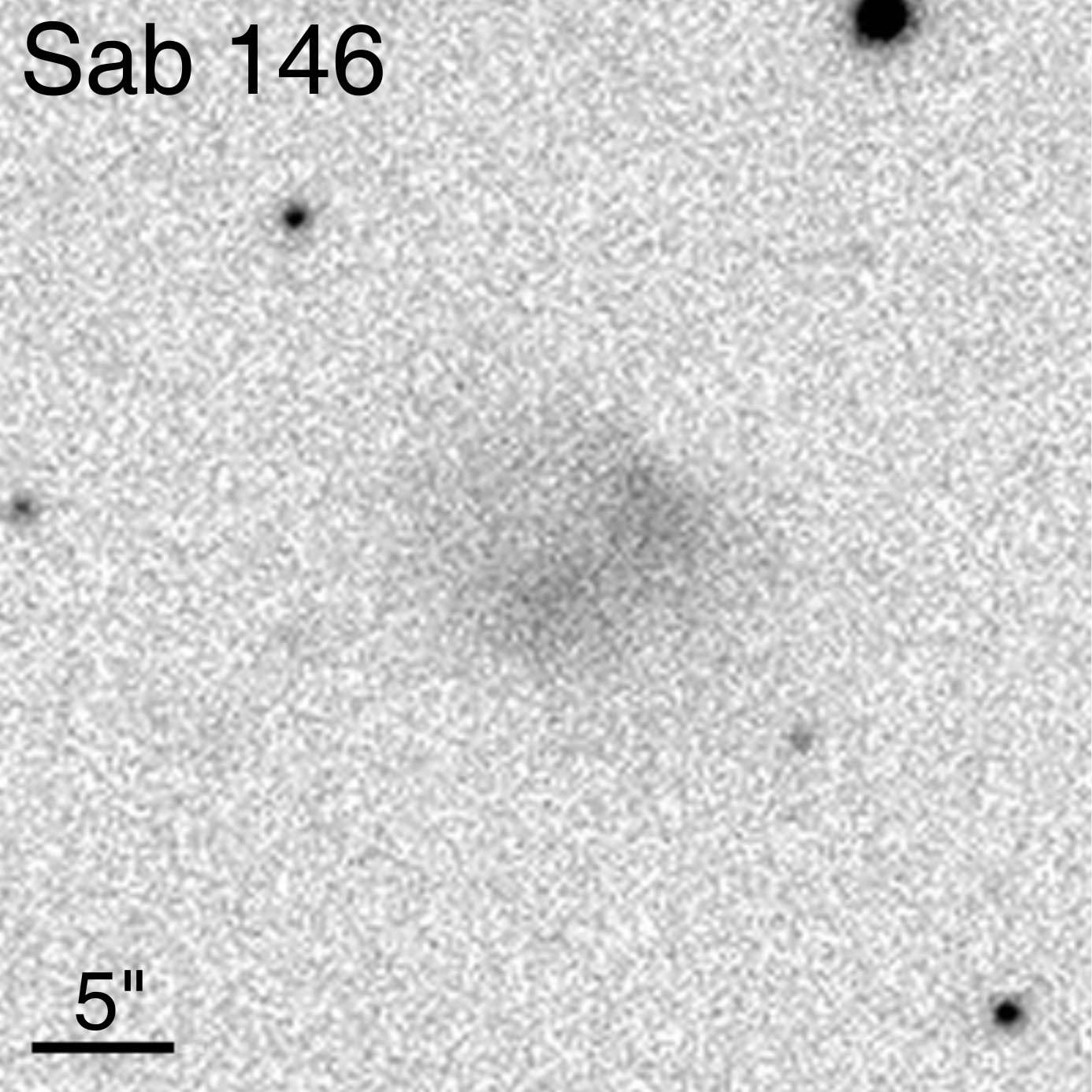} 
\includegraphics[height=1.7in]{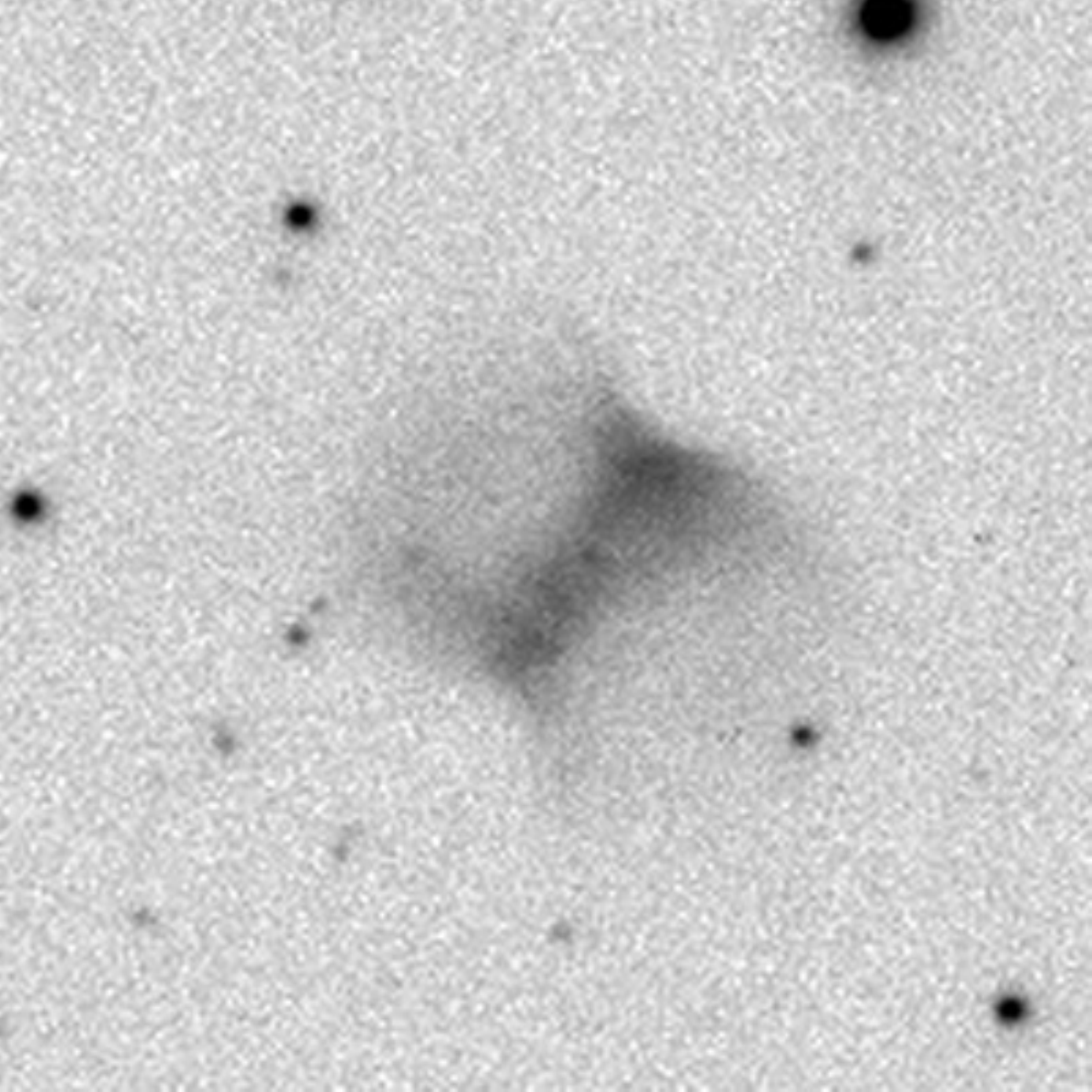}
\includegraphics[height=1.7in]{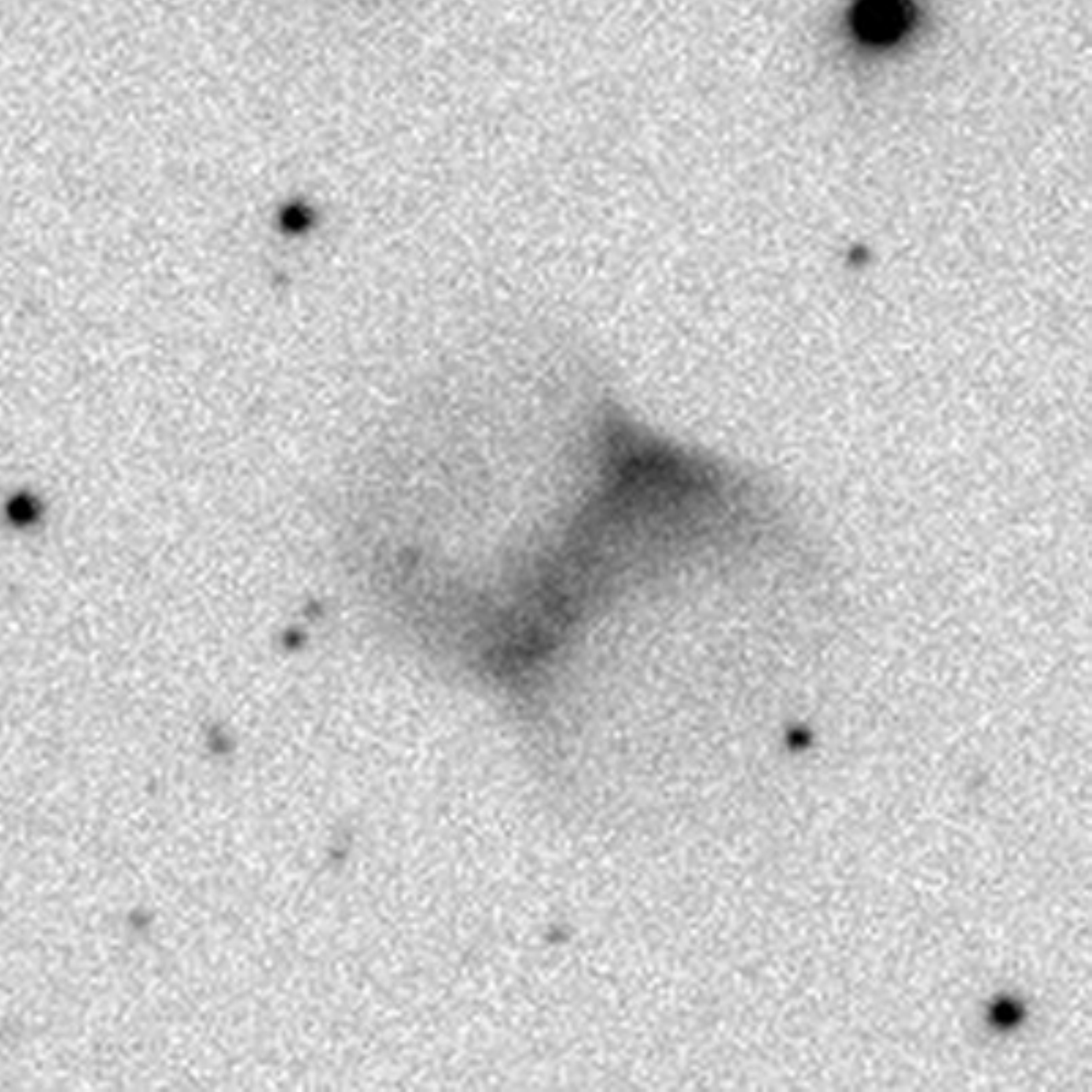}
\includegraphics[height=1.7in]{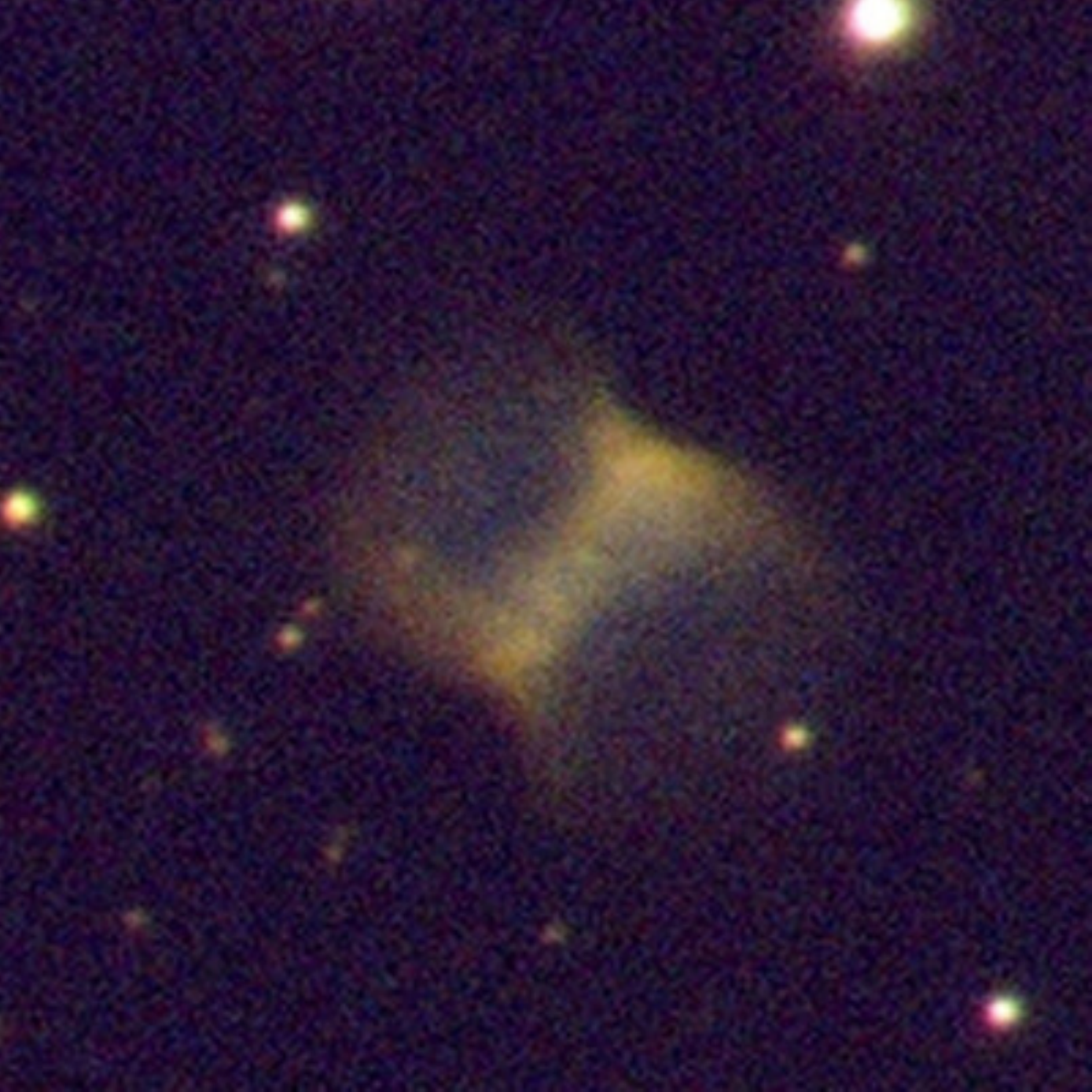}
\caption{Same as Figure~\ref{1.img}. } 
\label{10.img} 
\end{figure*}

\begin{figure*} 
\centering 
\includegraphics[height=1.7in]{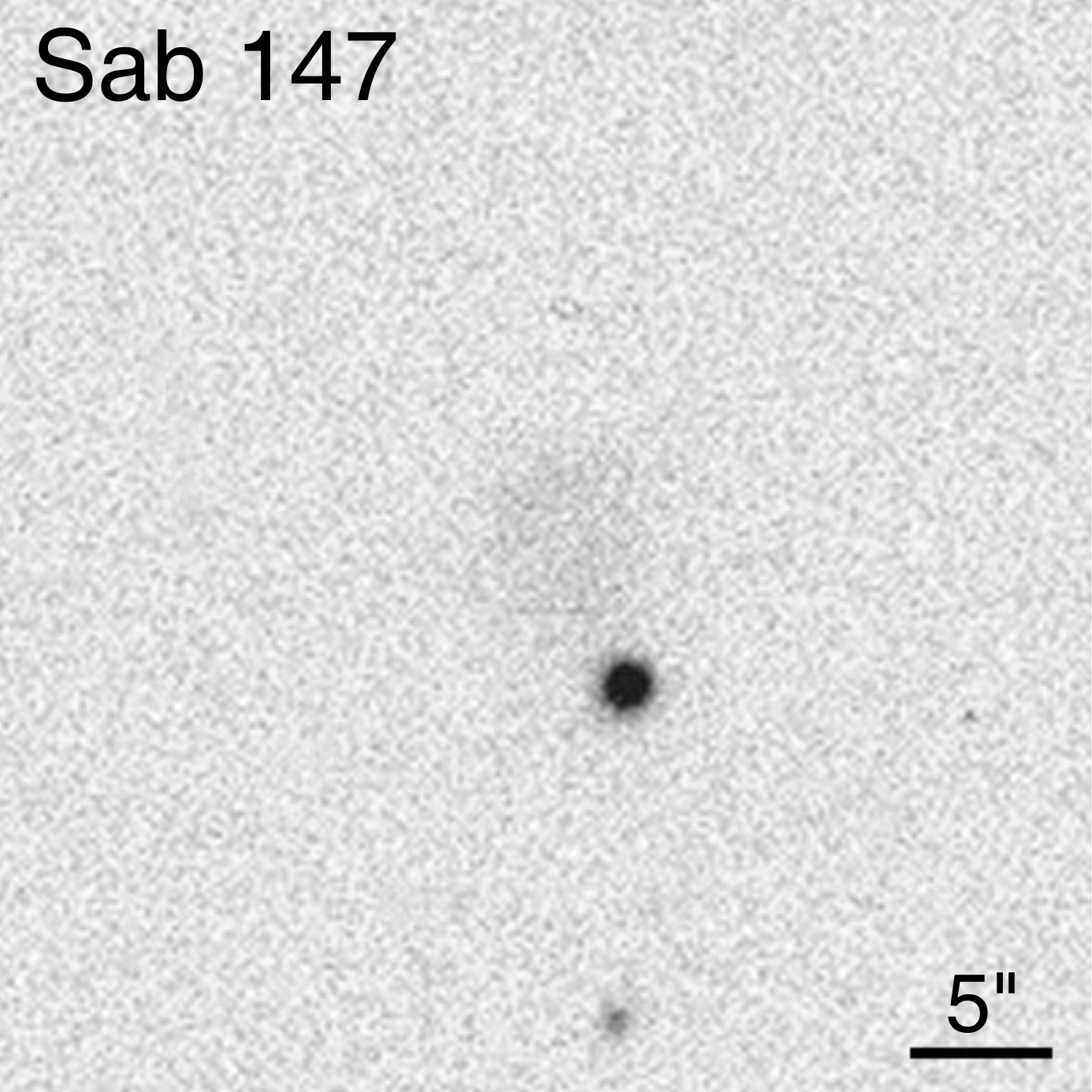} 
\includegraphics[height=1.7in]{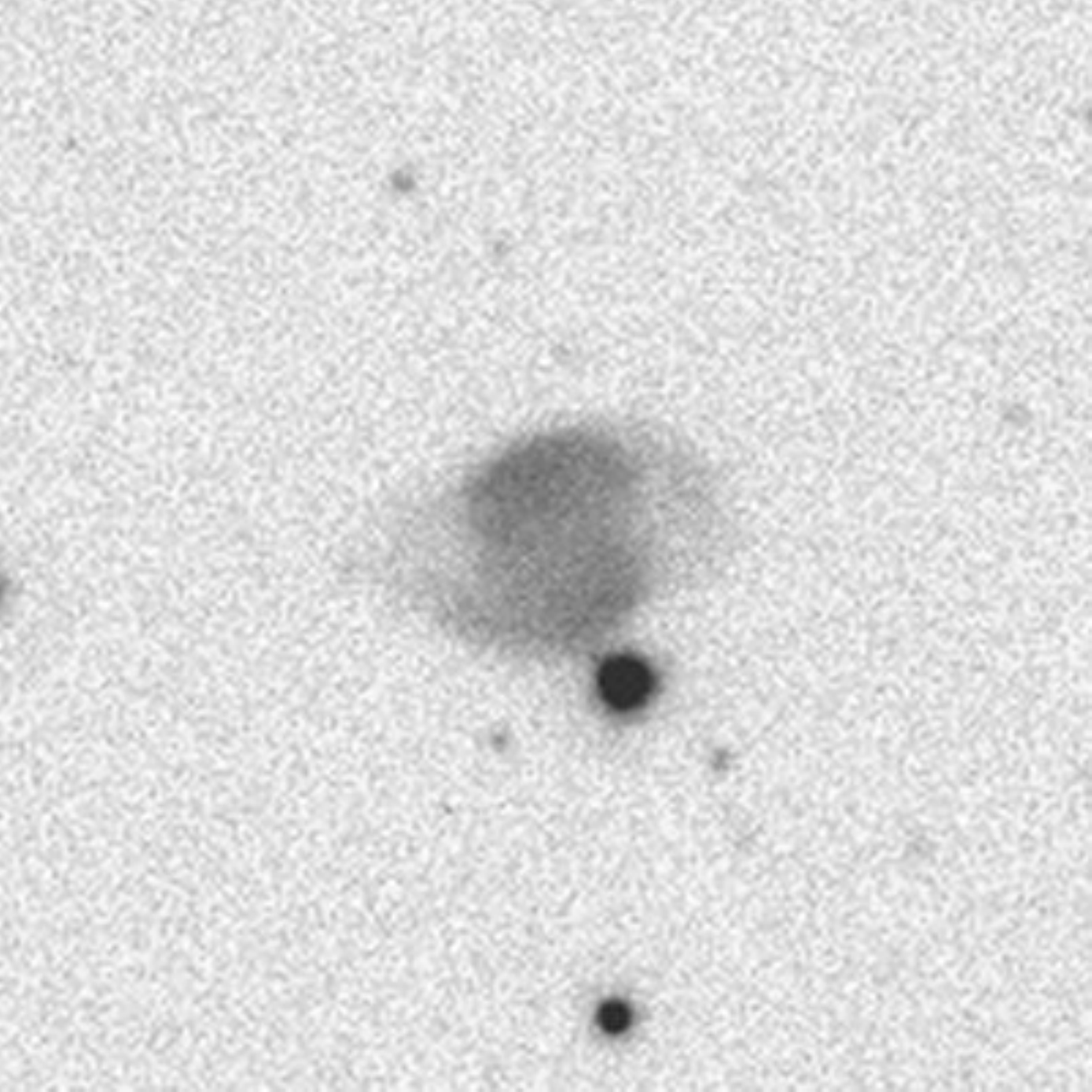}
\includegraphics[height=1.7in]{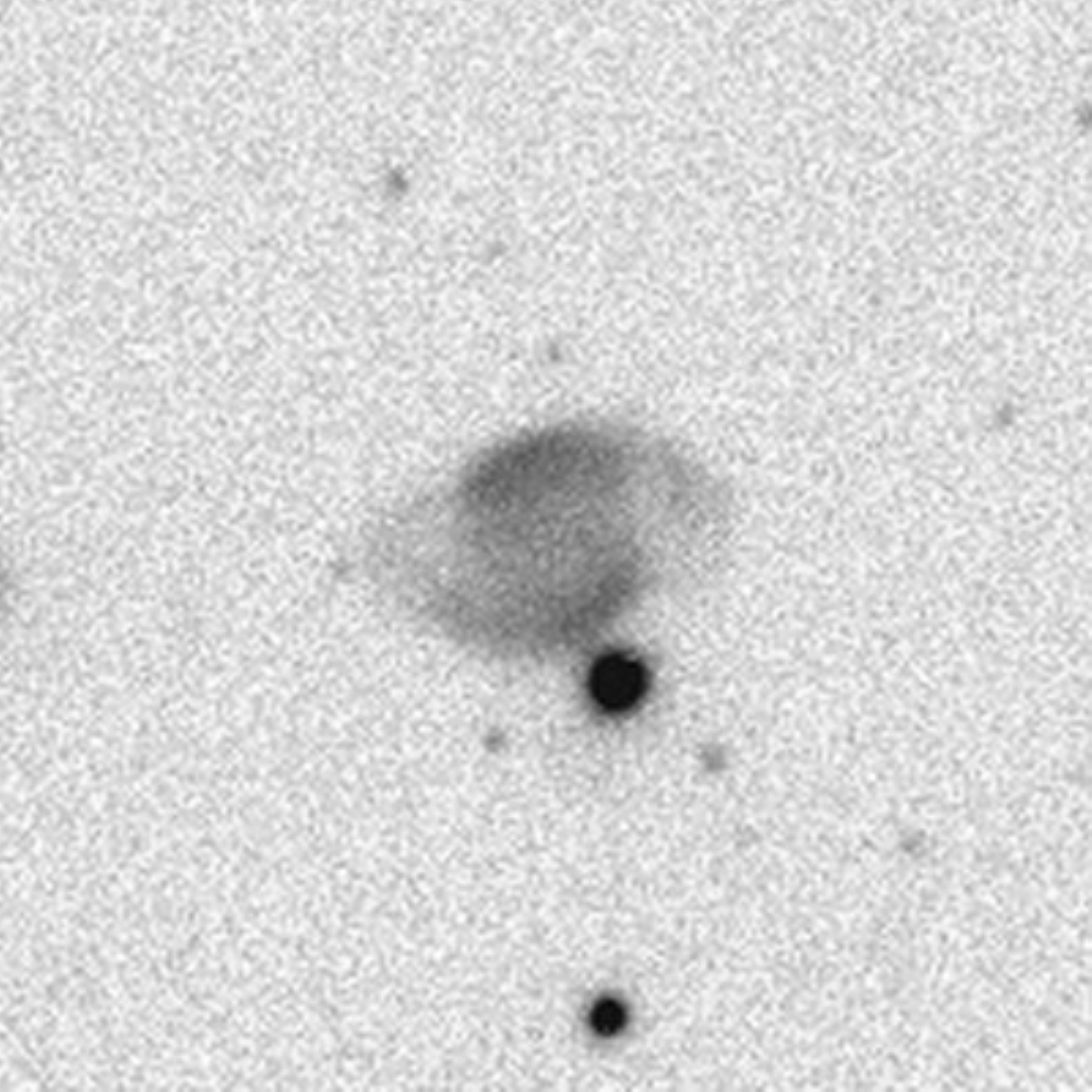}
\includegraphics[height=1.7in]{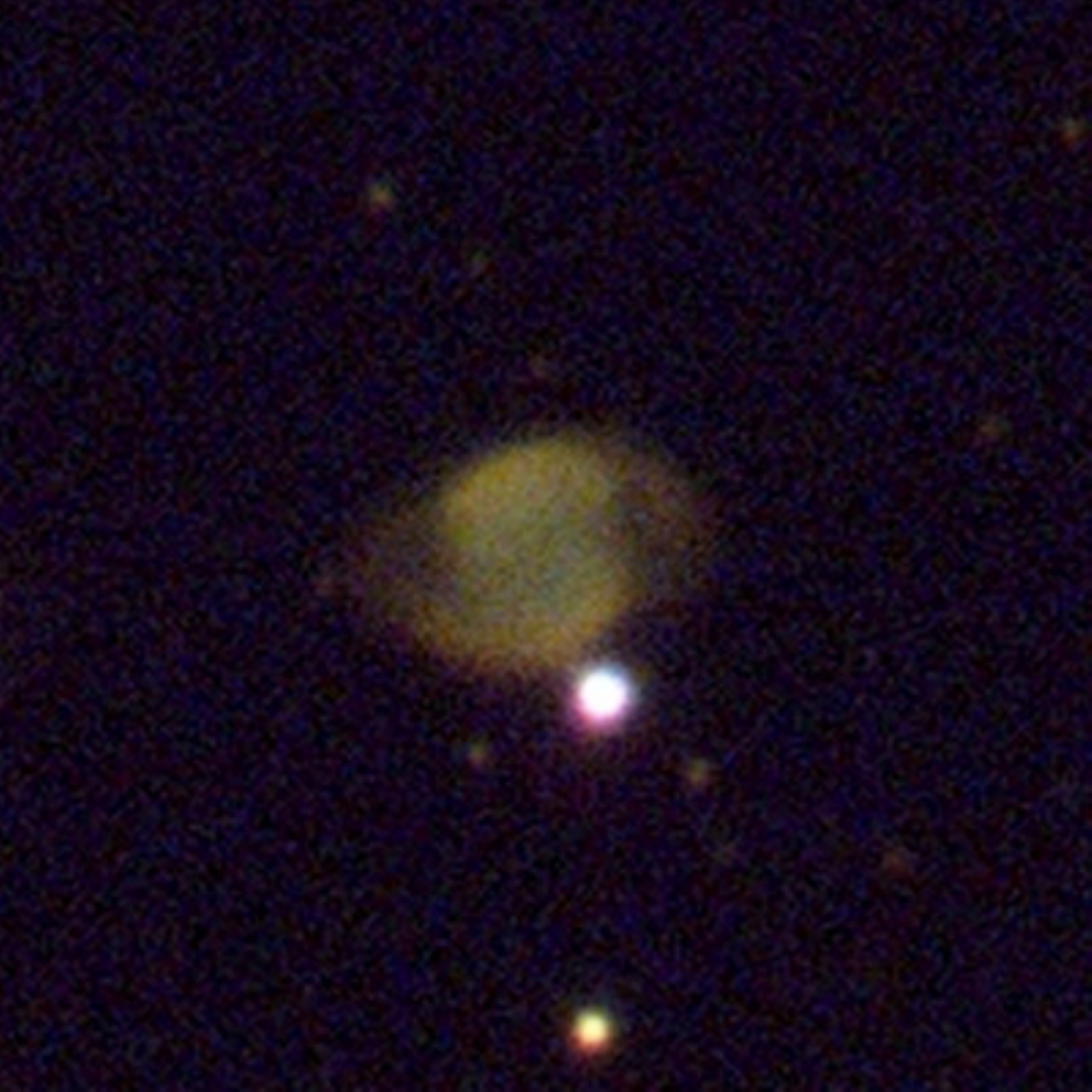}
\vskip .1in 
\includegraphics[height=1.7in]{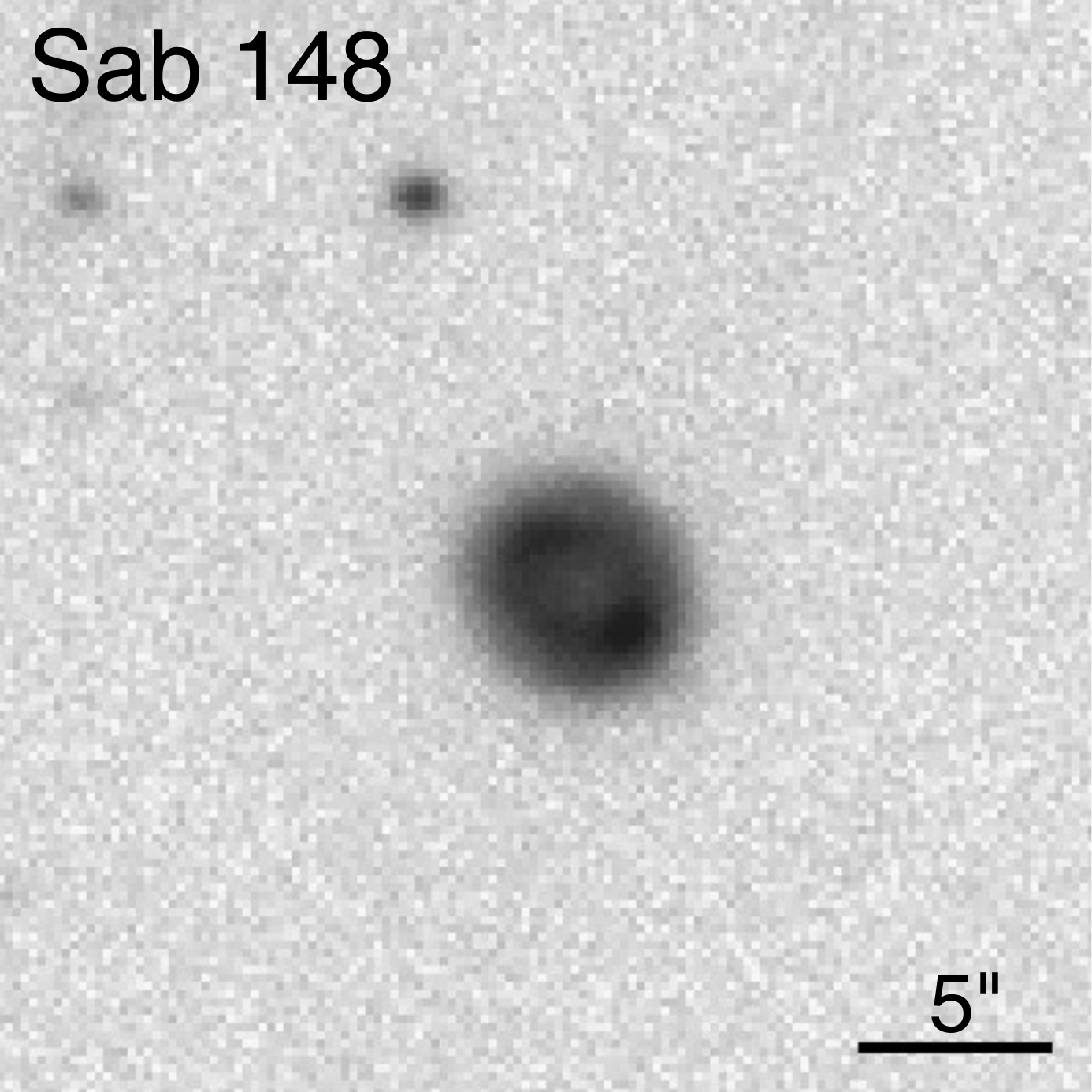} 
\includegraphics[height=1.7in]{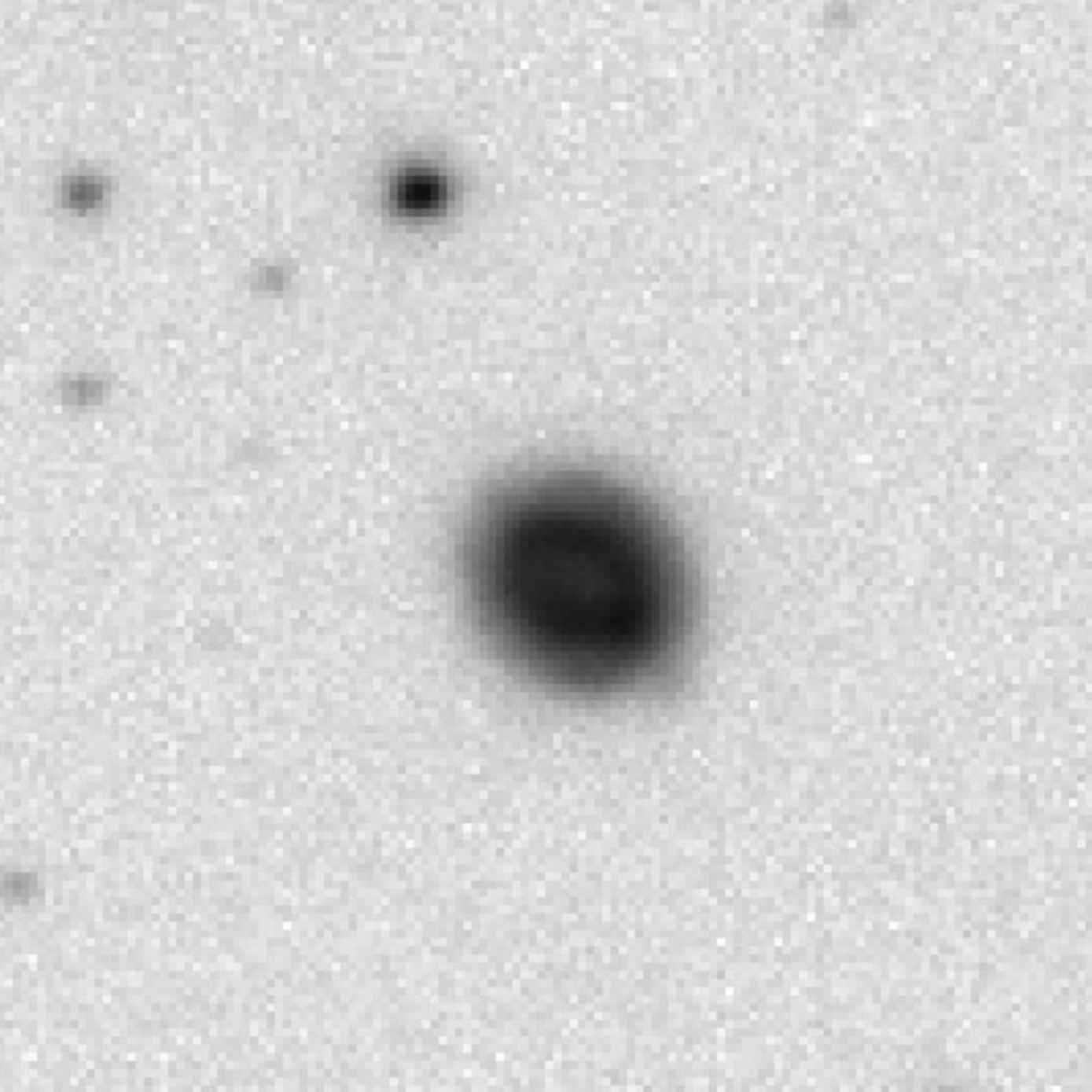}
\includegraphics[height=1.7in]{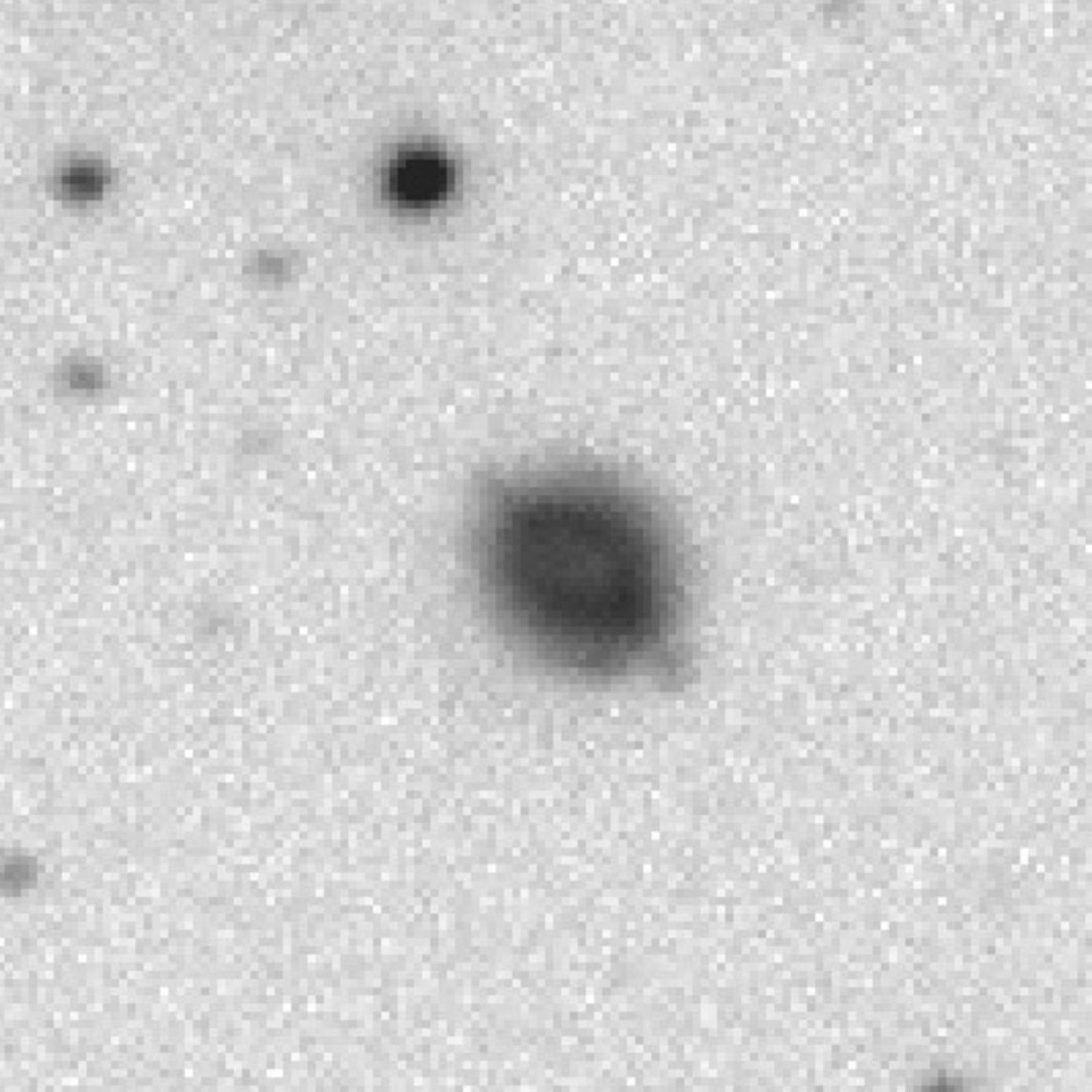}
\includegraphics[height=1.7in]{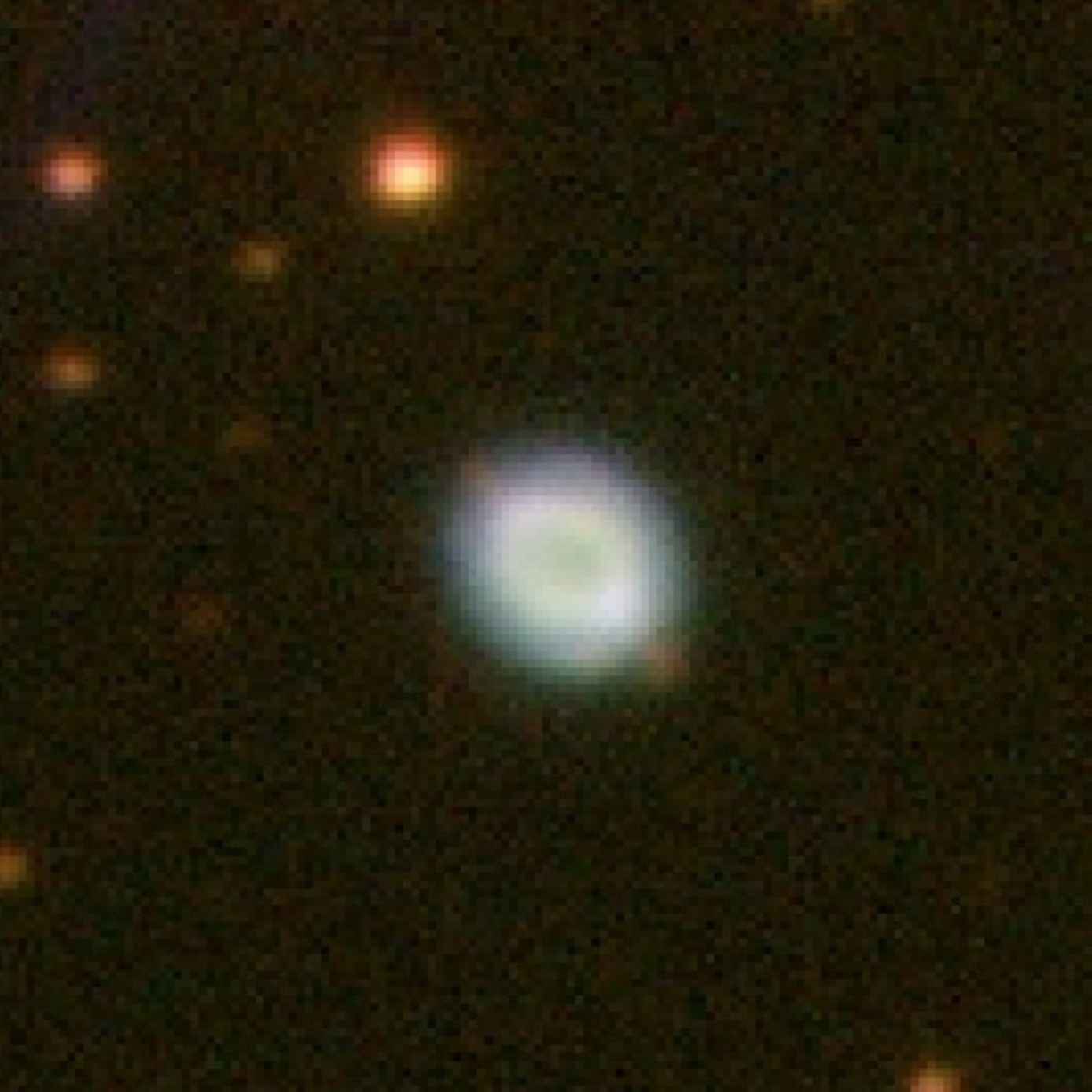}
\vskip .1in 
\includegraphics[height=1.7in]{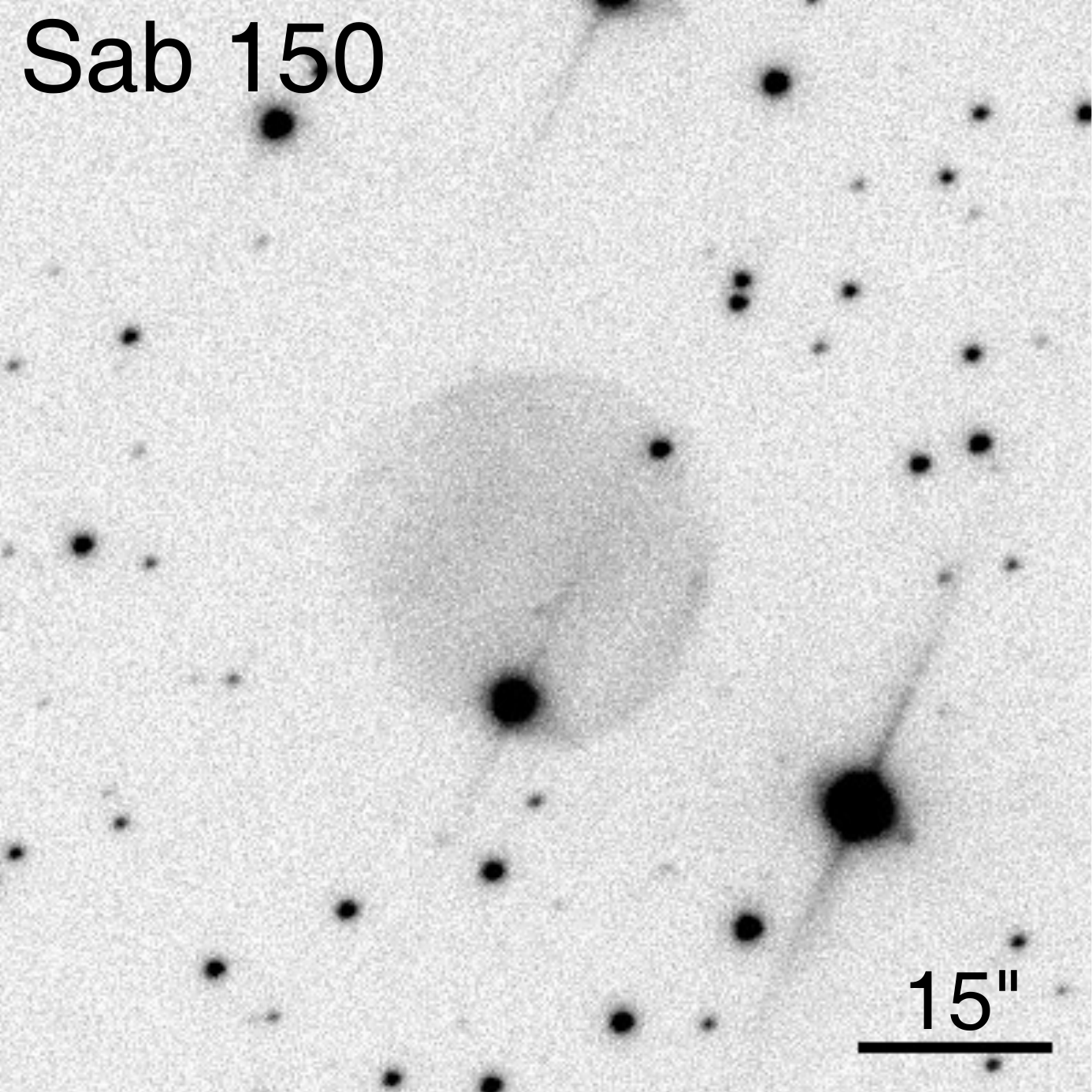} 
\includegraphics[height=1.7in]{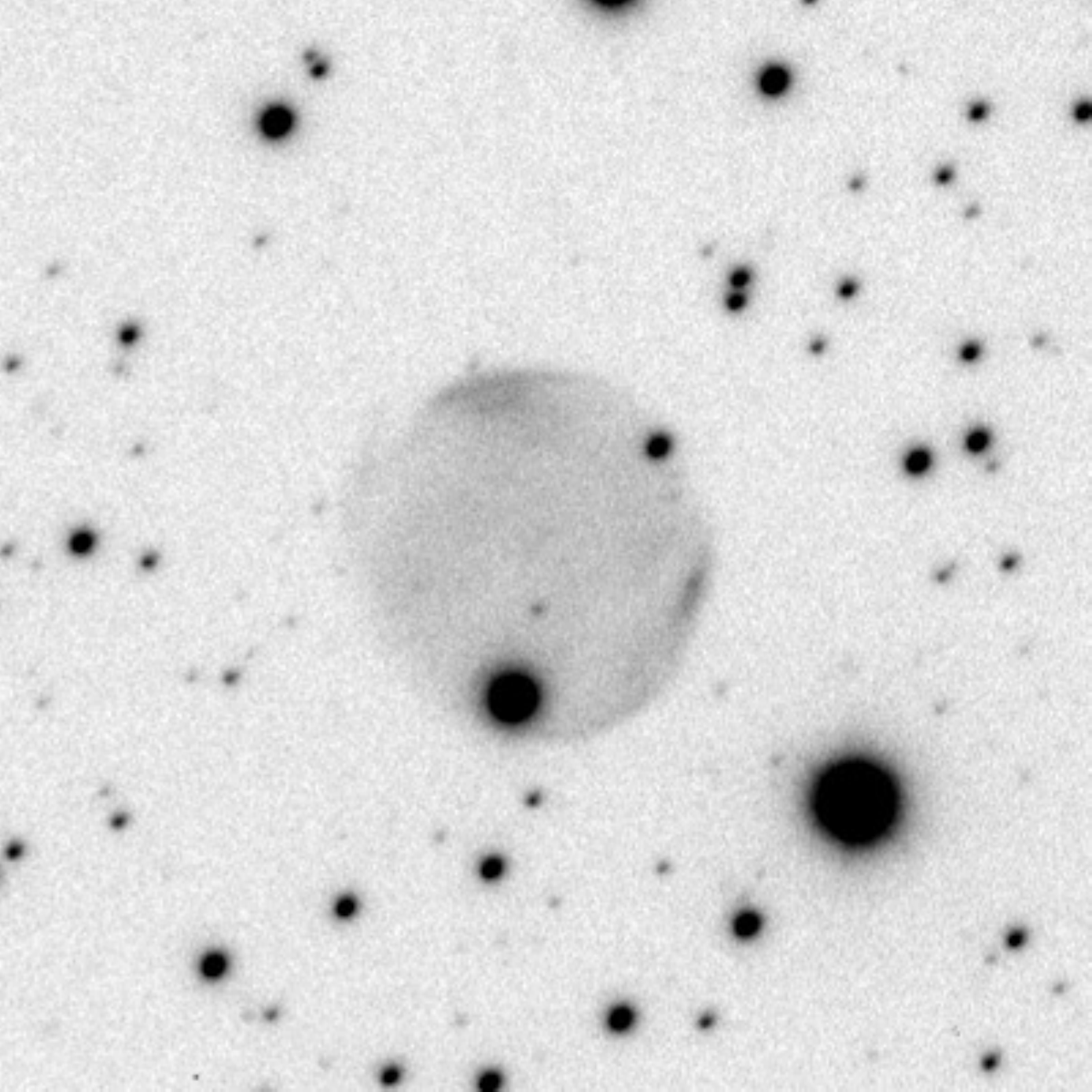}
\includegraphics[height=1.7in]{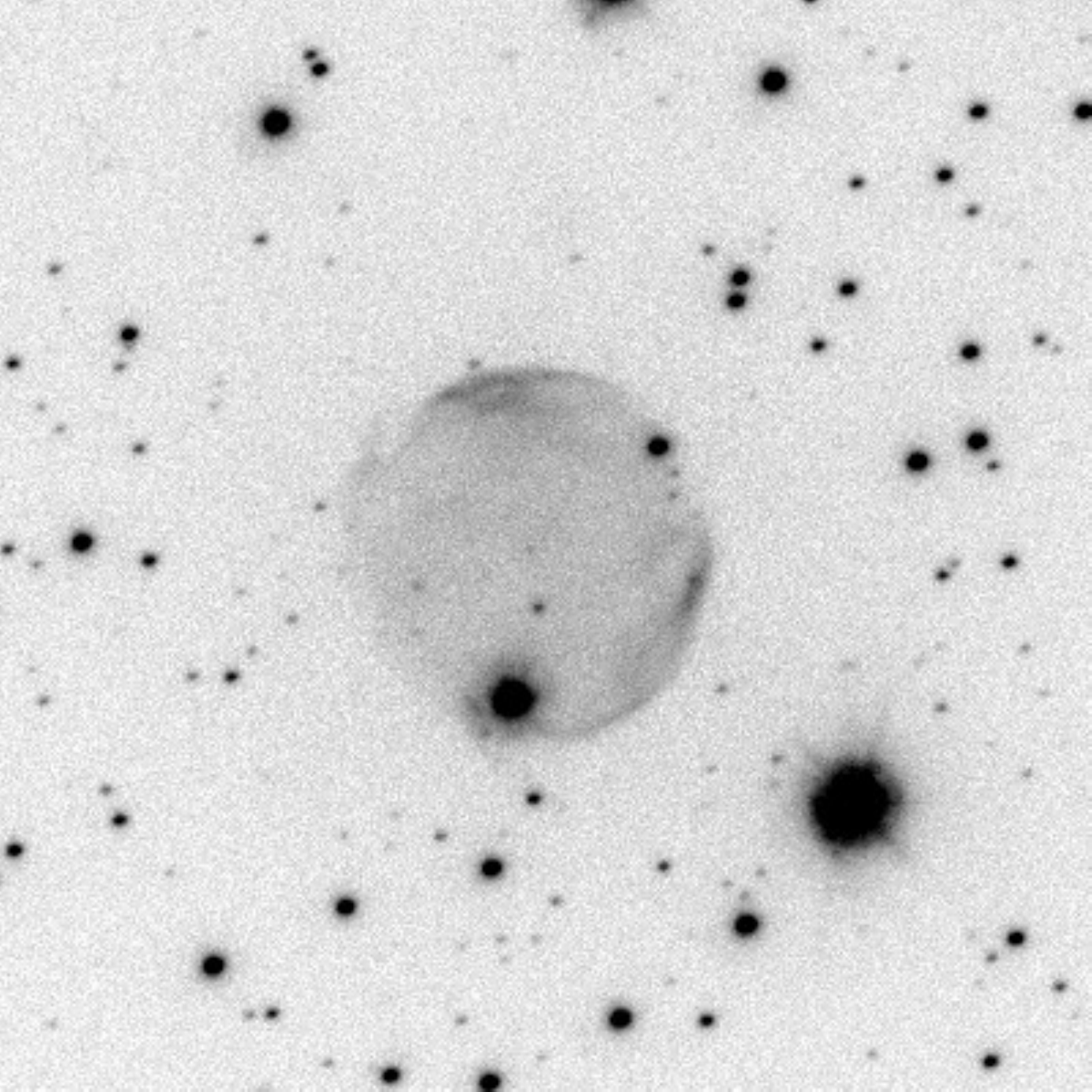}
\includegraphics[height=1.7in]{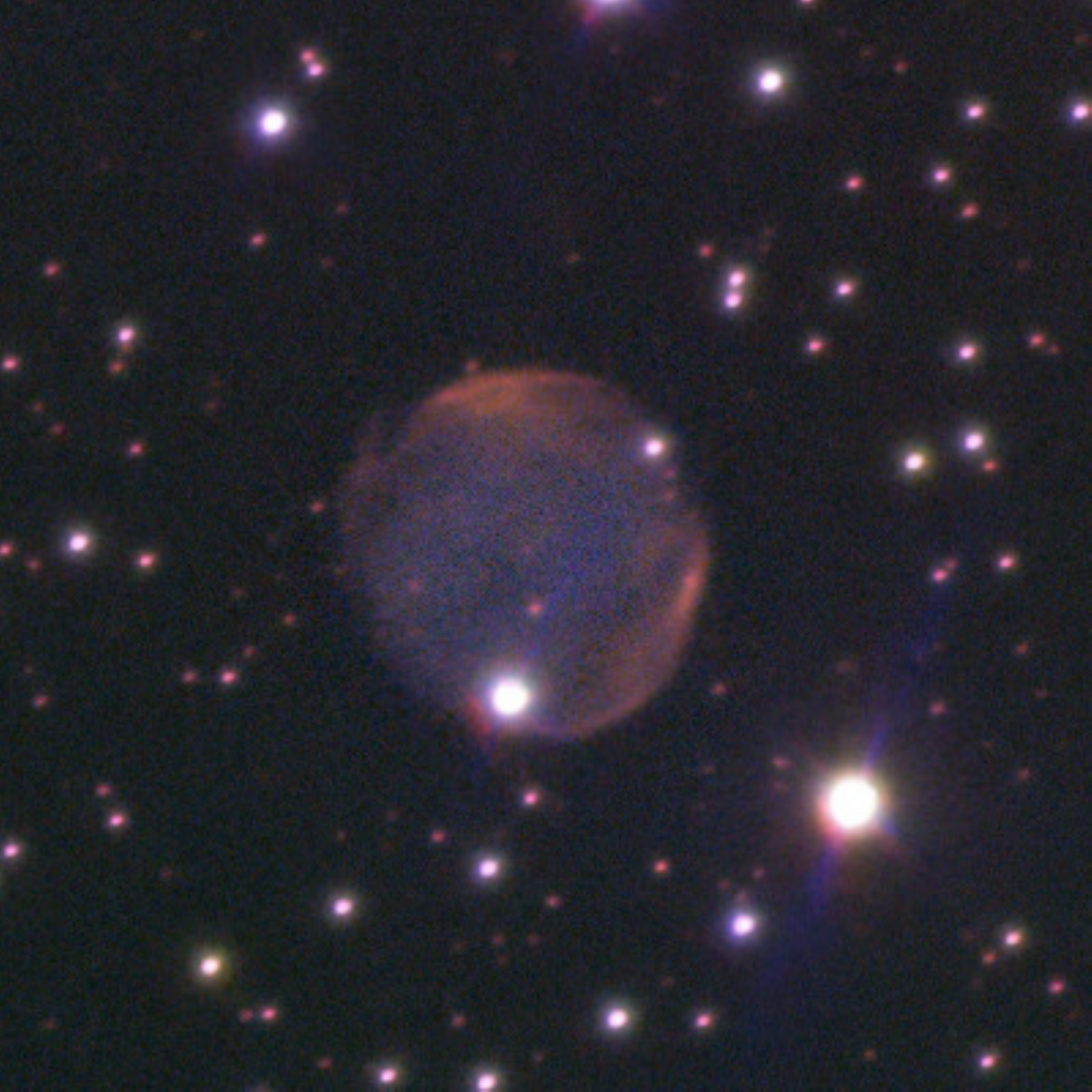}
\vskip .1in 
\includegraphics[height=1.7in]{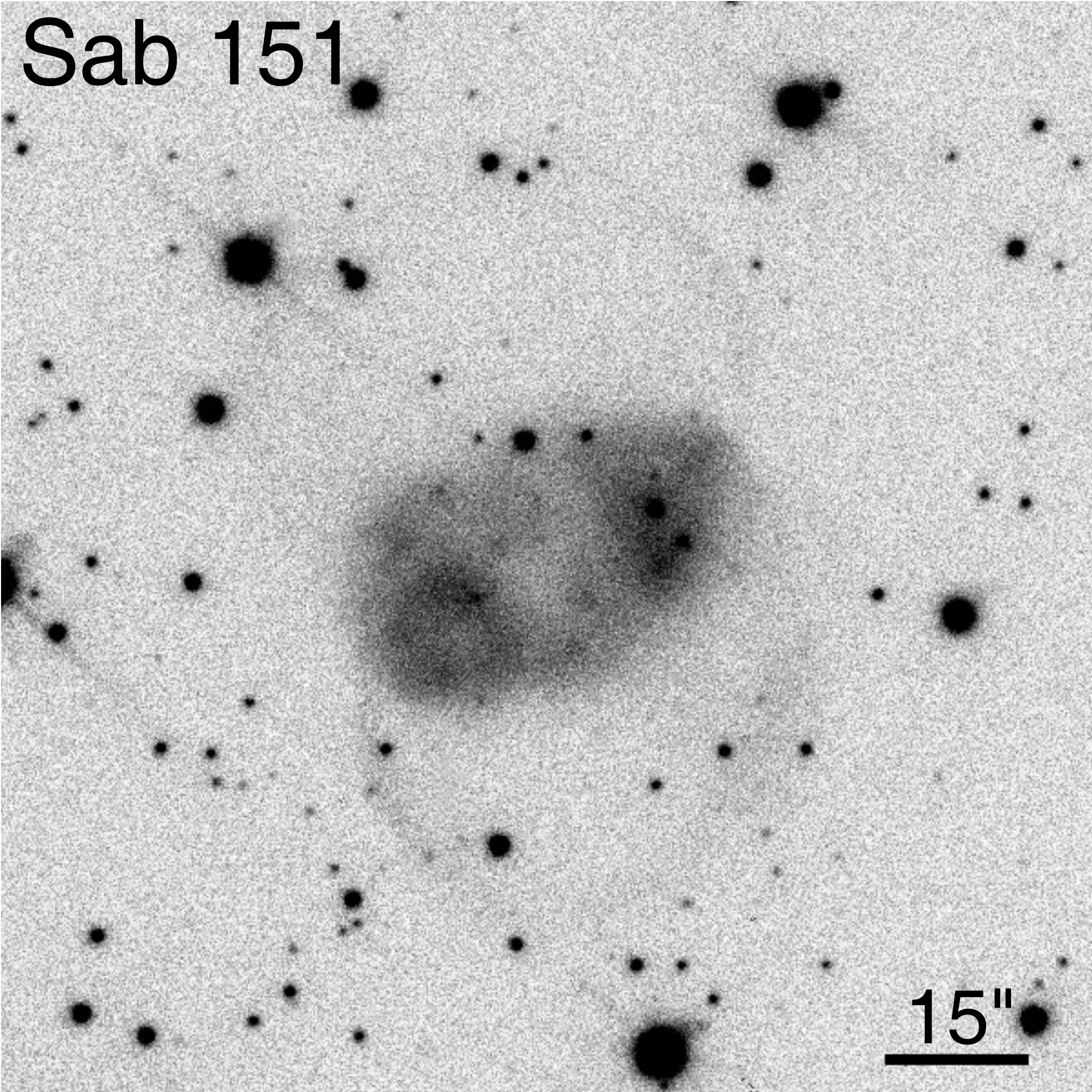} 
\includegraphics[height=1.7in]{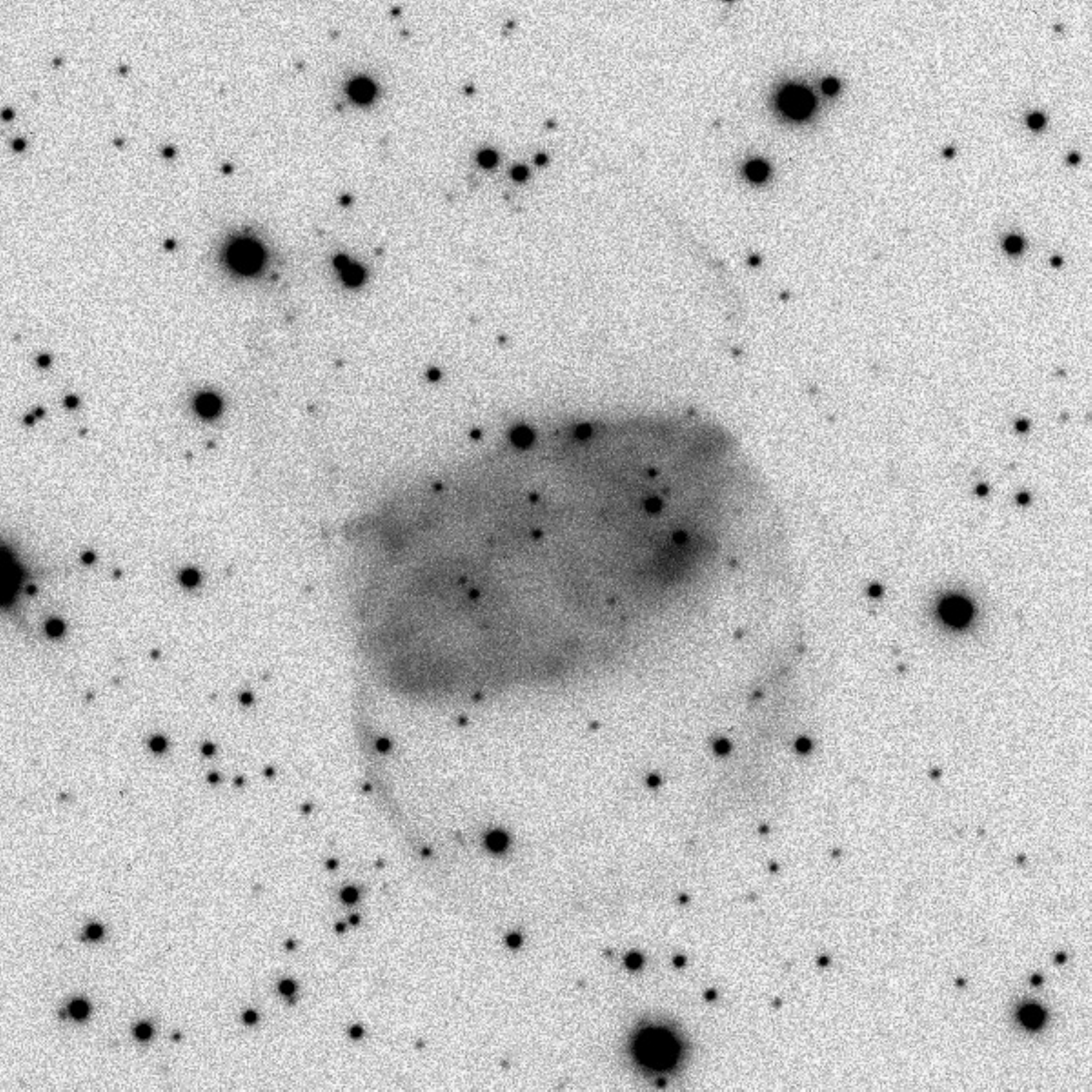}
\includegraphics[height=1.7in]{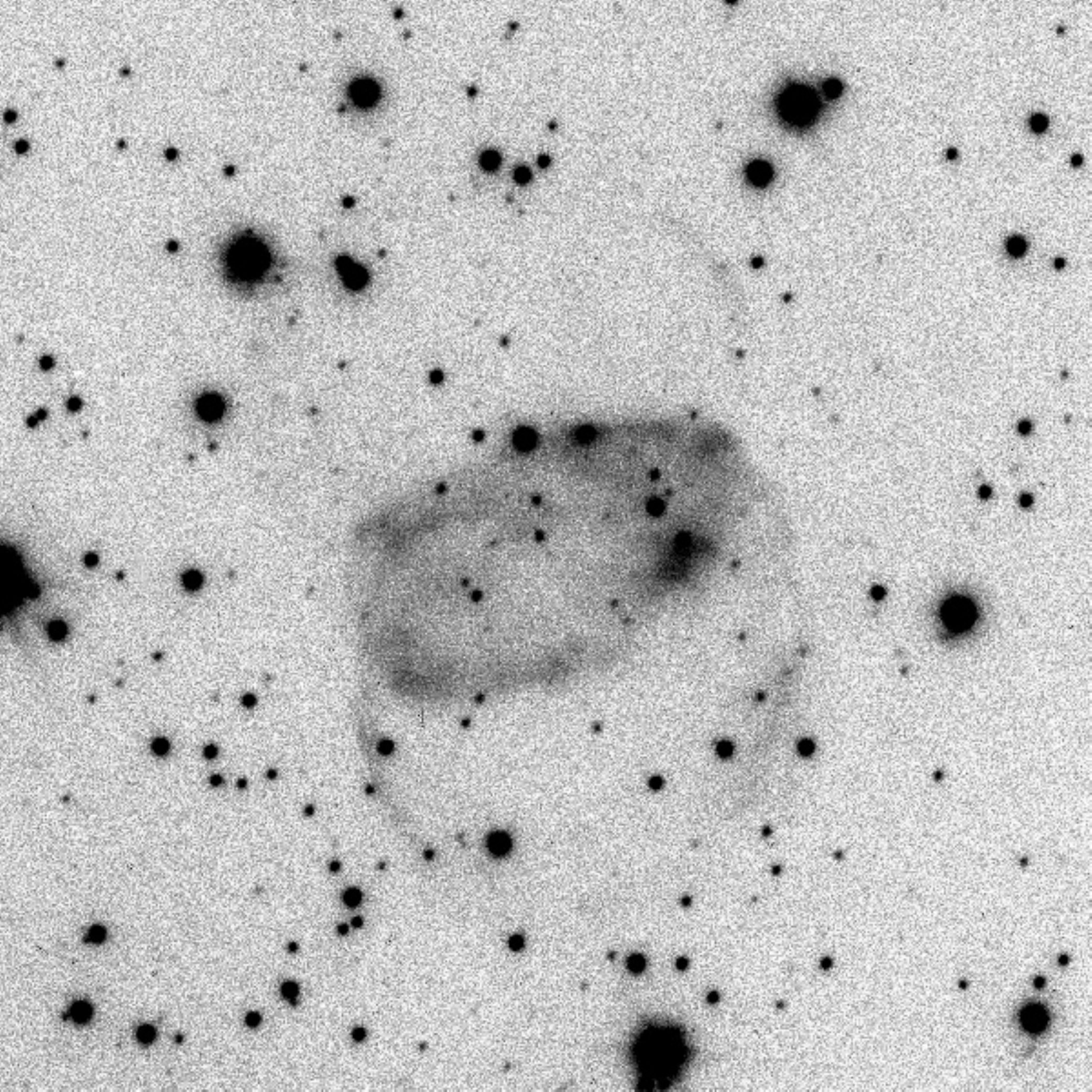}
\includegraphics[height=1.7in]{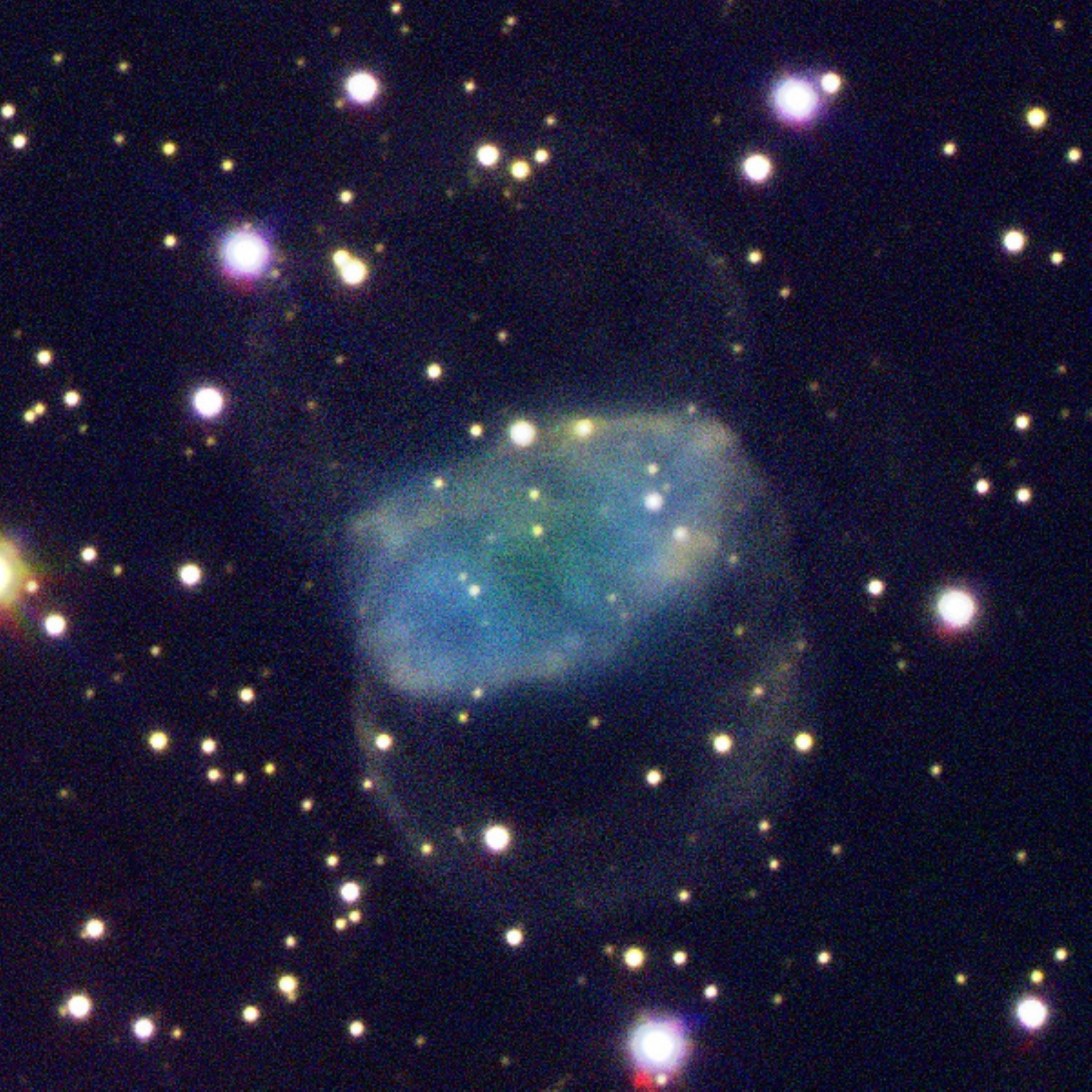}
\vskip .1in
\includegraphics[height=1.7in]{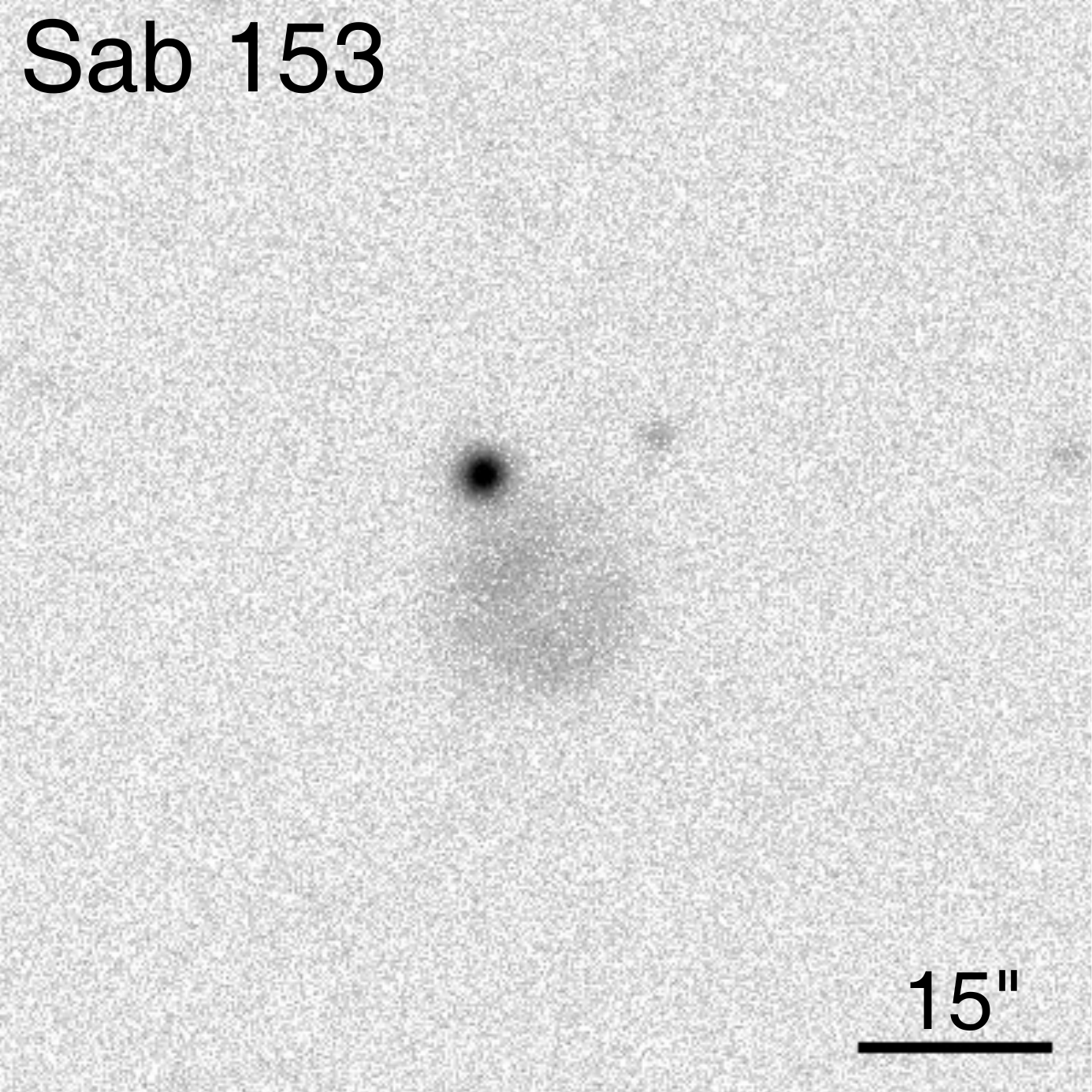} 
\includegraphics[height=1.7in]{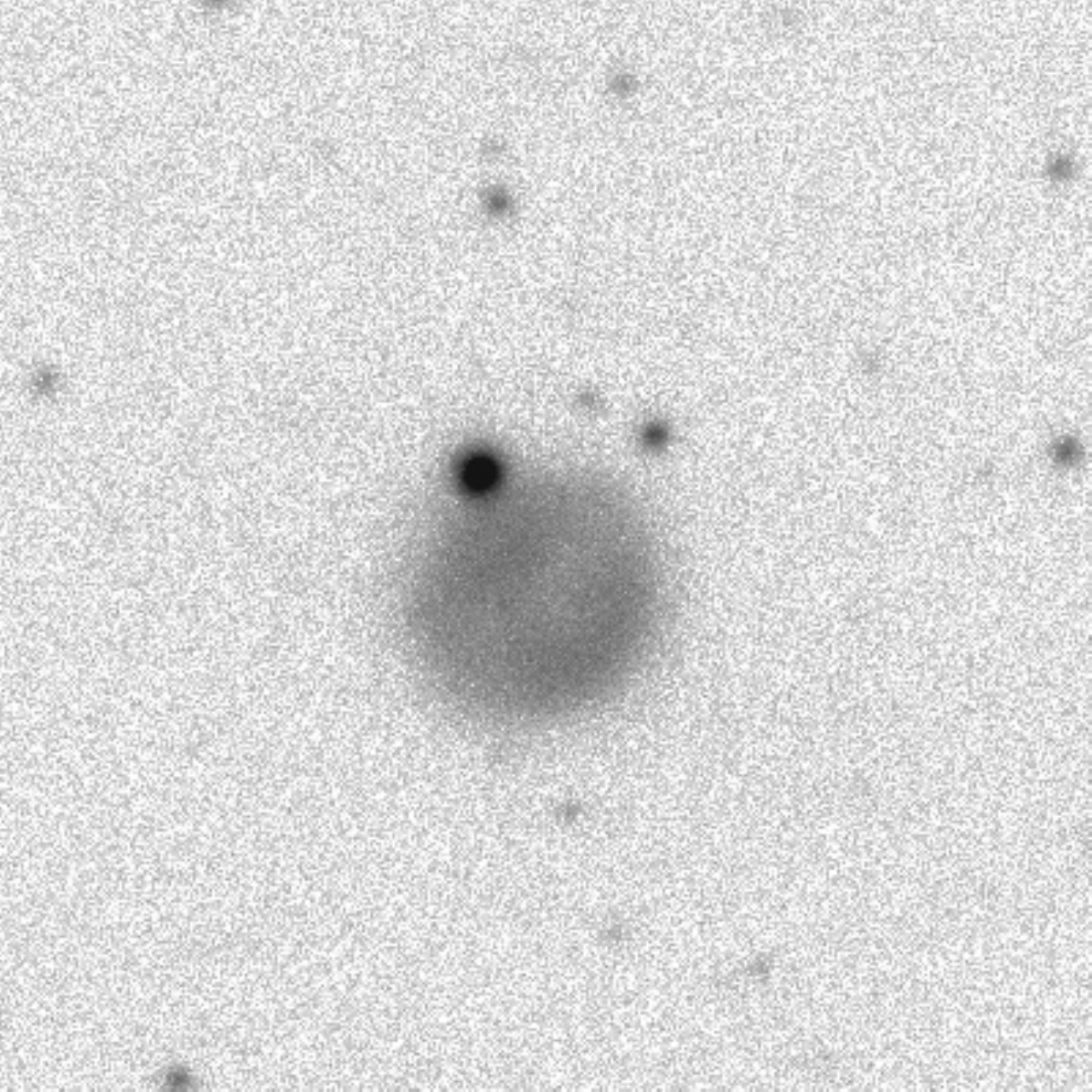}
\includegraphics[height=1.7in]{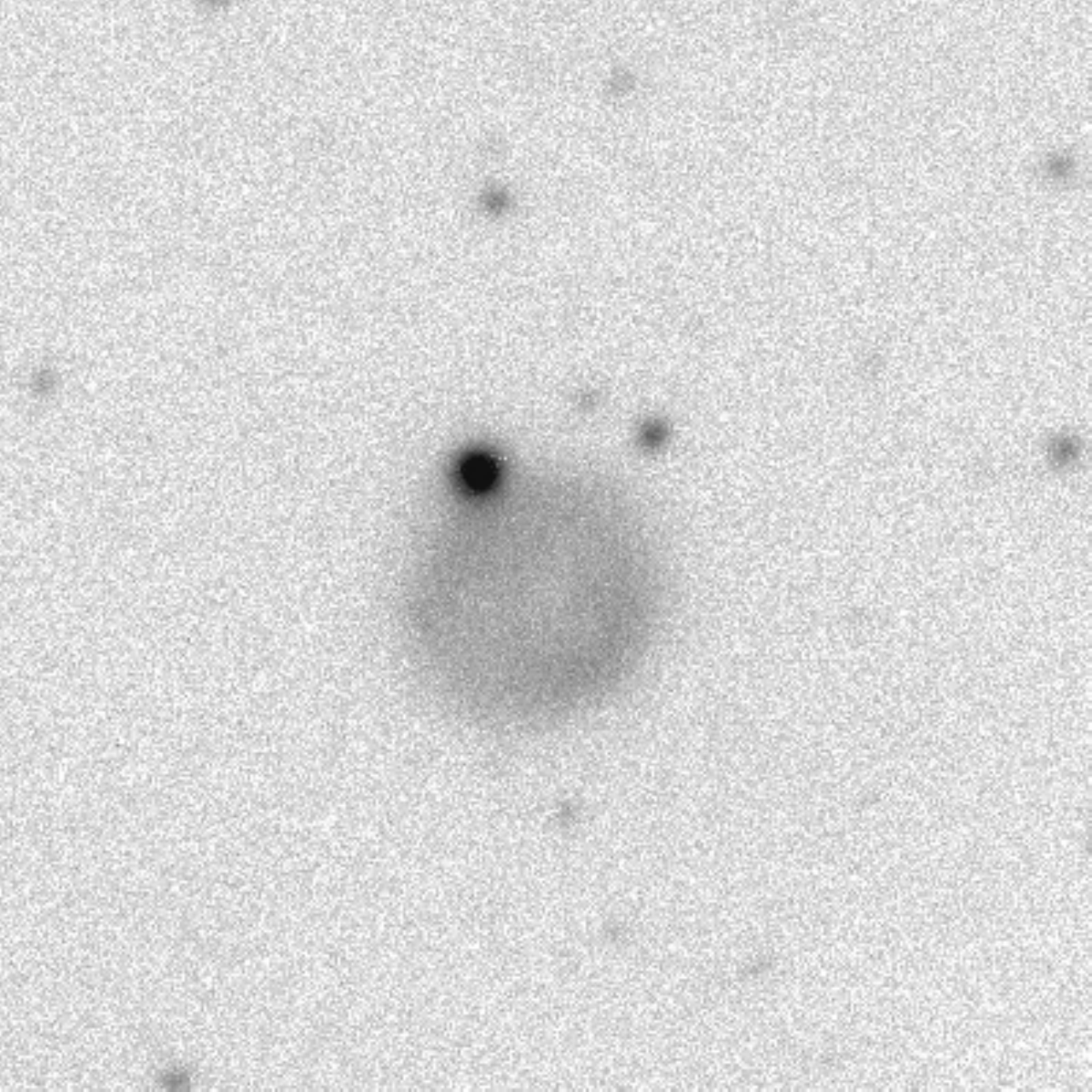}
\includegraphics[height=1.7in]{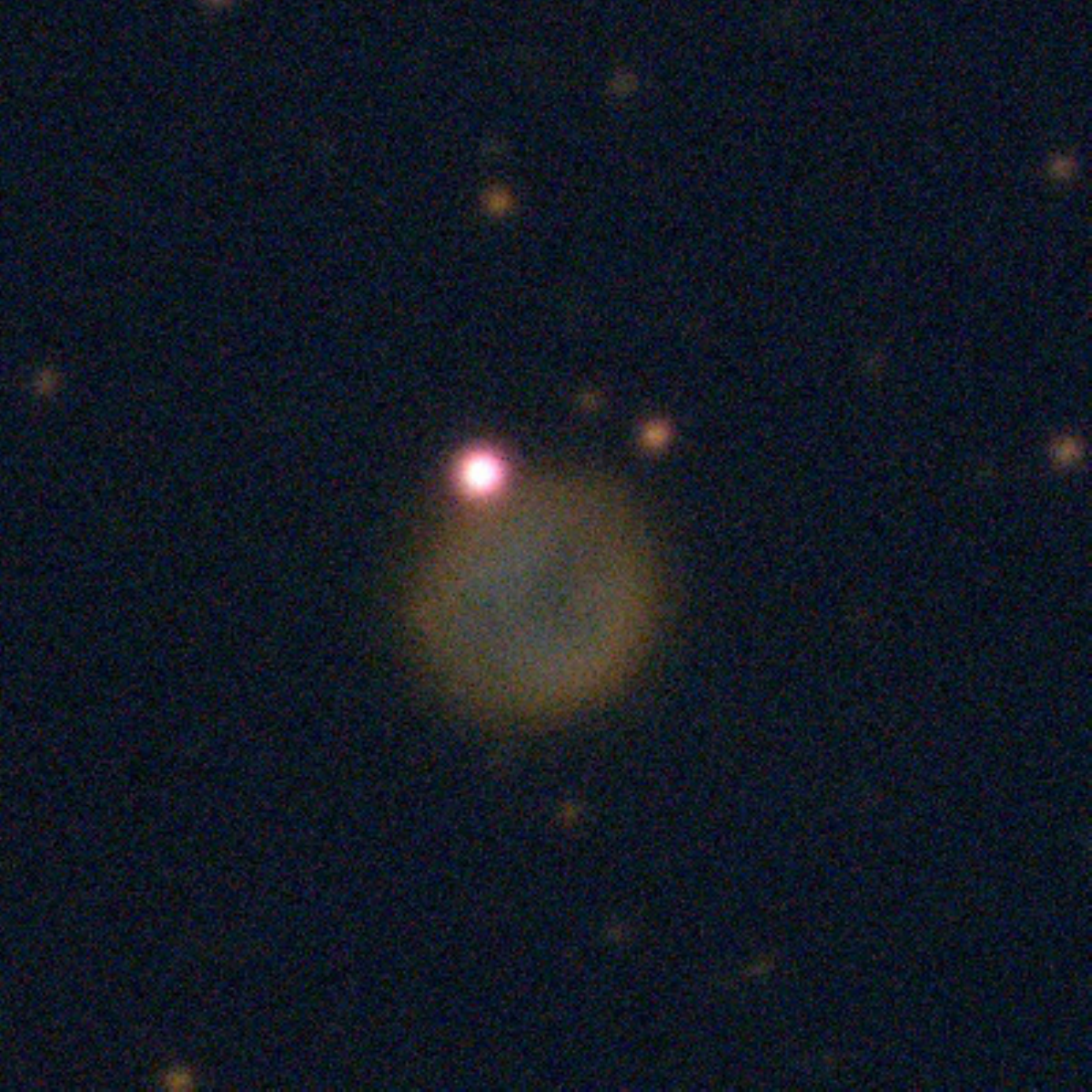}
\caption{Same as Figure~\ref{1.img}. } 
\label{11.img} 
\end{figure*}

\begin{figure*} 
\centering
\includegraphics[height=1.7in]{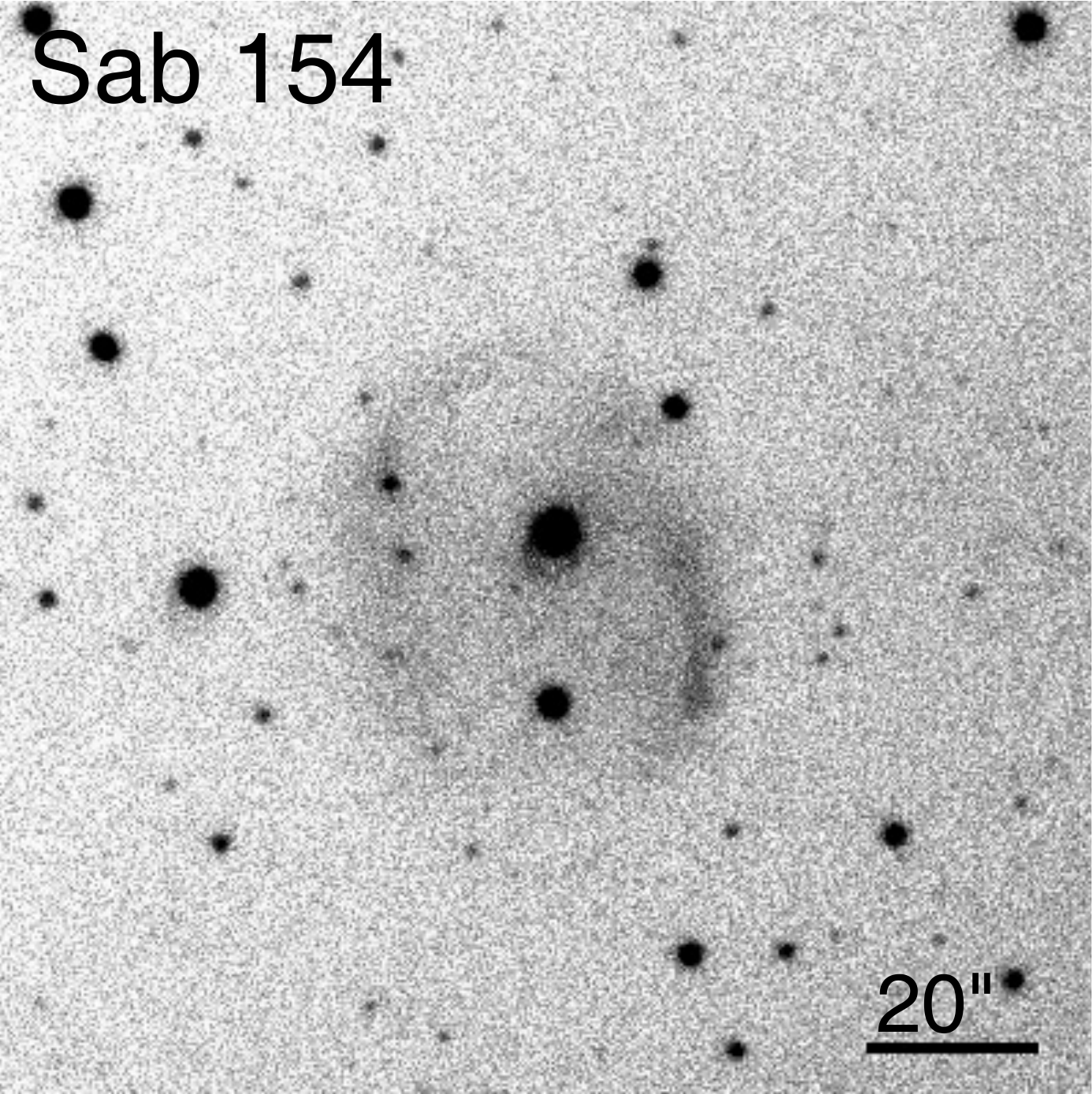} 
\includegraphics[height=1.7in]{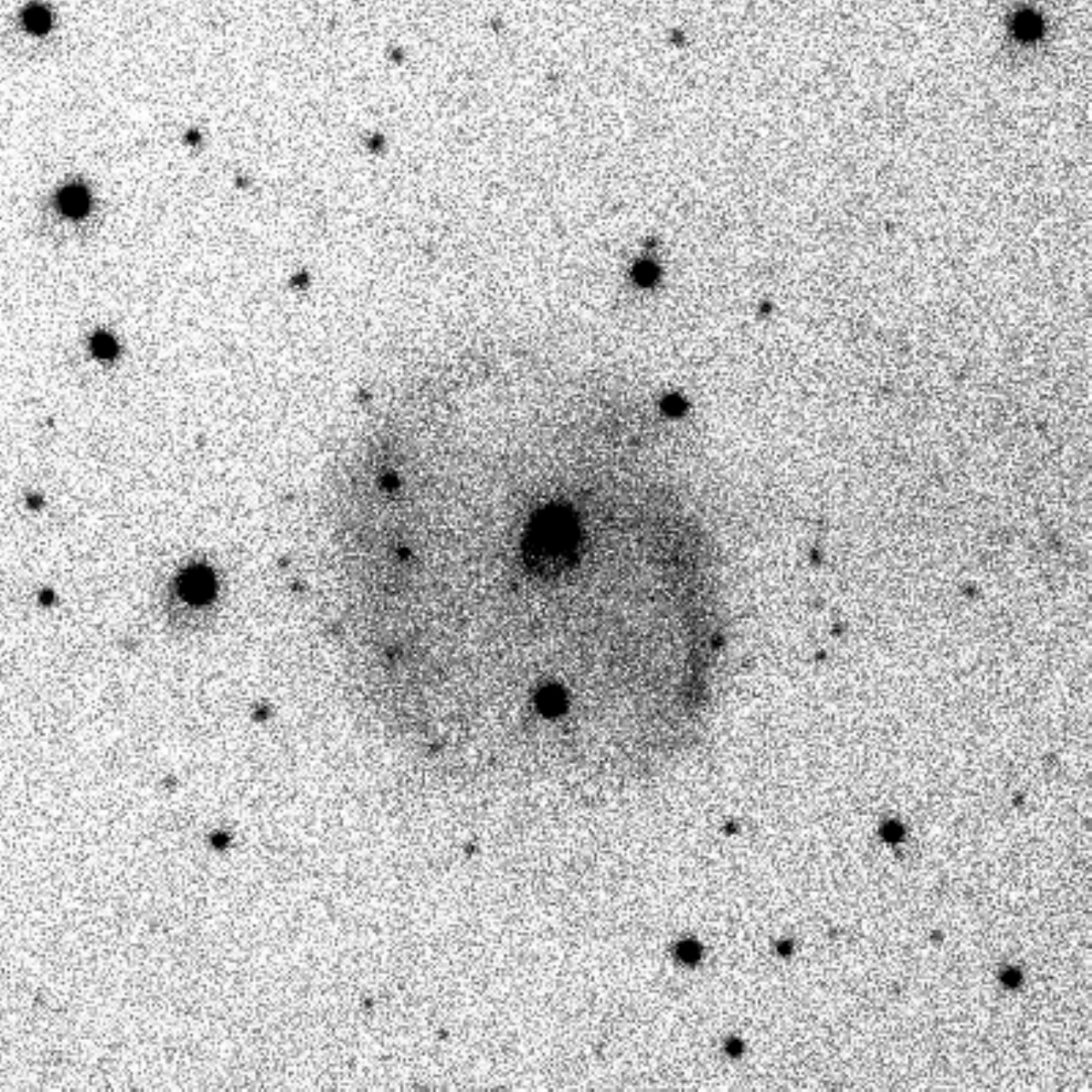}
\includegraphics[height=1.7in]{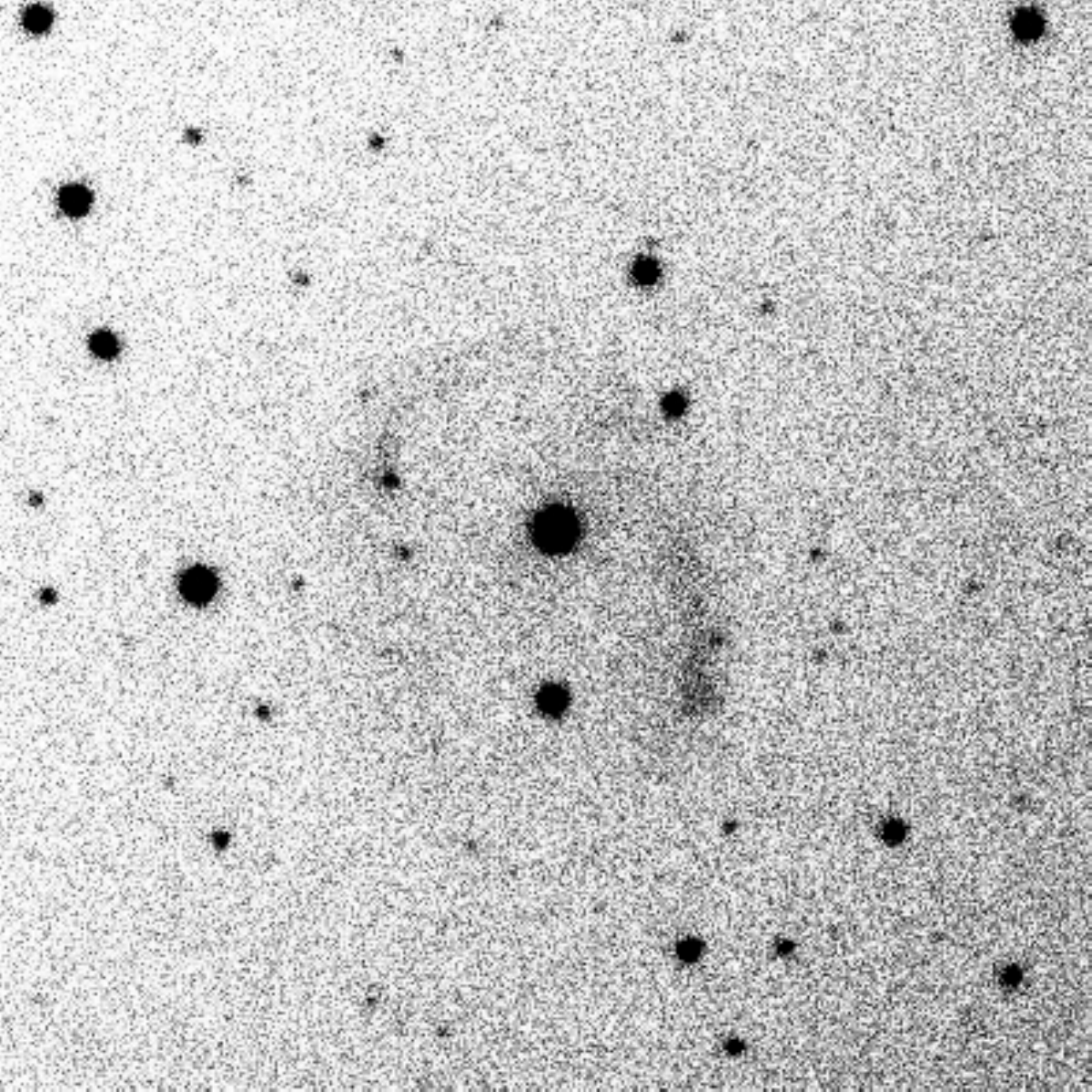}
\includegraphics[height=1.7in]{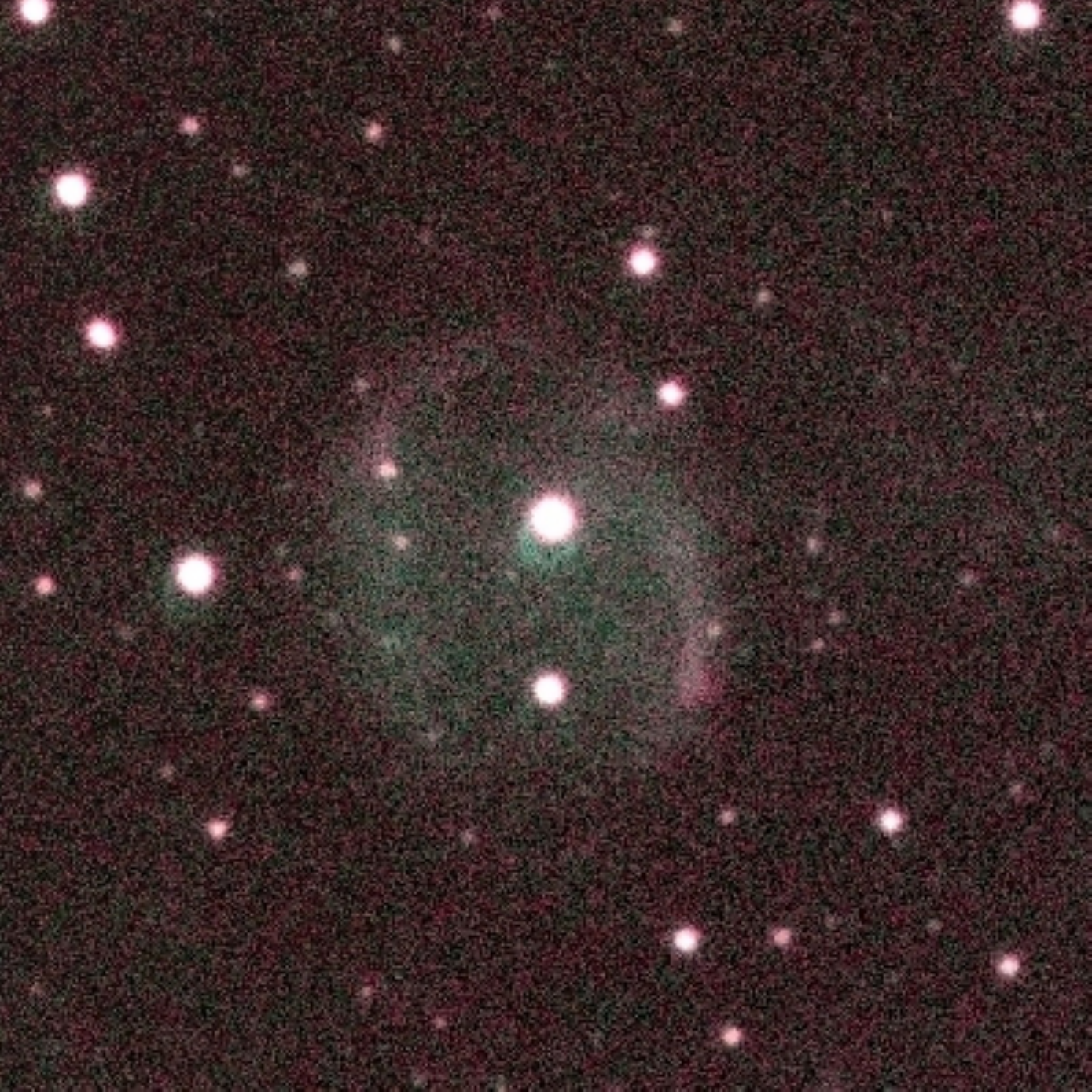}
\vskip .1in 
\includegraphics[height=1.7in]{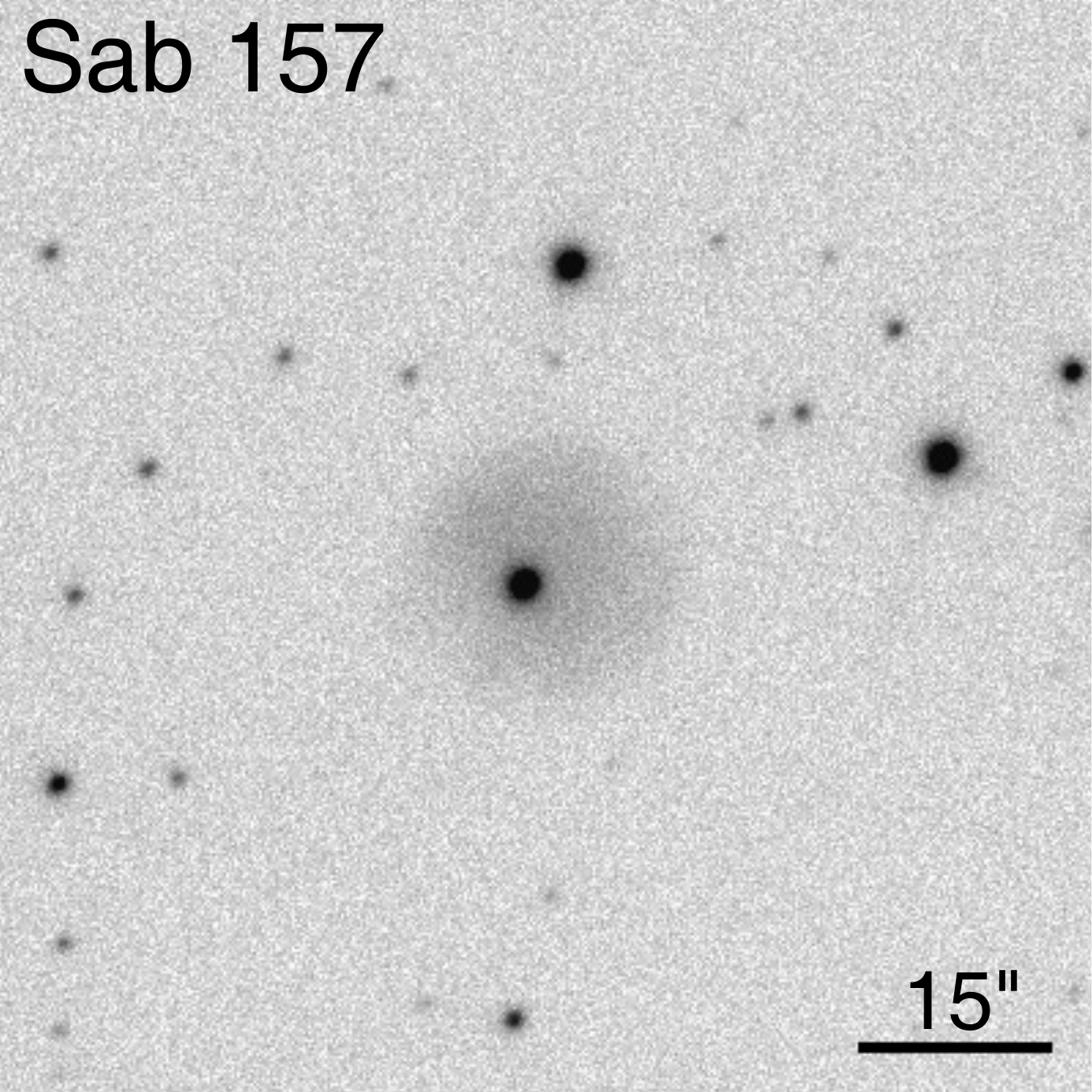} 
\includegraphics[height=1.7in]{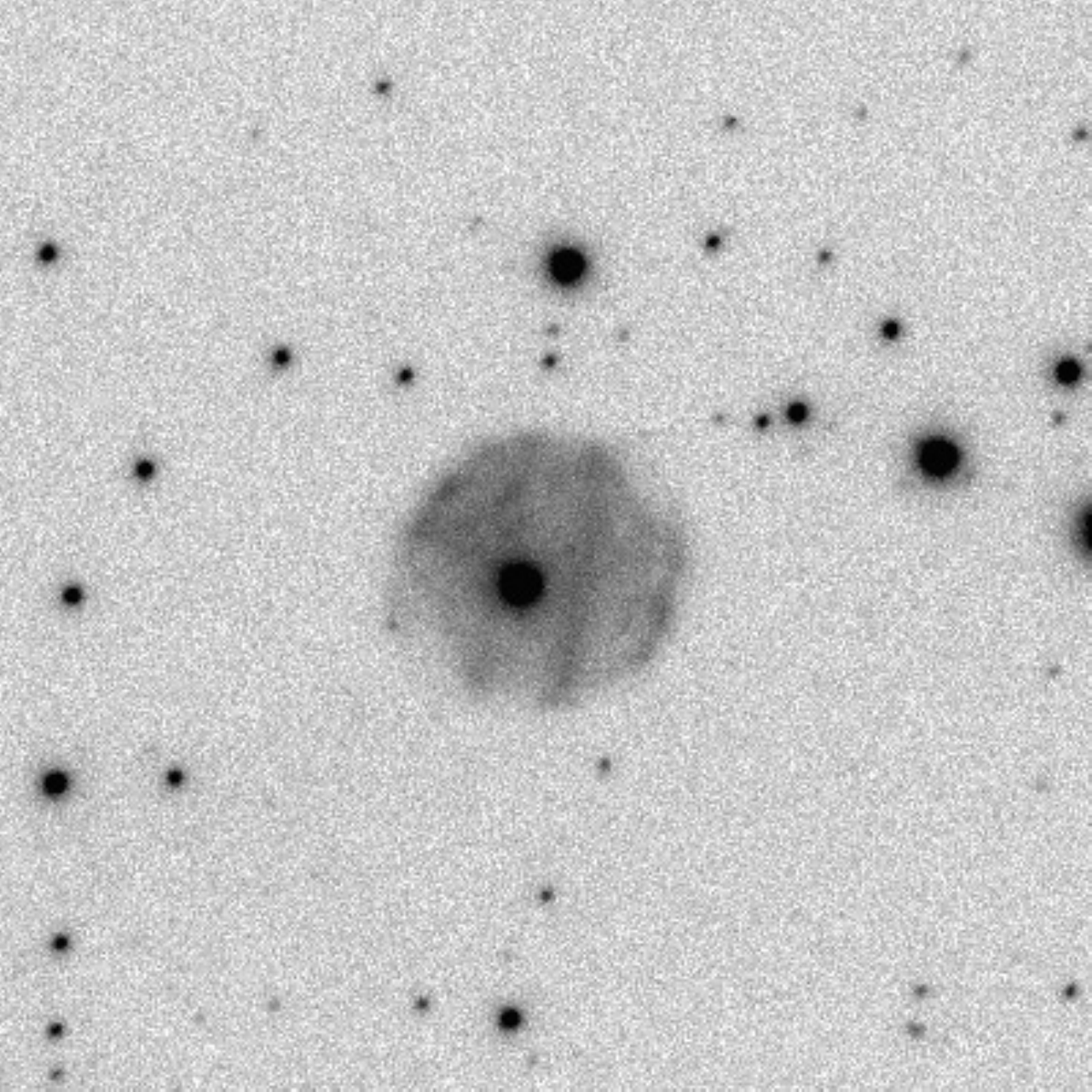}
\includegraphics[height=1.7in]{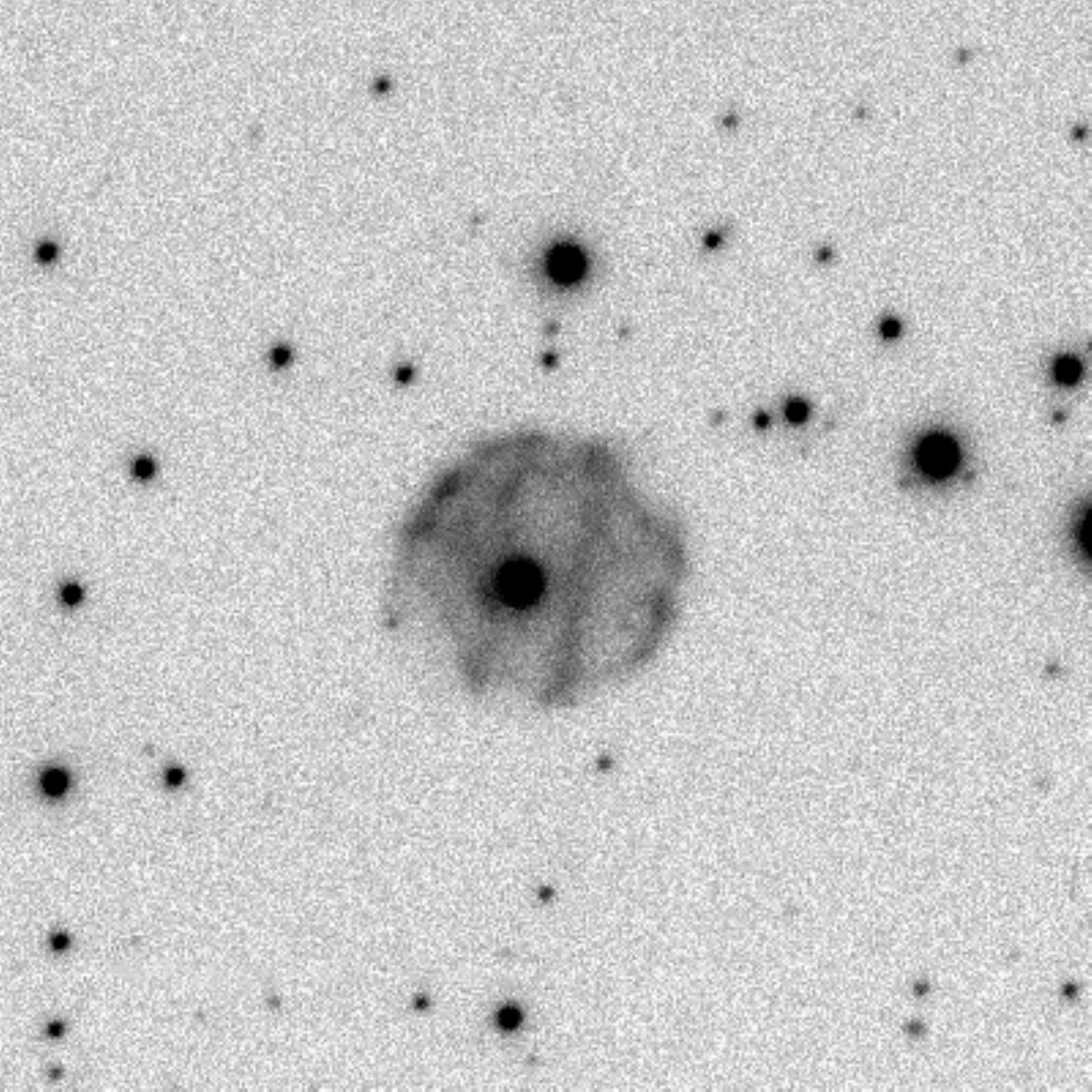}
\includegraphics[height=1.7in]{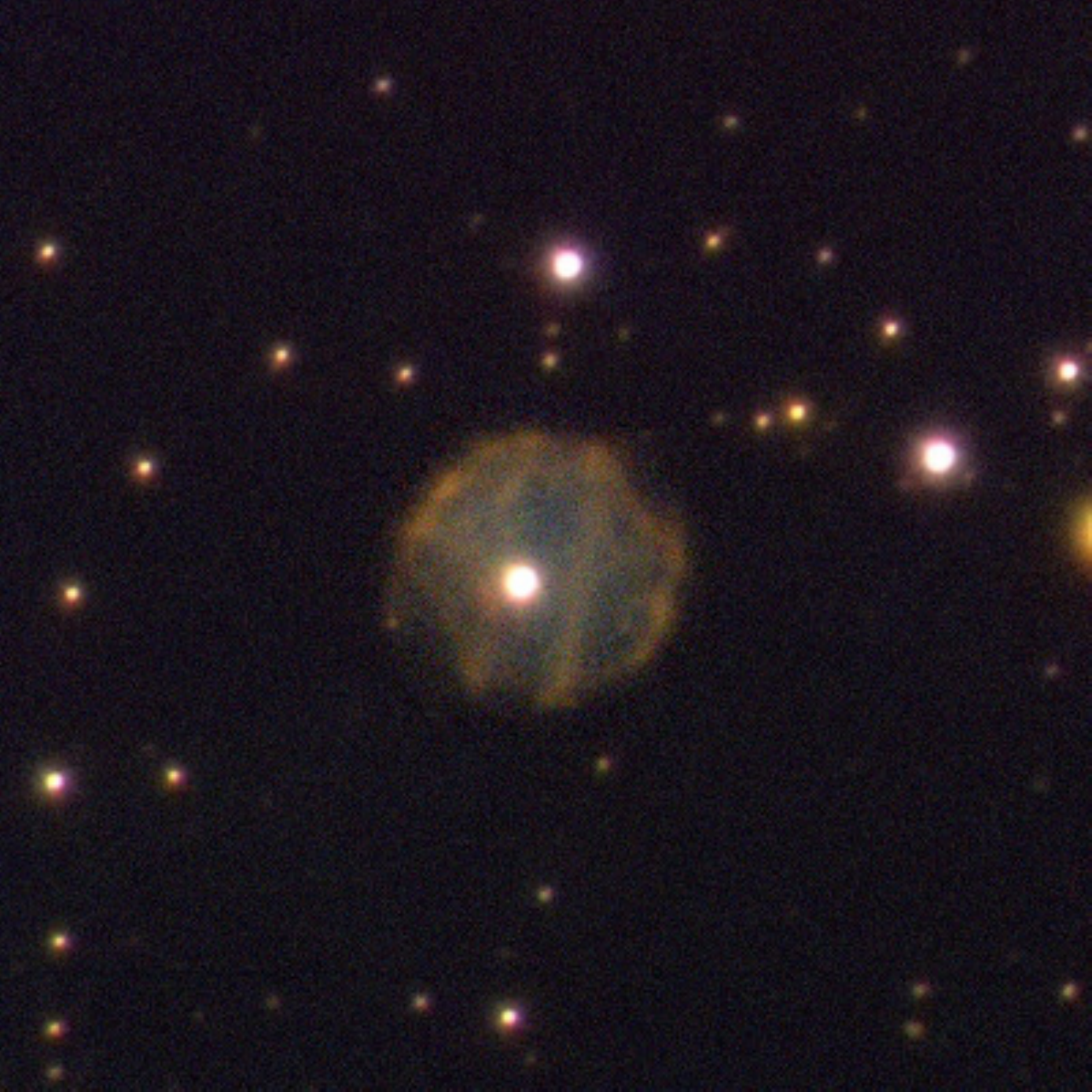}
\vskip .1in 
\includegraphics[height=1.7in]{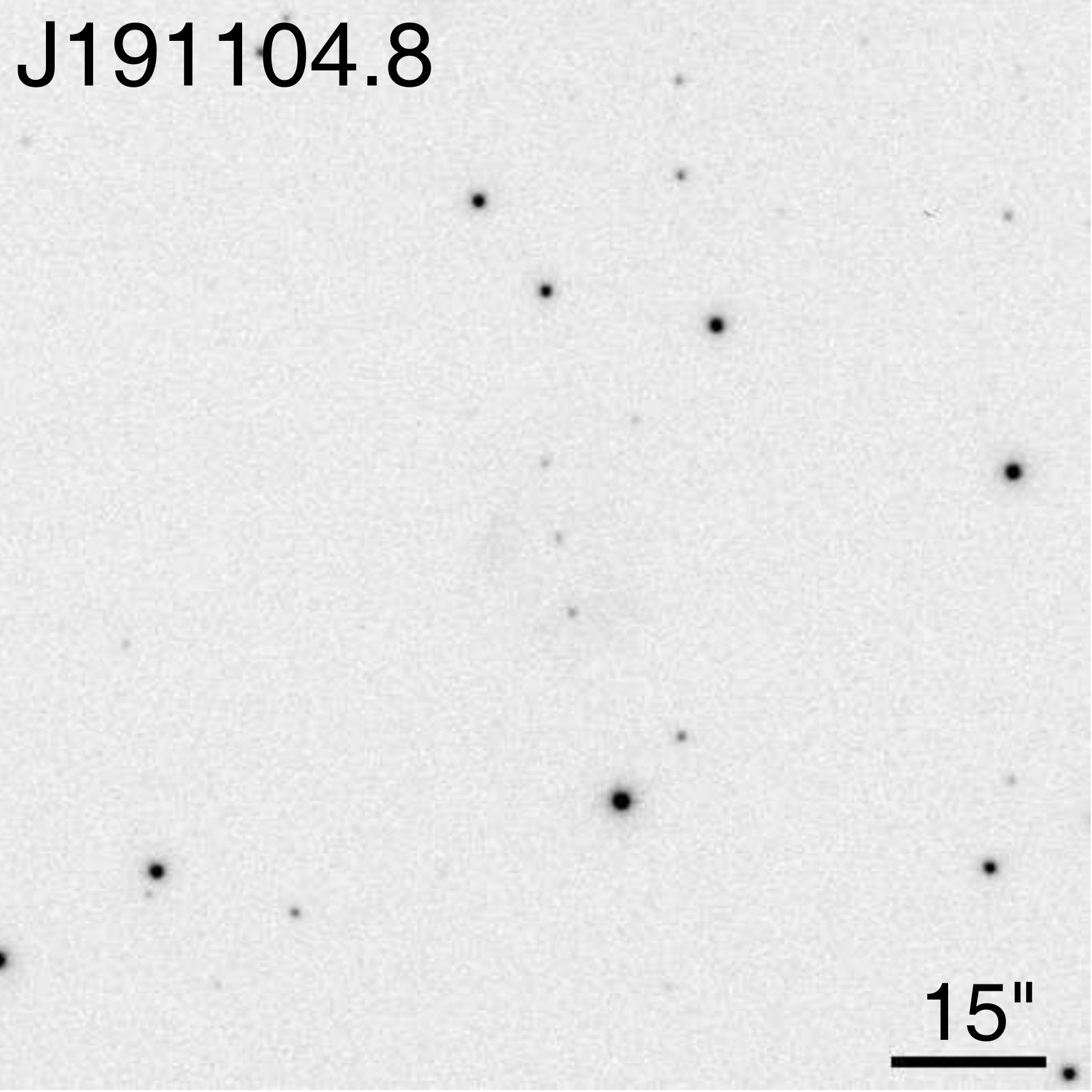} 
\includegraphics[height=1.7in]{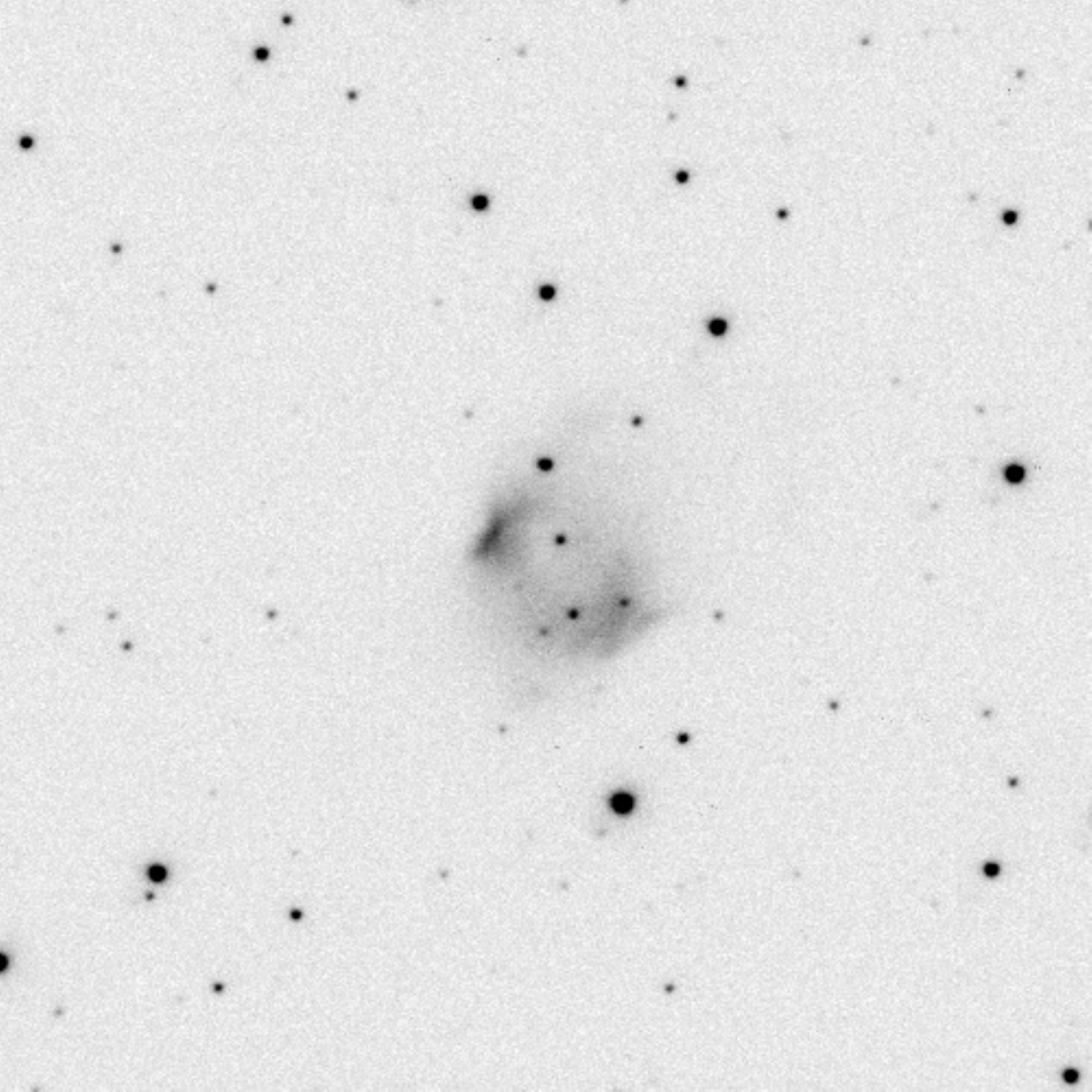}
\includegraphics[height=1.7in]{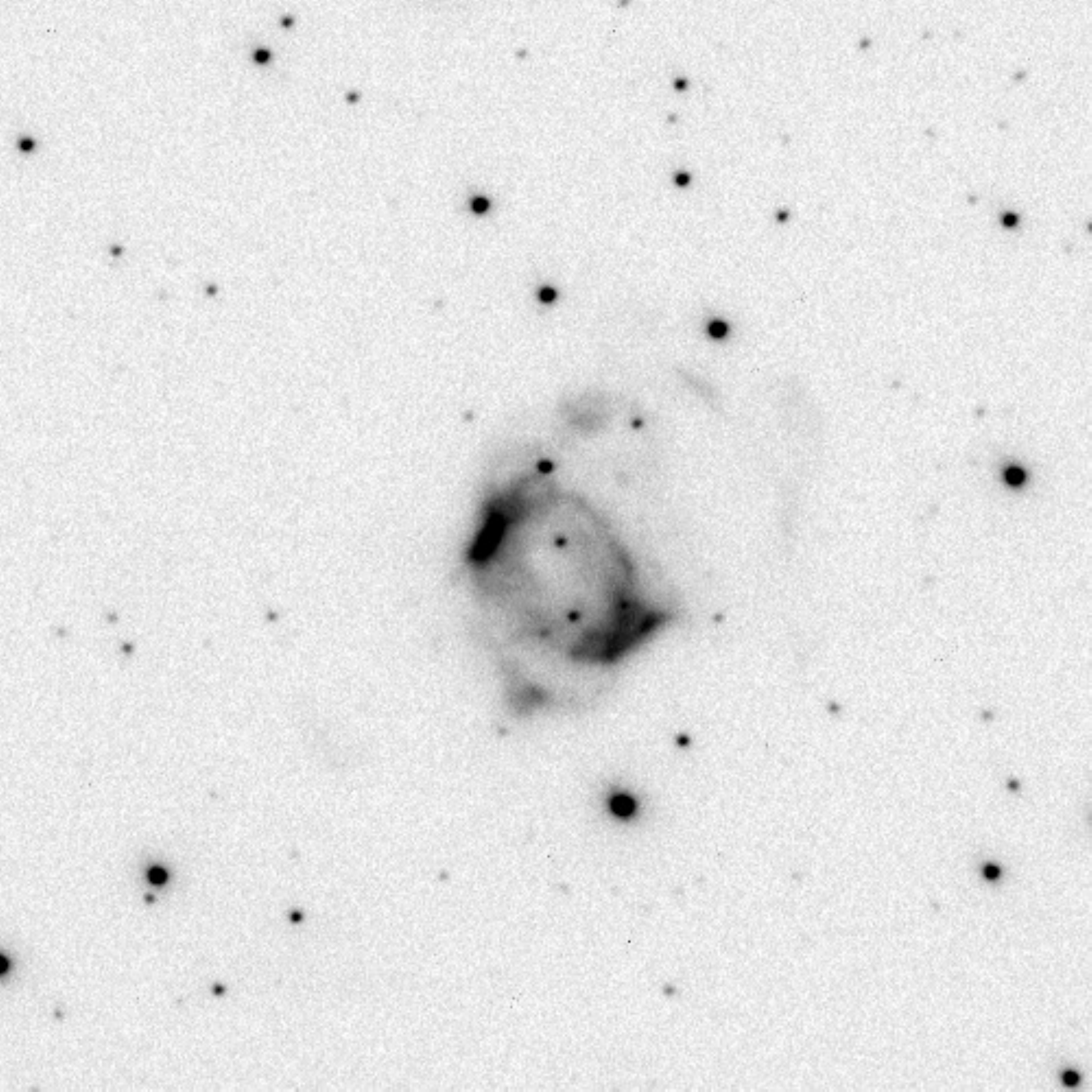}
\includegraphics[height=1.7in]{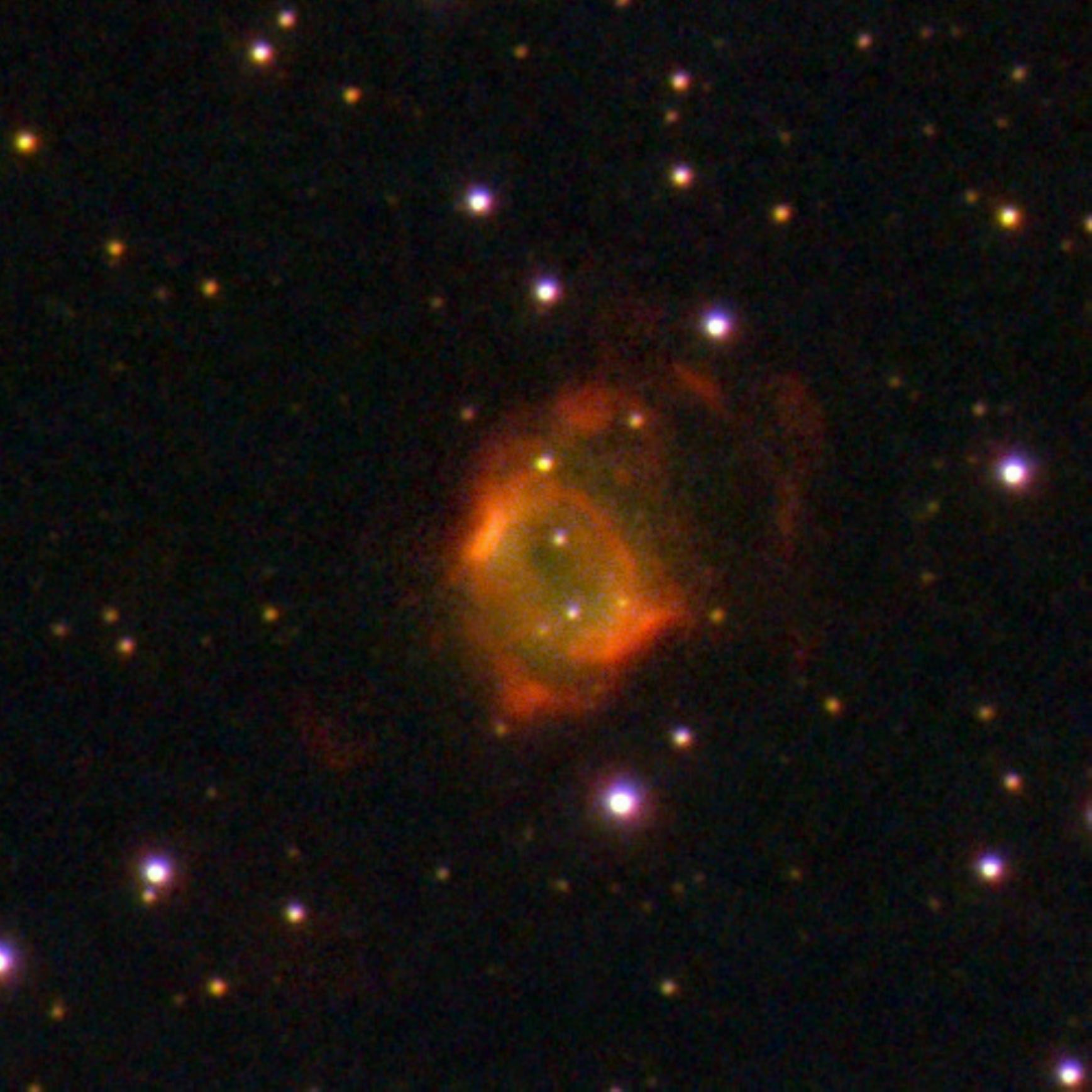}
\caption{Same as Figure~\ref{1.img}. } 
\label{12.img} 
\end{figure*}

In this appendix we present all the H$\alpha$, [N\,{\sc ii}], [O\,{\sc iii}] narrow-band images of all objects studied in this paper. In addition a colour-composite image combining these three filters is also presented for each object. The images are presented in Figures~\ref{1.img}--\ref{12.img}. We note that the images of J191104.8 (Fig.~\ref{12.img}), Sab~19 (Fig.~\ref{2.img}) and Sab~99 (Fig.~\ref{6.img}) were presented in \citet{Rodriguez2021}, \citet{Guerrero2021} and \citet{Sabin2021}, respectively.


\end{document}